%% file: hadronColliderTestsofSeesaws.tex
\documentclass[11pt,final,phd]{pittetd}
\usepackage{amsmath,amsfonts,amssymb}
\usepackage{array}
\usepackage{bm,bbm,bbm,bbold}
\usepackage{cite}
\usepackage{dsfont}
\usepackage{enumerate}
\usepackage{float}
\usepackage{graphicx}
\usepackage{lineno,lipsum}
\usepackage{multirow,multicol}
\usepackage{subfigure}
\usepackage[usenames]{color}
\usepackage{hyperref}
\def\etmiss{\slashchar{E}_{T}}
\def\lamDIS{\Lambda_{ \gamma }^{ \rm DIS  } }
\def\lamIn{\Lambda_{ \gamma}^{ \rm Inel   } }
\def\lamEl{\Lambda_{ \gamma}^{ \rm El } }

\def\GeV{{\rm ~GeV}}
\def\TeV{{\rm ~TeV}}

\def\invfb{{\rm ~fb^{-1}}}
\def\invab{{\rm ~ab^{-1}}}
\def\fb{{\rm ~fb}}

\def\lsim{\mathrel{\raise.3ex\hbox{$<$\kern-.75em\lower1ex\hbox{$\sim$}}}}
\def\gsim{\mathrel{\raise.3ex\hbox{$>$\kern-.75em\lower1ex\hbox{$\sim$}}}}

\def\wpri{W^\prime}

\def\lamDIS{ \Lambda_{ \gamma }^{ \rm DIS  } }
\def\lamIn{\Lambda_{ \gamma}^{ \rm Inel   } }
\def\lamEl{\Lambda_{ \gamma}^{ \rm El } }

\def\wpri{W^\prime}

\def\etmiss{\slashchar{E}_{T}}
\newcolumntype{x}[1]{%
{\centering\hspace{0pt}}p{#1}}%
\def\etmiss{E\!\!\!\!\slash_{T}}

\def\BR{{\rm BR}}
\newcommand{\coD}{\!\!\not\!\!D}

\def\mwpri{M_{W'}}

\def\beq{\begin{equation}}
\def\eeq{\end{equation}}
\def\bea{\begin{eqnarray}}
\def\eea{\end{eqnarray}}
\def\bmat{\begin{pmatrix}}
\def\emat{\end{pmatrix}}
\def\to{\rightarrow}

\def\GeV{{\rm ~GeV}}
\def\TeV{{\rm ~TeV}}

\def\invfb{{\rm ~fb^{-1}}}
\def\invab{{\rm ~ab^{-1}}}
\def\pb{{\rm ~pb}}
\def\fb{{\rm ~fb}}

\def\lsim{\mathrel{\raise.3ex\hbox{$<$\kern-.75em\lower1ex\hbox{$\sim$}}}}
\def\gsim{\mathrel{\raise.3ex\hbox{$>$\kern-.75em\lower1ex\hbox{$\sim$}}}}

\def\mwpri{M_{W'}}

\def\beq{\begin{equation}}
\def\eeq{\end{equation}}
\def\bea{\begin{eqnarray}}
\def\eea{\end{eqnarray}}
\def\bmat{\begin{pmatrix}}
\def\emat{\end{pmatrix}}
\def\to{\rightarrow}

\def\lsim{\mathrel{\raise.3ex\hbox{$<$\kern-.75em\lower1ex\hbox{$\sim$}}}}
\def\gsim{\mathrel{\raise.3ex\hbox{$>$\kern-.75em\lower1ex\hbox{$\sim$}}}}
\providecommand{\tabularnewline}{\\}

\newcommand{\xSlash}[1]{{\not\!\! #1}}

\newcommand{\I}{\rm 1\kern-.24em l} 
\newcommand{\Tr}{\mathop{\rm Tr}}
\def\etmiss{E\!\!\!\!\slash_{T}}

\newcommand{ \slashchar }[1]{\setbox0=\hbox{$#1$}   
   \dimen0=\wd0                                     
   \setbox1=\hbox{/} \dimen1=\wd1                   
   \ifdim\dimen0>\dimen1                            
      \rlap{\hbox to \dimen0{\hfil/\hfil}}          
      #1                                            
   \else                                            
      \rlap{\hbox to \dimen1{\hfil$#1$\hfil}}       
      /                                             
   \fi}                                             %

\subject{High Energy Physics}
\begin{document}

\title[Hadron Collider Tests of Neutrino Mass-Generating Mechanisms]{Hadron Collider Tests of Neutrino\\ Mass-Generating Mechanisms}
\author{Richard Efrain Ruiz}
\year{2015}

\degree{
B.S. in Mathematics, University of Chicago, 2010\\
B.A.~(Honors) in Physics, University of Chicago, 2010\\
M.S.~in Physics (Particle Theory and Collider Phenomenology),\\ University of Wisconsin - Madison, 2012}
\school{Kenneth P.~ Dietrich School of Arts and Sciences\\}
\maketitle

\date{April 24, 2015}
\committeemember{Tao Han, Ph.D., Distinguished Professor, University of Pittsburgh}
\committeemember{Ayres Freitas, Ph.D., Associate Professor, University of Pittsburgh}
\committeemember{Richard Holman, Ph.D., Professor, Carnegie Mellon University}	
\committeemember{Arthur Kosowsky, Ph.D., Professor, University of Pittsburgh}
\committeemember{Vladimir Savinov, Ph.D., Professor, University of Pittsburgh}

\makecommittee

\copyright

\begin{abstract}[Keywords:]
The Standard Model of particle physics (SM) is presently the best description of nature at small distances and high energies.
However, with tiny but nonzero neutrino masses, a Higgs boson mass unstable under radiative corrections, 
and little guidance on understanding the hierarchy of fermion masses, the SM remains an unsatisfactory description of nature. 
Well-motivated scenarios that resolve these issues exist but also predict extended gauge  (e.g., Left-Right Symmetric Models),
scalar (e.g., Supersymmetry), and/or  fermion sectors (e.g., Seesaw Models). 
Hence, discovering such new states would have far-reaching implications. 

After reviewing basic tenets of the SM and collider physics, several beyond the SM (BSM) scenarios that alleviate these shortcomings are investigated.
Emphasis is placed on the production of a heavy Majorana neutrinos at hadron colliders in the context of low-energy, 
effective theories that simultaneously explain the origin of neutrino masses and 
their smallness compared to other elementary fermions, the so-called Seesaw Mechanisms.
As probes of new physics, rare top quark decays to Higgs bosons in the context of the SM, the Types I and II Two Higgs Doublet Model (2HDM),
and the semi-model independent framework of Effective Field Theory (EFT) have also been investigated.
Observation prospects and discovery potentials of these models at current and future collider experiments are quantified.
\end{abstract}

\tableofcontents
\listoftables
\listoffigures

\phantomsection
\preface
In the process of writing this text, I came to the realization that
the most important conclusions contained in this document will have already been published in refereed journals~\cite{Han:2012vk,Han:2013sea,Alva:2014gxa}
by the time it has been defended.
Experts and students have already the resources available to obtain copies of referenced works, 
the capabilities to work out (unpublished) intermediate results, 
and it becomes redundant to simply recompile everything into a new document.
Therefore, very selfishly, I have decided to construct this dissertation for a single audience: myself;
and for a very specific reason: to have a mostly complete and standalone set of notes from which, 
given the appropriate number of scratch paper and computational resources, Refs.~\cite{Han:2012vk,Han:2013sea,Alva:2014gxa} can be readily reproduced.
While the contents of this book may not be sufficient to teach a graduate course on collider physics, I expect it to satisfy many of the requirements.

The text is laid out in the following fashion: 
In the first chapter, the Standard Model of Particle Physics (SM) is introduced.
Fundamentals of collider physics is introduced after this.
Three nontrivial extensions to the SM are then studied in great detail:
(i) an additional scalar doublet gauged under the SM symmetries (Two Higgs Doublet Model);
(ii) a right-handed neutrino with a Majorana mass (Seesaw Model Type I); and 
(iii) extended gauge sector with heavy Majorana neutrino (Seesaw Model Type I+II).

At this point, it is very much tradition to acknowledge and thank the many people who and institutions that have contributed to the completion of this thesis.
This is a tradition I am very pleased to continue. 
First and foremost, in my completely biased and subjective opinion, 
I would like to thank Tao Han for his guidance, undervalued criticisms that I hope to one day appreciate, mentorship, support, and words of wisdom.
To Efrain and Gilda, I am unsure how to appropriately acknowledge all that you have done. 
To write proportionally would fill libraries. Instead, I think acknowledging you two {\it inversely} to your importance is more appropriate,
so to summarize: thanks.

There are many from around the world I wish to thank. To minimize omissions, I resort to the alphabet:
Daniel Alva,
Marguerite Braun,
Helen Brown,
Terry and Joey Carlos,
Cindy Cercone,
Neil Christensen,
Andrew Cox,
Samuel Ducatman,
Anna Elder and Kevin Sapp,
Ayres Freitas,
Athena Frost,
Ayriole Frost,
Caitlyn Hunter,
Hiroyuki Ishida,
LJ,
Ali Kahler,
Rachel Landau-Lazerus, 
Bryce Lanham,
Ian Lewis,
Zhen Liu,
Amy Lowitz,
Phil Luciano,
Adriano Maron,
Thomas McLaughlan,
Ben Messerly,
Danny and Debbie Molina,
Jim Mueller,
Shannon Patterson,
Kevin Pedro,
Zachary Pierpoint,
Kara Ponder,
Zhuoni Qian,
my editors at Quantum Diaries,
Angela Ruiz,
Vladimir Savinov,
Josh Sayre,
Jared Schmitthenner,
Ben Smart,
Brittany Stalker,
Brock Tweedie,
Marina Tyquiengco,
Xing Wang,
Ying Wang,
C\'edric Weiland,
Adam Weingarten, 
Kristen Welke,
Daniel Wiegand,
Krispi Williams,
Ricky Williams,
and to the many, many people on whose couches I have slept during my travels.
Special appreciation and gratitude are owed to Zong-guo Si and Young-Kee Kim.
Much of the research presented in this text was made possible through the generous assistance of the Graduate Dean's Office at the University of Pittsburgh,
and the personal support of Kathleen Blee  and Philippa Carter.

As of this writing, the relevant publications and working group reports can be found in Refs.~\cite{Han:2012vk,Han:2013sea,Alva:2014gxa}
and Refs.~\cite{Dawson:2013bba,Agashe:2013hma}.
Reproduction of Refs.~\cite{Han:2012vk,Han:2013sea,Alva:2014gxa} is authorized under license numbers 
3610261016932 and 3610240521805 issued by Physical Review D, as well as 3604430001427 issued by the Journal of High Energy Physics.
Apologies are owed for typographical errors, which, of course, are my responsibility.
The realization (or lack thereof) of Seesaw Mechanisms in nature, on the other hand, is nature's responsibility. 

\hfill\\

Heavy Majorana neutrinos is a particularly favorite topic because it is an ideal, one-particle laboratory,
similar to the top quark, where rich physics can be enjoyed and studied.
Furthermore, the existence of tiny but nonzero neutrino masses indicate the existence of new particles and interactions not predicted by the Standard Model
of Particle Physics. It goes without saying that the physics presented here is quite interesting.

\hfill R.~Ruiz

\hfill Pittsburgh, Spring 2015

\include{./00_Preliminary/dedication}
\include{./01_Introduction/introduction}
\include{./02_ColliderPhysics/colliderPhysics}
\include{./03_HiggsFromTop/higgsTop}
\include{./04_NeutrinoFromWA/waHeavyN}
\include{./05_WprimeHeavyN/wprimeHeavyN}


\newpage

\end{document}

%% file: 00_Preliminary/dedication.tex
\vspace*{\fill}
\begingroup
\centering

\textit{To Armando Pacheco and Michael Hattar.}

\endgroup
\vspace*{\fill}

%% file: 01_Introduction/introduction.tex
\chapter{The Standard Model of Particle Physics}

\section{Introduction}
The Standard Model of Particle Physics (SM) is the quantum field theoretic model that, to date, 
best describes the interactions matter at small distances and large energies.
Though incredibly successful, as we will discuss later, the SM does remain an incomplete description of nature.
Before studying the particle content and forces of the SM, 
we begin by first considering what lies at the heart of the SM and physics in general: symmetries.

\section{Symmetries}
A symmetry exists when a quantity, for example: linear momentum or electric charge, 
does not change (remains invariant) while a system undergoes a transformation, 
such as a spatial translation or a local U$(1)$ phase shifts.
Here we discuss continuous, both global and local, symmetries, 
emphasizing along the way the consequences of their spontaneous breakdown.

The spontaneous breakdown of a symmetry through the acquiring of a nonzero vacuum expectation value, or vev, by a scalar field is an interesting topic
with subtle consequences. Typically in Quantum Field Theories (QFT), expectation values of fields, both bosonic and fermionic, are zero:
\begin{equation}
 \langle 0 \vert \phi(x),~ \psi(x),~ A^\mu(x) \vert 0 \rangle = 0.
\end{equation}
As fermions and vector bosons are nontrivial representations of the Lorentz group, they carry spinor and Lorentz indices.
A nonzero vev for these fields would imply that the vacuum itself carries corresponding indices, indicating a preferred state, 
thereby breaking Lorentz invariance. 
Scalars, on the other hand, being trivial representations of the Lorentz group, whether elementary or composite, are allowed to form a condensate and 
acquire a nonzero vev without violating Lorentz invariance.
However, whatever symmetries that are respected by scalars before acquiring a vev are not guaranteed to be preserved.

We now consider our first case study: continuous global symmetries.

\section{Continuous Symmetries I: Global Symmetries}
\textit{Global continuous symmetries} are transformations that remain independent of spacetime coordinates.
A familiar example intrinsic to all quantum mechanical processes is the invariance to an overall phase shift of the amplitude
that leaves the physical probability density 
unchanged\footnote{Throughout this text we employ \textit{active} transformations $U^{-1} = e^{-i\theta}$, 
which differs from some texts, e.g., \textit{Peskin~\&~Schroeder}~\cite{Peskin:1995ev}, which use \textit{passive} transformations.}:
\begin{eqnarray}
\mathcal{M}  &\rightarrow& \mathcal{M}' = e^{-i\theta} \mathcal{M},\quad \theta\in[0,2\pi)\\
\vert\mathcal{M}\vert^2 &\rightarrow& \vert\mathcal{M}'\vert^2 = \vert\mathcal{M}\vert^2
\end{eqnarray}
Though moving the amplitude $\mathcal{M}$ along the edge of a circle in the complex plane, such phase shifts are  unobservable.
Since the rotation holds for an arbitrary angle, it holds for all angles. 
The collection of all such transformation is the multiplicative group U$(1)$:
\begin{equation}
 U(1) = \{ e^{i\theta} \vert \theta \in [0,2\pi)\}
\end{equation}
For infinitesimal rotations, we have
\begin{equation}
 U(1) = 1 + i\theta
\end{equation}
The spacetime independence of $\theta$ means that $ \partial_\mu \exp[i\theta]=0$.
Hence, we say that physical probabilities derived from quantum mechanical amplitudes are symmetric (or invariant) 
under global U$(1)$ symmetries (or transformations).

We study global symmetries by considering a Lagrangian density, or Lagrangian for short,  at dimension-four 
consisting of both a Dirac fermion $\psi$ and a complex scalar $\phi$:
\begin{eqnarray}
 \mathcal{L}	&=& \overline{\psi}i\slash\!\!\!\partial\psi + (\partial^\mu\phi^*)\partial_\mu\phi 
		- m_\psi \overline{\psi}\psi - m_\phi^2 \phi^*\phi -\lambda(\phi^*\phi)^2.
\label{globalLag.sm.EQ}
\end{eqnarray}
Rotating $\psi$ and $\phi$ under the same global U$(1)$ transformations
\begin{equation}
   \psi \rightarrow \psi' = U^\dagger \psi = e^{-i\theta}\psi,  \quad\quad   \phi \rightarrow \phi' = U^\dagger\phi=e^{-i\theta}\phi, 
\end{equation}
it is self-evident that Eq.~(\ref{globalLag.sm.EQ}) remains unchanged.
Global symmetries, however, are not limited to simple Abelian transformations.
Suppose that our $\psi$ and $\phi$ fields were instead multiplets under a larger group, e.g., were in the fundamental representation of SU$(n)$ or U$(n)$:
\begin{eqnarray}
 \psi^T = (\psi_i \dots \psi_n), &\quad\text{for}\quad& SU(n) ~\text{or} ~U(n)\\
 \phi^T = (\phi_i \dots \phi_n), &\quad\text{for}\quad& SU(n) ~\text{or} ~U(n).
\end{eqnarray}
The infinitesimal transformations now behave as
\begin{equation}
\quad U^\dagger = e^{-i\theta} = 1 - i\theta \equiv 1 - i \theta^a T^a, \quad a = 1, \dots, n, 
\end{equation}
where $T^a$ denotes the generators of SU$(n)$ or U$(n)$ and $\theta^a$ are the linearly independent,
infinitesimal rotations in the space of our group's fundamental representation. 
As global (and local) transformations acting on scalars and fermions are unitary, i.e., $U^{-1} = U^\dagger$, we have
\begin{equation}
U^{-1} = e^{-i \theta^a T^a} = U^\dagger = \left(e^{i \theta^a T^a}\right)^\dagger = e^{-i \theta^{a *} T^{a \dagger}},
\end{equation}
implying that the generator and its adjoint are related by the expression
\begin{equation}
 \theta^a T^a = \theta^{a *} T^{a\dagger}.
\end{equation}
However, generators of physical transformations are Hermitian, and so $\theta^a$ must be real:
\begin{equation}
 0  = \theta^a T^a - \theta^{a *} T^{a\dagger} = T^a (\theta^a - \theta^{a *}) \implies \theta^a = \theta^{a *}.
\end{equation}
Despite the complication of non-Abelian groups, our Lagrangian remains unchanged under infinitesimal rotations
\begin{eqnarray}
 \mathcal{L} \rightarrow \mathcal{L'} 
	&=& \overline{\psi'}i\slash\!\!\!\partial\psi' + (\partial^\mu\phi'^\dagger)\partial_\mu\phi' 
	  - m_\psi \overline{\psi'}\psi' - m_\phi^2 \phi'^\dagger\phi' -\lambda(\phi'^\dagger\phi')^2
\\
	&=& \overline{\psi}U i\slash\!\!\!\partial (U^\dagger\psi) + [\partial^\mu(\phi^\dagger U)]U^\dagger\partial_\mu\phi
\nonumber\\
	&-& m_\psi \overline{\psi}U U^\dagger\psi' - m_\phi^2 \phi^\dagger U U^\dagger\phi -\lambda(\phi^\dagger U U^\dagger\phi')^2
\\
	&=& \mathcal{L}.
\end{eqnarray}

\subsection{Spontaneously Broken Global Symmetries}\label{ssbGlobal.sm.sec}
On our yellow brick road toward the emerald city of spontaneously broken gauge theories, 
we come across the related case of spontaneously broken global symmetries.
Though sharing many mechanics with broken local transformations, the phenomenological outcomes radically differ.
Consider the interacting theory of a complex scalar and a massive vector boson $\rho$
\begin{eqnarray}
 \mathcal{L}	&=&  (\partial^\mu\phi)^*\partial_\mu\phi - \frac{1}{4}\rho^{\mu\nu}\rho_{\mu\nu} 
 - \frac{1}{2}M_\rho^2 \rho_\mu \rho^\mu - m_\phi^2 \phi^*\phi 
 -\lambda(\phi^*\phi)^2 - g_\rho \phi^*\phi \rho_\mu \rho^\mu,
\end{eqnarray}
where the field strength $\rho_{\mu\nu}$ is 
\begin{equation}
 \rho_{\mu\nu}  = \partial_\mu\rho_\nu - \partial_\nu\rho_\mu.
\end{equation}
The theory is invariant under global U$(1)$ transformations of $\phi$. Inspecting the scalar potential
\begin{equation}
 V(\phi^*\phi) = m_\phi^2 (\phi^*\phi) + \lambda(\phi^*\phi)^2 + g_\rho \phi^*\phi \rho_\mu \rho^\mu,
 \label{scalarPot.sm.EQ}
\end{equation}
one sees that it is simply a quadratic function in $(\phi^*\phi)$ with coefficients $m_\phi^2$ and $\lambda$.
We ignore the contribution of $\rho_\mu$ as minima of vector fields must be  zero to preserve Lorentz invariance.
Potentials must also be bounded from below in order to bar tunneling to a state with infinite energy, so we require $\lambda>0$.
For $m_\phi^2 > 0$, the potential's minimum is zero at the origin
\begin{eqnarray}
 \frac{\partial V}{\partial \phi} \Bigg\vert_{\min} 	&=& 
 \frac{\partial V}{\partial (\phi^*\phi)} \frac{\partial (\phi^*\phi)}{\partial \phi} \Bigg\vert_{\min} 
 \\
							&=&  
\left( m_\phi^2 + 2\lambda (\phi^*\phi)\right)\phi^*  \Big\vert_{\min} = 0 \implies \phi^*(x)\Big\vert_{\min} = 0
 \\ 
 \frac{\partial V}{\partial \phi^*} \Bigg\vert_{\min} &=&  
\left( m_\phi^2  + 2\lambda (\phi^*\phi)\right)\phi	  \Big\vert_{\min} = 0 
\implies \phi(x)\Big\vert_{\min} = 0
\end{eqnarray}
However, curiously, when $m_\phi^2 < 0$, we discover a global minimum \textit{away} from the origin
\begin{eqnarray}
 \frac{\partial V}{\partial \phi} \Bigg\vert_{\rm extrema} 	
 = & \left(-\vert m_\phi^2\vert + 2\lambda (\phi^*\phi)\right)\phi^*  \Big\vert_{\rm extrema}&
\implies \phi^*(x)\Big\vert_{\rm extrema} = 0,~\sqrt{\frac{\vert m_\phi\vert^2}{2\lambda}}
 \\ 
 \frac{\partial V}{\partial \phi^*} \Bigg\vert_{\rm extrema} 
 = &\left(-\vert m_\phi^2\vert  + 2\lambda (\phi^*\phi)\right)\phi	  \Big\vert_{\rm extrema}&
\implies \phi(x)\Big\vert_{\rm extrema} = 0,~\sqrt{\frac{\vert m_\phi\vert^2}{2\lambda}}
\label{globalMin.sm.eq}
\end{eqnarray}
That is to say, the scalar $\phi$ possesses a nonzero vacuum expectation value (vev) given by
\begin{eqnarray}
\langle \phi \rangle  \equiv \frac{v}{\sqrt{2}} = \sqrt{\frac{\vert m_\phi^2\vert}{2\lambda}} > 0\quad 
\implies v = \sqrt{2} \langle \phi \rangle = \sqrt{\frac{\vert m_\phi^2\vert}{\lambda}} > 0.
\label{globalVEV.sm.eq}
\end{eqnarray}
The factor of $\sqrt{2}$ accounts for the normalization of $\phi$ as it can be expanded in terms of its real and imaginary components,
$\phi = (\Re(\phi) + i\Im(\phi))/\sqrt{2}$. 
Following this convention the kinetic term of $\phi$ results in properly normalized kinetic terms for $\Re(\phi)$ and $\Im(\phi)$.

We explore the consequences of quadratic (in $\phi^*\phi$) potential $V$ and, effectively, the tachyonic mass $m_\phi$ 
by considering small perturbations of $\phi$ around $v$.
We justify this by counting the degrees of freedom (states) in the theory before $\phi$ acquires a vev: 
two from the complex field $\phi$ and three from $\rho$ (two transverse and one longitudinal polarization).
Whether or not $\phi(x)$ is in a particular location, minimum or elsewhere, should not change the total number of physical states in the theory. 
So while their manifestations may depend on dynamics and momentum transfer scales, we expect to always have five physical states in our model.
The seemingly missing degrees can be traced back to $\phi$.
As fields with zero vevs, e.g., $\rho$, are fluctuations about classical minima, 
we should expect to have fluctuations around $\langle\phi\rangle.$
Therefore, expanding $\phi$ about its vev we have
\begin{equation}
 \phi \rightarrow \phi \approx \frac{v+h + ia}{\sqrt{2}}, \quad \langle h \rangle = \langle a \rangle = 0,
\end{equation}
where $h$ and $a$ are real scalar fields with zero vevs.
In passing, we note that the purely imaginary nature of $ia$ implies that its interactions are odd under charge conjugation unlike $h$, which is $C$-even.
Making the replacement, we see explicitly 
\begin{eqnarray}
 \mathcal{L}\rightarrow \mathcal{L'} &=& \frac{1}{2}\partial^\mu(v+h+ ia)^*\partial_\mu(v+h+ ia)	
		  + \frac{\vert m_\phi\vert^2}{2} (v+h+ ia)^*(v+h+ ia) - \frac{1}{2}M_\rho^2 \rho_\mu \rho^\mu
\nonumber\\
 		&-& \frac{\lambda}{4}((v+h+ ia)^*(v+h+ ia))^2 - g_\rho \rho_\mu \rho^\mu (v+h+ ia)^*(v+h+ ia) 
\\
		&=& \frac{1}{2}\partial^\mu h \partial_\mu h +\frac{1}{2}\partial^\mu a \partial_\mu a
		+ \frac{\vert m_\phi^2\vert }{2}(v^2 +2vh + h^2 + a^2)	 - \frac{1}{2}M_\rho^2 \rho_\mu \rho^\mu
 \nonumber\\
		&-&\frac{\lambda}{4}(v^4 + 4v^3 h + 6v^2 h^2  + 2v^2a^2 + 4vh^3 + h^4 + 2 h^2 a^2  + 4vha^2 + a^4)
\nonumber\\
		&-& \frac{g_\rho}{2} \rho_\mu \rho^\mu(v^2 +2vh + h^2 + a^2)
\end{eqnarray}
Regrouping terms in powers of $h$ and $a$, we get
\begin{eqnarray}
\mathcal{L}	&=& \frac{1}{2}\partial^\mu h \partial_\mu h +\frac{1}{2}\partial^\mu a \partial_\mu a	- \frac{1}{4}\rho^{\mu\nu}\rho_{\mu\nu}
\\
		&-&\frac{1}{2}\underset{(h~{\rm  mass})^2}{\underbrace{\left(3\lambda v^2  -\vert m_\phi^2\vert\right)}}h^2
		 - \frac{1}{2}\underset{{\rm zero}~a~{\rm mass}}{\underbrace{\left(\lambda v^2  -\vert m_\phi^2\vert\right)}}a^2
		 - \frac{1}{2}\underset{(\rho~{\rm  mass~shift})^2}{\underbrace{\left(M_\rho^2 + g_\rho v^2 \right)}}\rho_\mu\rho^\mu		  
\\
		&-& \frac{\lambda}{2} h^2 a^2  \underset{U(1){\rm -violating}}{\underbrace{-\lambda v h^3-\lambda v ha^3}}  
		-\frac{\lambda}{4}h^4 -\frac{\lambda}{4}a^4		
		 \underset{U(1){\rm -violating}}{\underbrace{-g_\rho v \rho_\mu\rho^\mu h}}  
		- \frac{g_\rho}{2} \rho_\mu \rho^\mu  h^2 - \frac{g_\rho}{2} \rho_\mu \rho^\mu  a^2		
\\		 
		&-& v\underset{0^2}{\underbrace{\left(\lambda v^2 - \vert m_\phi^2\vert\right)}}h
		 +\underset{(\frac{1}{2}\vert m_\phi\vert v)^2}{\underbrace{\frac{1}{2}\left(\vert m_\phi^2\vert -\frac{\lambda}{2}v^2\right)v^2}}
\end{eqnarray}
As one may expect, expanding $\phi$ into real and imaginary components has the effect of making $\phi$'s two degrees of freedom
manifest in the form of two real states, $h$ and $a$.
It also gives rise to the four-point couplings $h^4,~a^4,~\rho^2a^2$, etc.
Setting $v$ to zero has no impact on the existence of these couplings. Indeed, the original global U$(1)$ symmetry is still respected by the vertices.

A nonzero vev gives not-so-expected results. There are four immediate effects that merit our attention: 
(i) It endows the $C$-even component $h$ with positive definite mass of larger in magnitude than $m_\phi$
\begin{equation}
 m_h = \sqrt{2\lambda}v = \sqrt{3\lambda v^2  -\vert m_\phi^2\vert} = \sqrt{2}\vert m_\phi\vert;
\end{equation}
(ii) renders massless the $C$-odd component of $\phi$, $a$,
\begin{equation}
 m_a = \sqrt{\lambda v^2  -\vert m_\phi^2\vert} = 0;
\end{equation}
(iii) it makes a positive definite shift to $M_\rho$
\begin{equation}
 \tilde{M}_\rho = \sqrt{M_\rho^2 + g_\rho v^2} \geq M_\rho;
\end{equation}
(iv) and introduces global symmetry-violating, three-point interactions proportional to $v$
\begin{equation}
 -\lambda v h^3 \quad -\lambda v h a^2 -g_\rho v \rho_\mu\rho^\mu h 
 = -\sqrt{\frac{\lambda}{2}} m_h h^3 \quad -\sqrt{\frac{\lambda}{2}} m_h h a^2 -\frac{g_\rho m_h}{\sqrt{2\lambda}} \rho_\mu\rho^\mu h.
\end{equation}
The discrete charge conjugation symmetry \textit{protects} (forbids) the Lagrangian from spontaneously generating interaction terms with odd powers of $a$.
An imaginary vev, however, would generate such interactions.
A generalization of this presentation to arbitrary global group symmetries with a countable number of group generators is known as 
\textbf{Goldstone's Theorem} \cite{Goldstone:1961n}. It states that 
\textit{for each broken continuous global symmetry; equivalently, for each broken generator of a  continuous global symmetry, a massless scalar appears.}
These massless scalars, such as $a$ in our case, are called \textit{Nambu-Goldstone} (NG) bosons.
In the case of a spontaneously breakdown of an inexact global symmetry, 
the ``inexactness'' being controlled by some parameter $M$, 
the NG bosons acquire a mass proportional to $(M\times\text{vev})$ that vanishes in the limit that the global symmetry becomes exact.
In that case, the NG bosons are called \textit{pseudo-Nambu-Goldstone} (PNG) bosons, e.g., pions $(\pi^{0,\pm})$ in QCD after chiral symmetry breaking.

\section{Continuous Symmetries II: Local Symmetries}
As the name suggests, local symmetries, also known as \textit{gauge symmetries}, differ from global ones in that
\textit{continuous local symmetries} are infinitesimal transformations that are dependent on spacetime coordinates.
In a sense, they are a generalization of global transformations.
However, only derivatives acting on symmetry-respecting fields present a concern.
Operators without derivatives but respect a global symmetry, e.g., $\phi^\dagger\phi$, also respect their local analogs
\begin{equation}
 \phi^\dagger\phi \rightarrow \phi'^\dagger\phi' = \phi^\dagger U U^\dagger \phi = \phi^\dagger\phi.
\end{equation}
Thus our attention focuses on derivative operators, and in particular kinetic terms.

Lets consider a theory of a Dirac fermion $\psi$ and a complex scalar $\phi$
\begin{eqnarray}
 \mathcal{L}	&=& \overline{\psi}i\slash\!\!\!\partial\psi + (\partial^\mu\phi^\dagger)\partial_\mu\phi 
		- m_\psi \overline{\psi}\psi - m_\phi^2 \phi^\dagger\phi -\lambda(\phi^\dagger\phi)^2.
\label{localLag.sm.EQ}
\end{eqnarray}
We assume $\psi$ and $\phi$ satisfy some non-Abelian local symmetry but
delay a discussion of the sign of $\lambda$ and $m_\phi$ until the next section
However, the scalar potential's resemblance to Eq.~(\ref{scalarPot.sm.EQ}) is not coincidental.
A \textit{global} transformation on $\psi$ (or $\phi$) leaves kinetic terms unchanged since
\begin{equation}
 \not\!\partial \psi\rightarrow \not\!\partial \psi' = \not\!\partial (e^{-i\theta}\psi)
 =  \underset{0}{\underbrace{(\not\!\partial e^{-i\theta})}}\psi +  e^{-i\theta} (\not\!\partial\psi) = e^{-i\theta} \not\!\partial \psi.
\end{equation}
A spacetime-dependent symmetry rotation on the other hand, such as
\begin{equation}
\Theta(x)\equiv \Theta^a(x)T^a,
\end{equation}
where $T^a$ are the generators of the corresponding group and $\Theta^a(x)$ are spacetime-dependent rotations in the space of the group representation,
leads to an additional term:
\begin{eqnarray}
  \partial_\mu \psi\rightarrow \partial_\mu \psi' &=& \partial_\mu (e^{-i\Theta(x)}\psi)
 =  \underset{\neq0}{\underbrace{[-ie^{i\Theta(x)}\partial_\mu \Theta(x)]}}\psi  +  e^{-i\Theta} (\partial_\mu\psi)
 \\
 &=&  e^{-i\Theta(x)}\left( \partial_\mu -  i\partial_\mu \Theta(x)\right) \psi.
\end{eqnarray}
Suppressing $\Theta$'s $x^\mu$-dependence, the $\psi$ and $\phi$ kinetic terms under local transformations are 
\begin{eqnarray}
 \overline{\psi}i\not\!\partial\psi \rightarrow \overline{\psi'}i\not\!\partial\psi' 
 &=& 
 \overline{\psi}i\not\!\partial\psi + (\partial_\mu\Theta)\overline{\psi}\gamma^\mu\psi
 \nonumber
 \\
 \nonumber
 (\partial_\mu \phi)^* \partial^\mu\phi \rightarrow (\partial_\mu \phi')^* \partial^\mu\phi' 
 &=& 
 \partial_\mu \phi^* \partial^\mu\phi + (\partial_\mu\Theta)(\partial^\mu\Theta)\phi^\dagger\phi
 - \left[i(\partial_\mu\phi^*)\phi(\partial^\mu\Theta) + \text{H.c}\right].
\end{eqnarray}
The existence of terms linear and quadratic in $\partial_\mu\Theta$ very much violate our notion of invariance under infinitesimal transformations.
Therefore, if we must insist on such a symmetry, then additional terms must be introduced to cancel the $\partial_\mu\Theta$ terms.
The Lorentz and group indices on $\partial_\mu\Theta = \partial_\mu\Theta^a T^a$ provide us with much guidance.

As spin-zero and spin-half fields do not carry Lorentz four-vector indices, 
our symmetry-rescuing terms must come from modifications to our derivatives operators.
This considerably constraints our options. A modification must then take the form 
\begin{equation}
 i\partial_\mu \rightarrow iD_\mu \overset{\rm preliminarily}{=} i\partial_\mu - g A_\mu(x), \quad g> 0.
 \label{covDdef1.sm.EQ}
\end{equation}
where $A_\mu(x)$ is a quantum field.
The resemblance of Eq.~(\ref{covDdef1.sm.EQ}) to obtaining the 
Hamiltonian of a particle with electric charge $e > 0$ in classical electrodynamics using the substitution 
\begin{equation}
 p^\mu \rightarrow p^{'\mu} = p^\mu - e A^\mu(x),
\label{covDdef2.sm.EQ}
\end{equation}
where $A^\mu(x^\nu) = (\Phi(x^\nu),\vec{A}(x^\nu))$ is the classical electrodynamic vector potential, is motivational. 
In the case of a local U$(1)$ transformation with $g=e$, 
we can identify $A_\mu$ in Eq.~(\ref{covDdef1.sm.EQ}) as the quantized version of $A_\mu$ in Eq.~(\ref{covDdef2.sm.EQ}),
familiarly known as the photon.
For this reason, we take $g>0$.

As the infinitesimal rotation $\Theta^a$ is actually a vector in the space spanned by the group generators $T^a$,
it possesses as many independent components that spoil our symmetry as there are generators.
For SU$(N)$ and U$(N)$ theories, respectively, there are 
\begin{equation}
 n=N^2-1 \quad\text{and}\quad n=N^2
\end{equation}
generators. 
So to systematically cancel these terms, we must introduce not just one $A_\mu$ but as many as there are  $\Theta^a$.
To do this, $A_\mu$, like $\Theta^a$, must be a vector in the space spanned by $T^a$. 
Therefore, our derivative of Eq.~(\ref{covDdef1.sm.EQ}) is more completely written as
\begin{equation}
\boxed{ i\partial_\mu \rightarrow iD_\mu = i\partial_\mu - g A_\mu, \quad A_\mu \equiv A_\mu^a T^a, \quad \Tr[T^a T^b] = \frac{1}{2}\delta^{ab}}.
 \label{covDdef3.sm.EQ}
\end{equation}
The last equality represents the generator normalization convention we are adopting. 

A local transformation on $D_\mu\psi$ is given by
\begin{eqnarray}
 D_\mu\psi\rightarrow D_\mu\psi' &=& (\partial_\mu + igA_\mu)(e^{-i\Theta}\psi) = (\partial_\mu + ig A_\mu^a T^a)(\psi -i\Theta^b T^b \psi)
 \\
&=&  \partial_\mu\psi -i\Theta^b T^b (\partial_\mu \psi)   
+ ig A_\mu^a T^a \psi  -i(\partial_\mu\Theta^b) T^b \psi + g A_\mu^a \Theta^b  T^a  T^b \psi. 
\label{localTran2.sm.EQ}
\end{eqnarray}
Generically, the commutator for a non-Abelian field is expressed as
\begin{eqnarray}
 [A_{\mu},A_{\nu}] \equiv [A_{\mu}^{b}T^{b},A_{\nu}^{c}T^{c}] = A_{\mu}^{b}A_{\nu}^{c}[T^{b},T^{c}] = i f_{abc}  A_{\mu}^{b}A_{\nu}^{c} T^{a}.
 \label{commutator.sm.EQ}
\end{eqnarray}
where $f_{abc}$ is the \textit{structure constant} of the group. This allows us to rewrite Eq.~(\ref{localTran2.sm.EQ}) as
\begin{eqnarray}
  D_\mu\psi' &=& e^{-i\Theta}(\partial_\mu\psi)   
+ ig A_\mu^a T^a \psi -i(\partial_\mu\Theta^b) T^b \psi + g A_\mu^a \Theta^b T^b  T^a \psi + i g f_{abc} A_\mu^a \Theta^b T^c \psi
\\
&=& e^{-i\Theta}(\partial_\mu\psi)  + g A_\mu^a \Theta^b T^b
+ ig\left( A_\mu^a T^a    -\frac{1}{g}(\partial_\mu\Theta^b) T^b   +  f_{abc} A_\mu^a \Theta^b T^c \right)\psi
\end{eqnarray}
As the group indices within the parentheses are not external, the labels $a,b,c$ are dummy indices.
Using the cyclic properties of $f_{abc}$, we rewrite the parenthetical term as
\begin{eqnarray}
 (\dots) = A_\mu^a T^a    -\frac{1}{g}(\partial_\mu\Theta^a) T^a   +  f_{cba} A_\mu^c \Theta^b T^a
         = \left( A_\mu^a -\frac{1}{g}(\partial_\mu\Theta^a)   -  f_{abc} \Theta^b A_\mu^c \right)T^a.
\end{eqnarray}
Therefore, if $A_\mu^a$ \textit{also} rotates under infinitesimal transformations such that
\begin{equation}
\boxed{ A_\mu^a \rightarrow A^{'a}_\mu = A_\mu^a  +\frac{1}{g}(\partial_\mu\Theta^a)   +f_{abc} \Theta^b A_\mu^c},
\end{equation}
then $D_\mu\psi$ transforms under local rotations (keeping terms at most linear in $\Theta$) as
\begin{eqnarray}
 D'_\mu\psi'  &=& 
 e^{-i\Theta}(\partial_\mu\psi)  + g A^{'a}_\mu \Theta^b T^b T^a \psi
+ ig\left(A^{'a}_\mu   -\frac{1}{g}(\partial_\mu\Theta^a)   - f_{abc} \Theta^b A_\mu^c\right)  T^a\psi
\\
&=&  e^{-i\Theta}(\partial_\mu\psi)  + g A^{a}_\mu \Theta^b T^b T^a \psi + ig A^{a}_\mu  T^a\psi
\\
&=&  e^{-i\Theta}(\partial_\mu\psi)  + ig (1 -i  \Theta^b T^b)A^{a}_\mu  T^a \psi 
\\
&=& e^{-i\Theta} (D_\mu\psi).
\end{eqnarray}

Our fermion and scalar kinetic terms now transform satisfactorily as 
\begin{eqnarray}
\overline{\psi}i\not\!\!D\psi \rightarrow \overline{\psi'}i\not\!\!D'\psi' &=& \overline{\psi}\left(e^{i\Theta}e^{-i\Theta}\right)i\not\!\!D\psi =
 \overline{\psi}i\not\!\!D\psi 
 \\
(D_\mu\phi)^\dagger(D^\mu\phi) \rightarrow (D'_\mu\phi')^\dagger(D^{'\mu}\phi') &=& (D_\mu\phi)^\dagger(e^{i\Theta}e^{-i\Theta})(D^\mu\phi) =
(D_\mu\phi)^\dagger(D^\mu\phi)
\end{eqnarray}
Thus, the replacement in our theory of derivatives $\partial_\mu$ with \textit{covariant derivatives} $D_\mu$:
\begin{equation}
 \partial_\mu \rightarrow D_\mu \equiv \partial_\mu + i g A_\mu^a T^a,
\end{equation}
where $\phi$, $\psi$, and $A_\mu$ transform under the local symmetry as
\begin{eqnarray}
 \psi 		\rightarrow \psi' &=& \psi e^{-i\Theta(x)}\\
\phi 		\rightarrow \phi' &=& \phi e^{-i\Theta(x)}\\
A_\mu^a		\rightarrow A^{'a}_\mu &=& A^a_\mu +\frac{1}{g}(\partial_\mu\Theta^a), \quad\text{for Abelian symmetries}\\
A_\mu^a		\rightarrow A^{'a}_\mu &=& A^a_\mu +\frac{1}{g}(\partial_\mu\Theta^a)   +f_{abc} \Theta^b A_\mu^c \quad\text{for non-Abelian symmetries},
\end{eqnarray}
renders the entire Lagrangian given by Eq.~(\ref{localLag.sm.EQ}) invariant under local transformations.
The gauge fields $A_\mu^a$, as they carry Lorentz vector indices, and thus are in the vector boson representation of the Lorentz group.
We refer to such objects that correspond to gauge transformations as \textit{gauge bosons}.
Colloquially, we also say that such a theory described
\begin{eqnarray}
 \mathcal{L}	&=& \overline{\psi}i\slash\!\!\!\!D\psi + (D^\mu\phi)^* D_\mu\phi 
		- m_\psi \overline{\psi}\psi - m_\phi^2 \phi^*\phi -\lambda(\phi^*\phi)^2.
\label{sunLag.sm.EQ}
\end{eqnarray}
is \textit{gauge invariant}. 
However, Eq.~(\ref{sunLag.sm.EQ}) is incomplete. 
As we have introduced the gauge field $A_\mu$, or collection of gauge fields $A_\mu^a$, we must also specify how it propagates through spacetime.
In other words: its mass and kinetic terms.

Mass terms for vector bosons take the form
\begin{equation}
 \mathcal{L}_{\rm Mass} = \frac{1}{2}M_A^2 A_\mu^a A^{a\mu}.
\end{equation}
Under gauge transformations, however, we have
\begin{eqnarray}
  \frac{1}{2}M_A^2 A_\mu^a A^{a\mu} &\rightarrow& 
  \frac{1}{2}M_A^2 A_\mu^{'a} A^{'a\mu} 
  \\
  &=&
  \frac{1}{2}M_A^2
  \left(A_\mu^a  +\frac{1}{g}(\partial_\mu\Theta^a)   +f_{abc} \Theta^b A_\mu^c\right)
  \left(A^{\mu a}  +\frac{1}{g}(\partial^\mu\Theta^a)   +f_{ade} \Theta^d A^{\mu e}\right).
\end{eqnarray}
Keeping terms that are no more than linear in $\Theta^a$ and permuting the indices of $f$, we obtain
\begin{eqnarray}
 \frac{1}{2}M_A^2 A_\mu^{'a} A^{'a\mu} &=&
   \frac{1}{2}M_A^2
  \Bigg[
  A_\mu^a A^{\mu a}  +\frac{1}{g}(\partial_\mu\Theta^a)A^{\mu a}   + \frac{1}{g}(\partial^\mu\Theta^a)A_\mu^a
  \nonumber\\
  & & \qquad f_{abc} \Theta^b A_\mu^c A^{\mu a} +f_{cba} \Theta^b A_\mu^c A^{\mu a}
  \Bigg]
  \\
  &=&
  \frac{1}{2}M_A^2
  \left[  A_\mu^a A^{\mu a}  +\frac{1}{g}(\partial_\mu\Theta^a)A^{\mu a}   +\frac{1}{g}(\partial^\mu\Theta^a)A_\mu^a  \right],
  \label{gaugeViolation.sm.EQ}
\end{eqnarray}
which violates our gauge symmetry. Therefore, to preserve it, gauge bosons must be massless.

Kinetic terms for gauge bosons are constructed from the \textit{field strength} tensor
\begin{eqnarray}
 A_{\mu\nu} &=& \partial_\mu A_\nu - \partial_\nu A_\mu + i g [A_\mu,A_\nu] \\ 
	    &\equiv& A_{\mu\nu}^a T^a = \left(\partial_\mu A_\nu^a - \partial_\nu A_\mu^a -  g f_{abc}  A_{\mu}^{b}A_{\nu}^{c}\right) T^{a},
\end{eqnarray}
where we have used the commutator relationship of Eq.~(\ref{commutator.sm.EQ}).
Alternatively, we can construct the field strength tensor by evaluating the commutator of the covariant derivate:
\begin{eqnarray}
   A_{\mu\nu} &=& \frac{1}{i g}[D_\mu,D_\nu] 
   = \frac{1}{i g}(\partial_\mu + i g A_\mu)(\partial_\nu + i g A_\nu)-\frac{1}{i g}(\partial_\nu + i g A_\nu)(\partial_\mu + i g A_\mu)
  \\
  &=& \frac{1}{i g}\partial_\mu\partial_\nu + (\partial_\mu A_\nu) + A_\nu \partial_\mu + A_\mu\partial_\nu + i g A_\mu A_\nu
  \nonumber\\
  &-& \frac{1}{i g}\partial_\nu\partial_\mu - (\partial_\nu A_\mu) - A_\mu \partial_\nu - A_\nu\partial_\mu - i g A_\nu A_\mu
  \\
  &=& [(\partial_\mu A_\nu) -(\partial_\nu A_\mu)] +  i g (A_\mu A_\nu-  A_\nu A_\mu).
\label{fieldStr.sm.eq}
\end{eqnarray}
In this notation, it is easy to see that non-Abelian field strengths transform under gauge a transformation $U$ 
in the same manner as covariant derivatives do:
\begin{eqnarray}
 D_\mu \rightarrow D'_\mu &=& U^\dagger D_\mu U \\
 A_{\mu\nu} \rightarrow A'_{\mu\nu} &=& \frac{1}{i g}\left( U^\dagger D_\mu U U^\dagger D_\nu U - U^\dagger D_\nu U U^\dagger D_\mu U \right)
 = \frac{1}{i g}U^\dagger [D_\mu, D_\nu] U 
 \nonumber\\
 &=& U^\dagger A_{\mu\nu} U.
\end{eqnarray}
However, because of this property, we are not allowed to write a kinetic term simply as $A_{\mu\nu}A^{\mu\nu}$ if we require that 
it be independently gauge invariant from all other Lagrangian terms. We resolve this by taking only its diagonal elements, its trace.
Physically, we can understand this as needing to match a gauge boson's gauge charge (group index) with itself to form a gauge singlet state.
Subsequently,
\begin{equation}
 \Tr[A_{\mu\nu}A^{\mu\nu}] \rightarrow \Tr[A'_{\mu\nu}A^{'\mu\nu}] = \Tr[U^\dagger A_{\mu\nu}U U^\dagger A^{\mu\nu}U] = \Tr[A_{\mu\nu}A^{\mu\nu}]
 = \frac{1}{2}A^a_{\mu\nu}A^{a\mu\nu},
\end{equation}
there the factor of one-half came from the normalization of the generators, as given in Eq.~(\ref{covDdef3.sm.EQ}).
Scaling this trace by $1/2$ is necessary for correct field normalization for identical particles.
For Abelian gauge bosons, we have the additional property the field strengths are also individually gauge invariant:
\begin{eqnarray}
 A_{\mu\nu} \rightarrow A'_{\mu\nu} &=& \partial_\mu A'_\nu - \partial_\nu A'_\mu 
 =  \partial_\mu \left(A_\nu +\frac{1}{g}(\partial_\nu\Theta) \right) - \partial_\nu \left(A_\mu +\frac{1}{g}(\partial_\mu\Theta)\right)
 \\
 &=& \partial_\mu A_\nu - \partial_\nu A_\mu 
 +  \frac{1}{g}\underset{0}{\left( \underbrace{\partial_\mu\partial_\nu\Theta - \partial_\nu\partial_\mu\Theta}\right )}
 =  A_{\mu\nu}.
\end{eqnarray}
Thus, Abelian kinetic terms do take the simple form $A_{\mu\nu}A^{\mu\nu}$.
With this, we write
\begin{eqnarray}
 \mathcal{L}	&=& \overline{\psi}i\slash\!\!\!\!D\psi + (D^\mu\phi)^\dagger D_\mu\phi 
		- m_\psi \overline{\psi}\psi - m_\phi^2 \phi^\dagger\phi -\lambda(\phi^\dagger\phi)^2
 ~\underset{\rm Abelian}{\underbrace{-\frac{1}{4}A_{\mu\nu}A^{\mu\nu}}}
 ~\underset{\rm non-Abelian}{\underbrace{-\frac{1}{4}A^{a}_{\mu\nu}A^{a~\mu\nu}}}
\end{eqnarray}
and our local gauge theory is now complete.

As a concrete example, we briefly consider a simple U$(1)$ gauge theory: QED.

\subsection{Quantum Electrodynamics}\label{qed.sm.sec}
Quantum Electrodynamics (QED) is the theory of light at microscopic distances 
and, as the name suggests, is simply electrodynamics after second quantization, 
i.e., expansion of classical field in creation and annihilation operators.
We begin by supposing a massive fermion $\psi$ and complex scalar $\phi$ that transform according to a local U$(1)_{EM}$
with generator $\hat{Q}$ 
\begin{eqnarray}
\psi	\rightarrow \psi' = \psi e^{-i\Theta(x)\hat{Q}}, &\quad& 
\overline{\psi} \rightarrow \overline{\psi'} = \overline{\psi} e^{i\Theta(x)\hat{Q}}\\
\phi	\rightarrow \phi' = \phi e^{-i\Theta(x)\hat{Q}}, &\quad&
\phi^* 	  \rightarrow \phi'^* = \phi^* e^{i\Theta(x)\hat{Q}}.
\end{eqnarray}
When $\hat{Q}$ operates on $\psi~(\phi)$ gives its electric charge $q_\psi~(q_\phi)$ 
in units of $e>0$.
Following to our procedure above, our gauge invariant theory is given by the Lagrangian 
\begin{eqnarray}
 \mathcal{L}_{\rm QED} &=& \overline{\psi}i\slash\!\!\!\!D\psi + (D^\mu\phi)^\dagger D_\mu\phi -\frac{1}{4}A_{\mu\nu}A^{\mu\nu}
	      - m_\psi \overline{\psi}\psi - m_\phi^2 \phi^\dagger\phi
\end{eqnarray}
Expanding the covariant gauge derivatives gives one
\begin{eqnarray}
\overline{\psi}i \not\!\!D\psi &=&  \overline{\psi}i(\not\!\partial + i e  \not\!\!A \hat{Q})\psi =  
\overline{\psi}i\not\!\partial\psi -  e q_\psi \overline{\psi} \not\!\!A \psi 
\\
(D^\mu\phi)^\dagger D_\mu\phi &=& (\partial^\mu\phi + i e A^\mu \hat{Q}\phi)^\dagger (\partial_\mu\phi + i e A_\mu \hat{Q}\phi)
\\
\nonumber
&=& (\partial^\mu\phi^*)(\partial_\mu\phi) +  e^2 q^2_\phi   A^\mu A_\mu \phi^*  \phi
+  \left[ i e q_\phi (\partial^\mu\phi^*) A_\mu \phi  + \text{H.c} \right],
\end{eqnarray}
Shuffling terms, we have
\begin{eqnarray}
 \mathcal{L}_{\rm QED} &=& \mathcal{L}_{\rm Kinetic} + \mathcal{L}_{\rm Mass} + \mathcal{L}_{\rm Int.},
 \\
 \mathcal{L}_{\rm Kinetic} &=& \overline{\psi}i\not\!\partial\psi + (\partial^\mu\phi^*)(\partial_\mu\phi) -\frac{1}{4}A_{\mu\nu}A^{\mu\nu}
  \\
 \mathcal{L}_{\rm Mass} &=& - m_\psi \overline{\psi}\psi - m_\phi^2 \phi^\dagger\phi
 \\
 \mathcal{L}_{\rm Int.} &=& -e q_\psi \overline{\psi} \not\!\!A \psi +  e^2 q^2_\phi g_{\mu\nu}  A^\mu A^\nu \phi^*  \phi
+  \left[ i e q_\phi (\partial^\mu\phi^*) A_\mu \phi  + \text{H.c} \right]
\end{eqnarray}

\begin{figure}
\centering
\includegraphics[width=0.9\textwidth]{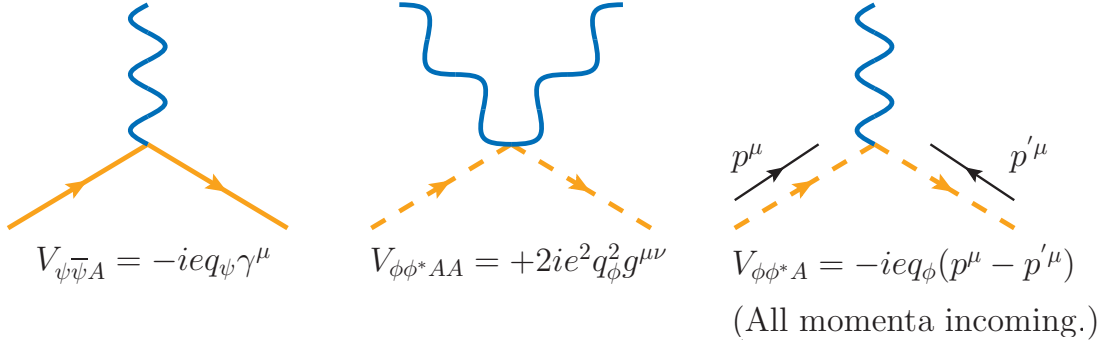} \vspace{0cm}
\caption{Interaction Feynman Rules for QED.} 
\label{qedFeynRules.fig}
\end{figure}

To efficiently obtain the \textit{Feynman Rules} for QED from the interaction Lagrangian $\mathcal{L}_{\rm Int.}$, 
we Fourier decompose the field operators $\psi$, $\phi$, and $A_\mu$ under its Action.
For example:
\begin{eqnarray}
\mathcal{S}_{\overline{\psi}\psi A} &=& i \int d^4x \mathcal{L}_{\overline{\psi}\psi A} = \int d^4x (-ie q_\psi) \overline{\psi} \not\!\!A \psi
\\
&\propto&  \int d^4x \left[d^3 p\right]^3 \sum_{\rm d.o.f.}
\left[a_s^\dagger(p') \overline{u}_s(p') e^{-ip'\cdot x} + b_s(p')\overline{v}_s(p')e^{ip'\cdot x}\right]
\nonumber\\
&\times&  (-ie q_\psi \gamma^\mu)\left[a_\lambda(q) \varepsilon_{\lambda\mu}(q)e^{iq\cdot x} +
a^{c\dagger}(q,\lambda) \varepsilon^{*}_{\mu}(q,\lambda)e^{-iq\cdot x}\right]
\nonumber\\
&\times& \left[a_s(p) u_s(p) e^{ip\cdot x} + b^\dagger_s(p)v_s(p)e^{-ip\cdot x}\right]. 
\end{eqnarray}
This represents every permutation of incoming/outgoing fermions/bosons that is allowed under QFT and gauge invariance
for the $\psi-\overline{\psi}-A$ coupling. The common factor is the term
\begin{equation}
 V_{\overline{\psi}\psi A}  \quad:\quad -ie q_\psi \gamma^\mu.
\end{equation}
We may do the same for the $\phi-\phi^*-A-A$ interaction vertex.
Keeping track of a multiplicity factor of $2$ that originates from having identical bosons, i.e., $A_\mu,~A_\nu$, we obtain 
\begin{equation}
 V_{\phi\phi A A}  \quad:\quad +2ie^2 q^2_\phi g_{\mu\nu}.
\end{equation}
Last, we have the $\phi-\phi^*-A$ vertex. It is marginally more complicated since one must keep track of 
incoming/outgoing momenta. For incoming $p$ and outgoing $p'$, we have
\begin{eqnarray}
\mathcal{S}_{\phi\phi A} &=& i \int d^4x \mathcal{L}_{\phi\phi A} = \int d^4x 
 (-e q_\phi)\left[(\partial^\mu\phi^*) A_\mu \phi   - \phi^* A_\mu (\partial^\mu \phi)\right]
\\
&\propto&  \int d^4x \left[d^3 p\right]^3 \sum_{\rm d.o.f.}
\left[a^\dagger(p')  \partial^\mu e^{-ip'\cdot x} + \underset{{\rm Outgoing:}~ip^{'\mu}}{\underbrace{a^{c}(p')\partial^\mu e^{ip'\cdot x}}}\right]
\nonumber\\
&\times&  (-e q_\phi )\left[a_\lambda(q) \varepsilon_{\lambda\mu}(q)e^{iq\cdot x} +
a^{c\dagger}(q,\lambda) \varepsilon^{*}_{\mu}(q,\lambda)e^{-iq\cdot x}\right]
\nonumber\\
&\times& \left[a(p)  e^{ip\cdot x} + 
\underset{{\rm Incoming}: -ip^\mu}{\underbrace{a^{c\dagger}(p) e^{-ip\cdot x}}}\right] - \cdots (\text{Incoming})
\end{eqnarray}
Collecting terms gives us the coupling vertex
\begin{equation}
 V_{\phi\phi A} \quad:\quad -ie q_\phi(p^\mu + p^{'\mu}) \quad\text{for incoming~(outgoing)~} p~(p').
\end{equation}
We summarize these Feynman rules in Fig.~\ref{qedFeynRules.fig}.

\subsection{The Higgs Mechanism: Spontaneously Broken Local Symmetries}\label{ssbLocal.sm.sec}
At last, we turn to the topic of spontaneously broken gauge symmetries.
Several of the intermediate steps here have been derived in Section~\ref{ssbGlobal.sm.sec}, where spontaneously broken global symmetries are studied.
We consider a model containing three complex scalars $\phi,$ $\Phi$, and $H$.
We let $\phi$ and $\Phi$ be respectively gauged under U$(1)_A$ and U$(1)_B$, and normalize their couplings to unity.
The field $H$ is charged under both U$(1)_A$ and U$(1)_B$, also with unity charges. 
Notationally, it is often stated that under the gauge group 
\begin{equation}
 \mathcal{G} = {\rm U}(1)_{A}\times{\rm U}(1)_B 
\end{equation}
the fields $\phi,~\Phi,~H$ are charged as follows:
\begin{equation}
 \phi : (+1,0), \quad \Phi : (0,+1), \quad H : (+1,+1).
\end{equation}
Assuming that only $H$ has a nonzero mass, our gauge invariant Lagrangian is
\begin{eqnarray}
\mathcal{L} &=& (D^\mu\phi)^\dagger D_\mu\phi + (D^\mu\Phi)^\dagger D_\mu\Phi + (D^\mu H)^\dagger D_\mu H  
-\frac{1}{4}A_{\mu\nu}A^{\mu\nu}- \frac{1}{4}B_{\mu\nu}B^{\mu\nu} 
\nonumber\\
	    & & - V_1 - V_2, \\
V_1 &=& m_H^2 H^* H + \lambda_{HH} (H^* H)^2 + \lambda_{H\phi} H^* H \phi^*\phi + \lambda_{H\Phi} H^*H \Phi^* \Phi, \\
V_2 &=& \lambda_{\phi\phi} (\phi^* \phi)^2 + \lambda_{\Phi\Phi} (\Phi^* \Phi)^2 + \lambda_{\Phi\phi} \Phi^*\Phi \phi^*\phi,
\end{eqnarray}
where $X_{\mu\nu}$ is the field strength of gauge boson $X=A,B$.
The covariant derivatives are
\begin{eqnarray}
 D_\mu\phi	&=& (\partial_\mu + ig_A A_\mu \hat{Q}_A)\phi = (\partial_\mu + (+1)ig_A A_\mu)\phi,\\
 D_\mu\Phi	&=& (\partial_\mu + ig_B B_\mu \hat{Q}_B)\Phi = (\partial_\mu + (+1)ig_B B_\mu)\Phi,\\
 D_\mu H 	&=& (\partial_\mu + ig_A A_\mu \hat{Q}_A + ig_B B_\mu \hat{Q}_B)H =
		    (\partial_\mu + (+1)ig_A A_\mu + (+1)ig_B B_\mu)H,
\end{eqnarray}
where $\hat{Q}_X$ denotes the charge generator of gauge interaction $X$.
The potentials $V_1$ and $V_2$ have been written in such a way that strictly $(H^* H)$-dependent terms are contained in $V_1$.

For the sake of avoiding a nonsensical theory, we require all four-point couplings $\lambda_{XY}$ to be positive-definite.
We require that neither  $\phi$ nor $\Phi$ carry nonzero vevs,
\begin{equation}
 \langle \phi \rangle = \langle \Phi \rangle = 0.
\end{equation}
The quartic potential in $(H^* H)$, $V_1$, gives rise to a nonzero vev for $H$ if $m_H^2 < 0$.
We may ignore $\phi$ and $\Phi$ in solving for the minimum of $H$ since their vevs are (by hypothesis) zero.
In this case, the extrema solutions of $H$ are the same as those given in Eq.~(\ref{globalMin.sm.eq}), 
and therefore $H$ possesses a vev given by
\begin{equation}
 v_H = \sqrt{2}\langle H \rangle = \sqrt{\frac{\vert m_H^2 \vert}{\lambda_{HH}}}.
\end{equation}

We now consider the covariant derivate acting on $\langle H\rangle$, which is
the qualitatively new feature in spontaneously broken \textit{local} symmetries.
Setting $H$ to its vev, we see for $D_\mu H$
\begin{eqnarray}
 D_\mu H &=& \left(\partial_\mu + ig_A A_\mu + ig_B B_\mu\right) \frac{v_H}{\sqrt{2}} 
	  = i\frac{v_H}{\sqrt{2}}\left(g_A A_\mu + g_B B_\mu\right).  
\end{eqnarray}
Pairing it with its conjugate, we obtain
\begin{eqnarray}
 (D^\mu H)^\dagger(D_\mu H) &=& \frac{v_H^2}{2}\left(g_A A^\mu + g_B B^\mu\right)\left(g_A A_\mu + g_B B_\mu\right).
 \label{dHdH.sm.EQ}
\end{eqnarray}
Without the loss of generality, we assume $g_A > g_B$ and define the quantities
\begin{equation}
\cos\theta_{A} \equiv \frac{g_A}{\sqrt{g_A^2 + g_B^2}}, \quad
g_Z \equiv \sqrt{g_A^2 + g_B^2} = \frac{g_A}{\cos\theta_A}, \quad\text{and}\quad   
g_\gamma = g_A \sin\theta_A.
\end{equation}
With these definitions, we may write Eq.~(\ref{dHdH.sm.EQ}) as
\begin{eqnarray}
  (D^\mu H)^\dagger(D_\mu H) &=& \frac{v_H^2}{2}g_Z^2 \left(\cos\theta_A A_\mu + \sin\theta_A B_\mu\right)^2.
\end{eqnarray}
However, one may recognize that the quantity in the parentheses is nothing more than a field redefinition in $g_A-g_B$ space
given by the rotation
\begin{equation}
 \begin{pmatrix}   Z \\ \gamma  \end{pmatrix}
  = 
 \begin{pmatrix}     \cos\theta_A & \sin\theta_A \\ -\sin\theta_A & \cos\theta_A   \end{pmatrix}
 \begin{pmatrix}   A \\ B  \end{pmatrix}.
 \label{gaugeMixing.sm.EQ}
\end{equation}
The field $\gamma$ is identified as the orthogonal state of $Z$ as $A,B$ are rotated by mixing angle $\theta_A$.
Expressing the kinetic term for $H$ at its minimum in terms of this $Z_\mu$ vector boson gives us a remarkable result:
a vector boson mass term.
\begin{equation}
   (D^\mu H)^\dagger(D_\mu H) = \frac{v_H^2}{2}g_Z \underset{Z^\mu Z_\mu}{\underbrace{\left(\cos\theta_A A_\mu + \sin\theta_A B_\mu\right)^2}} 
   = \frac{M_Z}{2} Z_\mu Z^\mu,
\end{equation}
where the mass of the $Z_\mu$ boson is 
\begin{equation}
 M_Z = g_Z v_H.
\end{equation}
As no such term for $\gamma_\mu$ materializes, it  remains massless.

These results are notable because in Eq.~(\ref{gaugeViolation.sm.EQ}) we showed that a gauge boson if forbidden 
to have mass as it would otherwise violate gauge invariance.
However, as $H$ was charged under the $A$ and $B$ gauge groups and has since acquired a vev, 
the vacuum too has acquired charges under $A$ and $B$, breaking the local symmetry. 
The massless field $\gamma$, on the other hand, is free to make gauge transformations with respect to a new (Abelian) generator:
\begin{equation}
 \hat{Q}_\gamma = \hat{Q}_A - \hat{Q}_B.
\end{equation}
The fields $\phi$ and $\Phi$, respectively, possess charges $q_\gamma^\phi = +1$ and $q_\gamma^\Phi = -1$ since
\begin{eqnarray}
 \hat{Q}_\gamma \phi &=& \hat{Q}_A \phi - \hat{Q}_B \phi = (+1)\phi - (0) \phi
 \\
 \hat{Q}_\gamma \Phi &=& \hat{Q}_A \Phi - \hat{Q}_B \Phi = (0) \Phi - (+1) \Phi.
\end{eqnarray}
Their couplings to the $Z$ and $\gamma$ fields are discovered by applying the field rotation Eq.~(\ref{gaugeMixing.sm.EQ}) to their 
covariant derivatives:
\begin{eqnarray}
  D_\mu\phi	= (\partial_\mu + ig_A A_\mu \hat{Q}_A)\phi 
  &=& \left[\partial_\mu + ig_A \left(\cos\theta_A Z_\mu - \sin\theta_A \gamma_\mu\right) \left(\hat{Q}_\gamma + \hat{Q}_B \right)\right]\phi
  \\
		&=& \left(\partial_\mu +  ig_Z\cos\theta^2_A Z_\mu - ig_\gamma \gamma_\mu \right)\phi,
  \\
 D_\mu\Phi	= (\partial_\mu + ig_B B_\mu \hat{Q}_B)\Phi 
 &=& \left[\partial_\mu + ig_B\left(\sin\theta_A Z_\mu + \cos\theta_A \gamma_\mu \right)\left( \hat{Q}_A - \hat{Q}_\gamma\right)\right]\Phi
 \\
		&=& \left(\partial_\mu + i g_Z\sin\theta^2_A Z_\mu + i g_\gamma \gamma_\mu \right)\Phi. 
\end{eqnarray}
Being a gauge interaction, $\gamma$ couples to $\phi$ and $\Phi$ with equal strength and proportionally to their charge.
$Z_\mu$, however, couples non-universally: for $g_A > g_B$, $\phi$ interacts more strongly with $Z_\mu$ than $\Phi$.
This is partly due to  $Z_\mu$ aligning more $(\theta_A > \pi/4$) with the gauge state $A_\mu$ than with $B_\mu$,
which is aligned more closely with $\gamma$.
Altogether, this is the crux of spontaneously broken gauge symmetries: the generation of vector boson masses and the emergence of a ``hidden'' local symmetry.
The hidden symmetry refers to  $H$ being charged under 
\begin{equation}
 U(1)_A \times U(1)_B,
\end{equation}
and so the associated gauge bosons are aware of its symmetry-breaking vev, but the subgroup 
\begin{equation}
 U(1)_{A-B}\subset  U(1)_A \times U(1)_B,
 \label{hiddenSymm.sm.eq}
\end{equation}
whose generator is given by $ \hat{Q}_\gamma = \hat{Q}_A - \hat{Q}_B$, 
is unknown to $H$ since it is \textit{neutral} under this local symmetry.
Being neutral, $H$ is unable to charge the vacuum under this gauge group, and therefore it remains unbroken after the spontaneously breakdown 
of its component generators.

In analog to the global symmetry case, we now expand $H$ about its minimum
\begin{equation}
 H \approx \frac{v_H + h(x) + i \xi(x)}{\sqrt{2}}.
\end{equation}
Writing this to lowest order in $h$ and $\xi$, however, gives us
\begin{equation}
 H = \frac{1}{\sqrt{2}}\left(v_H + h(x)\right)\left(1 + i \frac{\xi(x)}{v_H}\right) = \frac{1}{\sqrt{2}}\left(v_H + h(x)\right)e^{i \frac{\xi(x)}{v_H}},
 \label{qedGhost.sm.EQ}
\end{equation}
which we recognize as a gauge transformation of the form
\begin{equation}
 H \rightarrow H' = H e^{-i\Theta(x)}.
\end{equation}
This also indicates that our gauge field transforms locally as
\begin{equation}
Z_\mu	\rightarrow Z'_\mu = Z_\mu +\frac{1}{g_\gamma v}(\partial_\mu\xi).
\end{equation}
The field $\xi$ is an unphysical degree of freedom that represents our ability to make gauge transformations, 
in contrast to the global symmetry case, where $\xi$ was a real, massless scalar.
The gauge choice of removing the unphysical fields by explicitly setting $\xi(x)=0$ is known as the \textit{unitary gauge}.

Continuing, the covariant derivate on $H$ is now given by
\begin{eqnarray}
  D_\mu H &=& \left(\partial_\mu + ig_A A_\mu + ig_B B_\mu\right) \frac{1}{\sqrt{2}}(v_H + h)
          = \frac{1}{\sqrt{2}}\partial_\mu h + \frac{i}{\sqrt{2}}Z_\mu (M_Z + g_Z h).
\end{eqnarray}
Pairing $D_\mu H$ with its conjugate, we obtain
\begin{eqnarray}
 (D^\mu H)^\dagger(D_\mu H) &=& 
 \frac{1}{2}
 \left[(\partial^\mu h) - i Z_\mu (M_Z + g_Z h)\right]
 \left[(\partial_\mu h) + i Z_\mu (M_Z + g_Z h)\right]	  
\\
&=&  \frac{1}{2}(\partial^\mu h)(\partial_\mu h) + \frac{1}{2} Z^\mu  Z_\mu (M_Z + g_Z h)^2 
\\
&=&  \frac{1}{2}(\partial^\mu h)(\partial_\mu h) + 
\frac{1}{2} M_Z^2 Z^\mu  Z_\mu + g_Z M_Z Z^\mu  Z_\mu h + \frac{g_Z^2}{2} Z^\mu  Z_\mu h h,
\end{eqnarray}
which gives us three-point $ZZh$ interactions proportional to $M_Z$ and four-point $ZZhh$ couplings 
that is suppressed by two powers of $g_Z$.
For completeness, we turn to the potential $V_1$.
We observe the emergence of positive definite masses for $h$, $\phi$ and $\Phi$
as well as interactions that are linear and quadratic in $h$ in the same manner that we witnessed for the global case:
\begin{eqnarray}
 V_1 &=& m_H^2 H^* H + \lambda_{HH} (H^* H)^2 + \lambda_{H\phi} H^* H \phi^*\phi + \lambda_{H\Phi} H^*H \Phi^* \Phi,
 \\
 &=& \frac{m_H^2}{2}(v_H^2 + h^2 + 2v_H h  ) + \lambda_{HH}(v_H^2 + h^2 + 2v_H h)^2 
 \nonumber\\
 &+& \lambda_{H\phi} (v_H^2 + h^2 + 2v_H h ) \phi^*\phi + \lambda_{H\Phi}(v_H^2 + h^2 + 2v_H h)\Phi^* \Phi
 \\
 &=& 
 \frac{1}{2}\underset{(h~{\rm  mass})^2}{	\underbrace{\left(3\lambda_{HH}   v_H^2  -\vert m_H^2\vert\right)}}hh+
 \frac{1}{2}\underset{(\phi~{\rm  mass})^2}{	\underbrace{\left(2\lambda_{H\phi}v_H^2\right)}}\phi\phi+ 
 \frac{1}{2}\underset{(\Phi~{\rm  mass})^2}{	\underbrace{\left(2\lambda_{H\Phi}v_H^2\right)}}\Phi\Phi 
 \nonumber\\
 &+& \text{cubic interaction terms of the form~}~hhh,~h\phi\phi,~h\Phi\Phi,
  \nonumber\\
 &+& \text{quartic interaction terms of the form~}~hhhh,~hh\phi\phi,~hh\Phi\Phi.
\end{eqnarray}

This mechanism, proposed in 1964 first by Brout \& Englert~\cite{Englert:1964et}, Higgs~\cite{Higgs:1964ia,Higgs:1964pj}, 
and Guralnik, et. al.~\cite{Guralnik:1964eu}, later to be reviewed in Refs.~\cite{Higgs:1966ev,Kibble:1967sv}, 
is known as the \textit{Brout-Englert-Higgs {\rm (BEH)} Mechanism}, or more commonly, the the \textbf{Higgs Mechanism}.
It is a subtle caveat of Goldstone's Theorem stating that
\textit{for each broken continuous local symmetry, the gauge boson associated with the broken generator of the continuous local symmetry
acquires a mass.} The difference being that if the continuous symmetry is global (local), a massless scalar (massive vector boson) appears in the theory.
With this framework in place, we now move onto our Standard Model adventure.

\section{Introduction to the Standard Model of Particle Physics}

The Standard Model of particle physics, commonly denoted simply as SM, represents to-date our best understanding of 
matter and its interactions at energy scales on the order of 1 TeV and below.
In terms of distance, this corresponding to scales as small as $10^{-19}$ meters.
Though unsatisfactory, for example its prediction of massless neutrinos, 
the impressive agreement between high precision predictions and experimental observations demonstrate that most
any theory that supersedes it will contain the SM as its ``low energy'' effective field theory limit.
In this section, we introduce the ingredients of the SM and derive some of its most fundamental properties.
In doing so, we will be able to appreciate some of the more subtle aspects of SM extensions that alleviate its incompleteness.

Formally speaking, the SM contains a renormalizable Yang-Mills theory~\cite{'tHooft:1972fi} together with of chiral spin-half fermions and spin-zero bosons (scalars) 
with varying charges under strongly coupled~\cite{GellMann:1964nj,Zweig:1981pd,Zweig:1964jf,Gross:1973id,Politzer:1973fx}
and weakly coupled~\cite{Glashow:1961tr,Weinberg:1967tq,Salam:1968rm} gauge symmetries.
Respecting both Abelian and non-Abelian transformations,
the SM scalar sector breaks the weakly coupled gauge sector through the Higgs Mechanism.
The remaining unbroken gauge symmetries, color (non-Abelian) and electromagnetism (Abelian), 
possess somewhat interesting dynamics and eventually give rise to atoms,
which, to speak technically, are electronic bound states of light elementary fermions (electrons) and heavy hadronic bound states (nucleons).
The applicability and utility of atoms are (presumably) familiar to the reader.

The dimension-four Lagrangian of the SM is given as
\begin{equation}
 \mathcal{L}_{\rm SM} = \mathcal{L}_{\rm Gauge} + \mathcal{L}_{\rm Higgs} + \mathcal{L}_{\rm Fermion},
 \label{smLagFull.sm.eq}
\end{equation}
representing the gauge, scalar (or Higgs), and matter (or fermion) sectors of the models. 
We will now discuss each part in detail.

\section{Gauge Sector  of the Standard Model}\label{gauge.sm.sec}
The gauge sector of the SM is categorized into two parts: 
(i) a strongly coupled  (at low momentum transfers) but still asymptotically free (vanishing coupling at infinite momentum transfers)
sector obeying an SU$(3)$ symmetry, known as quantum chromodynamics (QCD);
and (ii) a weakly coupled SU$(2)\times$U$(1)$ symmetry, known as the electroweak (EW) sector, 
that spontaneously breaks to a weakly coupled U$(1)$ gauge symmetry.

The gauge field content of each symmetry constitute an adjoint representation of the group,
meaning that there are as many gauge bosons of a local symmetry as there are generators of the group.
For SU$(N)$ theories, there are $N^2-1$ generators (bosons). 
Similarly for U$(N)$ theories, there are $N^2$ generators (bosons).
Thus, there is a total of 12 $(8+3+1)$ massless spin-one (vector) bosons in the SM gauge group
\begin{equation}
 \mathcal{G}_{\rm SM} = SU(3)_C \times SU(2)_L \times U(1)_Y.
\end{equation}
The labels $C$, $L$, and $Y$ denote color, left-handed weak isospin, and weak hypercharge, 
the names for the respectively conserved charges. 

Defining indices $a,b,c=1,\dots,8$ to denote color degrees of freedom and $i,j,k=1,\dots,3$ to denote weak isospin degrees of freedom,
and with a summation implied for repeated indices, the gauge sector Lagrangian is given by the vector boson kinetic terms
\begin{eqnarray}
 \mathcal{L}_{\rm Gauge} &=& 
  ~-~\underset{\rm Color}{\underbrace{\frac{1}{4}\Tr[G_{\mu\nu}G^{\mu\nu}] }}
  ~-~\underset{\rm Weak~Isospin}{\underbrace{\frac{1}{4}\Tr[W_{\mu\nu}W^{\mu\nu}] }}
  ~- \underset{\rm Weak~Hypercharge}{\underbrace{\frac{1}{4}F_{\mu\nu}F^{\mu\nu}}}
  \\
  &=& 
  ~-~\frac{1}{4}G^{a}_{\mu\nu}G^{a~\mu\nu}
  ~-~\frac{1}{4}W^{i}_{\mu\nu}W^{i~\mu\nu}
  ~-~\frac{1}{4}F_{\mu\nu}F^{\mu\nu},
  \label{gaugeLag.sm.EQ}
\end{eqnarray}
For the SU$(3)$ color gauge bosons $G_\mu^a$, known as \textit{gluons}, the field strength [See Eq.~(\ref{fieldStr.sm.eq})] is
\begin{eqnarray}
 	G^{a}_{\mu\nu}	&=& \partial_{\mu}G_{\nu}^{a} - \partial_{\nu}G_{\mu}^{a} + ig_{s}[G_{\mu},G_{\nu}]^{a}\\
			&=& \partial_{\mu}G_{\nu}^{a} - \partial_{\nu}G_{\mu}^{a} - g_{s}f^{abc}G_{\mu}^{b}G_{\nu}^{c}.
\end{eqnarray}
A most striking feature of non-Abelian theories is the appearance of three-point $\partial G^{a}G^{b}G^{c}$ and 
four-point $G^{b}G^{c}G^{d}G^{e}$ interaction vertices among the gauge bosons. 
These self-interactions are proportional to both the coupling (quadratically in the four-gluon case) and the structure constant 
(product of structure constants in the four-gluon case). 
The reason for this self-coupling is due to the fact that the bosons in non-Abelian theories also carry gauge charges.
In the gauge sector Lagrangian [Eq.~(\ref{gaugeLag.sm.EQ})], 
this is why a trace over the non-Abelian generators to pair fields accordingly is required.
For QCD and its SU$(3)_{C}$ symmetry, its generators (in the fundamental representation) are proportional to the  Gell-Man (or color) matrices $\lambda^a$.
Explicitly, the generators are given by 
\begin{eqnarray}
 \hat{T}^{a} = \frac{1}{2}\lambda^{a}, &~& a=1,\dots,8,\label{colorMatrices.EQ}\\
\lambda^1  =  \begin{pmatrix} 0 & 1 & 0 \\ 1 & 0 & 0 \\ 0 & 0 & 0 \end{pmatrix},\quad
\lambda^2  =  \begin{pmatrix} 0 &-i & 0 \\ i & 0 & 0 \\ 0 & 0 & 0 \end{pmatrix},\quad
\lambda^3 &=& \begin{pmatrix} 1 & 0 & 0 \\ 0 &-1 & 0 \\ 0 & 0 & 0 \end{pmatrix},\quad
\lambda^4  =  \begin{pmatrix} 0 & 0 & 1 \\ 0 & 0 & 0 \\ 1 & 0 & 0 \end{pmatrix},
\nonumber\\
\lambda^5  =  \begin{pmatrix} 0 & 0 &-i \\ 0 & 0 & 0 \\-i & 0 & 0 \end{pmatrix},\quad
\lambda^6  =  \begin{pmatrix} 0 & 0 & 0 \\ 0 & 0 & 1 \\ 0 & 1 & 0 \end{pmatrix},\quad
\lambda^7 &=& \begin{pmatrix} 0 & 0 & 0 \\ 0 & 0 &-i \\ 0 & i & 0 \end{pmatrix},\quad
\lambda^8  =  \frac{1}{\sqrt{3}}\begin{pmatrix} 1 & 0 & 0 \\ 0 & 1 & 0 \\ 0 & 0 &-2 \end{pmatrix}.
\nonumber
\end{eqnarray}
The nonzero, antisymmetric color structure constants are
\begin{eqnarray}
 f_{123}=1, \quad f_{458}=f_{678}=\frac{\sqrt{3}}{2} \quad
 f_{147}=f_{246}&=&f_{257}=f_{345}=f_{516}=f_{637}=\frac{1}{2}.
\end{eqnarray}

\begin{table}[!t]
\caption{Bosons of the Standard Model Before Electroweak Symmetry Breaking}
 \begin{center}
\begin{tabular}{|c|c|c|c|}
\hline\hline
Interaction	& Symbol  	& Spin	& $SU(3)_{C}\times SU(2)_{L}\times U(1)_{Y}$  Charge \tabularnewline\hline\hline
Strong		& $G^{a}_{\mu}$	& 1	& $(\textbf{3},1,0)$ \tabularnewline\hline
Weak	     	& $W^{a}_{\mu}$	& 1	& $(1,\textbf{2},0)$ \tabularnewline\hline
Hypercharge  	& $B_{\mu}$	& 1	& $(1,1,0)$ \tabularnewline\hline
Yukawa		& $\Phi$	& 0	& $(1,\textbf{2},+1)$ (Complex)  \tabularnewline\hline
\hline
\end{tabular}
\label{smBosonsBeforeEWSB.TB}
\end{center}
\end{table}

Turning to the SU$(2)$ weak isospin gauge bosons $W_\mu^i$, the field strength is 
\begin{eqnarray}
	W^{i}_{\mu\nu}	&=& \partial_{\mu}W_{\nu}^{i} - \partial_{\nu}W_{\mu}^{i} + ig_{W}[W_{\mu},W_{\nu}]^{i}\\
			&=& \partial_{\mu}W_{\nu}^{i} - \partial_{\nu}W_{\mu}^{i} - g_W\epsilon^{ijk}W_{\mu}^{j}W_{\nu}^{k}.
\end{eqnarray}
As in the QCD case, we find three-point $\partial W^i W^j W^k $ and four-point $W^i W^j W^k W^j$ interaction vertices arising from the kinetic term.
In the triplet (adjoint) representation of SU$(2)$, the rotation matrices are equivalent to the SO$(3)$ spatial rotations for angular momentum $j=1$,
\begin{equation}
\hat{T}_{L}^{1}=\frac{1}{\sqrt{2}}\begin{pmatrix}0 & 1 & 0 \\ 1 & 0 & 1 \\ 0 & 1 & 0 \end{pmatrix}, \quad
\hat{T}_{L}^{2}=\frac{1}{\sqrt{2}}\begin{pmatrix}0 &-i & 0 \\ i & 0 &-i \\ 0 & i & 0 \end{pmatrix}, \quad
\hat{T}_{L}^{3}=\begin{pmatrix}1 & 0 & 0 \\ 0 & 0 & 0 \\ 0 & 0 &-1 \end{pmatrix}.
\end{equation}
Immediately, we read off from $\hat{T}_L^3$ that the isospin charges of the weak bosons are
\begin{equation}
 \hat{T}_L^3 \vert W^i \rangle = \pm1,0\vert W^i\rangle.
\end{equation}
In this relatively simple case, the SU$(2)_L$ generators (in the fundamental representation)
are proportional to the Pauli spin matrices (hence the label ``isospin''):
\begin{eqnarray}
\hat{T}_L^i = \frac{\sigma^i}{2}, &~& i=1,\dots,3, \quad \text{where} \quad 
\\
\sigma^1  =  \begin{pmatrix} 0 &  1 \\ 1  &  0 \end{pmatrix}, \quad
\sigma^2 &=& \begin{pmatrix} 0 & -i \\ i  &  0 \end{pmatrix}, \quad
\sigma^3  =  \begin{pmatrix} 1 &  0 \\ 0  & -1 \end{pmatrix}.
\label{pauliMatrices.EQ}
\end{eqnarray}
The structure constant 	 is the antisymmetric, three-dimensional Levi-Civita tensor $\epsilon_{ijk}$
\begin{equation}
  \left[W_{\mu},W_{\nu}\right]
= W_{\mu}^{j}W_{\nu}^{k}[T^{j},T^{k}] = W_{\mu}^{j}W_{\nu}^{k}\left[\frac{\sigma^{j}}{2},\frac{\sigma^{k}}{2}\right]
= W_{\mu}^{j}W_{\nu}^{k}~\frac{i\sigma^{i}}{2}\epsilon_{ijk} .
\end{equation}

Lastly, for the U$(1)$ weak hypercharge gauge boson $B_\mu$, which carries zero hypercharge, we have
\begin{eqnarray}
	F_{\mu\nu}	&=& \partial_{\mu}B_{\nu} - \partial_{\nu}B_{\mu}.
\end{eqnarray}
A special property of field strengths for Abelian gauge theories stems from its linear dependence on boson fields.
Namely, that linear transformations acting on a gauge field also hold for its field strength.
If we can express an Abelian gauge field by the following linear combination
\begin{equation}
 B_\mu = \sum_i c_i A_\mu^i,
\end{equation}
then we have that the field strength obeys the analogous linear decomposition:
\begin{equation}
 B_{\mu\nu} = \partial_\mu B_\nu - \partial_\nu B_\mu = \sum_i c_i \partial_\mu A_\nu^i - c_i \partial_\nu A_\mu^i = \sum_i c_i A_{\mu\nu}^i.
\end{equation}
Together, the isospin and hypercharge charge fields are the \textit{unbroken electroweak} gauge bosons.

The gauge boson content of the SM is summarized in the first three rows of Table~\ref{smBosonsBeforeEWSB.TB}.

\section{The Standard Model Higgs Sector and Electroweak Symmetry Breaking}\label{higgs.sm.sec}
The SM contains a single colorless, complex scalar field  $\Phi$ that is gauged under the electroweak sector.
Transforming as a doublet under SU(2)$_{L}$, i.e., under the fundamental representation, 
it possesses hypercharge $Y=+1$, mass $\mu$, and is expressible as
\begin{equation}
\Phi 
= 
\frac{1}{\sqrt{2}} \begin{pmatrix} \phi^1 + i\phi^2 \\  \phi^0 + i\phi^3 \end{pmatrix}
=
\begin{pmatrix} \phi^+ \\ \frac{1}{\sqrt{2}}(\phi^0 + i\phi^3) \end{pmatrix},
\quad
\phi^\pm \equiv \frac{1}{\sqrt{2}}(\phi^1 \pm i\phi^2). 
\label{higgsFieldDef.sm.EQ}
\end{equation}
From the requirements that it be charged under isospin (2 d.o.f.) and complex ($\times2$ d.o.f.), 
$\Phi$ actually comprises four real scalar fields $\phi^0, \dots, \phi^4.$
The fields $\phi^1,\phi^2$, are written as a linear combination for reasons that will become clear shortly.
It suffices for the moment to say  that the two isospin components of $\Phi$ separately respect the a second U$(1)$ gauge group with generator
\begin{equation}
 \hat{Q}\Phi = \left(\hat{T}_L^3 + \frac{1}{2}\hat{Y} \right)\Phi = \frac{1}{2}\left(\sigma^3 + \mathds{1}_2\right)\Phi
 = \begin{pmatrix} 1 & 0 \\ 0 & 0 \end{pmatrix} \Phi,
 \label{qedGenerator.sm.eq}
\end{equation}
indicating that $\phi^\pm$ have a charge $Q=\pm1$ and $\phi^0,\phi^1$ are charge zero.
In Eq.~(\ref{qedGenerator.sm.eq}), we expanded $\hat{T}^3_L$ and $\hat{Y}$ in SU$(2)_L$ space, 
in which case $\hat{Y}=\mathds{1}_2 Y_\Phi =(+1)\mathds{1}_2$.

Ignoring fermions, the most general Lagrangian at dimension-four we can write for $\Phi$ is
\begin{eqnarray}
\mathcal{L}_{\rm Higgs} = \left(D^\mu \Phi\right)^\dagger D_\mu\Phi - V(\Phi), \quad V(\Phi) = -\mu^2\Phi^\dagger\Phi +\lambda(\Phi^\dagger\Phi)^2,
\label{higgsLag.sm.EQ}
\end{eqnarray}
where the covariant derivative is given by
\begin{eqnarray}
 D_\mu\Phi	&=& \left[\partial_\mu + i g_W \hat{T}_L^{i}W_\mu^i 		+ i \frac{g_Y}{2} \hat{Y}B_\mu \right]\Phi \label{higgsCovD1.sm.EQ}
\\
		&=& \left[\partial_\mu + i \frac{g_W}{2}\sigma^i W_\mu^i 	+ i\frac{g_Y}{2} B_\mu \right]\Phi,
\end{eqnarray}
and $g_W~ (g_Y)$ denotes the weak isospin (hypercharge) coupling strength.
$\hat{Y}$ and $g_Y$ are normalized such that a factor of $1/2$ appears in the covariant derivative,
but it is sometimes absorbed in to the definition of $\hat{Y}$.
Though Eq.~(\ref{higgsCovD1.sm.EQ}) appears harmless, its present form does not show its utility.
Let us rewrite Eq.~(\ref{higgsCovD1.sm.EQ}) by taking advantage of familiar results of SU$(2)$ algebras. 
The raising and lowering ladder operators of SU$(2)$ are canonically given by
\begin{equation}
 \hat{T}^\pm_L = \hat{T}^1_L \pm i\hat{T}^2_L = \frac{1}{2}(\sigma^1 \pm i \sigma^2) 
 =
\left\{\begin{matrix}
\begin{pmatrix}0 & 1\\ 0 & 0\end{pmatrix} & \text{for~}+\\ 
\begin{pmatrix}0 & 0\\ 1 & 0\end{pmatrix} & \text{for~}-
\end{matrix}\right.
.
\end{equation}
We now write the SU$(2)_L$ gauge fields in terms of raising and lowering operators
\begin{eqnarray}
 \hat{T}_L^{i}W_\mu^i &=& \frac{1}{2}(\hat{T}_L^+ + \hat{T}_L^-)W^1_\mu - \frac{i}{2}(\hat{T}_L^+ - \hat{T}_L^-)W^2_\mu + W^3_\mu \hat{T}^3_L
 \\
 &=& \frac{1}{2}(W^1_\mu -i W^2_\mu)\hat{T}_L^+ + \frac{1}{2}(W^1_\mu + i W^2_\mu)\hat{T}_L^- + W^3_\mu \hat{T}^3_L,
\end{eqnarray}
and suggests the following linear field redefinitions
\begin{equation}
 \boxed{W_\mu^\pm = \frac{W_\mu^1 \mp i W_\mu^2}{\sqrt{2}} }.\label{wpwmDef.sm.EQ}
\end{equation}
In this form, the $W^+,W^-,W^3$ gauge bosons can be identified as increasing, decreasing, or leaving unchanged the weak isospin of a system that absorbs it.
Alternatively, the three respectively lower, raise, or leave unchanged isospin when radiated.
Equation~(\ref{higgsCovD1.sm.EQ}) becomes
\begin{equation}
  D_\mu\Phi	= \left[\partial_\mu 
  + \frac{i g_W}{\sqrt{2}}W_\mu^+\hat{T}_L^+ + \frac{i g_W}{\sqrt{2}}W_\mu^-\hat{T}_L^- 
  + i g_W W^3_\mu \hat{T}^3_L+ i \frac{g_Y}{2} \hat{Y}B_\mu \right]\Phi.
  \label{higgsCovD2.sm.eq}
\end{equation}
However, by decomposing $\hat{T}_L^a$ in this manner, Eq.~(\ref{higgsCovD2.sm.eq}) suggests an additional action: a redefinition of $W^3$ and $B$. 
As $W^3$ and $B$ share the same spacetime (massless, spin-1) and internal quantum numbers (zero isospin and hypercharge), they in principle can mix. 
This is very much like the gauge-mixing case we witnessed in Eq.~(\ref{gaugeMixing.sm.EQ}).

To construct the appropriate field redefinitions, 
we first recognize that this will require at most a $2\times2$ matrix, where the level of mixing is controlled by an angle $\theta_W$.
Expanding $\hat{T}_L^3$ and $\hat{Y}$ in SU$(2)_L$ space gives us such an object:
\begin{eqnarray}
 g_W W^3_\mu \hat{T}^3_L+  \frac{g_Y}{2} \hat{Y}B_\mu  &=& \frac{1}{2}
\begin{pmatrix}g_W W_\mu^3 + g_Y B_\mu  & 0 \\ 0 & -g_W W_\mu^3 + g_Y B_\mu \end{pmatrix}. 
\end{eqnarray}
Being diagonal, it is easy to read off the eigenstates, which we label $A_\mu$ and $Z_\mu$,
\begin{eqnarray}
 A_\mu &\sim& g_W W_\mu^3 + g_Y B_\mu \sim \sin\theta_W W_\mu^3 + \cos\theta_W B_\mu 
 \\
 Z_\mu &\sim& g_W W_\mu^3 - g_Y B_\mu \sim \cos\theta_W W_\mu^3 - \sin\theta_W B_\mu.
\end{eqnarray}
The interaction states then relate to gauge states $W^3_\mu,~B_\mu$ by the SU$(2)_L-$U$(1)_Y$ rotation
\begin{equation}
\boxed{
 \begin{pmatrix}   A_\mu \\ Z_\mu  \end{pmatrix}
  = 
 \begin{pmatrix}     \cos\theta_W & \sin\theta_W \\ -\sin\theta_W & \cos\theta_W   \end{pmatrix}
 \begin{pmatrix}   B_\mu \\ W_\mu^3  \end{pmatrix}}.
 \label{smGaugeMixing.sm.EQ}
\end{equation}
with coupling and mixing parameters defined by
\begin{equation}
\boxed{\sin\theta_W \equiv \frac{g_Y}{\sqrt{g_W^2 + g_Y^2}} \quad\text{and}\quad g_Z \equiv\sqrt{g_W^2 + g_Y^2} = \frac{g}{\cos\theta_W}}.
\end{equation}
Conventionally, we take the $g_Y\rightarrow0$ limit to be the decoupling regime where the SU$(2)_L$ and U$(1)_Y$ gauge bosons do not mix.
We now rewrite the last terms of $D_\mu\Phi$ in Eq.~(\ref{higgsCovD2.sm.eq}) as
\begin{eqnarray}
 g_W W^3_\mu \hat{T}^3_L+ \frac{g_Y}{2} \hat{Y}B_\mu &=& 
   g_W (\sin\theta_W A_\mu + \cos\theta_W Z_\mu)\hat{T}_L^3 
 + \frac{g_Y}{2} (\cos\theta_W A_\mu - \sin\theta_W Z_\mu)\hat{Y}
  \nonumber\\
 &=& g_W\sin\theta_W\underset{\hat{Q}\text{~of~Eq.}~(\ref{qedGenerator.sm.eq})}{\underbrace{(\hat{T}_L^3 + \frac{1}{2} \hat{Y})}}A_\mu 
   + \frac{g_W}{\cos\theta_W}(\cos^2\theta_W\hat{T}_L^3 - \sin^2\theta_W \underset{\hat{Q}-\hat{T}_L}{\underbrace{\frac{1	}{2}\hat{Y}}})Z_\mu
 \nonumber\\
 &=&
 e\hat{Q}A_\mu + g_Z (\hat{T}_L^3 - \sin^2\theta_W \hat{Q})Z_\mu,
\end{eqnarray}
where 
\begin{equation}
 \boxed{e \equiv g_W \sin\theta_W}.
\end{equation}
Finally, we have
\begin{equation}
   D_\mu	= \left[\partial_\mu 
  + \frac{i g_W}{\sqrt{2}}W_\mu^+\hat{T}_L^+ + \frac{i g_W}{\sqrt{2}}W_\mu^-\hat{T}_L^- 
  + ieA_\mu \hat{Q} + ig_Z Z_\mu (\hat{T}_L^3 - \sin^2\theta_W \hat{Q})\right].
  \label{higgsCovD3.sm.eq}
\end{equation}

Equation~(\ref{higgsCovD3.sm.eq}) is a rather dense expression, so we take time and explore the consequences of our successive field definitions.
The second and third terms, $W^+$ and $W^-$, 
being mixtures of pure SU$(2)_L$ fields that still carry $T_L^3=\pm1$ isospin, remain the gauge bosons of the local symmetry.
As previously mentioned, they transmute lower and upper components of $\Phi$ into each other. 
With the ladder operators $\hat{T}^\pm_L$, it is clear that terms proportional to 
\begin{equation}
 W^+ \phi^+ \quad\text{or}\quad W^-\phi^-,
\end{equation}
which would violate weak isospin charge conservation, do not exist.

The third term, $A_\mu \hat{Q}$, is interesting because we will eventually identify these objects as the photon field and the electric charge generator of QED. 
The term is also interesting because only half the Higgs doublet is aware of its existence;
see the discussion of ``Hidden Local Symmetries'' above Eq.~(\ref{hiddenSymm.sm.eq}).
As the sum of the hypercharge and third weak isospin generators, the (electric) charges $Q$ for the four $\phi^i$ fields are
\begin{eqnarray}
 Q_{\phi^{1,2}} &=& T_{L~\phi^{1,2}}^3 + \frac{1}{2}Y_{\phi^{0,3}} = \frac{+1}{2} + \frac{1}{2} = 1,
 \\
  Q_{\phi^{0,3}} &=& T_{L~\phi^{0,3}}^3 + \frac{1}{2}Y_{\phi^{0,3}} = \frac{-1}{2} + \frac{1}{2} = 0.
\end{eqnarray}
The relevant portion of the Higgs doublet's covariant derivate then simplifies to
\begin{eqnarray}
 D_\mu \Phi \ni \left(\partial_\mu + ieA_\mu \hat{Q}\right)\Phi 
 &=&   \partial_\mu\Phi + ie A_\mu
 \begin{pmatrix} 1 & 0 \\ 0 & 0 \end{pmatrix}
 \begin{pmatrix} \phi^+ \\ \frac{1}{\sqrt{2}}(\phi^0 + i\phi^3) \end{pmatrix} 
\\
 &\ni& 
 \left(\partial_\mu + ie A_\mu\right) \phi^+,
\end{eqnarray}
which, as we studied in Section~\ref{qed.sm.sec}, is the covariant derivative for scalar QED. 
However, as $W^\pm$ carry nonzero isospin but zero hypercharge, they too carry a net electric charge $Q_W = \pm1$.
These interaction terms are not present in $\mathcal{L}_{\rm Higgs}$
because they emerge after applying the field redefinitions $W^\pm_\mu$ in Eq.~(\ref{wpwmDef.sm.EQ}) and $Z_\mu,A_\mu$ in Eq.~(\ref{smGaugeMixing.sm.EQ})
to the $W^a_\mu$ and $B_\mu$ field strengths.

The last term of Eq.~(\ref{higgsCovD3.sm.eq}) is notable because its gives the appearance of predicting 
deviations from universal gauge couplings, even as the lower components of $\Phi$ have zero electromagnetic charge.
Of course, as $Z_\mu$ is neither an isospin or hypercharge gauge boson, gauge coupling universality is not actually violated.
It is enlightening to see the origin of slight coupling difference by considering the $\phi^0\phi^0 VV$ coupling for $V=W^\pm,Z$.
From the kinetic term in Eq.~(\ref{higgsLag.sm.EQ}), we have
\begin{eqnarray}
 \left(D_\mu\Phi\right)^\dagger D^\mu \Phi 
 &\ni&  
 \frac{g_W^2}{2}W_\mu^-W^{\mu+}\Phi^\dagger\left(\hat{T}_L^- \hat{T}_L^+\right)\Phi
 +
 g_Z^2 Z_\mu Z^\mu \Phi^\dagger \left(\hat{T}_L^3 - \sin^2\theta_W \hat{Q}\right)^2\Phi
 \\
 &\ni&
  \frac{g_W^2}{2}\left(\frac{1}{\sqrt{2}}\right)^2 W_\mu^-W^{\mu+}\phi^0\phi^0 
 +
 g_Z^2 \left(\frac{1}{2\sqrt{2}}\right)^2 Z_\mu Z^\mu  \phi^0\phi^0
\end{eqnarray}
\begin{figure}
\centering ~~
\includegraphics[width=0.7\textwidth]{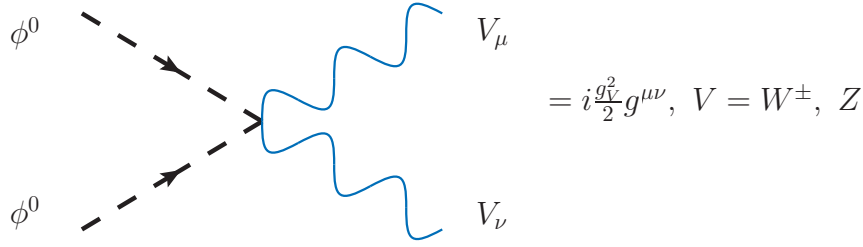} \vspace{0cm}
\caption{Feynman vertex rules for $\phi^0-\phi^0-V-V$, $V=W^\pm,~Z$, in the SM before EWSB.} 
\label{phiphiVV_Vertex.fig}
\end{figure}
Accounting for the symmetry factors from identical pairs $\phi^0\phi^0$ and $ZZ$,
the Feynman vertex rules for the four-point interactions, and shown in Fig.~\ref{phiphiVV_Vertex.fig}, are
\begin{eqnarray}
 \phi^0\phi^0 W^+W^- 	&:& ig{\phi\phi W W} = i\frac{g_W^2}{2} 
 \\
 \phi^0\phi^0 ZZ 	&:& ig{\phi\phi Z Z} = i\frac{g_Z^2}{2} =  i\frac{g_W^2}{2\cos^2\theta_W},
\end{eqnarray}
and become identical in the decoupling limit of $g_Y\rightarrow0$.
At leading order, we therefore have
\begin{equation}
 \lambda_{WZ} \overset{\rm Preliminarily}{=} 
 \frac{g_{\phi\phi W W }}{g_{\phi\phi Z Z}\cos^2\theta_W} = \frac{g_W^2}{g_Z^2\cos^2\theta_W} = 1,
\label{custodialTest1.sm.eq}
 \end{equation}
which is actually quite stable under radiative corrections~\cite{Willenbrock:2004hu}.
Equation (\ref{custodialTest1.sm.eq}) is very related to the notion of \textit{custodial symmetry}, and will be visited shortly.

The scalar boson content of the SM is summarized in the last row of Table~\ref{smBosonsBeforeEWSB.TB}.
We now turn our focus to the potential $V$ in $\mathcal{L}_{\rm Higgs}$ and the topic of \textit{electroweak {\rm(EW)} symmetry breaking} (EWSB).

\subsection{Electroweak Symmetry Breaking I: Massive Gauge Bosons}
From Section~\ref{ssbLocal.sm.sec}, we learned that if the $\Phi$ field  mass $\mu^2$ and the self-coupling $\lambda$ are both positive-definite, 
then its potential $V$,
\begin{equation}
 V(\Phi) = -\mu^2\Phi^\dagger\Phi +\lambda(\Phi^\dagger\Phi)^2,
\end{equation}
has a minimum at the origin and the Higgs field's ground state expectation value is zero.
However, for $\mu^2<0$ and positive $\lambda$, the minimum is away from the origin, leading to a nonzero vev,
triggering the Higgs Mechanism.
The EW symmetries under which the Higgs transforms are broken spontaneously, 
and the associated EW gauge bosons of these now-broken symmetries generate masses proportional to the size of the vev,
a process called \textit{electroweak {\rm(EW)} symmetry breaking} (EWSB).

We now apply the Higgs Mechanism to the SM and denote  the vev of $\Phi$ by
\begin{equation}
 v \equiv \sqrt{2}\langle \Phi \rangle = \sqrt{\frac{\vert \mu^2 \vert}{\lambda}}.
\end{equation}
Empirically, $v$ is measured from the muon lifetime~\cite{Halzen:1984mc}
\begin{equation}
\tau_\mu = \frac{1}{\Gamma_\mu} = \frac{192\pi^3}{G_F^2 m_\mu^2} \approx 2.2\time10^{-6}s, 
\end{equation}
where $G_F$ is Fermi's constant and
\begin{equation}
 \sqrt{\sqrt{2} G_F} = v \approx 246 ~\text{GeV}.
\end{equation}
Letting $\Phi$ settle at the minimum of its potential and take on the value of its vev, i.e., 
\begin{equation}
 \Phi =  \frac{1}{\sqrt{2}} \begin{pmatrix} 0 \\ v\end{pmatrix},
\end{equation}
its covariant derivate is then
\begin{eqnarray}
   D_\mu\Phi	= \Bigg	[
   \underset{\partial_\mu v=0}{\underbrace{\partial_\mu} }
&+& \frac{i g_W}{\sqrt{2}}W_\mu^+\hat{T}_L^+ 
+ \frac{i g_W}{\sqrt{2}}\underset{\hat{T}_L^-\Phi=0}{\underbrace{W_\mu^-\hat{T}_L^-} }
\nonumber\\
&+& ie\underset{\hat{Q}\Phi=0}{\underbrace{A_\mu\hat{Q}}}
+ ig_ZZ_\mu \left(\hat{T}_L^3 - \sin^2\theta_W \hat{Q}\right)\Bigg]
  \frac{v}{\sqrt{2}} \begin{pmatrix}0\\ 1\end{pmatrix}
  \\
  &=&
  \frac{iv}{\sqrt{2}} 
  \left[\frac{g_W}{\sqrt{2}}  W_\mu^+ \begin{pmatrix}1\\0\end{pmatrix}
             +\frac{g_Z}{2}Z_\mu \begin{pmatrix}0\\1\end{pmatrix} \right]   
  =
  \frac{iv}{2} \begin{pmatrix}g_W  W_\mu^+\\ \cfrac{-g_Z}{\sqrt{2}} Z_\mu\end{pmatrix}
\end{eqnarray}
The kinetic term of $\Phi$ at the bottom of the Higgs' potential then simplifies to 
\begin{eqnarray}
(D_\mu \Phi)^\dagger D^\mu\Phi &=& \frac{v^2}{2^2}
\begin{pmatrix}g_W  W_\mu^- &  \cfrac{-g_Z}{\sqrt{2}} Z_\mu\end{pmatrix}
\begin{pmatrix}g_W  W_\mu^+ \\ \cfrac{-g_Z}{\sqrt{2}} Z_\mu\end{pmatrix}
\\
&=& M_W^2 W_\mu^-W_\mu^+  +  \frac{1}{2}M_Z^2 Z_\mu Z^\mu
\label{wzMassLag.sm.EQ}
\end{eqnarray}
where we have define the mass terms
\begin{equation}
\boxed{ M_W^2 \equiv \frac{g_W^2 v^2}{4} \quad\text{and}\quad M_Z^2 \equiv \frac{g_Z^2v^2}{4} = \frac{(g_W^2 + g_Y^2)v^2}{4}}.
\end{equation}

\begin{table}[!t]
\caption{Bosons of the Standard Model After Electroweak Symmetry Breaking}
 \begin{center}
\begin{tabular}{|c|c|c|c|c|c|}
\hline\hline 
Name	& Gauge Interaction	 & Symbol  & Spin	& Mass [GeV]~\cite{Agashe:2014kda}	& $SU(3)_{C}\times U(1)_{EM}$ 
\tabularnewline\hline\hline
Gluon	& Strong		 & $G^{a}_{\mu}$	& 1	& 0 	     	& $(\textbf{3},0)$ \tabularnewline\hline
W	& Weak	     	 & $W^{\pm}_{\mu}$	& 1	& 80.385$\pm$ 0.015 	& $(1,\pm1)$ \tabularnewline\hline
Z	& Weak	     	 & $Z_{\mu}$		& 1	& 91.1876$\pm$0.0021 	& $(1,0)$ \tabularnewline\hline
Photon	& Electromagnetism & $A_{\mu}/\gamma$	& 1	& 0 		& $(1,0)$ \tabularnewline\hline
Higgs	& Yukawa (Not Gauged)	 & $h$		& 0	& $125.09\pm0.21$	 	& $(1,0)$ \tabularnewline\hline
\hline
\end{tabular}
\label{smBosonsAfterEWSB.TB}
\end{center}
\end{table}

The massive gauge bosons, $W^\pm$ and $Z$, have been measured at many experiments since their first direct production at CERN's
Super Proton Synchrotron by the UA1 and UA2 experiments in 1983. Presently, the world's best average for these masses
are~\cite{Agashe:2014kda}
\begin{equation}
 M_W^{\rm World~Avg.} = 80.385\pm 0.015~\text{GeV}
 \quad\text{and}\quad
 M_Z^{\rm World~Avg.} = 91.1876\pm0.0021~\text{GeV}.
\end{equation}
As tempted as we are to comment on the similarity of the masses, we continue with EWSB.

Counting degrees of freedom before EWSB, we had four fields from $\Phi$, three SU$(2)_L$ gauge fields, and one U$(1)_Y$ gauge field.
Since each (massless) gauge boson possess two transverse polarizations, this gives us 12 total degrees of freedom.
Presently, we have recovered only nine from the massive $W^\pm,~Z$ bosons (two transverse and one longitudinal polarization).
We saw in Section \ref{ssbLocal.sm.sec} that vector boson masses break gauge invariance, 
and thus $M_W$ an $M_Z$ in Eq.~(\ref{wzMassLag.sm.EQ}) ruin the generators
\begin{equation}
 \hat{T}^\pm, ~ \hat{T}^3_L - \sin^2\theta_W \hat{Q}.
\end{equation}
However, $\Phi$ does not carry a charge associated with generator
\begin{equation}
 \hat{Q} =  \hat{T}^3_L + \frac{1}{2}\hat{Y},
\end{equation}
and $A_\mu$ remains massless:
\begin{equation}
 \boxed{m_\gamma = 0}.
\end{equation}
Two more physical degrees of freedom are thus recovered as transverse polarizations.

The last physical state comes from fluctuations of $\Phi$ around $\langle\Phi\rangle$.
We define the field $h$ with a vanishing vev such that
\begin{equation}
 \Phi(x) \approx \frac{1}{\sqrt{2}} \begin{pmatrix} 0 \\ v + h(x)\end{pmatrix}, \quad \langle h(x)\rangle = 0. 
\end{equation}
Recall from Eq.~(\ref{qedGhost.sm.EQ}) that $\Im[\Phi(x)]$ around $v$ is an unphysical field that represents 
the ability of $A_\mu$ to make a gauge transformation.
The covariant derivate acting on $\Phi$ now takes the form
\begin{eqnarray}
   D_\mu\Phi  &=&
  \left[
  \cfrac{1}{\sqrt{2}} 		\begin{pmatrix}0\\\partial_\mu h\end{pmatrix}
  +i\cfrac{g_W}{2}  W_\mu^+ 	\begin{pmatrix}v+h\\0\end{pmatrix}
  +i\cfrac{g_Z}{2\sqrt{2}}Z_\mu \begin{pmatrix}0\\v+h\end{pmatrix} \right]   
  \\
  &=&	
  \begin{pmatrix}i \cfrac{g_W}{2}  W_\mu^+(v+h)\\ \cfrac{1}{\sqrt{2}}\partial_\mu h - i\cfrac{g_Z}{2\sqrt{2}} Z_\mu(v+h)\end{pmatrix}.
\end{eqnarray}
This gives rise to the three-point and four-point interaction terms
\begin{eqnarray}
 (D_\mu \Phi)^\dagger D^\mu\Phi &=& 
 \begin{pmatrix}\cfrac{-ig_W}{2}  W_\mu^- (v+h),& \cfrac{\partial_\mu h}{\sqrt{2}} + \cfrac{ig_Z}{\sqrt{2}} Z_\mu(v+h)\end{pmatrix}
 \begin{pmatrix}\cfrac{ ig_W}{2}  W^{\mu+}(v+h)\\ \cfrac{\partial^\mu h}{\sqrt{2}} - \cfrac{ig_Z}{2\sqrt{2}} Z^\mu(v+h)\end{pmatrix}
\nonumber\\
\\
&=& \frac{1}{2}\partial_\mu h\partial^\mu h + M_W^2 W_\mu^-W^{\mu+} + M_Z^2 Z_\mu Z^\mu
+ \underset{hVV~\text{Coupling}~\propto~M_V}{\underbrace{g_W M_W W_\mu^-W^{\mu+} h + g_Z M_Z Z_\mu Z^\mu h}}
\nonumber\\
& &\quad +\underset{hhVV~\text{Coupling}}{\underbrace{\cfrac{g_W^2}{4}W_\mu^-W^{\mu+}hh + \cfrac{g_Z^2}{8}Z_\mu Z^\mu hh}}.
\end{eqnarray}

Expending $\Phi$ in the potential $V$, we obtain the mass and self-interaction terms for $h$: 
\begin{eqnarray}
 V(\Phi) &=& -\mu^2\left(\frac{1}{\sqrt{2}}\right)^2 \begin{pmatrix}   0, &  v+h \end{pmatrix}\begin{pmatrix}   0, \\  v+h \end{pmatrix}
+
\lambda \left(\frac{1}{\sqrt{2}}\right)^2 \left[\begin{pmatrix}   0, &  v+h \end{pmatrix}\begin{pmatrix}   0, \\  v+h \end{pmatrix}\right]^2
\\
&=&
\frac{1}{2} m_H^2  h^2  + \sqrt{\frac{\lambda}{2}}m_H hhh + h^4 + \frac{\lambda}{4}hhhh + \frac{\lambda v^4}{2},
\end{eqnarray}
where the mass of the \textit{Higgs bosons}, $h$, is given by
\begin{equation}
 m_H \equiv  v \sqrt{2 \lambda} = \sqrt{2} \vert \mu \vert  
\end{equation}
Discovered only recently by the ATLAS and CMS experiments at the CERN's Large Hadron Collider~\cite{Aad:2012tfa,Chatrchyan:2012ufa}, 
the discovery of $h$ represent the completion of the SM as its last unknown parameter.
Presently, the best combination measurement of the Higgs mass is~\cite{Aad:2015zhl}
\begin{equation}
 m_H^{\rm ATLAS+CMS} = 125.09 \pm 0.21 ~(\text{stat.}) \pm 0.11 (\text{syst.}).
\end{equation}
Direct measurements of the Higgs self-coupling have not been achieved at the time of this writing.
Taking the central value of $m_H$, it is predicted to be
\begin{equation}
\lambda = \frac{m_H^2}{2 v^2} = \frac{1}{\sqrt{2}}m_H^2 G_F \approx  0.129 \approx \frac{1}{8}.
\end{equation}
The elementary boson content of the SM after EWSB is summarized in Table~\ref{smBosonsAfterEWSB.TB}.

Before introducing the fermionic content of the SM, we return to the similarity of $M_W$ and $M_Z$.		

\begin{figure}
\centering ~~
\includegraphics[width=0.9\textwidth]{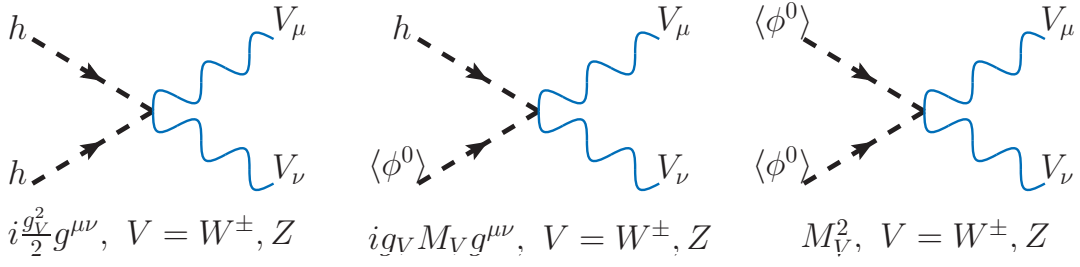} \vspace{0cm}
\caption[Feynman rules for $h-h-V-V$, $h-V-V$, $V=W^\pm,~Z$ after EWSB.]{Feynman vertex rules for $h-h-V-V$ (L) and $h-V-V$ (C), $V=W^\pm,~Z$, in the SM after EWSB.
(R) $W/Z$ mass.} 
\label{hhVV_hVV_Vertex.fig}
\end{figure}

\subsection{Custodial Symmetry}
Recalling the definitions $M_W$ and $M_Z$, we have
\begin{eqnarray}
  M_W^2 &\equiv& \frac{g_W^2 v^2}{4} 
  \\
  M_Z^2 &\equiv& \frac{g_Z^2v^2}{4} = \frac{(g_W^2 + g_Y^2)v^2}{4},
\end{eqnarray}
which means that the mass ratio of the two is a measure of how much, or how little,
the isospin and hypercharge groups rotate into each other during EWSB:
\begin{equation}
 \frac{M_W^2}{M_Z^2} = \frac{g_W^2}{g_W^2 + g_Y^2} = \cos\theta_W^2.
\end{equation}
In the zero mixing limit, which arises from either $Y_\Phi=0$ or a negligibly small $g_Y$, the $W$ and $Z$ bosons masses converge.
From this, the observable $\rho$, also called the $\rho$-parameter, can be constructed.
At tree-level in the SM, the $\rho$-parameter is defined to be
\begin{eqnarray}
 \rho \equiv \cfrac{M_W^2}{M_Z^2 \cos^2\theta_W} = 1,
 \label{rhoDef.sm.EQ}
\end{eqnarray}
changes very little under radiative corrections~\cite{Willenbrock:2004hu}.
This stability is due to \textit{custodial symmetry}.
In the SM, the Higgs field obeys an approximate global SU$(2)_L\times$SU$(2)_R$ symmetry, and is exact in the zero hypercharge limit.
After EWSB, the (approximate) left-right symmetry breaks down to an (approximate) vector symmetry, i.e.,
\begin{equation}
 \text{SU}(2)_L\times\text{SU}(2)_R \rightarrow \text{SU}(2)_V,
\end{equation}
thereby ensuring a near mass degeneracy among the gauge bosons.

However, it is straightforward to see how drastically $\rho$ can change in the presence of new scalars participating in EWSB.
Supposing it were the case that many complex scalars, all gauged under SU$(2)_L\times$U$(1)_Y$, acquire various vevs.
For such a scalar $\Phi_i$ in weak isospin representation $\hat{T}_i$ with weak isospin charge $T_i$,
only its electrically neutral component, $\phi^0$ can acquire a nonzero vev $v_i \equiv \langle \phi_i \rangle$ in order to preserve electromagnetism. 
As the electric charge is given by $Q_\phi = T_i^3 + Y_i/2$, it goes to show that $Y_i= -2T_i^3$ for each participating scalar, 
where $T^3_i$ is the isospin of the electrically neutral component $\phi^0_i$.
For reference, in the SM, the Higgs field $\Phi$ is an SU$(2)_L$ double with isospin charge $T = 1/2$; 
its electrically neutral component $\phi^0$ has isospin $T^3 = -1/2$ and vev
\begin{equation}
 v =  \sqrt{2}\langle\Phi\rangle =  \langle\phi^0\rangle,
\end{equation}
where
\begin{equation}
 \Phi = \begin{pmatrix}         \phi^+ \\ \frac{1}{\sqrt{2}}\left(\phi^0 + i\phi^3\right)        \end{pmatrix}.
\end{equation}

The covariant derivate acting on a generic $\Phi_i$ at its minimum is then
\begin{eqnarray}
D_\mu \Phi_i \rightarrow D_\mu v_i &=& \left[  \frac{i g_W}{\sqrt{2}}\left(W_\mu^+\hat{T}_i^+ + W_\mu^-\hat{T}_i^- \right)
  + ig_Z Z_\mu \hat{T}_i^3 \right] v_i,
\end{eqnarray}
which implies a kinetic term
\begin{eqnarray}
(D^\mu \Phi_i)^\dagger(D_\mu \Phi_i)\rightarrow (D^\mu v_i)^\dagger(D_\mu v_i) &=&
\left[  \frac{g_W^2}{2}W^{\mu -}W_\mu^+\left(\hat{T}_i^-\hat{T}_i^+ + \hat{T}_i^+\hat{T}_i^- \right)
+ g_Z^2   \left(\hat{T}_i^3\right)^2 Z^\mu Z_\mu \right]v_i^2
\nonumber
\end{eqnarray}
Following the usual ladder operator algebra, we have
\begin{eqnarray}
 \hat{T}_i^\pm\hat{T}_i^\mp\vert v_i\rangle &=& \hat{T}_i^\pm\hat{T}_i^\mp \vert T_i, T_i^3\rangle
 \\
 &=&  \sqrt{T_i(T_i+1) - T_i^3(T_i^3\mp1)}\hat{T}_i^\pm \vert T_i, T_i^3\mp1\rangle
 \\
 &=& \sqrt{T_i(T_i+1) - T_i^3(T_i^3\mp1)} \sqrt{T_i(T_i+1) - (T_i^3\mp1)T_i^3}\vert T_i, T_i^3\rangle
  \\
 &=& \left[T_i(T_i+1) - T_i^3(T_i^3\mp1)\right] \vert T_i, T_i^3\rangle,
\end{eqnarray}
indicating that for the $W$ boson, we have
\begin{equation}
 \left(\hat{T}_i^-\hat{T}_i^+ + \hat{T}_i^+\hat{T}_i^- \right)v_i = 2\left[T_i(T_i+1) - (T_i^3)^2\right]v_i.
\end{equation}
Similarly, 
\begin{equation}
 (\hat{T}_i^3)^2 v_i = (T_i^3)^2 v_i.
\end{equation}
The kinetic term for $\phi$ now simplifies to
\begin{eqnarray}
 (D^\mu v_i)^\dagger(D_\mu v_i) &=&
\left[g_W^2v_i^2\left(T_i(T_i+1) - (T_i^3)^2\right)W^{\mu -}W_\mu^+ + g_Z^2 v_i  (T_i^3)^2  Z^\mu Z_\mu \right].
\end{eqnarray}
Summing over all vev-acquiring fields $\Phi_i$, the total kinetic term gives
\begin{eqnarray}
 \mathcal{L}_{\rm Kinetic} &=& \sum_i (D^\mu \Phi_i)^\dagger(D_\mu \Phi_i) \ni \sum_i (D^\mu \phi^0_i)^\dagger(D_\mu \phi^0_i)
 \rightarrow \sum_i (D^\mu v_i)^\dagger(D_\mu v_i)
 \\
 &=&  \underset{M_W^2}{\underbrace{\sum_i\left[g_W^2v_i^2(T_i(T_i+1) - (T_i^3)^2)\right]}}W^{\mu -}W_\mu^+ 
 + 
 \frac{1}{2}\underset{M_Z^2}{\underbrace{2\sum_i \left[g_Z^2 v_i^2  (T_i^3)^2\right]}}  Z^\mu Z_\mu,
\end{eqnarray}
giving us expressions for $M_W$ and $M_Z$ in terms of the various $v_i$ and isospins
\begin{eqnarray}
 M_W^2 &=& \sum_i g_W^2v_i^2\left(T_i(T_i+1) - (T_i^3)^2\right),
 \\
 M_Z^2 &=& 2\sum_i g_Z^2 v_i^2  (T_i^3)^2.
\end{eqnarray}
The $\rho$-parameter for an arbitrary number of Higgs doublets is then given by
\begin{eqnarray}
 \rho \equiv \cfrac{M_W^2}{M_Z^2 \cos^2\theta_W} &=& 
 \cfrac{\sum_i\left[g_W^2v_i^2(T_i(T_i+1) - (T_i^3)^2)\right]}{2\sum_i \left[g_Z^2 v_i  (T_i^3)^2\right]\cos^2\theta_W}
 \\
 &=&
  \cfrac{\sum_i v_i^2\left[ T_i(T_i+1) - (T_i^3)^2\right]}{2\sum_i v_i^2  (T_i^3)^2}
\end{eqnarray}

As $M_W$, $M_Z$, and $\cos\theta_W$ in Eq.~(\ref{rhoDef.sm.EQ}) can be measured independently, 
$\rho$ represents a high-precision into the EW sector and the origin of EWSB.
Accounting for smalls radiative corrections, labeled by $\hat{\rho}$, the best measurement for $\rho$ is given by~\cite{Agashe:2014kda}
\begin{equation}
 \rho_0 \equiv \frac{\rho}{\hat{\rho}} = \frac{M_W^2}{M_Z^2 \cos^2\theta_W \hat{\rho}} = 1.00040\pm 0.00024,
\end{equation}
and is consistent with the SM at $1.67\sigma$.
An proxy test of custodial symmetry is measuring the branching fraction ratios of Higgs boson decays to weak bosons.

As we saw in Eq.~(\ref{custodialTest1.sm.eq}) as well as in Fig.~\ref{hhVV_hVV_Vertex.fig}, 
the three-point $hVV$ and four-point $hhVV$ couplings are proportional to the amount of mixing between the isospin and hypercharge bosons.
In the vanishing $g_Y$ limit, the two couplings for $WW$ and $ZZ$ become identical.
\begin{eqnarray}
 \lambda_{WZ} &=& 
 \cfrac{\text{BR}\left(H\rightarrow WW\right)}{\text{BR}^{\rm SM}\left(H\rightarrow WW\right)}
 \times
 \cfrac{\text{BR}^{\rm SM}\left(H\rightarrow ZZ\right)}{\text{BR}\left(H\rightarrow ZZ\right)}
 \sim   \cfrac{g_{hWW}^2}{g_{hWW}^{2~\rm SM}} \times\cfrac{g_{hZZ}^{2~\rm SM}}{g_{hZZ}^2} 
\end{eqnarray}
Measurements by ATLAS~\cite{Aad:2013wqa} and CMS~\cite{Khachatryan:2014jba} find it consistent with SM prediction of $1$
\begin{eqnarray}
 \lambda^{\rm ATLAS}_{WZ} 	&=& 0.81^{+0.16}_{-0.15},
 \\
 \lambda^{\rm CMS}_{WZ} 	&=& 0.94^{0.22}_{-0.18}.
\end{eqnarray}

\section{Fermion Sector of the Standard Model}\label{matter.sm.sec}
The SM is a theory of massless, chiral fermion that are coupled through Yukawa interactions and interact via the exchange of gauge bosons. 
We now introduce the SM fermionic sector, their gauge and Yukawa interactions, and their spontaneous generation of mass.

\subsection{Fermion Content}
We first denote the LH (RH) components of a Dirac fermion $\psi$ by
\begin{equation}
 \psi_{L(R)} \equiv P_{L(R)} \psi \equiv \frac{1}{2} (1_4 \mp \gamma^5)\psi,
\end{equation}
where $P_{L(R)}$ is the chiral projection operator, and under charge conjugation one has
\begin{equation}
\psi_L^c \equiv (\psi^c)_L = (\psi_R)^c.
\end{equation}
All known (anti)fermionic states that are gauged under a non-Abelian group are charged in a (anti)fundamental representation of the gauge group.
The absoluteness of this statement is of much interest and speculation.

The fermionic content of the SM consists of the LH states
\begin{eqnarray}
Q_L^{\alpha I} =  \begin{pmatrix} u_L^{\alpha I} \\ d_{L}^{\alpha I}\end{pmatrix}, 
 \quad
L_L^{I} =  \begin{pmatrix} \nu_L^{I} \\ e_{L}^{I}\end{pmatrix},
\label{lhDoublets.sm.eq}
\end{eqnarray}
and the RH states
\begin{eqnarray}
u_R^{\alpha I}, \quad d_{R}^{\alpha I}, \quad  e_{R}^{I}.
\end{eqnarray}
The LH objects are arranged to make manifest that they satisfy an SU$(2)$ (weak isospin) gauge symmetry.
The lowercase Greek index $\alpha=1,\dots,N_c=3$ denote SU$(3)$ (color) indices.
The capital Roman index $I=1,\dots3$ represent that there are three copies of these fields called \textit{generations},
or sometimes \textit{families}.
The ordering is such that generation-$n$ fields have smaller Yukawa couplings (masses) than generation-$(n+1)$ fields.
Despite the wide body of literature, and despite its suggestive structure, presently there is no confirmed ``theory of generations''. 
Measurements of Higgs boson properties indicate that additional generations, 
if they exist and obtain their from the Higgs fields, must be very massive~\cite{Chatrchyan:2013sfs}.
For a fixed generation, each of the seven fields posses a unique charge under U$(1)_Y$ hypercharge.
However, for a fixed generation, the sum of all hypercharges is identically zero, thereby rendering it ``anomaly free''~\cite{Peskin:1995ev}.
As gauge  quantum number assignments are independent of generation, this cancellation holds for each generation.

\begin{table}[!t]
\caption{Matter Content of the Standard Model}
\begin{center}
\begin{tabular}{|c|c|c|c|c|}
\hline\hline
Species & Symbol  & SU$(3)_{C}\times \text{SU}(2)_{L}\times \text{U}(1)_{Y}$ Rep. & U$(1)_{EM}$ Charge [Units of $e>0$] \tabularnewline\hline\hline
Quark   & $Q_L =  \begin{pmatrix} u_L \\ d_{L} \end{pmatrix}$ & $(\textbf{3},\textbf{2},+\frac{1}{3})$ &     
$\begin{pmatrix} +2/3 \\ -1/3 \end{pmatrix}$
\tabularnewline\hline
Quark   & $u_{R}$ & $(\textbf{3},\textbf{1},+\frac{4}{3})$    &	$+2/3$   \tabularnewline\hline
Quark   & $d_{R}$ & $(\textbf{3},\textbf{1},-\frac{2}{3})$    &	$-1/3$   \tabularnewline\hline
Lepton  & $L_L =  \begin{pmatrix} \nu_L \\ e_{L}\end{pmatrix}$ & $(\textbf{1},\textbf{2},-1)$   &
$\begin{pmatrix} 0 \\ -1 \end{pmatrix}$
\tabularnewline\hline
Lepton  & $e_{R}$ & $(\textbf{1},\textbf{2},-2)$  & $-1$   \tabularnewline\hline
\hline
\end{tabular}
\label{smMatter.TB}
\end{center}
\end{table}

The SU$(3)$-colored fields $Q_L$, $u_R$, $d_R$ are called \textit{quarks},
and the SU$(3)$-neutral fermions $L_L$, $e_R$ are the \textit{leptons}.
Leptons are further categorized into (electrically) \textit{charged leptons}
\begin{equation}
 e^I_L, \quad e^I_R,
\end{equation}
and (electrically) \textit{neutral leptons} or \textit{neutrinos}
\begin{equation}
 \nu^I_L.
\end{equation}
Though not used there, the notation
\begin{equation}
q_L^{\alpha i I}, \quad \ell_L^{i I},
\end{equation}
is very often found to denote LH quark and lepton doublets with  SU$(2)_L$ index $i=1,~2$.

The identity of each of the four LH fields in Eq.~(\ref{lhDoublets.sm.eq}) is referred to as \textit{flavor}.
Accounting for three generations, there are 12 flavors in total.
The RH analogs of Eq.~(\ref{lhDoublets.sm.eq}), if they exist, have the same flavor, e.g.,
$e_L^{I=1}$ is the LH electron and $e_R^{I=1}$ is the RH electron.
When speaking of a particular particle species across generations, the qualifier \textit{type} is used,
e.g., a $u$-type or up-type quark represents 
\begin{equation}
 u_L^{\alpha,I}, ~u_{R}^{\alpha I}\quad\text{for}\quad I=1,\dots,3.
\end{equation}
The SM elementary fermion content is summarized in Table~\ref{smMatter.TB}.

There are no RH neutrinos, $N_R^I$, in the SM.

\subsection{Fermion Lagrangian}
The last piece of the SM Lagrangian [Eq.~(\ref{smLagFull.sm.eq})] is the fermionic contribution, given by
\begin{eqnarray}
 \mathcal{L}_{\rm Fermion} &=& \mathcal{L}_{\rm Fermion~Kin.} - V_{\rm Yukawa}
 \label{smFermionLag.sm.eq}
\end{eqnarray}
where the kinetic term is given by
\begin{eqnarray}
 \mathcal{L}_{\rm Fermion~Kin.} &=&
 \overline{Q}_L^{\beta I}i\not\!\!D^{\beta\alpha} Q_L^{\alpha I} +  \overline{L}_L^{I}i\not\!\!D L_L^{I} 
\nonumber\\
& & \quad + 
\overline{u}_R^{\beta I}i\not\!\!D^{\beta\alpha} u_R^{\alpha I} + 
\overline{d}_R^{\beta I}i\not\!\!D^{\beta\alpha} d_R^{\alpha I} +
\overline{e}_R^{I}i\not\!\!D e_R^{I}
\end{eqnarray}
and Yukawa potential by
\begin{eqnarray}
V_{\rm Yukawa} &=&
  y_{u}^{JI}\overline{Q}_{L}^{\alpha J}	\tilde{\Phi}	u_{R}^{\alpha I} +
  y_{d}^{JI}\overline{Q}_{L}^{\alpha J}	\Phi		d_{R}^{\alpha I} +
  y_{e}^{JI}\overline{L}_{L}^{J}	\tilde{\Phi}	e_{R}^{I} + 
  \text{H.c.}
  \label{fermionYukawaLag.sm.EQ}
\end{eqnarray}
We unpack Eq.~(\ref{smFermionLag.sm.eq}) by first listing the covariant derivatives explicitly using Eq.~(\ref{higgsCovD3.sm.eq}):
\begin{eqnarray}
\not\!\!D^{\beta\alpha} Q_L^\alpha &=&  \Bigg[ \delta^{\beta\alpha}\partial_\mu  + ig_s  G^a_\mu (\hat{T}^a)^{\beta\alpha}
+ \delta^{\beta\alpha}\frac{i g_W}{\sqrt{2}}\left(W_\mu^+\hat{T}_L^+ +W_\mu^-\hat{T}_L^- \right)
  \nonumber\\
  & & \qquad + \delta^{\beta\alpha}ieA_\mu \hat{Q} 
  + \delta^{\beta\alpha}ig_Z Z_\mu \left(\hat{T}_L^3 - \sin^2\theta_W \hat{Q}\right)\Bigg] \gamma^\mu Q_L^\alpha
\nonumber\\
\not\!\!D L_L &=& \left[ \partial_\mu  + \frac{i g_W}{\sqrt{2}}\left(W_\mu^+\hat{T}_L^+ +W_\mu^-\hat{T}_L^- \right)
 + ieA_\mu \hat{Q} + ig_Z Z_\mu \left(\hat{T}_L^3 - \sin^2\theta_W \hat{Q}\right)\right] \gamma^\mu L_L
\nonumber\\
\not\!\!D^{\beta\alpha} u_R^{\alpha} &=&  \left[ \delta^{\beta\alpha}\partial_\mu  + ig_s  G^a_\mu (\hat{T}^a)^{\beta\alpha}
+ \delta^{\beta\alpha}ieA_\mu \hat{Q} - \delta^{\beta\alpha}ig_Z \sin^2\theta_W Z_\mu \hat{Q} \right] \gamma^\mu u_R^{\alpha}
\nonumber\\
\not\!\!D^{\beta\alpha} d_R^{\alpha} &=& \left[ \delta^{\beta\alpha}\partial_\mu  + ig_s  G^a_\mu (\hat{T}^a)^{\beta\alpha}
+ \delta^{\beta\alpha}ieA_\mu \hat{Q} - \delta^{\beta\alpha}ig_Z \sin^2\theta_W Z_\mu \hat{Q} \right] \gamma^\mu d_R^{\alpha}
\nonumber\\
\not\!\!D e_R &=& \left[ \partial_\mu  + ieA_\mu \hat{Q} - ig_Z \sin^2\theta_W  Z_\mu  \hat{Q} \right] \gamma^\mu e_R
 \nonumber
\end{eqnarray}
The index $a=1,\dots (N_c^2 - 1) = 8$ denotes the SU$(3)$ color generator (in the adjoint representation).
In the Yukaway potential, $\Phi$ is the scalar doublet introduced in Section~\ref{higgs.sm.sec}.
The field $\tilde{\Phi}$ is its ``isospin-hypercharge'' conjugate, defined by
\begin{eqnarray}
 \tilde{\Phi} \equiv i\sigma^{2}\Phi^{*} = 
 \begin{pmatrix}0 & 1 \\ -1 & 0 \end{pmatrix}
\begin{pmatrix} \phi^+ \\ \frac{1}{\sqrt{2}}(\phi^0 + i\phi^3) \end{pmatrix}
=
\begin{pmatrix} \frac{1}{\sqrt{2}}(\phi^0 - i\phi^3) \\ -\phi^-  \end{pmatrix}.
\end{eqnarray}
$y_f^{JI}$ are the $3\times3$ Yukawa coupling matrices for a fermion species $f$ in generation space, e.g., $b$- or $t$-type quarks.
Equation~(\ref{smFermionLag.sm.eq}) will be explored in considerable depth throughout the remaining chapters.
For now, we focus on how $\mathcal{L}_{\rm Fermion}$ changes as $\Phi$ acquires its vev.

\subsection{Electroweak Symmetry Breaking II: Fermion Masses}
To break EW symmetry in the fermion sector with the Higgs field, we follow the (by now) standard procedure of 
setting $\Phi$ equal to its vev and considering perturbative fluctuations $(h)$ around it.
We consider the up-type interaction as an example and see
\begin{eqnarray}
   y_{u}^{JI}\overline{Q}_{L}^{\alpha J}\tilde{\Phi}u_{R}^{\alpha I} 
   &=&
   y_{u}^{JI}
   \begin{pmatrix}    \overline{u}_L^{\alpha J} & \overline{d}_L^{\alpha J}    \end{pmatrix}
   \begin{pmatrix} \frac{1}{\sqrt{2}}(v+h) \\ 0  \end{pmatrix}u_{R}^{\alpha I}
   = 
\frac{y_{u}^{JI}}{\sqrt{2}} \overline{u}_L^{\alpha J} (v+h)u_{R}^{\alpha I}.
\label{fermionEWSB.sm.eq}
\end{eqnarray}
In the last term of Eq.~(\ref{fermionEWSB.sm.eq}) we see two very interesting terms: 
(i) a three-point point coupling between left- and right-handed fields of the same species type mediated by a Higgs radiation, and 
(ii) a two-point coupling between  left- and right-handed fields proportional to a dimensionful parameter.
The second term should be identified as a  \textit{fermion mass} term that has been spontaneously by the Higgs field.
Making the definition 
\begin{equation}
 \boxed{m_f^{JI} \equiv y_{f}^{JI}\langle \Phi \rangle  = \frac{y_{f}^{JI}}{\sqrt{2}}v },
\end{equation}
we now have
\begin{eqnarray}
   y_{u}^{JI}\overline{Q}_{L}^{\alpha J}\tilde{\Phi}u_{R}^{\alpha I} + \text{H.c} = 
   m_u^{JI} \overline{u}_L^{\alpha J}u_{R}^{\alpha I} + \frac{m_u^{JI}}{v} \overline{u}_L^{\alpha J}u_{R}^{\alpha I} h + \text{H.c}.
\end{eqnarray}
Applying this systematically, we obtain masses for all SM fermions with RH partners:
\begin{eqnarray}
 V_{\rm Yukawa} &=&   
 m_u^{JI} \overline{u}_L^{\alpha J}	u_{R}^{\alpha I} + 
 m_d^{JI} \overline{d}_L^{\alpha J}	d_{R}^{\alpha I} +
 m_e^{JI} \overline{e}_L^{J}		e_{R}^{I}
 \nonumber\\
 &+&
 \frac{m_u^{JI}}{v} \overline{u}_L^{\alpha J}	u_{R}^{\alpha I} h +
 \frac{m_d^{JI}}{v} \overline{d}_L^{\alpha J}	d_{R}^{\alpha I} h +
 \frac{m_e^{JI}}{v} \overline{e}_L^{J}		e_{R}^{I} h  + \text{H.c}
 \label{yukawaAfterEWSB.sm.eq}
\end{eqnarray}

Having broken EW symmetry in the fermion sector, we find ourself at another interesting junction. 
Our fermion Lagrangian $\mathcal{L}_{\rm Fermion}$ was written in terms of massless chiral/gauge eigenstates. 
However, as the broken Lagrangian now only respects
\begin{equation}
 \text{SU}(3)_c \times \text{U}(1)_{EM},
\end{equation}
we may conclude that our massive fermion states as they are presently written may no longer be aligned with their mass eigenstates.

\subsection{Quark and Lepton Mass Mixing}
Like in the gauge sector, fermionic gauge states and mass eigenstates before EWSB were aligned.
We no long have this luxury and must rotate our states out of the gauge basis in order to obtain mass eigenstates,
which are necessary to discuss particle scattering.

Generically, we may  decompose our LH and RH chiral fields into mass eigenstates with a unitary transformation:
\begin{eqnarray}
\underset{\rm Chiral~Basis}{\underbrace{\begin{pmatrix}u^1\\u^2 \\ u^3\end{pmatrix}_{L(R)}}}
=
U_{L(R)}
\underset{\rm Mass~Basis}{\underbrace{\begin{pmatrix}u\\c \\ t\end{pmatrix}_{L(R)}}},
&\quad&
\underset{\rm Chiral~Basis}{\underbrace{\begin{pmatrix}d^1\\d^2 \\ d^3\end{pmatrix}_{L(R)}}}
=
D_{L(R)}
\underset{\rm Mass~Basis}{\underbrace{\begin{pmatrix}d\\s \\ b\end{pmatrix}_{L(R)}}},
\\
\underset{\rm Chiral~Basis}{\underbrace{\begin{pmatrix}e^1\\e^2 \\ e^3\end{pmatrix}_{L(R)}}}
&=&
E_{L(R)}
\underset{\rm Mass~Basis}{\underbrace{\begin{pmatrix}e\\ \mu \\ \tau\end{pmatrix}_{L(R)}}},
\end{eqnarray}
Our mass and Yukawa matrices $m_f^{JI}$ and $y_f^{JI}$ can now be diagonalized
\begin{eqnarray}
 \mathcal{M}_u &=& U^{-1}_L m_u U_R = 
\begin{pmatrix} m_u & 0 & 0 \\ 0 & m_c & 0 \\ 0 & 0 & m_t \end{pmatrix}
= \frac{v}{\sqrt{2}}
\begin{pmatrix} y_u & 0 & 0 \\ 0 & y_c & 0 \\ 0 & 0 & y_t \end{pmatrix}
\\
 \mathcal{M}_d &=& D^{-1}_L m_d D_R = 
\begin{pmatrix} m_d & 0 & 0 \\ 0 & m_s & 0 \\ 0 & 0 & m_b \end{pmatrix}
= \frac{v}{\sqrt{2}}
\begin{pmatrix} y_d & 0 & 0 \\ 0 & y_s & 0 \\ 0 & 0 & y_b \end{pmatrix}
\\
 \mathcal{M}_e &=& E^{-1}_L m_e E_R = 
\begin{pmatrix} m_e & 0 & 0 \\ 0 & m_\mu & 0 \\ 0 & 0 & m_\tau \end{pmatrix}
= \frac{v}{\sqrt{2}}
\begin{pmatrix} y_e & 0 & 0 \\ 0 & y_\mu & 0 \\ 0 & 0 & y_\tau \end{pmatrix}.
\end{eqnarray}
This allows us to rewrite the Yukawa interactions and mass terms of Eq.~(\ref{yukawaAfterEWSB.sm.eq}) compactly as
\begin{eqnarray}
 V_{\rm Yukawa} &=&   
 m_f  \overline{f}_L	f_{R} +  \frac{m_f}{v} \overline{f}_L f_{R} h + \text{H.c}, \quad m_f = \frac{y_f v}{\sqrt{2}},
 \label{yukawaAfterMixing.sm.eq}
\end{eqnarray}
for $f=u,~d,~c,~s,~t,~b,~e,~\mu,~\tau$, and repeated color indices are implicit for quarks.
In Table~\ref{smMatterMasses.TB}, we summarize the SM fermion content after EWSB, in the mass eigenbasis, 
along with the most precise measurements of their masses presently available.

\begin{table}[!t]
\caption[Masses of the Elementary Standard Model Fermions]{Masses of the Elementary Standard Model Fermions~\cite{Agashe:2014kda}}
 \begin{center}
\begin{tabular}{|c|c|c|c|c|c|}
\hline\hline
\multicolumn{6}{|c|}{ Quark Generation}\tabularnewline\hline
\multicolumn{2}{|c|}{I} & \multicolumn{2}{|c|}{II} & \multicolumn{2}{|c|}{III} \tabularnewline\hline\hline
Species 	& Mass [MeV]		& Species	& Mass [GeV]		& Species	& Mass [GeV] \tabularnewline\hline
Up $(u)$	& $2.3^{+0.7}_{-0.5}$ 	& Charm $(c)$	& $1.275\pm0.025$ 	& Top $(t)$	& $173.21\pm0.51$ \tabularnewline\hline
Down $(d)$ 	& $4.8^{+0.5}_{-0.3}$  	& Strange $(s)$	& $0.95\pm0.05$  	& Bottom $(b)$  & $4.18\pm0.03$  \tabularnewline\hline
\hline
\multicolumn{6}{|c|}{ Lepton Generation}\tabularnewline\hline
\multicolumn{2}{|c|}{I} & \multicolumn{2}{|c|}{II} & \multicolumn{2}{|c|}{III} \tabularnewline\hline\hline
Species 	& Mass 		& Species	& Mass 			& Species	& Mass  \tabularnewline\hline
Electron-	& \multirow{2}{*}{$<2$ eV} & 
Muon-		& \multirow{2}{*}{$<0.19$ MeV} & 
Tau- 		& \multirow{2}{*}{$<18.2$ MeV}\tabularnewline
Neutrino $(\nu_e)$ & & Neutrino $(\nu_\mu)$ & & Neutrino $(\nu_\tau)$ & \tabularnewline\hline
\multirow{2}{*}{Electron $(e)$}	& $510.998928\pm$ 	& 
\multirow{2}{*}{Muon $(\mu)$} 	& $105.6583715\pm$ & 
\multirow{2}{*}{Tau $(\tau)$} 	& $1.77682\pm$ 	 \tabularnewline
				& $0.000011$ 	KeV &  
				& $0.0000035$	MeV &  
				& $0.00016$	GeV\tabularnewline\hline
\hline

\end{tabular}
\label{smMatterMasses.TB}
\end{center}
\end{table}

At this point, it is worth noting that as the Higgs-vector boson couplings originate from the Higgs kinetic term
and as Higgs-fermion couplings originate from the Yukawa potential,
which are subtracted from kinetic terms in the Lagrangian formalism.
Thus, the $hVV$ and $hff$ couplings differ by a relative minus sign.
An analysis of $h\rightarrow \gamma\gamma$ decays,
which is mediated at LO by the interference between a $W$ boson and $t$ quark loop, and therefore sensitive to this sign difference,
concludes that Higgs boson data is consistent with the SM description~\cite{Ellis:2013lra}.

For QCD, QED, and $Z$ interactions, as the coupling vertex have the structure
\begin{equation}
 \overline{u}^{\alpha}_{L} \gamma^\mu u_L =  \overline{u}^{' \alpha}_{L} U_L^{-1} \gamma^\mu U_L u_L^{'} = \overline{u}^{' \alpha}_{L} \gamma^\mu u_L^{'},
\label{neutralCurrents.sm.eq}
 \end{equation}
we see that the currents are \textit{flavor-conserving}, and that the interaction basis is still aligned with the mass basis.
Thus, they need not be discussed further.
We now explore what consequences this rotation in \textit{flavor space} has on our charged current interactions.
Recalling the covariant derivatives from above,  we have
\begin{eqnarray}
\overline{Q}_L^{\beta}i \not\!\!D^{\beta\alpha I} Q_L^\alpha &\ni& 
\overline{Q}_L^{\alpha}i 
\left[\delta^{\beta\alpha}\frac{i g_W}{\sqrt{2}}\left(W_\mu^+\hat{T}_L^+ +W_\mu^-\hat{T}_L^- \right)\right]\gamma^\mu Q_L^{\alpha}
\\
&=&
\delta^{\beta\alpha}\frac{- g_W}{\sqrt{2}}
\left(W_\mu^+ \overline{u}_L^{\beta} \gamma^\mu d_L^{\alpha} + W_\mu^- \overline{d}_L^{\beta} \gamma^\mu u_L^{\alpha} \right).
\end{eqnarray}
Rotating into the mass basis, we get
\begin{eqnarray}
\overline{Q}_L^{\beta}i \not\!\!D^{\beta\alpha I} Q_L^\alpha &\ni& 
\delta^{\beta\alpha}\frac{- g_W}{\sqrt{2}}
\left(W_\mu^+ \overline{u'}_L^{\beta}  \gamma^\mu \left( U_L^{-1} D_L \right) d_L^{'\alpha} 
+ W_\mu^- \overline{d'}_L^{\beta} \gamma^\mu \left(D_L^{-1} U_L\right) u_L^{'\alpha} \right).
\end{eqnarray}
Defining the matrix $V$ such that
\begin{equation}
\boxed{ V \equiv U_L^{-1} D_L},
\end{equation}
our \textit{flavor-changing charged currents}  with (small) intergenerational mixing are governed by
\begin{eqnarray}
\overline{Q}_L^{\beta}i \not\!\!D^{\beta\alpha I} Q_L^\alpha &\ni& 
\delta^{\beta\alpha}\frac{- g_W}{\sqrt{2}}
\left(W_\mu^+ \overline{u}_L^{\beta}  \gamma^\mu V_{ud} d_L^{\alpha} 
+ W_\mu^- \overline{d}_L^{\beta} \gamma^\mu V_{ud}^\dagger u_L^{\alpha} \right),
\end{eqnarray}
where $u_L$ and $d_L$ now represents mass eigenstates.
$V$ is the \textit{Cabbibo-Kobayashi-Maskawa}  (CKM) matrix~\cite{Cabibbo:1963yz,Kobayashi:1973fv}.
As a $3\times3$ unitary matrix, it expressible by three angles and a phase:
\begin{eqnarray}
V^{\rm CKM} &=& 
 \begin{pmatrix}
  V_{ud} & V_{us} & V_{ub} \\
  V_{cd} & V_{cs} & V_{cb} \\
  V_{td} & V_{ts} & V_{tb}
 \end{pmatrix} 
 \nonumber\\
 &=&
  \begin{pmatrix}  1 & 0 & 0 \\   0 & \cos\theta_{23} & \sin\theta_{23} \\   0 & -\sin\theta_{23} & \cos\theta_{23}  \end{pmatrix}
  \begin{pmatrix}  \cos\theta_{13}& 0 & \sin\theta_{13} e^{-i\delta_{13}}\\
  0 & 1 & 0 \\  -\sin\theta_{13} e^{-i\delta_{13}} & 0 & \cos\theta_{13} \end{pmatrix}
  \begin{pmatrix}  \cos\theta_{12} & \sin\theta_{12} & 0 \\  -\sin\theta_{12} & \cos\theta_{12}& 0 \\ 0 & 0 & 1  \end{pmatrix}.
  \nonumber
\end{eqnarray}
The presence of the complex phase $\delta_{13}\neq0$ in $^{\rm CKM}$ indicates $CP$ violation in weak interactions.
The best measurements available at the time of this writing of $V^{\rm CKM}$ are given in Table~\ref{ckmMatrix.TB}~\cite{Agashe:2014kda}.
Other useful parameterizations can also be found in Ref.~\cite{Agashe:2014kda} and references within.

Gauge invariance bars gluons and photons to undergo \textit{flavor changing neutral currents} (FCNCs), even at the higher orders of perturbation.
However, as the $Z$ is not associated with a good local symmetry, 
off-diagonal elements of $U^{-1}_L U_L$ may be generated at the loop-level,
a process known as the Glashow-Iliopoulos-Maiani (GIM) mechanism~\cite{Glashow:1970gm}.
These FCNCs processes, however, face both coupling and phase space suppression.

\begin{table}[!t]
\caption{Components of CKM Matrix}
\begin{center}
\begin{tabular}{|l l l|}
\hline\hline 
$\vert V_{ud}\vert=0.97425\pm0.00022$ & 
$\vert V_{us}\vert=0.2253\pm0.0008$ 		& 
$\vert V_{ub}\vert=(4.13\pm0.49)\times10^{-3} $	\tabularnewline
$\vert V_{cd}\vert=0.225\pm0.008$   & 
$\vert V_{cs}\vert=0.986\pm0.016 $ & 
$\vert V_{cb}\vert=(41.1\pm1.3)\times10^{-3} $	\tabularnewline
$\vert V_{td}\vert=(8.4\pm0.6)\times10^{-3} $ & 
$\vert V_{ts}\vert=(40.0\pm2.7)\times10^{-3} $ & 
$\vert V_{tb}\vert=1.021\pm0.032 $ 		\tabularnewline\hline
\multicolumn{3}{|c|}{Jarlskog invariant $J=3.06^{+0.21}_{-0.20}\times10^{-5}$}  \tabularnewline\hline
\hline
\end{tabular}
\label{ckmMatrix.TB}
\end{center}
\end{table}

\section{Beyond the Standard Model}\label{bsm.sm.sec}
In this chapter we have introduced global and local continuous symmetries as well as their spontaneous breakdown 
via the acquiring of a nonzero vacuum expectation value by a scalar field. 
Building on these principles, we constructed the Standard Model of particle physics.
However, despite the SM's  experimental success, it remains an unsatisfactory description of nature.
The existence of nonzero neutrino masses, dark matter, a large hierarchy among fermion masses,
and a Higgs boson whose mass is unstable under radiative corrections highlight the theory's shortcomings.
Extensions of the SM that alleviate these issues vary in size and scope,
but commonly predict, among new principles and symmetries, 
the existence of new gauge bosons (e.g., Left-Right Symmetry), new scalars (e.g., Supersymmetry), and new fermions (e.g., Seesaw Mechanisms).
In the following chapters, we will explore several such models, including more phenomenological, semi-model-independent approaches,
and derive testable  consequences. 

%% file: 02_ColliderPhysics/colliderPhysics.tex
\chapter{Principles of Collider Physics}

\section{Introduction}
\label{sec:intro}
Collider physics and phenomenology explore the manifestation of the SM in high energy and high momentum transfer scattering experiments.
It is a deeply rich and enjoyable subject that incorporates perturbative, 
non-perturbative (all-orders summed and effective field theory), and computational techniques
in order to simulate with a reasonably high degree of accuracy the results of lepton-lepton, lepton-hadron, and hadron-hadron collisions.
In this chapter, we introduce many fundamental topics of collider physics. 
Many excellent texts on the topic are available, in particular the classic Barger~\&~Phillips~\cite{Barger:1987nn} 
as well as lectures by Han~\cite{Han:2005mu} and Willenbrock~\cite{Willenbrock:1990ei}.
The texts Halzen~\&~Martin~\cite{Halzen:1984mc} and Thomson~\cite{Thomson:2013zua} provide an excellent introduction to the field, 
providing an inordinately large number of useful examples.

\section{Helicity Amplitudes}
We start our study of collider phenomenology with the introduction of helicity amplitudes
and helicity eigenstates for representations of the Lorentz group.
The theory of scalars, spin one-half fermions, and vector bosons as irreducible representations of the Lorentz group is a very important topic.
A rigorous construction from first principles can be found in Weinberg~\cite{Weinberg:1995mt}. 
We now  briefly review spin one-half fermions and spin-one bosons. 
As spin-zero bosons are a trivial representations of the Lorentz group, they transform as scalars; no review  of their properties is needed.

\subsection{Spin One-Half Fermions}
To construct the explicit forms of Dirac spinors in the helicity basis,
we suppose a fermion propagating in the direction $\hat{p}_0$ relative to its quantized spin axis. 
Along this direction, the helicity operator $\Sigma$ is defined as 
\begin{equation}
 \Sigma \equiv \sigma\cdot \hat{p}_0 
  = \begin{pmatrix} \hat{p}^3_0 & \hat{p}^1_0 -i\hat{p}^2_0 \\ \hat{p}^1_0 +i\hat{p}^2_0 & -\hat{p}^3_0 \end{pmatrix}
  = \begin{pmatrix} \cos\theta & e^{-i\phi}\sin\theta  \\ e^{i\phi}\sin\theta & -\cos\theta \end{pmatrix}.
\end{equation}
The corresponding helicity eigenstates are the two-component solutions $\chi_\lambda(\hat{p}_0)$ to the relationship
\begin{eqnarray}
  \Sigma~\chi_\lambda(\hat{p}_0) = \lambda \chi_\lambda(\hat{p}_0),
  \label{helicityDef.col.EQ}
\end{eqnarray}
the eigenvalues of which are $\lambda=\pm1$ and twice the fermion's actual helicity. 
Conventionally, when the direction of propagation is (anti-)parallel to the spin axis, 
which results in the  eigenvalue $\lambda=(-)1$, we refer to the state as being in its (left-) right-handed helicity eigenstate.

Fixing the spin quantization axis with a definite direction, say $\hat{z}$, the fermion is aligned with its spin axis when $\hat{p}_0=\pm\hat{z}$.
In such a situation, the four solutions to Eq.~(\ref{helicityDef.col.EQ}) are
\begin{equation}
\chi_{\lambda=+1}( \hat{z}) = \begin{pmatrix} 1  \\ 0 \end{pmatrix},\quad
\chi_{\lambda=-1}( \hat{z}) = \begin{pmatrix} 0  \\ 1 \end{pmatrix},\quad
\chi_{\lambda=+1}(-\hat{z}) = \begin{pmatrix} 0  \\ 1 \end{pmatrix},\quad
\chi_{\lambda=-1}(-\hat{z}) = \begin{pmatrix} -1 \\ 0 \end{pmatrix}.
\end{equation}
Boosting our fermion to an arbitrary reference frame
\begin{equation}
 p^\mu = (E_0,0,0,\vert \vec{p}_0\vert) \rightarrow 
 p'^\mu = p^\mu = (E,\vert\vec{p}\vert \sin\theta\cos\phi,\vert\vec{p}\vert\sin\theta\sin\phi,\vert\vec{p}\vert\cos\theta),
 \quad E^2 = \vert\vec{p}\vert^2 + m^2
\end{equation}
the two-component eigenstates are 
\begin{eqnarray}
 \chi_{\lambda=+1}(\hat{p})&=&\frac{1}{\sqrt{2\vert\vec{p}\vert(\vert\vec{p}\vert+p_z)}}
\begin{pmatrix} \vert\vec{p}\vert+p_z \\ p_x + ip_y \end{pmatrix}
=\begin{pmatrix} \cos\frac{\theta}{2} \\ e^{i\phi}\sin\frac{\theta}{2} \end{pmatrix}
\\
\chi_{\lambda=-1}(\hat{p})&=&\frac{1}{\sqrt{2\vert\vec{p}\vert(\vert\vec{p}\vert+p_z)}}
\begin{pmatrix} -p_x + ip_y \\ \vert\vec{p}\vert+p_z \end{pmatrix}
=\begin{pmatrix} -e^{-i\phi}\sin\frac{\theta}{2}  \\ \cos\frac{\theta}{2} \end{pmatrix}
\\
\chi_\lambda(-\hat{p}) &=& -\lambda e^{i\lambda\phi}\chi_{-\lambda}(\hat{p}).
\end{eqnarray}
The four-component Dirac spinors for a fermion $(u_\lambda)$ and antifermion $(v_\lambda)$ can now be constructed:
\begin{eqnarray}
 u_\lambda(p) = \left(\begin{matrix}\sqrt{E-\lambda\vert\vec{p}\vert} \chi_\lambda(\hat{p})
\\ \sqrt{E+\lambda\vert\vec{p}\vert} \chi_\lambda(\hat{p})\end{matrix}\right),
\quad
v_\lambda(p) = \left(\begin{matrix}-\lambda\sqrt{E+\lambda\vert\vec{p}\vert} \chi_{-\lambda}(\hat{p})
\\ \lambda\sqrt{E-\lambda\vert\vec{p}\vert} \chi_{-\lambda}(\hat{p})\end{matrix}\right).
\end{eqnarray}

\subsubsection{Properties of Dirac Spinors}

In the high-energy limit, when $E\gg m$, degrees of freedom are decoupled and the Dirac spinors simplify to
\begin{eqnarray}
u_{\lambda=+1}(p) \approx \sqrt{2E} \left(\begin{matrix} 0 \\  \chi_{\lambda=+1}(\hat{p})\end{matrix}\right)
&\quad&
u_{\lambda=-1}(p) \approx \sqrt{2E} \left(\begin{matrix} \chi_{\lambda=-1}(\hat{p}) \\ 0 \end{matrix}\right)
\\
v_{\lambda=+1}(p) \approx -\sqrt{2E} \left(\begin{matrix} \chi_{\lambda=-1}(\hat{p}) \\ 0 \end{matrix}\right).
&\quad&
v_{\lambda=-1}(p) \approx  \sqrt{2E} \left(\begin{matrix} 0 \\ \chi_{\lambda=+1}(\hat{p}) \end{matrix}\right).
\end{eqnarray}
In this limit, $u_\lambda$ and $v_\lambda$ are found to be eigenstates of the chiral projection operators
\begin{equation}
 P_L = \frac{1}{2}(1-\gamma^5), \quad\text{and}\quad P_R = \frac{1}{2}(1+\gamma^5)
\end{equation}
In particular, the LH fermion and RH antifermion helicity states are LH chiral states
\begin{equation}
 P_L u_{\lambda=-1},~v_{\lambda=+1} = u_{\lambda=-1},~v_{\lambda=+1};
\end{equation}
and  the RH fermion and LH antifermion helicity states are RH chiral states
\begin{equation}
   P_R u_{\lambda=+1},~v_{\lambda=-1} = u_{\lambda=-1},~v_{\lambda=+1}.
\end{equation}
It is in this limit that chirality and helicity become equivalent.

\subsection{Spin-One Vector Bosons}
Massive and massless vector bosons are of central importance to broken and unbroken gauge theories, and QFTs in general.
Indeed, a non-Abelian gauge theory without additional fermions or scalars represents 
an entirely nontrivial, self-consistent and self-contained theory with predictive scattering rates.
Consider a vector boson with mass $M_V$ and momentum
\begin{eqnarray}
 k^\mu = (E, k_x, k_y, k_z) &=& (E, \vert\vec{k}\vert \sin\theta\cos\phi, \vert\vec{k}\vert \sin\theta\sin\phi, \vert\vec{k}\vert \cos\theta)
 \\
 &=& (E, k_T \cos\phi, k_T \sin\phi, \vert\vec{k}\vert \cos\theta), \quad E^2 = M_V^2 + \vert\vec{k}\vert^2.
\end{eqnarray}
The transverse momentum is defined by 
\begin{equation}
 k_T = \sqrt{k_x^2 + k_y^2} = \vert\vec{k}\vert \sin\theta.
\end{equation}
In the $M_V\rightarrow0$ limit, we have
\begin{eqnarray}
 k^\mu = (E, k_x, k_y, k_z) &=& E (1, \sin\theta\cos\phi, \sin\theta\sin\phi, \cos\theta)
 \\
&=& (E, E_T \cos\phi, E_T \sin\phi, E \cos\theta),
\quad
E_T = E \sin\theta,
\end{eqnarray}
with \textit{transverse energy} $E_T$.

The polarization vectors in the Cartesian representation are given by
\begin{eqnarray}
 \varepsilon^\mu (k,x) &=& \frac{1}{\vert\vec{k}\vert k_T} (0, k_x k_y, k_y k_z, -k_T^2 )
 \\
 \varepsilon^\mu (k,y) &=& \frac{1}{ k_T} (0, - k_y, k_x, 0 )
 \\
 \varepsilon^\mu (k,z) &=& \frac{E}{M_V\vert\vec{k}\vert^2 } \left( \frac{\vert\vec{k}\vert}{E}, k_x , k_y , k_z \right)
\end{eqnarray}
Checking, we have that the expected orthogonal relationships
\begin{eqnarray}
 k_\mu \varepsilon^\mu (k,x) &=& \frac{1}{\vert\vec{k}\vert k_T} (0 \underset{-k_T^2 k_z}{\underbrace{- k_x^2 k_y - k_y^2 k_z }} + k_z k_T^2  )=0
 \\
 k_\mu \varepsilon^\mu (k,y) &=& \frac{1}{ k_T}  \left(0 + k_x k_y - k_y k_x + 0 \right) =0
 \\
 k_\mu \varepsilon^\mu (k,z) &=& \frac{E}{M_V\vert\vec{k}\vert } \left(  \frac{\vert\vec{k}\vert^2}{E} E - k_x^2 - k_y ^2 + k_z^2  \right) =0
\end{eqnarray}

In the polar representation, the right- $(\lambda=+1)$, left- $(\lambda=-1)$, and longitudinal $(\lambda=0)$ polarization vectors are
\begin{eqnarray}
 \varepsilon^\mu (k,\lambda=\pm) &=& \frac{1}{\sqrt{2}}\left( \mp\varepsilon^\mu(k,x) - i\varepsilon^\mu(k,y)\right),
 \\
 \varepsilon^\mu (k,\lambda=0) &=& \varepsilon^\mu (k,z).
\end{eqnarray}
As these are (at most) linear redefinitions of the Cartesian polarization vectors, inner product relationships hold.
For massless vector bosons, there are no longitudinal polarization states.

\subsection{Decay of Heavy Fermionic Top Quark Partner}
Hypothetical, TeV-scale top quark partners represent an excellent example that highlights the differences between chirality and helicity, 
as well as the interesting roles played by the transverse and longitudinal polarizations of gauge bosons.
Such particles are proposed to cancel the large quadratic corrections the SM Higgs' mass receives from the top quark at 1-loop.
In these models, the SM gauge state $t_L$ is decomposed into light $(\sim 173\GeV)$ and heavy $(\gtrsim 1\TeV)$ mass eigenstates, 
denoted as $t$ and $T$, respectively. 
The alignment of the top quark gauge state and the mass eigenstates can be (phenomenologically) parameterized by the angle $\theta_t:$
\begin{equation}
\underset{\rm Gauge~Basis}{\underbrace{t_L}} \simeq ~\underset{\rm Mass~Basis}{\underbrace{\cos\theta_t t + \sin\theta_t T}}.
\end{equation}
The corresponding charged current Feynman rules are then given by
\begin{eqnarray}
 tWb &:& \frac{-ig}{\sqrt{2}}V_{tb}^{*}\cos\theta_t \label{tWbFeynRule.col.EQ}\\
 TWb &:& \frac{-ig}{\sqrt{2}}V_{tb}^{*}\sin\theta_t \label{TWbFeynRule.col.EQ}
\end{eqnarray}
Qualitatively, $\theta_t \sim m_t / m_T$, and thus in the large $m_T$ limit $T$ decouples from the SM.

For the heavy top partner decay into a massive SM $W$ boson and bottom quark,
\begin{equation}
 T_{\tau}(p_T) \rightarrow W_\lambda^+(p_W) + b_{\tau'}(p_b),
\end{equation}
with helicities $\tau,\tau'=L,R$ and polarizations $\lambda=\pm,0$, the helicity amplitudes are generically
\begin{equation}
 \mathcal{M}_{\lambda\tau'\tau} = 
 \frac{-ig}{\sqrt{2}}V^{*}_{tb}\cos\theta_t~
[\overline{u}_{\tau'}(p_b) \xSlash{\varepsilon}_\lambda^{*}(p_W)P_L u_\tau(p_T)].
\end{equation}
In the rest frame of $T$, the momenta are
\begin{eqnarray}
 p_T &=& (m_T,\vec{0})\\
 p_b &=& (E_b, \vert\vec{p}_b\vert \sin\theta\cos\phi,  \vert\vec{p}_b\vert \sin\theta\sin\phi,  \vert\vec{p}_b\vert \cos\theta)\\
 p_W &=& (E_W, -\vert\vec{p}_b\vert \sin\theta\cos\phi,-\vert\vec{p}_b\vert \sin\theta\sin\phi, -\vert\vec{p}_b\vert \cos\theta)\\
 E_b &=& \frac{m_T}{2}(1+r_b - r_W),\quad \vert\vec{p}_b\vert = \frac{m_T}{2}(1-r_b - r_W),\quad E_W = \frac{m_T}{2}(1+r_W - r_b),
\end{eqnarray}
where $r_X = m_X^2 / m_T^2.$ 
Omitting a universal factor of $\left(-igV^{*}_{tb}\cos\theta_t/\sqrt{2}\right)$, the orthogonal helicity amplitudes for transverse $W$ bosons are 
\begin{eqnarray}
\mathcal{M}_{-LL} =   m_T \sqrt{2(1-r_W)} \sin\frac{\theta}{2}, &\quad& \mathcal{M}_{-LR} =  m_T \sqrt{2(1-r_W)} e^{i\phi} \cos\frac{\theta}{2} 
\\
\mathcal{M}_{+RL} =  - mT \sqrt{2r_b} e^{-i\phi} \cos\frac{\theta}{2} &\quad& \mathcal{M}_{+RR} =  m_T \sqrt{2r_b} \sin\frac{\theta}{2} 
\\
\mathcal{M}_{+LL} = \mathcal{M}_{-LR} &=& \mathcal{M}_{+LR} = \mathcal{M}_{-RR} = 0 	
\end{eqnarray}
Several appreciable lessons can be learned from these expressions.
In the helicity-conserving cases $(\lambda\tau'\tau) = (-LL)$ and $(+RR)$, 
zero angular momentum can be carried away by the transversely polarized $W$ when the bottom quark is aligned with its parent fermion. 
Thus, the amplitudes vanish as $\theta$ tends toward zero. 
Conversely, this is precisely when the helicity-flipping amplitudes $(\lambda\tau'\tau) = (-LR)$ and $(+RL)$ are maximal.
As $W$ radiation is a purely LH chiral coupling, the $P_L$ projection operator in Eq.~(\ref{tWbFeynRule.col.EQ}) 
collects terms proportional to $m_b = m_T\sqrt{r_b}$ from RH (helicity) bottom quarks.
Therefore as the bottom quark is taken massless, its chiral and helicity states align, 
and contributions from RH bottom quarks, i.e.,  $(\lambda\tau'\tau) = (+RL)$ and $(+RR)$, turn off.
By angular momentum conservation, the remaining amplitudes for transverse $W$ bosons are identically zero.

Omitting the same coupling factor, the helicity amplitudes for a longitudinally polarized $W$ are
\begin{eqnarray}
\mathcal{M}_{0LL} &=&   \frac{m_T}{2} \sqrt{\frac{1-r_W}{r_W}}
\left(1-rb+rW +\lambda^{1/2}(1,r_b,r_W) \right) \cos\frac{\theta}{2}
\\
\mathcal{M}_{0LR} &=&  -\frac{m_T}{2} \sqrt{\frac{1-r_W}{r_W}}
\left(1 - rb + rW +     \lambda^{1/2}(1,r_b,r_W) \right) e^{i\phi} \sin\frac{\theta}{2} 
\\
\mathcal{M}_{0RL} &=&  -\frac{m_T}{2} \sqrt{\frac{r_b}{r_W}}
\left(1 - rb + rW -    \lambda^{1/2}(1,r_b,r_W) \right) e^{-i\phi}\sin\frac{\theta}{2}
\\
\mathcal{M}_{0RR} &=& -\frac{m_T}{2} \sqrt{\frac{r_b}{r_W}}
\left(1 - rb + rW -    \lambda^{1/2}(1,r_b,r_W) \right) \cos\frac{\theta}{2},
\end{eqnarray}
where $\lambda(x,y,z)$ is the usual kinematic function and simplifies to
\begin{equation}
 \lambda(1,r_b,r_W) = 1 - 2 (r_b + r_W) + (r_b - r_W)^2.
\end{equation}
The most striking feature of these amplitudes is the inverse dependence on the $W$ boson mass,
which leads to a quadratic growth with respect to $m_T$ in the case of LH (helicity) bottom quarks.
Its origin is in the zeroth component of the $W$ polarization vector, 
\begin{equation}
 \varepsilon_{\mu=0}^{\lambda=0} = E_W/M_W \sim m_T/2M_W.
\end{equation}
If $m_T$ is the result of some Higgs-like mechanism, then it can generically be written as as the product of a Yukawa coupling and scalar vev: 
$m_T \sim y_T v_T$. We see now that $\varepsilon_{\mu=0} \sim m_T/2M_W \sim y_T v_T / g v$. 
In other words, for a fixed ratio of vevs, 
fermionic decays to longitudinally polarized gauge bosons is a measure of the relative coupling strength to their respective Higgs sectors.
In the large $m_T$ limit, these amplitudes become the dominant contributions to the T quark decay, 
a phenomenon known as \textit{longitudinal polarization enhancement}, and has observed in SM top quark decays~\cite{Aaltonen:2012tk}.

\section{Phase Space}

Phase space, abbreviated by $PS$, is far-reaching concept in physics. 
It is the set of all allowed configurations in which a system may exist and not forbidden by a symmetry (conservation law).
The volume of phase space is a measure of how many unique configurations a system possesses: 
more available states correspond to a larger phase space volume.

In momentum space, the $n-$body differential phase space with a total momentum $P_{\rm Tot.}$ is
\begin{equation}
  dPS_n(P_{\rm Tot}; p_1, p_2 \dots p_n) =  (2\pi)^{4}\delta^{4}\left(P_{\rm Tot} - \sum_{k=1}^{n} p_k \right) 
  \prod_{k=1}^{n} \frac{d^{3} p_k}{(2\pi)^{3} 2E_k}
\label{phaseSpaceDef.col.EQ}
\end{equation}
In~Eq.~(\ref{phaseSpaceDef.col.EQ}), the Dirac function enforces momentum conservation.
For the $1-$, $2-$ and $3-$body configurations, the number of d.o.f.~are sufficiently constrained by momentum conservation that 
the differential phase space can be reasonably expressed analytically. 
For situations with weakly coupled (narrow width) particles propagating intermediately as well as $n\geq 4$-body systems, 
it is helpful to apply the phase space recursion relationship, given in Section~\ref{psRecursion.col.sec}.
When the former is coupled with the narrow width approximation (NWA), on-shell factorization can be applied.

For two-body processes, the \textit{K\"allen kinematic function}, also called the ``$\lambda$'' function, is of considerable use and is given by
\begin{eqnarray}
 \lambda(x,y,z) &=& (x - y -z)^2 -4yz =  x^2 + y^2 + z^2 - 2xy - 2xz - 2yz.
 \label{kallen.col.eq}
\end{eqnarray}
Quite often, we deal with arguments normalized to the leading variable
\begin{eqnarray}
 \lambda\left(1,\frac{y}{x},\frac{z}{x}\right) &=& \left(1 - \frac{y}{x} -\frac{z}{x}\right)^2 - \frac{4yz}{x^2}
 = \frac{1}{x^2}\left[(x - y - z)^2 - 4yz\right]
 \\
 &=& \frac{1}{x^2}\lambda(x,y,z)
 \\
 \implies \lambda^{1/2}(x,y,z) &=& x \lambda^{1/2}\left(1,\frac{y}{x},\frac{z}{x}\right).
\end{eqnarray}
Physically, for momentum $p_i$ and  $r_i  = p_i^2 / P_{\rm Tot.}^2$ is the mass ratio (squared) of momentum $i=1,2$ and c.m.~mass $\sqrt{P_{\rm Tot.}^2}$,
$\lambda^{1/2}(1,r_1,r_2)$ can be interpreted as speed of $i=1$ and $i=2$ in the parent $P_{\rm Tot.}$ frame. 
It is then the case that
\begin{equation}
 \vert \vec{p}_i \vert = \frac{E_{\rm Tot}}{2}\beta = \frac{E_{\rm Tot}}{2}\lambda^{1/2}(1,r_1,r_2).
 \label{lamSpeed.col.eq}
\end{equation}

\subsection{One-Body Phase Space}\label{dPS1.col.sec}
The one-body final state  is a very special scenario because the invariant mass of the final-state momentum must equal the total c.m.~energy
by momentum conservation. Consider a state with 4-momentum $p_1$ and mass $m_1$. It follows that
\begin{eqnarray}
\int dPS_{1}  
  &=& (2\pi)^{4} ~ \int \frac{d^{3}p_1}{(2\pi)^{3}2E_1}~\delta^{4}(P_{\rm Tot.} - p_1) \\
  &=& (2\pi)     ~ \int d^{4}p ~\delta(p_1^2-m_{1}^{2})~\delta^{4}(P_{\rm Tot.} - p_1) \\
  &=&  2\pi~\delta(\hat{s}-m_{1}^{2}),
\end{eqnarray}
where we have applied the relationship
\begin{equation}
 \int \frac{d^{3}p_1}{2E_1} = \int d^{4}p_1 ~\delta(p_1^2-m_{1}^{2}).
\end{equation}
and introduced the Mandelstam collider variable
\begin{equation}
\hat{s} = P_{\rm Tot.}^2, 
\end{equation}
Quantities labeled by the caret 	$(\hat{~})$, colloquially called ``hat'', denote \textit{partonic} quantities within composite system.
Most often this is applied to parton scattering in hadron-hadron and hadron-lepton collisions,
it also applied to parton scattering in lepton-lepton collisions when objects are convolved about distribution functions;
for example, see Section~\ref{isPhotons.col.sec}.

\subsection{Solid Angle in $d$ Dimension}
The derivations of compact expressions for two- and three-body phase space volume elements differ little from their $d$-dimensional analogs.
By introducing this slight but nonetheless additional complexity, we can greatly reap the benefits of having results applicable to higher order calculations.
The solid angle volume element for a $k$-sphere in $d$-dimensions is given by
\begin{eqnarray}
 d^d\Omega_k &=& \left[ d\theta_1 \dots d\theta_{k-1}\right] \left[\sin^{k-2}\theta_{k-1}\dots \sin\theta_2\right], \quad \theta_i\in (0,\pi).
\end{eqnarray}
We separate the volume element into $2$-angle and $(k-2)$-angle orientations to obtain
\begin{eqnarray}
 d^d\Omega_k &=& \left(d\theta_{k-1} \sin^{k-2}\theta_{k-1}\right)\left(d\theta_{k-2}\sin^{k-3}\theta_{k-2}\right) 
 \left[ d\theta_1 \dots d\theta_{k-3}\right] \left[\sin^{k-4}\theta_{k-3}\dots \sin\theta_2\right]
\\
 &=&
 \left(d\theta_{k-1} \sin^{k-2}\theta_{k-1}\right)\left(d\theta_{k-2}\sin^{k-3}\theta_{k-2}\right)  \times d\Omega_{k-2}.
\end{eqnarray}
For a $k$-sphere, the integrated solid angle is
\begin{equation}
 \Omega_k = \frac{2(\pi)^{k/2}}{\Gamma\left(\frac{k}{2}\right)} \overset{\rm E.g.}{=}
 \left\{\begin{matrix}2, & \text{for}~ k=1\\ 2\pi, & \text{for}~ k=2\\ 4\pi, & \text{for}~ k=3\end{matrix}\right.
\end{equation}
Integrating over the $(k-2)$ space and relabeling our variables, we get
\begin{equation}
\boxed{
 d\Omega_k = \left(d\theta \sin^{k-2}\theta\right)\left(d\phi\sin^{k-3}\phi\right)
 \frac{2(\pi)^{\frac{k-2}{2}}}{\Gamma\left(\frac{k-2}{2}\right)},  \quad \theta, \phi\in(0,\pi). } 
\end{equation}

\subsection{Two-Body Phase Space in $d$ and $4$ Dimensions}\label{dPS2.col.sec}
For a 2-body phase in $d$ dimensions, we have
\begin{eqnarray}
 d^dPS_{2}(P_{\rm Tot};p_{1},p_{2})&=&(2\pi)^{d}\delta^{d}\left(P_{\rm Tot}-p_1 - p_2\right)
 \frac{d^{d-1}p_{1}}{(2\pi)^{d-1}2E_{1}}\frac{d^{d-1}p_{2}}{(2\pi)^{d-1}2E_{2}},
 \label{dPS2Prelim.col.EQ}
\end{eqnarray}
Integrating (without the loss of generality) over the $p_2$ momentum, we have
\begin{eqnarray}
 d^dPS_{2}(P_{\rm Tot};p_{1},p_{2})&=& 
 \frac{\delta(u)du}{2^2(2\pi)^{d-2}}\delta\left(E_{\rm Tot}-E_1 - E_2\right)
 \frac{d\vert\vec{p}_1\vert~\vert\vec{p}_1\vert^{d-2}}{E_1 E_2} d\Omega_{d-1}^{p_1}.
\end{eqnarray}
Now, defining $u = E_1 + E_2 - E_{\rm Tot}$ and taking note that $\vert\vec{p}_1\vert = \vert\vec{p}_2\vert$, we get
\begin{equation}
 du = \frac{\vert\vec{p}_1\vert d\vert\vec{p}_1\vert}{E_1} + \frac{\vert\vec{p}_1\vert d\vert\vec{p}_1\vert}{E_2} - 0 
 = \frac{\vert\vec{p}_1\vert d\vert\vec{p}_1\vert (E_{\rm Tot} + u)}{E_1 E_2}.
\end{equation}
Plugging this into Eq.~(\ref{dPS2Prelim.col.EQ}) and using the momentum-$\lambda$ relationship of Eq.~(\ref{lamSpeed.col.eq}), we find
\begin{eqnarray}
 d^dPS_{2}(P_{\rm Tot};p_{1},p_{2})&=&  \frac{d\Omega_{d-1}^{p_1}}{2^2(2\pi)^{d-2}}
 \frac{\vert\vec{p}_1\vert^{d-3}}{(E_{\rm Tot} + u)}
 =
  \frac{d\Omega_{d-1}^{p_1}}{2^2(2\pi)^{d-2}}
 \frac{\vert\vec{p}_1\vert^{d-3}}{E_{\rm Tot}}
 \\
 &=&
  \frac{d\Omega_{d-1}^{p_1}}{2^2(2\pi)^{d-2}}
 \frac{E_{\rm Tot}^{d-4}}{2^{d-3}} \lambda^{\frac{d-3}{2}}(1,r_1,r_2), \quad r_i = \frac{p_i^2}{P_{\rm Tot}^2}.
\end{eqnarray}
For the $d=4$ case, this simplifies to
\begin{equation}
 \boxed{dPS_{2}(P_{\rm Tot};p_{1},p_{2}) =  
 \frac{d\Omega_{3}^{p_1}}{2(4\pi)^{2}} \lambda^{1/2}(1,r_1,r_2), \quad r_i = \frac{p_i^2}{P_{\rm Tot}^2}, \quad d\Omega_{3}^{p_1} = d\cos\theta_1 d\phi_1}.
 \label{dPS2.col.eq}
\end{equation}
Useful, equivalent expressions include
\begin{eqnarray}
dPS_{2}(P_{\rm Tot};p_{1},p_{2})&=&\frac{d\Omega_{3}^{p_1}}{2(4\pi)^{2}}\sqrt{1-2(r_1+r_2)+(r_1-r_2)^{2}}
\\
&=&\frac{d\Omega_{3}^{p_1}}{2(4\pi)^{2}}\sqrt{\left[1-(\sqrt{r_1}+\sqrt{r_2})^{2}\right]\left[1-(\sqrt{r_1}-\sqrt{r_2})^{2}\right]}.
\end{eqnarray}
In the $r_2\rightarrow0$ limit,
\begin{equation}
dPS_{2}(P_{\rm Tot};p_{1},p_{2})=\frac{d\Omega_{3}^{p_1}}{2(4\pi)^{2}}(1-r_{1});
\end{equation}
and when $r_{1}=r_{2}$, we have
\begin{equation}
dPS_{2}(P_{\rm Tot};p_{1},p_{2})=\frac{d\Omega_{3}^{p_1}}{2(4\pi)^{2}}\sqrt{1-4r_1}.
\end{equation}	
In four dimensions, the momenta and energies of the final-state particles in the $P_{\rm Tot}$~frame are
\begin{equation}
 \vert \vec{p}_1 \vert = \vert\vec{p}_2\vert = \frac{\sqrt{P_{\rm Tot}^2}}{2}\lambda^{1/2}(1,r_1,r_2), \quad 
 E_1 = \frac{\sqrt{P_{\rm Tot}^2}}{2} (1 + r_1 - r_2), \quad
 E_2 = \frac{\sqrt{P_{\rm Tot}^2}}{2} (1 - r_1 + r_2).
\end{equation}

To generate the two-body phase space via Monte Carlo integration, we define $y_1,~y_2$ such that
\begin{eqnarray}
 \cos\theta = 2y_1 - 1, \quad \phi = 2\pi y_2, \quad y_1,~y_2 \in [0,1].
\end{eqnarray}
Equation ~(\ref{dPS2.col.eq}) then simplifies to
\begin{eqnarray}
 dPS_{2}(P_{\rm Tot};p_{1},p_{2}) =   \frac{dy_1~ dy_2}{(8\pi)} \lambda^{1/2}(1,r_1,r_2)
\approx  \Delta y_1 \Delta y_2 ~  \frac{1}{(8\pi)} \lambda^{1/2}(1,r_1,r_2).
\end{eqnarray}
In practice, $y_1,~y_2$ are randomly generated, which are used to build $\cos\theta$ and $\phi$.
$\Delta y_1,~ \Delta y_2$ represent the finite volume element from which $y_1,~y_2$ are generated.

\subsection{Three-Body Phase Space in $d$ and $4$ Dimensions}\label{dPS3.col.sec}
For the three-body case, we follow a procedure similar to the two-body situation.
The phase space in $d$ dimensions is given by
\begin{eqnarray}
 d^dPS_{3}(P_{\rm Tot};p_{1},p_{2},p_3) &=& (2\pi)^{d}\delta^{d}\left(P_{\rm Tot}-p_1 - p_2 - p_3\right)
 \nonumber\\
 & & \times 
 \frac{d^{d-1}p_{1}}{(2\pi)^{d-1}2E_{1}}\frac{d^{d-1}p_{2}}{(2\pi)^{d-1}2E_{2}}\frac{d^{d-1}p_{3}}{(2\pi)^{d-1}2E_{3}}.
 \label{dPS3Prelim.col.EQ}
\end{eqnarray}
As $d^{d-1} p_i / E_i$ is a boost-invariant quantity, we rotate $p_2$ and $p_3$  into the $p_{(23)}=p_{2}+p_{3}$ rest frame.
Also integrating out (without the loss of generality) $p_3$ by use of the $\delta$-function, we get
\begin{eqnarray}
 d^dPS_{3}(P_{\rm Tot};p_{1},p_{2},p_3) &=& 
 \frac{1}{2^3(2\pi)^{2d-3}}\delta\left(E_{\rm Tot}-E_1 - (E_2 - E_3)\right)
\nonumber\\ 
& & \times
 \frac{d\vert\vec{p}_1\vert \vert\vec{p}_1\vert^{d-2} d\Omega_{d-1}^{p_1}}{E_{1}}
 \frac{d\vert\vec{p}_2^{(23)}\vert \vert\vec{p}_2^{(23)}\vert^{d-2} d\Omega_{d-1}^{p_2^{(23)}}}{E_{2}^{(23)}E_{3}^{(23)}}
 \label{dPS3Prelim2.col.EQ}
\end{eqnarray}
In the $p_{23}$-frame, we have the relationship
\begin{equation}
 m_{23} \equiv \sqrt{p_{23}^2} = E_{\rm Tot}^{(23)} =  E_{2}^{(23)} +  E_{3}^{(23)},
\end{equation}
which implies
\begin{eqnarray}
 dE_{\rm Tot}^{(23)} &=&  dE_{2}^{(23)} +  dE_{3}^{(23)} = 
 \frac{d\vert\vec{p}_2^{(23)}\vert \vert\vec{p}_2^{(23)}\vert }{E_{2}^{(23)}} +  
 \frac{\vert\vec{p}_2^{(23)}\vert \vert\vec{p}_2^{(23)}\vert}{E_{3}^{(23)}}
\\
 &=&
 d\vert\vec{p}_2^{(23)}\vert \vert\vec{p}_2^{(23)}\vert \left(\frac{m_{23}}{E_2^{(23)} E_3^{(23)}}\right).
\end{eqnarray}
We also note that in the $p_{23}$ frame the momentum-$\lambda$ relationship gives us 
\begin{eqnarray}
 \vert\vec{p}_2^{(23)}\vert = \frac{m_{23}}{2}\lambda^{1/2}\left(1, \frac{p_2^2}{m_{23}^2},\frac{p_3^2}{m_{23}^2}  \right),
 \label{dPS3mom23.col.eq}
\end{eqnarray}
and more generally
\begin{eqnarray}
 \frac{\vert\vec{p}_2^{(23)}\vert^{d-3}}{m_{23}}
 = \frac{m_{23}^{d-4}}{2^{d-3}}\lambda^{\frac{d-3}{2}}\left(1, \frac{p_2^2}{m_{23}^2},\frac{p_3^2}{m_{23}^2}  \right).
\end{eqnarray}
Now making the appropriate substitutions in Eq.~(\ref{dPS3Prelim2.col.EQ}) gives us
\begin{eqnarray}
 d^dPS_{3}(P_{\rm Tot};p_{1},p_{2},p_3) &=& 
 \frac{d\Omega_{d-1}^{p_1}d\Omega_{d-1}^{p_2^{(23)}}}{2^3(2\pi)^{2d-3}}\delta\left(E_{\rm Tot}-E_1 - (E_2 - E_3)\right)
\nonumber\\ 
& & \times
 \underset{2\vert\vec{p}_1\vert d\vert\vec{p}_1\vert  = 2E_1 dE_1}{\underbrace{\frac{d\vert\vec{p}_1\vert \vert\vec{p}_1\vert^{d-2} }{E_{1}}}}
 \left[    dE_{\rm Tot}^{(23)}\left(\frac{m_{23}^{d-4}}{2^{d-3}}\right)
  \lambda^{\frac{d-3}{2}}\left(1, \frac{p_2^2}{m_{23}^2},\frac{p_3^2}{m_{23}^2}  \right)\right]
  \\
  &=&
  \frac{d\Omega_{d-1}^{p_1}d\Omega_{d-1}^{p_2^{(23)}}}{2^3(2\pi)^{2d-3}}\delta\left(E_{\rm Tot}-E_1 - (E_2 - E_3)\right)dE_{\rm Tot}^{(23)}
\nonumber\\ 
& & \times
 dE_1  \vert\vec{p}_1\vert^{d-3}
 \left[    \left(\frac{m_{23}^{d-4}}{2^{d-3}}\right)
  \lambda^{\frac{d-3}{2}}\left(1, \frac{p_2^2}{m_{23}^2},\frac{p_3^2}{m_{23}^2}  \right)\right].
\end{eqnarray}
In the last line, we made the change of variable from $d\vert\vec{p}_1\vert$ to $dE_1$.
Simplifying, regrouping, and integrating over the final $\delta$-function gives us
\begin{eqnarray}
 d^dPS_{3}(P_{\rm Tot};p_{1},p_{2},p_3) &=& 
 \frac{m_{23}^{d-4}}{2^{d+1}(2\pi)^{2d-3}}2\vert\vec{p}_1\vert^{d-3}dE_1d\Omega_{d-1}^{p_1}d\Omega_{d-1}^{p_2^{(23)}}
  \lambda^{\frac{d-3}{2}}\left(1, \frac{p_2^2}{m_{23}^2},\frac{p_3^2}{m_{23}^2}  \right)
\end{eqnarray}
In the $d=4$ limit, we have
\begin{eqnarray}
\boxed{
 dPS_{3}(P_{\rm Tot};p_{1},p_{2},p_3) =
 \frac{1}{(4\pi)^5}2\vert\vec{p}_1\vert dE_1d\Omega_{3}^{p_1}d\Omega_{3}^{p_2^{(23)}}
  \lambda^{1/2}\left(1, \frac{p_2^2}{m_{23}^2},\frac{p_3^2}{m_{23}^2}  \right).}
  \label{dPS3.col.eq}
\end{eqnarray}
And in the $P_{\rm Tot}$~frame, the maximum momentum and energy of $p_1$  are
\begin{equation}
 \vert\vec{p}_1^{\max}\vert = 
 \frac{\lambda^{1/2}\left(P_{\rm Tot}^2, m_1^2, (m_2 + m_3)^2 \right)}{2\sqrt{P_{\rm Tot}^2}},
 \quad 
 E_1^{\max} = \frac{P_{\rm Tot}^2 + m_1^2 - (m_2 + m_3)^2}{2\sqrt{P_{\rm Tot}^2}}
\end{equation}

To generate this phase space via Monte Carlo integration, we define $y_1,\dots, y_5$ such that
\begin{eqnarray}
 E_1 = (E_1^{\max} - m_1) y_1 + m_1, &\quad& \cos\theta_1 = 2y_2 - 1, \quad \phi_1 = 2\pi y_3, \\
 \cos\theta_2^{(23)} = 2y_4 - 1, &\quad& \phi_2^{(23)} = 2\pi y_5, \quad y_1,\dots, y_5 \in [0,1].
\end{eqnarray}
Equation (\ref{dPS3.col.eq}) then simplifies to
\begin{eqnarray}
 dPS_{3}(P_{\rm Tot};p_{1},p_{2},p_3) &=&
 dy_1 dy_2 dy_3 dy_4 dy_5
 \frac{(E_1^{\max} - m_1)}{(4\pi)^3}2\vert\vec{p}_1\vert
  \lambda^{1/2}\left(1, \frac{p_2^2}{m_{23}^2},\frac{p_3^2}{m_{23}^2}  \right)
  \\
  &\approx&
   [\Delta y_i]^5
 \frac{(E_1^{\max} - m_1)}{(4\pi)^3}2\vert\vec{p}_1\vert
  \lambda^{1/2}\left(1, \frac{p_2^2}{m_{23}^2},\frac{p_3^2}{m_{23}^2}  \right),
\end{eqnarray}
where $y_1, ~i=1,\dots,5$ are randomly generated and $\Delta y_i$ represent the finite volume element from which the $y_i$ are generated.
After constructing $p_1$, the Lorentz invariant $m_{23}$ can be built:
\begin{equation}
 m_{23}^2 = (p_2 + p_3)^2 = (P_{\rm Tot} - p_1)^2 = P_{\rm Tot}^2 + m_1^2 - 2\sqrt{P_{\rm Tot}^2} E_1.
\end{equation}
Then using Eq.~(\ref{dPS3mom23.col.eq}), $p_2$ and $p_3$ can be constructed in the $(23)$-frame using
the procedure outlined in section~\ref{dPS2.col.sec}. Finally, $p_2$ and $p_3$ are boosted from the $(23)$-frame into the $P_{\rm Tot}$~frame.

\subsection{Phase Space Decomposition}
\label{psRecursion.col.sec}
A powerful property of phase space for an arbitrary number of final states is the ability to decompose it into the product of smaller phase spaces.
For automated event generator packages, the \textit{phase space recursion relationship} forms the basis of their phase space integration modules.
Formally, the relationship states that
{\it the $n$-body phase space volume of $P_{\rm Tot}$ is equivalent to the volume enclosed by 
(i) an $(n-1)$-body phase space made by combining two final-state momenta $p_i$ and $p_j$ into $p_{ij}$, 
(ii) the corresponding $ij\rightarrow i + j$ $2$-body phase space, and (iii) the allowed virtuality of $p_{ij}^2$}:
\begin{eqnarray}
 dPS_n(P_{\rm Tot}; p_1, \dots, p_i, \dots, p_j, \dots p_n) &=&  
 dPS_{n-1}(P_{\rm Tot}; p_1, \dots, p_{i-1}, p_{i+1}, \dots, p_{j-1}, p_{j+1}, \dots p_n, p_{ij}) \nonumber\\ 
 &\times&
 dPS_2(p_{ij}; p_i, p_j)
 \times
 \frac{d~m_{ij}^2}{2\pi}, 
 \quad
 m_{ij}^2 = p_{ij}^2 = (p_i + p_j)^2.
\end{eqnarray}
The proof of decomposition is quite general, so we present it $d$-dimensions.
We start by factoring the $n$-body phase space into $(n-2)$- and $2$-particle momentum integrals and factors of $1$:
\begin{eqnarray}
  dPS_n(P_{\rm Tot}; p_1, \dots p_n) &=&  (2\pi)^{d}\delta^{d}(P_{\rm Tot} -  p_1 \dots - p_i  \dots - p_j \dots -p_n) 
  \prod_{f=1}^{n} \frac{d^{d-1} p_f}{(2\pi)^{d-1} 2E_f} \\
  &=& 
  (2\pi)^{d}\delta^{d}(P_{\rm Tot} -  p_1 \dots -p_{i-1} -p_{i+1} \dots -p_{j-1} -p_{j+1} \dots -p_n -p_{ij}) \nonumber \\
  &\times&    \left(\prod_{f=1, ~\neq i,j}^{n} \frac{d^{d-1} p_f}{(2\pi)^{d-1} 2E_f} \right)
  \times \frac{d^{d-1} p_i}{(2\pi)^{d-1} 2E_i}  ~ \frac{d^{d-1} p_j}{(2\pi)^{d-1} 2E_j} \nonumber \\
  &\times& 
  \underset{=1}{\underbrace{\delta^{d}(p_{ij}-p_i - p_j)~d^d p_{ij}}} \times
  \frac{(2\pi)^{d} 2E_{ij}}{(2\pi)^{d} 2E_{ij}} \times
  \underset{=1}{\underbrace{\delta(p_{ij}^2-m_{ij}^2)~d m_{ij}^2}}
\end{eqnarray}
Shuffling around terms, we can construct a differential $n-1$-body phase space volume element:
\begin{eqnarray}
dPS_n(P_{\rm Tot}; p_1, \dots p_n) &=&
  (2\pi)^{d}\delta^{d}(P_{\rm Tot} -  p_1 \dots -p_{i-1} -p_{i+1} \dots -p_{j-1} -p_{j+1} \dots -p_n -p_{ij}) \nonumber \\
  &\times&   \underset{(n-1)\text{\rm-momentum integrals}}{\underbrace{\left(\prod_{f=1, ~\neq i,j}^{n} \frac{d^{d-1} p_f}{(2\pi)^{d-1} 2E_f} \right)  \times \frac{d^{d-1} p_{ij}}{(2\pi)^{d-1} 2E_{ij}}}}  
  ~ \times \frac{d E_{ij}}{2\pi} ~ 2E_{ij}
  \nonumber \\
  &\times&  (2\pi)^d \delta^{d}(p_{ij}-p_i - p_j) \frac{d^{d-1} p_i}{(2\pi)^{d-1} 2E_i}  ~ \frac{d^{d-1} p_j}{(2\pi)^{d-1} 2E_j} \nonumber \\
  &\times&    \delta(p_{ij}^2-m_{ij}^2)~d m_{ij}^2
  \\
    &=& 
   dPS_{n-1}(P_{\rm Tot}; p_1, \dots, p_{i-1}, p_{i+1}, \dots, p_{j-1}, p_{j+1}, \dots p_n, p_{ij}) \nonumber\\ 
  &\times&  dPS_2(p_{ij}; p_i, p_j) \nonumber\\
  &\times&  d E_{ij}^2 ~ \delta(p_{ij}^2-m_{ij}^2)~\frac{d m_{ij}^2}{2\pi},
\end{eqnarray}
giving us our desired expression
\begin{eqnarray}
dPS_n(P_{\rm Tot}; p_1, \dots  p_n) &=&
 dPS_{n-1}(P_{\rm Tot}; p_1, \dots, p_{i-1}, p_{i+1}, \dots, p_{j-1}, p_{j+1}, \dots p_n, p_{ij}) \nonumber\\ 
 &\times&
 dPS_2(p_{ij}; p_i, p_j)  \times  \frac{d~m_{ij}^2}{2\pi}.
\end{eqnarray}
As alluded, two useful applications phase space decomposition are (i) in automated phase space integration packages and (ii) resonant decays. 
In the first, the phase space for an arbitrarily high number of final states can be reduced to successive boosts of two final-state particles $p_i$ and $p_j$ 
into their total momentum frame $(p_i+p_j)$, followed by simple and efficient integration over two-body or three-body solid angles.
The explicit formulae for these are given in Sections ~\ref{dPS2.col.sec} and ~\ref{dPS3.col.sec}.
In the latter case, the invariant mass integral $d~m_{ij}^2$ can be interpreted as the virtuality integral for an intermediate resonance.
For example: in on-shell top quark decays into a bottom quark, muon, and neutrino, the leading contribution occurs through 
$W$ boson radiation from the top quark that then splits into leptons. The phase space decomposition
\begin{equation}
 dPS_3(t;\mu^+,\nu_\mu, b) = dPS_2(t; p_{\mu+\nu}, b) \times dPS(p_{\mu+\nu}; \mu^+, \nu_\mu) \times \frac{d~q_{\mu+\nu}^2}{2\pi}
 \label{dPSTop.col.eq}
\end{equation}
has a physical interpretation as the $t\rightarrow W^* b$ two-body phase space, 
the $W^*\rightarrow \mu\nu_\mu$ two-body phase space, 
and the virtuality integral for $W^*$ can spans the entire spectrum of invariant masses that are allowed by conservation of momentum.

\subsection{One-Particle Phase Space Splitting}
It is convenient to write explicitly the one-body phase space for collinear splittings, and similar situations.
Consider the process 
\begin{equation}
 A(p_A) + B(p_B) \rightarrow A(p_1) + X
\end{equation}
that is mediated by the subprocess
\begin{equation}
 g(p_g) + B(p_B) \rightarrow X, \quad p_g = p_A - p_1
\end{equation}
where $X$ is some arbitrary $n$-body final state, 
$g$ originates from an $A\rightarrow A g$ splitting,
and $A$ is an otherwise a spectator in the entire process.
Momentum conservation tells us
\begin{equation}
 p_A + p_B = \underset{p_A}{\underbrace{p_g + p_1}} + p_B = p_1 + p_2 \cdots + p_n,
\end{equation}
and so we may write in $d$ dimensions the $\delta$-function
\begin{equation}
 \delta^d(p_A + p_B - p_1 - p_2 -\cdots - p_n) = \delta^d(p_g + p_1 + p_B - p_1  - p_2 -\cdots - p_n) = \delta^d(p_g + p_B - p_2 -\cdots - p_n). 
\end{equation}
Therefore, factoring out the momentum integral for $A(p_1)$, we have 
\begin{eqnarray}
 dPS_{n}(A+B\rightarrow A+X)
 &=&
 (2\pi)^{4}\delta^{4}\left(p_A + p_B - p_1 - p_2 -\cdots - p_n\right)\prod_{k=1}^{n}\frac{d^{3}p_{k}}{(2\pi)^{3}2E_{k}}
\\
 &=&
 \left[(2\pi)^{4}\delta^{4}\left(p_g + p_B - p_2 -\cdots - p_n\right)\prod_{k=2}^{n}\frac{d^{3}p_{k}}{(2\pi)^{3}2E_{k}} \right]
 \frac{d^{3}p_{1}}{(2\pi)^{3}2E_{1}}
 \\
 &=& dPS_{n}(A+B\overset{A\rightarrow Ag}{\longrightarrow} X) \times \frac{d^{3}p_{1}}{(2\pi)^{3}2E_{1}}, \quad p_1 = p_A - p_g.
 \label{dPSsplit.sm.col}
\end{eqnarray}

\section{Partial Width}
The \textit{partial width} of an unstable, unpolarized particle $A$ with mass $m_A$, spin states $(2s_A+1)$, 
and SU$(3)_c$ color multiplicity $N_c^A$ , decaying into an $n$-body final-state $f$ is given by formula
\begin{equation}
\Gamma(A\rightarrow f)=\frac{1}{2m_{A}}\frac{1}{(2s_{A}+1)N_{c}^A}\sum\vert\mathcal{M}\vert^{2}\cdot dPS_{n}(p_A;p_{1},\dots,p_{n}).
\end{equation}
Here, $\mathcal{M}$ is usual $A\rightarrow f$ amplitude that can be calculated perturbatively using Feynman Rules.
The sum over all partial widths, is the \textit{total width},
\begin{equation}
 \Gamma_{\rm Tot}^A \equiv \sum_f \Gamma(A\rightarrow f).
\end{equation}
The fraction of times $A$ will decay into a particular final state $X$ is called the \textit{branching fraction}, and is given as the 
ratio of the partial and total widths
\begin{equation}
 \text{BR}(A\rightarrow X) \equiv \frac{\Gamma(A\rightarrow X)}{\Gamma_{\rm Tot}^A} = \frac{\Gamma(A\rightarrow X)}{\sum_f \Gamma(A\rightarrow f)}.
 \label{brDef.col.EQ}
\end{equation}
The total width is also related to $A$'s mean lifetime, $\tau$, by the expression
\begin{equation}
 \tau = \frac{\hbar}{\Gamma_{\rm Tot}^A}.
\end{equation}
We make explicit the conversion of units from $\Gamma$ [GeV] to $\tau$ [s] for clarity.
Hence, partial and total widths are simultaneous estimations of how strongly $A$ couples to its final states (larger coupling, larger width), 
and its likelihood to decay (larger width, smaller decay time).
To appreciate this observable better, we consider the {Optical Theorem} as derived from unitarity of the $S$-matrix.

\subsection{The Optical Theorem and Breit-Wigner Propagators}
The $S$-matrix in QFT can be decomposed into its trivial non-scattering and scattering component, $T$, by the relationship
\begin{equation}
 S = 1 + i T.
\end{equation}
To leading order in scattering amplitudes, the unitarity of $S$ tells us
\begin{eqnarray}
 1 = S^\dagger S = (1- iT^\dagger)(1+ iT) = 1 + T^\dagger T + i(T - T^\dagger),
\end{eqnarray}
or that the squared norm of the transition operator is equal to its imaginary part: 
\begin{equation}
 T^\dagger T  =  -i(T - T^\dagger).
 \label{unitarity1.col.EQ}
\end{equation}
For initial $i$, final states $f$, and arbitrary intermediate $n$-body state $k$, this implies
\begin{eqnarray}
 \langle f \vert T^\dagger T \vert i\rangle &=& \sum_{ k}\int dPS_n \langle f \vert T^\dagger\vert k \rangle \langle k \vert T \vert i\rangle
\label{unitarity2.col.EQ}
 \\
 &=& \sum_{k}\int dPS_n ~\mathcal{M}^*(f\rightarrow k) \mathcal{M}(i\rightarrow k),
\end{eqnarray}
where by the completeness relationship we sum/integrate over all discrete degrees of freedom and phase space configurations.
In words, the result states that the matrix elements of the squared norm transition operator $T^\dagger T$ is 
equal to the sum of transition amplitudes, $\mathcal{M}$, to all intermediate states.
The unitarity condition of Eq.~(\ref{unitarity1.col.EQ}) also tells use that 
the matrix element of the imaginary part of the transition operator is 
the imaginary part of the transition operator matrix elements, i.e., 
\begin{equation}
 -i\langle f\vert (T - T^\dagger)\vert i\rangle = -i\left[\mathcal{M}(i\rightarrow f) - \mathcal{M}^*(f\rightarrow i)\right],
\end{equation}
which ultimately follows from linearity.
When combined, we obtain the \textbf{Optical Theorem}, 
which states that \textit{imaginary part of a scattering amplitude is equivalent to the sum of all its intermediate states}:
\begin{equation}
 -i\left[\mathcal{M}(i\rightarrow f) - \mathcal{M}^*(f\rightarrow i)\right]
 =
  \sum_{k}\int dPS_n ~\mathcal{M}^*(f\rightarrow k) \mathcal{M}(i\rightarrow k).
\end{equation}

In the special case of $1\rightarrow 1$ scattering, i.e., particle propagation, of particle $A$ with mass $m_A$,
the initial and final states $i$ and $f$ are equivalent.
The Optical Theorem then stipulates 
\begin{eqnarray}
 -i\left[\mathcal{M}(A\rightarrow A) - \mathcal{M}^*(A\rightarrow A)\right]
 &=&
 2\Im[\mathcal{M}(A\rightarrow A)]
 \\
 &=&
  \frac{2m_A}{2m_A}\sum_{\rm k}\int dPS_n ~\mathcal{M}^*(A\rightarrow k) \mathcal{M}(A\rightarrow k).
   \\
 &=&
  2m_A \sum_{\rm k}\underset{\text{Partial~Width}~\Gamma_k}{\underbrace{\frac{1}{2m_A}\int dPS_n ~\vert\mathcal{M}^*(A\rightarrow k)\vert^2}}.
\end{eqnarray}
We recognize the last expression as the definition of the total width. 
In other words, the imaginary part of the $1\rightarrow1$ scattering amplitude proportional to the total width of the propagating particle
\begin{equation}
 \Im[\mathcal{M}(A\rightarrow A)] = m_A \Gamma_{\rm Tot}^A,
\end{equation}
with the constant of proportionality being the object's mass.

More significantly is that the $1\rightarrow1$ amplitudes are precisely the one-particle irreducible (1PI) 
correlation function diagrams that constitute the self-energy of $A$.
We denote the self-energy generically (whether $A$ is a scalar, fermion, or vector boson) by $\Pi(q^2)$, where $q^2$ is the virtuality of $A$.
As $A$ comes on-shell, its inverse propagator (again, generically written)
\begin{equation}
 \Delta^{-1}_A(q^2) = q^2 - m_A^2 + \Pi(q^2),
\end{equation}
takes the form
\begin{eqnarray}
\lim_{q^2\rightarrow m_A^2} \Delta^{-1}(q^2) &\approx& m_A^2 - m_A^2 + i\Im[\Pi(m_A^2)] = +  i m_A \Gamma_{\rm Tot}^A,
\end{eqnarray}
indicating that when $A$ is on mass-shell, the imaginary part of its self-energy is given by its mass and its total width:
\begin{equation}
 \boxed{\Im[\Pi(m_A^2)] = m_A \Gamma_{\rm Tot}^A}.
\end{equation}

Furthermore, transition amplitudes are dominantly populated by regions of phase space where particles are close to being on-shell. 
In other words: in the neighborhood of a pole in the $S$-matrix.
Thus, for intermediate, resonant states with momentum $q$, mass $M$, and a well-defined, on-shell self-energy, 
its all-orders summed propagator is well-modeled in matrix element calculations by making the substitution 
\begin{equation}
 \frac{i}{(q^2 - m^2) + \Pi(q^2)} \rightarrow \frac{i}{(q^2 - M^2) + i M\Gamma}.
\end{equation}
This is the \textit{Breit-Wigner {\rm(BW)} propagator}. 
As a distribution function, its normalization is set by
\begin{eqnarray}
1 &=&  \int_{-\infty}^{\infty} \frac{d~q^2 ~N}{(q^2 - M^2)^2 + (M\Gamma)^2} = \frac{N}{M\Gamma}\int_{-\infty}^{\infty}\frac{dx}{1+x^2} 
\\
 &=& N\frac{\arctan x}{M\Gamma} \Bigg\vert_{-\infty}^{\infty} = N\frac{\pi}{M\Gamma}, \quad x = \frac{q^2 - M^2}{M\Gamma},
\end{eqnarray}
implying
\begin{equation}
 N = \frac{M\Gamma}{\pi}.
\end{equation}
In application, phase space integration over BW propagators can be make more efficient 
by making the change of variable
\begin{equation}
q^{2}-M^{2}\equiv M\Gamma\tan\theta. 
\end{equation}
This has the action of \textit{smoothening} the Breit-Wigner resonance distribution 
\begin{equation}
\int_{q_{min}^{2}}^{q_{max}^{2}}\frac{dp^{2}}{(q^{2}-M^{2})^{2}+(\Gamma M)^{2}}=
\int_{\theta_{\min}}^{\theta_{\max}}\frac{d\theta}{\Gamma M}=\frac{(\theta_{\max}-\theta_{\min})}{\Gamma M}\int_{0}^{1}dy,
\end{equation}
where
\begin{eqnarray}
\theta_{i}&=&\tan^{-1}\left[\frac{q_{i}^{2}-M^{2}}{\Gamma m}\right],\quad i\in\{\min,\max\}
\\
\theta&=&(\theta_{\max}-\theta_{\min})y+\theta_{\min}.
\end{eqnarray}
This has great utility in Monte Carlo or other sampling-based integration techniques, 
which are adversely affected by sharp peaks in integrands.

\subsection{Narrow Width Approximation}
Generally, widths scale like $\Gamma \sim g^2 ~ M.$ Thus, for weakly coupled objects one finds
\begin{equation}
 \frac{\Gamma}{M} \sim g^2 \ll 1.
\end{equation}
We label such particles as ``narrow'' resonances, in reference to their narrow BW distributions.
A few examples of SM particles with narrow widths include~\cite{Agashe:2014kda}
\begin{eqnarray}
 W^{\pm}&:& 	\frac{\Gamma}{M} \approx  0.026 \\
 Z&:&  		\frac{\Gamma}{M} \approx  0.027 \\
 \eta&:&	\frac{\Gamma}{M} \approx  2.34\times 10^{-06}\\
 t&:&		\frac{\Gamma}{m} \lesssim 0.012.
\end{eqnarray}

Formally, in the zero width (infinitely stable) limit, the BW distribution function approaches the normal distribution function
that in turns approaches a $\delta$-function~\cite{Plehn:2009nd}
\begin{eqnarray}
 \lim_{\Gamma\rightarrow 0}\frac{1}{\pi}\frac{M\Gamma}{(q^2 - M^2)^2 + (M\Gamma)^2} &=& 
 \lim_{\Gamma\rightarrow 0}\frac{1}{(\alpha M\Gamma)\sqrt{\pi}}e^{-(q^2 - M^2)^2/(\alpha M\Gamma)^2}
 \\
 &=&  \int \frac{d p}{2\pi} e^{-i(q^2 - M^2)p} = \delta(q^2 - M^2),
\end{eqnarray}
where $\alpha^{-1}\approx 1.177 M$ is the conversion between total width $\Gamma$ and the standard deviation $\sigma$ in normal distributions.
In particularly extreme situations where the the total width of an intermediate resonance is much smaller than its mass,
e.g., top quarks and $W$ bosons at the few percent accuracy~\cite{Campbell:2012uf}, 
it is often sufficient  to approximate BW distributions as $\delta$-functions:
\begin{equation}
 \frac{1}{(p^2 - M^2)^2 + (M\Gamma)^2} \rightarrow \frac{\pi}{M\Gamma}\delta(p^2-M^2).
\end{equation}
This is the {\it narrow width approximation} (NWA). 
Physically, the NWA says that an intermediate particle is sufficiently longed live that its intermediate 
production can be well-approximated as its on-shell production and subsequent on-shell decay.
When used in conjunction with the phase space recursion relationship, the NWA is a very powerful tool 
that greatly simplifies calculations of cascade decays into on-shell particles.
In the top quark decay example of Eq.~(\ref{dPSTop.col.eq}), applying the NWA has the affect of putting the $W$ boson on-shell at all times
\begin{eqnarray}
\frac{dPS_3(t;\mu^+,\nu_\mu, b)}{(p^2 - M^2)^2 + (M\Gamma)^2}   &=& 
\frac{\pi}{M\Gamma}\delta(p^2-M^2)
dPS_2(t; p_{\mu+\nu}, b) \times dPS(p_{\mu+\nu}; \mu^+, \nu_\mu) \times \frac{d~q_{\mu+\nu}^2}{2\pi}
\nonumber\\
&=&
\frac{\pi}{M\Gamma} dPS_2(t;W^+, b) \times dPS(W^+; \mu^+, \nu_\mu).
\end{eqnarray}
As an example, we now carry out the full $t\rightarrow W^+b\rightarrow \mu^+\nu_\mu b$ calculation with the NWA.

\subsection{Example: NWA Applied to Leptonic Decays of Top Quarks}
We consider the decay of a top quark into a bottom quark and a pair of massless leptons
\begin{equation}
 t ~\rightarrow ~W^{+*} ~b ~\rightarrow ~\ell^+ ~\nu_\ell ~b,
\end{equation}
and model the intermediate $W$ boson propagation using a BW propagator with a momentum transfer $q^2$ and total with $\Gamma_W$.
The $t \rightarrow \ell^+ \nu_\ell b$ matrix element can then be expressed as
\begin{equation}
 \mathcal{M}(t\rightarrow \ell^+ \nu_\ell b) = 
 \mathcal{M}^\mu(t\rightarrow W^{*}b) ~ \cfrac{-i \left[g_{\mu\nu} + (\xi-1)\frac{q_\mu q_\nu}{(q^2 - \xi M_W^2)} \right]}{q^2 - M_W^2 + iM_W\Gamma_W} ~ \mathcal{M}^{\nu}(W^{*}\rightarrow \ell^+ \nu_\ell).
\label{topFullME.col.EQ}
\end{equation}

Physical observables, e.g., cross sections and partial widths, are independent of the gauge parameter $\xi$,
which is sometimes written as $\eta$ instead of $(\xi-1)$.
A wise choice of $\xi$ can greatly simplify a calculation but at the potential cost of increasing the number of subprocesses (Feynman diagrams) 
contributing to the process. Common choices of the \textit{gauge fixing} parameter are
\begin{equation}
 \xi=\left\{\begin{matrix}
0, & \text{Landau Gauge} \\ 
1, & \text{Feynman Gauge}\\ 
\infty, & \text{Unitary Gauge for Massive Bosons}
\end{matrix}\right.
\end{equation}
However, the gauge-term $q_\mu q_\nu$ in Eq.~(\ref{topFullME.col.EQ}) is unimportant and does not contribute to the final result.
This follows from the $W$ propagator contracting with a vector current of massless, external fermions.
This can be understood from two semi-independent arguments:
(i) Since the $W$ momentum is the sum of the massless lepton momenta
\begin{equation}
q_W = p_\mu + p_\nu,  
\end{equation}
by the Dirac equation we have (ignoring factors of $-ig/\sqrt{2}$)
\begin{eqnarray}
 q_\nu \mathcal{M}^\nu(W^*\rightarrow \ell^+ \nu_\ell) &=& 
 \overline{u}(p_\nu) \left( \not\!\!p_\mu + \not\!\!p_\nu \right)P_L v(p_\mu)
\\
& =&
 \overline{u}(p_\nu)P_R  \underset{=m_\mu v(p_\mu)=0}{\underbrace{\not\!\!p_\mu  v(p_\mu)}}
+
\underset{=\overline{u}(p_\nu)m_\nu=0}{\underbrace{\overline{u}(p_\nu)  \not\!\!p_\nu}} P_L v(p_\mu) =0.
\end{eqnarray}
(ii) By virtue of being on-shell, massless isospin partners, the leptons respect an unbroken SU$(2)_L$ symmetry 
and therefore do not couple to the $W$ boson's longitudinal polarizations, i.e., $q_\mu q_\nu$.

Using the completeness relationship in the Unitarity gauge,
\begin{equation}
 \sum_{\lambda,\lambda'} \varepsilon_{\mu,\lambda}^{*}(q)\varepsilon_{\nu,\lambda'}(q) = 
 -g_{\mu\nu} + q_\mu q_\nu / M_W^2,
\label{vectorCompleteness.col.EQ}
 \end{equation}
we can express Eq.~(\ref{topFullME.col.EQ}) as the product of matrix elements for two independent processes
\begin{eqnarray}
  \mathcal{M}(t\rightarrow \ell^+ \nu_\ell b) &=& \sum_{\lambda,\lambda'}~
  \mathcal{M}^\mu(t\rightarrow W^{*}b) 
  ~ \frac{i\varepsilon_{\mu,\lambda}^{*}(q)\varepsilon_{\nu,\lambda'}(q)}{q^2 - M_W^2 + iM_W\Gamma_W} 
  ~ \mathcal{M}^{\nu}(W^{*}\rightarrow \ell^+ \nu_\ell)
  \\
  &=& \sum_{\lambda,\lambda'}~
  \mathcal{M}_{\lambda}(t\rightarrow W^{*}b) ~   ~ \frac{i}{q^2 - M_W^2 + iM_W\Gamma_W}  ~ \mathcal{M}_{\lambda'}(W^{*}\rightarrow \ell^+ \nu_\ell).
\end{eqnarray}
Squaring the amplitudes and summing over all degrees of freedom, i.e., spins/colors, gives us
\begin{equation}
\sum_{\rm d.o.f.}
 \vert \mathcal{M}(t\rightarrow \ell^+ \nu_\ell b)\vert^2 
 = \frac{1}{(2s_W+1)N_c^W}
 ~\sum_{\rm d.o.f.}
 ~\frac{\vert \mathcal{M}_{\lambda\tilde{\lambda}}(t\rightarrow W^{*}b) \vert^2  ~ \vert \mathcal{M}_{\lambda'\tilde{\lambda'}}(W^{*}\rightarrow \ell^+ \nu_\ell)\vert^2}
 {(q^2 - M_W^2)^2 + (M_W\Gamma_W)^2}.
\end{equation}
In order to prevent double-counting of $W$ boson spins, we must average over the number of spin states.
Though trivial in this instance, one must also average over the intermediate messenger's color multiplicity in order to avoid double-counting color states.
This sees innocuous but systematic practice allows us to write intermediate subprocesses, e.g., $W$ boson splitting, 
in terms of unpolarized widths and cross sections.
We are now in position to make the NWA approximation:
\begin{equation}
 \sum_{\rm dof}
 \vert \mathcal{M}(t\rightarrow \ell^+ \nu_\ell b)\vert^2 
 \approx \frac{1}{(2s_W+1)N_c^W} \sum_{\rm dof}~
 \vert \mathcal{M}_{\lambda\tilde{\lambda}}(t\rightarrow W^{*}b) \vert^2  ~ \vert \mathcal{M}_{\lambda'\tilde{\lambda'}}(W^{*}\rightarrow \ell^+ \nu_\ell)\vert^2 ~
 ~\frac{\pi\delta(q^2 - M_W^2)}{M_W\Gamma_W}.
 \nonumber
\end{equation}
Finally, we evaluate the few remaining steps needed to compute the $t\rightarrow W^{*}b \rightarrow \ell^+ \nu_\ell b$ partial width.
Averaging over the top quark's spin states and colors, as well as integrating over the 3-body phase space, 
which is immediately decomposed into two two-body spaces, we obtain
\begin{eqnarray}
 \Gamma(t\rightarrow \ell^+ \nu_\ell b) &=& 
 \int dPS_3(t;\mu^+,\nu_\mu,b) ~\frac{1}{2m_t (2s_t+1)N_c^t} \sum_{\rm dof} \vert \mathcal{M}(t\rightarrow \ell^+ \nu_\ell b)\vert^2 
 \\
 &=&
 \int dPS_2(t;W^{*},b)~dPS_2(W^{*};\mu^+,\nu_\mu)~\frac{d q^2}{2\pi} \times
   \frac{1}{2m_t (2s_t+1)N_c^t} ~ \frac{1}{(2s_W+1)N_c^W}  \nonumber\\
  &\times& \sum_{\rm dof}~
 \vert \mathcal{M}_{\lambda\tilde{\lambda}}(t\rightarrow W^{*}b) \vert^2  ~ \vert \mathcal{M}_{\lambda'\tilde{\lambda'}}(W^{*}\rightarrow \ell^+ \nu_\ell)\vert^2 ~
 \frac{\pi}{M_W\Gamma_W}\delta(q^2 - M_W^2)
  \\
 &=&
 \int dPS_2(t;W^{*},b)~ \frac{1}{2m_t (2s_t+1)N_c^t}   \sum_{\rm dof} \vert \mathcal{M}_{\lambda\tilde{\lambda}}(t\rightarrow W^{*}b) \vert^2  \nonumber\\
 &\times& 
  \frac{1}{(2s_W+1)N_c^W} ~\frac{\pi}{M_W\Gamma_W} ~\int \frac{d q^2}{2\pi} \delta(q^2 - M_W^2) \nonumber\\
 &\times& 
 \int dPS_2(W^{*};\mu^+,\nu_\mu) ~  \sum_{\rm dof} \vert \mathcal{M}_{\lambda'\tilde{\lambda'}}(W^{*}\rightarrow \ell^+ \nu_\ell)\vert^2
\label{topWidthNWA.col.EQ}
 \end{eqnarray}
Combining the $\delta$-function from the NWA and the virtuality integral from the Recursion Theorem together require that the $W$ be on-shell at all times.
With this, we immediately recognize the first line of Eq.~(\ref{topWidthNWA.col.EQ}) as the partial width for $t\rightarrow W b$ decay,
and the last two lines of Eq.~(\ref{topWidthNWA.col.EQ}) provide us the ingredients for the $W\rightarrow \ell \nu_\ell$ partial width:
 \begin{eqnarray}
 \Gamma(t\rightarrow \ell^+ \nu_\ell b) &=&  \Gamma(t\rightarrow W b) ~\times~ \frac{1}{\Gamma_W} \nonumber\\
 &\times& 
 \frac{1}{2M_W(2s_W+1)N_c^W}  \int dPS_2(W^{*};\mu^+,\nu_\mu) ~  \sum_{\rm dof} \vert \mathcal{M}_{\lambda'\tilde{\lambda'}}(W^{*}\rightarrow \ell^+ \nu_\ell)\vert^2
 \\
 &=& \Gamma(t\rightarrow W b) ~\times~ \frac{1}{\Gamma_W} \Gamma(W\rightarrow \ell \nu_\ell) 
 \\
 &=&
 \Gamma(t\rightarrow W b) ~\times~  {\rm BR}(W\rightarrow \ell \nu_\ell) \label{topDecayNWA.col.EQ}
\end{eqnarray}
Thus, by combining the NWA and phase space recursion theorem, 
we can approximate a top quark decay rate to leptons as the rate of a top quark decaying into
an on-shell $W$ boson scaled by the probability that an on-shell $W$ will decay to leptons. 
This probability is given by the branching fraction of $W$ into leptons, and is defined in Eq.~(\ref{brDef.col.EQ}).

Before moving on, it is worth reflecting what we have sacrificed in order to obtain Eq.~(\ref{topDecayNWA.col.EQ}).
The first is that we have considered only a single slice of phase space, namely when the $W$ boson is on-shell.
An entire continuum of phase space for $q^2 \neq M_W^2$ has been neglected and thus Eq.~(\ref{topDecayNWA.col.EQ}) 
is an underestimation, albeit a very good one, of the actual LO $t\rightarrow \ell^+ \nu_\ell b$ partial width.
As the stipulation that $\Gamma/M \ll 1$ breaks down, 
larger regions of phase space where $q^2 \neq M_W^2$ become increasingly important,
and lead to worse estimates of the partial width.

More subtle is the fact that spin correlation between the initial-state top quark and final-state leptons has been lost due to the use of the 
completeness relationship in Eq.~(\ref{vectorCompleteness.col.EQ}), which acts to decouple the initial-state and final-state fermion currents.
Though the total scattering and decay rates remain unaffected when the NWA holds, 
angular distributions and ``fiducial'' rates that are obtained by imposing phase space cuts will be inaccurate.
The exception of course being the case of a narrow scalar mediator, for example: a charged Higgs in the 2HDM.
In the top quark decay $t\rightarrow H^+ b, ~ H^+ \rightarrow \tau^+ \nu_\tau$, 
the completeness relationship for scalars can be imposed without any loss of spin correlation because, as a scalar, 
the scalars carry no such information.

An alternative procedure that preserves spin correlation would be to forgo the use of Eq.~(\ref{vectorCompleteness.col.EQ}).
Starting from the $t\rightarrow W^{*}b \rightarrow \ell^+ \nu_\ell b$ matrix element given in Eq.~(\ref{topFullME.col.EQ}), we have
\begin{eqnarray}
 \mathcal{M}(t\rightarrow \ell^+ \nu_\ell b) &=& 
 \mathcal{M}^\mu(t\rightarrow W^{*}b) ~ \frac{-i g_{\mu\nu}}{q^2 - M_W^2 + iM_W\Gamma_W} ~ \mathcal{M}^{\nu}(W^{*}\rightarrow \ell^+ \nu_\ell)
 \\
 &=& \mathcal{M}^\mu(t\rightarrow W^{*}b)\mathcal{M}_{\mu}(W^{*}\rightarrow \ell^+ \nu_\ell)~ \frac{-i}{q^2 - M_W^2 + iM_W\Gamma_W} 
\end{eqnarray}
Repeating the above procedure will give us
\begin{equation}
 \sum_{\rm dof} \vert \mathcal{M}(t\rightarrow \ell^+ \nu_\ell b) \vert^2 
 \approx
 \sum_{\rm dof} \vert \mathcal{M}^\mu(t\rightarrow W^{*}b)\mathcal{M}_{\mu}(W^{*}\rightarrow \ell^+ \nu_\ell)\vert^2
 \times \frac{\pi}{M_W\Gamma_W}\delta(q^2 - M_W^2). 
\end{equation}
Averaging and integrating over a decomposed phase space gives us
\begin{eqnarray}
  \Gamma(t\rightarrow \ell^+ \nu_\ell b) &=& 
 \int dPS_2(t;W^{*},b)~dPS_2(W^{*};\mu^+,\nu_\mu)~\frac{d q^2}{2\pi} \times
   \frac{1}{2m_t (2s_t+1)N_c^t} ~   \nonumber\\
  &\times& 
   \sum_{\rm dof} \vert \mathcal{M}^\mu(t\rightarrow W^{*}b)\mathcal{M}_{\mu}(W^{*}\rightarrow \ell^+ \nu_\ell)\vert^2
 \times \frac{\pi}{M_W\Gamma_W}\delta(q^2 - M_W^2). 
  \\
  &=& \frac{1}{2^2 m_t M_W\Gamma_W(2s_t+1)N_c^t} ~    \int dPS_2(t;W,b)~dPS_2(W;\mu^+,\nu_\mu)~
   \\
  &\times&    \sum_{\rm dof} \vert \mathcal{M}^\mu(t\rightarrow Wb)\mathcal{M}_{\mu}(W\rightarrow \ell^+ \nu_\ell)\vert^2.
  \end{eqnarray}
  Inserting the closed expressions for each of the two-body phase spaces and grouping together factors of $2\pi$, we obtain the spin-correlated expression
  \begin{eqnarray}
  \Gamma(t\rightarrow \ell^+ \nu_\ell b) 
  &=& \frac{\sqrt{1-2(r_W+r_b)+(r_W-r_b)^2}}{2^{12}\pi^3 m_t M_W\Gamma_W N_c^t} ~\int d\cos\theta_{b} ~d\Omega_{\mu}
   \\
  &\times&    \sum_{\rm dof} \vert \mathcal{M}^\mu(t\rightarrow Wb)\mathcal{M}_{\mu}(W\rightarrow \ell^+ \nu_\ell)\vert^2,
\end{eqnarray}
where $r_X = m_X / m_t^2$.
Despite its apparent bulkiness, the expression above can be evaluated analytically or numerically with little additional effort.
The key point is that numerical integration of the original three-body phase space (4 integrals) over a BW propagator is inefficient and
can be approximated well by one fewer integrals over zero propagators.

\section{Partonic Level Cross Section}
The statistical nature of quantum mechanics lends itself to counting experiments to test predictions made by models.
In colliders, antiparallel particle beams are focused onto each other in order to reproduce a type of Rutherford scattering.
For a given flux, or luminosity, $\mathcal{L}$ of particles transversing 
through an effective scattering area, or cross section, $\sigma$,
the number of scattering events is given schematically by
\begin{equation}
 {\rm Number~of~events} = 
  \underset{\rm Luminosity,~\mathcal{L}}{\underbrace{\rm (Number~of~particles~per~beam~area)}} 
~\times~ 
  \underset{\rm Cross~section,~\sigma}{\underbrace{\rm (Effective~target~area)}}
\end{equation}
For a fixed beam luminosity, we can interpret the cross section as a measure of the likelihood for a particular scattering to occur.
Again, schematically, this is given by
\begin{eqnarray}
  {\rm (Scattering~cross~section)} &=& \frac{\rm Number~of~events}{\rm Incoming~particle~flux} \\
  &=& \frac{\rm (Scattering~likelihood) ~\times~ (Scattering~configurations)}{\rm Incoming~particle~flux}
\end{eqnarray}
From Fermi's Golden Rule, we identify the numerator of this expression as simply the squared matrix element 
summed over discrete final-state degrees of freedom and integrated over continuous ones, i.e., phase space.
For randomly polarized and charged initial states, symmetry factors must be introduced to average over initial-state degrees of freedom.
And as is typical for scattering experiments, flux can be factored into the product of number densities of each beam and the relative velocity of the two,
implying its invariance under longitudinal boosts (along the beam line).
The scattering cross section can now be written (schematically) as
\begin{eqnarray}
  {\rm (Scattering~cross~section)} &=&   \frac{1}{\rm (Number~density)\times(Relative~velocity)\times(Symmetry~factors)}\nonumber\\
  &\times&   \sum_{\rm Discrete~dof} ~\int d{\rm [Phase~space]}~ {\rm (Probability~density)}
\end{eqnarray}
Formally, for incoming particles $A$ and $B$, with masses $m_A,~m_B$ and c.m.~energy
\begin{equation}
\sqrt{\hat{s}}=\sqrt{(p_A + p_B)^2}, 
\end{equation}
the $2\rightarrow n$ scattering rate is given by the formula 
\begin{eqnarray}
\sigma(A+B	&\rightarrow& X+ {\rm anything~else}) = \int dPS_n ~ \frac{d\sigma}{dPS_n},\\
 \frac{d\sigma}{dPS_n} &=& 
 \frac{1}{2\hat{s}\lambda^{1/2}(1,r_{A},r_{B})} \frac{1}{(2s_{A}+1)(2s_{B}+1)N_{c}^A N_{c}^B} \sum_{\rm dof}\vert\mathcal{M}\vert^{2}, 
\end{eqnarray}
where, for $X=A,B$, $r_X = m_X^2/\hat{s}$, 
$\lambda$ is the kinematic K\"allen function of Eq.~(\ref{kallen.col.eq}), 
$(2s_X+1)$ represents the number of spin states possessed by particle $X$,
$N_c^X$ is the SU$(3)_c$ color factor of $X$,
$dPS_n$ denotes the Lorentz-invariant $n$-body differential phase space as defined in Eq.~(\ref{phaseSpaceDef.col.EQ}), and
$\mathcal{M}$ is the Lorentz-invariant matrix element for scattering process 
\begin{equation}
 A + B \rightarrow X.
\end{equation}
In the $r_{B}\rightarrow0$ limit, $\lambda^{1/2}(1,r_{A},0)=(1-r_{A})$; and for $r_{A},r_{B}\rightarrow0$, $\lambda\rightarrow 1$.

\subsection{Example: $Zh$ Production at Electron Colliders}
Lepton collider-based Higgs factories are premised on the fact that Z bosons couple directly to both electrons and the Higgs boson,
and so Higgs bosons can be produced in the $2\rightarrow 2$ process
\begin{equation}
 e^{-}_\tau(p_A) + e^{+}_{\tau'}(p_B) \rightarrow Z_\lambda(p_Z) + h(p_h),
\end{equation}
where $\tau=L,R$ and $\lambda=0,\pm$ denote the helicity takes of electrons and the $Z$.
For massless electrons, the matrix element is given by
\begin{eqnarray}
 \mathcal{M}^{Zh}_{\lambda\tau'\tau} &=&  
 \kappa_Z \left[\overline{v}_{\tau'}(p_B)\xSlash{\varepsilon}_\lambda^{*}(p_Z)\left(c_R^e P_R+ c_L^e P_L\right)u_\tau(p_A)\right]~D_Z(s),
 \\
 c_R^e &=& g_V^e + g_A^e = \frac{1}{2}(T_3^e)_L - Q^e\sin^2\theta_W + \left(\frac{-1}{2}\right)(T_3^e)_L  = \sin^2\theta_W,
 \\
 c_L^e &=& g_V^e-g_A^e = \frac{1}{2}(T_3^e)_L - Q^e\sin^2\theta_W - \left(\frac{-1}{2}\right)(T_3^e)_L = \sin^2\theta_W - \frac{1}{2},
 \\
 D_X(p^2) &=& \frac{1}{(p^2 - M_X^2) + iM_X\Gamma_X }, \quad \kappa_Z = \frac{g^2 M_Z}{\cos^2\theta_W}
\end{eqnarray}
where the helicities of $e^-$, $e^+$, and $Z$ are denoted, respectively, by $\tau,~\tau',~\lambda$.
In the center of mass frame, the 4-momenta can be expressed as
\begin{eqnarray}
 p_A &=& \frac{\sqrt{s}}{2}~(1,0,0,1), \quad p_B = \frac{\sqrt{s}}{2}~(1,0,0,-1)
 \\
 p_Z &=& (E_Z, \vert\vec{p}_Z\vert\sin\theta\cos\phi, \vert\vec{p}_Z\vert\sin\theta\sin\phi, \vert\vec{p}_Z\vert\cos\theta),
 \\
 p_h &=& (E_h, -\vert\vec{p}_Z\vert\sin\theta\cos\phi, -\vert\vec{p}_Z\vert\sin\theta\sin\phi, -\vert\vec{p}_Z\vert\cos\theta),
 \\
  E_Z &=& \frac{\sqrt{s}}{2}(1+r_Z-r_h), 
  \quad \vert\vec{p}_Z\vert = \frac{\sqrt{s}}{2}\lambda^{1/2}(1,r_Z,r_h),
  \quad E_h = \frac{\sqrt{s}}{2}(1+r_h-r_Z),
\end{eqnarray}
and for $X=Z,h$  we define $r_X = m_X^2  / s$.
The nonzero amplitudes are given by
\begin{eqnarray}
\mathcal{M}_{0RL} &=&  \frac{-c_L}{2}\kappa_Z D_Z(s)\sqrt{\frac{s}{r_Z}} ~ (1 - r_h + r_Z)  e^{-i\phi}\sin\theta
\\
\mathcal{M}_{0LR} &=&  \frac{-c_R}{2}\kappa_Z D_Z(s)\sqrt{\frac{s}{r_Z}} ~   (1 - r_h + r_Z)  e^{i\phi}\sin\theta
\\
\mathcal{M}_{+RL} &=&  -c_L~\kappa_Z D_Z(s)\sqrt{\frac{s}{2}}  e^{-i\phi}  (1 - \cos\theta)
\\
\mathcal{M}_{+LR} &=&  c_R~\kappa_Z D_Z(s)\sqrt{\frac{s}{2}}  e^{i\phi}  (1 + \cos\theta)
\\
\mathcal{M}_{-RL} &=& -c_L~\kappa_Z D_Z(s)\sqrt{\frac{s}{2}}  e^{-i\phi}  (1 + \cos\theta)
\\
\mathcal{M}_{-LR} &=&  c_R~\kappa_Z D_Z(s)\sqrt{\frac{s}{2}}  e^{i\phi}  (1 - \cos\theta).
\end{eqnarray}
As the vector couplings are helicity conserving, the $(\lambda,\tau',\tau)=(\lambda,L,L)$ and $(\lambda,R,R)$ contributions are zero.
We observe longitudinal enhancement in the $(\lambda,\tau',\tau) = (0RL)$ and $(0RL)$ helicity amplitudes.
The squared and summed amplitude is then
\begin{eqnarray}
\sum\vert\mathcal{M}_{Zh}\vert^2 &=& \frac{g^4(c_L^2 + c_R^2)s^2}{8 c_W^4} \vert D_Z(s)\vert^2
\nonumber\\
&\times&
 \left[ (1  - r_h)^2 - 2 r_h r_Z +    r_Z (14 + r_Z) + \left(1 + 2 r_h r_Z -(1  - r_h)^2  - (1 - r_Z)^2   \right) \cos(2\theta) \right],
 \nonumber\\
 \vert D_Z(s)\vert^2 &=& \frac{1}{(s - M_Z^2)^2 + (M_Z\Gamma_Z)^2 }.
\end{eqnarray}
The corresponding 2-body phase space is
\begin{eqnarray}
dPS_{2}(P_{\rm Tot};p_{Z},p_{h})&=&\frac{d\cos\theta d\phi}{2(4\pi)^{2}} \lambda^{1/2}(1,r_Z,r_h)
\\
&=& \frac{d\cos\theta}{2^4\pi} \sqrt{1-2(r_Z+r_h)+(r_Z - r_h)^{2}}. 
\end{eqnarray}
Integrating and averaging over the quantum numbers of the initial states, 
the $Zh$ production cross section at $\sqrt{s}=250\GeV$ for $m_h = 125\GeV$ is
\begin{eqnarray}
\sigma(e^- e^+	\rightarrow Z h)  &=& 
\underset{\rm GeV^{-2}~to~fb~conversion~factor}{\underbrace{(\hbar c)^2} ~ \times}
~
\underset{\rm Flux}{\underbrace{\left(\frac{1}{2^{3}s}\right) }}
\times \nonumber\\
&\times&
 \frac{g^4 s^2 \left(c_L^2+c_R^2\right)}{3 c_W^4}
 \frac{\left(1 + 10r_Z - 2r_h + (r_Z-r_h)^2 \right)}{(s - M_Z^2)^2 + (M_Z\Gamma_Z)^2 }
\nonumber\\
&\times& \underset{\rm Phase~space}{\underbrace{\left(\frac{\lambda^{1/2}(1,r_Z,r_h)}{2^4\pi}\right)}}\approx 240~\fb.
\end{eqnarray}

\section{Initial-State Photons from Electron-$X$ Scattering}\label{isPhotons.col.sec}
On-shell factorization of scattering amplitudes into a product of \textit{universal}, i.e., process-independent, terms
and (usually) much simpler (though process dependent) hard scattering  matrix element calculations for 
participants whose masses are much smaller than momentum transfers scales is a central tenet of perturbative QCD and collider experiments.
Here we derive $\gamma X$ scattering of a quasi-real photon that originates from a nearly collinear splitting with an electron in high energy $eX$ collisions.
It should be emphasized that this is a property of gauge theories, not unique to photons or electrons,
and holds with very minor modifications, though with various physics interpretations, for $gg$ or $gq$ splitting.

\begin{figure}[!t]
\centering
\includegraphics[width=.7\textwidth]{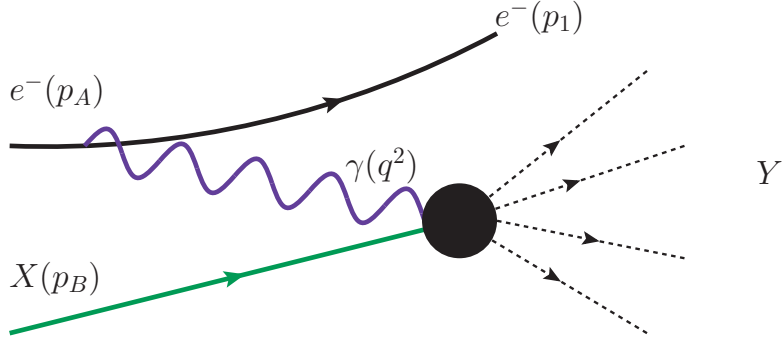}
\caption{Diagrammatic representation of quasi-real, initial-state photon from $e\gamma$ splitting.} 
\label{eX_eAX.FIG}
\end{figure}

As shown in Fig.~(\ref{eX_eAX.FIG}), we consider the $eX$ scattering process
\begin{equation}
 e^-(p_A) ~+~ X(p_B) \rightarrow e^-(p_1) ~+~ \gamma^*(q) ~+~ X(p_B) \rightarrow e^-(p_1) ~+~ Y
\end{equation}
that is mediated by the subprocess
\begin{equation}
 \gamma^*(q) + X(p_B) \rightarrow  Y, \quad q = p_A - p_1.
 \label{aXScatt.col.eq}
\end{equation}
The total $eX$ and subprocess $\gamma^* X$ center of mass energies are respectively denoted by
\begin{equation}
 s = (p_A + p_B)^2, \quad \hat{s} = (q + p_B)^2.
\end{equation}
In essence, the initial-state electron radiates a photon that participates in the $2\rightarrow 1$ scattering but otherwise acts as a spectator 
of the process.
The matrix element is given by
\begin{eqnarray}
 \mathcal{M} &=& \left[\overline{u}(p_1) (-ie q_e)\gamma^\mu u(p_A)\right] 
 \frac{(-i)\left(g_{\mu\nu} - (\xi-1)q_\mu q_\nu/q^2\right)}{q^2}\mathcal{M}^\nu(\gamma^* X\rightarrow Y)
 \\
 &=& \frac{(-i e q_e)}{q^2} \left[\overline{u}(p_1)\gamma^\mu u(p_A)\right] 
 \left(\sum_{\lambda\lambda'}\varepsilon_\mu^*(q,\lambda) \varepsilon_\nu(q,\lambda')\right)
 \mathcal{M}^\nu(\gamma^* X\rightarrow Y).
  \\
 &=& \frac{(-i e q_e)}{q^2}\sum_{\lambda\lambda'}
 \left[\overline{u}(p_1)\not\!\varepsilon u(p_A)\right] 
 \left[\varepsilon\cdot \mathcal{M}(\gamma X\rightarrow Y)\right]
 \\
 &\equiv& \frac{(-i e q_e)}{q^2}\sum_{\lambda\lambda'} \mathcal{M}(e\rightarrow e\gamma^*) \mathcal{M}(\gamma^*X\rightarrow X).
\end{eqnarray}
The quantity $\mathcal{M}^\nu(\gamma X\rightarrow Y)$ is defined such that the scattering process Eq.~(\ref{aXScatt.col.eq}) 
for an on-shell (massless), initial-state photon is given by
\begin{equation}
 \mathcal{M}(\gamma_\lambda X\rightarrow Y) = \varepsilon_\mu(q,\lambda) \cdot \mathcal{M}^\nu(\gamma_\lambda X\rightarrow Y).
\end{equation}
We keep explicit the electron charge as $q_e$. 
In the second line, we applied the completeness relationship of Eq.~(\ref{vectorCompleteness.col.EQ})
and find that we can express the entire matrix element as the product of two subprocess matrix elements:
$e\rightarrow\gamma$ splitting and $\gamma X$ scattering.
Squaring and summing over final state degrees of freedom gives us
\begin{eqnarray}
 \sum_{\rm d.o.f.} \vert M\vert^2  &=& 
 \frac{(e^2 q_e^2)}{q^4}\sum_{\lambda\lambda'}\sum_{\rm d.o.f.}
 \vert\mathcal{M}(e\rightarrow e\gamma^*)\vert^2 
 \vert\mathcal{M}(\gamma X\rightarrow X)\vert^2.
\end{eqnarray}
However, this is not quite correct as we are inadvertently double counting degrees of freedom of our  intermediate $\gamma^*$.
We must introduce spin-state and (in principle) color-state averaging factors, which effectively gives us a recipe for an 
unpolarized intermediate photon with, for now, an arbitrary virtuality.
Therefore, we should instead have
\begin{eqnarray}
 \sum_{\rm d.o.f.} \vert M\vert^2  &=& 
 \frac{(e^2 q_e^2)}{q^4}\sum_{\lambda\lambda'}\sum_{\rm d.o.f.}
 \vert\mathcal{M}(e\rightarrow e\gamma^*)\vert^2 
 \left(\frac{1}{(2s_\gamma+1)N_c^\gamma}\right)
 \vert\mathcal{M}(\gamma X\rightarrow X)\vert^2.
\end{eqnarray}
The obviously trivial color factor $N_c^\gamma$ is present for completeness.
We are also free to introduce the K\"allen function
\begin{equation}
 \hat{s}\lambda^{1/2}(1,\tilde{r}_\gamma,\tilde{r}_B), \quad \tilde{r}_\gamma =  \frac{q^2}{\hat{s}}, \tilde{r}_B =  \frac{p_B^2}{\hat{s}},
\end{equation}
which represents the energy available in $\gamma^* X$ scattering.
The spin- and color-averaged squared amplitude is then
\begin{eqnarray}
  \overline{\vert M\vert^2}  &=& 
 \frac{(e^2 q_e^2)}{q^4}
 \sum_{\lambda\lambda'}\sum_{\rm d.o.f.}
  \frac{ \hat{s}\lambda^{1/2}(1,\tilde{r}_\gamma,\tilde{r}_B)}{(2s_e + 1)N_c^e}
 \vert\mathcal{M}(e\rightarrow e\gamma^*)\vert^2 
 \nonumber\\
 & & \times 
 \left(\frac{1}{\hat{s}\lambda^{1/2}(1,\tilde{r}_\gamma,\tilde{r}_B)(2s_\gamma+1)(2s_B+1)N_c^\gamma B_c^b}\right)
 \vert\mathcal{M}(\gamma^*X\rightarrow X)\vert^2.
\end{eqnarray}
The $eX$ cross section is then given by
\begin{eqnarray}
 \sigma(e^-X\rightarrow e^-Y) &=& \int \frac{d^{3}p_{1}}{(2\pi)^{3}2E_{1}}
 \sum_{\lambda, \rm d.o.f.}
  \frac{ \hat{s}\lambda^{1/2}(1,\tilde{r}_\gamma,\tilde{r}_B)}{s\lambda^{1/2}(1,r_e,r_B)}
  \frac{\vert\mathcal{M}(e\rightarrow e\gamma^*)\vert^2 }{(2s_e + 1)N_c^e}
 \nonumber\\
 &\times & \frac{(4\pi\alpha q_e^2)}{q^4} \times  \hat{\sigma}(\gamma X\rightarrow Y),
\end{eqnarray}
where $r_i = m_i^2/s$, $e^2 = 4\pi\alpha$, we have split the $n$-body phase space into two using Eq.~(\ref{dPSsplit.sm.col}), 
and $\hat{\sigma}$ is the subprocess $\gamma X$ cross section.
Assuming that the scatting scales $s$ and $\hat{s}$ are much larger than any mass relevant initial-state mass,
this further refines down to
\begin{eqnarray}
 \sigma(e^-X\rightarrow e^-Y) &=& \int \frac{d^{3}p_{1}}{(2\pi)^{3}2E_{1}}
 \sum_{\lambda, \rm d.o.f.}
  \frac{ \hat{s}}{2s}
  \vert\mathcal{M}(e\rightarrow e\gamma^*)\vert^2
\frac{(4\pi\alpha q_e^2)}{q^4}
 \hat{\sigma}(\gamma X\rightarrow Y).
\end{eqnarray}
It now remains to evaluate the $e\rightarrow\gamma$ splitting matrix elements.

\subsection{$e\rightarrow \gamma$ Splitting}
For $e^-(p_A) \rightarrow e^-(p_1) \gamma^*(q)$ splitting, 
where $\gamma^*$ with small transverse momentum $q_T\ll E_A$ carries away an energy fraction $z$ from its parent electron,
we assign the following momenta
\begin{eqnarray}
  p_A &=& (E_A, 0, 0, E_A)\\
  p_1 &=& (1-z)E_A (1, \sin\theta_1 ,0, \cos\theta_1) \nonumber\\
      &=&\left((1 - z) E_A, -q_T, 0, (1 - z) E_A - \frac{q_T^2}{2(1-z) E_A} \right)\\
  q   &=&  (z E_A, \vert\vec{q}\vert \sin\theta_\gamma , 0,\vert\vec{q}\vert \cos\theta_\gamma)
  \\
      &=&\left(z E_A, q_T, 0, z E_A + \frac{q_T^2}{2 (1-z) E_A} \right).
\end{eqnarray}
The corresponding (complex conjugated) photon polarization vectors in the helicity basis are
\begin{eqnarray}
 \varepsilon^{\lambda=+*}_\mu &=& \frac{1}{\sqrt{2}}\left(0, -\cos\theta_\gamma, i, -\sin\theta_\gamma\right)
 \\
 \varepsilon^{\lambda=-*}_\mu &=& \frac{1}{\sqrt{2}}\left(0, \cos\theta_\gamma, i, \sin\theta_\gamma\right).
\end{eqnarray}
Then, neglecting terms higher than of $\mathcal{O}\left(\frac{q_T^2}{E_A^2}\right)$, we have
\begin{eqnarray}
 p_1^2 &=& (1-z)^2 E_A^2 - q_T^2 - (1 - z)^2 E_A^2 +  \frac{2(1-z)E_A q_T^2}{2(1-z) E_A} + \mathcal{O}\left(\frac{q_T^4}{E_A^4}\right) = 0
 \\
 q^2 &=& z^2 E_A^2 - q_T^2 - z^2 E_A^2 - \frac{2z E_A q_T^2}{2(1- z) E_A} + \mathcal{O}\left(\frac{q_T^4}{E_A^4}\right) = \frac{-q_T^2}{(1-z)},
\end{eqnarray}
that is: a massless final-state $e^-$ and internal photon with virtuality proportional to its transverse momentum.

As photon radiation is helicity-conserving, the only nonzero fermion currents are the $e_L^- \rightarrow e_L^- \gamma$
$e_R^- \rightarrow e_R^-\gamma$ channels, given by
\begin{eqnarray}
 J_{LL}^\mu &=&  
2 E_A \sqrt{1-z}\left[
 \cos \left(\frac{\theta_1}{2}\right),  \sin \left(\frac{\theta_1}{2}\right),- i   \sin \left(\frac{\theta_1}{2}\right), \cos \left(\frac{\theta_1}{2}\right)
 \right]
 \\
 J_{RR}^\mu &=& 
 2 E_A \sqrt{1-z}\left[ 
 \cos \left(\frac{\theta_1}{2}\right), \sin \left(\frac{\theta_1}{2}\right), i   \sin \left(\frac{\theta_1}{2}\right), \cos \left(\frac{\theta_1}{2}\right) \right].
\end{eqnarray}
The permutation of helicity amplitudes (in the small $q_T$ limit) is therefore
\begin{eqnarray}
\mathcal{M}_{LL-} &=& 2 E_A \sqrt{2-2 z} \cos \left(\frac{\theta_\gamma }{2}\right) \sin \left(\frac{\theta_1+\theta_\gamma }{2}\right)\\
		&\approx&  E_A \sqrt{2-2 z}  \left(\theta_1+\theta_\gamma\right)
\\
\mathcal{M}_{LL+} &=& -2 E_A \sqrt{2-2 z} \sin \left(\frac{\theta_\gamma }{2}\right) \cos \left(\frac{\theta_1+\theta_\gamma }{2}\right)\\
 &\approx& - E_A \sqrt{2-2 z}  \left(\theta_\gamma\right)
 \\
\mathcal{M}_{RR-} &=& 2 E_A \sqrt{2-2 z} \sin \left(\frac{\theta_\gamma }{2}\right) \cos \left(\frac{\theta_1+\theta_\gamma }{2}\right)\\
&\approx&  E_A \sqrt{2-2 z}  \left(\theta_\gamma\right) 
 \\ 
\mathcal{M}_{RR+} &=& -2 E_A \sqrt{2-2 z} \cos \left(\frac{\theta_\gamma }{2}\right) \sin \left(\frac{\theta_1+\theta_\gamma }{2}\right)\\
&\approx& - E_A \sqrt{2-2 z}   \left(\theta_1+\theta_\gamma\right)
\end{eqnarray}
Squaring the amplitudes, making the replacements
\begin{equation}
\theta_\gamma \rightarrow  \frac{q_T}{z E_A},\quad \theta_1 \rightarrow  \frac{q_T}{(1-z) E_A},
\end{equation}
and summing, we obtain
\begin{equation}
 \sum\vert\mathcal{M}(e\rightarrow e\gamma^*)\vert^2 = \frac{4 q_T^2}{z(1-z)}P_{\gamma e}(z),
\end{equation}
where $P_{\gamma e}(z)$ is the universal Altarelli-Parisi splitting function
\begin{equation}
 \boxed{P_{\gamma e}(z) = \frac{1+(1-z)^2}{z}}.
\end{equation}
Its pole represents the soft divergence that appears when $q_T$, and hence $\gamma^*$'s virtuality goes to zero.
The function $P_{\gamma e}(z)$ is universal 
in the sense that it holds for all spin-half-to-internal spin-one bosons splittings in the small transverse momentum limit.
Accounting for color factors, the QCD equivalent is in fact given by
\begin{equation}
 P_{g q}(z) = C_F \frac{1+(1-z)^2}{z}.
\end{equation}
Had we considered instead an internal electron and on-shell photon, neglecting terms higher than $\mathcal{O}\left(\frac{q_T^2}{E_A^2}\right)$,
the $e_i\rightarrow e_f$ splitting function, where $e_f$ carries a momentum fraction $z$ from $e_i$, is
\begin{equation}
\boxed{ P_{ee}(z) = \frac{1+z^2}{1-z}},
\end{equation}
and possesses a collinear divergence. It is interesting to note that the $\gamma\rightarrow e$ splitting function
\begin{equation}
\boxed{ P_{e\gamma}(z) = z^2 + (1-z)^2}
\end{equation}
does not have a pole as massless fermion currents turn off (vanish) in the zero fermion energy limit.

\subsection{Weizs\"acker-Williams Approximation}
We can now assemble our final result. We start by expressing our one-body phase space in cylindrical coordinates,
and ultimately photon virtuality and momentum fraction $z$
\begin{eqnarray}
 \frac{d^{3}p_{1}}{(2\pi)^{3}2E_{1}} &=& \frac{d\phi_1~ dp_z ~d q_T^2}{(2\pi)^{3}2^2 (1-z)E_A}  = \frac{dz ~d q_T^2}{(2^4\pi^2)(1-z)}
 =
 \frac{dz~ d q^2}{(2^4\pi^2)},
\end{eqnarray}
where we made use of the fact that
\begin{equation}
 p_z \simeq E_1 = (1-z)E_A \quad\text{and}\quad  q^2 = \frac{- q_T^2}{(1-z)}.
\end{equation}
Under the working assumptions that masses are negligible compared to the scattering scale and that we are in the 
collinear (small $q_T$) $e\rightarrow\gamma$ splitting regime, we also have
\begin{equation}
 \hat{s} = (q + p_B)^2 = 2 q\cdot p_B = z p_A \cdot p_B = z s.
\end{equation}
Making the appropriate substitutions, we then have
\begin{eqnarray}
 \sigma(e^-X\rightarrow e^-Y) &=&
 \int \frac{dz~ d q^2}{(2^4\pi^2)}
 \frac{z}{2}
 \frac{4(1-z) q^2}{z(1-z)}P_{\gamma e}(z)
\frac{(4\pi\alpha q_e^2)}{q^4}
 \hat{\sigma}(\gamma X\rightarrow Y)
 \\
 &=&
 \int  dz ~
 \frac{\alpha q_e^2}{2\pi} ~P_{\gamma e}(z)
\int
\frac{d q^2}{q^2}
 \hat{\sigma}(\gamma X\rightarrow Y)
\end{eqnarray}
A pause is necessary to to address the limits for the virtuality integrations.
At zero momentum transfer $(q^2=0)$, the cross section diverges but only artificially.
We made the assumption that masses are negligible compared to the hard scattering energies,
but at zero momentum transfer this is no longer true. Strictly speaking, the photon virtuality is given by
\begin{equation}
 q^2 = (p_A - p_1)^2 = 2m_e^2 - E_A E_1(1 - \beta_A \beta_1\cos\theta_1),
\end{equation}
indicating that the supposed collinear divergence is actually regulated by $\beta_{A,1}<1$, or in other words, the electron mass.
For quark-gluon splitting, this is regulated analogously by the bare quark mass.
The electron mass then sets the scale for momentum transfers and we \textit{evolve} our momentum transfer scale starting from $q^2 = m_e^2$.
The upper limit of integration must be chosen based on its type of calculation that is being performed.
For inclusive cross section calculations, and despite contradicting the small $q_T$ assumption, 
evolving the integral upwards to $\hat{s}$ is a reasonable estimation~\cite{Peskin:1995ev}.
However, as we will discuss shortly later and much detail in later chapters, 
this can be matched with the deeply inelastic process, rendering the \textit{sum} of the two components
relatively scale-independent~\cite{Alva:2014gxa}. For now, we will evolve our integral upwards to $q^2 = Q^2$.
The evolution scale $Q^2$ is also called the \textit{factorization scale}.
Doing so gives us  
  \begin{eqnarray}
   \sigma(e^-X\rightarrow e^-Y)
  &=&
 \int  dz ~
 \frac{\alpha q_e^2}{2\pi} ~P_{\gamma e}(z)
\log\left(\frac{Q^2}{m_e^2}\right)~
 \hat{\sigma}(\gamma X\rightarrow Y).
\end{eqnarray}
This result, also known as the Weizs\"acker-Williams~\cite{vonWeizsacker:1934sx,Williams:1934ad} approximation,
is expressed more commonly in the form
\begin{eqnarray}
 \sigma(e^-X\rightarrow e^-Y) =  \int^1_{z_{\min}}  dz ~ f_{\gamma/e}(z,Q^2)  ~  \hat{\sigma}(\gamma X\rightarrow Y),
\end{eqnarray}
where the \textit{photon distribution function} is given by
 \begin{eqnarray}
\boxed{ f_{\gamma/e}(z,Q^2) = \frac{\alpha q_e^2}{2\pi} ~P_{\gamma e}(z)\log\left(\frac{Q^2}{m_e^2}\right)}.
\label{photonPDF.col.eq}
\end{eqnarray}
Written in this form, we interpret $f_{\gamma/e}(z,Q^2)$ as the likelihood of 
observing a photon in an electron, possessing an energy fraction $z$ of the electron's total energy at a momentum transfer scale $Q^2$.
The limits of integration are derived from the relation $\hat{s} = z s$, which tells us that
\begin{eqnarray}
 \max(z) &=& \max \left(\frac{\hat{s}}{s}\right) = \frac{\max (\hat{s})}{s} = \frac{s}{s} =1\\
 \min(z) &=& \min \left(\frac{\hat{s}}{s}\right) = \frac{\min (\hat{s})}{s},
\end{eqnarray}
and $\min(\hat{s})$ is the minimum invariant mass required for $\gamma B\rightarrow X$ to kinematically proceed.

\subsection{Weak Boson Distribution Functions}
Following the identical procedure with $W$ and $Z$ bosons will yield similar results.
The polarization-dependent distributions functions for Weak bosons 
carrying energy fraction $z$ from fermion $f$ evolved to a scale $Q_V\gg M_W$ are given by~\cite{Dawson:1984gx,Barger:1987nn}
\begin{eqnarray}
 P_{V/f}(z,Q^2,\lambda=\pm) &=& \frac{C}{16\pi^2 z}\left[(g_V^f \mp g_A^f)^2  + (g_V^2 \pm g_A^f)^2(1-z)^2\right]\log\frac{Q^2}{M_V^2}
 \\
 P_{V/f}(z,Q^2,\lambda=0)   &=& \frac{C}{4\pi}\left[(g_V^f)^2 + (g_A^f)^2\right]\left(\frac{1-z}{z}\right),
\end{eqnarray}
where for $V=W^\pm$ we have
\begin{equation}
 C=\frac{g^2}{8}, \quad g_V=-g_A=1,
\end{equation}
and for $V = Z$
\begin{equation}
 C=\frac{g^2}{\cos^2\theta_W}, \quad g_V = \frac{1}{2}(T_f^3) - Q_f \sin^2\theta_W, \quad g_A = \frac{-1}{2}(T_f^3).
\end{equation}
Summing and averaging the transverse $W^\pm$ distributions gives
\begin{eqnarray}
 f_{W_T/q}(z,Q_V^2) &=& \frac{C}{8\pi^2}\frac{\left[1 + (1-z)^2\right]}{z} \log\left(\frac{Q_V^2}{M_W^2}\right), 
\end{eqnarray}

\subsection{Beyond Leading Logarithm}
The above \textit{leading logarithm} (LL) result, gives us  a result of the form
\begin{equation}
 eX\text{-scattering} = e\gamma~\text{-splitting} \quad\otimes\quad \gamma X\text{-scattering}.
\end{equation}
However, the $e\gamma$-splitting function is universal and will reappear for each successive splitting.
This is particularly important for when the momentum transfer is very large, in which case 
\begin{equation}
 \alpha(Q^2)\log\left(\frac{Q^2}{m_f^2}\right) \sim 1,
\end{equation}
and  our perturbative treatment breakdowns. 
Though $\alpha(M_Z)\sim 1/128$ is quite small, is becomes a considerable problem from QCD where $\alpha_s(M_Z)\sim 0.1$.
The solution is actually to consider summing over an arbitrary number of parton splittings
\begin{equation}
 qX\text{-scattering} = \sum_{k=1}^{n} \underset{k\text{-splittings}}{\underbrace{
qg~\text{-splitting} \otimes \cdots qg~\text{-splitting}}} \otimes g X\text{-scattering}.
\end{equation}
The result is an expression that can be exponentiated 
\begin{equation}
 \sum \alpha^k(Q^2)\log^k\left(\frac{Q^2}{m_f^2}\right) \sim \exp\left[\alpha(Q^2)\log\left(\frac{Q^2}{m_f^2}\right) \right].
 \label{resum.col.eq}
\end{equation}
This process, only given schematically here, is called \textit{resummation} and is an all-orders, hence non-perturbative, result.
For the the case of QCD, collinear radiation is ``resummed'' using the 
Dokshitzer-Gribov-Lipatov-Altarelli-Parisi (DGLAP) equations,
resulting in what are known as the \textit{parton distribution functions} (PDFs).
These functions give the likelihood of observing a particular parton species, e.g., anti-strange quark or gluon, in a hadron, e.g., proton or Pb nuclei,
possessing a fraction $x$ of the hadron's energy at a momentum transfer of $Q^2$.
The distribution function $f_{\gamma/e}$ in Eq.~(\ref{photonPDF.col.eq}) is another example of a PDF.
By fixing the PDFs at a particular momentum transfer and energy fraction, 
in say deeply inelastic scattering (DIS) $ep$ experiments, the DGLAP equations are used to evolve the PDFs to a different scale, 
such as those observed in LHC collisions or potentially at a future 100 TeV collider.

\subsection{Elastic Photon PDF}\label{AppEl.col.sec}
It is worth noting that our previous results were reliant on the parton model and dealt with point-particles.
As the proton is charged, at momentum transfers below a couple GeV, it too can give rise initial-state photons in $pX$ collisions.
The elastic photon PDF for a proton is given analytically by~\cite{Budnev:1974de}
\begin{eqnarray}
 \label{elEPA.EQ}
 f_{\gamma/p}^{\rm El }(\xi) &=& \frac{\alpha_{\rm EM}}{\pi}\frac{(1-\xi)}{\xi}
 \left[
 \varphi\left(\frac{{\lamEl}^2}{Q_{0}^2}\right)
 -
  \varphi\left(\frac{Q_{\min}^2}{Q_{0}^2}\right)
 \right],
 \quad \alpha_{\rm EM} \approx 1/137,
 \label{modWWApprox.EQ}
 \\
 Q_{\min}^{2} &=& m_{p}^{2}y, 
 \quad y = \frac{\xi^{2}}{(1-\xi)},
 \quad Q_{0}^2 = 0.71~\text{GeV}^{2},
 \quad m_{p} = 0.938~\text{GeV},
 \\
 \varphi(x) &=& (1+a y) \left[-\log\left(1 + \frac{1}{x}\right) + \sum_{k=1}^{3}\frac{1}{k(1+x)^k}\right]
 + \frac{y(1-b)}{4x(1+x)^3}
 \nonumber\\
 &+& c\left(1+\frac{y}{4}\right)\left[\log\left(\frac{1+x-b}{1+x}\right) + \sum_{k=1}^{3} \frac{b^{k}}{k(1+x)^{k}}\right], 
 \label{varphi.EQ}
 \\
 a&=& \frac{1}{4}(1+\mu_{p}^{2})+\frac{4m_{p}^{2}}{Q_{0}^{2}} \approx 7.16,
 \quad b = 1 -\frac{4m_{p}^{2}}{Q_{0}^{2}} \approx -3.96, 
 \quad c = \frac{\mu_{p}^{2}-1}{b^{4}}\approx 0.028. 
 \end{eqnarray}
 Here, $\lamEl$ is a upper limit on elastic momentum transfers such that $f_{\gamma/p}^{\rm El }=0$ for $Q_{\gamma} >\lamEl$.
In Eq.~(\ref{modWWApprox.EQ}), and later in Eq.~(\ref{WWApprox.EQ}), since $Q_\gamma \ll m_Z$, 
$\alpha(\mu=Q_\gamma)\approx \alpha_{\rm EM} \approx 1/137$ is used.
In the hard scattering matrix elements, $\alpha(\mu = M_Z)$ is used. See Ref.~\cite{Drees:1994zx} for further details.
 
Equation (\ref{elEPA.EQ}) has been found to agree well with data from TeV-scale collisions at $Q_\gamma\sim m_\mu$~\cite{Chatrchyan:2011ci}.
However, applications to cases with larger momentum transfers and finite angles lead to large errors and increase scale sensitivity.
Too large a choice for $\lamEl$ will lead to overestimate of cross sections~\cite{Budnev:1974de}.
However, we observe negligible growth in $f^{\rm El}_\gamma$ at scales well above $\lamEl=1-2\GeV$, in agreement with Ref.~\cite{Sahin:2010zr}.

 Briefly, we draw attention to a typo in the original manuscript that derives Eq.~(\ref{elEPA.EQ}).
This has been only scantly been mentioned in past literature~\cite{deFavereaudeJeneret:2009db,Chapon:2009hh}. 
The sign preceding the ``$y(1-b)$'' term of $\varphi$ in Eq.~(\ref{varphi.EQ}) is erroneously flipped in Eq.~(D7) of Ref.~\cite{Budnev:1974de}.
Both CalcHEP~\cite{Pukhov:1999gg,Pukhov:2004ca,Belyaev:2012qa} and MG5\_aMC@NLO~\cite{Alwall:2014hca} have the correct sign in their default PDF libraries.
 
At these scales, the gauge state $\gamma$ is a understood to be a linear combination of 
discrete states: the physical (massless) photon and (massive) vector mesons $(\omega,\phi,...)$, and a continuous mass spectrum,
a phenomenon known as generalized vector meson dominance (GVMD)~\cite{Sakurai:1972wk}.
An analysis of ZEUS measurements of the $F_2$ structure function at $Q^2_\gamma < m_p^2$ and Bjorken-$x\ll1$ concludes that GMVD effects are 
included in the usual dipole parameterizations of the proton's electric and magnetic form factors $G_{\rm E}$ and $G_{\rm M}$~\cite{Alwall:2004wk}.
Thus, the radiation of vector mesons by a proton that are then observed as photon has been folded into Eq.~(\ref{elEPA.EQ}).

\section{Factorization Theorem, Parton Luminosities, and Hadronic Cross Section}
We are now in position to introduce hadronic level cross sections and the Factorization theorem. 
We start by considering some partonic-level process
\begin{equation}
 a + b \rightarrow X.
\end{equation}
We suppose that both $a$ and $b$ are massless and possess proton PDFs, denoted by $f_{a/p}(\xi_1,\mu^2)$ and $f_{b/p}(\xi_2,\mu^2)$,
where $\xi_i$ is the energy fraction of proton $P_i$, and $f_{i/p}$ are evolved to factorization scale $\mu^2$.
The partonic center of mass is denoted as $\hat{s} = (p_a + p_b)^2$, 
and the minimal invariant mass needed for the process to proceed is denoted by $\hat{s}_X$.
The $pa\rightarrow X+Y$ scattering rate  is then 
\begin{equation}
 \sigma(pa \rightarrow X +Y ) = \int_{\xi_{2}^{\rm min}}^{1}d\xi_{2}~ f_{b/p}(\xi_2)~ \hat\sigma(ab\rightarrow X),
\end{equation}
where 
\begin{equation}
 \hat{s} = (p_{a}+p_{b})^{2} = 2p_{a}p_{b} = 2P_{2}p_{a}\xi_{2} = s_{pa}\xi_{2}
\end{equation}
is the relation between the partonic and the $p-a$ system's c.m.~energies. This last line implies
\begin{equation}
 \xi_{2}^{\rm min} = \min\left( \frac{\hat{s}}{s_{pa}} \right)= \frac{\hat{s}^{\rm min}}{s_{pa}} = \frac{m_X^2}{s_{pa}},
 \quad\text{and}\quad  s_{pb}^{\rm min} = \min\left( \frac{\hat{s}}{\xi_{a}} \right) = m_X^2.
\end{equation}
Similarly, we can construct the $pp\rightarrow X+Y''$ scattering rate from the semi-partonic $pb$ initial-states.
The corresponding limits of integration for the splitting function integrals are
\begin{eqnarray}
 \xi_{1}^{\rm min} &=& \min\left( \frac{s_{pa}}{s_{pp}} \right) = \frac{s^{\rm min}_{pa}}{s_{pp}} = \frac{m_X^2}{s_{pp}} \equiv \tau_{\rm min}
 \\
 \xi_{2}^{\rm min} &=& \frac{m_X^2}{s_{pb}} = \frac{m_X^2}{s_{pp}\xi_{1}} = \tau_{\rm min}/\xi_1.
\end{eqnarray}
This gives us
\begin{eqnarray}
 \sigma(pp\rightarrow X+Y'') 
 &=& \int_{\xi_{1}^{\rm min}}^{1}d\xi_{1}~ f_{a/p}(\xi_1)~ \hat\sigma(pb\rightarrow X + Y),\\
 &=& \int_{\xi_{1}^{\rm min}}^{1}d\xi_{1}\int_{\xi_{2}^{\rm min}}^{1}d\xi_{2}~   f_{a/p}(\xi_1)f_{b/p}(\xi_2)~ \hat\sigma(ab\rightarrow X),\\
 &=& \int_{\tau_{\rm \min}}^{1}  d\xi_{1}\int_{\tau_{\rm \min}/\xi_{1}}^{1}d\xi_{2}~ f_{a/p}(\xi_1)f_{b/p}(\xi_2)~ \hat\sigma(ab\rightarrow X).
\end{eqnarray}
However, our assignment of $a$ to the first proton and $b$ to the second proton was arbitrary and indistinguishable from the reverse assignment.
Thus, we obtain the \textbf{Factorization Theorem}
\begin{equation}
\label{factTheorem.col.EQ}
\boxed{
  \sigma(pp\rightarrow X+Y'') 
 = \int_{\tau_{\rm min}}^{1}d\xi_{1}\int_{\tau_{\rm min}/\xi_{1}}^{1}d\xi_{2}~ 
 \left[ f_{a/p}(\xi_1,\mu^2)f_{b/p}(\xi_2,\mu^2) \hat\sigma(ab\rightarrow X) + (a\leftrightarrow b) \right]},
 \end{equation}
 where $\tau_{\rm}$ is the minimal energy fraction required for the process to be kinematically allowed
 \begin{equation}
   \tau \equiv \frac{\hat{s}}{s} = \xi_1\xi_2, \quad \tau_{\min} = \frac{\hat{s}_{\min}}{s}, 
 \end{equation}
 and states that \textit{a sufficiently inclusive hadronic-level scattering at sufficiently large momentum transfers
 can be expressed as a convolution of the partonic-level scattering with the probability (PDFs)
of observing the participating partons in the hadron.} 
In other words, the likelihood of observing a particular process in hadron collisions can be  obtained by ``multiplying'' (convolving)
the probabilities of partons reproducing the desired final-state and the likelihood of finding said probabilities in the scattering hadrons.

Using the relationship $\tau = \xi_1\xi_2$, we can make the change of variable
 \begin{equation}
  \int_{\xi_{2}^{\rm min}}^{1}d\xi_{2} = \int_{\tau_{\rm min}}^{1} \frac{d\tau}{\xi_1},
 \end{equation}
allowing us to write
\begin{eqnarray}
  \sigma(pp\rightarrow X+Y'') 
 =  \int_{\tau_{\rm min}}^{1}d\tau \int_{\tau}^{1}\frac{d\xi_1}{\xi_1}~ 
 \left[ f_{a/p}(\xi_1)f_{b/p}(\xi_2) \hat\sigma(ab\rightarrow X) + (1\leftrightarrow 2) \right].
 \end{eqnarray}
 For $2\rightarrow 1$ processes, this readily simplifies to
 \begin{eqnarray}
  \sigma(pp\rightarrow X+Y'') 
  &=&  \int_{\tau_{\rm min}}^{1}d\tau \int_{\tau}^{1}\frac{d\xi_1}{\xi_1}~ 
 \left[ f_{a/p}(\xi_1)f_{b/p}(\xi_2)  \frac{d\hat\sigma}{dPS_{1}} \int dPS_{1} + (1\leftrightarrow 2) \right]\\
  &=&  \int_{\tau_{\rm min}}^{1}d\tau \int_{\tau}^{1}\frac{d\xi_1}{\xi_1}~ 
 \left[ f_{a/p}(\xi_1)f_{b/p}(\xi_2)  \frac{d\hat\sigma}{dPS_{1}} \frac{2\pi}{s}\delta(\tau-\tau_{\rm min}) + (1\leftrightarrow 2) \right]\\ 
 &=&  \frac{2\pi}{s} \int_{\tau_{\rm min}}^{1}\frac{d\xi_1}{\xi_1}~ 
 \left[ f_{a/p}(\xi_1)f_{b/p}(\xi_2)  \frac{d\hat\sigma}{dPS_{1}}  + (1\leftrightarrow 2) \right].
 \end{eqnarray}
 
\subsection{Inelastic Photon PDF}\label{sec:AppIn}
Following the methodology of Ref.~\cite{Drees:1994zx}, 
we can extend our discussion on initial-state photons from electrons and protons in Section~\ref{isPhotons.col.sec}
to initial state photons from quarks in protons.
The inelastic cross section for producing final-state $X$ is given explicitly by
\begin{eqnarray}
  \sigma_{\rm Inel}(pp\rightarrow X &+& \text{anything}) = 
  \sum_{q,q'}
  \int^{1}_{\tau_0}d\xi_{1}\int^{1}_{\tau_0/\xi_1}d\xi_{2}\int^{1}_{\tau_0/\xi_1/\xi_2}dz
  \nonumber\\
  &\times&\left[
  f_{q/p}\left(\xi_1,Q_{f}^2\right) ~  f_{\gamma/q'}\left(z,Q_\gamma^2\right)  ~ f_{q'/p}\left(\xi_2,Q_{f}^2\right)
  \hat{\sigma}\left(q_1\gamma_2\right) + (1\leftrightarrow2)\right],
   \\
 & & \tau_0=m_{X}^{2}/s,	
 \quad	
 \tau = \hat{s} / s = \xi_1\xi_2 z. \nonumber
   \label{inelepa.EQ}
\end{eqnarray}
The Weizs\"acker-Williams photon structure function~\cite{Williams:1934ad,vonWeizsacker:1934sx} is given by
\begin{eqnarray}
 f_{\gamma/q}(z,Q_\gamma^2) = \frac{\alpha_{\rm EM} ~ e_{q}^{2}}{2\pi} 
 \left(\frac{1+(1-z)^2}{z} \right)
 \log\left(\frac{Q_{\gamma}^{2}}{\lamIn}\right),\quad
 \alpha_{\rm EM}\approx 1/137,	
 \label{WWApprox.EQ}
\end{eqnarray}
where $e_{q}^{2} = 4/9 ~ (1/9)$ for up-(down-)type quarks and $\lamIn$ is a low-momentum transfer cutoff.
In DGLAP-evolved photon PDFs~\cite{Martin:2004dh}, $\lamIn$ is taken as the mass of the participating quark.
Ref.~\cite{Drees:1994zx} argues a low-energy cutoff $\mathcal{O}(1-2)$ GeV  so that the 
associated photon is sufficiently off-shell for the parton model to be valid.
Taking $\lamIn = \lamEl = \mathcal{O}(1-2)\GeV$ 
allows for the inclusion of non-perturbative phenomena without worry of double counting of phase space~\cite{Alva:2014gxa}.

Fixing $z$ and defining $\xi_\gamma \equiv \xi_2 z$, we have the relationships
\begin{equation}
 \tau_0 = \min\left(\xi_1 \xi_2 z\right) = \min\left(\xi_1 \xi_\gamma \right)
 \implies \min(\xi_\gamma) = \frac{\tau_0}{\xi_1} \text{ for fixed } \xi_1.
\end{equation}
Physically, $\xi_\gamma$ is the fraction of proton energy carried by the initial-state photon.
Eq.~(\ref{inelepa.EQ}) can be expressed into the more familiar two-PDF factorization theorem, i.e., Eq.~(\ref{factTheorem.col.EQ}),
by grouping together the convolutions about $f_{q'/p}$ and $f_{\gamma/q'}$:
\begin{eqnarray}	
\sum_{q'}
 \int^{1}_{\tau_0/\xi_1}d\xi_{2}\int^{1}_{\tau_0/\xi_1/\xi_2}dz 
 ~ f_{\gamma/q'}(z ) 
 ~ f_{q'/p}(\xi_2 )
 &=& 
 \sum_{q'}
 \int^{1}_{\tau_0/\xi_1}\frac{d\xi_{\gamma}}{z}\int^{1}_{z_{\min}}dz 
~ f_{\gamma/q'}(z )
 ~ f_{q'/p}\left(\frac{\xi_\gamma}{z} \right)
 \\
 &=& 
 \int^{1}_{\tau_0/\xi_1} d\xi_\gamma ~f_{\gamma/p}^{\rm Inel}(\xi_\gamma)
 \\
 f_{\gamma/p}^{\rm Inel}\left(\xi_\gamma,Q_\gamma^2,Q_{f}^2\right)
 &\equiv& 
 \sum_{q'} \int^{1}_{z_{\min} = \xi_\gamma}\frac{dz}{z}  
 ~ f_{\gamma/q'}\left(z, Q_\gamma^2 \right)  
 ~ f_{q'/p}\left(\frac{\xi_\gamma}{z},Q_f^2 \right).~~~~~
\end{eqnarray}
The minimal fraction $z$ of energy that can be carried away by the photon from the quark corresponds to when the quark has 
the maximum fraction $\xi_2$ of energy from its parent proton. Thus, for a fixed $\xi_\gamma$, we have 
\begin{equation}
 1 = \max(\xi_2) = \max\left(\frac{\xi_\gamma}{z}\right) = \frac{\xi_\gamma}{\min(z)} \implies \min(z) = \xi_\gamma.
\end{equation}
The resulting expression is
\begin{eqnarray}
  \sigma_{\rm Inel}(pp\rightarrow N\ell^{\pm}X) = 
  \sum_{q}
  \int^{1}_{\tau_0}d\xi_{1} \int^{1}_{\tau_0/\xi_1} d\xi_2 
  \left[
  ~f_{q/p}\left(\xi_1,Q_{f}^2\right)
  ~f_{\gamma/p}^{\rm Inel}\left(\xi_2,Q_\gamma^2,Q_{f}^2\right) 
  \hat{\sigma}\left(q_1\gamma_2\right) + (1\leftrightarrow2)\right]\nonumber
\end{eqnarray}

Real, initial-state photons from inelastic quark emissions can be studied in MG5 by linking the appropriate Les Houches accord PDFs (LHAPDF) 
libraries~\cite{Whalley:2005nh} and using the MRST2004 QED~\cite{Martin:2004dh} or NNPDF QED~\cite{Ball:2013hta} PDF sets.
With this prescription, sub-leading (but important) photon substructure effects~\cite{Drees:1984cx}, 
e.g., $P_{g\gamma}$ splitting functions, are included in evolution equations.

  \begin{figure}[!t]
\centering
\subfigure[]{	\includegraphics[width=0.48\textwidth]{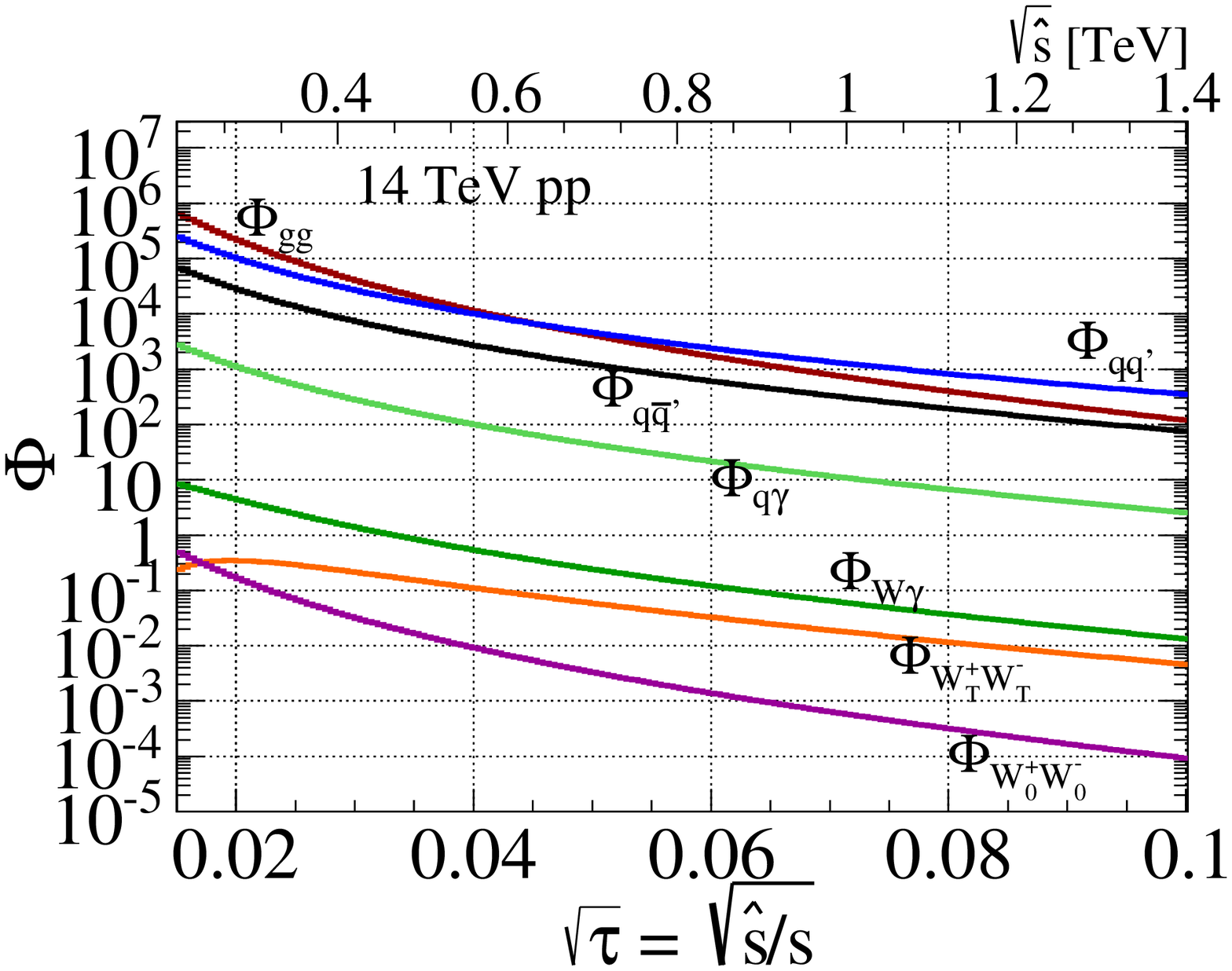}	}
\subfigure[]{	\includegraphics[width=0.48\textwidth]{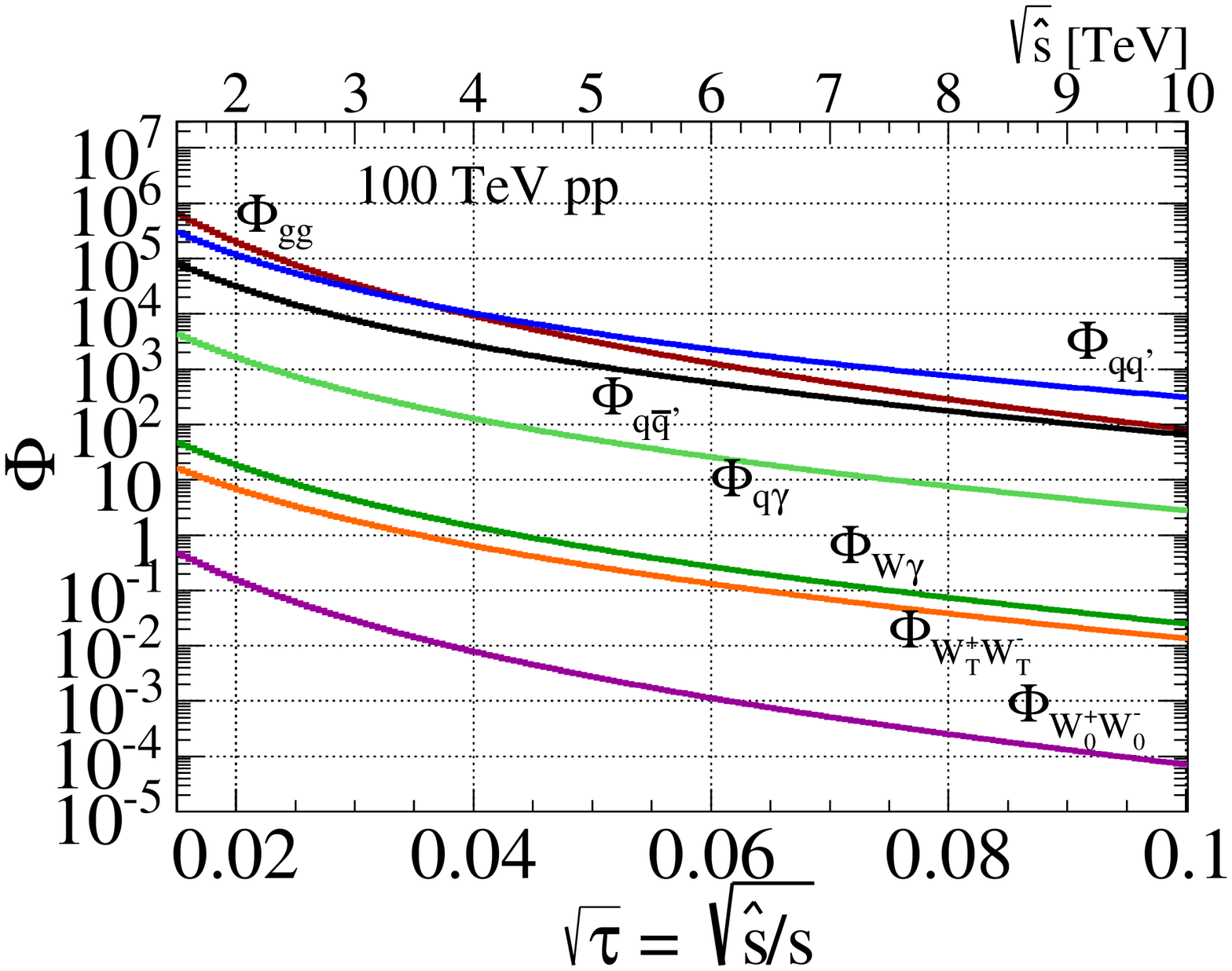}	}
\caption{Parton luminosity as a function of $\sqrt{\tau}$ at (a) 14 TeV and (b) 100 TeV.}
\label{lumi.col.FIG}
\end{figure}

\subsection{Parton Luminosities} 
From the Factorization Theorem, we can extract the \textit{parton luminosity} $\mathcal{L}$, 
which is a measure of the parton-parton flux in hadron collisions.
Parton luminosities are given in terms of the PDFs $f_{i,j/p}$ by the expression
\begin{equation}
\Phi_{ij}(\tau) \equiv  \frac{d\mathcal{L}_{ij}}{d\tau}  = 
  \frac{1}{1+\delta_{ij}}  \int^{1}_{\tau} \frac{d\xi}{\xi} 
 \left[  f_{i/p}(\xi, Q_{f}^{2})f_{j/p}\left(\frac{\tau}{\xi},Q_{f}^{2} \right) + (i \leftrightarrow j) \right],
 \label{partonLumi.EQ}
\end{equation}
where for a process 
\begin{equation}
 i + j \rightarrow X,
\end{equation}
we have
\begin{eqnarray}
  \sigma(pp\rightarrow X+Y) &=& 
  \sum_{i,j}
  \int^{1}_{\tau_{0}}d\xi_{a}\int^{1}_{\tau_{0}\over \xi_a}d\xi_{b} 
  \left[f_{i/p}(\xi_a,\mu^{2})f_{j/p}(\xi_b,\mu^{2}) \hat{\sigma}(i j \rightarrow X) + (i\leftrightarrow j)\right]~~
    \label{factTheoremLumi.col.EQ}
  \\
   &=& \int_{\tau_{0}}^{1} d\tau  \sum_{ij}\frac{d\mathcal{L}_{ij}}{d\tau}\  \hat{\sigma}(ij\rightarrow X).
\end{eqnarray}

In Fig.~\ref{lumi.col.FIG}, we plot the parton luminosities for various initial-state pairs in $\sqrt{s}=$14 and 100 TeV $pp$ collisions. 
We include the light quarks $(u,d,c,s)$ and adopt the 2010 update of the CTEQ6L PDFs\cite{Pumplin:2002vw}.  
We evolve the quark PDFs to half the total partonic energy,
\begin{equation}
 Q_f = \frac{\sqrt{\hat{s}}}{2}.
\end{equation}

\subsection{Parton-Vector Boson Luminosities}
We can extend the definition of parton luminosities to quark-$V$-scattering where $V$ is a spin-1 vector boson 
that collinearly splits from initial parton $i$ by making the replacement in Eq.~(\ref{factTheoremLumi.col.EQ})
\begin{eqnarray}
 f_{i/p}(\xi,Q_f^2) &\rightarrow& f_{V/p}(\xi,Q_V^2,Q_f^2),
 \\
 f_{V/p}(\xi,Q_V^2,Q_f^2) &=& \sum_{q}\int^1_\xi \frac{dz}{z} ~ f_{V/q}(z,Q_V^2) ~ f_{q/p}\left(\frac{\xi}{z},Q_f^2\right)
\end{eqnarray}
resulting in the following $qV$ luminosity formula
\begin{equation}
 \Phi_{qV}(\tau) = \int^1_\tau\frac{d\xi}{\xi} ~\int^1_{\tau/\xi}\frac{dz}{z} ~ \sum_{q'}
 \left[
 f_{q/p}(\xi)f_{V/q'}(z)f\left(\frac{\tau}{\xi z}\right) + f_{q/p}\left(\frac{\tau}{\xi z}\right)f_{V/q'}(z)f_{q'/p}(\xi)
 \right].
\end{equation}
We plot the $qV$ parton luminosities at 14 and 100 TeV $pp$ collisions in Fig.~\ref{lumi.col.FIG}
and observe that the luminosities are typically $\sim\alpha$ smaller than the $qq$ rates.

\subsection{Vector Boson Scattering: Double Initial-State Parton Splitting}
We further extend luminosities to initial-state $VV'$ scattering by making a substitution of initial-state parton $j$ in Eq.~(\ref{factTheoremLumi.col.EQ}):
\begin{equation}
  f_{j/p}(\xi,Q_f^2) \rightarrow f_{V/p}(\xi,Q_V^2,Q_f^2).
\end{equation}
The resulting luminosity expression is
\begin{eqnarray}
 \Phi_{VV'}(\tau) &=& 
 \frac{1}{(\delta_{VV'}+1)}
 \int^1_\tau\frac{d\xi}{\xi} ~\int^1_{\tau/\xi}\frac{dz_1}{z_1}~\int^1_{\tau/\xi/ z_1}\frac{dz_2}{z_2} ~ 
 \sum_{q,q'}
 \\
 &\times&
 \left[
 f_{V/q}(z_2)f_{V'/q'}(z_1)~f_{q/p}(\xi)f_{q'/p}\left(\frac{\tau}{\xi z_1 z_2}\right) + 
 f_{V/q}(z_2)f_{V'/q'}(z_1)~f_{q/p}\left(\frac{\tau}{\xi z_1 z_2}\right)f_{q'/p}(\xi)
 \right]
 \nonumber
\end{eqnarray}
We plot the $WW$ parton luminosities at 14 and 100 TeV $pp$ collisions in Fig.~\ref{lumi.col.FIG}.

\section{Statistics}
\label{sec:stats}
\subsection{Poisson Statistics}
To determine the discovery potential at a particular significance, 
we first translate significance into a corresponding confidence level (CL),\footnote{We use $\sigma$-sensitivity and CL interchangeably in the text.} e.g.,
\begin{equation}
 2\sigma \leftrightarrow 95.45\%~\text{CL},\quad
  3\sigma \leftrightarrow 99.73\%~\text{CL},\quad
   5\sigma \leftrightarrow 99.9999\%~\text{CL}.
\end{equation}
Given an given integrated luminosity $\mathcal{L}$, SM background rate $\sigma_{\rm SM}$, and CL, say 95.45\% CL,
we solve for the maximum number of background-only events, denoted by $n^{b}$, using  the Poisson distribution:
\begin{equation}
 0.9545 = \sum_{k=0}^{n^b} P\left(k \vert \mu^b = \sigma_{\rm SM}\mathcal{L} \right)
	= \sum_{k=0}^{n^b} \frac{(\sigma_{\rm SM}\mathcal{L})^k}{k!} e^{-\sigma_{\rm SM}\mathcal{L}}.
\end{equation}
The requisite number of signal events at a 95.45\% CL (or $2\sigma$ significance) 
is obtained by solving for the mean number of signal events $\mu^s$ such that a mean number of total expected events $(\mu^s + \mu^b)$
will generate $n^{b}$ events only $4.55\%(=100\%-95.45\%)$ of the time, i.e., find $\mu^s$ such that
\begin{equation}
 P\left(k\geq n^b \vert \mu = \mu^s + \mu^b \right) = \frac{(\mu^s + \mu^b)^{n^b}}{(n^b)!} e^{-(\mu^s + \mu^b)} = 0.455.
 \label{poiDis.EQ}
\end{equation}
The 2$\sigma$ sensitivity to nonzero $S_{\ell\ell}$ is then
\begin{equation}
 S_{\ell \ell'}^{2\sigma} = \frac{\mu^s}{\mathcal{L} \times \sigma_{\rm Tot~0}}.
\end{equation}
For fixed signal $\sigma_{s}$ and background $\sigma_{\rm SM}$ rates, $\mu^s + \mu^b = (\sigma_{s}+\sigma_{\rm SM})\times\mathcal{L}$.
The required luminosity for a $2\sigma$ discovery can then be obtained by solving Eq.~(\ref{poiDis.EQ}) for $\mathcal{L}$.

\subsection{Gaussian Statistics}
In the large event limit, we approach Gaussian statistics and the uncertainty greatly simplifies.
For given number of expected SM background events $N_b$ and number of actual observed events $N_o$,
the significance estimator is given by
\begin{equation}
 \sigma = \frac{N_o - N_b}{\sqrt{N_o}}.
\end{equation}
For a new physics signal, we may replace $N_o-N_b$ by $N_s$, the number of signal events after a luminosity $\mathcal{L}$.
Both the signal and background processes then have corresponding cross sections, labeled by $\sigma_s$ and $\sigma_b$.
The significance can then expressed as
\begin{equation}
 \sigma = \frac{N_s}{\sqrt{N_s + N_b}} = \frac{\sigma_s \mathcal{L}}{\sqrt{\sigma_s \mathcal{L} + \sigma_b \mathcal{L}}} = 
 \frac{\sigma_s}{\sqrt{\sigma_s + \sigma_b }}\sqrt{\mathcal{L}},
\end{equation}
indicating a power-law growth in significance as a function of data.

%% file: 03_HiggsFromTop/higgsTop.tex
\chapter{Higgs Bosons from the Top Decays}

\section{Introduction}
\label{intro.SEC}

The discovery of a light, Standard Model (SM)-like Higgs boson at the CERN Large Hadron Collider (LHC)~\cite{Aad:2012tfa,Chatrchyan:2012ufa} 
is a tremendous step towards understanding the  underlying mechanism of electroweak (EW) symmetry breaking (EWSB). 
The observed signal, consistent with the leading production mechanism $gg \to h$,
indicates the existence of the Higgs boson coupling to the top-quark~\cite{Chatrchyan:2013lba}. 
Ultimately, the $t \bar t h$ coupling may be determined at the LHC luminosity upgrade and at a high energy $e^{+}e^{-}$ linear collider~\cite{Dawson:2013bba}.
Regardless of their rarity, a Higgs boson that is less massive than the top quark implies that $t\to h$ transitions exist.
With an annual luminosity of $\mathcal{L}=100~\text{fb}^{-1}/\text{yr}$, the 14 TeV LHC will produce over 90 million $t\overline{t}$ pairs a year~\cite{Czakon:2013goa}.
Thus, searches for $t\to h$ transitions that are sensitive to new physics scenarios are an essential part of the LHC program.
For example: the rare decay involving the Flavor Changing Neutral Current (FCNC) 
\begin{equation}
t\rightarrow ch. 
\label{eq:ch}
\end{equation}
This process is particularly interesting for several reasons. 
At leading order, it is induced at one-loop in the SM and,
due to GIM suppression \cite{Glashow:1970gm,Eilam:1990zc,Mele:1998ag}, its branching fraction is very small, about  $10^{-14}$.
NLO QCD contributions increase this by 10\% \cite{Zhang:2013xya}.
However, new physics beyond the SM (BSM), such as an extended Higgs sector~\cite{Eilam:1990zc,Hou:1991un,Atwood:1996vj,AguilarSaavedra:2004wm,Chen:2013qta} or Supersymmetry (SUSY) \cite{Yang:1997dk,Cao:2007dk,Eilam:2001dh}, can significantly enhance this decay, making it a very sensitive channel to new physics.

\begin{figure}[ptb]
\centering{\includegraphics[width=0.9\textwidth]{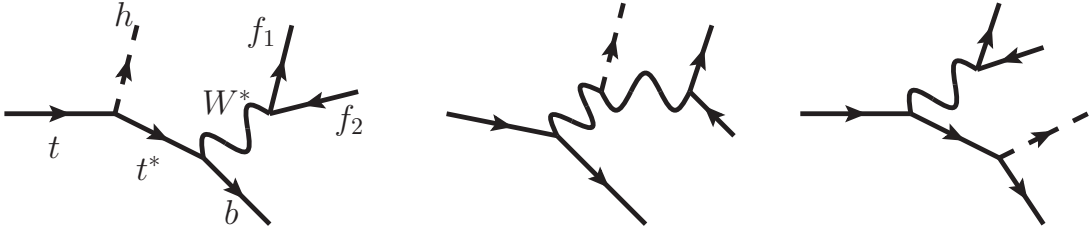}}
\caption[Feynman diagrams representing the leading transition $t\to H$.]{Feynman diagrams representing the leading transition $t\to H$ in Eqs.~(\ref{tWbh.EQ}) and (\ref{tWbH.EQ}).
Drawn using the package JaxoDraw~\cite{Binosi:2003yf}.
}
\label{feynman.FIG}
\end{figure}

In this study, we consider another $t\to h$ transition:
\begin{equation}
 t\rightarrow W^{*}b\ h,
\label{tWbh.EQ}
\end{equation}
where the off-shell $W^{*}$ decays to a pair of light fermions. We now know that this is kinematically allowed in the SM. 
Proceeding at tree-level through the diagrams depicted in Fig.~\ref{feynman.FIG}, Eq.~(\ref{tWbh.EQ}) has been previously evaluated \cite{Rizzo:1986sh,Barger:1987bv,Barger:1989ur,Mahlon:1994us,Decker:1991cz,Decker:1992wz,Iltan:2002am}. 
Both the $t\overline{t}h$ and $WWh$ interactions are simultaneously involved, resulting in a certain subtle, but accidental, cancellation. The predicted branching fraction in the SM is about $10^{-9}$. 
Though still small, the rate is significantly larger than that of Eq.~(\ref{eq:ch}), thereby representing the leading $t\to h$ transition in the SM. 
Subsequently, we are  motivated to investigate how sensitive Eq.~(\ref{tWbh.EQ}) is to new physics.

To systematically quantify this sensitivity in a model-independent fashion, we first employ the approach of  Effective Field Theory (EFT). 
In particular, we consider the effects of gauge invariant, dimension-six operators that can alter the $t\overline{t}h$ interaction 
and take into account constraints on anomalous $t\overline{t}h$ couplings imposed by data.

It is highly probable that the scalar sector responsible for the EWSB extends well beyond a solitary Higgs boson. 
For example: in the Two Higgs Doublet Model (2HDM), one of the best motivated SM extensions, an additional scalar $SU(2)_{L}$ doublet is introduced to facilitate EWSB. 
We extend our study into leading $t\rightarrow h$ transitions by considering CP-conserving variants of the so-called Type I and Type II 2HDM,
denoted by 2HDM(I) and 2HDM(II), respectively.
The corresponding decay channel is 
\begin{equation}
 t\rightarrow W^{*}bH\rightarrow 
 f_{1} \bar f_{2}\  bH,
\label{tWbH.EQ}
\end{equation}
where $H$ is generically either one of the two CP-even ($h,~H$) or the CP-odd ($A$) Higgs bosons, and $f_{1}, f_{2}$ are the light fermions in the SM. 
For $h/H$, Eq.~(\ref{tWbH.EQ}) proceeds identically though Fig.~\ref{feynman.FIG}.
For $A$, the middle diagram is absent.

The remainder of this analysis proceeds as follows: 
In section~\ref{th.SEC}, we introduce our theoretical framework and comment on current experimental constraints for each new physics scenario.
We then present in section~\ref{br.SEC} the SM, EFT, 2HDM(I), and 2HDM(II) predictions for the top quark branching fraction of 
Eq.~(\ref{tWbH.EQ}) over respective parameter spaces.
Observation prospects at present and future colliders are briefly addressed in section~\ref{LHC.SEC}.
Finally in section~\ref{conc.SEC}, we summarize our results and conclude.

\section{Theoretical Framework}
\label{th.SEC}
The theoretical frameworks under consideration include the effective field theory (EFT) for $t\overline{t} h$ interactions up to dimension-six operators (Section~\ref{eft.SEC}), the two Higgs doublet model of Type I [2HDM(I)] (Section~\ref{2hdmI.SEC}), and Type II [2HDM(II)] (Section~\ref{2hdmII.SEC}). 
Current experimental constraints on the model parameters are also presented. 

\subsection{The SM as an Effective Field Theory}
\label{eft.SEC}
To systematically search for new physics beyond the reach of present-day experiments,
we employ Effective Field Theory (EFT) to model new physical phenomena and linearly realize the SM gauge symmetries\cite{Burges:1983zg,Whisnant:1997qu,Grzadkowski:2010es}. 
After integrating out heavy degrees of freedom at a scale $\Lambda$, the low energy effects can be parameterized by 
\begin{equation}
 \mathcal{L} = \mathcal{L}_{\rm SM}+\mathcal{L}_{\rm Eff.},\quad \mathcal{L}_{\rm Eff.}=\sum_{i,j}\frac{f_{i,j}}{\Lambda^{i}}\mathcal{O}_{i,j},
 \label{formalEffLag.EQ}
\end{equation}
where $\mathcal{L}_{\rm SM}$ is the SM Lagrangian, the $f_{i,j}$ are real, dimensionless ``anomalous couplings'' naturally of order $1\sim 4\pi$, 
and $\mathcal{O}_{i,j}$ represent $SU(3)_{c}\times SU(2)_{L}\times U(1)_{Y}$-invariant, dimension-$(4+i)$ operators constructed solely from SM fields. 
When $f_{i,j}\rightarrow 4\pi$, however, one is likely in the strong coupling regime
and the EFT approach breaks down. Here, $f_{i,j}$ is assumed to be $\mathcal{O}(1)$. 
For the remainder of the text, we consider only the next-to-leading interactions at dimension-six and drop the $i=2$ subscript.

\subsubsection{EFT framework and parameters}
Many linearly independent dimension-six operators can affect the $t\overline{t}h$, $WWh$, $b\overline{b}h$, $tWb$, $htc/u$, or 4-point $tWbh$ vertices
~\cite{Burges:1983zg,Whisnant:1997qu,Gounaris:1996yp,Yang:1997iv,Han:1999xd,DeRujula:1991se,AguilarSaavedra:2008zc,
AguilarSaavedra:2009mx,Grzadkowski:2010es,Einhorn:2013kja,Einhorn:2013tja}.
Results from the ATLAS and CMS experiments indicate that the $WWh$ coupling is close to its SM value~\cite{Aad:2012tfa,Chatrchyan:2012ufa,Chatrchyan:2013lba},
and evidence suggest that the $b\overline{b}h$ coupling cannot be much larger than the SM prediction~\cite{Chatrchyan:2013zna,ATLASHbb}.
As dimension-six $tWbh$ verticies originate from terms of the form $tWb(v+h)$~\cite{Whisnant:1997qu,Yang:1997iv},
the size of anomalous 4-point $tWbh$ couplings are restricted to be small by 
the stringent limits on anomalous $tWb$ couplings~\cite{Abazov:2010jn,Abazov:2012uga,Aaltonen:2012lua,Aad:2012ky,Gounaris:1996yp,Han:1999xd}. 
Anomalous $htc/u$ couplings are constrained to be small~\cite{Chen:2013qta,Craig:2012vj,Atwood:2013ica,TheATLAScollaboration:2013nia}.

As we are interesting in the next-to-leading contribution to the $t\rightarrow W^{*}bh$ transition, 
we consider those operators affecting the weakly constrained $t\overline{t}h$ vertex only.
In the basis of Ref.~\cite{Whisnant:1997qu}, 
the most general $t\overline{t}h$ interaction Lagrangian one can construct using linearly independent dimension-six operators 
requires only two operators~\cite{AguilarSaavedra:2009mx} (one CP-even and one CP-odd):
\begin{eqnarray}
 \mathcal{O}_{t1}=\left(\Phi^\dagger\Phi -\frac{v^{2}}{2}\right)\left(\overline{q_{L}}t_{R}\tilde{\Phi}+\tilde{\Phi}^\dagger\overline{t_{R}}q_{L}\right),
 \quad
 \overline{\mathcal{O}}_{t1}=i\left(\Phi^\dagger\Phi -\frac{v^{2}}{2}\right)\left(\overline{q_{L}}t_{R}\tilde{\Phi}-\tilde{\Phi}^\dagger\overline{t_{R}}q_{L}\right),
 \label{GIOp1.EQ}
 \end{eqnarray}
 where $\Phi$ is the SM Higgs $SU(2)_{L}$ doublet with $U(1)_{Y}$ hypercharge $+1$, 
\begin{equation}
 v=\sqrt{2}\langle\Phi\rangle\approx246~\text{GeV},\quad
 \overline{q_{L}}=(\overline{t_L},\overline{b_{L}}),\quad
 \tilde{\Phi}=i\sigma_{2}\Phi^{*},\quad
 t_{L/R}=P_{L/R}t,
\end{equation}
and $P_{L/R}=\frac{1}{2}(1\mp\gamma^{5})$ is the left/right-handed (LH/RH) chiral projection operator. 
These respectively lead to anomalous scalar- and pseudoscalar-type interactions and 
correspond to the operator $Q_{u\varphi}$ in Refs.~\cite{Grzadkowski:2010es,Einhorn:2013kja}, which assume \textit{complex} Wilson coefficients.
To investigate the sensitivity of operators that select out different kinematic features from those listed above,
we consider also the two redundant\footnote{
Using integration by parts and the appropriate equations of motion, e.g.,
$i\overrightarrow{\not\!\! D} q_{L} = y_{u} u_{R} \tilde{\Phi} + y_{d} d_{R} \Phi$, 
one finds that the operator $\overline{\mathcal{O}}_{t2}$ is linearly dependent on $\mathcal{O}_{t1}$ and $\overline{\mathcal{O}}_{t1}$ plus the bottom quark analogues. Similarly, $\overline{\mathcal{O}}_{\Phi q}^{(1)}$ is linearly dependent on $\mathcal{O}_{t1}$ and $\overline{\mathcal{O}}_{t1}$ \cite{Grzadkowski:2010es}.} (CP-odd) operators
\begin{eqnarray}
  \overline{\mathcal{O}}_{\Phi q}^{(1)}=\left[\Phi^\dagger(D_{\mu}\Phi)+(D_{\mu}\Phi)^\dagger\Phi\right](\overline{q_{L}}\gamma^{\mu}q_{L}),
   \quad
  \overline{\mathcal{O}}_{t2}=\left[\Phi^\dagger(D_{\mu}\Phi)+(D_{\mu}\Phi)^\dagger\Phi\right](\overline{t_{R}}\gamma^{\mu}t_{R}), 
 \label{GIOp2.EQ}
\end{eqnarray}
which respectively lead to anomalous left/right-handed (LH/RH) chiral couplings. 
We do not consider other operators that can affect the $t\rightarrow W^{*}bh$ decay because their Wilson coefficients are strongly constrained by data.

After EWSB, the $t\overline{t}h$ interaction Lagrangian contains four\footnote{
The anomalous LH chiral $b\overline{b}h$ coupling from $\overline{\mathcal{O}}_{\Phi q}^{(1)}$ 
is ignored as its contribution suffers from kinematic and helicity suppression. 
See the discussion in Sections~\ref{br.SEC} and ~\ref{EffBR.SEC}.
} 
new independent terms:
\begin{equation}
 \mathcal{L}_{tth}= - \frac{1}{\sqrt{2}}\overline{t}\left(y_{t}-g^{S}-ig^{P}\gamma^{5}\right)th + \left(\frac{\partial_{\mu}h}{v}\right)\overline{t}\gamma^{\mu}\left(g^{L}P_{L}+g^{R}P_{R}\right)t, 
 \label{eftLag.EQ}
\end{equation}
where $y_{t}$ is the SM top quark Yukawa coupling,
\begin{equation}
y_{t}  = \frac{gm_t}{ \sqrt 2 M_W} \simeq 1,
\end{equation}
and  the anomalous couplings $g^{X}$ beyond the SM (BSM) are
\begin{equation}
 g^{S}=f_{t1}\frac{v^{2}}{\Lambda^{2}},\quad
 g^{P}=\overline{f}_{t1}\frac{v^{2}}{\Lambda^{2}},\quad
 g^{L}=\overline{f}_{\Phi q}^{(1)}\frac{v^{2}}{\Lambda^{2}},\quad
 g^{R}=\overline{f}_{t2}\frac{v^{2}}{\Lambda^{2}}.
 \label{eftCoup.EQ}
\end{equation}
The relative minus signs between $y_{t}$ and $g^{X}$ are arbitrary due to the unknown couplings $f$. 
To better understand the influence of $g^{S}$ and $g^{P}$ on Eq.~(\ref{tWbh.EQ}), it is useful to rewrite the relevant parts of Eq.~(\ref{eftLag.EQ}) as
\begin{equation}
y_{t}-g^{S}-ig^{P}\gamma^{5} =  g^{\rm Eff.}\left( e^{-i\delta_{\rm CP}}P_{R} + e^{i\delta_{\rm CP}} P_{L}\right),
  \label{cpvLag.EQ} 
\end{equation}
 where the  effective coupling, $g^{\rm Eff.}$, and the CP-violating (CPV) phase, $\delta_{\rm CP}$, are 
 \begin{equation}
  g^{Eff.} \equiv \sqrt{(y_{t}-g^{S})^{2}+g^{P~2}},\quad
  \delta_{CP} \equiv \sin^{-1}\left[ \frac{g^{P}}{\sqrt{(y_{t}-g^{S})^{2}+g^{P~2}}} \right].
 \end{equation}

\subsubsection{EFT Constraints}

\begin{table}
\caption{Bounds on EFT couplings}
 \begin{center}
\begin{tabular}{|c|c|c|}
\hline \hline
Operator & $g^{X}$ Bound & $\Lambda/\sqrt{\vert f_{\mathcal{O}}\vert}$ [GeV]  \tabularnewline\hline\hline 
\multirow{2}{*}{$\mathcal{O}_{t1}$} & $-0.72<g^{S}<0.21$   	& $>537$    \tabularnewline 
				    & $~~1.77<g^{S}<2.70$   	& $150 - 185$    \tabularnewline\hline
$\overline{\mathcal{O}}_{t1}$       & $-1.4<g^{P}<1.4$ 		& $>208$ \tabularnewline\hline
\end{tabular}
\label{eftBound.TB}
\end{center}
\end{table}

 Independent of deviations in the $h\rightarrow\gamma\gamma$ channel and with no assumption on the Higgs boson's total width,  
 ATLAS has measured the gluon-gluon fusion (ggF) scale factor to be~\cite{Chatrchyan:2013lba}
 \begin{equation}
  \kappa_g = 1.08^{+0.32}_{-0.14},\quad 
    \kappa_{g}^{2}\equiv \sigma(gg\rightarrow h)/\sigma^{\text{SM}}(gg\rightarrow h).
  \label{tthAnom.EQ}
  \end{equation}
Since ggF is dominated by a top quark loop, we can approximate an anomalous $g^{S}$ contribution to the observed rate by
\begin{equation}
 \sigma(gg\rightarrow h)= \kappa_{g}^{2}\times\sigma^{SM}(gg\rightarrow h)\approx \frac{(y_{t}-g^{S})^{2}}{y_{t}^{2}} \times\sigma^{SM}(gg\rightarrow h),
 \label{ggFApprox1.EQ}
\end{equation}
implying
\begin{equation}
 g^{S} \in [-0.72,0.21]\cup [1.77,2.70]~\text{at}~2\sigma.
 \label{gS2Sig.EQ}
\end{equation}
Similarly, we can relate Eq.~(\ref{tthAnom.EQ}) to $g^{P}$ by
 \begin{equation}
   \sigma(gg\rightarrow h)= \kappa_{g}^{2}\times\sigma^{SM}(gg\rightarrow h)\approx \frac{y_{t}^{2}+(g^{P})^{2}}{y_{t}^{2}} \times\sigma^{SM}(gg\rightarrow h),
   \label{ggFApprox2.EQ}
 \end{equation}
indicating
 \begin{equation}
 g^{P} \in [-1.41,1.41]~\text{at}~2\sigma.
 \label{gP2Sig.EQ}
\end{equation}
We next translate measurements of $\kappa_{g}$ into bounds on the cutoff scale of new physics involving operators 
$\mathcal{O}_{t1}$ and $\overline{\mathcal{O}}_{t1}$. 
The bounds on new physics scales $\Lambda/\sqrt{\vert f_{\mathcal{O}}\vert}$ are given in Table~\ref{eftBound.TB}. 
With the Naive Dimensional Analysis (NDA)~\cite{Manohar:1983md,Luty:1997fk} and $f_{\mathcal{O}}\sim \mathcal{O}(1)$, 
the new physics scale is pushed to about $\cal{O}$(1 TeV). 
Translating limits on $\kappa_{g}$ into bounds on $g^{L/R}$, and hence on $\overline{\mathcal{O}}_{\Phi q}^{(1)}$ and $\overline{\mathcal{O}}_{t2}$, 
is a nontrivial procedure due to the derivative coupling. 
Subsequently, such results are not presently available.

\subsection{Linear Dependence of EFT Operators}
Reference ~\cite{Einhorn:2013kja,Einhorn:2013tja} argue that the operators
 \begin{eqnarray}
 \mathcal{O}_{t1}=\left(\varphi^\dagger\varphi -\frac{v^{2}}{2}\right)\left(\overline{q_{L}}t_{R}\tilde{\varphi}+\tilde{\varphi}^\dagger\overline{t_{R}}q_{L}\right),&\quad&
 \overline{\mathcal{O}}_{t1}=i\left(\varphi^\dagger\varphi -\frac{v^{2}}{2}\right)\left(\overline{q_{L}}t_{R}\tilde{\varphi}-\tilde{\varphi}^\dagger\overline{t_{R}}q_{L}\right)\nonumber\\
 \overline{\mathcal{O}}_{t2}=\left[\varphi^\dagger(D_{\mu}\varphi)+(D_{\mu}\varphi)^\dagger\varphi\right](\overline{t_{R}}\gamma^{\mu}t_{R}),&\quad&
 \overline{\mathcal{O}}_{\varphi q}^{(1)}=\left[\varphi^\dagger(D_{\mu}\varphi)+(D_{\mu}\varphi)^\dagger\varphi\right](\overline{t_{L}}\gamma^{\mu}t_{L}) ,
 \label{GIOp.EQ}
\end{eqnarray}
 are linearly dependent with respect to each other.
 Following the notation of Ref.~\cite{Grzadkowski:2010es}, $\varphi$ is the Higgs $SU(2)_L$ doublet,
 $\tilde{\varphi}=i\sigma^{2}\varphi^{*}$,
 $\overline{q_{L}} = (\overline{t_{L}},\overline{b_{L}})$, and $t_{L/R}=P_{L/R}t$, 
 where $P_{L/R} = \frac{1}{2}(1\mp\gamma^{5})$ is the LH/RH chiral projection operator. 
 In this basis, all the operators above have real Wilson coefficients.
 These effective operators introduce (clockwise beginning from $\mathcal{O}_{t1}$) anomalous scalar, pseudoscalar, LH vector couplings, and RH vector couplings.
 The bottom two operators introduce derivative couplings of the form 
  \begin{equation}
    (\partial_\mu h)\gamma^\mu P_{R/L}.
  \end{equation}
  We will show that $\overline{\mathcal{O}}_{t2}$ is equivalent to the top two operators, up to an overall coefficient;
 we will also show that $\overline{\mathcal{O}}_{\varphi q}^{(1)}$ can be expressed in terms of the first two operators and the $b_{R}$ analogue.

 To demonstrate this, the equations of motion (EoM) for quarks will be necessary. 
 They can be obtained from the SM Lagrangian:
 \begin{eqnarray}
 \mathcal{L}_{Quarks} &=& i \overline{q_L}\coD q_{L} + i\overline{u_R}\coD u_{R} +i\overline{d_R}\coD d_{R} \\
		    &-& \Gamma_{u} \overline{q_{L}}u_{R}\tilde{\varphi} -  \Gamma_{d} \overline{q_{L}}d_{R}\varphi
		    ~
		    -\Gamma_{u}^\dagger \tilde{\varphi}^\dagger \overline{u_{R}}q_{L} 
		    -\Gamma_{d}^\dagger \varphi^\dagger \overline{d_{R}}q_{L},				    
\end{eqnarray}
where $\Gamma_{f}$ represent Yukawa couplings and, for $T^{A}=\frac{1}{2}\lambda^{A}$ and $S^{I} = \frac{1}{2}\sigma^{I}$,
\begin{equation}
 \left(D_\mu q_{L}\right)^{\alpha j}= 
 \left( 
 \partial_\mu  + ig_{s}T^{A}_{\alpha\beta}G_{\mu}^{A} + igS^{I}_{jk}W_{\mu}^{I} + ig'Y_{q}B_{\mu}
 \right)q^{\beta k},
\end{equation}
with weak isospin and color indices $j,k=1,2$ and $\alpha,\beta=1,2,3$.

 Taking the appropriate functional derivatives, the EoMs can be obtained.
 For $q_L$, the RH up-type quark $u_{R}$ and the RH down-type quark $d_{R}$ these are
\begin{eqnarray}
 i\overrightarrow{\not\!\! D} q_{L} = \Gamma_{u} u_{R} \tilde{\varphi} + \Gamma_{d} d_{R} \varphi
 &~\Longleftrightarrow~&
 i\overline{q_L}~\overleftarrow{\coD} = -i \overline{\left(\overrightarrow{\coD}q_{L}\right)}
 = -\Gamma_{u}^\dagger \tilde{\varphi}^\dagger \overline{u_{R}} - \Gamma_{d}^\dagger  \varphi^\dagger \overline{d_{R}}
\\
 i\overrightarrow{\not\!\! D} u_{R} = \Gamma_{u}^\dagger \left(\tilde{\varphi}^\dagger q_{L}\right)
 &~\Longleftrightarrow~&
   i\overline{u_R}~\overleftarrow{\coD} = -i \overline{\left(\overrightarrow{\coD}u_{R}\right)}
   = - \Gamma_{u} \left(\overline{q_{L}}\tilde{\varphi} \right)
 \\
 i\overrightarrow{\not\!\! D} d_{R} = \Gamma_{d}^\dagger \left(\varphi^\dagger q_{L}\right) 
 &~\Longleftrightarrow~&
    i\overline{d_R}~\overleftarrow{\coD} = -i \overline{\left(\overrightarrow{\coD}d_{R}\right)}
   = - \Gamma_{d} \left(\overline{q_{L}}\varphi \right)
\end{eqnarray}

\subsubsection{ $\mathcal{O}_{t1}$ and $\overline{\mathcal{O}}_{t1}$}
As noted above, $\mathcal{O}_{t1}$ and $\overline{\mathcal{O}}_{t1}$ possess \textit{real} Wilson coefficients. 
The operators in Ref.~\cite{Grzadkowski:2010es} possess \textit{complex} coefficients, and so the operator
\begin{equation}
    Q_{u\varphi} = \left(\varphi^\dagger\varphi\right)\left(\overline{q_L}d_{R}\tilde{\varphi}\right)
\end{equation}
in Ref.~\cite{Grzadkowski:2010es} maps with a one-to-one correspondence to $\mathcal{O}_{t1}$ and $\overline{\mathcal{O}}_{t1}$.

\subsubsection{ $\overline{\mathcal{O}}_{t2}$}
For the operator $\overline{\mathcal{O}}_{t2}$, we see
\begin{eqnarray}
     \overline{\mathcal{O}}_{t2}&=&\left[\varphi^\dagger(D_{\mu}\varphi)+(D_{\mu}\varphi)^\dagger\varphi\right](\overline{t_{R}}\gamma^{\mu}t_{R}) 
    \\
     &=& \left[\varphi^\dagger(\partial_{\mu}\varphi)+(\partial_{\mu}\varphi)^\dagger\varphi\right](\overline{t_{R}}\gamma^{\mu}t_{R}) 
    \\
     &=& \left[\partial_{\mu}(\varphi^\dagger\varphi)\right](\overline{t_{R}}\gamma^{\mu}t_{R}) 
    \\
    &\overset{\rm IBP}{=}& 
    \left[\int\dots \right] \quad +\quad     (\varphi^\dagger\varphi) D_\mu \left(\overline{t_{R}}\gamma^{\mu}t_{R}\right)
    \\
    &=& \left[\int \dots\right]\quad + \quad (\varphi^\dagger\varphi)
    \left[ \left(\overline{t_{R}}\overleftarrow{\not\!\!D}t_{R}\right) + \left(\overline{t_{R}}\overrightarrow{\not\!\!D} t_{R}\right) \right]
    \\
    &=& \left[\int \dots\right]\quad + \quad (\varphi^\dagger\varphi)
    \left[ 
    i\Gamma_{t} \left(\overline{q_{L}} \tilde{\varphi}\right)t_{R} 
    +i\Gamma_{t}^\dagger \overline{t_{R}}\left(\tilde{\varphi}^\dagger q_{L}\right)
    \right]       
    \\
    &=& \left[\int \dots\right]\quad+\quad i\left[ \Gamma_{t}Q_{u\varphi} + \text{H.c.} \right],
 \end{eqnarray}
  where $\left[\int\dots \right]$ denotes a total derivative and has no observable effect on the physical amplitude.
  Subsequently, we see that this operator is proportional to $Q_{u\varphi}$ and its Hermitian conjugate,
  and hence is a linear combination of $\mathcal{O}_{t1}$ and $\overline{\mathcal{O}}_{t1}$.

 \subsubsection{ $\overline{\mathcal{O}}_{\varphi q}^{(1)}$} 
 For the operator $\overline{\mathcal{O}}_{\varphi q}^{(1)}$, we obtain
 \begin{eqnarray}
  \overline{\mathcal{O}}_{\varphi q}^{(1)}&=&\left[\varphi^\dagger(D_{\mu}\varphi)+(D_{\mu}\varphi)^\dagger\varphi\right](\overline{q_{L}}\gamma^{\mu}q_{L})
  \\
  &=& \left[\partial_{\mu}(\varphi^\dagger\varphi)\right](\overline{q_{L}}\gamma^{\mu}q_{L}) 
  \\
      &\overset{\rm IBP}{=}& 
    \left[\int\dots \right] \quad +\quad     (\varphi^\dagger\varphi) D_\mu \left(\overline{q_{L}}\gamma^{\mu}q_{L}\right)
    \\
        &=& \left[\int \dots\right]\quad + \quad (\varphi^\dagger\varphi)
    \left[ \left(\overline{q_{L}}\overleftarrow{\not\!\!D}q_{L}\right) + \left(\overline{q_{L}}\overrightarrow{\not\!\!D} q_{L}\right) \right]
    \\
    &=& \left[\int \dots\right]\quad + \quad (\varphi^\dagger\varphi)
    \left[ 
    i\left(\Gamma_{t}^\dagger ~\tilde{\varphi}^\dagger\overline{t_{R}}   + \Gamma_{b}^\dagger~ \varphi^\dagger\overline{b_{R}} \right)q_{L} 
    -i\overline{q_{L}}\left(\Gamma_{t}~ t_{R}\tilde{\varphi} + \Gamma_{b}~   b_{R}\varphi\right) 
    \right]
    \\
        &=& \left[\int \dots\right]\quad + \quad i(\varphi^\dagger\varphi)
    \left[ 
    \Gamma_{t}^\dagger(\tilde{\varphi}^\dagger\overline{t_{R}}q_{L})
    +
    \Gamma_{b}^\dagger(\varphi^\dagger\overline{b_{R}}q_{L}) 
    -\Gamma_{t}(\overline{q_{L}}t_{R}\tilde{\varphi}) 
    -\Gamma_{b}(\overline{q_{L}}b_{R}\varphi) 
    \right]
    \\
    &=& \left[\int \dots\right]\quad + \quad i\left[ \text{H.c.} - \Gamma_{t}Q_{u\varphi} - \Gamma_{b}Q_{d\varphi} \right]
 \end{eqnarray}
   where $Q_{d\varphi}$ is a Ref.~\cite{Grzadkowski:2010es} operator and  
 \begin{equation}
      Q_{d\varphi} = \left(\varphi^\dagger\varphi\right)\left(\overline{q_L}u_{R}\varphi\right).
 \end{equation}
   In this case, $\overline{\mathcal{O}}_{\varphi q}$ is linearly independent  only because we do not include an operator analogous to $Q_{d\varphi}$.
   However, Ref.~\cite{AguilarSaavedra:2009mx} points out that the most general $tth$ Lagrangian constructed from the minimal set of dimension-6 operators
   has the form
   \begin{equation}
 \mathcal{L}_{tth}= - {1\over\sqrt{2}}\overline{t}\left(Y_{t}^{V}+iY^{P}_{t}\gamma^{5}\right)th 
 \label{eftLag.EQ}
\end{equation}
  for real $Y_{t}^{V,A}$ because the derivative coupling terms identically vanish due to equations of motion.

  The dimension-six operators used here are taken from from Whisnant, et al.~\cite{Whisnant:1997qu}.
  However, the issue of redundant operators reported by Grzadkowski~\cite{Grzadkowski:2010es}, Aguilar-Saavedra~\cite{AguilarSaavedra:2009mx}, 
  and Einhorn \& Wudka~\cite{Einhorn:2013kja,Einhorn:2013tja} appeared more than a decade after the Whisnant, et al.


\subsection{Type I Two Higgs Doublet Model}
\label{2hdmI.SEC}
In the generic CP-conserving 2HDM, EWSB is facilitated by two $SU(2)_{L}$ doublets, $\Phi_{i}$, for $i\in\{1,2\}$, 
each with $U(1)_{Y}$ hypercharge $+1$ and a nonzero vacuum expectation value (vev) $v_{i}$. 
A $\mathcal{Z}_{2}$ symmetry is applied for $\Phi_1 \leftrightarrow \Phi_{2}$ to eliminate tree-level FCNC but may be softly broken at loop-level. 
After EWSB, there are five physical spin-0 states: $h,~H,~A,$ and $H^{\pm}$, 
which are respectively the two CP-even, single CP-odd, and $U(1)_{EM}$ charged Higgs bosons with masses $m_{h},m_{H}$, $m_{A}$, and $m_{H^\pm}$.
By convention, we fix the ordering of $h$ and $H$ by taking $$m_{h}<m_{H}.$$
Two angles, $\alpha$ and $\beta$, remain as free parameters.
$\alpha$ measures the mixing between the two CP-even Higgs fields to form the mass eigenstates ($h,\ H$)
and spans $\alpha\in[-\pi/2,\pi/2]$.
$\beta$ represents the relative size of $\langle\Phi_{i}\rangle$ and is defined by
\begin{equation}
 \tan\beta\equiv \langle\Phi_{2}\rangle / \langle\Phi_{1}\rangle= v_{2}/v_{1},  \quad \beta\in[0,\pi/2].
 \label{tB.EQ}
\end{equation}
Reviews of various 2HDMs and their phenomenologies can be found in Refs.~\cite{Gunion:1989we,Gunion:2002zf,Branco:2011iw}. 

\subsubsection{Type I 2HDM framework and parameters}
\label{2hdmITh.SEC}
In the 2HDM(I), much like in the SM, only one Higgs doublet is responsible for generating fermion masses and couples accordingly;
the second CP-even Higgs boson interacts with fermions through mixing.
The interaction Lagrangian relevant to this study is
\begin{eqnarray}
\mathcal{L}\ni
&-&\frac{gm_{u}}{2M_{W}}\overline{u}\left(h\frac{\cos\alpha}{\sin\beta}+H\frac{\sin\alpha}{\sin\beta}-i\gamma^{5}A\cot\beta\right)u\nonumber\\
&-&\frac{gm_{d}}{2M_{W}}\overline{d}\left(h\frac{\cos\alpha}{\sin\beta}+H\frac{\sin\alpha}{\sin\beta}+i\gamma^{5}A\cot\beta\right)d\nonumber\\
&+&gM_{W}W_{\mu}W^{\mu}\left[h\sin(\beta-\alpha)+H\cos(\beta-\alpha)\right]. 
\label{typeIlag.EQ}
\end{eqnarray} 
In Eq.~(\ref{typeIlag.EQ}), $u_{L(R)}$ is the LH (RH) up-type quark spinor, $d_{L(R)}$ is the down-type quark analogue, and
$g$ is the weak coupling constant in the SM.

\begin{table}
\caption{Neutral Scalar Boson Couplings in the 2HDM(I) Relative to the SM Higgs Couplings}
 \begin{center}
\begin{tabular}{|c|c|c|c|c|}
\hline \hline
Vertex & SM & 2HDM I & $\sin(\beta-\alpha)=1-\Delta_{V}$ \tabularnewline
\hline\hline 
$hu\overline{u}/d\overline{d}$ & $0$ or $1$ & $\frac{\cos\alpha}{\sin\beta}$  & $1-\Delta_{V}+\sqrt{2\Delta_{V}-\Delta_{V}^{2}}\cot\beta $ \tabularnewline\hline 
$hW^{+}W^{-}$    & $0$ or $1$ & $\sin(\beta-\alpha)$            & $1-\Delta_{V}$ \tabularnewline\hline 
$Hu\overline{u}/d\overline{d}$ & $0$ or $1$ & $\frac{\sin\alpha}{\sin\beta}$  & $(\Delta_{V}-1)\cot\beta+\sqrt{2\Delta_{V}-\Delta_{V}^{2}}$ \tabularnewline\hline 
$HW^{+}W^{-}$    & $0$ or $1$ & $\cos(\beta-\alpha)$            & $\sqrt{2\Delta_{V}-\Delta_{V}^{2}}$  \tabularnewline\hline 
$Au\overline{u}$ & - & $\cot\beta$                     & $\cot\beta$\tabularnewline\hline 
$Ad\overline{d}$ & - & $-\cot\beta$                    & $-\cot\beta$\tabularnewline\hline 
\end{tabular}
\label{2hdmICouple.TB}
\end{center}
\end{table}

Discovering a Higgs boson with SM-like couplings greatly impacts the 2HDM.
In particular, the measured couplings to weak bosons\cite{Aad:2012tfa,Chatrchyan:2012ufa,Chatrchyan:2013lba} imply either
\begin{eqnarray}
&& 
\sin(\beta-\alpha)\approx1\quad \text{for $h$ to be SM-like,}\\
\text{or}\ \ &&
\cos(\beta-\alpha)\approx1\quad \text{for $H$ to be SM-like.}
\label{sinBA.EQ}
\end{eqnarray}
Generally, we may parameterize how far $\sin(\beta-\alpha)$ is away from one and define $\Delta_{V}$ such that
\begin{equation}
\sin(\beta-\alpha)\equiv1-\Delta_{V},\quad0\leq\Delta_{V}\leq1.
\label{DVDef.EQ}
\end{equation}
We restrict the couplings to have the same sign as those of the SM \cite{Ellis:2013lra} and limit $\Delta_{V}$ up to one.
Eq.~(\ref{DVDef.EQ}) maps to the parameterization used by the SFitter Collaboration~\cite{Klute:2012pu} 
by taking $\Delta_{V}\rightarrow -\Delta_{V}$ and allowing $\Delta_{V}<0$.
After substituting $\alpha$ by $\Delta_{V}$ in Eq.~(\ref{typeIlag.EQ}), we have

{\small
\begin{eqnarray}
\mathcal{L}\ni
&-&\frac{gm_{u}}{2M_{W}}\overline{u}\left[h\left(1-\Delta_{V}+\sqrt{2\Delta_{V}-\Delta_{V}^{2}}\cot\beta\right)+H\left((\Delta_{V}-1)\cot\beta+\sqrt{2\Delta_{V}-\Delta_{V}^{2}}\right)\right]u\nonumber\\
&-&\frac{gm_{d}}{2M_{W}}\overline{d}\left[h\left(1-\Delta_{V}+\sqrt{2\Delta_{V}-\Delta_{V}^{2}}\cot\beta\right)+H\left((\Delta_{V}-1)\cot\beta+\sqrt{2\Delta_{V}-\Delta_{V}^{2}}\right)\right]d\nonumber\\
&+&\frac{gm_{u}}{2M_{W}}\overline{u}\left[i\gamma^{5}A\cot\beta\right]u - \frac{gm_{d}}{2M_{W}}\overline{d}\left[i\gamma^{5}A\cot\beta\right]d\nonumber\\
&+&gM_{W}W_{\mu}W^{\mu}\left[h(1-\Delta_{V})+H\sqrt{2\Delta_{V}-\Delta_{V}^{2}}\right].
\label{typeIlagDV.EQ}
\end{eqnarray} }
Table \ref{2hdmICouple.TB} summarizes the bosonic and fermionic couplings to the neutral scalar in the 2HDM(I) relative to those in the SM, i.e., the 2HDM(I) coupling coefficient divided by the SM coupling coefficient.
In the small (large) $\Delta_{V}$ limit, $h~(H)$ becomes SM-like and $H~(h)$ becomes non-SM-like.
At $\Delta_{V}=0~(\Delta_{V}=1)$, $H~(h)$ decouples from the gauge bosons. 
The relevant tree-level couplings to $A$ are independent of $\Delta_{V}$ as they are initially independent of $\alpha$. 
In the large $\tan\beta$ limit, $A$ decouples from the theory.
For all parameter scenarios considered, we identify the SM-like Higgs as the one with stronger couplings to $WW,\ ZZ$, and having a mass of $125.5$ GeV.

\subsubsection{Type I 2HDM Constraints}
\label{2hdmIConst.SEC}
Since the Higgs boson's discovery, many reports have appeared investigating the 2HDMs' compatibility with 
data~\cite{Ellis:2013lra,Klute:2012pu,Baak:2011ze,Gorczyca:2011he,Ferreira:2012my,Altmannshofer:2012ar,Azatov:2012qz,Chen:2013kt,Chiang:2013ixa,Chang:2013aya,Barroso:2013zxa,Chen:2013rba,Craig:2013hca,Barger:2013mga,Mahmoudi:2012ej}.
We list here constraints relevant to the 2HDM(I) and note when a result is applicable to other types.
The following bounds assume one SM-like Higgs boson at approximately $126$ GeV.

\begin{enumerate}[(i)]
 \item{\it $\cos(\beta-\alpha)-\tan\beta$ Parameter Space}:
    A global fit of available LHC data, in particular from $h \to \gamma\gamma, \ VV,\ b\bar b,\ \tau^{+}\tau^{-}$,  
    has set stringent bounds \cite{Craig:2013hca}. Representative values at $95\%$CL are
    \begin{equation}
     \cos(\beta-\alpha)< 0.3~(0.40)~[0.42] \quad\text{for}\quad \tan\beta=2.4~(10)~[100].
     \label{2hdmICostB.EQ}
    \end{equation}
    Similar conclusions have been reached by Refs.~\cite{Azatov:2012qz,Chiang:2013ixa,Barger:2013mga,Barroso:2013zxa,Chen:2013rba}.    
   
  \item{\it $m_{H^{\pm}}-\tan\beta$ Parameter Space}:
  For all 2HDMs, flavor observables exclude at 95\%~CL~\cite{Barberio:2008fa,Mahmoudi:2009zx}
  \begin{equation}
   \tan\beta < 1 \quad\text{for}\quad  m_{H^{\pm}}<500~\text{GeV}.
  \end{equation}
  Values of $\tan\beta<1$ are allowed given a sufficiently heavy $H^{\pm}$~\cite{Barberio:2008fa,Mahmoudi:2009zx,Mahmoudi:2012ej,Chen:2013kt}.
  Due to the particular $\tan\beta$ dependence, 
  no absolute lower bound on $m_{H^{\pm}}$ from flavor constraints exists in the 2HDM(I)~\cite{Mahmoudi:2009zx}.
  An observation of excess $B\rightarrow D^{*}\tau\nu$ decays~\cite{Lees:2013uzd} has yet to be confirmed and is not considered.
  
  \item {\it Additional Higgs Masses}: 
  For both 2HDM(I) and (II), additional CP-even scalars below LEP bounds~\cite{Barate:2003sz,Abbiendi:2004gn,Abdallah:2004wy} are allowed 
  given sufficiently decoupled $H^{\pm}$ and $A$~\cite{Ferreira:2012my}.
  A second CP-even Higgs is incompatible with LHC data for mass
  \begin{equation}
  180~\text{GeV}< m_{H} < 350~\text{GeV},
  \end{equation}
  but allowed outside this range~\cite{Altmannshofer:2012ar}.
  Direct searches for $H^{\pm}$ and $A$ exclude ~\cite{Abbiendi:2004gn,Aubert:2009cp,Abdallah:2004wy,Abbiendi:2013hk}
   \begin{equation}
   m_{H^\pm},~m_{A} \lesssim 80~\text{GeV}.
   \end{equation}

\end{enumerate}

 Additional considerations include the compatibility of a  
SM-like Higgs boson with EW precision data in general 2HDMs \cite{Baak:2011ze},
the perturbative unitarity limits on the heavy Higgs masses in a general, CP-conserving 2HDM 
\cite{Kanemura:1993hm,Akeroyd:2000wc,Chang:2013aya}, 
and perturbative unitarity limits on $\tan\beta$ in an exact $\mathcal{Z}_{2}$-symmetric, CP-conserving 2HDM~\cite{Gorczyca:2011he,Barroso:2013zxa}.
Since FCNC do exist in nature and the SM, it is unnecessary to impose the severe constraints on $\tan\beta$ associated with an exact  $\mathcal{Z}_{2}$ symmetry.


\subsection{Type II Two Higgs Doublet Model}
\label{2hdmII.SEC}

\subsubsection{Type II 2HDM framework and parameters}
\label{2hdmIITh.SEC}
In the 2HDM(II), one Higgs doublet is assigned a hypercharge $+1$, giving masses to fermions with weak isospin $T^{3}_{L}=+\frac{1}{2}$, 
and the second is assigned a hypercharge $-1$, giving masses to $T^{3}_{L}=-\frac{1}{2}$ fermions.
The doublets are denoted respectively by $\Phi_{u}$ and $\Phi_{d}$, and $\beta$ is written as
\begin{equation}
\tan\beta\equiv\langle \Phi_{u}\rangle / \langle \Phi_{d}\rangle  = v_{u}/v_{d}.
\end{equation}
After EWSB, the CP-conserving interaction Lagrangian relevant to Eq.~(\ref{tWbH.EQ}) is similar to Eq.~(\ref{typeIlag.EQ}), 
with the only difference being the down-type quark Yukawa couplings:
\begin{eqnarray}
\mathcal{L}\ni
&-&\frac{gm_{d}}{2M_{W}}\overline{d}\left(-h\frac{\sin\alpha}{\cos\beta}+H\frac{\cos\alpha}{\cos\beta}-i\gamma^{5}A\tan\beta\right)d.
\label{typeIILag.EQ}
\end{eqnarray}
The notation used in Eq.~(\ref{typeIILag.EQ}) is the same as the 2HDM(I) Lagrangian Eq.~(\ref{typeIlag.EQ}).
Using Eq.~(\ref{DVDef.EQ}), and similar to Eq.~(\ref{typeIlagDV.EQ}), the preceding line becomes 
{\small
\begin{eqnarray}
\mathcal{L}\ni 
&-&\frac{gm_{d}}{2M_{W}}\overline{d}\left[h\left(1-\Delta_{V}-\sqrt{2\Delta_{V}-\Delta_{V}^{2}}\tan\beta\right)+H\left((1-\Delta_{V})\tan\beta+\sqrt{2\Delta_{V}-\Delta_{V}^{2}}\right)\right]d\nonumber\\
&+& i \frac{gm_{d}}{2M_{W}}\overline{d}\gamma^{5}d \ A\tan\beta .
\label{lagDV.EQ}
\end{eqnarray} }
Table \ref{2HDMII.TB} summarizes the bosonic and fermionic couplings to the neutral scalars in the 2HDM(II) relative to those in the SM.
Like the 2HDM(I), in the small (large) $\Delta_{V}$ limit, $h~(H)$ becomes SM-like and $H~(h)$ becomes non-SM-like.
At $\Delta_{V}=0~(\Delta_{V}=1)$, $H~(h)$ decouples from the gauge bosons. 
In this same limit, the $h~(H)$ Yukawa couplings become independent of $\tan\beta$. 
Unlike the 2HDM(I), $A$ only  decouples from the theory if taken to be infinitely heavy.

An important feature for the Higgs couplings to fermions is that 
the down-type quark couplings are enhanced at higher values of $\tan\beta$, while 
the up-type quark couplings are suppressed. For the charged Higgs however, there is an interplay between the two and the particular value $\tan\beta =\sqrt{m_{t}^{\overline{\text{MS}}}(m_{t})/m_{b}^{\overline{\text{MS}}}(m_{t})}\ \approx 7.6$ minimizes the decay $t\rightarrow H^{+}b$.
Though no such minima occur in the 2HDM(I), sensitivity to $\tan\beta=7.6$ will be investigated in both 2HDM scenarios.

\begin{table}
\caption{Neutral Scalar Boson Couplings in the 2HDM(II) Relative to the SM Higgs Couplings}
 \begin{center}
\begin{tabular}{|c|c|c|c|c|}
\hline \hline
Vertex & SM & 2HDM II & $\sin(\beta-\alpha)=1-\Delta_{V}$ \tabularnewline
\hline\hline 
$hu\overline{u}$ & $0$ or $1$ & $\frac{\cos\alpha}{\sin\beta}$  & $1-\Delta_{V}+\sqrt{2\Delta_{V}-\Delta_{V}^{2}}\cot\beta $ \tabularnewline\hline 
$hd\overline{d}$ & $0$ or $1$ & $-\frac{\sin\alpha}{\cos\beta}$ & $1-\Delta_{V}-\sqrt{2\Delta_{V}-\Delta_{V}^{2}}\tan\beta $ \tabularnewline\hline 
$hW^{+}W^{-}$    & $0$ or $1$ & $\sin(\beta-\alpha)$            & $1-\Delta_{V}$ \tabularnewline\hline 
$Hu\overline{u}$ & $0$ or $1$ & $\frac{\sin\alpha}{\sin\beta}$  & $(\Delta_{V}-1)\cot\beta+\sqrt{2\Delta_{V}-\Delta_{V}^{2}}$ \tabularnewline\hline 
$Hd\overline{d}$ & $0$ or $1$ & $\frac{\cos\alpha}{\cos\beta}$  & $(1-\Delta_{V})\tan\beta+\sqrt{2\Delta_{V}-\Delta_{V}^{2}}$ \tabularnewline\hline 
$HW^{+}W^{-}$    & $0$ or $1$ & $\cos(\beta-\alpha)$            & $\sqrt{2\Delta_{V}-\Delta_{V}^{2}}$  \tabularnewline\hline 
$Au\overline{u}$ & $-$ & $\cot\beta$                     & $\cot\beta$\tabularnewline\hline 
$Ad\overline{d}$ & $-$ & $\tan\beta$                     & $\tan\beta$\tabularnewline\hline 
  \end{tabular}
\label{2HDMII.TB}
\end{center}
\end{table}

\subsubsection{Type II 2HDM Constraints}
\label{2hdmIIConst.SEC}
Constraints relevant to the 2HDM(II) are listed here.
See Section~\ref{2hdmI.SEC} for generic 2HDM bounds.

\begin{enumerate}[(i)]

\item{\it $\cos(\beta-\alpha)-\tan\beta$ Parameter Space}:
    A global fit of available LHC data, in particular from $h \to \gamma\gamma, \ VV,\ b\bar b,\ \tau^{+}\tau^{-}$,  
    has set stringent bounds \cite{Craig:2013hca}. Representative values at $95\%$~CL are
    \begin{equation}
     \cos(\beta-\alpha)< 0.06~(0.01) \quad\text{for}\quad \tan\beta=2.4~(10).
     \label{2hdmIICostB.EQ}
    \end{equation}
    Similar conclusions have been reached by Refs.~\cite{Azatov:2012qz,Chiang:2013ixa,Barger:2013mga,Barroso:2013zxa,Chen:2013rba}.

  \item {\it $m_{H^{\pm}}-\tan\beta$ Parameter Space}: 
  Flavor observables, and in particular $\BR(B\rightarrow X_{s}\gamma)$, exclude at 95\%~CL~\cite{Misiak:2006zs,Barberio:2008fa,Lees:2012ym}
  \begin{equation}
    m_{H^{\pm}}<327~\text{GeV}\quad\text{for all}~\tan\beta
  \end{equation}
   From $\BR(B\rightarrow\tau\nu)$ measurements, the {\bf UT}{\it fit} Collaboration~\cite{Bona:2009cj} has determined the absolute bound
 \begin{equation}
  \tan\beta < 7.4 \frac{m_{H^\pm}}{100~\text{GeV}}.
 \end{equation}

 \end{enumerate}

\section{Branching Ratios}
\label{br.SEC}
The discovery of a SM-like Higgs boson at 126 GeV\cite{Aad:2012tfa,Chatrchyan:2012ufa,Chatrchyan:2013lba} implies that
\begin{equation}
 t\rightarrow W^{+*}bh,\quad
  W^{+*}\rightarrow f_{1}\bar f_{2}
 \label{tWbhWmv.EQ}
\end{equation}
is kinematically allowed and proceeds through the diagrams given in Fig.~\ref{feynman.FIG}. 
Following Ref.~\cite{Mahlon:1994us}, we define the $t\rightarrow W^{*}bh$ partial width as
\begin{equation}
 \Gamma(t\rightarrow Wbh)= \frac{\Gamma(t\rightarrow \mu^{+}\nu_{\mu}bh)}{\BR(W\rightarrow\mu\nu_{\mu})},
 \label{GWbh.EQ}
\end{equation}
and the $t\rightarrow W^{*}bh$ branching ratio by
\begin{equation}
 \BR(t\rightarrow Wbh)= \frac{\Gamma(t\rightarrow Wbh)}{\Gamma_{\rm Tot.}},\quad \Gamma_{\rm Tot.}\equiv \Gamma(t\rightarrow Wb).
 \label{brWbh.EQ}
\end{equation}

With CalcHEP 3.4.2~\cite{Pukhov:1999gg,Pukhov:2004ca,Belyaev:2012qa}, we find excellent agreement with Ref.~\cite{Mahlon:1994us}. 
With updated parameters~\cite{Aad:2012tfa,Chatrchyan:2012ufa,Chatrchyan:2013lba,Beringer:1900zz}:
{\small
\begin{eqnarray}
 m_{t}^{\overline{\text{MS}}}(m_{t})=173.5~\text{GeV},\quad
 m_{b}^{\overline{\text{MS}}}(m_{t})=3.01~\text{GeV},\quad
 m_{h}&=&125.5~\text{GeV},\quad
 m_{\mu}=0~\text{GeV},
 \nonumber\\
 M_{W}=80.385~\text{GeV}, \quad
 M_{Z}=91.1876~\text{GeV},  \quad
 G_{F}&=&1.1663787\times10^{-5}~\text{GeV}^{-1},
 \nonumber\\  
 \Gamma_{W}=2.085~\text{GeV},\quad
 \BR(W\rightarrow\mu\nu)&=&0.1057,
 \label{input.EQ}
\end{eqnarray} }
we calculate $\Gamma_{\rm Tot.}$ at leading order to be
\begin{equation}
 \Gamma_{\rm Tot.} = 1.509~\text{GeV},
 \label{totWidth.EQ}
\end{equation}
and find that the SM predicts
\begin{equation}
 \BR^{\rm SM}(t\rightarrow Wbh) = 1.80\times10^{-9}.
 \label{smBR.EQ}
 \end{equation} 
The smallness of this branching fraction 
falls from several features, including phase space suppression associated with the three-body final state, kinematic suppression due to the off-shell $W$ boson, 
and an accidental cancellation between the leading $t\overline{t}h$ and subleasing $WWh$ diagrams.
Nevertheless, this decay rate is $\mathcal{O}(10^{5})$ larger than the well-studied~\cite{Eilam:1990zc,Mele:1998ag} two-body $t\rightarrow ch$ transition.
This is due to the GIM suppression for the FCNC. 
 
In the remainder of this section, we investigate how the branching fraction can change in the context of EFT, 2HDM(I), and 2HDM(II).
  

\subsection{EFT BR$(t\rightarrow Wbh)$}
\label{EffBR.SEC}

 \begin{table}[!t]
\caption{$\BR(t\rightarrow Wbh)$ for Benchmark Values of Anomalous $t\overline{t}h$ Couplings}
 \begin{center}
\begin{tabular}{|c|c|c|}
\hline \hline
 \multicolumn{2}{|c|}{$g^{X}$} & $\BR(t\rightarrow Wbh)$  \tabularnewline\hline\hline 
 $g^{S}$& $~0.5$ & $1.075\times 10^{-9}$ \tabularnewline\hline
  & $-0.5$ & $3.078\times 10^{-9}$  \tabularnewline\hline
 $g^{P}$& $~0.5$ & $1.929\times 10^{-9}$  \tabularnewline\hline
  & $-0.5$ & $1.928\times 10^{-9}$  \tabularnewline\hline
 $g^{L}$& $~0.5$ & $1.812\times 10^{-9}$  \tabularnewline\hline 
  & $-0.5$ & $1.812\times 10^{-9}$  \tabularnewline\hline
 $g^{R}$& $~0.5$ & $1.927\times 10^{-9}$  \tabularnewline\hline
  & $-0.5$ & $1.928\times 10^{-9}$  \tabularnewline\hline\hline
\end{tabular}
\label{eftBench.TB}
\end{center}
\end{table}

\begin{figure}[!t]
\centering
\subfigure[]{	\includegraphics[width=0.46\textwidth]{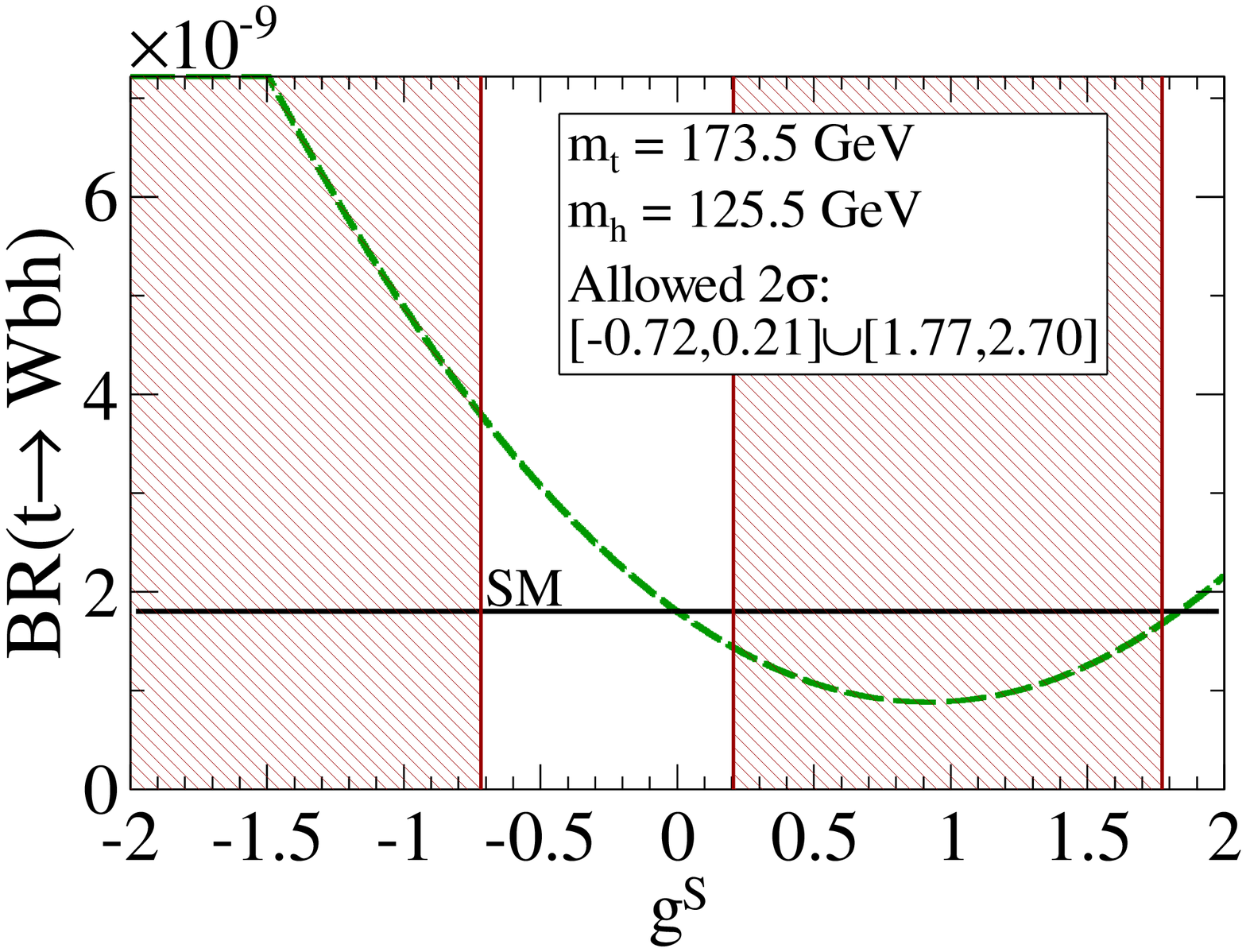}	\label{gs.FIG}	}
\subfigure[]{	\includegraphics[width=0.46\textwidth]{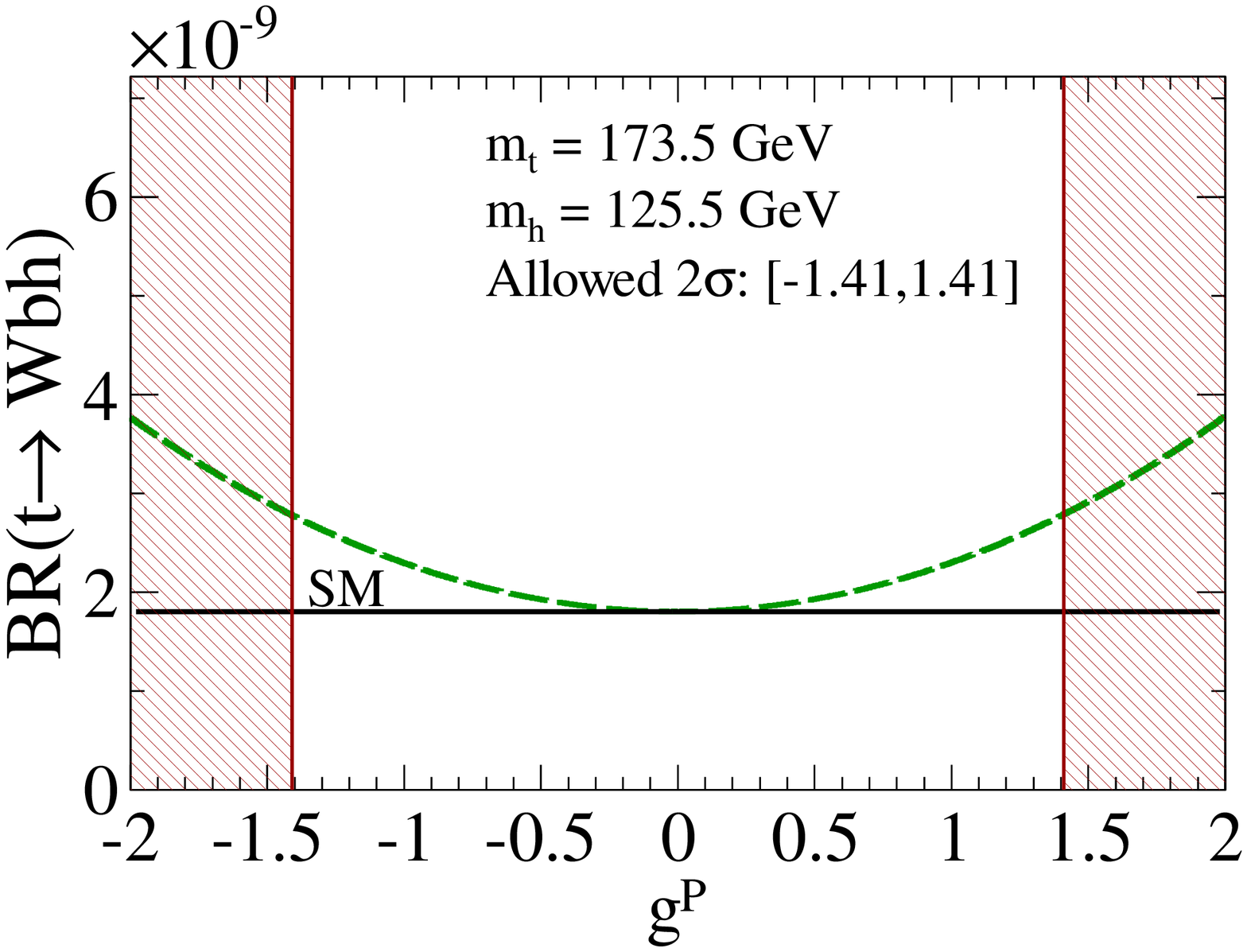}	\label{gp.FIG}	}
\\
\subfigure[]{	\includegraphics[width=0.46\textwidth]{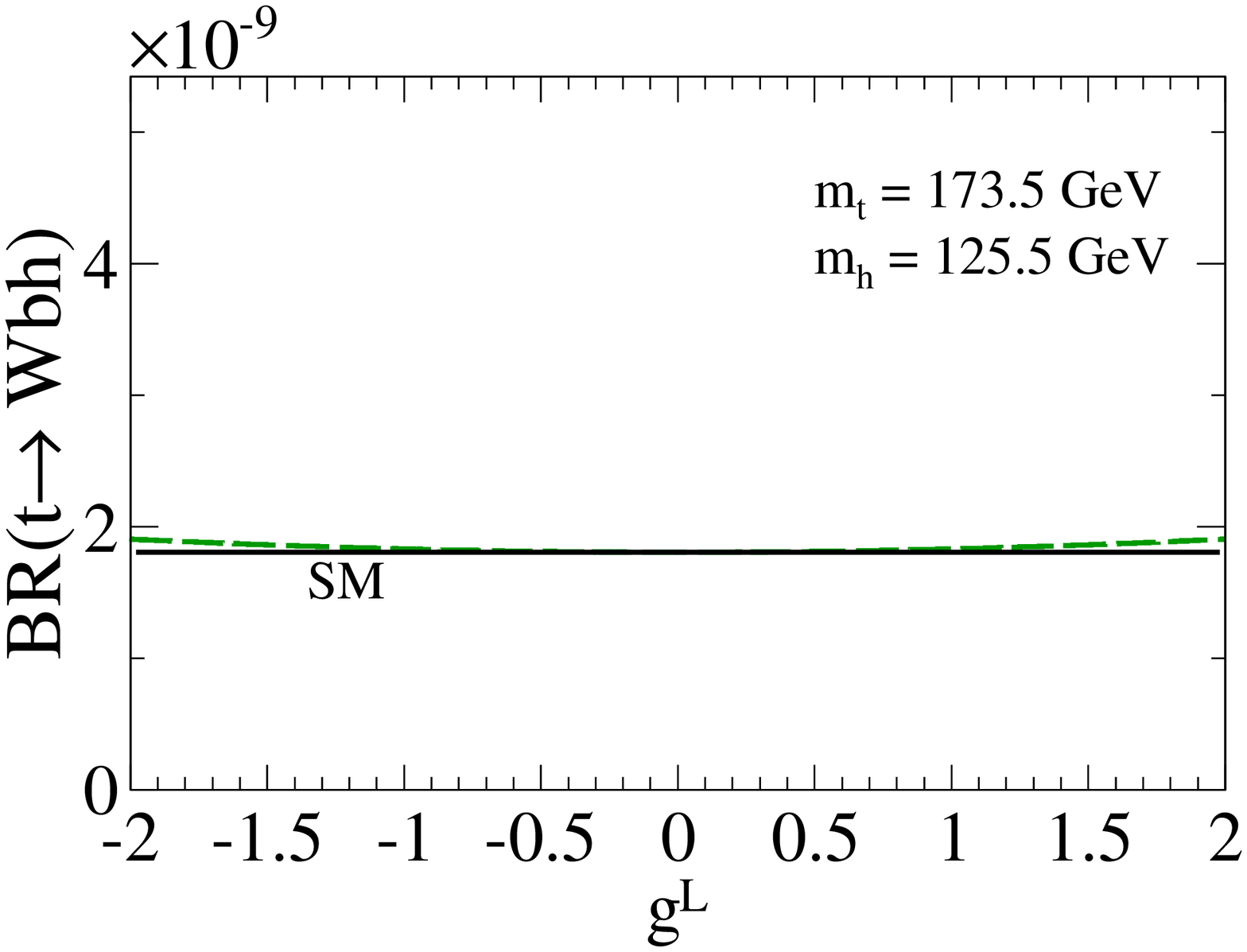}	\label{gl.FIG}	}
\subfigure[]{	\includegraphics[width=0.46\textwidth]{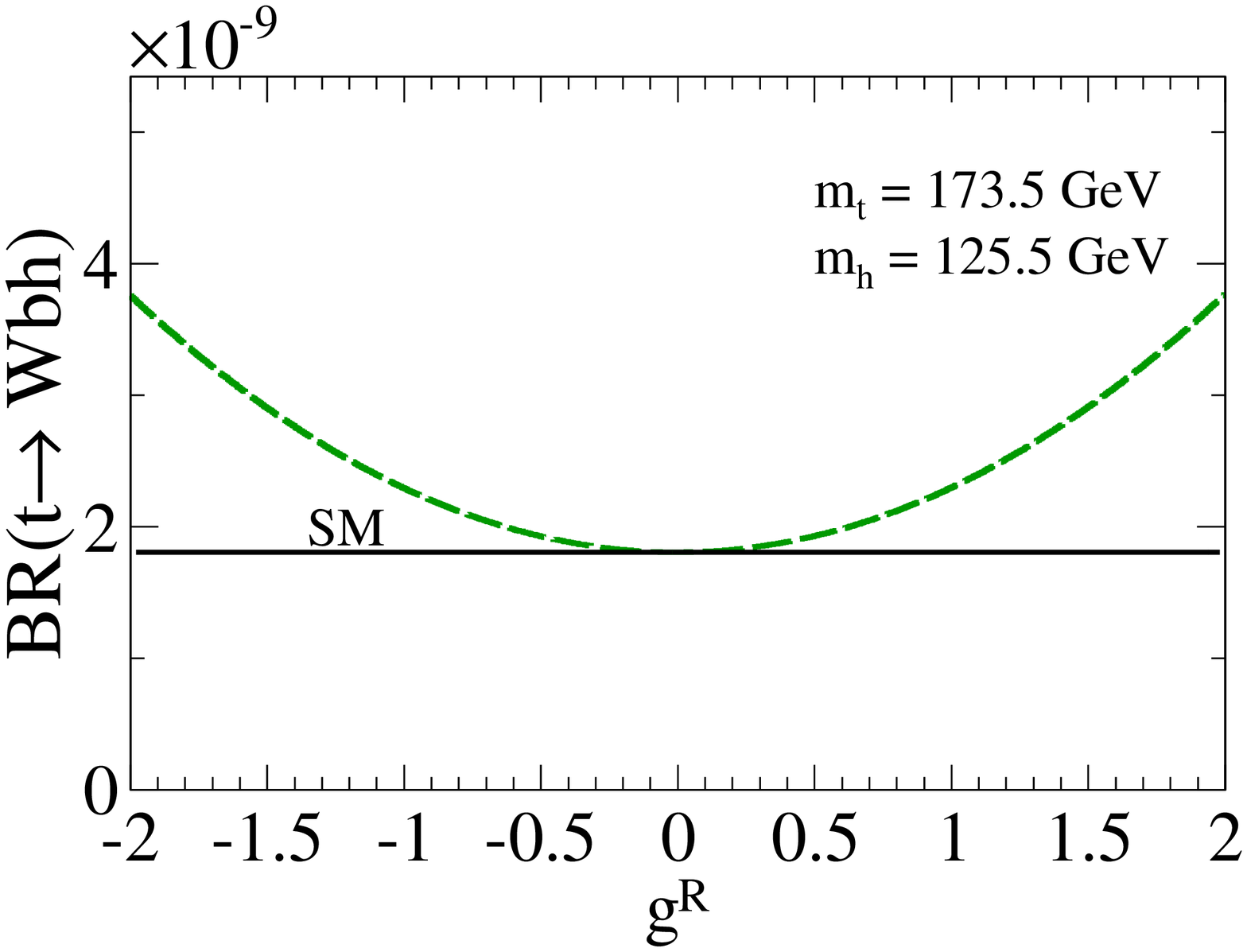}	\label{gr.FIG}}
\caption[$\BR(t\rightarrow Wbh)$ as a function of EFT coupling]{
$\BR(t\rightarrow Wbh)$ as a function of
(a) $g^{S}$, (b) $g^{P}$, (c) $g^{L}$, (d) $g^{R}$.
The solid line denotes the SM prediction, Eq.~(\ref{smBR.EQ}).
The shaded region is excluded at 95\% C.L.
}
\label{effBR.FIG}
\end{figure}

We present first the behavior of  $\BR(t\rightarrow Wbh)$ as a function of anomalous $t\overline{t}h$ couplings.
For one non-zero anomalous coupling from Eq.~(\ref{eftLag.EQ}) at a time, we calculate the branching fraction over the domain $g^{X}\in[-2,+2]$ 
and set all other anomalous couplings to zero.
Bounds on $g^{S}$ and $g^{P}$, Eqs.~(\ref{gS2Sig.EQ}) and (\ref{gP2Sig.EQ}) respectively, are applied. 
The results are shown in Fig.~\ref{effBR.FIG}.
To investigate the sensitivity of operators that select out different kinematic features, 
we include the redundant operators listed in Eq.~(\ref{GIOp2.EQ}), which give rise to anomalous $g^{L}$ and $g^{R}$. 
In all plots, the SM prediction  as in Eq.~(\ref{smBR.EQ}) is shown as a (black) solid line labeled by ``SM''.
Table~\ref{eftBench.TB} lists values of the branching fraction for various benchmark values of $g^{X}$.

In Fig.~\ref{gs.FIG}, $\BR(t\rightarrow Wbh)$ as a function of the anomalous scalar coupling $g^{S}$ is shown. 
From the Lagrangian in Eq.~(\ref{cpvLag.EQ}), it is clear that $(y_{t}-g^{S})$ acts as an  effective Yukawa coupling.
For $g^{S}<0$, the anomalous coupling enhances the already dominant top-Higgsstrahlung diagram.
For $g^{S}>0$, an accidental cancellation among the anomalous scalar, Yukawa, and gauge terms results in a minimum at $g^{S}\approx0.92$.
When $g^{S}\gtrsim0.92$, the quadratic term takes over and causes the branching fraction to grow.
An observed transition rate smaller than the SM prediction thus implies that $g^{S}>0$.
Indirect measurements of the $t\overline{t}h$ coupling, as seen in Fig.~\ref{gs.FIG}, indicate that 
\begin{equation}
\BR(t\rightarrow Wbh) = (0.8\sim2.1) \times \BR^{\rm SM}(t\rightarrow Wbh).
\label{eftBRLimitI.EQ}
\end{equation}

Figure~\ref{gp.FIG} shows the influence of an anomalous pseudoscalar coupling, $g^{P},$ on $\BR(t\rightarrow Wbh)$.
From the Lagrangian in Eq.~(\ref{cpvLag.EQ}), similar to the discussions in the previous session, the $t\rightarrow h$ transition is symmetric with respect to $g^{P}$ due to the dominance of the quadratic term.
Both couplings contribute greatest when the intermediate, off-shell top quark propagates in its RH helicity state, which gives an $m_{t}$ enhancement over other diagrams.
The CPV associated with $\delta_{\rm CP}$ is unobservable here because the asymmetry is proportional to interference terms, which are small. 

The  linear dependence on $g^{S}$ in interference terms from the previous case and the strict quadratic dependence on $g^{P}$ here implies that that branching fraction is less sensitive to small values of $g^{P}$ than it is to small values of $g^{S}$.
The rate therefore grows more slowly as a function of  $g^{P}$ than $g^{S}$.
As seen in Figure~\ref{gp.FIG}, the bounds on $g^{P}$ allow
\begin{equation}
\BR(t\rightarrow Wbh) = (1\sim 1.5) \times \BR^{\rm SM}(t\rightarrow Wbh).
\end{equation}

In Fig.~\ref{gl.FIG}, we see the branching fraction as a function of an anomalous LH vector current with coupling $g^{L}$.
Over the domain investigated, the contribution is rather small.
We turn to kinematics to elucidate this behavior.
First, the anomalous contribution is proportional to $k_\mu/v$, where $k_{\mu}$ is the momentum of the Higgs. 
Since the energy budget for this process is fixed at $m_{t}$, and since we require a final state Higgs $(E_{h}\gtrsim m_{h})$, 
$k_\mu/v\sim E_{h}/v$ ranges between $0.5\sim0.6$, leading to kinematic suppression of anomalous contributions.
Second, note that a fermion participating in two sequential LH chiral interactions necessarily propagates in its LH helicity state.
Hence, the anomalous contribution is proportional to the internal, off-shell top quark momentum and leads to helicity suppression of anomalous contributions. 
We consequently expect and observe very small growth in the branching fraction over the range of $g^{L}$.

Figure~\ref{gr.FIG} displays the results for $\BR(t\rightarrow Wbh)$ as a function of anomalous RH vector current with coupling $g^{R}$.
Unlike the LH case, the anomalous contribution has a large effect over the domain considered, comparable to $g^{S}$ and $g^{P}$.
As in the previous case, there is kinematic suppression; however, there is no longer helicity suppression.
A massive fermion participating in a RH chiral interaction followed by a LH chiral interaction propagates in its RH helicity state.
Hence, as in the $g^{P}$ case, the anomalous contribution is proportional to $m_{t}$. 
Comparatively, there is a faster rise in the transition rate as a function of $g^{R}$ than $g^{L}$.

 \begin{table}
\caption{$\BR(t\rightarrow WbH)$ for Benchmark Values of Higgses in the 2HDM(I)}
 \begin{center}
\begin{tabular}{|c|c|c|c|c|}
\hline \hline
 $H~(125.5~{\rm GeV})$ &  $\Delta_{V}$ & $\tan\beta$ & $\BR^{2HDM(I)}(t\rightarrow WbH)$  \tabularnewline\hline\hline 
 $h$ &  $0.05$ & $3$   & $1.840 \times 10^{-9}$     \tabularnewline\hline
 $h$ &         & $7.6$ & $1.714 \times 10^{-9}$     \tabularnewline\hline
 $H$ &  $0.7$  & $3$   & $1.460 \times 10^{-9}$     \tabularnewline\hline 
 $H$ &         & $7.6$ & $1.567 \times 10^{-9}$     \tabularnewline\hline
 $h$ &  $0.7$  & $3$   & $4.643 \times 10^{-10}$    \tabularnewline\hline
 $h$ &         & $7.6$ & $2.573 \times 10^{-10}$    \tabularnewline\hline
 $A~(100~{\rm GeV})$ &      & $3$   & $1.814 \times 10^{-9}$    \tabularnewline\hline
 $A~(100~{\rm GeV})$ &      & $7.6$ & $2.829 \times 10^{-10}$  \tabularnewline\hline\hline\end{tabular}
\label{2hdmIBench.TB}
\end{center}
\end{table}

\begin{figure}[ptb]
\centering
\subfigure[]{	\includegraphics[width=0.47\textwidth]{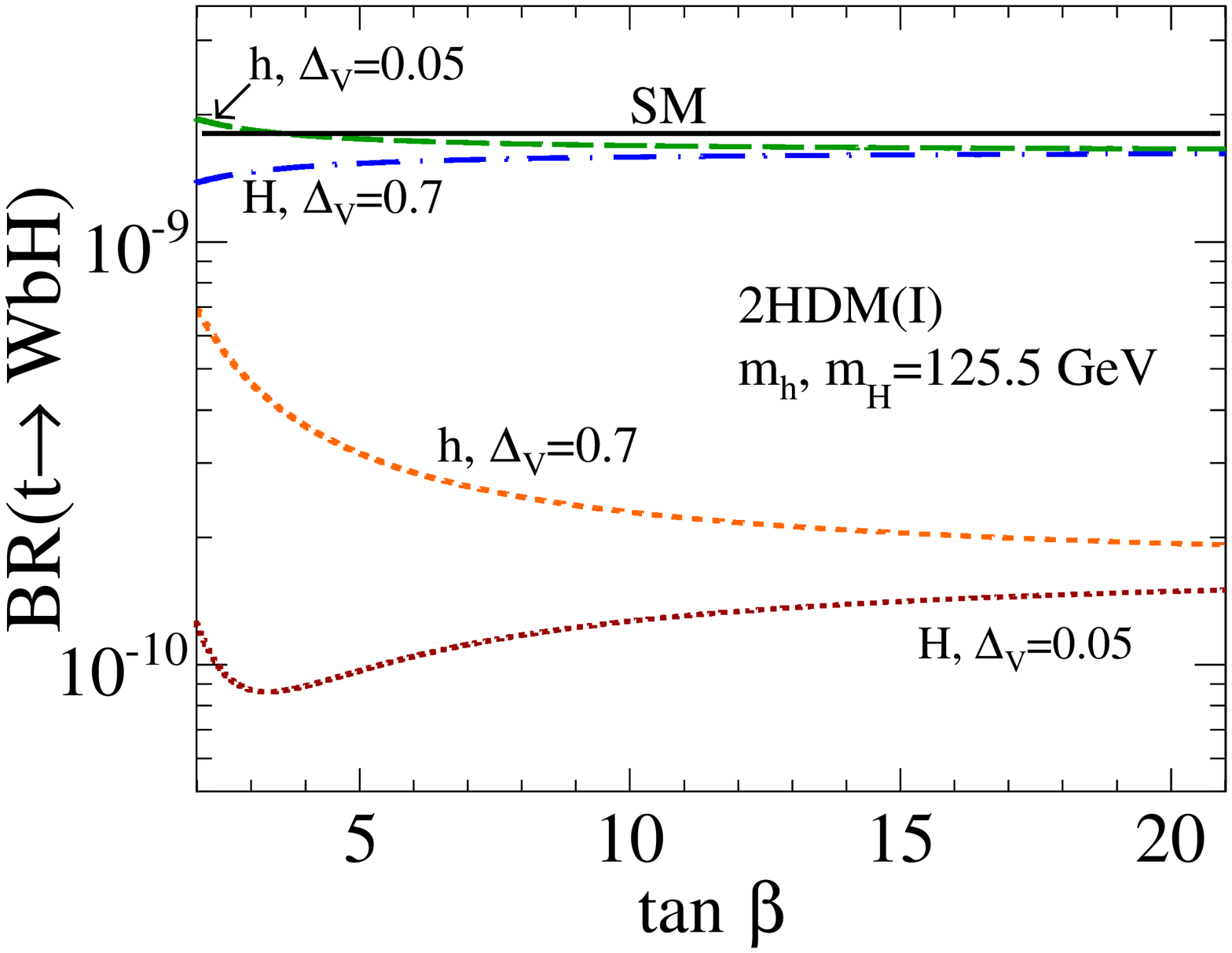}	\label{2hdmI_brVstB.FIG}}
\subfigure[]{	\includegraphics[width=0.47\textwidth]{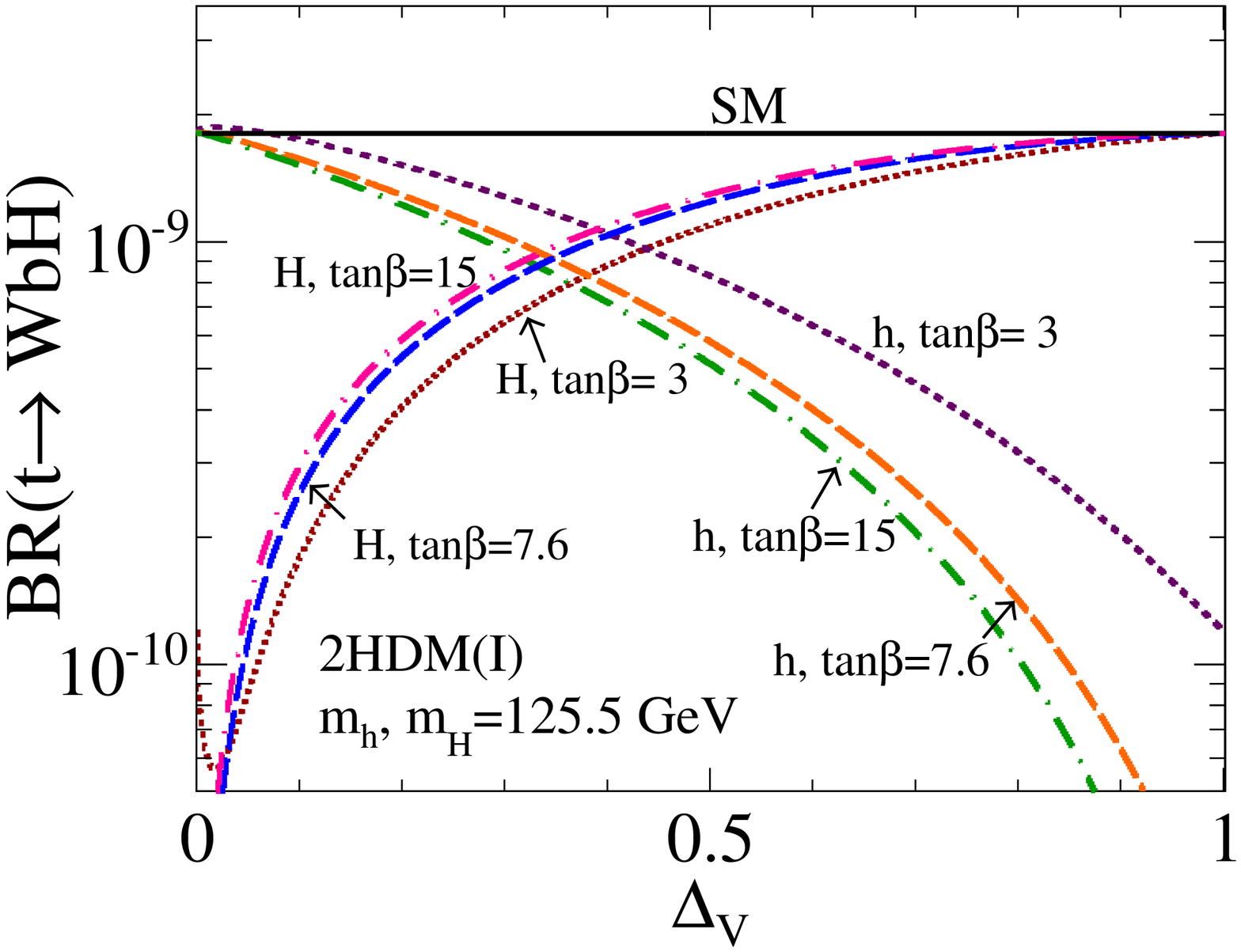}	\label{2hdmI_brVsDV.FIG}}
\caption[The 2HDM(I) $\BR(t\rightarrow WbH)$ as a function of $\tan\beta$ and $\Delta_V$]{
The 2HDM(I) $\BR(t\rightarrow WbH)$ as a function of 
(a) $\tan\beta$ for SM-like $h$ (long dash), $H$ (dash-dot), and non-SM-like $h$ (short dash), $H$ (dot);
(b) $\Delta_{V}$ for $h$ at $\tan\beta = 3~7.6,~15$ (short dash, long dash, dash-dot), and for $H$ (dot, long dash, dash-dot).
The solid line denotes the SM prediction, Eq.~(\ref{smBR.EQ}).
}
\label{2hdmI_HvsInput.FIG}
\end{figure}

\subsection{Type I 2HDM BR$(t\rightarrow WbH)$}
\label{2hdmIBR.SEC}
The behavior of $\BR(t\rightarrow WbH)$, where $H$ represents $h,~H,$ or $A$ in the 2HDM(I), is presented in this section.
To explore sensitivity to the anomalous $WWH$ coupling, $\Delta_{V}$, we consider
\begin{equation}
 \tan\beta=3,~7.6,~15\quad\text{for}~\Delta_{V}\in[0,1].
\end{equation}
For these values of $\tan\beta$, 
the largest deviation in the $WWH$ coupling allowed by present data corresponds to a light SM-like Higgs with $\cos(\beta-\alpha)=0.3$, i.e.,
\begin{equation}
  \Delta_{V}=0.05~(0.7)\quad\text{for}\quad h~(H)\approx h^{SM}.
\end{equation}
To determine the mass sensitivity, we focus on the mass windows 
\begin{eqnarray}
  m_{h}\in[95~\text{GeV},126~\text{GeV}]\label{mhMass.EQ},\ \
  m_{H}\in [126~\text{GeV},155~\text{GeV}]\label{mHMass.EQ},\ \
  m_{A}\in [95~\text{GeV},155~\text{GeV}]\label{mAMass.EQ}.
\end{eqnarray}
Below 95 GeV, the SM $Z$ boson background becomes relevant, making observation of the transition very difficult;
above $155$ GeV the kinematic suppression of $t\rightarrow H/A$ becomes too great for practical purposes.
However, it is straightforward to extrapolate these results in the event of a neutral scalar's discovery in these peripheral ranges.

Table~\ref{2hdmIBench.TB} lists values of $\BR(t\rightarrow WbH)$ for several Higgses and benchmark parameter values.

\subsubsection{BR$(t\rightarrow Wbh, H)$ vs $\tan\beta$, $\Delta_{V}$}

\begin{figure}[!t]
\centering
\subfigure[]{\includegraphics[width=0.47\textwidth]{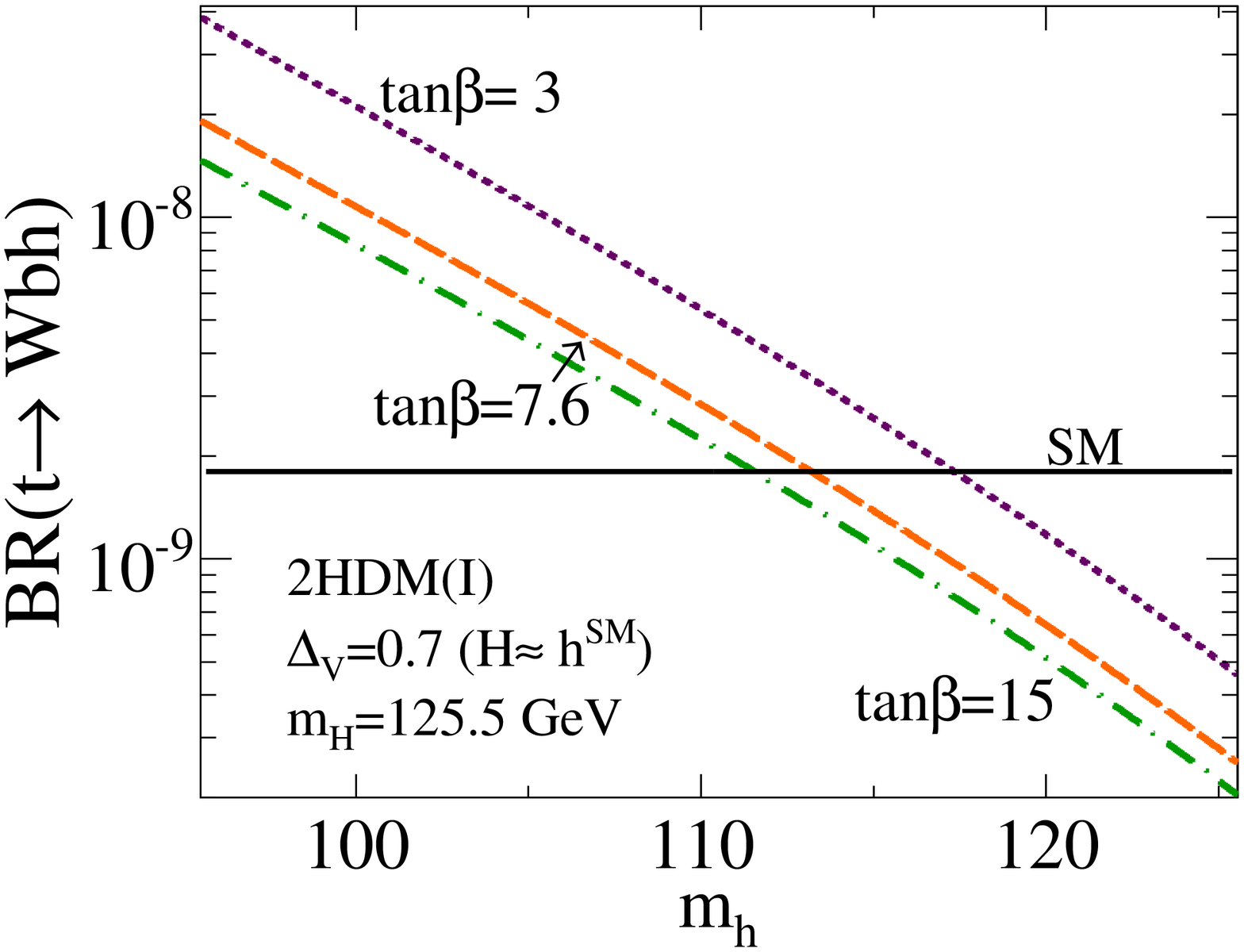}	 \label{2hdmI_brVsMH1.FIG}}
\subfigure[]{\includegraphics[width=0.47\textwidth]{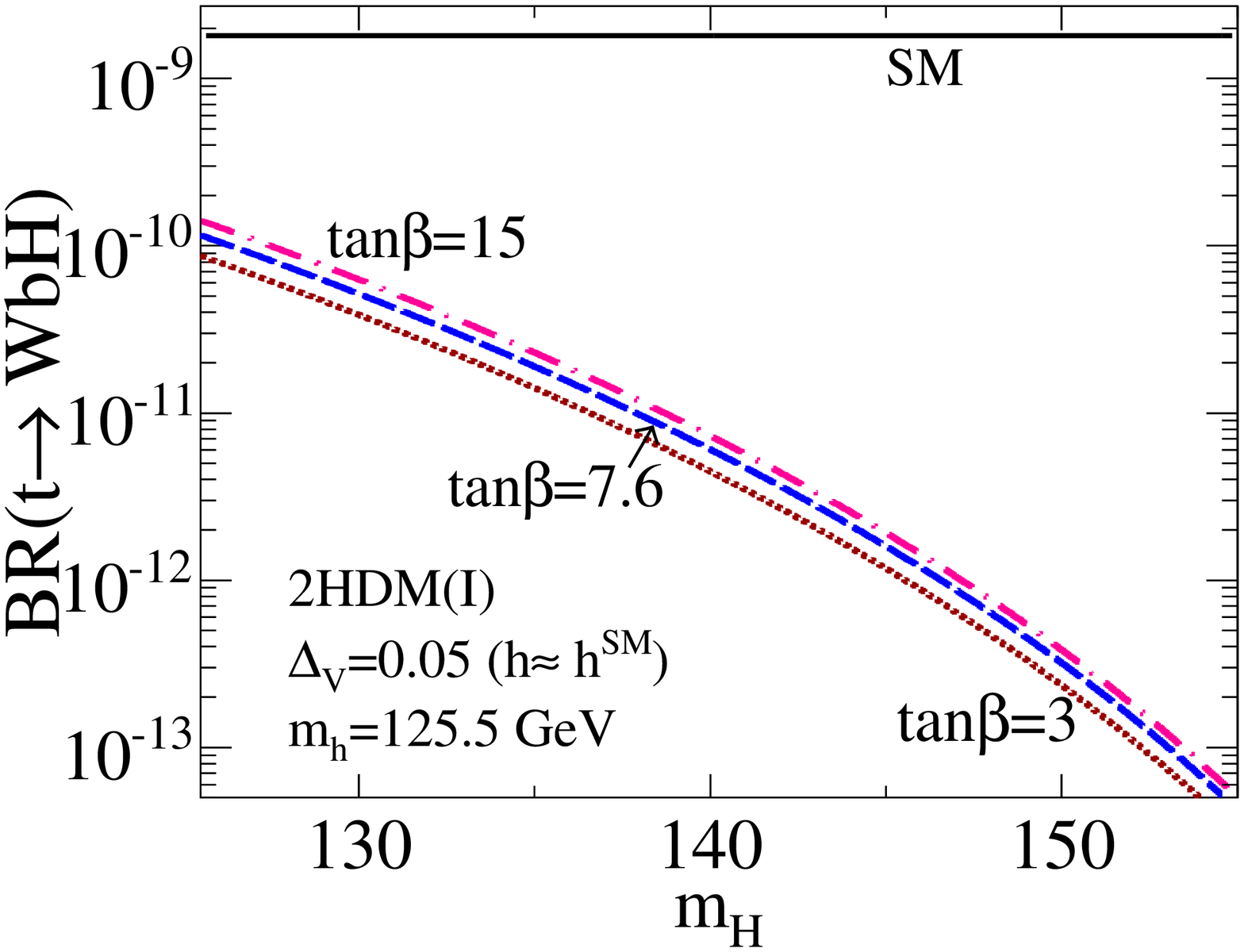}	\label{2hdmI_brVsMH2.FIG}}
\vspace{.15in}\\
\caption[2HDM(I) $\BR(t\rightarrow WbH)$ as a function of mass for a non-SM-like Higgs]{
The 2HDM(I) $\BR(t\rightarrow WbH)$ as a function of mass for a non-SM-like 
(a) $h$ and (b) $H$ assuming $\tan\beta=3,~7.6,~15$ (short dash, long dash, dash-dot).
The solid line denotes the SM prediction, Eq.~(\ref{smBR.EQ}).
}
\label{2hdmI_HvsMH.FIG}
\end{figure}

\begin{figure}[!t]
\centering
\subfigure[]{	\includegraphics[width=0.47\textwidth]{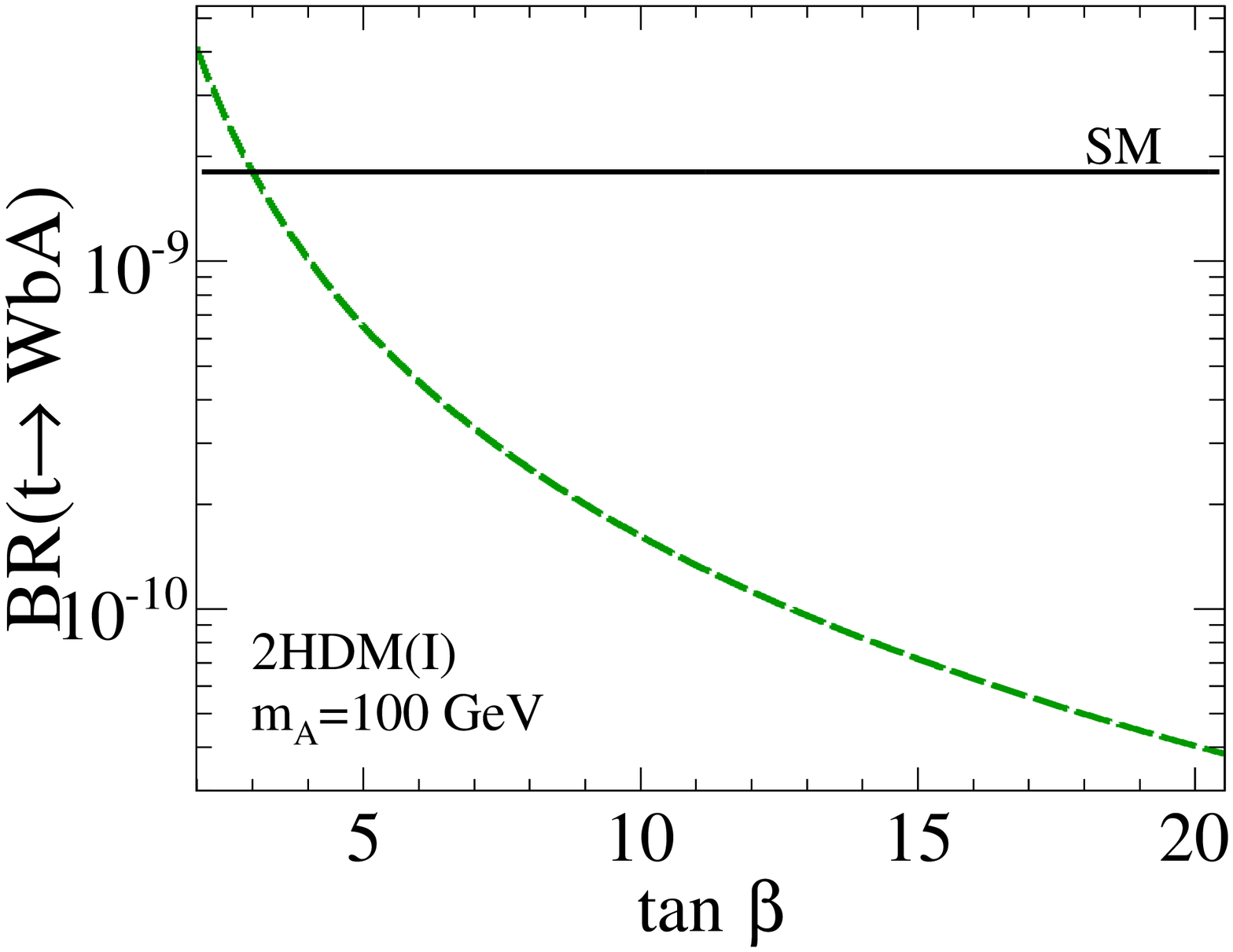}	 \label{2hdmIA0vstB.FIG}}
\subfigure[]{	\includegraphics[width=0.47\textwidth]{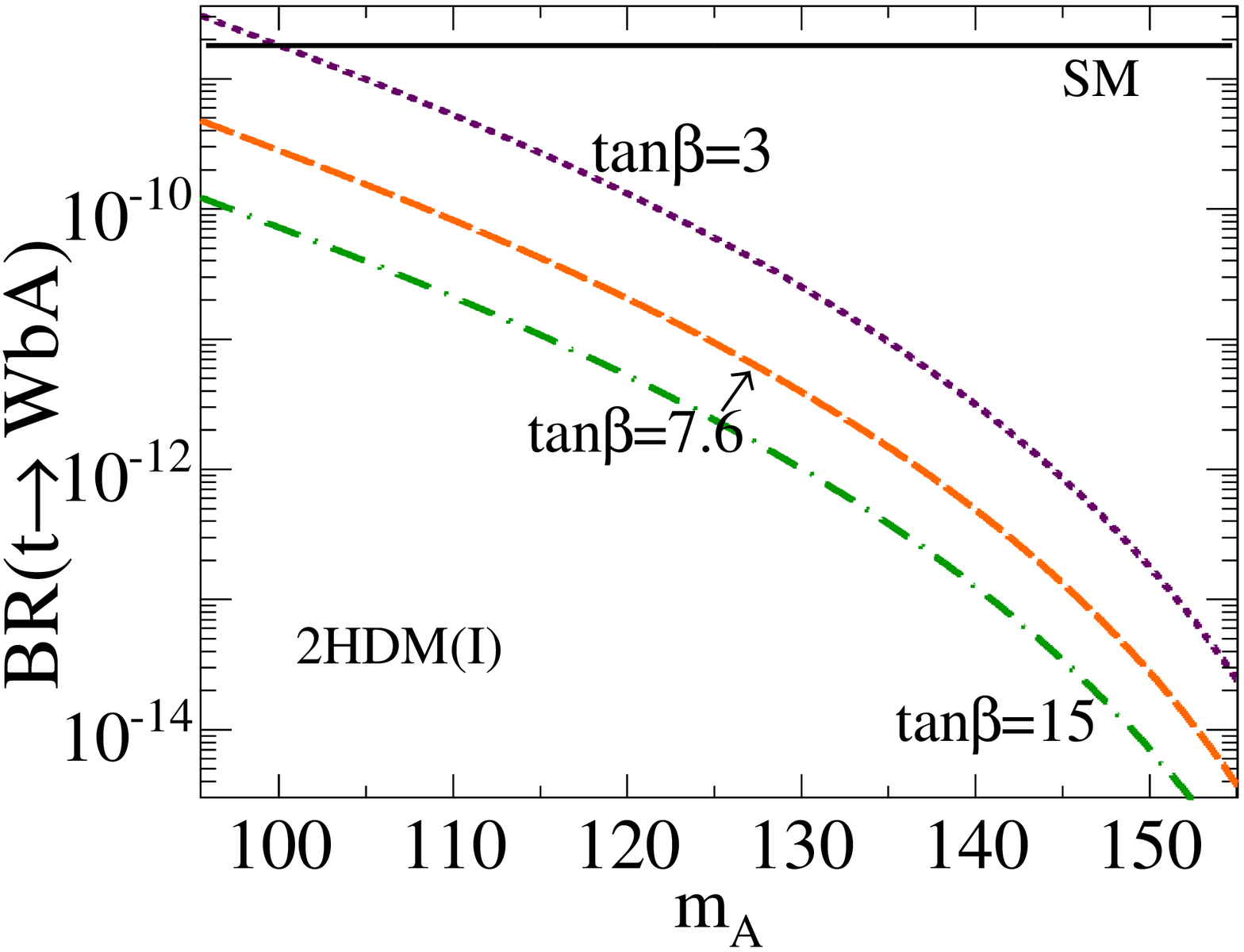}	\label{2hdmIA0vsMA.FIG}}
\caption[2HDM(I) $\BR(t\rightarrow WbA)$ as a function of  $\tan\beta$ and $m_{A}$]{
The 2HDM(I) $\BR(t\rightarrow WbA)$ as a function of 
(a) $\tan\beta$ and (b) $m_{A}$ for 
$\tan\beta=3,~7.6,~15$ (short dash, long dash, dash-dot).
The solid line denotes the SM prediction, Eq.~(\ref{smBR.EQ}).
}
\label{2hdmI_A0.FIG}
\end{figure}

The decay rates for $t\rightarrow W^{*}bh$ and $t\rightarrow W^{*}bH$ as a function of (a) $\tan\beta$ and (b) $\Delta_{V}$ are shown in Fig.~\ref{2hdmI_HvsInput.FIG}.
Except for low value of $\tan\beta <3$, the rates are always smaller than the SM rate.
Beyond $\tan\beta\approx3$, the SM-like CP-even Higgs rates become independent of $\tan\beta$ and converge to asymptotic values;
for the non-SM-like Higgses, this occurs at $\tan\beta\approx15$. 
To see how this happens, note that the Yukawa couplings in the 2HDM(I) (Table \ref{2hdmICouple.TB}) take the simple form
\begin{equation}
 c_{1} \cot\beta + c_{2},
\end{equation}
where $c_{1,2}$ are elementary functions of $\Delta_{V}$, as seen in Table~\ref{2hdmICouple.TB}.
In the large $\tan\beta$ limit, the $c_{1}$ part vanishes, leaving the asymptotic value $c_{2}$.
In the SM-like limit, the $c_{2}$ terms are larger than the $c_{1}$ contributions, whereas the reverse holds in the non-SM-like limit.
We extract asymptotic values by observing that for a given CP-even Higgs the $c_{2}$ terms and $WWH$ couplings are the identical.
Consequently,
\begin{eqnarray}
 \underset{\tan\beta\rightarrow\infty}{\lim} \BR^{2HDM(I)}(t\rightarrow Wbh) &=& (1-\Delta_{V})^{2}\BR^{SM}(t\rightarrow Wbh)\nonumber\\
									    &=& \sin^{2}\left(\beta-\alpha\right)\BR^{SM}(t\rightarrow Wbh)
  \label{asympI.EQ}\\
 \underset{\tan\beta\rightarrow\infty}{\lim} \BR^{2HDM(I)}(t\rightarrow WbH) &=& (2\Delta_{V}-\Delta_{V}^{2})\BR^{SM}(t\rightarrow Wbh)\nonumber\\
									    &=& \cos^{2}\left(\beta-\alpha\right)\BR^{SM}(t\rightarrow Wbh).
 \label{asympII.EQ}
 \end{eqnarray}
For our choices of $\Delta_{V}$, the asymptotic rates in Fig.~\ref{2hdmI_brVstB.FIG} are 
\begin{eqnarray}
\underset{\tan\beta\rightarrow\infty}{\lim} \BR^{2HDM(I)}_{\Delta_{V}=0.7}(t\rightarrow WbH) &=& 0.910 \times \BR^{SM}(t\rightarrow Wbh),
 \label{smlikeH1Rate.EQ}\\
\underset{\tan\beta\rightarrow\infty}{\lim} \BR^{2HDM(I)}_{\Delta_{V}=0.05}(t\rightarrow Wbh) &=& 0.903 \times \BR^{SM}(t\rightarrow Wbh),
 \label{smlikeH2Rate.EQ}\\
\underset{\tan\beta\rightarrow\infty}{\lim} \BR^{2HDM(I)}_{\Delta_{V}=0.05}(t\rightarrow WbH) &=& 0.098 \times \BR^{SM}(t\rightarrow Wbh),
 \label{smunlikeH1Rate.EQ}\\
\underset{\tan\beta\rightarrow\infty}{\lim} \BR^{2HDM(I)}_{\Delta_{V}=0.7}(t\rightarrow Wbh) &=& 0.090 \times \BR^{SM}(t\rightarrow Wbh),
 \label{smunlikeH2Rate.EQ}
\end{eqnarray}
and agree well with numerical calculations.

The $\Delta_{V}$ dependence in Fig~\ref{2hdmI_brVsDV.FIG} and the relationship between $h$ and $H$ 
is indicative of much broader behavior found in all 2HDM variants.
To saturate the sum rule for the electroweak symmetry breaking \cite{Gunion:1989we}, the $hWW$ coupling ($g_{hWW}$) and the $HWW$ coupling ($g_{HWW}$) obey 
\begin{equation}
  g^{2}_{hWW}+g^{2}_{HWW}=g^{2}_{h^{SM}WW},
 \label{2hdmSumRule.EQ}
\end{equation}
where $g_{h^{SM}WW}$ is the SM $hWW$ coupling. 
For $h$ and $H$ with degenerate masses, 
\begin{equation}
  BR(t\rightarrow Wbh) + BR(t\rightarrow WbH) = \BR^{SM}(t\rightarrow Wbh)+\mathcal{O}(\cot^{2}\beta).
 \label{2hdmSumRule2.EQ}
\end{equation}
Indeed, Eqs.~(\ref{asympI.EQ}) and (\ref{asympII.EQ}) satisfy this relationship. 
Furthermore, this can be extended to an arbitrary number of scalar $SU(2)_{L}$ doublets and singlets~\cite{Gunion:1989we}.
Though mass splittings, etc., will break this equality, it provides a useful estimate for processes involving transitions in models with additional scalar $SU(2)_{L}$ doublets and singlets.

\subsubsection{BR$(t\rightarrow Wbh/H)$ vs $m_{h/H}$}
As a function of mass, we plot in Fig.~\ref{2hdmI_HvsMH.FIG} the decay rates for $t\rightarrow W^{*}bH$ where $H$ is a non-SM-like CP-even Higgs;
the mass of the SM-like Higgs is taken to be $125.5$ GeV.
For a mass below (above) $110$ GeV, we observe that transition rate to a non-SM-like Higgs remains above (below) the SM rate.
As the scalar mass decreases and the $W^{*}$ comes closer to being on-shell, 
the availability of phase space greatly ameliorates the coupling suppression associated with $\Delta_{V}$.
However, despite this relief, the transition rate to a non-SM-like $H$ stays below the SM rate for much of the parameter space.
The insensitivity to large and moderate $\tan\beta$ seen in Fig.~\ref{2hdmI_HvsMH.FIG} is consistent with previous discussions.

\begin{table}
\caption{$\BR(t\rightarrow WbH)$ for Benchmark Values of Higgses in the 2HDM(II)}
 \begin{center}
\begin{tabular}{|c|c|c|c|}
\hline \hline
 $H~(125.5~{\rm GeV})$ &  $\Delta_{V}$ & $\tan\beta$ & $\BR^{2HDM(II)}(t\rightarrow WbH)$  \tabularnewline\hline\hline 
 $h$ &  $5\times10^{-5}$ & $3$   & $1.813 \times 10^{-9}$     \tabularnewline\hline
 $h$ &                   & $7.6$ & $1.809 \times 10^{-9}$     \tabularnewline\hline
 $H$ &  $0.99$           & $3$   & $1.798 \times 10^{-9}$     \tabularnewline\hline 
 $H$ &                   & $7.6$ & $1.802 \times 10^{-9}$     \tabularnewline\hline
 $h$ &  $0.99$           & $3$   & $1.440 \times 10^{-10}$    \tabularnewline\hline
 $h$ &                   & $7.6$ & $4.990 \times 10^{-11}$    \tabularnewline\hline
 $A~(100~{\rm GeV})$ &   & $3$   & $1.760 \times 10^{-9}$    \tabularnewline\hline
 $A~(100~{\rm GeV})$ &   & $7.6$ & $1.007 \times 10^{-9}$  \tabularnewline\hline\hline
\end{tabular}
\label{2hdmIIBench.TB}
\end{center}
\end{table}

\begin{figure}[ptb]
\centering
\subfigure[]{
	\includegraphics[width=0.47\textwidth]{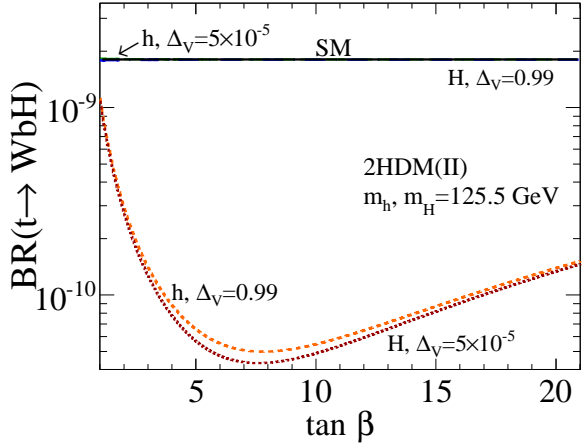}	
	\label{brVstB.FIG}
}
\subfigure[]{
	\includegraphics[width=0.47\textwidth]{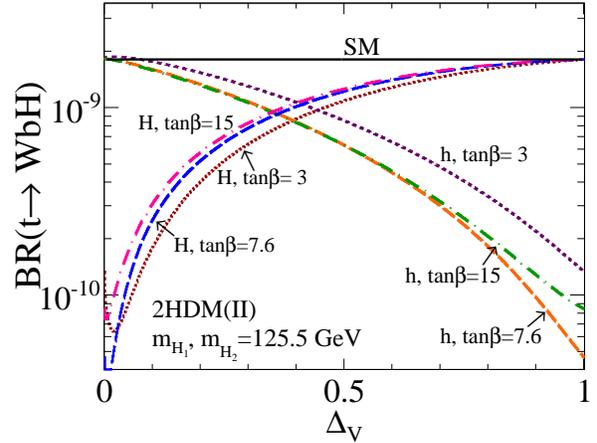}	
	\label{brVsDV.FIG}
}
\caption[2HDM(II) $\BR(t\rightarrow WbH)$ as a function of  $\tan\beta$ and $\Delta_{V}$]{
The 2HDM(II) $\BR(t\rightarrow WbH)$ as a function of 
(a) $\tan\beta$ for SM-like $h$ (long dash), $H$ (dash-dot), and non-SM-like $h$ (short dash), $H$ (dot);
(b) $\Delta_{V}$ for $h$ at $\tan\beta = 3~7.6,~15$ (short dash, long dash, dash-dot), and for $H$ (dot, long dash, dash-dot).
The solid line denotes the SM prediction, Eq.~(\ref{smBR.EQ}).
}
\label{HvsInput.FIG}
\end{figure}	

\subsubsection{BR$(t\rightarrow WbA)$ vs $\tan\beta$, $m_{A}$}
Here, we consider the decay rate to the CP-odd Higgs, $t\rightarrow W^{*}bA$.
Fig. \ref{2hdmI_A0.FIG} shows $\BR(t\rightarrow WbA)$ as a function of (a) $\tan\beta$ and (b) $m_{A}$.
Except for very low $\tan\beta$ and mass, the branching fraction remains well below the SM prediction for much of the parameter space,
approximately equaling it at $\tan\beta\simeq3$ for $m_{A}=100$ GeV.
Due to CP-invariance in the gauge sector there is no tree-level $AWW$ contribution.
And since the $f\overline{f}A$ couplings are independent of $\Delta_{V}$, the decay rate is fixed entirely by $m_{A}$ and $\tan\beta$. 
Destructive interference still exists, however, since the $t\overline{t}A$ and $b\overline{b}A$ vertices differ by a minus sign.
A quadratic dependence on $\cot\beta$ is the consequence the $f\overline{f}A$ coupling $(\propto \cot\beta)$.
See Table~\ref{2hdmICouple.TB}.
Despite this monotonic dependence on $\tan\beta$, which implies that $\BR(t\rightarrow WbA)$ is a direct measure of $\tan\beta$ were it to be measured, 
the recuperation of available phase space at low $m_{A}$ is unable to compensate for the $\cot^{2}\beta$ suppression.


\subsection{Type II 2HDM BR$(t\rightarrow WbH)$}
\label{2hdmIIBR.SEC}

\begin{figure}[ptb]
\centering
\subfigure[]{	\includegraphics[width=0.47\textwidth]{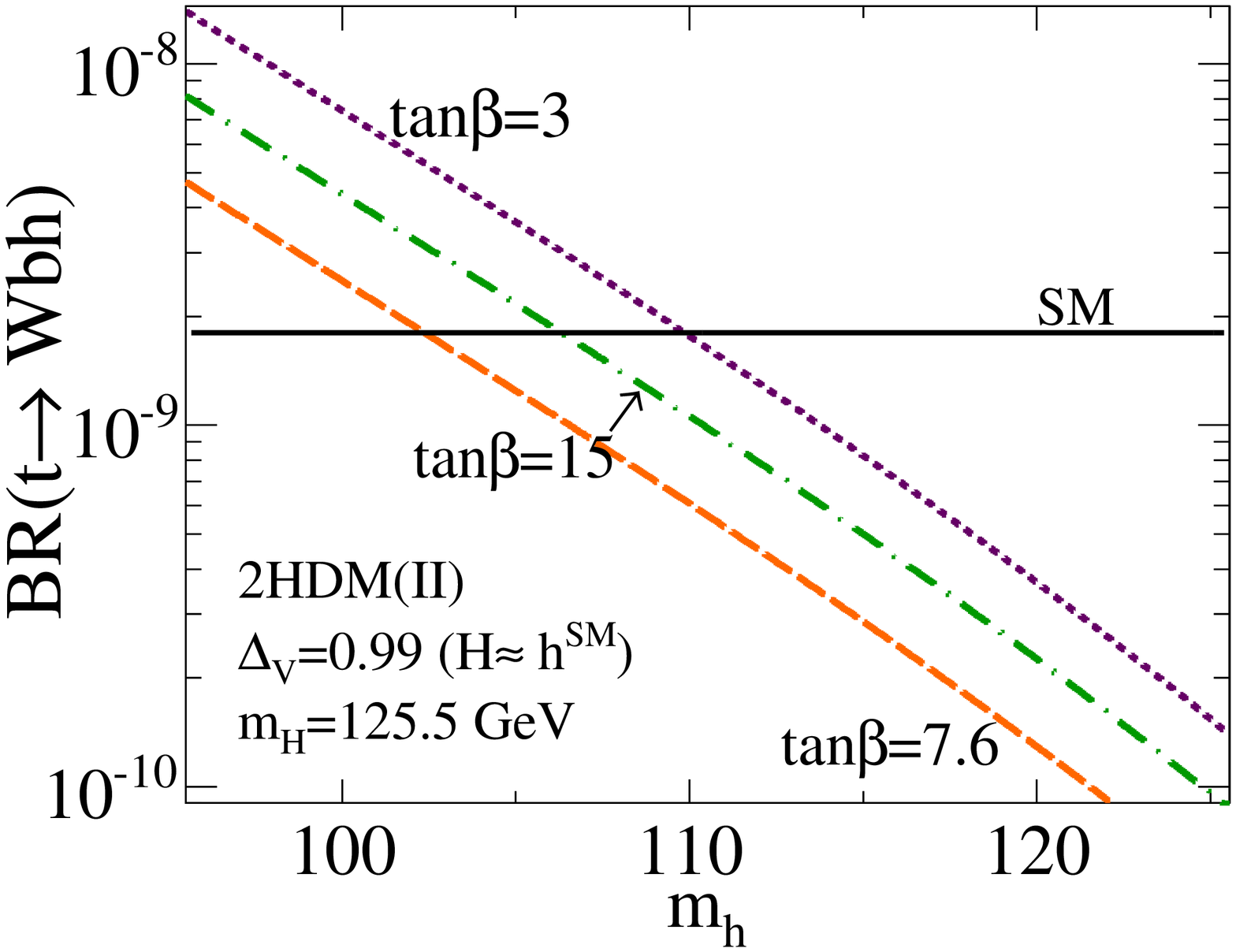}}
\subfigure[]{	\includegraphics[width=0.47\textwidth]{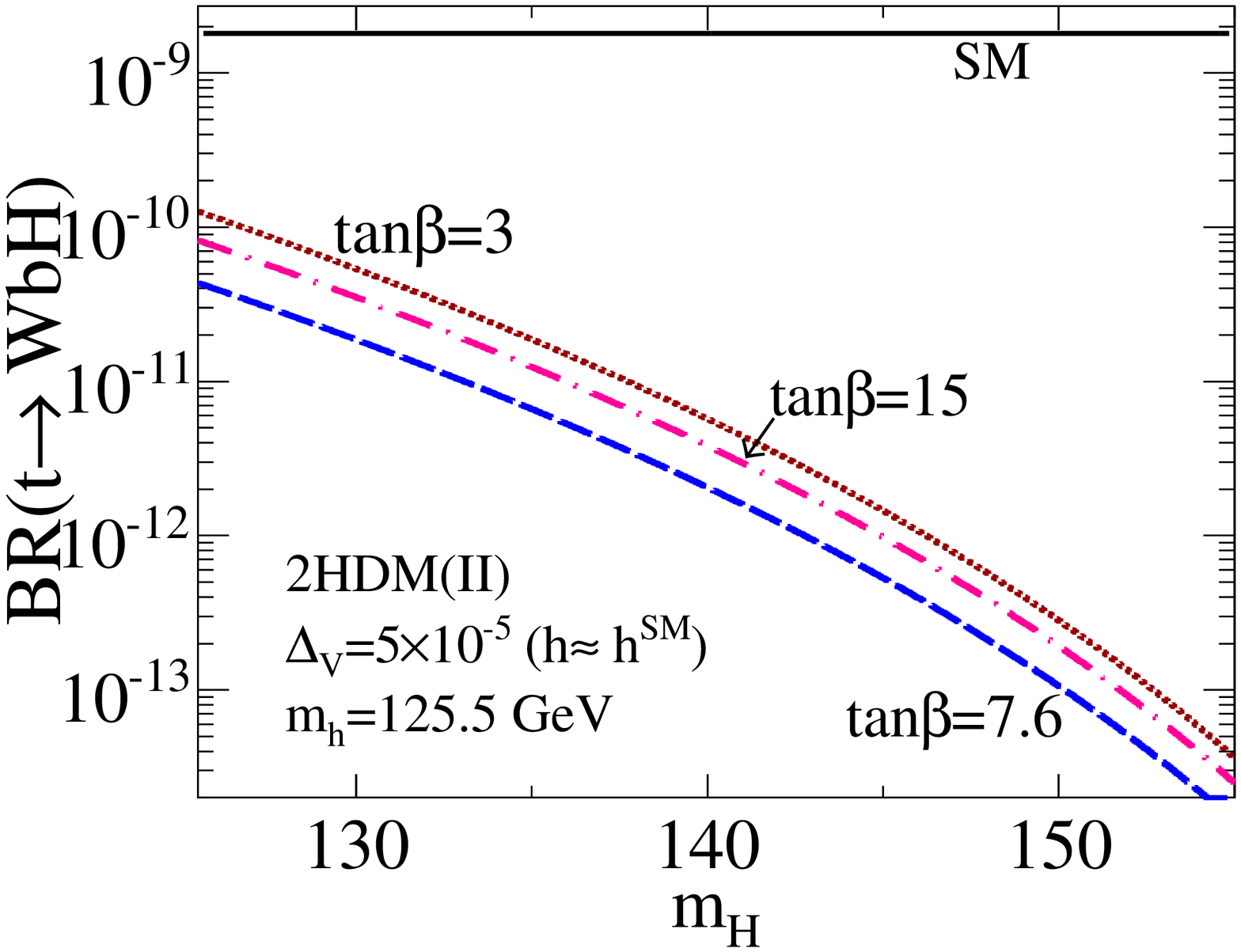}	}
\caption[2HDM(I) $\BR(t\rightarrow WbH)$ as a function of mass for a non-SM-like Higgs]{
The 2HDM(I) $\BR(t\rightarrow WbH)$ as a function of mass for a non-SM-like 
(a) $h$ and (b) $H$ assuming $\tan\beta=3,~7.6,~15$ (short dash, long dash, dash-dot).
The solid line denotes the SM prediction, Eq.~(\ref{smBR.EQ}).
}
\label{HvsMH.FIG}
\end{figure}

We report here the behavior of $\BR(t\rightarrow WbH)$, where $H$ represents $h,~H,$ or $A$ in the 2HDM(II). 
The same values of $\tan\beta$ are used here as the Type I case.
To avoid constraints, we choose a $\Delta_{V}$ that corresponds to a light Higgs with $\cos(\beta-\alpha)=0.01$, i.e.,
\begin{equation}
  \Delta_{V}=5.\times10^{-5}~(0.99)\quad\text{for}\quad h~(H)\approx h^{SM}.
\end{equation}
Table~\ref{2hdmIIBench.TB} lists values of the branching fraction for several Higgses and benchmark parameter values.
In the following figures, the predicted SM decay rate is shown as a black, solid line labeled by ``SM''.

\subsubsection{BR$(t\rightarrow Wbh, H)$ vs $\tan\beta$, $\Delta_{V}$}

Figure~\ref{HvsInput.FIG} depicts the branching ratio $\BR(t\rightarrow WbH)$ for both of the CP-even Higgses as a function of 
(a) $\tan\beta$ for SM-like and non-SM-like $h$ and $H$, and (b) $\Delta_{V}$ for small and large values of $\tan\beta$.

In Fig~\ref{brVstB.FIG}, for SM-like Higgses, the branching fraction is indistinguishable from the SM prediction as a function of $\tan\beta$;
for non-SM-like Higgses, however, the rates minimize for $\tan\beta=7\sim8$.
This dependence on $\tan\beta$ is indicative of a playoff between the $t\overline{t}H$ and the $b\overline{b}H$ couplings.
In the SM limit, this specific behavior is suppressed for SM-like Higgs bosons because the couplings to these bosons grow independent of $\tan\beta$. 
When $h$ is non-SM-like $(\Delta_{V}=0.99)$, sensitivity to $\tan\beta$ maximizes because the $\tan\beta$-independent parts of the fermionic Higgs couplings nearly cancel.
As $\tan\beta$ grows, the contribution from $t\overline{t}h~(\propto \cot\beta)$ 
runs $\BR(t\rightarrow Wbh)$ down until the $b\overline{b}h$ contribution $(\propto \tan\beta)$ takes over at $\tan\beta\approx 7.6$.
When $H$ is non-SM-like $(\Delta_{V}=5.\times 10^{-5})$ we expect and observe similar behavior as the non-SM-like $h$ case.

Much of the relationship between $h$ and $H$ observed in in Fig.~\ref{brVsDV.FIG} is type-independent and the discussion can be found in the Type I scenario.
For a light Higgs, we indeed see that at $\tan\beta=7.6$ transition rates are minimized for all values of $\Delta_{V}$.
For a heavy Higgs, however, this value of $\tan\beta$ only minimizes the rate in the $h\rightarrow h^{SM}$ limit, 
in which case the $t\rightarrow H$ transition rate vanishes.
The $t\rightarrow H$ rate minimum occurs at larger $\Delta_{V}$ with decreasing $\tan\beta$ because the $t\overline{t}H~(b\overline{b}H)$  contribution becomes numerically larger (smaller).


\subsubsection{BR$(t\rightarrow Wbh, H)$ vs $m_{H}$}

Figure~\ref{HvsMH.FIG} presents the $t\rightarrow W^{*}bH$ branching ratio for a non-SM-like Higgs boson as a function of mass.
For the mass window given in Eq.~(\ref{mAMass.EQ}), we find considerable enhancement in the decay rate relative to the SM rate due to the increase in available phase space, overcoming the coupling suppression associated with scalars that have non-SM-like coupling.

\subsubsection{BR$(t\rightarrow WbA)$ vs $\tan\beta$, $m_{A}$}
\begin{figure}[ptb]
\centering
\subfigure[]{	\includegraphics[width=0.47\textwidth]{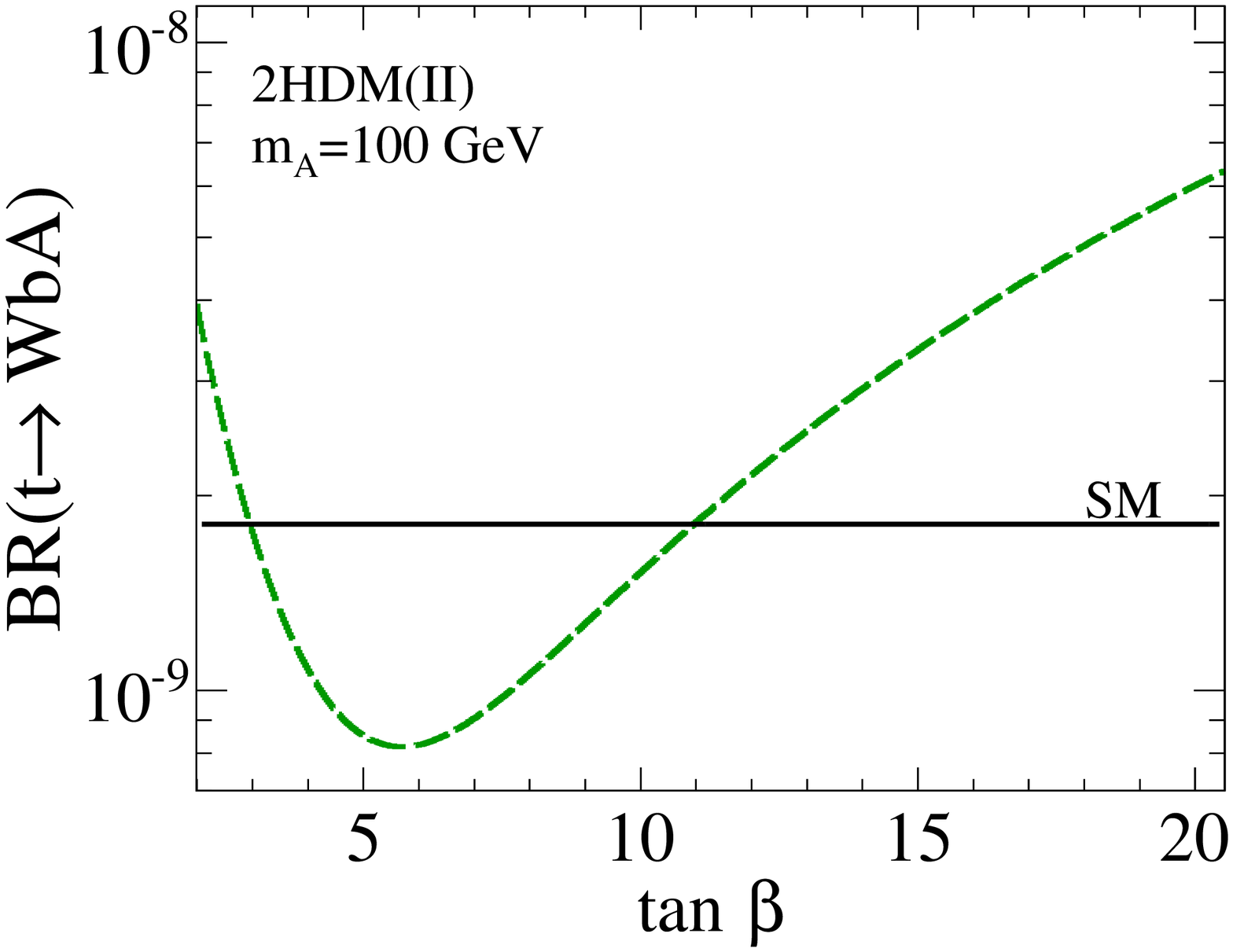}	 \label{A0vstB.FIG}}
\subfigure[]{	\includegraphics[width=0.47\textwidth]{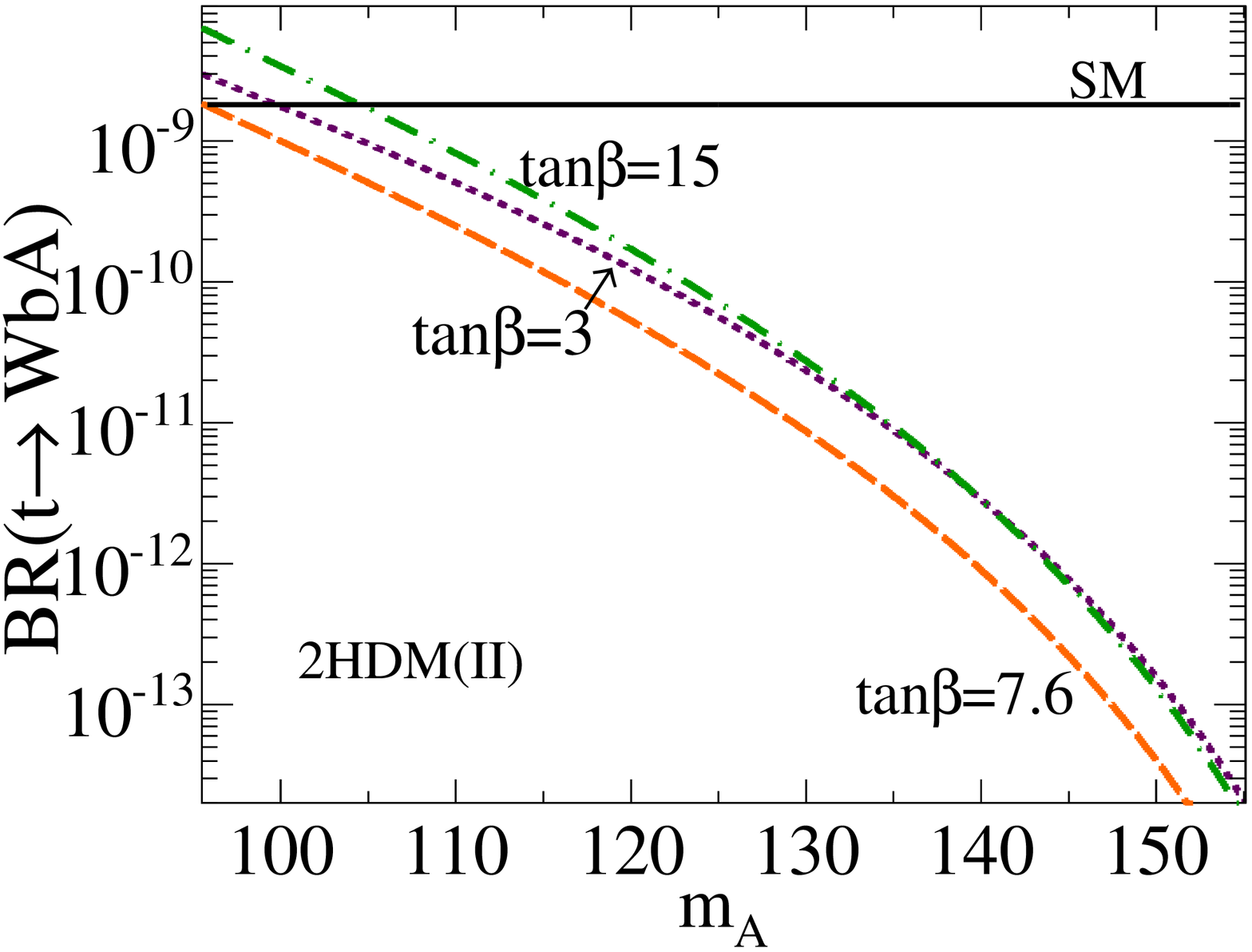}	\label{A0vsMA.FIG}	}
\caption[2HDM(II) $\BR(t\rightarrow WbA)$ as a function of  $\tan\beta$ and  $m_{A}$]{
The 2HDM(II) $\BR(t\rightarrow WbA)$ as a function of 
(a) $\tan\beta$ and (b) $m_{A}$ for $\tan\beta=3,~7.6,~15$ (short dash, long dash, dash-dot).
The solid line denotes the SM prediction, Eq.~(\ref{smBR.EQ}).
}
\label{A0vsX.FIG}
\end{figure}

Turning to the CP-odd Higgs decay channel, $t\rightarrow W^{*}bA$,
we note that many of the arguments made in the 2HDM(I) case carry over to this situation. 
Unlike the Type I scenario, however, there is only {\it constructive} interference between the fermion contributions.
Figure~\ref{A0vsX.FIG} shows $\BR(t\rightarrow WbA)$ as a function of (a) $\tan\beta$ and (b) $m_{A}$.

In Fig.~\ref{A0vstB.FIG}, due to an accidental cancellation, the branching fraction minimizes at $\tan\beta\approx 5.8$,
which is unsurprisingly close to the $t\rightarrow H^{+}b$ minimum at $\tan\beta=\sqrt{m_{t}/m_{b}}\approx 7.6$.
At $\tan\beta\approx 5.8$, 
the $t\overline{t}A$ coupling $(\propto\cot\beta)$ and the $b\overline{b}A$ coupling $(\propto\tan\beta)$ contribute equally.  
At smaller values of $\tan\beta$, $t\overline{t}A$ is the dominant term but is driven down by an increasing $\tan\beta$; 
and at larger values, $b\overline{b}A$ is the dominant term, which ramps up the rate.
In the large $\tan\beta$ limit, the $t\overline{t}A$ graph becomes negligible and the rate becomes quadratically with $\tan\beta$.

In Fig.~\ref{A0vsMA.FIG}, we observe a similarity between $A$ and the non-SM-like Higgs boson, $H_{X}$. 
We attribute this to a similarity of contributing diagrams. 
For example: the $WWA$ vertex does not exist because of CP-invariance, and by virtue of being {\it non}-SM-like, the $WWH_{X}$ vertex is considerably suppressed. 
In this domain, fermionic couplings to $A$ and $H_{X}$ also have the same dependence on $\tan\beta$.

\section{Observation Prospects at Colliders}
\label{LHC.SEC}
In this section, we estimate observation prospects at current and future colliders.
The 14 TeV LHC $t\overline{t}$ production cross section at NNLO in QCD has been calculated~\cite{Czakon:2013goa} to be
\begin{equation}
\sigma_{LHC14}^{NNLO}(t\overline{t})= 933~\text{pb}.
\end{equation}
The SM $pp\rightarrow t\overline{t}\rightarrow WW^{*}b\overline{b}h$ cross section at the LHC is thus estimated to be
\begin{equation}
 \sigma_{LHC14}(pp\rightarrow t\overline{t}\rightarrow WW^{*}b\overline{b}h)
 \approx2\times\sigma^{NNLO}_{LHC14}(t\overline{t})\times BR(t\rightarrow Wbh) 
 = {3.4}~\text{ab}.
\label{lhcppWbh.EQ}
 \end{equation}
The factor of two in the preceding line accounts for either top or antitop quark decaying into the Higgs.
To assure a clear trigger and to discriminate against the large SM backgrounds, we require at least one $W$ boson decaying leptonically ($\ell = e, \mu$), i.e.
\begin{equation}
 \BR(WW^{*}\rightarrow \ell^{+}\ell^{'-}\nu_{\ell}\overline{\nu}_{\ell'} + jj\ell^{\pm}\overset{(-)}{\nu_{\ell}})\approx 0.33.
\end{equation}
The total cross section for an arbitrarily decaying $h$ is therefore estimated to be 
\begin{eqnarray}
 \sigma_{LHC14}(pp\rightarrow t\overline{t}\rightarrow WW^{*}b\overline{b}h\rightarrow 
 h(\ell^{+}\ell^{'-}\nu_{\ell}\overline{\nu}_{\ell'} + jj\ell^{\pm}\overset{(-)}{\nu_{\ell}})) &\approx&  1.1~\text{ab}.
\label{lhcppWhbDecay.EQ}
 \end{eqnarray}
Higgs branching fractions and detector efficiencies will further suppress this rate.
Such a small cross section means that observing this SM process will be challenging. 
Following the same procedure, we estimate Eq.~(\ref{lhcppWhbDecay.EQ}) for several proposed colliders and collider upgrades; 
the results are given in Table~\ref{collider.TB}.

\begin{table}
\caption[Cross sections for $t\overline{t}$ and 
$t\overline{t}\rightarrow WW^{*}b\overline{b}h\rightarrow  h(\ell^{+}\ell^{'-}\nu_{\ell}\overline{\nu}_{\ell'} + jj\ell^{\pm}\nu_{\ell})$]{Cross sections for $t\overline{t}$ and 
$t\overline{t}\rightarrow WW^{*}b\overline{b}h\rightarrow  h(\ell^{+}\ell^{'-}\nu_{\ell}\overline{\nu}_{\ell'} + jj\ell^{\pm}\nu_{\ell})$
at 14~\cite{Czakon:2013goa}, 33~\cite{ttbar33TeV:2012}, and 100~\cite{Baur:2002ka} TeV $pp$, and 350 GeV $e^{+}e^{-}$~\cite{seidelTop:2012} Colliders.
}
 \begin{center}
\begin{tabular}{|c|c|c|c|c|}
\hline \hline
Process  & 14 TeV $pp$ & 33 TeV $pp$ & 100 TeV $pp$ & 350 GeV $e^{+}e^{-}$  \tabularnewline
\hline\hline
$\sigma(t\overline{t})$[pb]  & $933$ & $5410$ & $2.7\times10^{4}$  & $0.45$  \tabularnewline \hline
$\sigma(t\overline{t}\rightarrow 
 h(\ell^{+}\ell^{'-}\nu_{\ell}\overline{\nu}_{\ell'} + jj\ell^{\pm}\overset{(-)}{\nu_{\ell}}))$ [ab] & $1.1$ & $6.5$  & $32$ & $5\times10^{-4}$      
 \tabularnewline \hline
\hline
\end{tabular}
\label{collider.TB}
\end{center}
\end{table}

\section{Summary and Conclusion}
\label{conc.SEC}

Given the discovery of a SM-like Higgs boson, we have recalculated the rare top quark decay mode $t\rightarrow W^{*}bh$,  
where $h$ represents the SM Higgs boson. 
We have extended this calculation to include the effects of anomalous $t\overline{t}h$ couplings originating from effective operators 
as well as both CP-even and the single CP-odd scalars in the CP-conserving 2HDM Types I and II. 
The most updated model constraints have been reported. 
We summarize our results:
\begin{enumerate}[(i)]

 \item The SM predicts a $t\rightarrow W^{*}bh$ branching ratio of
 \begin{equation}
 \BR(t\rightarrow Wbh) = 1.80\times 10^{-9}\quad\text{for}\quad m_{h}=125.5~\text{GeV}.
 \label{conBR.EQ}
 \end{equation}
This is the leading $t\rightarrow h$ transition, five orders of magnitude larger than the next channel $t\rightarrow ch$.  See Eq.~(\ref{smBR.EQ}).

 \item Present LHC Higgs constraints on anomalous $t\overline{t}h$ couplings permit up to a factor of two enhancement of the $t\rightarrow W^{*}bh$ transition. See Eq.~(\ref{eftBRLimitI.EQ}).

 \item The operator $\overline{\mathcal{O}}_{t2}$, 
 which selects different kinematic features than either $\mathcal{O}_{t1}$ or $\overline{\mathcal{O}}_{t1}$,
 results in comparable enhancement of the $t\rightarrow W^{*}bh$ transition. See Fig.~\ref{effBR.FIG}.
 
  \item In the 2HDM(I), decays to CP-even Higgses do not decouple in the large $\tan\beta$ limit and their rates approach asymptotic values that are functions of the anomalous $WWh$ coupling. They are given in Eqs.~(\ref{asympI.EQ}) and (\ref{asympII.EQ}).

 \item In the Type I (II) 2HDM, due to the increase in available phase space, the branching ratio to a light, non-SM-like Higgs boson can as much as $2~(7)$ times larger than Eq.~(\ref{conBR.EQ}). 

 \item In the Type I (II) 2HDM, the branching ratio to a light, CP-odd Higgs can be as much as $1.6~(3)$ times larger than Eq.~(\ref{conBR.EQ}). 
 
 \item The $pp\rightarrow t\overline{t}\rightarrow WW^{*}b\overline{b}h\rightarrow 
 h(\ell^{+}\ell^{'-}\nu_{\ell}\overline{\nu}_{\ell'} + jj\ell^{\pm}\overset{(-)}{\nu_{\ell}})$ production cross section at the 14 TeV LHC and future colliders have been estimated [Eq.~(\ref{lhcppWhbDecay.EQ})]; a few $t\rightarrow W^{*}bh$ events over the full LHC lifetime. 
      Due to enhancements in gluon distribution functions, any increase in collision energies can greatly increase this rate.
\end{enumerate}

%% file: 04_NeutrinoFromWA/waHeavyN.tex
\chapter[Inclusive Heavy Majorana Neutrino Production at Hadron Colliders]{Inclusive Heavy Majorana Neutrino Production at Hadron Colliders}

\section{Introduction}
\label{sec:intro}

The discovery of the Higgs boson completes the Standard Model (SM). Yet, the existence of nonzero neutrino masses 
remains one of the clearest indications of physics beyond the Standard Model 
(BSM)~\cite{Mohapatra:1998rq,Gluza:2002vs,Fukugita:2003en,Barger:2003qi,Eidelman:2004wy,GonzalezGarcia:2007ib,Mohapatra:2006gs,Strumia:2006db}
The simplest SM extension that can simultaneously explain both the existence of neutrino masses and their smallness,
the so-called Type I seesaw 
mechanism~\cite{Minkowski:1977sc,Yanagida:1979as,VanNieuwenhuizen:1979hm,Ramond:1979py,Glashow:1979nm,Mohapatra:1979ia,GellMann:1980vs,Schechter:1980gr,Shrock:1980ct,Schechter:1981cv}, 
introduces a right handed (RH) neutrino $N_{R}$.
Via a Yukawa coupling $y_{\nu}$,
the resulting Dirac mass is $m_{D}= y_{\nu}\langle \Phi\rangle$, where $\Phi$ is the SM Higgs SU$(2)_{L}$ doublet.
As $N_{R}$ is a SM-gauge singlet, 
one could assign $N_{R}$ a Majorana mass $m_{M}$ without violating any fundamental symmetry of the model. 
Requiring that $m_{M}\gg m_{D}$, the neutrino mass eigenvalues are
\begin{equation}
 m_{1}\sim m_D \frac{m_D}{m_M} \quad\text{and}\quad m_2 \sim m_M.
\end{equation}
Thus, the apparent smallness of neutrino masses compared to other fermion masses is due to the suppression by a new scale above the EW scale. Taking the Yukawa coupling to be $y_{\nu}\sim \mathcal{O}(1)$, the Majorana mass scale must be of the order $10^{13}$ GeV to recover sub-eV light neutrinos masses.
However, if the Yukawa couplings are as small as the electron Yukawa coupling, i.e., $y_{\nu}\lesssim\mathcal{O}(10^{-5})$, 
then the mass scale could be at $\mathcal{O}(1)$ TeV or lower \cite{ArkaniHamed:2000bq,Borzumati:2000mc,deGouvea:2005er,deGouvea:2006gz}. 

\begin{figure}[!t]
\centering
\subfigure[]{\includegraphics[width=.65\textwidth]{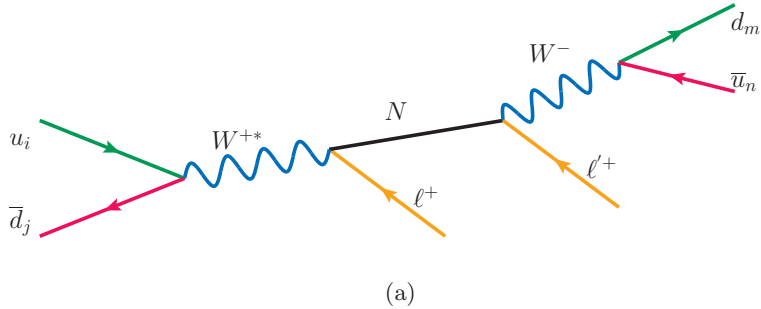}}
\caption[Diagram representing resonant heavy Majorana neutrino production]{Diagram representing resonant heavy Majorana neutrino production through the DY process and its decay into same-sign leptons and dijet.
All diagrams drawn using JaxoDraw~\cite{Binosi:2003yf}.} 
\label{feynDYDecay.fig}
\end{figure}

Given the lack of guidance from theory of lepton flavor physics, 
searches for Majorana neutrinos must be carried out as general and model-independent as possible. 
Low-energy phenomenology of Majorana neutrinos has been studied in detail
~\cite{Keung:1983uu,Dicus:1991fk,Pilaftsis:1991ug,Datta:1993nm,deGouvea:2005er,deGouvea:2006gz,Han:2006ip,
delAguila:2007em,Atre:2009rg,Chao:2009ef,AguilarSaavedra:2012fu,
Das:2012ii,AguilarSaavedra:2012gf,Han:2012vk,Chen:2013foz,Dev:2013wba,Davoudiasl:2014pya}.
Studied first in Ref.~\cite{Keung:1983uu} and later in 
Refs.~\cite{Dicus:1991fk,Pilaftsis:1991ug,Datta:1993nm,Han:2006ip,del Aguila:2007em,Atre:2009rg},
the production channel most sensitive to heavy Majorana neutrinos $(N)$ at hadron colliders is the resonant Drell-Yan (DY) process,
\begin{equation}
 p p \rightarrow W^{\pm*} \to N ~\ell^{\pm},\ \ 
\ \ {\rm with}\ \ N \rightarrow W^{\mp} ~\ell^{'\pm},  \quad W^{\mp} \rightarrow j ~j,
 \label{dy.EQ}
\end{equation}
in which the same-sign dilepton channel violates lepton number $L$ by two units $(\Delta L=2)$; see figure~\ref{feynDYDecay.fig}.
Searches for Eq.~(\ref{dy.EQ}) are underway at LHC experiments~\cite{Chatrchyan:2012fla,ATLAS:2012yoa,Aaij:2012zr}. 
Non-observation in the dimuon channel has set a lower bound on the heavy neutrino mass of 100 (300) GeV for mixing 
$\vert V_{\mu N}\vert^{2} = 10^{-2~(-1)}$~\cite{ATLAS:2012yoa}.
Bounds on mixing from $0\nu\beta\beta$ \cite{Belanger:1995nh,Benes:2005hn} and EW precision data \cite{Nardi:1994iv,Nardi:1994nw,delAguila:2008pw,Antusch:2014woa} indicate that 
the 14 TeV LHC is sensitive to Majorana neutrinos with mass between 10 and 375 GeV after 100 $\invfb$ of data~\cite{Han:2006ip}.
Recently renewed interest in a very large hadron collider (VLHC) with a center of mass (c.m.)~energy about 100 TeV,
which will undoubtedly extend the coverage, suggests a reexamination of the search strategy at the new energy frontier.

Production channels for heavy Majorana neutrinos at higher orders of $\alpha$ were systematically cataloged in Ref.~\cite{Datta:1993nm}. 
Recently, the vector boson fusion (VBF) channel $W\gamma \to N\ell^{\pm}$ was studied at the LHC, and its $t$-channel enhancement 
to the total cross section was emphasized~\cite{Dev:2013wba}. 
Along with that, they also considered corrections to the DY process by including the tree-level QCD contributions to $N\ell^{\pm} +$jets.
Significant enhancement was claimed over both the leading order (LO) DY signal \cite{Han:2006ip,Atre:2009rg} and the expected next-to-next-to-leading order (NNLO) in QCD-corrected DY rate~\cite{Hamberg:1990np}, prompting us to revisit the issue.

\begin{figure}[!t]
\centering
\subfigure[]{\includegraphics[width=.44\textwidth]{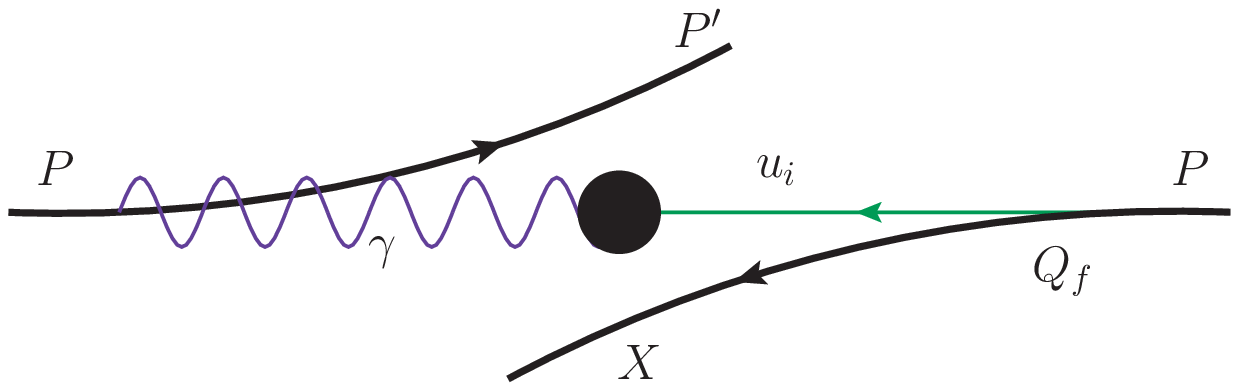}	\label{gamFromP.fig}}
\subfigure[]{\includegraphics[width=.52\textwidth]{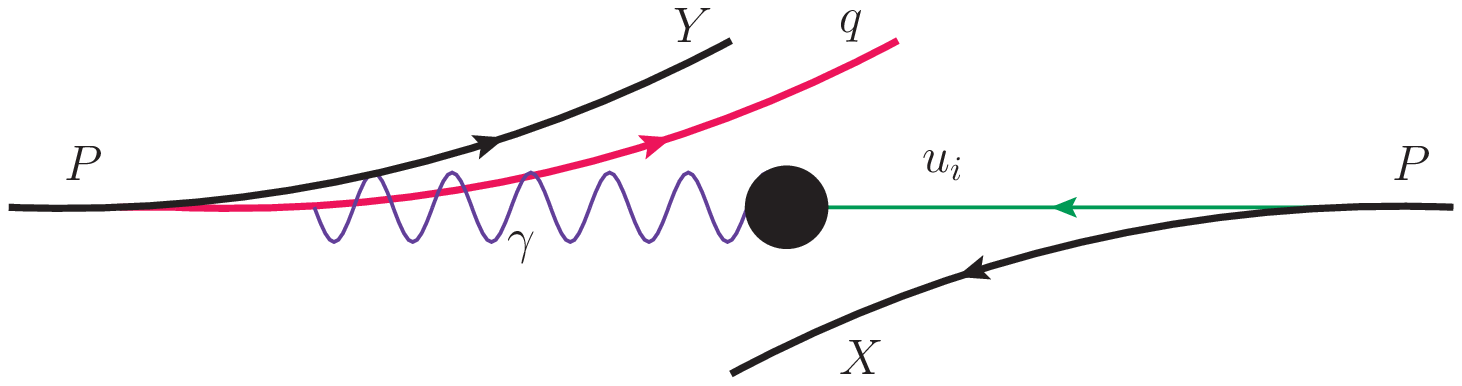}	\label{gamFromQ.fig}}
\caption[Diagrammatic description of elastic \& inelastic/deeply inelastic $\gamma p$ scattering.]{Diagrammatic description of (a) elastic and (b) inelastic/deeply inelastic $\gamma p$ scattering.} 
\label{pascatt.FIG}
\end{figure}

We carry out a systematic treatment of the photon-initiated processes. 
The elastic emission (or photon emission off a nucleon) at colliders, as shown in figure~\ref{pascatt.FIG}(a), 
is of considerable interest for both 
SM~\cite{Budnev:1974de,Kniehl:1990iv,Block:1998hu,Gluck:2002fi,Cox:2005if,deFavereaudeJeneret:2009db,d'Enterria:2013yra} 
and 
BSM processes~\cite{Drees:1984cx,Drees:1994zx,Khoze:2001xm,Han:2007bk,Chapon:2009hh,Sahin:2010zr,Gupta:2011be,Sahin:2012zm,Sahin:2012mz},
and has been observed at 
electron~\cite{Abreu:1994vp}, hadron~\cite{Chatrchyan:2011ci,Chatrchyan:2013foa}, and lepton-hadron~\cite{Aid:1996dn,Adloff:2000vm} colliders.
The inelastic (collinear photon off a quark) and deeply inelastic (large momentum transfer off a quark) channels, as depicted in figure~\ref{pascatt.FIG}(b), may take over at higher momentum transfers~\cite{Drees:1988pp,Gluck:2002fi,Martin:2004dh}.
Comparing with the DY production $q q'\to W^{*}\to N\ell^{\pm}$, we find that the $W\gamma$ fusion process becomes relatively more important at higher scales,
taking over the QCD-corrected DY mechanism at {$\gtrsim 1\TeV~ (770\GeV)$} at the 14-TeV LHC (100 TeV VLHC).
At {$m_{N} \sim 375$} GeV, a benchmark value presented in \cite{Atre:2009rg}, 
we find the $W\gamma$ contribution to be about {20\%\ (30\%)} of the LO DY cross section.

NNLO in QCD corrections to the DY processes are well-known \cite{Hamberg:1990np} and 
the K-factor for the inclusive cross sections are about {$1.2-1.4~(1.2-1.5)$} at LHC (VLHC) energies.
Taking into account all the contributions, we present the state-of-the-art results for the inclusive production of heavy neutrinos in 14 and 100 TeV $pp$ collisions. 
We further perform a signal-versus-background analysis for a 100 TeV collider of the fully reconstructible and $L$-violating final state in Eq.~(\ref{dy.EQ}).
With the currently allowed mixing {$\vert V_{\mu N}\vert ^2<6\times 10^{-3}$,
we find that the $5\sigma$ discovery potential of Ref.~\cite{Atre:2009rg} can be extended to
{$m_N = 530~(1070)$} GeV at the 14 TeV LHC (100 TeV VLHC) after 1 ab$^{-1}$.
Reversely, for $m_N = 500$ GeV and the same integrated luminosity, 
a mixing $\vert V_{\mu N}\vert^2$ of the order {$1.1\times10^{-3}~(2.5\times10^{-4})$} may be probed.}
Our results are less optimistic than reported in \cite{Dev:2013wba}. 
We attribute the discrepancy to their significant overestimate of the signal in the tree-level QCD calculations, as quantified in section~\ref{sec:totIsPhoton}.

The rest of paper is organized as follows: 
In section~\ref{sec:inclusive}, 
we describe our treatment of the several production channels considered in this study,
address the relevant scale dependence, 
and present the inclusive $N\ell^\pm$ rate at the 14 TeV LHC and 100 TeV VLHC.
In section~\ref{sec:100TeV}, we perform the signal-versus-background analysis at a future 100 TeV $pp$ collider and report the discovery potential.
Finally summarize and conclude in section~\ref{sec:summary}.

\section{Neutrino Mixing Formalism}
Our formalism and notation follow Ref.~\cite{Atre:2009rg,Han:2012vk}.
We assume that there are three left-handed (L.H.) neutrinos (denoted by $\nu_{aL}, a=1,2,3$) 
with three corresponding light mass eigenstates (denoted by $m$),
and $n$ right-handed (R.H.) neutrinos (denoted by $N_{a'R}, \  a'=1,\dots,\: n$)
with $n$ corresponding heavy mass eigenstates (denoted by $m'$).
The mixing between chiral states and mass eigenstates may then be parameterized~\cite{Atre:2009rg} by
\begin{equation}
\left(\begin{array}{c}
\nu_{L}\\
N_{L}^{c}
\end{array}\right)=\left(\begin{array}{cc}
U_{3\times3} & V_{3\times n}\\
X_{n\times3} & Y_{n\times n}
\end{array}\right)\left(\begin{array}{c}
\nu_{m}\\
N_{m'}^{c}
\end{array}\right), 
\label{appNuMix.EQ}
\end{equation}
where $\psi^{c}=\mathcal{C}\overline{\psi}^{T}$ denotes the charge conjugate of the spinor field $\psi$, 
with $\mathcal{C}$ labeling the charge conjugation operator,
and the chiral states satisfy $\psi_{L}^{c}\equiv(\psi^{c})_{L}=(\psi_{R})^{c}.$ 
Expanding the L.H. and R.H. chiral states, we obtain:
\begin{eqnarray}
\overline{\nu_{aL}}=\sum_{m=1}^{3}\overline{\nu_{m}}U_{ma}^{*}+\sum_{m'=4}^{n+3}\overline{N_{m'}^{c}}V_{m'a}^{*},
&\qquad&
\overline{N_{a'L}^{c}}=\sum_{m=1}^{3}\overline{\nu_{m}}X_{ma'}^{*}+\sum_{m'=4}^{n+3}\overline{N_{m'}^{c}}Y_{m'a'}^{*} \\
\overline{\nu_{aR}^{c}}=\sum_{m=1}^{3}\overline{\nu_{m}^{c}}U_{ma}+\sum_{m'=4}^{n+3}\overline{N_{m'}}V_{m'a},
&\qquad&
\overline{N_{a'R}}=\sum_{m=1}^{3}\overline{\nu_{m}^{c}}X_{ma'}+\sum_{m'=4}^{n+3}\overline{N_{m'}}Y_{m'a'}. 
\end{eqnarray}
Under this formalism,
one expects diagonal mixing of order $1$, 
\begin{equation}
UU^{\dagger}\:\text{and}\: YY^{\dagger}\sim\mathcal{O}(1); 
\end{equation}
and suppressed off-diagonal mixing,
\begin{equation}
VV^{\dagger}\:\text{and}\: XX^{\dagger}\sim\mathcal{O}(m_{m}/m_{m'}). 
\end{equation}
For simplicity, we consider only the lightest, heavy mass eigenstate neutrino $N$.
The SM $W$ coupling to heavy neutrino $N$ and charged lepton $\ell$ can now be written as 
\begin{eqnarray}
\mathcal{L}=
&-& \frac{g}{\sqrt{2}}\sum_{\ell=e}^{\tau}W_{\mu}^{+}
\left[\sum_{m=1}^{3}\overline{\nu_{m}}U_{\ell m}^{*} 
+ 
\overline{N^{c}}V_{\ell N}^{*}\right]\gamma^{\mu}P_{L}\ell^{-}+\text{H.c.}.
\end{eqnarray}

\section{Heavy $N$ Production at Hadron Colliders}
\label{sec:inclusive}


For the production of a heavy Majorana neutrino at hadron colliders, the leading channel is the DY process at order $\alpha^{2}$ (LO)~\cite{Keung:1983uu}
\begin{equation}
 q ~ \overline{q}' \rightarrow W^{\pm*} \to N ~\ell^\pm  .
 \label{dyDef.EQ}
\end{equation}
The QCD corrections to DY-type processes up to $\alpha_{s}^{2}$ (NNLO) are known \cite{Hamberg:1990np}, and will be included in our analysis. 
Among other potential contributions, the next promising channel perhaps is the VBF channel \cite{Datta:1993nm}
\begin{equation}
W\ \gamma \to N\ \ell^{\pm}, 
\label{eq:fuse}
\end{equation}
due to the collinear logarithmic enhancement from $t$-channel vector boson radiation. 
Formally of order $\alpha^{2}$, there is an additional $\alpha$ suppression from the photon coupling to the radiation source. 
Collinear radiation off charged fermions (protons or quarks) leads to significant enhancement but requires proper treatment.
In our full analysis, $W$s are not considered initial-state partons~\cite{Datta:1993nm} 
and all gauge invariant diagrams, including non-VBF contributions, are included.

We write the production cross section of a heavy state $X$ in hadronic collisions as 
\begin{eqnarray}
  \sigma(pp\rightarrow X+ \text{anything}) &=& 
  \sum_{i,j}
  \int^{1}_{\tau_{0}}d\xi_{a}\int^{1}_{\tau_{0}\over \xi_a}d\xi_{b} 
  \left[f_{i/p}(\xi_a,Q_{f}^{2})f_{j/p}(\xi_b,Q_{f}^{2}) \hat{\sigma}(i j \rightarrow X) + (i\leftrightarrow j)\right]~~
    \label{factTheorem.EQ}
  \\
   &=& \int_{\tau_{0}}^{1} d\tau  \sum_{ij}\frac{d\mathcal{L}_{ij}}{d\tau}\  \hat{\sigma}(ij\rightarrow X).
\end{eqnarray}
where 
$\xi_{a,b}$ are the fractions of momenta carried by initial partons $(i,j)$, $Q_{f}$ is the parton factorization scale,
and $\tau = \hat s/s$ with 
$\sqrt{s}\ (\sqrt{\hat{s}})$ the proton beam (parton) c.m.~energy. For heavy neutrino production, the threshold is 
$\tau_{0} = m_{N}^{2}/s$. 
Parton luminosities are given in terms of the parton distribution functions (PDFs) $f_{i,j/p}$ by the expression
\begin{equation}
\Phi_{ij}(\tau) \equiv  \frac{d\mathcal{L}_{ij}}{d\tau}  = 
  \frac{1}{1+\delta_{ij}}  \int^{1}_{\tau} \frac{d\xi}{\xi} 
 \left[  f_{i/p}(\xi, Q_{f}^{2})f_{j/p}\left(\frac{\tau}{\xi},Q_{f}^{2} \right) + (i \leftrightarrow j) \right].
 \label{partonLumi.EQ}
\end{equation}
We include the light quarks $(u,d,c,s)$ and adopt the 2010 update of the CTEQ6L PDFs\cite{Pumplin:2002vw}.  
Unless stated otherwise, all quark (and gluon) factorization scales are set to half the c.m.~energy:
\begin{equation}
 Q_{f} = \sqrt{\hat s} / 2 .
 \label{QfScale.EQ}
\end{equation}
For the processes with initial state photons $(\gamma)$, their treatment and associated scale choices are given in section~\ref{sec:isPhoton}.

\begin{figure}[!t]
\centering
\subfigure[]{\includegraphics[width=.46\textwidth]{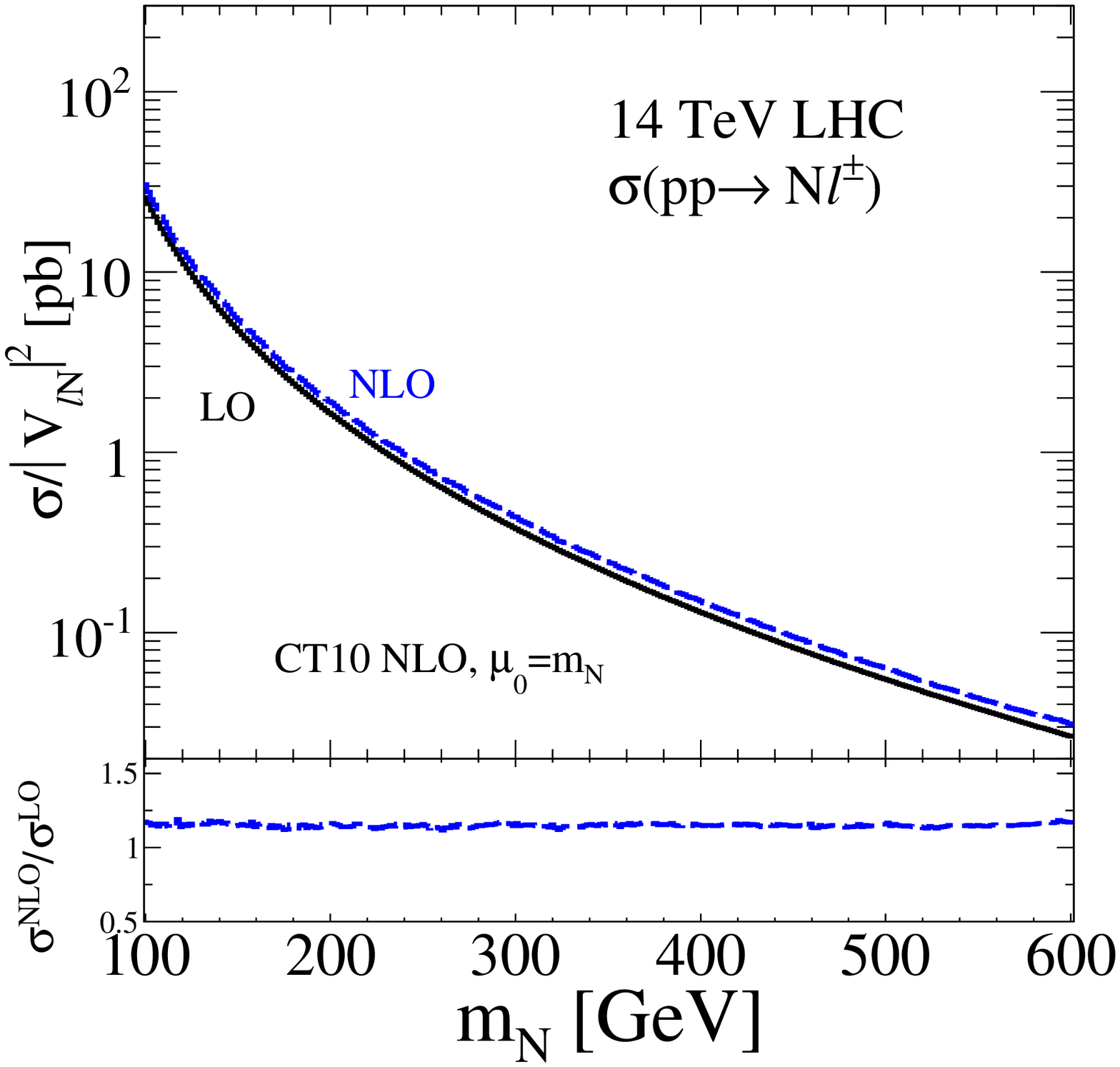} \label{ppNl_SeesawI_14TeV_XSec_mN.FIG}}
\subfigure[]{\includegraphics[width=.46\textwidth]{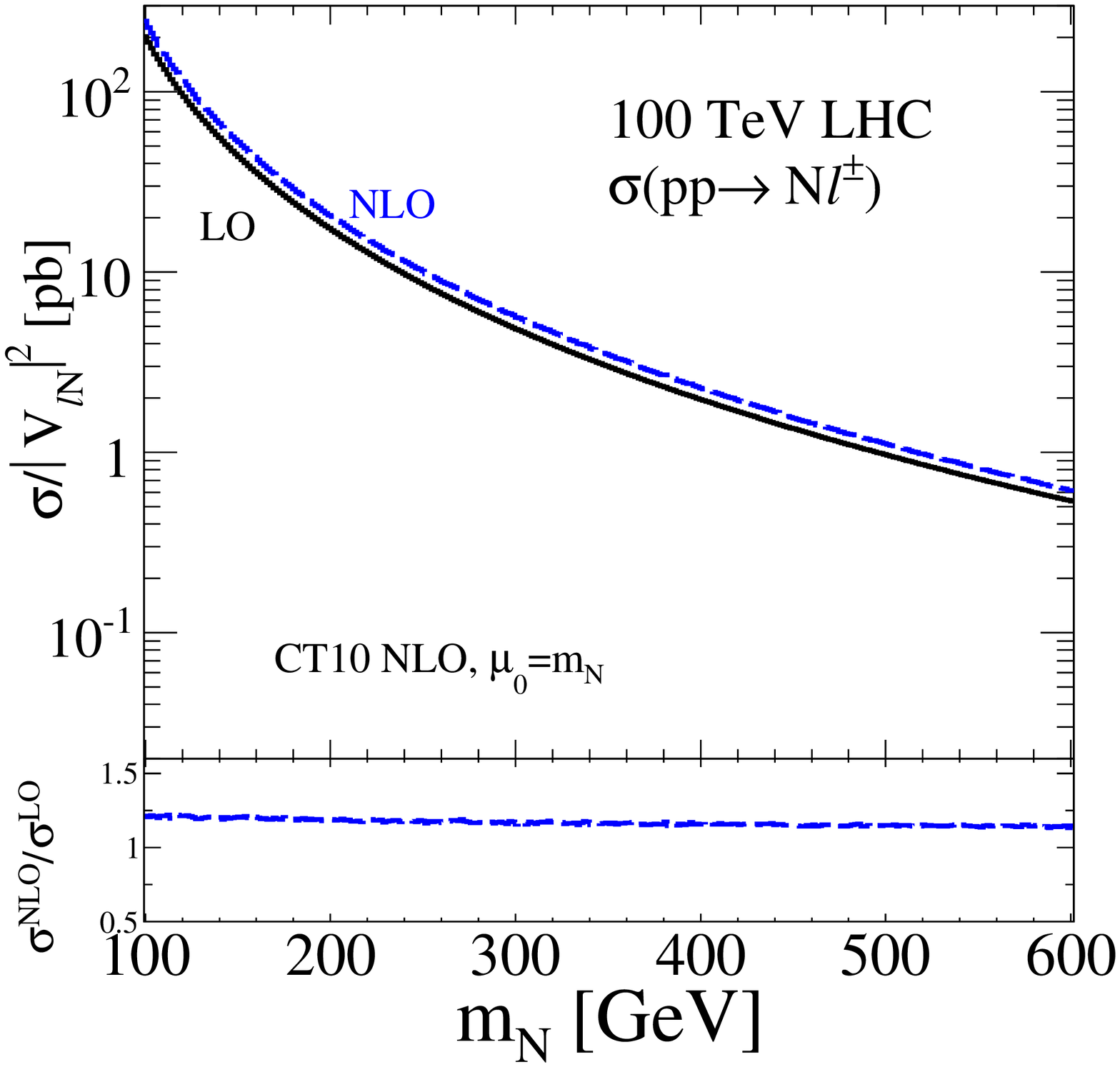}\label{ppNl_SeesawI_100TeV_XSec_mN.FIG}}
\caption[LHC and VLHC $N\ell^\pm$ cross section, divided by $\vert V_{\ell N}\vert^2$, at LO and NLO in QCD]{
(a) 14 TeV LHC (b) 100 TeV VLHC $N\ell^\pm$ cross section, divided by $\vert V_{\ell N}\vert^2$, and NLO $K$-factor as a function of $m_N$
at LO DY (solid) and NLO in QCD(bash).} 
\label{xsecNLO.FIG}
\end{figure}

For the heavy neutrino production via the SM charged current coupling, 
the cross section is proportional to the mixing parameter (squared) between the mass eigenstate $N$ and the charged lepton $\ell\ (e,\mu,\tau)$. 
Thus it is convenient to factorize out the model-dependent parameter $\vert V_{\ell N}\vert^2$
\begin{equation}
 \sigma(pp\rightarrow N\ell^{\pm}) \equiv  \sigma_{0}(pp\rightarrow N\ell^{\pm}) ~\times~  \vert V_{\ell N}\vert^2,
 \label{bareProd.EQ}
\end{equation}
where $\sigma_{0}$ will be called the ``bare cross section''.
Using the phase space slicing method~\cite{Fabricius:1981sx,Kramer:1986mc,Baer:1989jg,Harris:2001sx}, 
the heavy Majorana neutrino production can be evaluated at next-to-leading order (NLO) in QCD accuracy.
Using the 2012 update of the CT10 PDFs~\cite{Lai:2010vv} and factorization, renormalization scales $\mu_f=\mu_r = m_N$,
we plot in Fig.~\ref{xsecNLO.FIG} the LO and NLO bare cross section and 
NLO $K$-factor\footnote{The $N^nLO$ $K$-factor is defined as $K = \sigma^{N^nLO}(N\ell) / \sigma^{LO}(N\ell)$, 
where $\sigma^{N^nLO}(N\ell)$ is the N$^n$LO-corrected cross section and $\sigma^{LO}(N\ell)$ is the lowest order $(n=0)$, or  Born, cross section.}
as a function of Majorana neutrino mass $m_N$ at the (a) 14 TeV LHC and (b) 100 TeV VLHC. 
At 14 (100) TeV, for $m_N = 100-600\GeV$, the bare NLO rate ranges from $0.03-30\pb~(0.6-250\pb)$.
The corresponding $K$-factor spans $1.13-1.17~(1.15-1.2)$. 

The branching fraction of a heavy neutrino to a particular lepton flavor $\ell$ 
is proportional to $\vert V_{N\ell}\vert^{2} / \sum_{\ell'}\vert V_{N\ell'}\vert^{2}$.
Thus for neutrino production and decay into same-sign leptons with dijet, it is similarly convenient to factorize out this ratio~\cite{Han:2006ip}:
\begin{eqnarray}
 \sigma(pp\rightarrow \ell^{\pm} \ell^{'\pm}+2j) 
 &\equiv& \sigma_{0}(pp\rightarrow \ell^{\pm} \ell^{'\pm} +2j) ~\times ~ S_{\ell\ell'},
  \label{bareXSecDef.EQ}\\
  S_{\ell\ell'} & = & \frac{\vert V_{\ell N}\vert^2\vert V_{\ell' N}\vert^{2}}{\sum_{\ell''} \vert V_{\ell'' N}\vert^{2}}.
 \label{sellell.EQ}  
\end{eqnarray}
The utility of this approach is that all the flavor-model dependence is encapsulated into a single, measurable number. Factorization into a bare rate and mixing coefficient holds generally for QCD and EW corrections as well.


\subsection{Constraints on Heavy Neutrino mixing}
\label{sec:constraints}

As seen above in Eq.~(\ref{bareProd.EQ}), one of the most important model-dependent parameters to control the signal production rate is the neutrino mixing $V_{\ell N}$. 
Addressing the origin of lepton flavor is beyond the scope of this study, so 
masses and mixing factors are taken as independent, phenomenological parameters.
We consider only the lightest, heavy neutrino mass eigenstate and require it to be kinematically accessible.
Updates on heavy neutrino constraints can be found elsewhere~\cite{Atre:2009rg,Han:2012vk,Dev:2013oxa}.
Here we list only the most stringent bounds relevant to our analysis.
\begin{itemize}
 \item \textbf{Bounds from $0\nu\beta\beta$}: 
 For heavy Majorana neutrinos with $M_i \gg 1\GeV$, the absence of $0\nu\beta\beta$ 
 decay restricts the mixing between 
 heavy mass and electron-flavor eigenstates~\cite{Belanger:1995nh,Benes:2005hn}:
 \begin{equation}
  \sum_{m'} \frac{\vert V_{em'}\vert^{2} }{M_{m'}} < 5 \times 10^{-5} \TeV^{-1}.
 \end{equation}
 \item \textbf{Bounds from EW Precision Data}: Mixing between a SM singlet above a few hundred GeV in mass and lepton flavor eigenstates 
is constrained by EW data~\cite{delAguila:2008pw}:
\begin{equation}
 \vert V_{\mu N}\vert^2  < 3.2 \times 10^{-3}, \quad 
 \vert V_{\tau N}\vert^2 < 6.2 \times 10^{-3} \quad \text{at}\quad 90\%~\text{C.L.}
\end{equation}
\end{itemize}
We consider the existence of only the lightest heavy Majorana neutrino,
which is equivalent to the decoupling limit where heavier eigenstates are taken to have infinite mass.
Thus, for representative neutrino masses
\begin{equation}
 m_N = 300~(500)~[1000]\GeV,
 \label{massParam.EQ}
\end{equation}
we use the following mixing coefficients
\begin{equation}
 \vert V_{e N}\vert^{2}    = 1.5~(2.5)~[5]\times 10^{-5}, \qquad
 \vert V_{\mu N}\vert^{2}  = 3.2 \times 10^{-3}, \qquad
 \vert V_{\tau N}\vert^{2} = 6.2 \times 10^{-3},
 \label{mixingParam.EQ}
\end{equation}
corresponding to a total neutrino width of
\begin{equation}
 \Gamma_{N} = 0.303~(1.50)~[12.3]\GeV.
  \label{widthParam.EQ}
\end{equation}
As $\Gamma_t / m_{N} \approx 0.1\% - 1\%$, the heavy neutrino resonance is very narrow and application of the narrow width approximation (NWA) is justified.
For $S_{\ell\ell}$, these mixing parameters imply
\begin{eqnarray}
 S_{ee} 	=  2.4~(6.6)~[26] \times 10^{-8} &\quad\text{for}\quad& m_N= 300~(500)~[1000]\GeV
 \label{see.EQ}
 \\
 S_{e\mu} 	= S_{\mu e} = 5.1~(8.5)~[17] \times 10^{-6} &\quad\text{for}\quad& m_N= 300~(500)~[1000]\GeV 
 \label{semu.EQ}
 \\
 S_{\mu\mu} 	= 1.1 \times 10^{-3} &\quad\text{for}\quad& m_N\in[100,1000]\GeV
 \label{smumu.EQ}
\end{eqnarray}
Though the bound on $\vert V_{eN}\vert$ varies with $m_N$, 
$S_{\mu\mu}$ changes at the per mil level over the masses we investigate and is taken as constant.
The allowed sizes of $S_{e\mu}$, $S_{\mu\mu}$, and $S_{\tau\ell}$
demonstrate the complementarity to searches for $L$-violation at $0\nu\beta\beta$ experiments afforded by hadron colliders.
To make an exact comparison with Ref.~\cite{Atre:2009rg}, we also consider the bound~\cite{Nardi:1994iv,Nardi:1994nw}
\begin{equation}
 S_{\mu\mu} \approx \frac{\vert V_{\mu N} \vert^4}{\vert V_{\mu N} \vert^2 } = \vert V_{\mu N} \vert^2 = 6\times 10^{-3}
\end{equation}
However, bare results, which are mixing-independent, are presented wherever possible.


\subsection{$N$ Production via the Drell-Yan Process at NNLO}
\begin{figure}[!t]
\begin{center}
\subfigure[]{\includegraphics[scale=1,width=.48\textwidth]{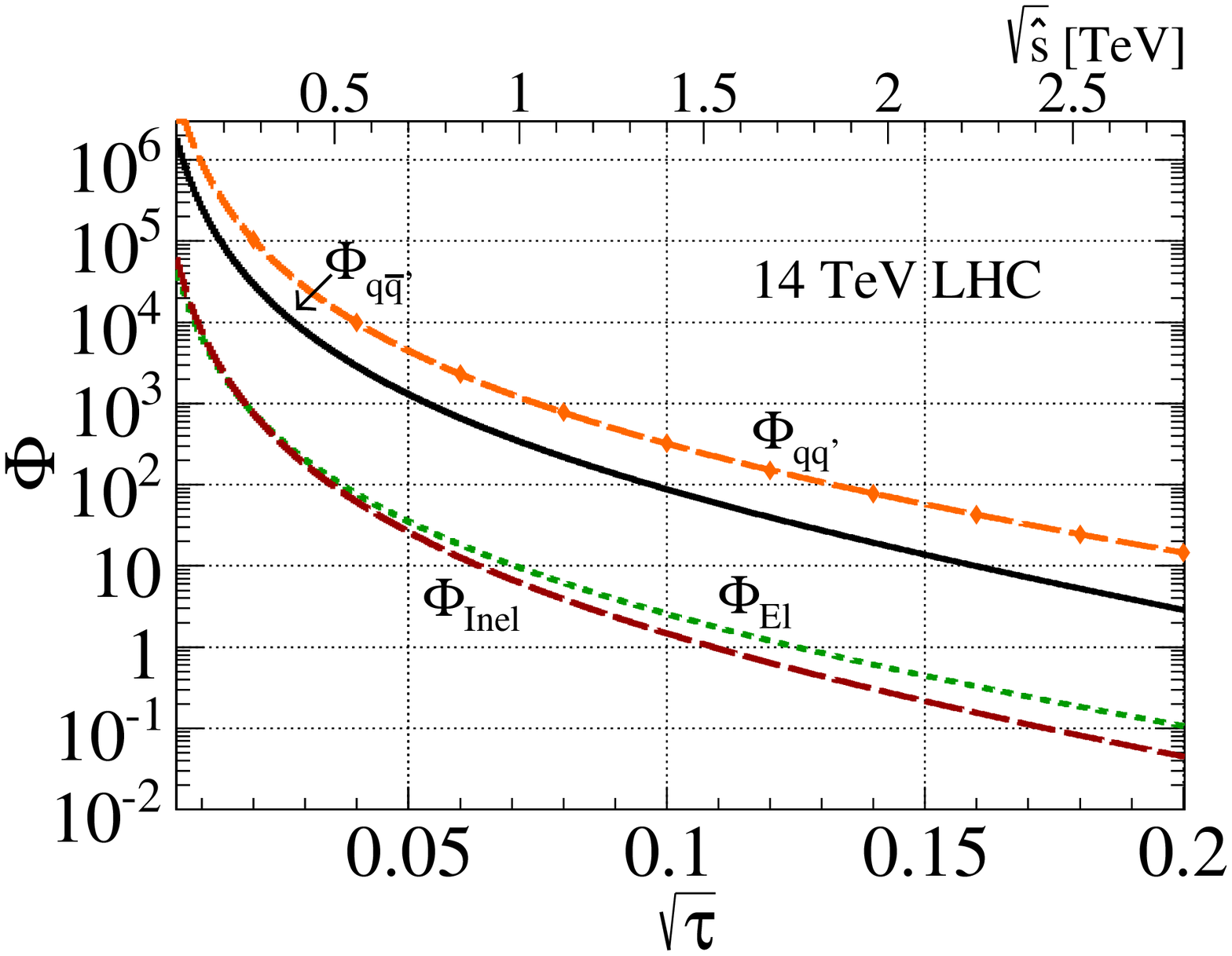}\label{lumilumi.fig}}
\subfigure[]{\includegraphics[scale=1,width=.48\textwidth]{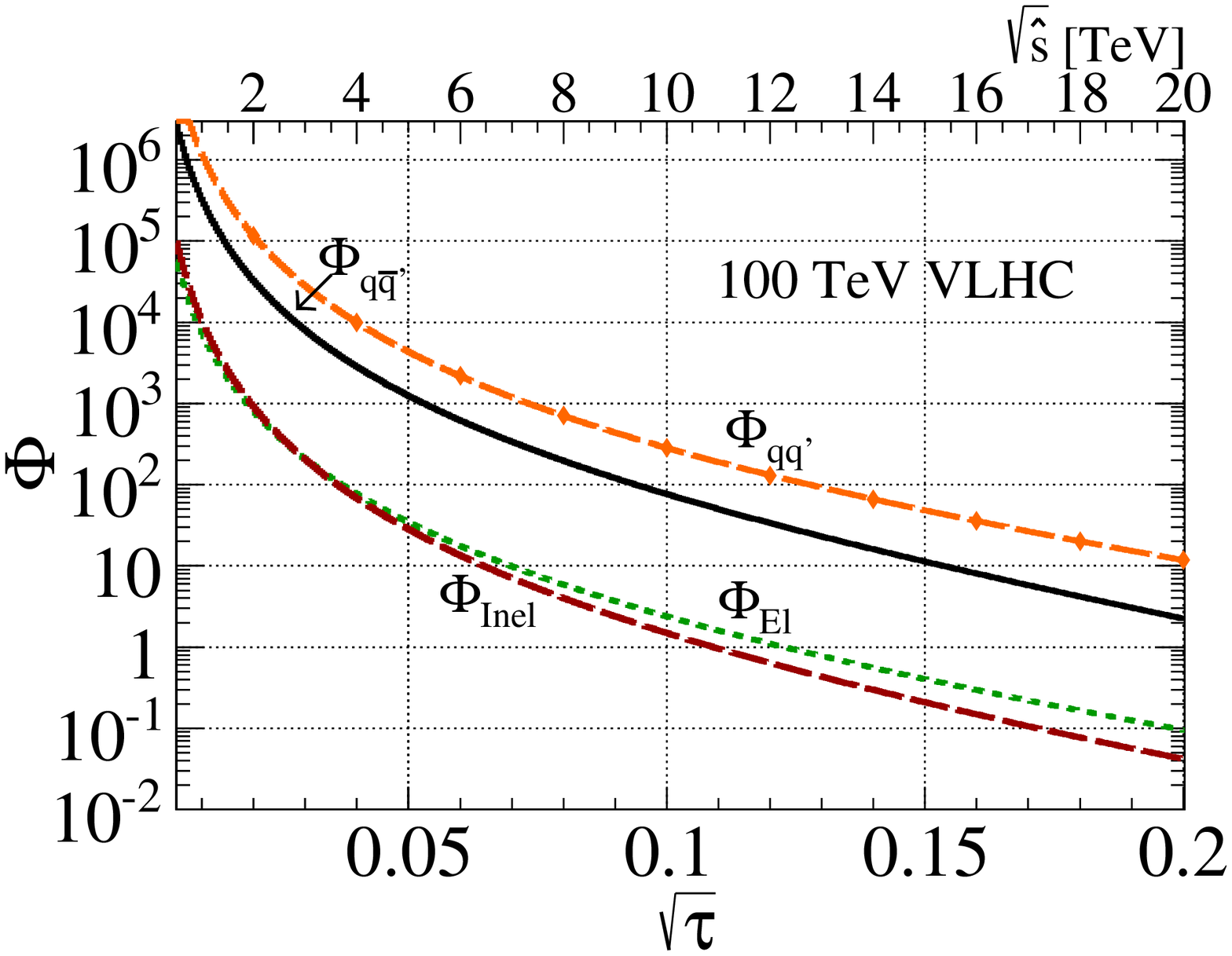}\label{lumilumi100TeV.fig}}
\vspace{.2in}\\
\subfigure[]{\includegraphics[scale=1,width=.48\textwidth]{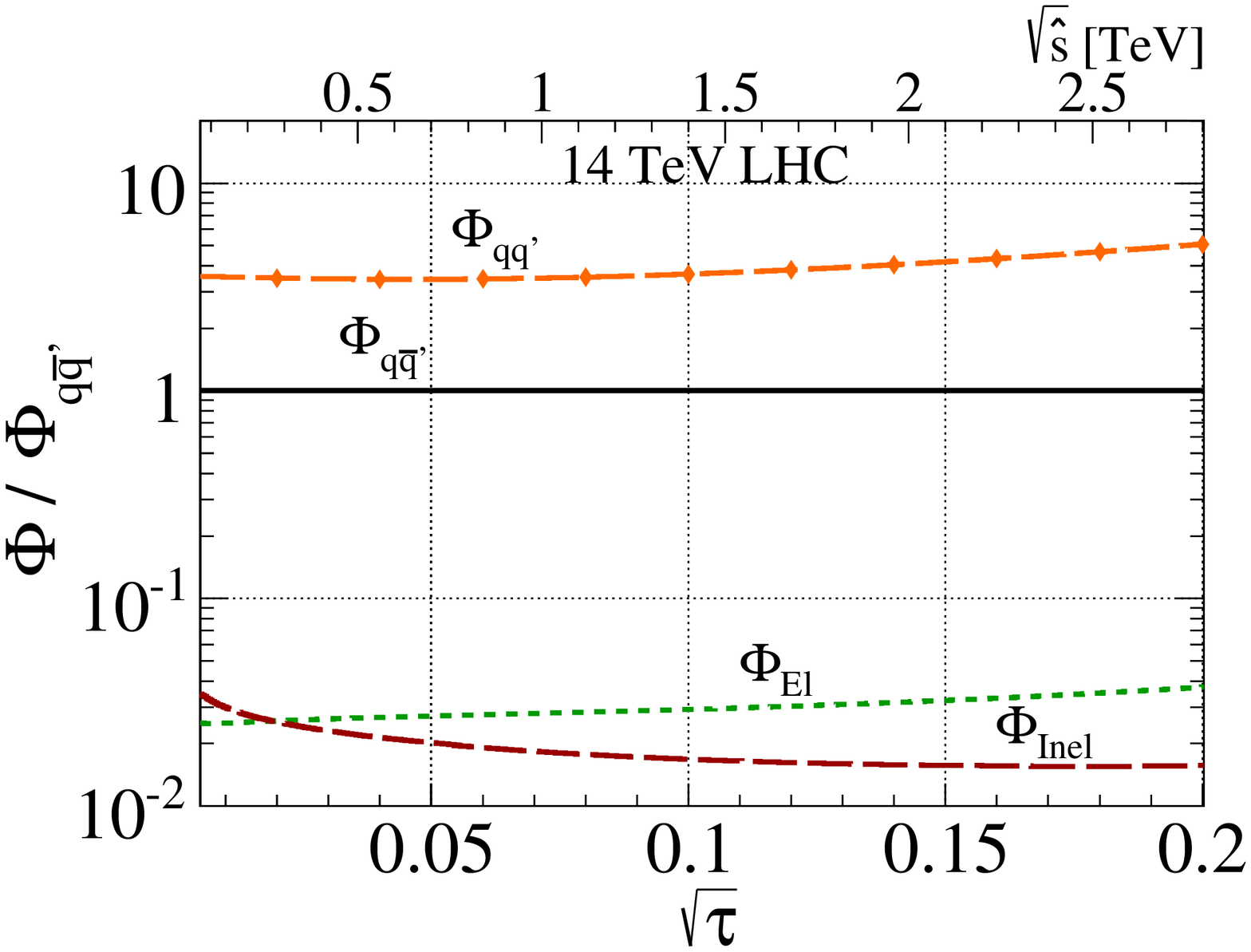}\label{lumiratio.fig}}
\subfigure[]{\includegraphics[scale=1,width=.48\textwidth]{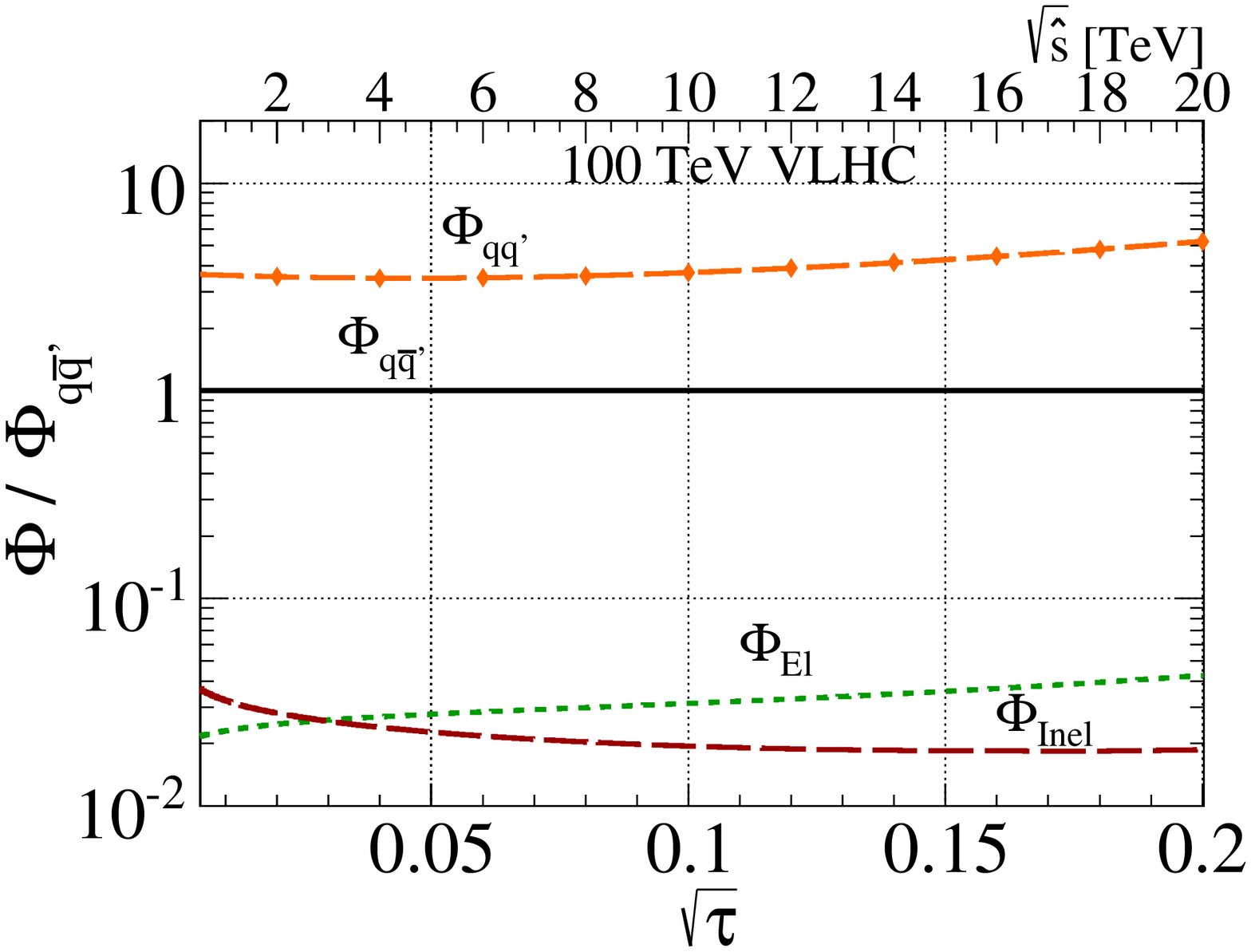}\label{lumiratio100TeV.fig}}
\caption[Parton luminosities at 14 and 100 TeV]{
Parton luminosities at (a) 14 TeV and (b) 100 TeV 
for the DY (solid), elastic (dot), inelastic (dash), and DIS (dash-diamond) $N\ell X$ processes; 
Ratio of parton luminosities to the DY luminosity in (c) and (d).
}
\label{lumi.fig}
\end{center}
\end{figure}

Before presenting the production cross sections, it is informative to understand the available parton luminosities $(\Phi_{ij})$ as defined in Eq.~(\ref{partonLumi.EQ}).
We show $\Phi_{\rm q\overline{q}'}$ versus $\sqrt \tau$ for $q\overline{q}'$ annihilation summing over light quarks ($u,d,c,s$) by the solid (black) curves in figures~\ref{lumilumi.fig} and \ref{lumilumi100TeV.fig} for the 14 TeV LHC and 100 TeV VLHC, respectively. 
The upper horizontal axis labels the partonic c.m.~energy $\sqrt{\hat{s}}$.
As expected, at a fixed $\sqrt{\hat{s}}$ the DY luminosity at 100 TeV significantly increases over that at 14 TeV. 
At $\sqrt{\hat{s}}\approx500\GeV~(2\TeV)$, the gain is a factor of {$600~(1.8\times10^3)$}, and the discovery potential of heavy Majorana neutrinos is greatly expanded.
Luminosity ratios with respect to $\Phi_{\rm q\overline{q}'}$ are given in figure~\ref{lumiratio.fig} and \ref{lumiratio100TeV.fig}, and will be discussed when appropriate.

Cross sections for resonant $N$ production via the charged current DY process in Eq.~(\ref{dy.EQ}) and shown in figure~\ref{feynDYDecay.fig}
are calculated with the usual helicity amplitudes at the LO $\alpha^2$. Monte Carlo integration is performed using CUBA~\cite{Hahn:2004fe}.
Results are checked by implementing the heavy Majorana neutrino model into FeynRules 2.0.6 \cite{Alloul:2013bka,Christensen:2008py} and MG5\_aMC@NLO 2.1.0 \cite{Alwall:2014hca} (MG5).
For simplicity, percent-level contributions from off-diagonal Cabbibo-Kobayashi-Maskawa (CKM) matrix elements are ignored and the diagonal elements are taken to be unity.
SM inputs $\alpha^{\rm \overline{MS}}(M_{Z})$, $M_{Z}$, and $\sin^{2}_{\rm \overline{MS}}(\theta_{W})$ 
are taken from the 2012 Particle Data Group (PDG)~\cite{Beringer:1900zz}.

\begin{table}[!t]
\caption[LO and NNLO cross sections for $pp\rightarrow W^{*}\rightarrow \mu^\pm \nu$ at 14 and 100 TeV]{LO and NNLO cross sections for $pp\rightarrow W^{*}\rightarrow \mu^\pm \nu$ at 14 and 100 TeV with successive invariant mass cuts
using MSTW2008LO and NNLO PDF Sets.}
 \begin{center}
\begin{tabular}{|c|c|c|c||c|c|c|}
\hline\hline
$\sqrt{\hat{s}^{\min}}$	& 14 TeV LO [pb] & NNLO [pb] & $K$ & 100 TeV LO [pb] & NNLO [pb] & $K$ 
\tabularnewline\hline\hline
100 GeV 	& 152	& 209 	& 1.38	& 1150	& 1420	& 1.23	\tabularnewline\hline
300 GeV 	& 1.54	& 1.90	& 1.23	& 17.0	& 25.6	& 1.50	\tabularnewline\hline
500 GeV 	& 0.248	& 0.304	& 1.22	& 3.56	& 4.97	& 1.40	\tabularnewline\hline
1  TeV 		& 17.0 $\times 10^{-3}$ & 20.5 $\times 10^{-3}$ & 1.20	& 0.380	& 0.485	& 1.28	\tabularnewline\hline
\hline
\end{tabular}
\label{kFactor.TB}
\end{center}
\end{table}

We estimate the 14 and 100 TeV $pp$ NNLO $K$-factor 
by using FEWZ 2.1~\cite{Gavin:2010az,Gavin:2012sy} to compute the equivalent quantity for the SM process
\begin{equation}
pp\rightarrow W^{*}\rightarrow \mu^\pm {\nu}, 
\label{smDY.EQ}
\end{equation}
and impose only an minimum invariant mass cut, $\sqrt{\hat{s}^{\min}}$.
Because LO $N\ell$ production and Eq.~(\ref{smDY.EQ}) are identical DY processes (up mass effects) with the same color structure, 
$K$-factors calculated with a fixed $\hat{s}$ are equal. 

\begin{figure}[!t]
\begin{center}
\subfigure[]{\includegraphics[scale=1,width=.47\textwidth]{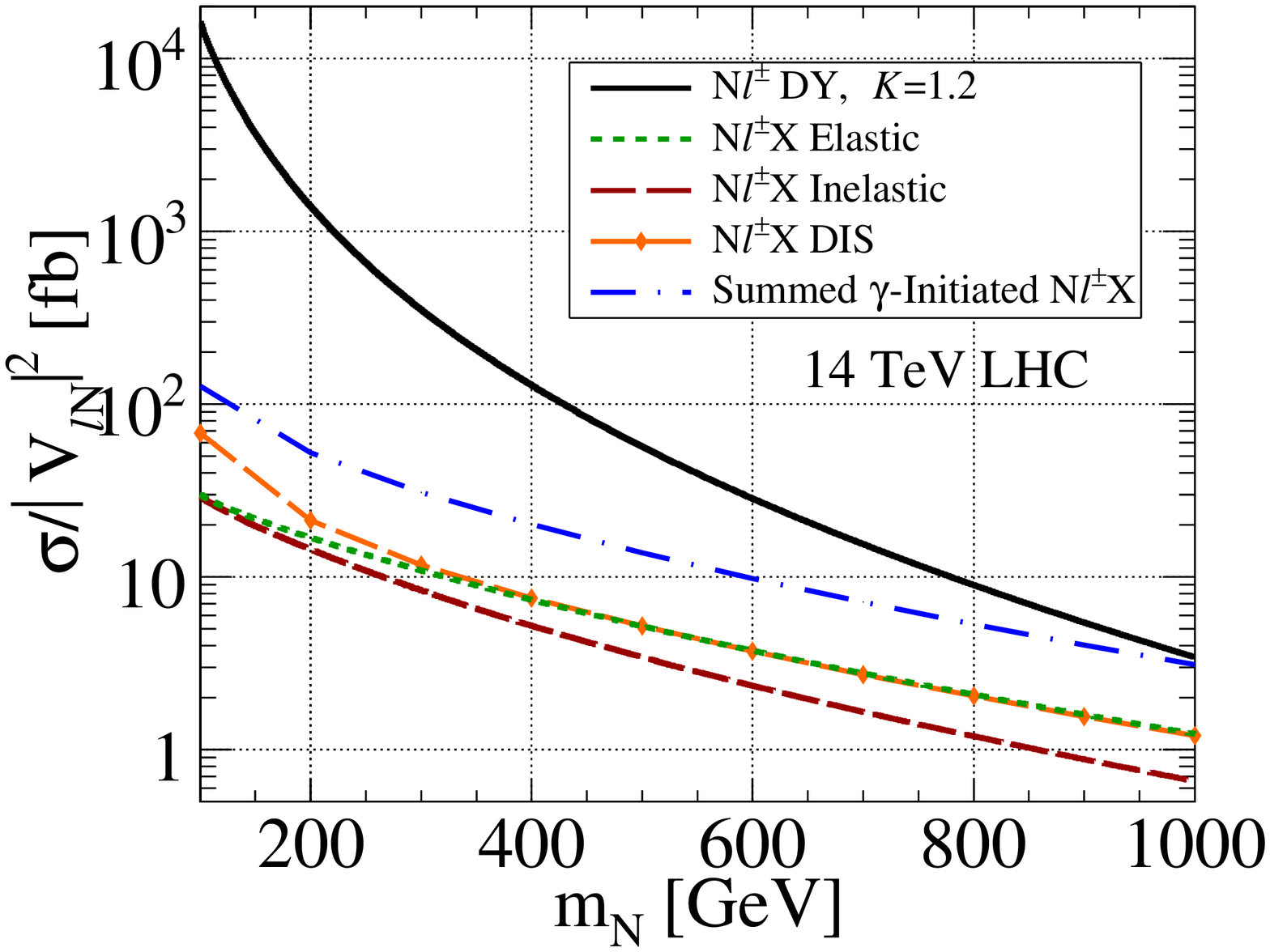}		\label{xsecComb.fig}}
\subfigure[]{\includegraphics[scale=1,width=.48\textwidth]{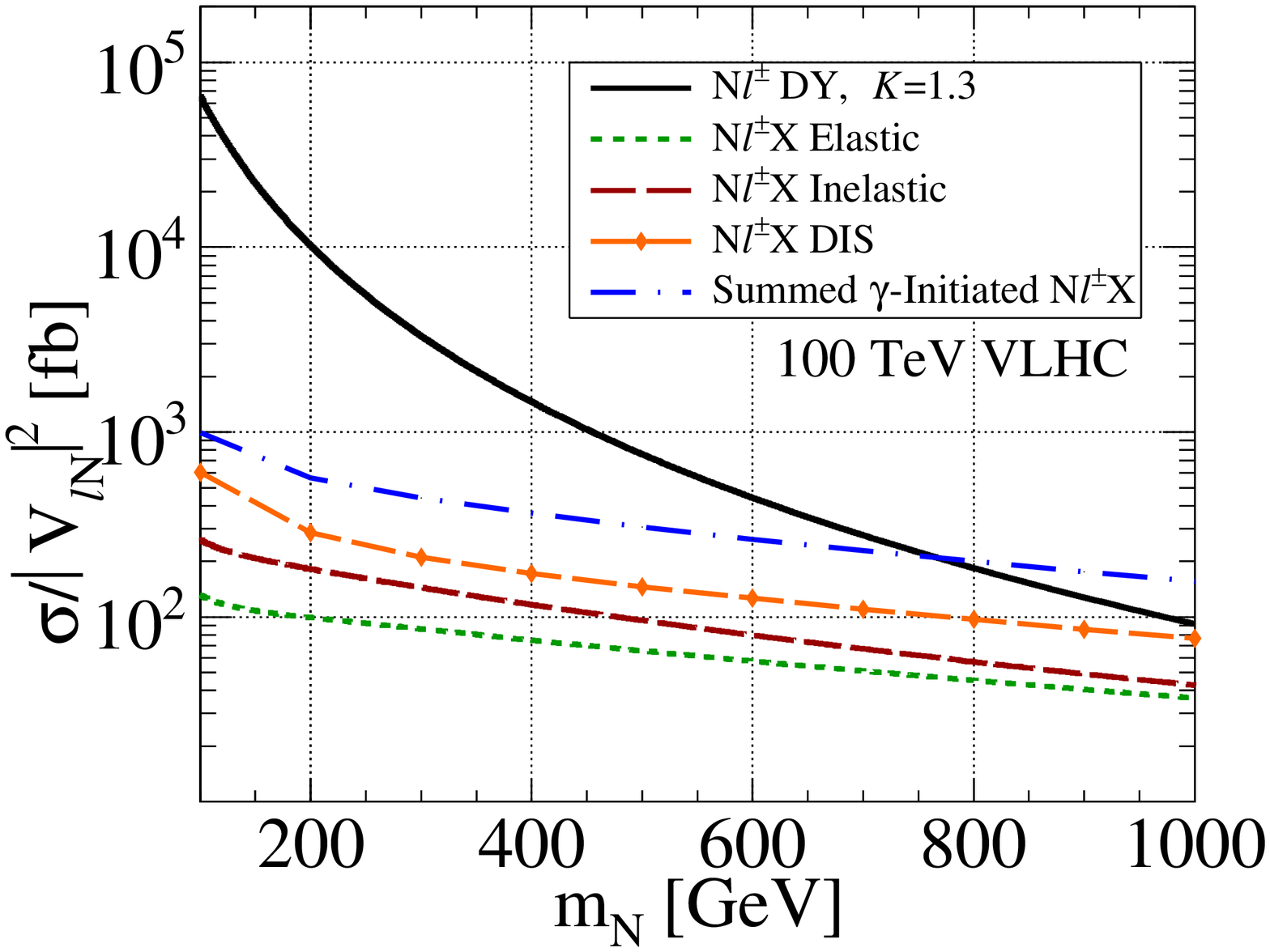}		\label{xsecComb100TeV.fig}}
\vspace{.2in}\\
\subfigure[]{\includegraphics[scale=1,width=.47\textwidth]{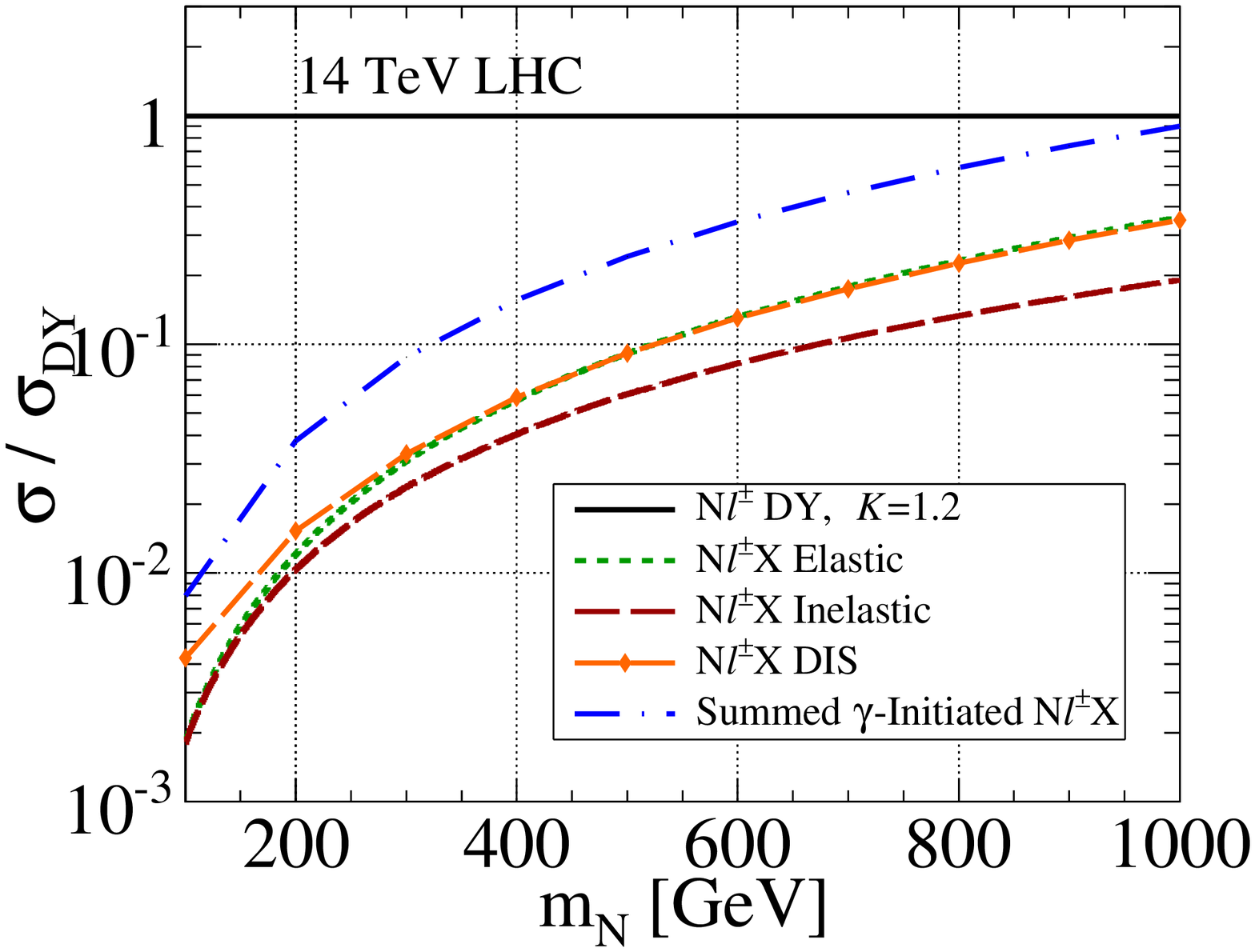}	\label{xsecRatio.fig}}
\subfigure[]{\includegraphics[scale=1,width=.48\textwidth]{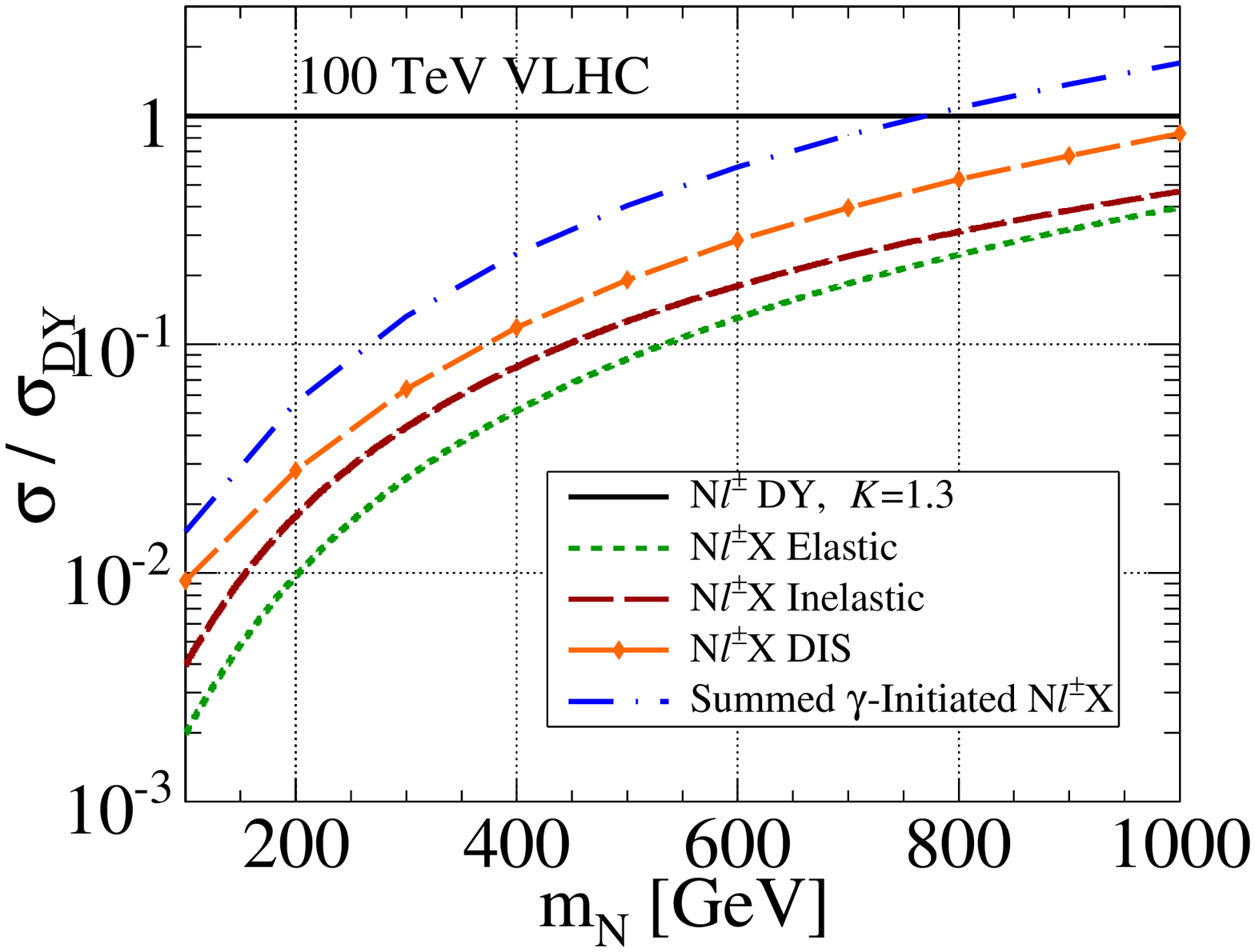}	\label{xsecRatio100TeV.fig}}
\end{center}
\caption[$N\ell X$ cross section, divided by $\vert V_{\ell N}\vert^2$]{(a) 14 TeV LHC (b) 100 TeV VLHC $N\ell X$ cross section, divided by $\vert V_{\ell N}\vert^2$, as a function of the $N$ mass for 
the NNLO DY (solid), elastic (dot), inelastic (dash), DIS (dash-diamond), and summed $\gamma$-initiated (dash-dot) processes.
(c,d) Ratio of cross sections relative to NNLO DY rate.
} 
\label{xsec.fig}
\end{figure}

Table~\ref{kFactor.TB} 
lists\footnote{As no NNLO CTEQ6L PDF set exists, we have adopted the MSTW2008 series to obtain a self-consistent estimate of the NNLO $K$-factor.} 
the LO and NNLO cross sections as well as the NNLO $K$-factors for several representative values of $\sqrt{\hat{s}^{\min}}$.
At $\sqrt{\hat{s}^{\min}} = 1\TeV$, the QCD-corrected charged current rate can reach tens (several hundreds) of fb at 14 (100) TeV.
Over the range from $\sqrt{\hat{s}^{\min}}=100\GeV - 1\TeV,$ 
\begin{eqnarray}
 K &=& 1.20-1.38 \ \ {\rm at\ 14\ TeV},  \label{K.EQ}\\
   &=& 1.23 -1.50 \ \ {\rm at\ 100\ TeV}.
\end{eqnarray}
This agrees with calculations for similar DY processes~\cite{Nemevsek:2011hz,Chatrchyan:2012meb}.
We see that the higher order QCD corrections to the DY channel are quite stable, which will be important for our discussions in section~\ref{sec:isPhoton}.
Throughout the study, independent of neutrino mass, we apply to the DY-process a $K$-factor of
\begin{equation}
 K = 1.2~(1.3) \quad\text{for}\quad 14~(100)~\TeV.
\end{equation}
Including the QCD K-factor, 
we show the NNLO total cross sections [called the ``bare cross section $\sigma_0$'' by factorizing out $\vert V_{\ell N}\vert^2$ as defined in  Eq.~(\ref{bareProd.EQ})] as a function of heavy neutrino mass in figures~\ref{xsecComb.fig} and \ref{xsecComb100TeV.fig} for the 14-TeV LHC and 100-TeV VLHC, respectively. The curves are denoted by the (black) solid lines.
Here and henceforth, we impose the following basic acceptance cuts on the transverse momentum and pseudorapidity of the charged leptons for 14 (100) TeV, 
\begin{equation}
p_{T}^{\ell} > 10~(30)\GeV,\quad	\vert \eta^{\ell}\vert < 2.4~(2.5).
\label{regCutsLep.EQ}
\end{equation}
The motive to include these cuts is two-fold. 
First,  they are consistent with the detector acceptance for our future simulations and the definition of ``fiducial'' cross section.
Second, they serve as kinematical regulators for potential collinear singularities, to be discussed next.
The $p_T$ and $\eta$ criteria at 100 TeV follow the 2013 Snowmass benchmarks~\cite{Avetisyan:2013onh}.


\subsection{Photon-Initiated Processes}
\label{sec:isPhoton}

\begin{figure}[!t]
\begin{center}
\includegraphics[scale=1,width=.96\textwidth]{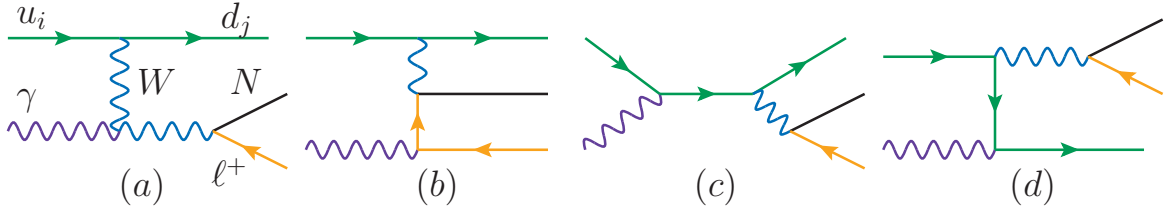}
\end{center}
\caption{Feynman diagrams for photon-initiated process $q\gamma \to N\ell^\pm q'$.} 
\label{feynQA.fig}
\end{figure}

After the dominant DY channel, VBF via $W\gamma$ fusion, as introduced in Eq.~(\ref{eq:fuse}), presents a promising additional contribution to the heavy $N$ production. We do not make any approximation for the initial state $W$ and treat its radiation off the light quarks with exact matrix element calculations.
In fact, we consistently treat the full set of diagrams, shown in figure~\ref{feynQA.fig}, for the photon-initiated process at order $\alpha^3$
\begin{equation}
 q ~ \gamma \rightarrow  N ~\ell^\pm ~q'.
 \label{pgammaDef.EQ}
\end{equation}
Obviously, diagrams  figure~\ref{feynQA.fig}(c) and (d) do not add to $W\gamma$ fusion and are just small QED corrections.\footnote{
Diagram \ref{feynQA.fig}(d) involves a collinear singularity from massless quark splitting.
It is unimportant for our current consideration since its contribution is simply a QED correction to the quark PDF.
For consistency and with little change to our results, $\lamDIS = 15\GeV$ [defined in Eq.~(\ref{qDISMinCut.EQ})] is applied as a regulator.}
Diagram figure~\ref{feynQA.fig}(b) involves a massless $t$-channel charged lepton. 
The collinear pole is regularized by the basic acceptance cuts in Eq.~(\ref{regCutsLep.EQ}). 
What is non-trivial, however, is how to properly treat initial-state photons across the different sources depicted in figure~\ref{pascatt.FIG}.
We now discuss the individual channels in detail. 

\subsubsection{Elastic Scattering: Intact Final-State Nucleons}
\label{sec:semi}
Here and henceforth, the virtuality for the incoming photon in $W\gamma$ fusion is denoted as $Q_\gamma>0$.
In the collinear limit that results in momentum transfers on the order of the proton mass or less, $Q^{2}_\gamma \lesssim m_{p}^{2}$,
initial-state photons are appropriately described as massless radiation by an elastic proton, 
i.e., does not break apart and remains as an on-shell nucleon, as indicated in figure~\ref{gamFromP.fig}.
To model this, we use the ``Improved'' Weizs\"acker-Williams approximation~\cite{Budnev:1974de}
and factorize the photon's collinear behavior into a structure function of the proton to obtain the elastic photon PDF $f_{\gamma/p}^{\rm El }$. 
In Eq.~(\ref{factTheorem.EQ}), this entails replacing one $f_{i/p}$ with $f_{\gamma/p}^{\rm El }$:
\begin{equation}
 f_{i/p}(\xi,Q_f^2) \rightarrow f_{\gamma/p}^{\rm El }(\xi).
 \label{elPDF.EQ}
\end{equation}
The expression for $f_{\gamma/p}^{\rm El }$, given in Section~\ref{AppEl.col.sec}, is dependent on a cutoff scale $\Lambda_{\gamma}^{\rm El }$, 
above which the description of elastic $p\rightarrow \gamma$ emission starts to break down.
Typically, the scale is taken to be $\mathcal{O}(m_{p}-2\GeV)$ 
~\cite{Budnev:1974de,deFavereaudeJeneret:2009db,d'Enterria:2013yra,Chapon:2009hh,Sahin:2010zr,Gupta:2011be,Sahin:2012zm,Sahin:2012mz}
but should be insensitive to small variations if an appropriate scale is chosen.
Based on analysis of $ep$ scattering at low $Q_\gamma$~\cite{Alwall:2004wk}, we take
\begin{equation}
 \Lambda_\gamma^{\rm El } = \sqrt{1.5 \GeV^{2}} \approx 1.22\GeV.
\label{lamEl.EQ}
\end{equation}
The scale dependence associated with $\Lambda_\gamma^{\rm El }$ is discussed in section~\ref{sec:scale}.

In figure~\ref{lumi.fig}, the elastic luminosity spectrum $(\Phi_{\rm El })$ is denoted by the (green) dot line.
For the range studied, $\Phi_{\rm El }$ is roughly {$2-4\%$} of the $q\bar q'$ DY luminosity at 14 and 100 TeV.

We calculate the matrix element for the diagrams in figure~\ref{feynQA.fig} in the same manner as the DY channel.
The results are checked with MG5 using the elastic, asymmetric $p\gamma$ beam mode.
In figures~\ref{xsecComb.fig} and \ref{xsecComb100TeV.fig}, 
we plot the bare cross section for the elastic process, denoted by a (green) dot line, as a function of neutrino mass.
The rate varies between {$1-30 ~(40-100)$} fb at 14 (100) TeV for $m_N = 100$ GeV$-$1 TeV.
As seen in figures~\ref{xsecRatio.fig} and \ref{xsecRatio100TeV.fig}, where the cross sections are normalized to the DY rate,
it reaches about {$30~(40)\%$} of the DY rate for large $m_N$.


\subsubsection{Inelastic Scattering: Collinear Photons From Quarks}
For momentum transfers above the proton mass, the parton model is valid.
When this configuration coincides with the collinear radiation limit, 
initial-state photons are appropriately described as being radiated by quark partons.
To model a quark splitting to a photon, we follow the methodology of Ref.~\cite{Drees:1994zx}
and use the (original) Weizs\"acker-Williams approximation~\cite{Williams:1934ad,vonWeizsacker:1934sx} 
to obtain the inelastic photon PDF $f_{\gamma/p}^{\rm Inel}$. 
Unlike the elastic case, factorization requires us to convolve about a splitting function.
The inelastic $N\ell^{\pm}X$ cross section is obtained by making the replacement in Eq.~(\ref{factTheorem.EQ})
\begin{eqnarray}
 f_{i/p}(\xi,Q_f^2) &\rightarrow & f_{\gamma/p}^{\rm Inel}(\xi,Q_\gamma^2,Q_f^2), 
 \\
 f_{\gamma/p}^{\rm Inel}(\xi,Q_\gamma^2,Q_f^2) &=& 
 \sum_{j}
 \int^{1}_{\xi} \frac{dz}{z} ~ f_{\gamma/j}(z,Q_\gamma^{2}) ~ f_{j/p}\left(\frac{\xi}{z},Q_{f}^2\right) ,
   \label{inelPDF.EQ}
\end{eqnarray}
where $f_{\gamma/j}$ is the Weizs\"acker-Williams $j\rightarrow \gamma$ distribution function,
with $Q_\gamma$ and $Q_f$ being the factorization scales for the photon and quark distributions, respectively. 
The summation is over all charged quarks.
Details regarding Eq.~(\ref{inelPDF.EQ}) can be found in Section~\ref{sec:AppIn}. 

Clearly, the scale for the photon momentum transfer should be above the elastic bound $Q_\gamma \ge \lamEl$. What is not clear, however, is how high we should evolve $Q_\gamma$. If we crudely consider the total inclusive cross section, we could simply choose the kinematical upper limit $Q_\gamma^2 \approx Q_f^2 \approx \hat s /4$ or $\hat{s}/4 - m^2_N$, which is a quite common practice in the literature \cite{Drees:1994zx}. 
However, we do not consider this a satisfactory treatment. Well below the kinematical upper limit,  
the photon virtuality $Q_\gamma$ becomes sufficiently large so that the collinear photon approximation as in figure~\ref{feynQA.fig} breaks down. Consequently, ``deeply inelastic scattering'' (DIS), as in figure~\ref{feynDIS.fig}, becomes the dominant feature. 
For a brief review of DIS, see Ref.~\cite{DeRujula:1979kh}.
Thus, a more reasonable treatment is to introduce an upper limit for the inelastic process $\lamDIS$, above which a full DIS calculation of figure~\ref{feynDIS.fig} should be applied. We adopt the following scheme
\begin{eqnarray}
Q_\gamma = \lamDIS =
\left\{\begin{matrix}
 15 \GeV & \text{for 14 TeV}\\ 
 25 \GeV & \text{for 100 TeV}
\end{matrix}\right.
 \label{disDef.EQ}
\end{eqnarray}
Sensitivity to variations $\lamDIS$  are discussed in section~\ref{sec:scale}.

Consistent with $\Phi_{ij}(\tau)$ in Eq.~(\ref{partonLumi.EQ}),  we define the inelastic $\gamma q$ parton luminosity $\Phi_{\rm Inel}$ to be
\begin{eqnarray}
 \Phi_{\rm Inel}(\tau) =
 \int^{1}_{\tau} \frac{d\xi}{\xi}  \int^{1}_{\tau/\xi}\frac{dz}{z} 
  ~ 
  \sum_{q,q'} 
  \left[
  f_{q/p}(\xi)f_{\gamma/q'}(z)f_{q'/p}\left(\frac{\tau}{\xi z}\right) +    f_{q/p}\left(\frac{\tau}{\xi z}\right)f_{\gamma/q'}(z)f_{q'/p}(\xi) 
 \right].
   \label{inelLumi.EQ}
\end{eqnarray}

In figure~\ref{lumi.fig}, we give the $\Phi_{\rm Inel}$ spectrum as a function of $\sqrt \tau$, denoted by the (red) dash curve, for 14 and 100 TeV.
For the range investigated, $\Phi_{\rm Inel}$ ranges between {$2-4\%$} of the DY luminosity.
Compared to its elastic counterpart, the smallness of the inelastic luminosity is attributed the limited $Q_\gamma^2$ evolution.

The inelastic matrix element is identical to the elastic case. 
In figures~\ref{xsecComb.fig} and \ref{xsecComb100TeV.fig}, 
we show the bare cross section for the inelastic process, denoted by the (red) dash line, as a function of the neutrino mass.
The rate varies between {$0.7-30 ~(40-260)$} fb at 14 (100) TeV for $m_N = 100\GeV-1\TeV$.
As seen in figures~\ref{xsecRatio.fig} and \ref{xsecRatio100TeV.fig}, where the cross sections are normalized to the DY rate,
it reaches about {$10~(50)\%$} of the DY rate at large $m_N$.


\subsubsection{Deeply Inelastic Scattering: High $p_{T}$ Quark Jet}
\begin{figure}[!t]
\begin{center}
\includegraphics[scale=1,width=.96\textwidth]{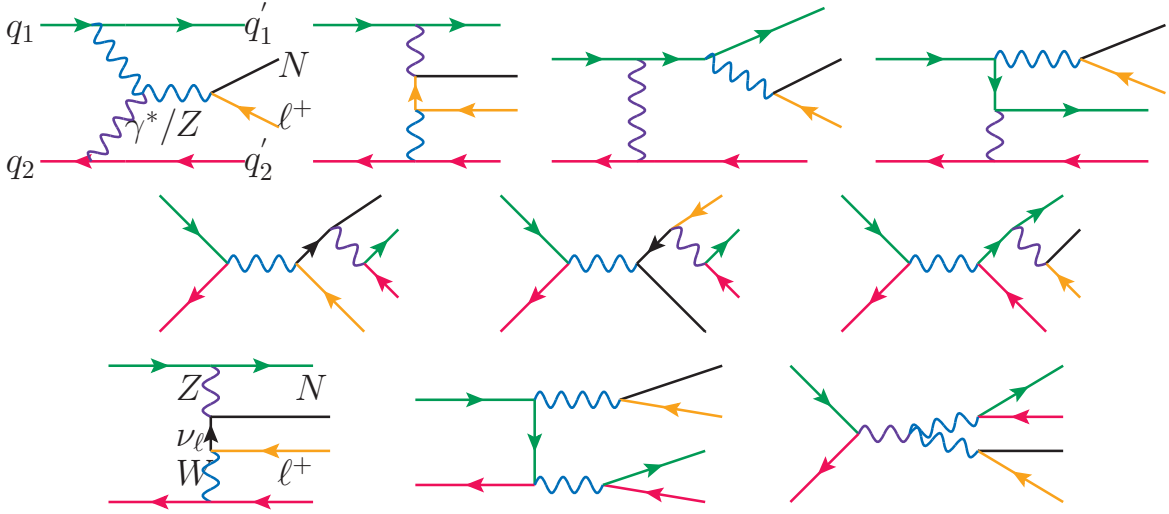}
\end{center}
\caption{Feynman diagrams for the DIS process $q_1 q_2 \rightarrow  N \ell^\pm q'_1 q'_2.$} 
\label{feynDIS.fig}
\end{figure}

As discussed in the previous section, at a sufficiently large momentum transfer
the collinear photon description breaks down and the associated final-state quark emerges as an observable jet. 
The electroweak process at $\alpha^4$
\begin{equation}
  q_1 ~ q_2 \rightarrow  N ~\ell^\pm ~q'_1 ~q'_2.
 \label{dis.EQ}
\end{equation}
becomes DIS, as shown by the Feynman diagrams in figure~\ref{feynDIS.fig}.
The top row of figure~\ref{feynDIS.fig} can be identified as the DIS analog of those diagrams in figure~\ref{feynQA.fig}.
Again, the first two diagrams represent the $W\gamma$ fusion with collinear log-enhancement from $t$-channel $W$ exchange. 
At these momentum transfers, the $WZ$ fusion channel~\cite{Datta:1993nm} turns on but is numerically smaller; 
see figure~\ref{feynDIS.fig}, bottom row, first diagram. 
The center row and two bottom-rightmost diagrams in figure~\ref{feynDIS.fig} represent on-shell $W/Z$ production at $\alpha^3$
with subsequent $W/Z\rightarrow q\overline{q}'$ decay.
Those processes, however, scale as $1/\hat s$ and are not log-enhanced.
A subset of these last diagrams also represent higher-order QED corrections to the DY process.

To model DIS, we use MG5 and simulate Eq.~(\ref{dis.EQ}) at order $\alpha^4$.
We impose\footnote{For consistency, we also require the lepton cuts given in Eq.~(\ref{regCutsLep.EQ})
and a jet separation $\Delta R_{jj} > 0.4$ to regularize irrelevant $\gamma^{*} \to q\overline{q}$ diagrams,
where $\Delta R \equiv \sqrt{\Delta\phi^2 + \Delta\eta^2}$ with $y = \eta \equiv -\log[\tan(\theta/2)]$ in the massless limit.}  
at the generator level a minimum on momentum transfers between initial-state and final-state quarks 
\begin{equation}
\underset{i,j=1,2}{\min}\sqrt{\vert (q_{i} - q'_{j})^2 \vert } > \lamDIS .
 \label{qDISMinCut.EQ}
\end{equation}
This requirement serves to separate the elastic and inelastic channels from  DIS.
Sensitivity to this cutoff is addressed in section~\ref{sec:scale}.

In figure~\ref{lumi.fig}, we show the quark-quark parton luminosity spectrum $\Phi_{\rm qq'}$, 
the source of the DIS processes, and represented by the (orange) dash-diamond  curves.
Though possessing the largest parton luminosity, the channel must overcome its larger coupling and phase space suppression.
At 14 and 100 TeV, $\Phi_{\rm qq'}$ ranges {$3-5$} times larger than $\Phi_{\rm q\overline{q}'}$. 
The difference in size between $\Phi_{\rm qq'}$ and $\Phi_{\rm El~(Inel)}$ is due to the additional coupling $\alpha_{\rm EM}$ in $f_{\gamma/p}^{\rm El~(Inel)}$.

In figures~\ref{xsecComb.fig} and \ref{xsecComb100TeV.fig}, we plot bare cross section as in Eq.~(\ref{bareProd.EQ}), denoted by the (orange) dash-diamond curve.
In figures~\ref{xsecRatio.fig} and \ref{xsecRatio100TeV.fig}, the same curves are normalized to the DY rate.
At 14 (100) TeV, the cross section ranges from {$1-60$ ($80-500$)} fb, reaching about {35\% (80\%)} of the DY rate.

To compare channels, we observe that the DIS (elastic) process increases greatest (least) with increasing collider energies.
This is due to the increase likelihood for larger momentum transfers in more energetic collisions.
A similar conclusion was found for elastic and inelastic $\gamma\gamma$ scattering at the Tevatron and LHC~\cite{Han:2007bk}.

\begin{table}[!t]
\caption[Total cross sections of various $pp\rightarrow N\ell^{\pm}X$ channels for representative values of $m_N$.]{
Total cross sections of various $pp\rightarrow N\ell^{\pm}X$ channels for representative values of $m_N$
after applying minimal acceptance cuts of Eqs.~(\ref{regCutsLep.EQ}).}
 \begin{center}
\begin{tabular}{|c|c|c|c|}
\hline\hline 
$\sigma_{\rm14\TeV~LHC}/\vert V_{\ell N}\vert^{2}$ [fb] & $m_{N}=300\GeV$ 	&  $m_{N}=500\GeV$ & $m_{N}=1\TeV$ \tabularnewline\hline\hline 
$pp\rightarrow N\ell^{\pm}$ LO DY $[K=1.2]$	&293 (352) 	&47.3 (56.8) 	&2.87 (3.44) \tabularnewline\hline
$pp\rightarrow N\ell^{\pm}X$ Elastic			&10.8971	&5.16756	&1.23693	\tabularnewline\hline
$pp\rightarrow N\ell^{\pm}X$ Inelastic			&8.32241	&3.44245	&0.65728	\tabularnewline\hline
$pp\rightarrow N\ell^{\pm}X$ DIS			&11.7 		&5.19 		&1.21 \tabularnewline\hline
$\sigma_{\gamma{\rm-Initiated}}$/$\sigma_{\rm DY}^{K=1.2}$ & 0.09  & 0.24   & 0.90 \tabularnewline\hline
\hline
\hline 
$\sigma_{\rm 100\TeV~VLHC}/\vert V_{\ell N}\vert^{2}$ [fb] & $m_{N}=300\GeV$ 	&  $m_{N}=500\GeV$ & $m_{N}=1\TeV$ \tabularnewline\hline\hline 
$pp\rightarrow N\ell^{\pm}$ LO DY  $[K=1.3]$	& 2540 (3300) 	& 583 (758) 	& 70.5 (91.6)\tabularnewline\hline
$pp\rightarrow N\ell^{\pm}X$ Elastic			& 85.8 		& 65.5 		& 36.4 \tabularnewline\hline
$pp\rightarrow N\ell^{\pm}X$ Inelastic 			& 144 		& 96.0		& 42.7 \tabularnewline\hline
$pp\rightarrow N\ell^{\pm}X$ DIS			& 210 		& 145 		& 76.7 \tabularnewline\hline
$\sigma_{\gamma{\rm-Initiated}}$/$\sigma_{\rm DY}^{K=1.3}$ & 0.13 	& 0.40  	& 1.7 \tabularnewline\hline
\hline
\end{tabular}
\label{xsec.TB}
\end{center}
\end{table}
\subsubsection{Total Neutrino Production from $\gamma$-Initiated Processes}
\label{sec:totIsPhoton}
The total heavy neutrino production cross section from $\gamma$-initiated processes may be obtained by summing the elastic, inelastic, and DIS channels \cite{Drees:1994zx,Han:2007bk}:
\begin{eqnarray}
 \sigma_{\rm \gamma-Initiated}(N\ell^\pm X) = \sigma_{\rm El }(N\ell^\pm X) + \sigma_{\rm Inel}(N\ell^\pm X) + \sigma_{\rm DIS}(N\ell^\pm X),
\label{gammaSum.EQ}
\end{eqnarray}
We plot Eq.~(\ref{gammaSum.EQ}) as a function of $m_N$ in figures~\ref{xsecComb.fig} and \ref{xsecComb100TeV.fig} at 14 and 100 TeV, denoted by the 
(blue) dash-dot curve. In figures~\ref{xsecRatio.fig} and \ref{xsecRatio100TeV.fig}, the same curves are normalized to the DY rate.
For $m_N = 100\GeV-1\TeV$, the total rate spans {$3-100$~($150-1000$) fb} at 14 (100) TeV,
reaching about {$90~(110)\%$} of the DY rate at large $m_N$.
We find that the $W\gamma$ fusion represents the largest heavy neutrino production mechanism for {$m_N>1\TeV~(770)\GeV$} at 14 (100) TeV.
We expect for increasing collider energy this crossover will occur earlier at lighter neutrino masses.
Cross sections for representative values of $m_N$ for all channels at 14 and 100 TeV are given in Table \ref{xsec.TB}.

\begin{figure}[!t]
\centering
\subfigure[]{\includegraphics[width=.48\textwidth]{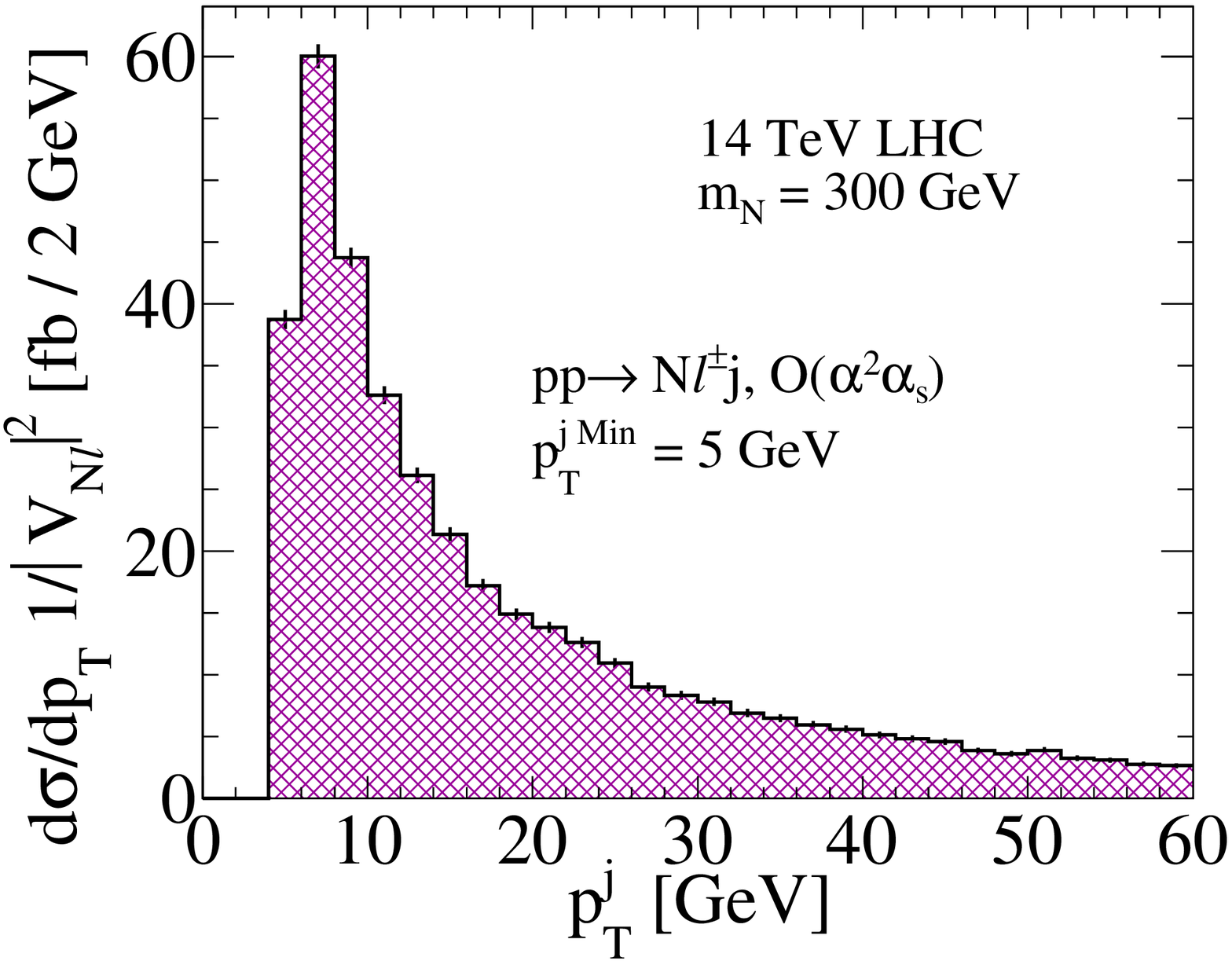}\label{ppNljQCD_dXSec_vs_pt.FIG}}
\subfigure[]{\includegraphics[width=.48\textwidth]{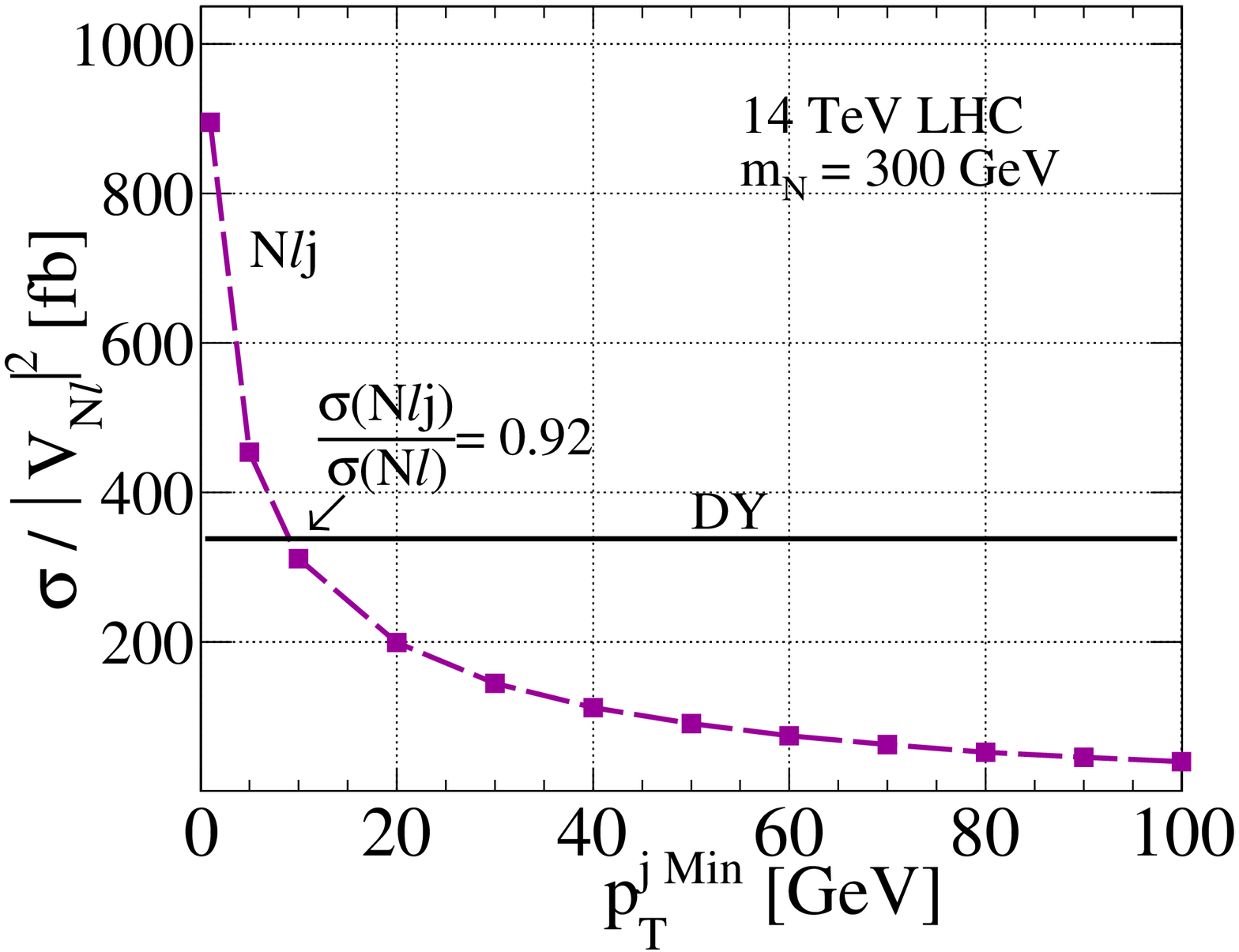}\label{ppNljQCD_XSec_vs_pTMinCut.FIG}}
\caption[Total and differential cross section for $N\ell^\pm j$]{
(a) The tree-level differential cross section for $N\ell^\pm j$ at $\alpha^2\alpha_{\rm s}$ with respect to $p_T^j$;
(b) Integrated cross section $\sigma(N\ell^\pm j)$ versus the minimum $p_T^{j}$ cutoff. The solid line denotes the LO DY rate.
} 
\label{ppNl1jQCD.FIG}
\end{figure}

Before closing the discussion for the heavy $N$ production at hadron colliders, an important remark is in order. 
We have taken into account the {\it inclusive} QCD correction at NNLO as a $K$-factor.
In contrast, Ref.~\cite{Dev:2013wba} included only the tree-level process at order $\alpha^2\alpha^2_{s}$ and  $\alpha^4$
\begin{equation}
 p p \rightarrow N \ell^\pm j j .
\end{equation}
When calculating the exclusive $N\ell^\pm jj$ cross section,  
kinematical cuts of $p_{Tj} > 10$ GeV and $\Delta R_{jj}>0.4$ were applied to regularize the cross section. 
For $m_N = 300\GeV$, the exclusive cross section was found to exceed the LO DY channel at 14 TeV, 
whereas we find that the NNLO correction to the inclusive cross section is only $20\%$ with DIS contributing $3\%$.
More recently~\cite{Das:2014jxa}, the tree-level rate for $N\ell j$ with $p_T^j>30\GeV$  was calculated to be $80\%$ of the LO DY rate at $m_N=500\GeV$; 
at NNLO, we find the inclusive correction to be only $20\%$.
We attribute these discrepancies to their too low a $p_T^j$ cut that overestimate the contribution of initial-state radiation based on a tree-level calculation.

To make the point concrete, we consider the tree-level QCD correction to the DY process at order $\alpha^2\alpha_{\rm s}$
\begin{equation}
 p ~ p \rightarrow N ~\ell^\pm ~j,
 \label{qcd.EQ}
\end{equation}
where the final-state jet originates from an initial-state quark or gluon.
MG5 is used to simulated Eq.~(\ref{qcd.EQ}).
In figure~\ref{ppNljQCD_dXSec_vs_pt.FIG}, the differential cross section of $p_T^j$ is shown for a minimal $p_T$ at $5\GeV$. 
The singularity at the origin is apparent. 
In figure~\ref{ppNljQCD_XSec_vs_pTMinCut.FIG}, the 14 TeV LHC cross section as a function of minimum $p_T$ cut on the jet is presented.
A representative neutrino mass of $m_N = 300\GeV$ is used; no additional cut has been imposed. At $p_{T}^{j \min} = 10\GeV$, as adopted in
Ref.~\cite{Dev:2013wba}, the $N\ell j$ rate is nearly equal to the DY rate, well above the NNLO prediction for the inclusive cross section \cite{Hamberg:1990np}.


\subsection{Kinematic Features of $N$ Production with Jets at 14 TeV}
\label{sec:kine}

To explore the kinematic distributions of the inclusive neutrino production, we fix $\sqrt{s} = 14\TeV$ and $m_N = 500\GeV$. 
At 100 TeV, we observe little change in the kinematical features and our conclusions remain the same. The most notable difference, however, is a broadening of rapidity distributions.
This is due an increase in longitudinal momentum carried by the final states,
which follows from the increase in average momentum carried by initial-state partons.
For $m_N \geq 100\GeV$, we observe little difference from the 500 GeV case we present.
Throughout this study, jets are ranked by $p_T$, namely, 
the jet with the largest (smallest) $p_T$ is referred to as hardest (softest).

\begin{figure}[!t]
\begin{center}
\subfigure[]{\includegraphics[scale=1,width=.48\textwidth]{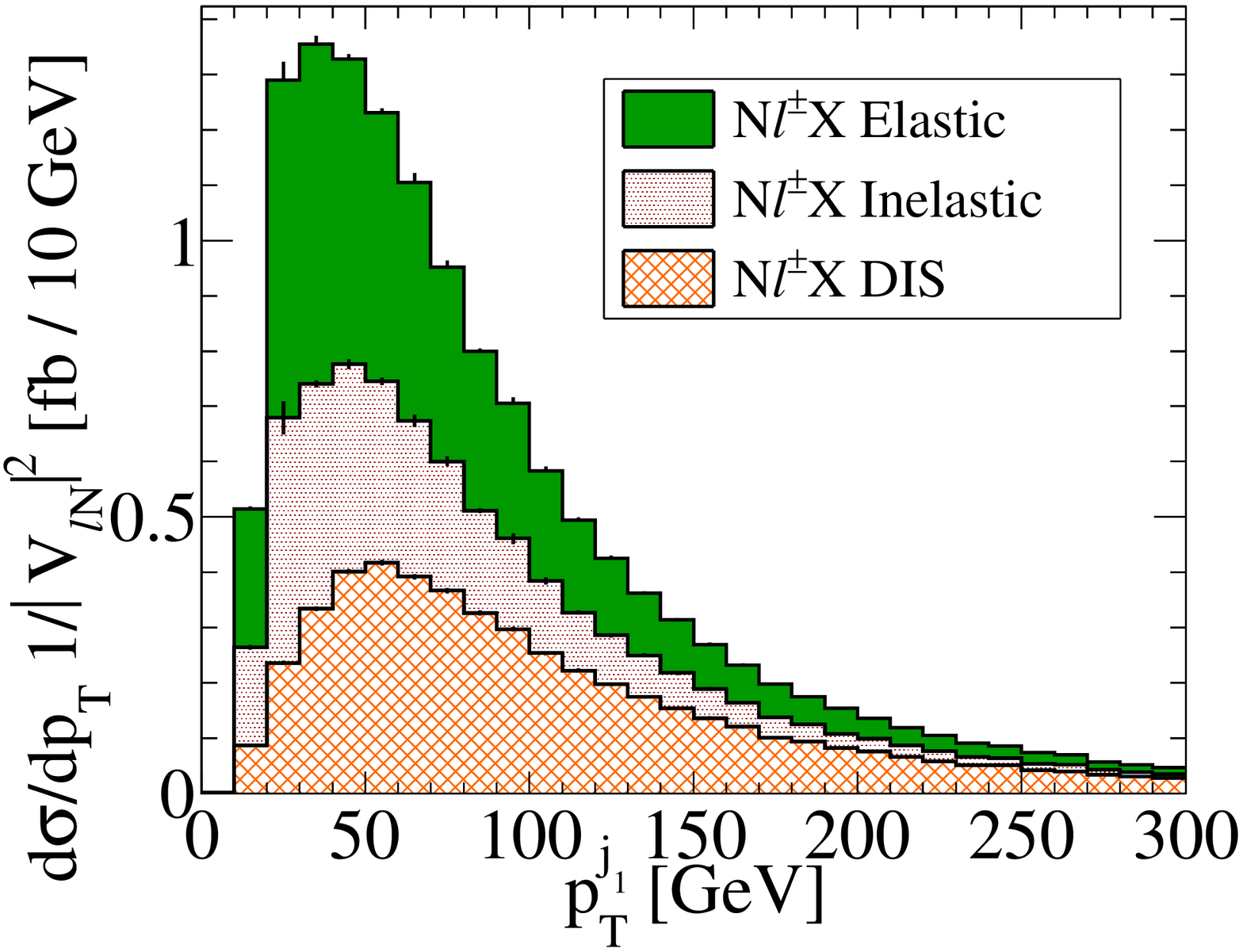}	\label{assocPTj1.FIG} }
\subfigure[]{\includegraphics[scale=1,width=.48\textwidth]{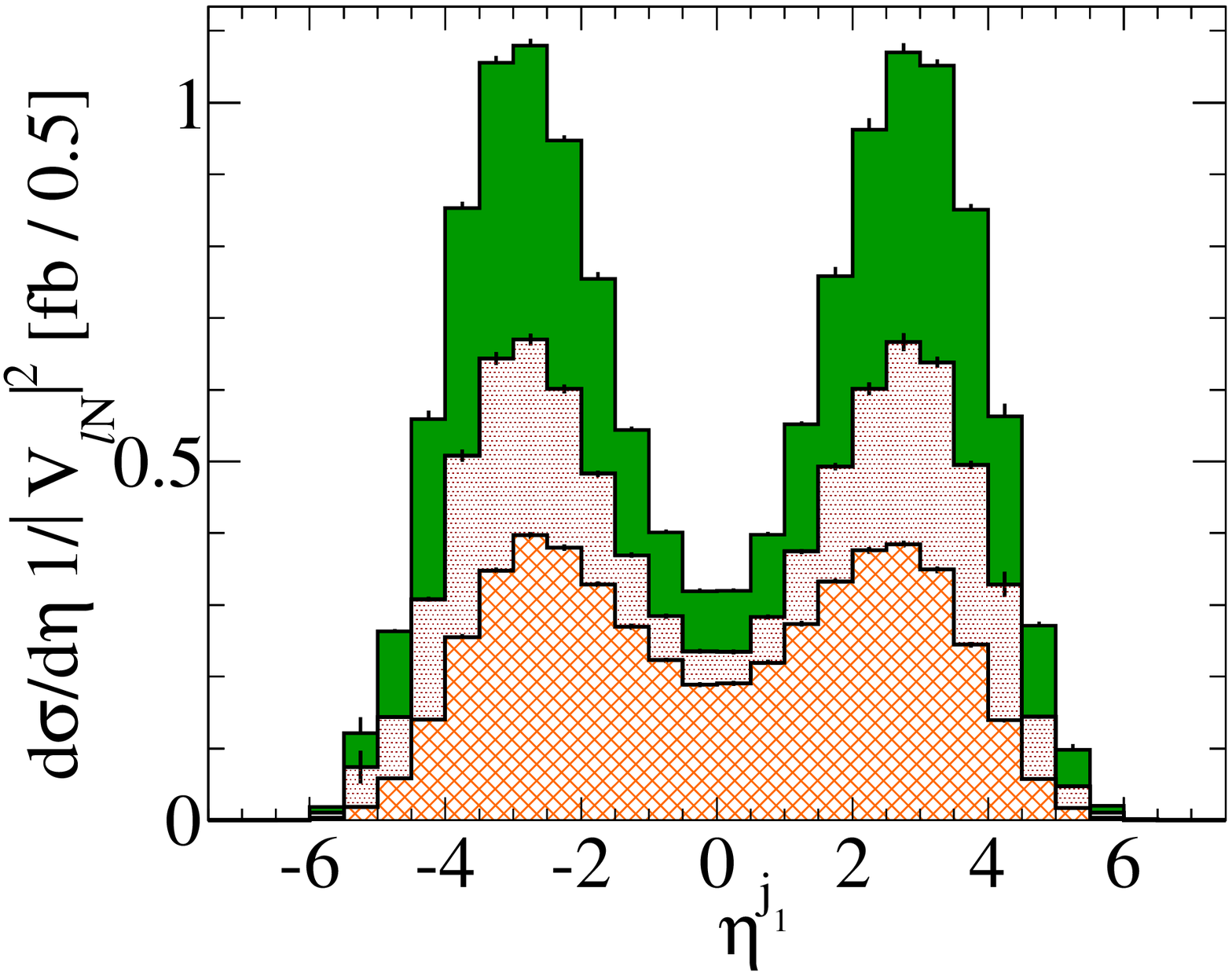}	\label{assocEtaj1.FIG} }
\vspace{.2in}\\
\subfigure[]{\includegraphics[scale=1,width=.48\textwidth]{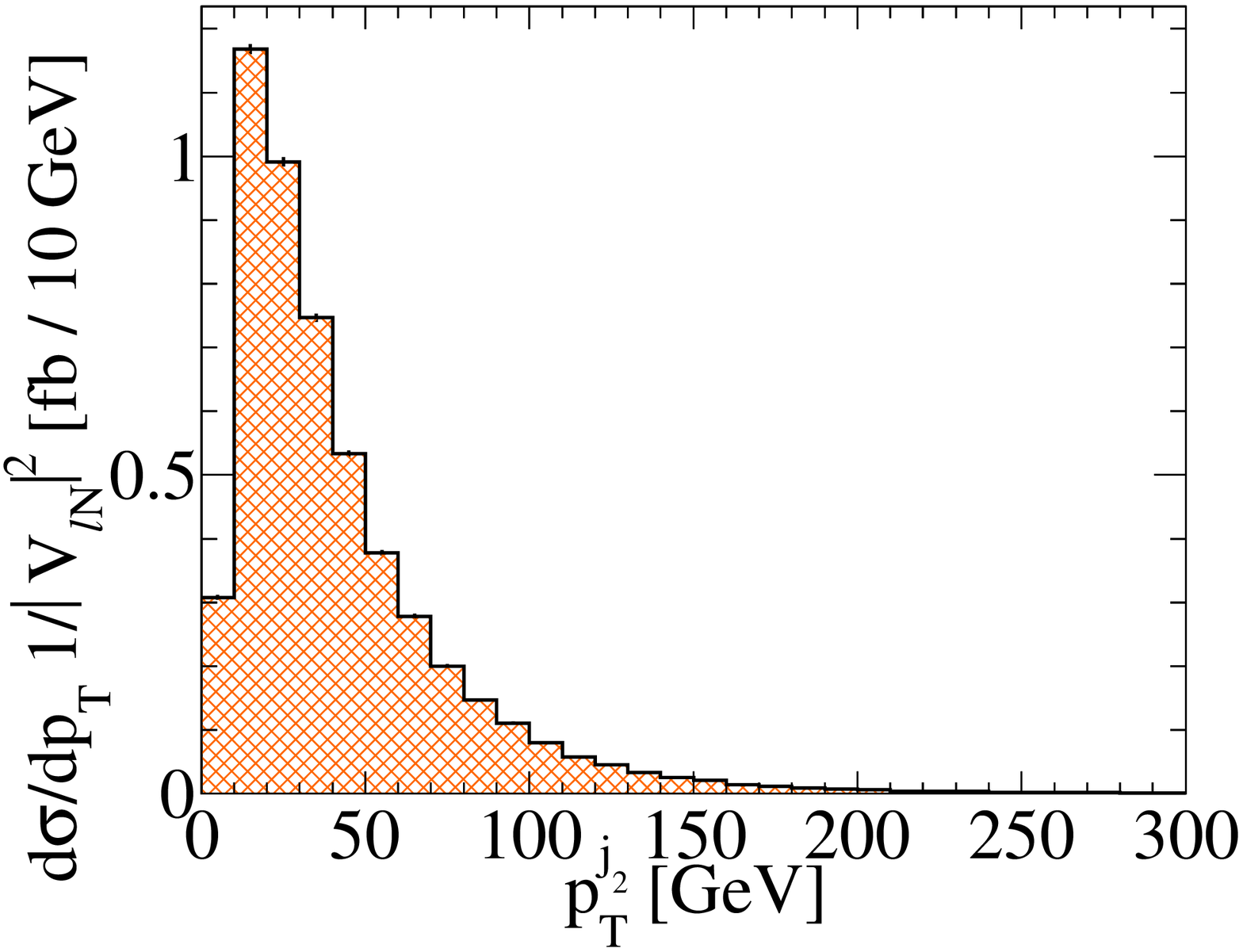}	\label{assocPTj2.FIG} }
\subfigure[]{\includegraphics[scale=1,width=.48\textwidth]{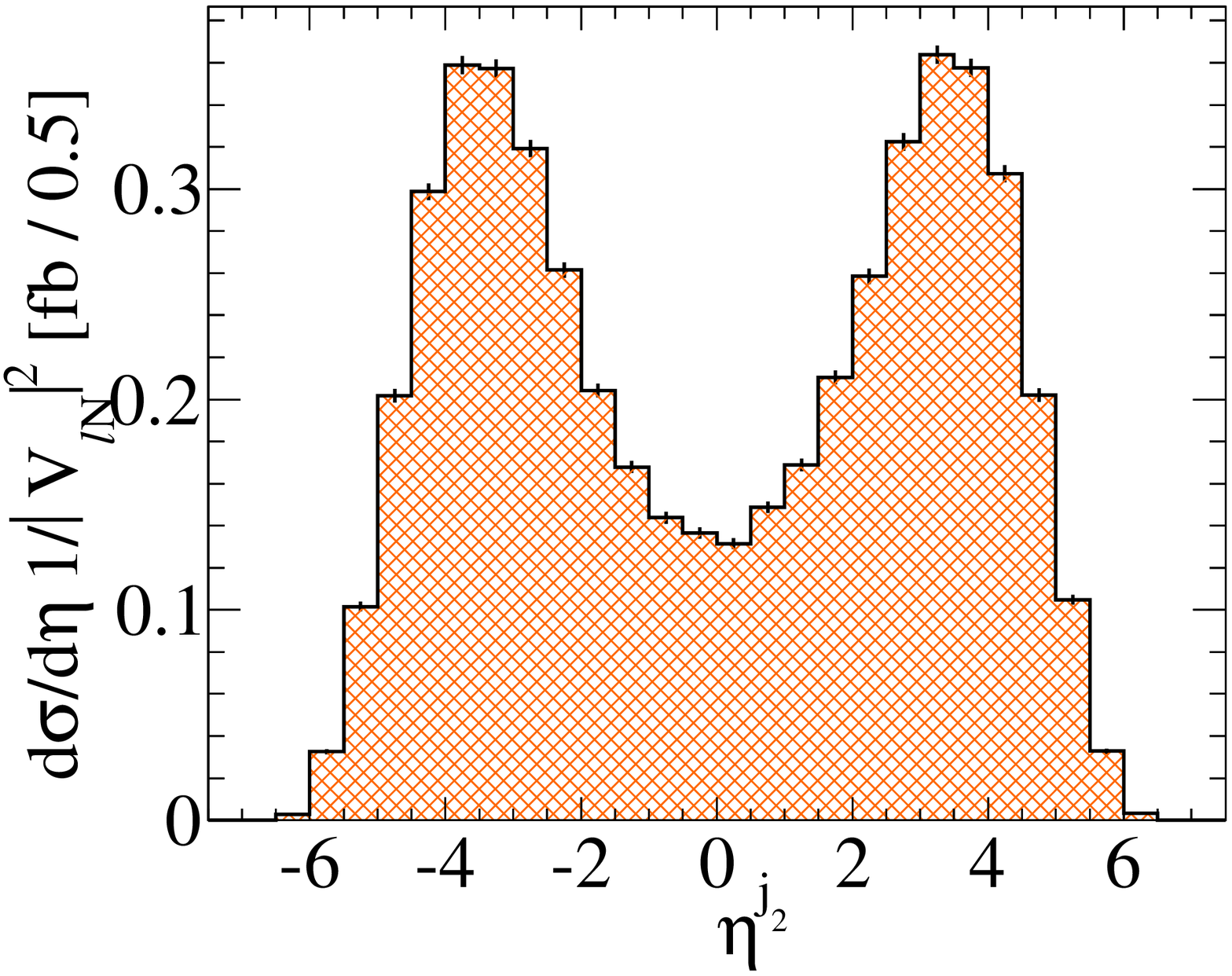}	\label{assocEtal2.FIG} }
\end{center}
\caption[Hadronic $p_{T}$ and $\eta$ differential distributions]{Stacked (a) $p_{T}$ and (b) $\eta$ differential distributions, divided by $\vert V_{\ell N}\vert^{2}$,  
at 14 TeV LHC of the leading jet in the elastic (solid fill), inelastic (dot fill), and DIS (crosshatch fill) processes. 
(c) $p_{T}$ and (d) $\eta$ of the sub-leading jet in DIS.}
\label{assocJet.FIG}
\end{figure}

In figure~\ref{assocJet.FIG}, we plot the (a) $p_T$ and (b) $\eta$ distributions of the hardest jet in $p_T$ produced in association with $N$ 
for the various $W\gamma$ fusion channels.
Also shown are (c) $p_{T}$ and (d) $\eta$ distributions of the sub-leading jet for the DIS channel.
For the hardest jet, we observe a plateau at $p_T \sim M_W / 2$ and a rapidity concentrated at $\vert\eta\vert\sim 3.5$,
suggesting dominance of $t$-channel $W$ boson emission.
For the soft jet, we observe a rise in cross section at low $p_T$ and a rapidity also concentrated at $\vert\eta\vert\sim 3.5$,
indicating $t$-channel emission of a massless vector boson.
We conclude that VBF is the driving contribution $\gamma$-initiated heavy neutrino production.

\begin{figure}[!t]
\begin{center}
\subfigure[]{\includegraphics[scale=1,width=.48\textwidth]{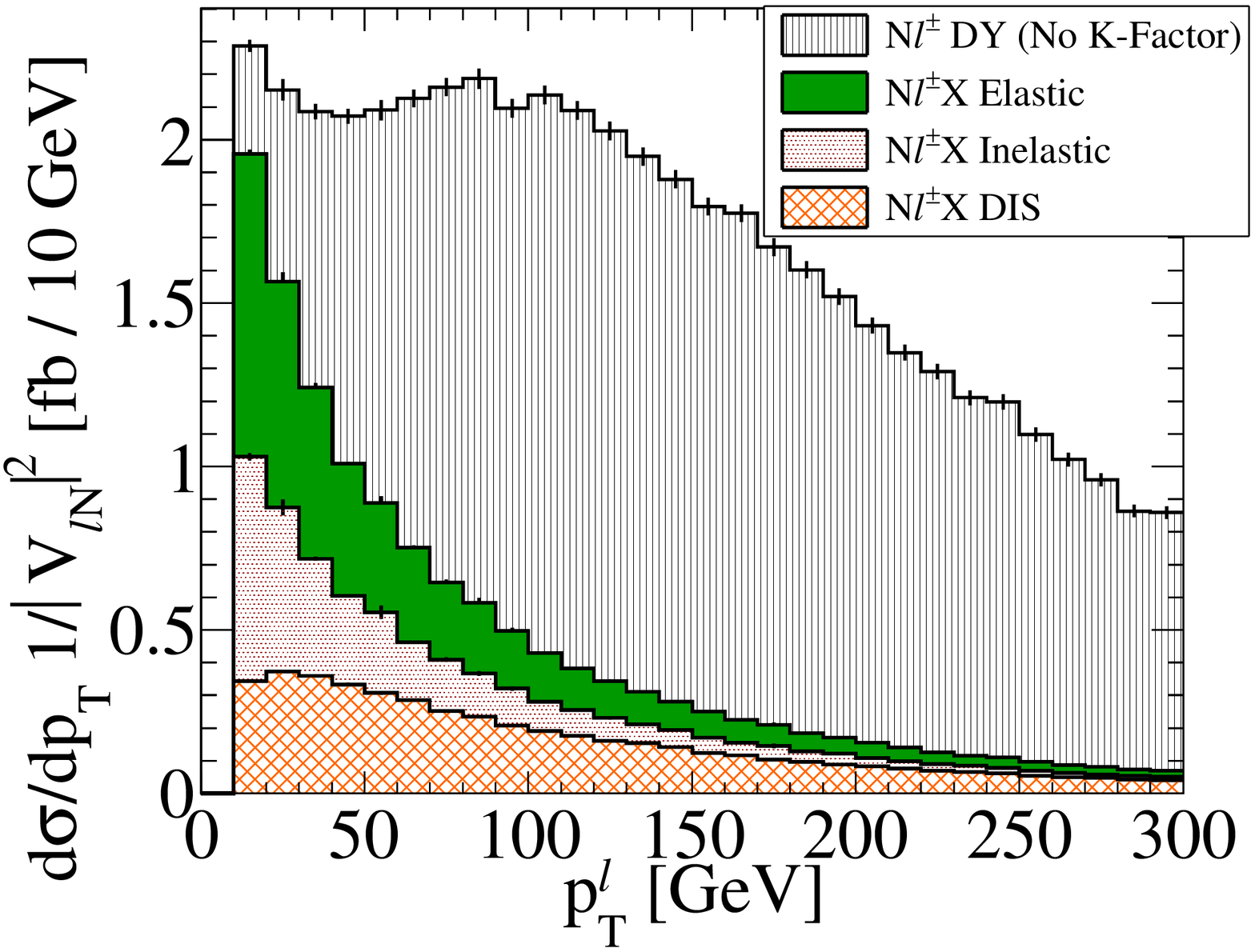}	\label{assocPTl.FIG} }
\subfigure[]{\includegraphics[scale=1,width=.48\textwidth]{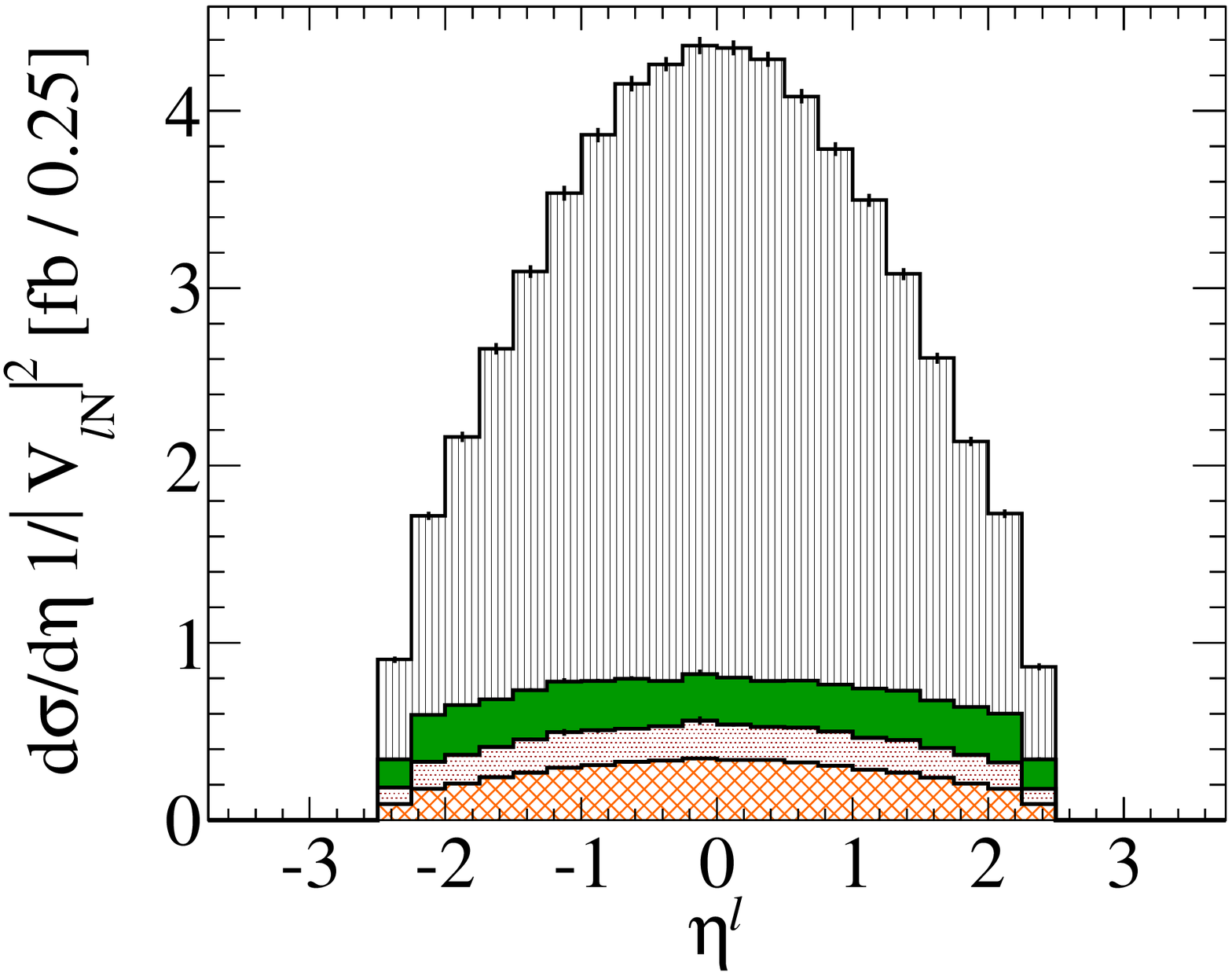}	\label{assocEtal.FIG} }
\vspace{.2in}\\
\subfigure[]{\includegraphics[scale=1,width=.48\textwidth]{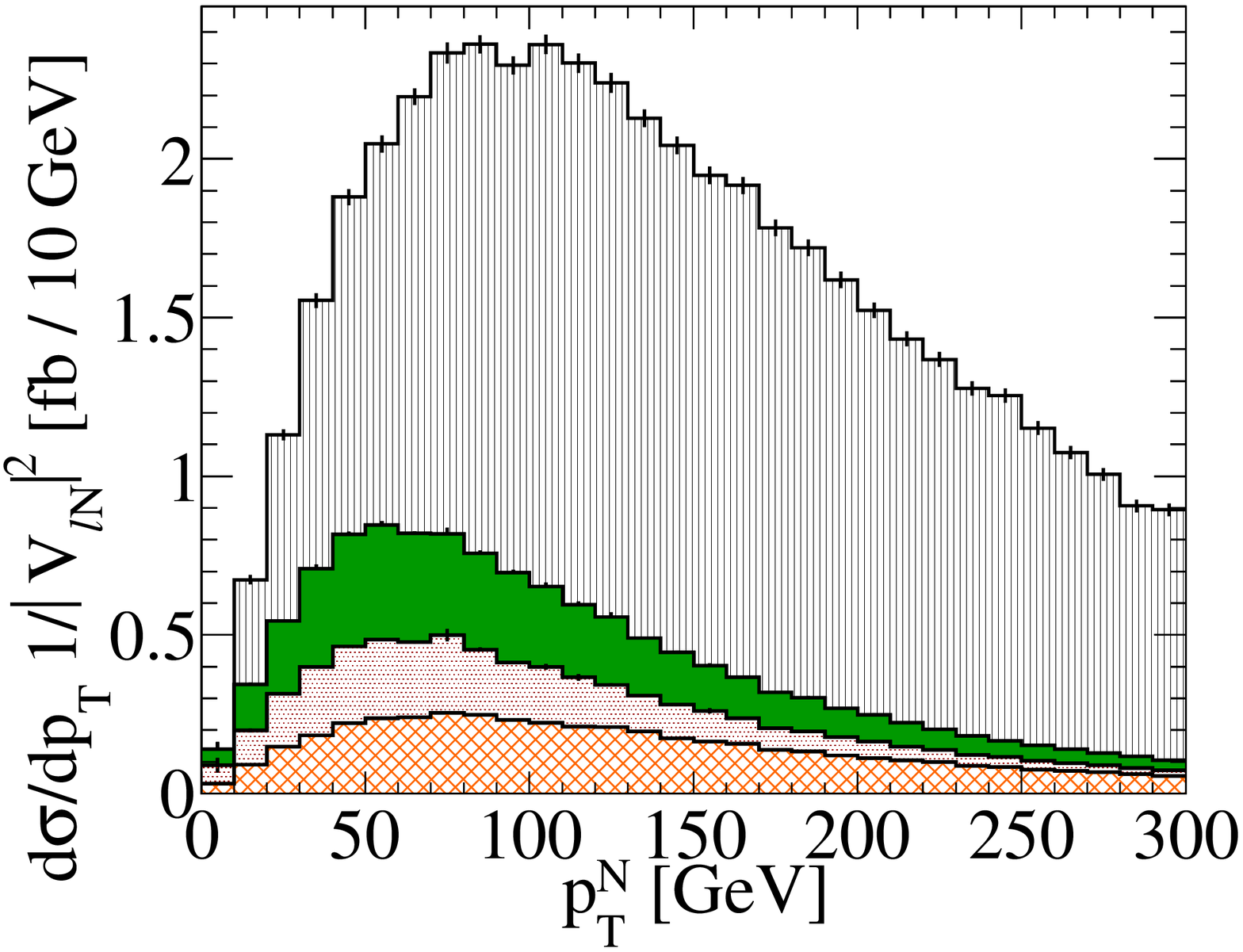}	\label{ptN.FIG} }
\subfigure[]{\includegraphics[scale=1,width=.48\textwidth]{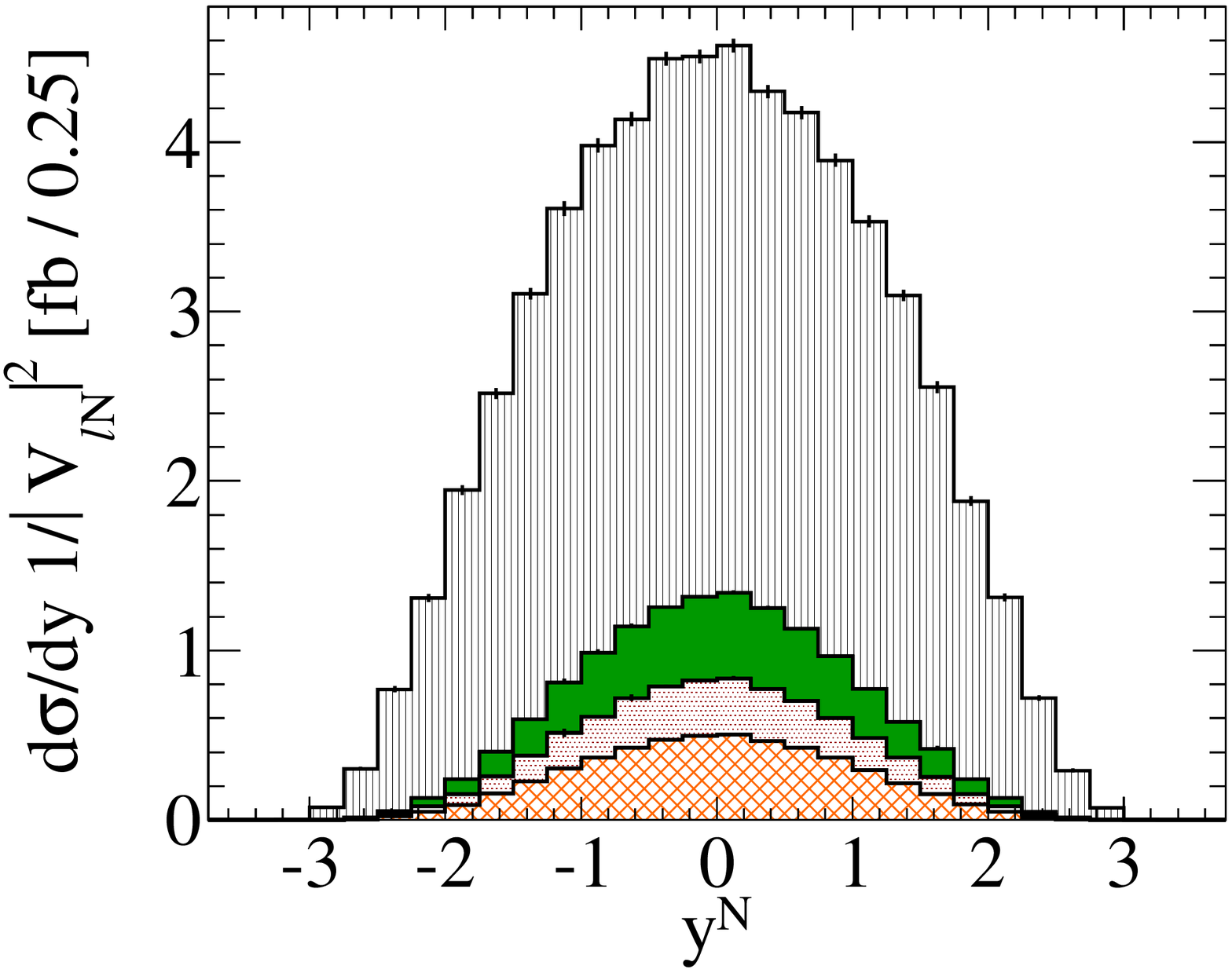}	\label{yN.FIG} }
\end{center}
\caption[Leptonic $p_{T}$ and $y/\eta$ differential distributions]{Stacked (a) $p_{T}$ and (b) $\eta$ differential distributions at 14 TeV LHC of the charged lepton produced in association with $N$ for the
DY (line fill), elastic, inelastic and DIS processes. (c) $p_{T}$ and (d) $y$ of $N$ for the same processes.
Fill style and normalization remain unchanged from figure~\ref{assocJet.FIG}.}
\label{leptonKin.FIG}
\end{figure}

In figure~\ref{leptonKin.FIG}, we plot the (a) $p_T$ and (b) $\eta$ distributions of the charged lepton produced in association with $N$
for all channels contributing to $N\ell$ production. Also shown are the (c) $p_T$ and (d) $y$ distribution of $N$.
For both leptons, we observed a tendency for softer $p_T$ and broader rapidity distributions in $\gamma$-initiated channels than in the DY channel.
As DY neutrino production proceeds through the $s$-channel, $N$ and $\ell$ possess harder $p_T$ than the $\gamma$-initiated states, 
which proceed through $t$-channel production and are thus more forward.


\subsection{Scale Dependence}
\label{sec:scale}

\begin{table}[!t]
\caption{Summary of scale dependence in $N\ell^\pm X$ production at 14 TeV and 100 TeV.}
 \begin{center}
\begin{tabular}{|c|c|c|c|c|}
\hline\hline
\multirow{2}{*}{Scale Parameter} & Default at & \multirow{2}{*}{Lower} & \multirow{2}{*}{Upper} & Variation \tabularnewline
				 & 14 (100) TeV &		& 				& at 14 (100) TeV	\tabularnewline\hline\hline
\multirow{2}{*}{$\Lambda_\gamma^{\rm El }$ [Eq.~(\ref{elPDF.EQ})]} & \multirow{2}{*}{ 1.22 GeV} &
		$m_p$	& 2.3 GeV	&	$\mathcal{O}(10\%)\ \  (12\%)$ \tabularnewline
	  & &	$m_p$	& 5 GeV		&	$\mathcal{O}(22\%)\ \  (28\%)$ \tabularnewline\hline
\multirow{2}{*}{$\lamDIS$ [Eq.~(\ref{disDef.EQ})]} & \multirow{2}{*}{15 GeV (25 GeV)} &
	        5 GeV	& 50 GeV	&	$\mathcal{O}(10\%)\ \ (15\%)$ \tabularnewline
	  & &   5 GeV	& 150 GeV	&	$\mathcal{O}(18\%)\ \ (27\%)$ \tabularnewline\hline\hline
$Q_f^{\rm DY}$  [Eq.~(\ref{factTheorem.EQ})]	& $\sqrt{\hat{s}}/2$	& $m_N/2$	& $\sqrt{\hat{s}}$	&$\mathcal{O}(10\%)\ \ (5\%)$\tabularnewline\hline
$Q_f^{\rm DIS}$ [Eq.~(\ref{factTheorem.EQ})]	& $\sqrt{\hat{s}}/2$	& $m_N/2$	& $\sqrt{\hat{s}}$	&$\mathcal{O}(15\%)\ \ (8\%)$\tabularnewline\hline
\hline
\end{tabular}
\label{scale.TB}
\end{center}
\end{table}

For the processes under consideration, namely DY and $W\gamma$ fusion, there are two factorization scales involved: $Q_{f}$ and $Q_{\gamma}$.
They characterize typical momentum transfers of the physical processes. 
For the $\gamma$-initiated channels, we separate the contributions into three regimes using $\lamEl$ and $\lamDIS$. 
Though the quark parton scale $Q_{f}$ is present in all channels, we assume it to be near $m_N^{}$ and set it as in Eq.~(\ref{QfScale.EQ}). 

\begin{figure}[!t]
\begin{center}
\subfigure[]{\includegraphics[scale=1,width=.45\textwidth]{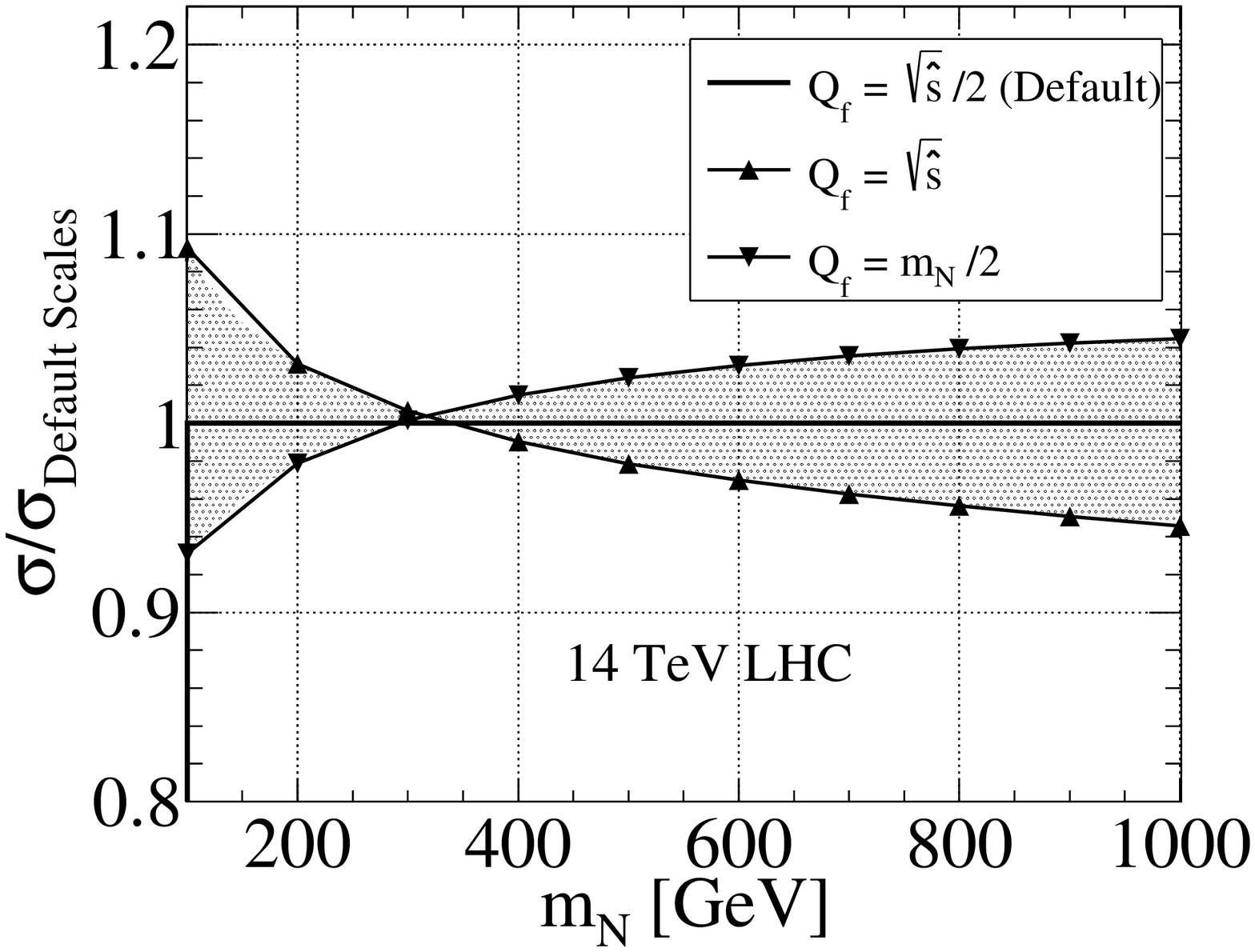}		\label{scale_DY.FIG} }
\subfigure[]{\includegraphics[scale=1,width=.45\textwidth]{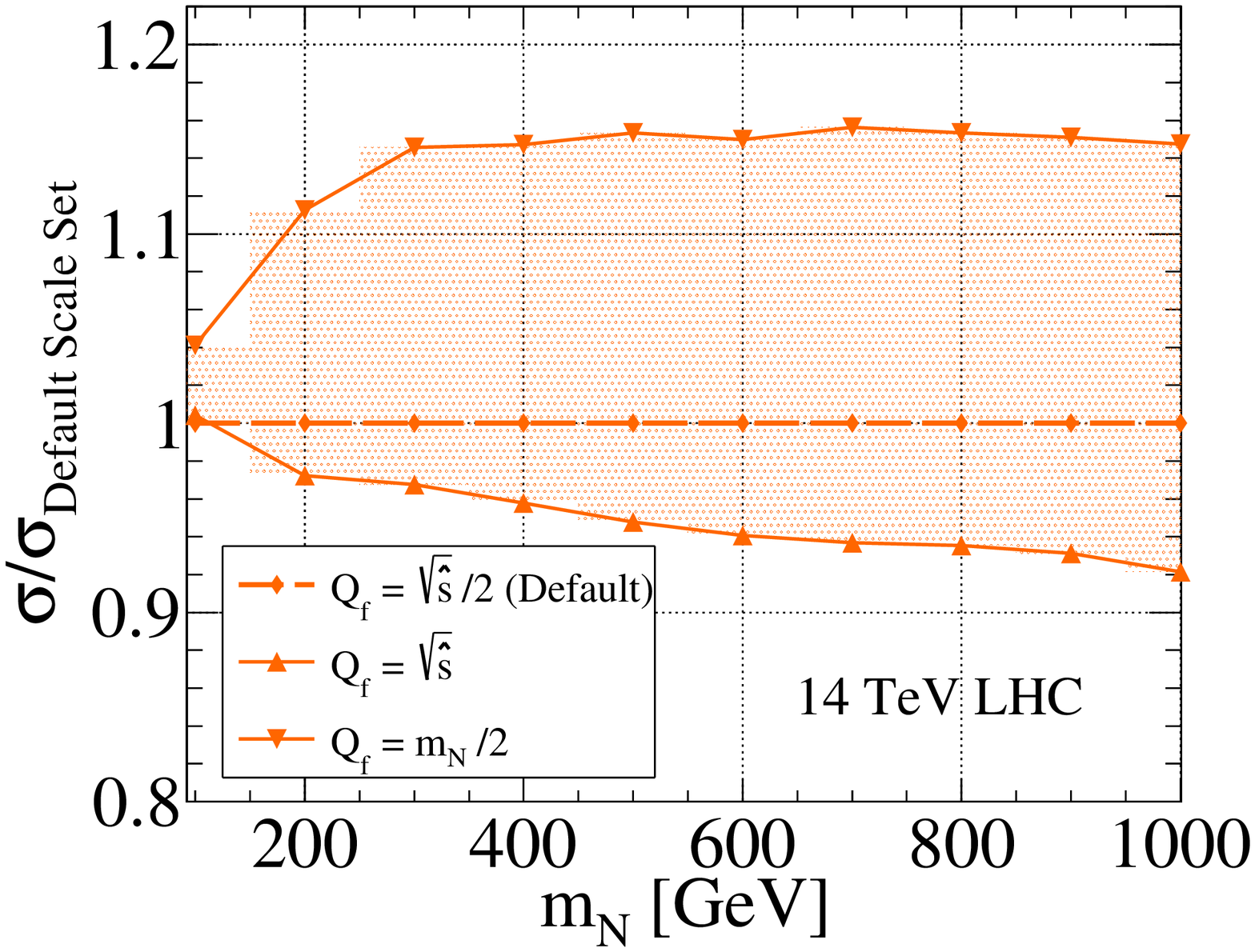}		\label{scale_DIS_qFact.FIG}}
\\
\subfigure[]{\includegraphics[scale=1,width=.45\textwidth]{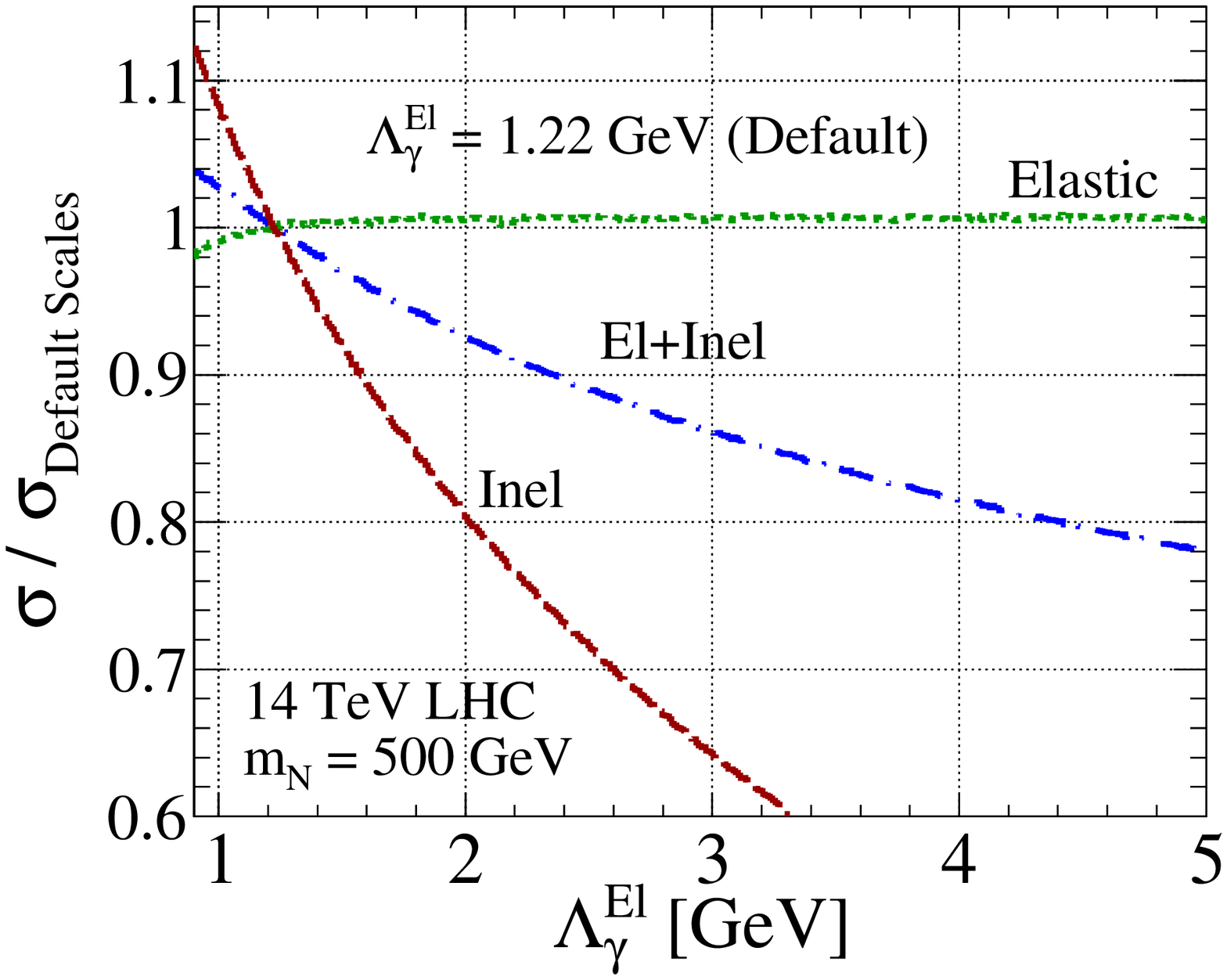}		\label{scale_ElMatch.FIG} }
\subfigure[]{\includegraphics[scale=1,width=.45\textwidth]{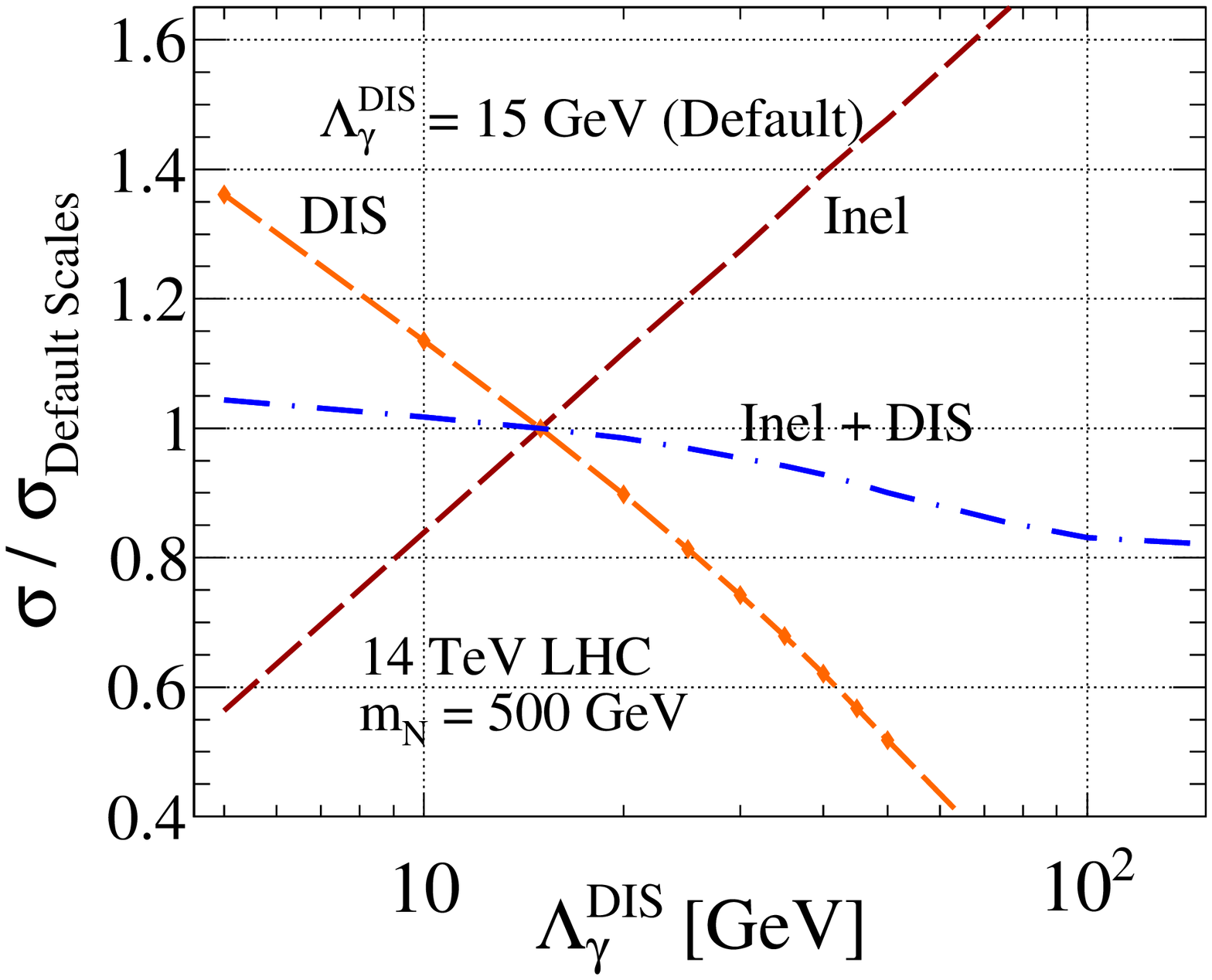}		\label{scale_DISMatch.FIG} }
\end{center}
\caption[Cross section ratios relative to the default scale scheme]{
Cross section ratios relative to the default scale scheme, as a function of $m_N$, 
for the high-scale (triangle) and low-scale (upside-down triangle) $Q_{f}$ scheme in (a) DY and (b) DIS.
The same quantity as a function of (c) $\lamEl$ in elastic (dot), inelastic (dash), elastic+inelastic (dash-dot) scattering;
(d) $\lamDIS$ in inelastic (dash), DIS (dash-diamond), and inelastic+DIS (dash-dot).}
\label{scale.fig}
\end{figure}

To quantify the numerical impact of varying these scales, 
each relevant cross section as a function of $m_N$ is computed with one scale varied while all other scales are held at their default values. 
The test ranges are taken as 
\begin{eqnarray}
  m_{p} \leq \Lambda_\gamma^{\rm El } \leq 5\GeV,
  \quad
  5\GeV \leq Q_\gamma = \lamDIS \leq  150\GeV,
  \quad
  \frac{m_N}{2}\leq Q_{f} \leq \sqrt{\hat{s}},
\label{scaleVar.EQ}
\end{eqnarray}
In figure~\ref{scale.fig}, we plot the variation band in each production channel cross section due to the shifting scale.
For a given channel, rates are normalized to the cross section using the default scale choices, as discussed in the previous sections and summarized in the first column of Table \ref{scale.TB}.
High-(low-) scale choices are denoted by a solid line with right-side (upside-down) up triangles.

For the 14 TeV LO DY process, we observe in figure~\ref{scale_DY.FIG} maximally a 9\% upward (7\% downward) variation for the range of $m_N$ investigated.
Below $m_N\approx 300\GeV$, the default scale scheme curve is below (above) the high (low) scale scheme curve.
The trend is reversed for above $m_N\approx 300\GeV$.
At 100 TeV, the crossover point shifts to much higher values of $m_N$.
Numerically, we observe a smaller scale dependence at the {5\%} level.

In figure~\ref{scale_DIS_qFact.FIG}, we plot scale variation associated with the factorization scale $Q_{f}$ for DIS.
Maximally, we observe a 16\% upward (8\% downward) shift.
We observe that the crossover between the high and low scale schemes now occurs at $m_N\lesssim 100\GeV$.
This is to be expected as $\hat{s}$ for the 4-body DIS at a fixed neutrino mass is much larger than that for the 2-body DY channel.
Similarly, as $\sqrt{\hat{s}}$ and $m_N$ are no longer comparable, as in the DY case, an asymmetry between the high- and low-scale scheme curves emerges. 
At 100 TeV, we observe a smaller dependence at the {10\%} level.

In figure~\ref{scale_ElMatch.FIG}, 
we show the dependence on $\lamEl$ in the elastic (dot) and inelastic (dash) channels, as well as the sum of the two channels (dash-dot). 
For the elastic channel we find very small dependence on $\lamEl$ between $m_p$ and $5\GeV$,
with the analytical expression for $f^{\rm El}_{\gamma/p}$ given in Section~\ref{AppEl.col.sec}.
For the inelastic channel, on the other hand, we find rather large dependence on $\lamEl$ between $m_p$ and $5\GeV$.
Since $\lamEl$ acts as the regulator of the inelastic channel's collinear logarithm, this large sensitivity is expected;
see Section~\ref{sec:AppIn} for details regarding $f^{\rm Inel}_{\gamma/p}$.
We find that the summed rate is slightly more stable.
In the region $m_p < \lamEl < 2.3\GeV$, the variation is below the 10\% level.
Over the entire range studied, this grows to {$20\%$}. 
At 100 TeV, similar behavior is observed and the dependence grows to the {$30\%$} level over the whole range.

In figure~\ref{scale_DISMatch.FIG}, for $m_N=500\GeV$, 
we plot the scale dependence on $\lamDIS$ in the inelastic (dash) and DIS (dash-diamond) channels, as well as the sum of the two channels (dash-dot).
Very large sensitivity on the scale is found for individual channels, ranging {$40\%-60\%$} over the entire domain.
However, as the choice of $\lamDIS$ is arbitrary, we expect and observe that their sum is considerably less sensitive to $\lamDIS$.
For $\lamDIS = 5-50~(5-150)$ GeV, we find maximally a {10\% (18\%)} variation.
The stability suggests the channels are well-matched for scales in the range of $5-50\GeV$. 
Results are summarized in Table \ref{scale.TB}.


\section{Heavy Neutrino Observability at Hadron Colliders}
\label{sec:100TeV}

\subsection[Kinematic Features at 100 TeV]{Kinematic Features of Heavy $N$ Decays to Same-Sign Leptons with Jets at 100 TeV}
We consider at a $100$ TeV $pp$ collider charged current production of a heavy Majorana neutrino $N$ in association with $n=0,~1~\text{or}~2$ jets,
and its decay to same-sign leptons and a dijet via the subprocess $N\to \ell W \to \ell jj$:
\begin{equation}
 p~p \rightarrow ~N ~\ell^{\pm} ~+~ nj 	\rightarrow 
   \ell^{\pm} ~\ell^{'\pm} ~+~ (n+2)j, \quad n = 0,~1,~2.
 \label{ppllnj.EQ}
\end{equation}
Event simulation for the DY and DIS channels was handled with MG5.
A NNLO $K$-factor of $K=1.3$ is applied to the LO DY channel; kinematic distributions are not scaled by $K$.
Elastic and inelastic channels were handled by extending neutrino production calculations to include heavy neutrino decay.
The NWA with full spin correlation was applied.
The elastic channel matrix element was again checked with MG5.

Detector response was modeled by applying a Gaussian smearing to jets and leptons.
For jet energy, the energy resolution is parameterized by \cite{Aad:2009wy} 
\begin{equation}
 \frac{\sigma_E}{E} = \frac{a}{\sqrt{E/\GeV}} \oplus b, 
 \label{jetSmear.EQ}
\end{equation}
with $a = 0.6~(0.9)$ and $b= 0.05~(0.07)$ for $\vert\eta\vert\leq3.2 ~(>3.2)$,  
and where the terms are added in quadrature, i.e., $x\oplus y = \sqrt{x^2 + y^2}$.
For muons, the inverse-$p_T$ resolution is parameterized by \cite{Aad:2009wy}
\begin{equation}
 \frac{\sigma_{1/p_T}}{(1/p_T)} = \frac{0.011\GeV}{p_T} \oplus 0.00017.
 \label{muSmear.EQ}
\end{equation}
We will eventually discuss the sensitivity to the $e^\pm\mu^\pm$ final state and thus consider electron $p_T$ smearing.
For electrons,\footnote{
For this group of exotic searches, the dominant lepton uncertainty stems  from $p_T$ mis-measurement.
The energy uncertainty is only 1\% versus a 20\% uncertainty in the electron $p_T$ resolution~\cite{Aad:2009wy}.} 
the $p_T$ resolution is parameterized by \cite{Aad:2009wy}
\begin{equation}
 \frac{\sigma_{p_T}}{p_T} = 0.66 \times \left( \frac{0.10}{\sqrt{p_T/\GeV}} \oplus 0.007 \right). 
 \label{eleSmear.EQ}
\end{equation}
Both the muon $1/p_T$ and electron $p_T$ smearing are translated into an energy smearing, keeping the polar angle unchanged.
We only impose the cuts on the charged leptons as listed in Eq.~(\ref{regCutsLep.EQ}).

\begin{figure}[!t]
\begin{center}
\subfigure[]{\includegraphics[scale=1,width=.48\textwidth]{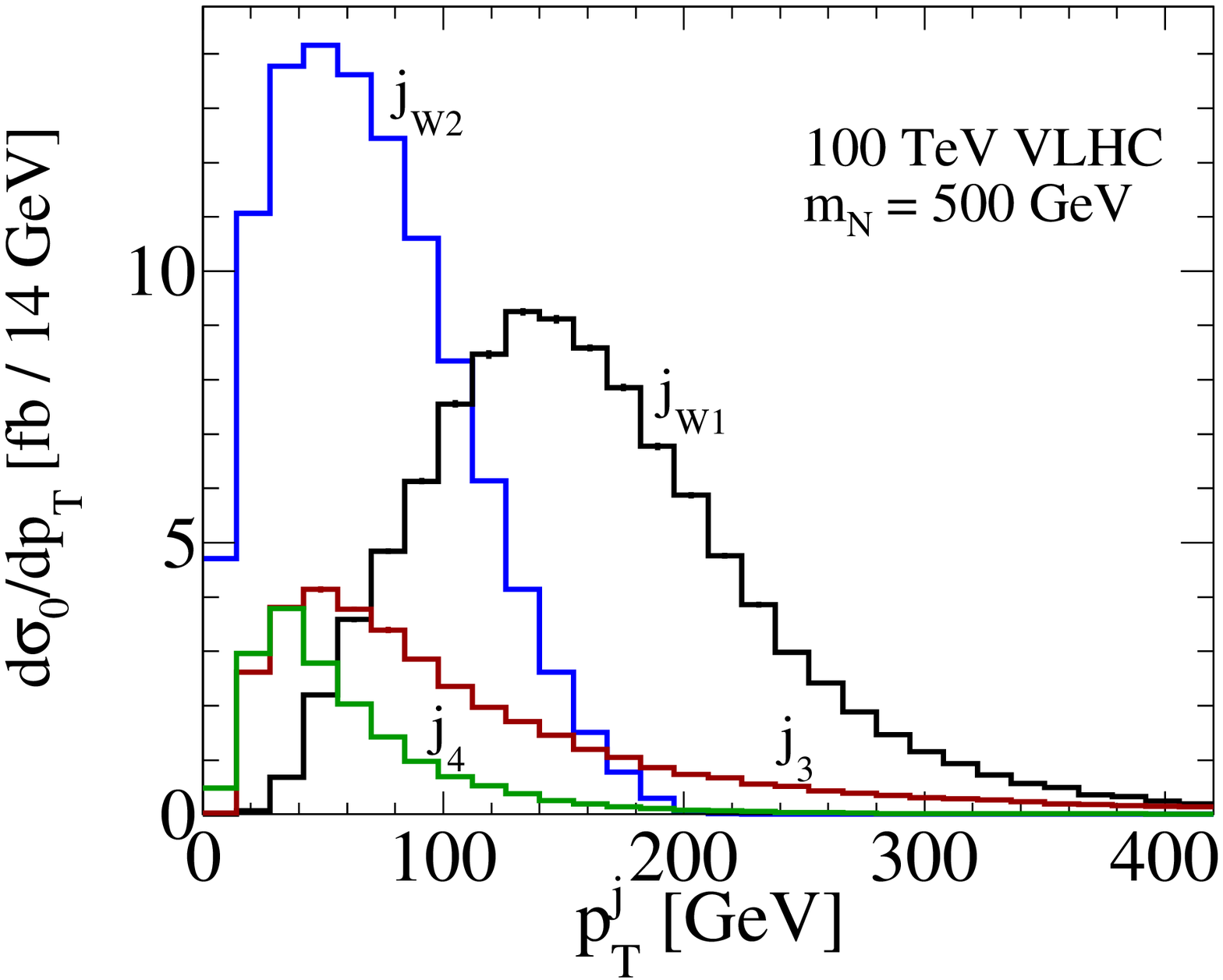}\label{ptj_100TeV_500GeV.fig}}
\subfigure[]{\includegraphics[scale=1,width=.48\textwidth]{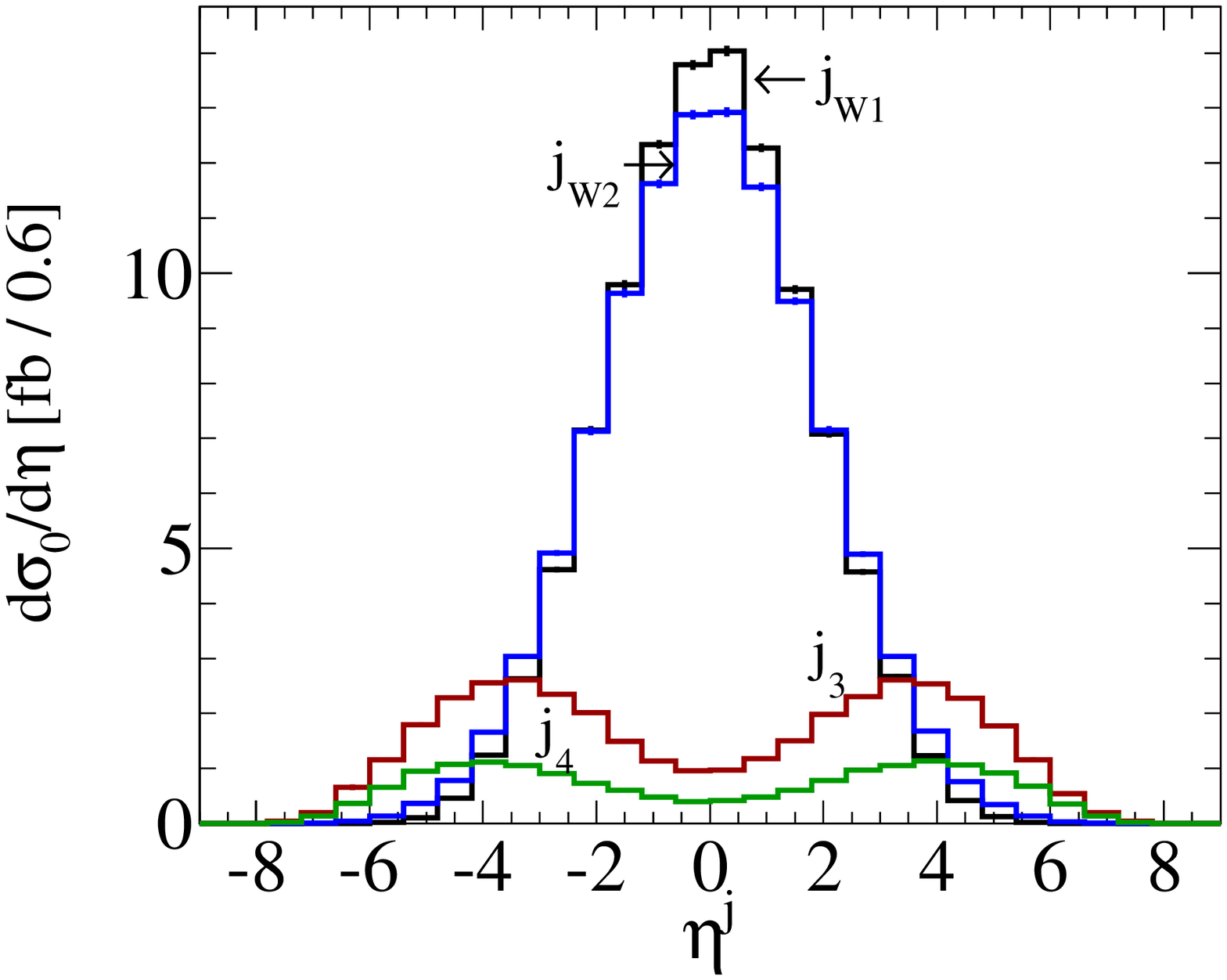}\label{etaj_100TeV_500GeV.fig}}
\\
\subfigure[]{\includegraphics[scale=1,width=.48\textwidth]{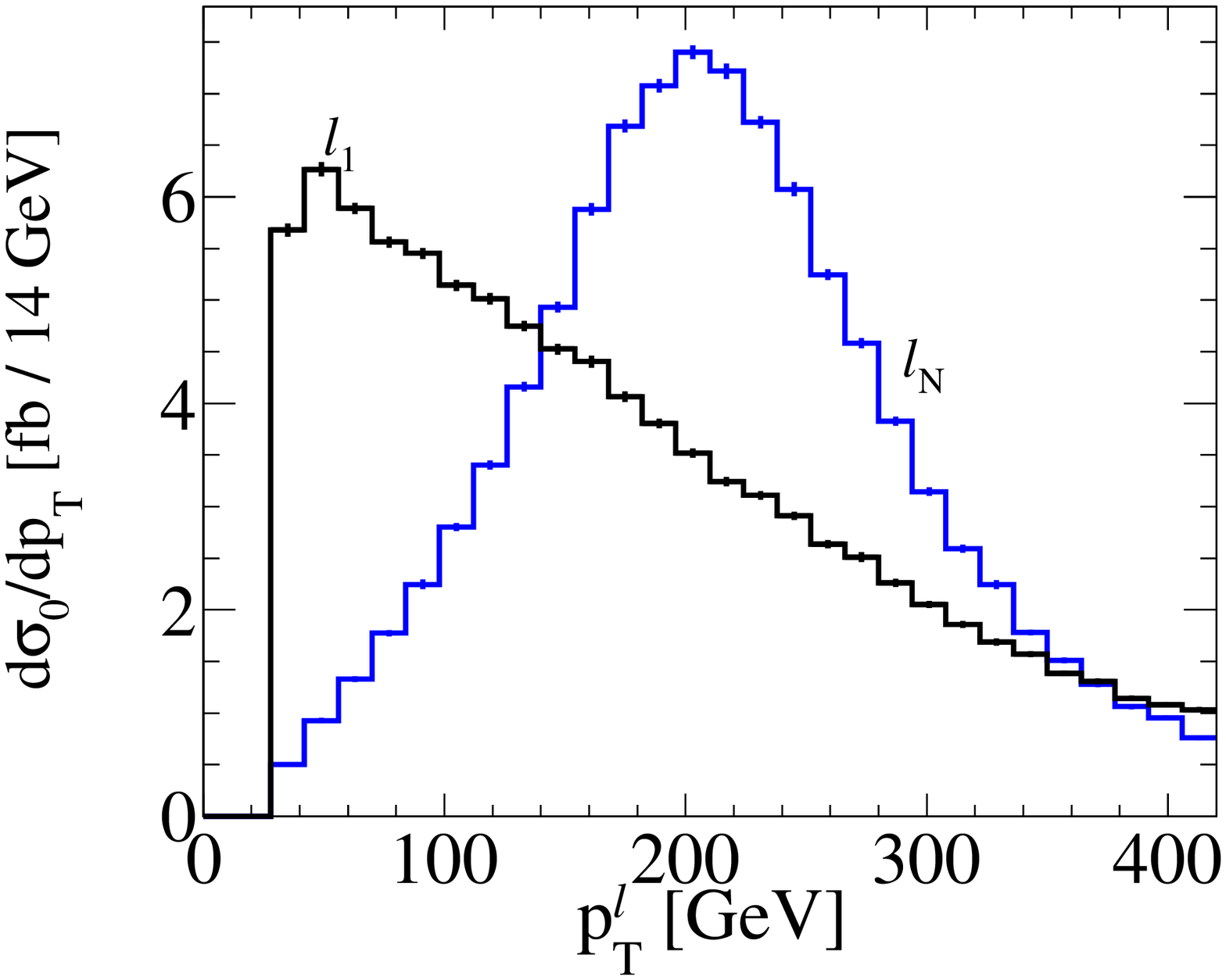}\label{ptl_100TeV_500GeV.fig}}
\subfigure[]{\includegraphics[scale=1,width=.48\textwidth]{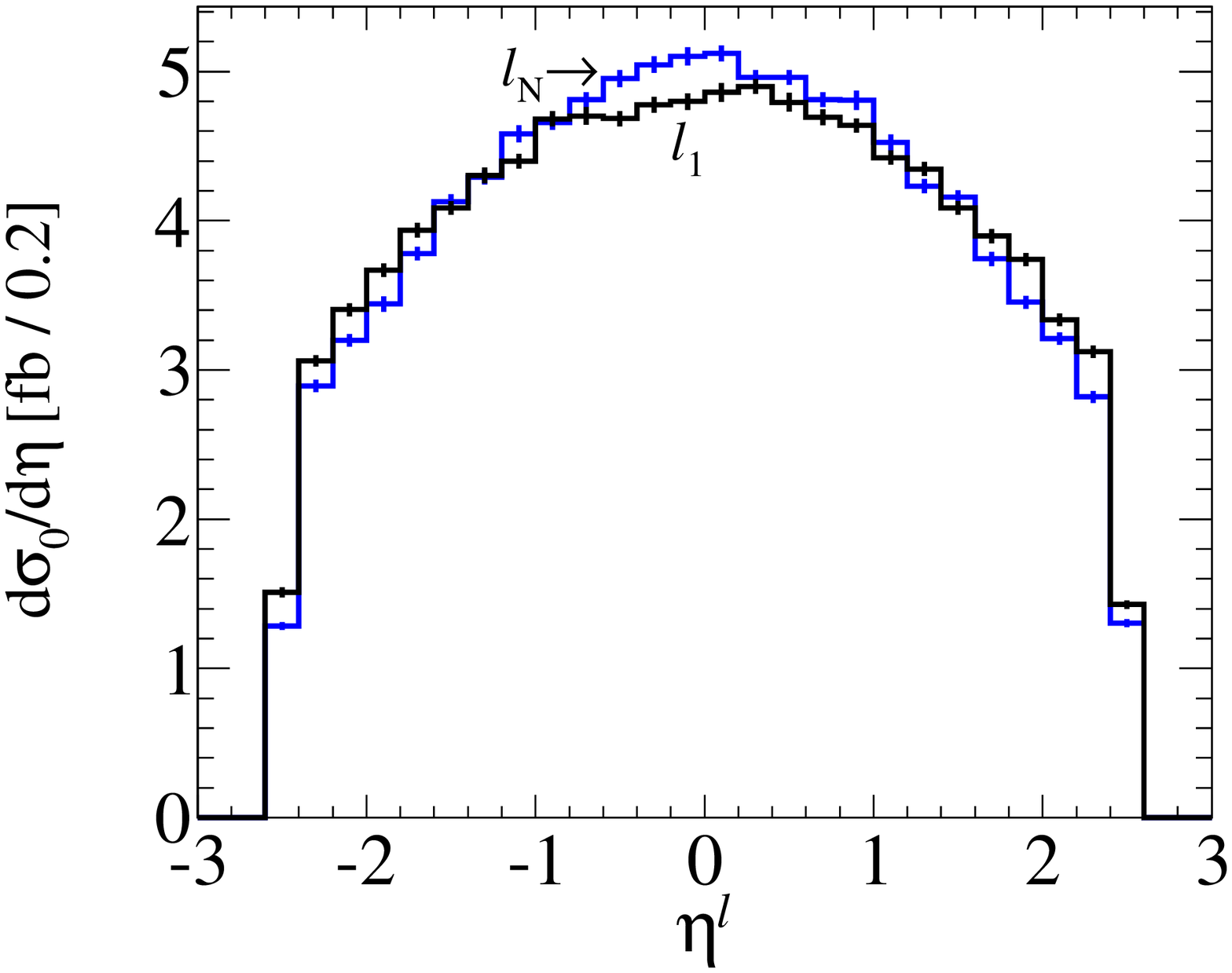}\label{etal_100TeV_500GeV.fig}}
\end{center}
\caption[Kinematic distributions of $pp\rightarrow \ell^{\pm}\ell^{'\pm}jjX$ II]{(a) $p_T$ and (b) $\eta$ differential distributions of the final-state jets for the processes in Eq.~(\ref{ppllnj.EQ}), for $m_N=500\GeV$;
(c,d) the same for final-state same-sign dileptons.}
\label{jets_leps_100TeV.fig}
\end{figure}

In figure~\ref{jets_leps_100TeV.fig}, we show the transverse momentum and pseudorapidity distributions of the final-state jets and same-sign dileptons for the processes in Eq.~(\ref{ppllnj.EQ}), for $m_N = 500$ GeV. 
Jets originating from $N$ decay are denoted by $j_{W_{i}}$, for $i=1,2$, and are ranked by $p_T~(p_{T}^{j_{W_1}}>p_{T}^{j_{W_2}})$. As the three-body $N\rightarrow \ell j j$ decay is preceded by the two-body $N\rightarrow \ell W$ process, $p_T^{j_W}$ scales like $m_N/4$,
as seen in figure~\ref{ptj_100TeV_500GeV.fig}.
The jets produced in association with $N$ are denoted by $j_{3}$ or $j_{4}$, and also ranked by $p_T$.
As VBF drives these channels, we expect $j_3$ (associated with $W^*$) and $j_4$ (associated with $\gamma^*$) to scale 
like $M_W/2$ and $\lamDIS$, respectively.
In figure~\ref{etaj_100TeV_500GeV.fig}, the $\eta$ distributions of all final-state jets are shown.
We see that $j_3$ and $j_4$ are significantly more forward than $j_{W1}$ and $j_{W2}$, consistent with jets participating in VBF.
The high degree of centrality of $j_{W1}$ and $j_{W2}$ follows from the central $W$ decay.

In figures~\ref{ptl_100TeV_500GeV.fig} and~\ref{etal_100TeV_500GeV.fig}, we plot the $p_T$ and $\eta$ distributions of the final-state leptons.
The charged lepton produced in association with $N$ is denoted by $\ell_1$ and the neutrino's child lepton by $\ell_N$. 
As a decay product, $p_T^{\ell_N}$ scales like $(m_N-M_W)/2$, 
whereas $p_{T}^{\ell_1}$ scales as $(\sqrt{\hat{s}}-m_N)/2$.
$\ell_1$ tends to be soft and more forward in the $\gamma$-initiated channels.

\begin{table}[!t]
\caption{Parton-level cuts on 100 TeV $\mu^\pm\mu^\pm jjX$ Analysis.}
 \begin{center}
\begin{tabular}{|c|c|c|}
\hline\hline
Lepton Cuts & Jet Cuts & Other Cuts \tabularnewline\hline\hline
 $\Delta R_{\ell\ell}>0.2$				&$\Delta R_{jj}>0.4$				& $\Delta R_{\ell j}^{\rm Min} > 0.6$	
 \tabularnewline
 $p_T^\ell ~(p_{T}^{\ell ~\rm Max})> 30~(60)\GeV$ 	&$p_T^j ~(p_{T}^{j~\rm Max})> 30~(40)\GeV$ 	& $\not\!\! E_T < 50\GeV$ \tabularnewline
$\vert \eta^\ell\vert<2.5$ 				&$\vert\eta^j\vert < 2.5$	 		& $\vert m_{N}^{\rm Candidate} - m_N \vert < \rm 20\GeV$
\tabularnewline
							&$\vert M_{W}^{\rm Candidate} - M_W \vert < 20\GeV$ 	& \tabularnewline	
							&$\vert m_{jjj} - m_t \vert < 20\GeV$ (Veto) 		&\tabularnewline\hline	
\hline
\end{tabular}
\label{100TeVCuts.TB}
\end{center}
\end{table}
\begin{table}[!t]
\caption[Acceptance rates for $\mu^\pm\mu^\pm jjX$ at 100 TeV VLHC.]{Acceptance rates and percentage efficiencies for the signal $\mu^\pm\mu^\pm jjX$ at 100 TeV VLHC.}
 \begin{center}
\begin{tabular}{|c|c|c|c|}
\hline \hline
 $\sigma_{0}$ [Eq.~(\ref{bareXSecDef.EQ})] [fb]	$\quad\backslash\quad$ $m_N$ [GeV] & $300$ & $500$ & $1000$ \tabularnewline\hline\hline  
Fiducial + Kin.~ + Smearing	[Eq.~(\ref{fidkinsmCut.EQ})]	&281~(41\%)	&83.9~(45\%)   	& 11.6~(28\%)  	\tabularnewline\hline
Leading $p_{T}$ Minimum			[Eq.~(\ref{leadPTCut.EQ})]	&278~(99\%)	&83.8~($>$99\%)	& 11.6~($>$99\%) \tabularnewline\hline
$\Delta R_{\ell j}$ Separation		[Eq.~(\ref{dRljCut.EQ})]    	&264~(95\%)	&79.3~(95\%)	& 10.7~(92\%)  \tabularnewline\hline
$\not\!\! E_{T}$ Maximum		[Eq.~(\ref{metCut.EQ})]   	&263~($>$99\%)	&78.1~(99\%) 	& 10.1~(95\%) \tabularnewline\hline
$M_{W}$ Reco.~ 				[Eq.~(\ref{mWCut.EQ})]	   	&252~(96\%)	&74.1~(95\%) 	& 9.51~(94\%) \tabularnewline\hline
$m_{t}$ Veto				[Eq.~(\ref{mtCut.EQ})]		&251~(99\%)	&73.5~(99\%) 	& 9.42~(99\%)\tabularnewline\hline
$m_{N}$ Reco.~ 				[Eq.~(\ref{mNCut.EQ})]		&244~(98\%)	&64.7~(88\%) 	 & 7.79~(83\%)  \tabularnewline\hline
\hline
Acceptance $[\mathcal{A}] = \sigma_{0}^{\rm ~All~Cuts} / \sigma_{0}^{\rm Fid.+Kin.+Sm.}$	& 87\%	& 77\%	& 67\%	\tabularnewline\hline
\hline
\end{tabular}
\label{acceptXSec.TB}
\end{center}
\end{table}

\subsection{Signal Definition and Event Selection: Same-Sign Leptons with Jets}

For simplicity,  we restrict our study to electrons and muons.
We design our cut menu based on the same-sign muon channel.
Up to detector smearing effects, the analysis remains unchanged for electrons.
A summary of imposed cuts are listed in Table~\ref{100TeVCuts.TB}.
Jets and leptons are identified by imposing an isolation requirement; we require
\begin{equation}
 \Delta R_{jj} > 0.4,\quad \Delta R_{\ell\ell}>0.2.
\label{regCutsDIS.EQ}
\end{equation}
We define our signal as two muons possessing the same electric charge and at least two jets satisfying the following fiducial and kinematic requirements:
\begin{equation}
 \vert \eta^\ell \vert < 2.5, \quad 
 p_{T}^\ell > 30\GeV,\quad  
 \vert \eta^j \vert < 2.5,\quad
 p_{T}^j > 30\GeV.
 \label{fidkinsmCut.EQ}
\end{equation}
The bare cross sections [defined by factorizing out $S_{\ell\ell}$ as defined in  Eq.~(\ref{bareXSecDef.EQ})]
after cuts listed in Eqs.~(\ref{fidkinsmCut.EQ}) and (\ref{regCutsDIS.EQ}) and smearing are given in the first row of Table~\ref{acceptXSec.TB},
for representative masses $m_N = 300,~500,$ and 1000 GeV.
Events with additional leptons are rejected. 
Events with additional jets are kept; we have not tried to utilize the VBF channel's high-rapidity jets.
About 30-45\% of all $\ell^\pm\ell^{'\pm} jjX$ events survive these cuts.
As learned from figure~\ref{jets_leps_100TeV.fig}, the $\eta$ requirement given in Ref.~\cite{Avetisyan:2013onh} considerably reduces selection efficiency.
Extending the fiducial coverage to $\eta^{\rm Max} = 3$ or larger, though technically difficult, can be very beneficial experimentally.

\begin{figure}[!t]
\begin{center}
\subfigure[]{\includegraphics[scale=1,width=.48\textwidth]{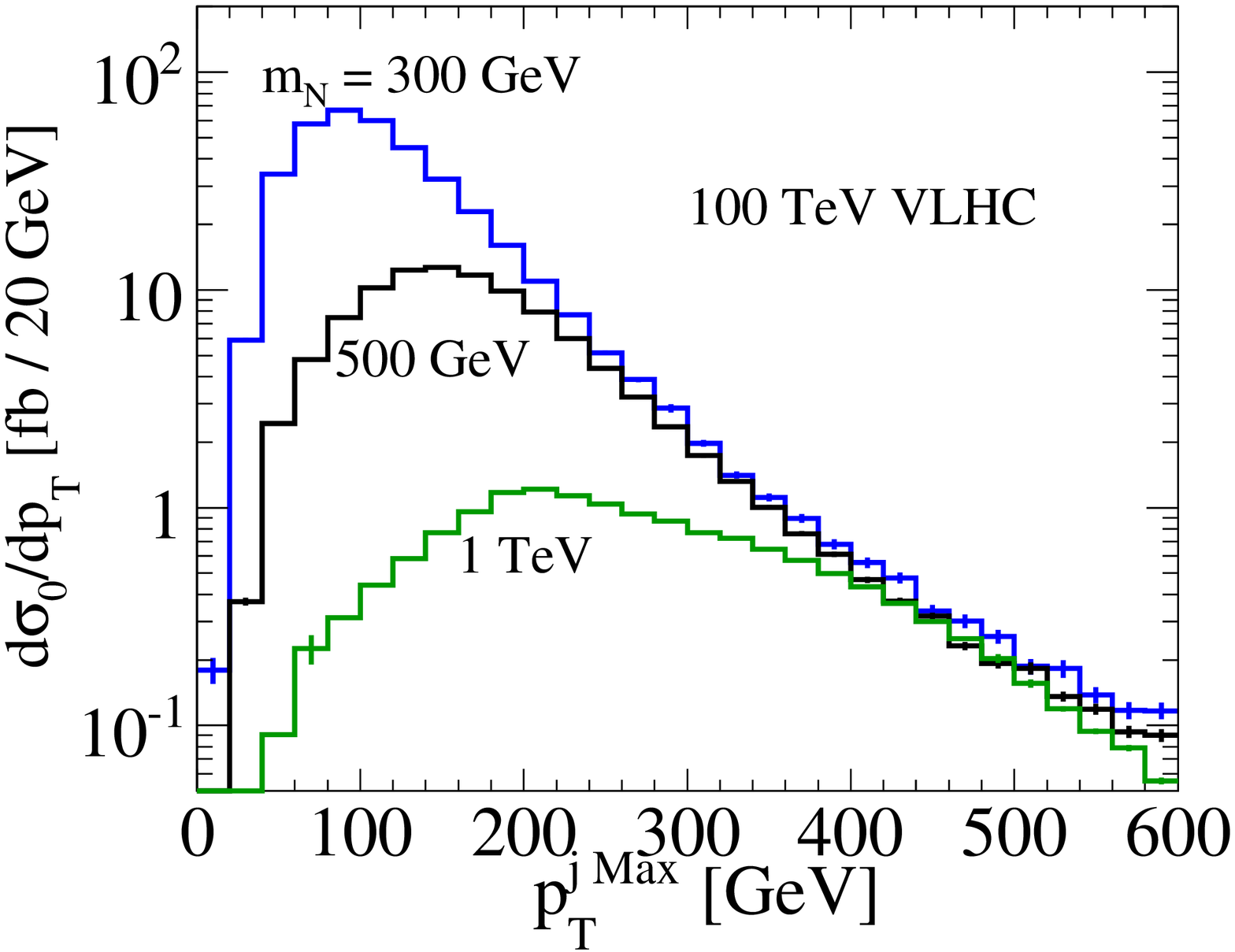}\label{ptjMax_100TeV_MultiGeV.fig}}
\subfigure[]{\includegraphics[scale=1,width=.48\textwidth]{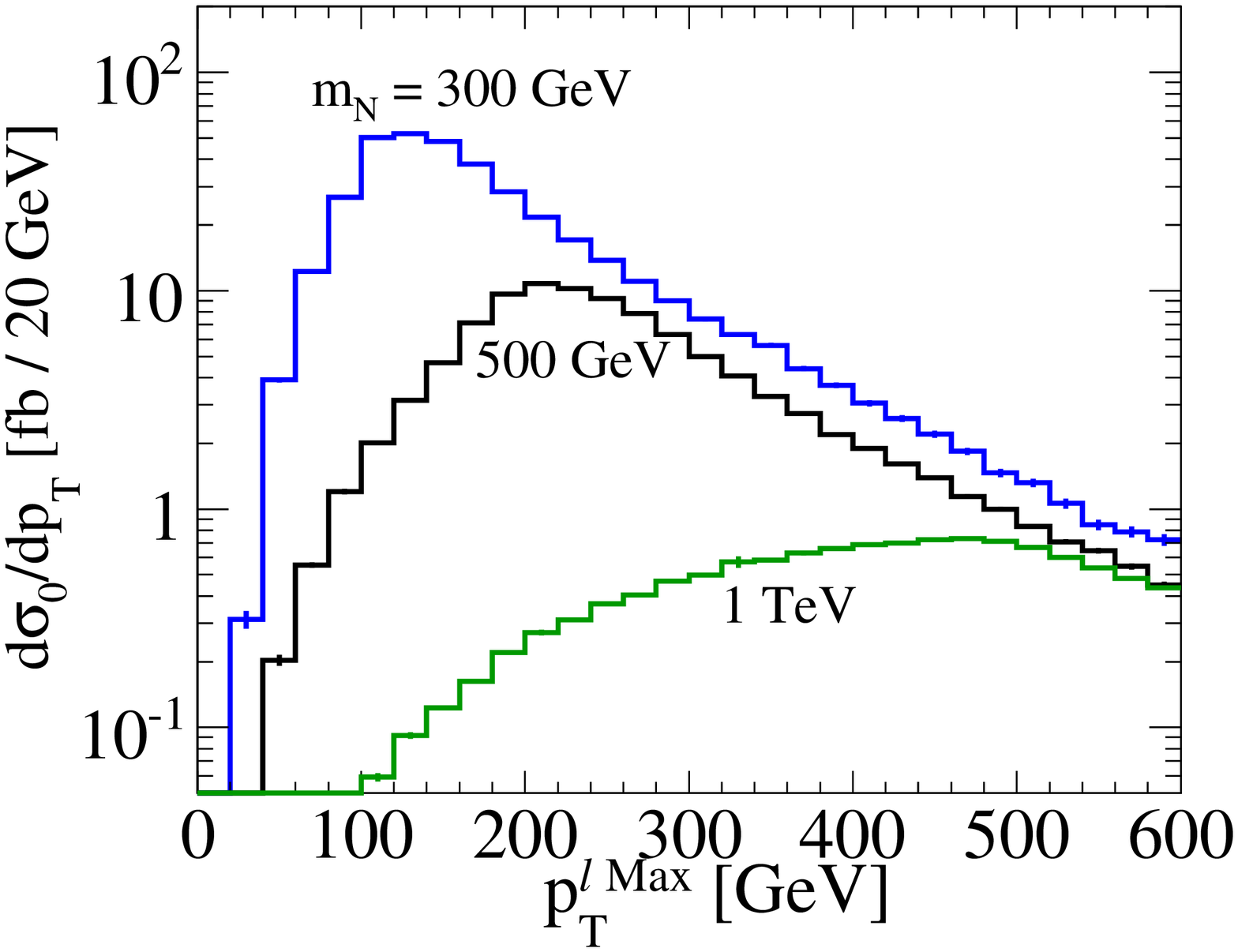}\label{ptlMax_100TeV_MultiGeV.fig}}
\vspace{.2in}\\
\subfigure[]{\includegraphics[scale=1,width=.48\textwidth]{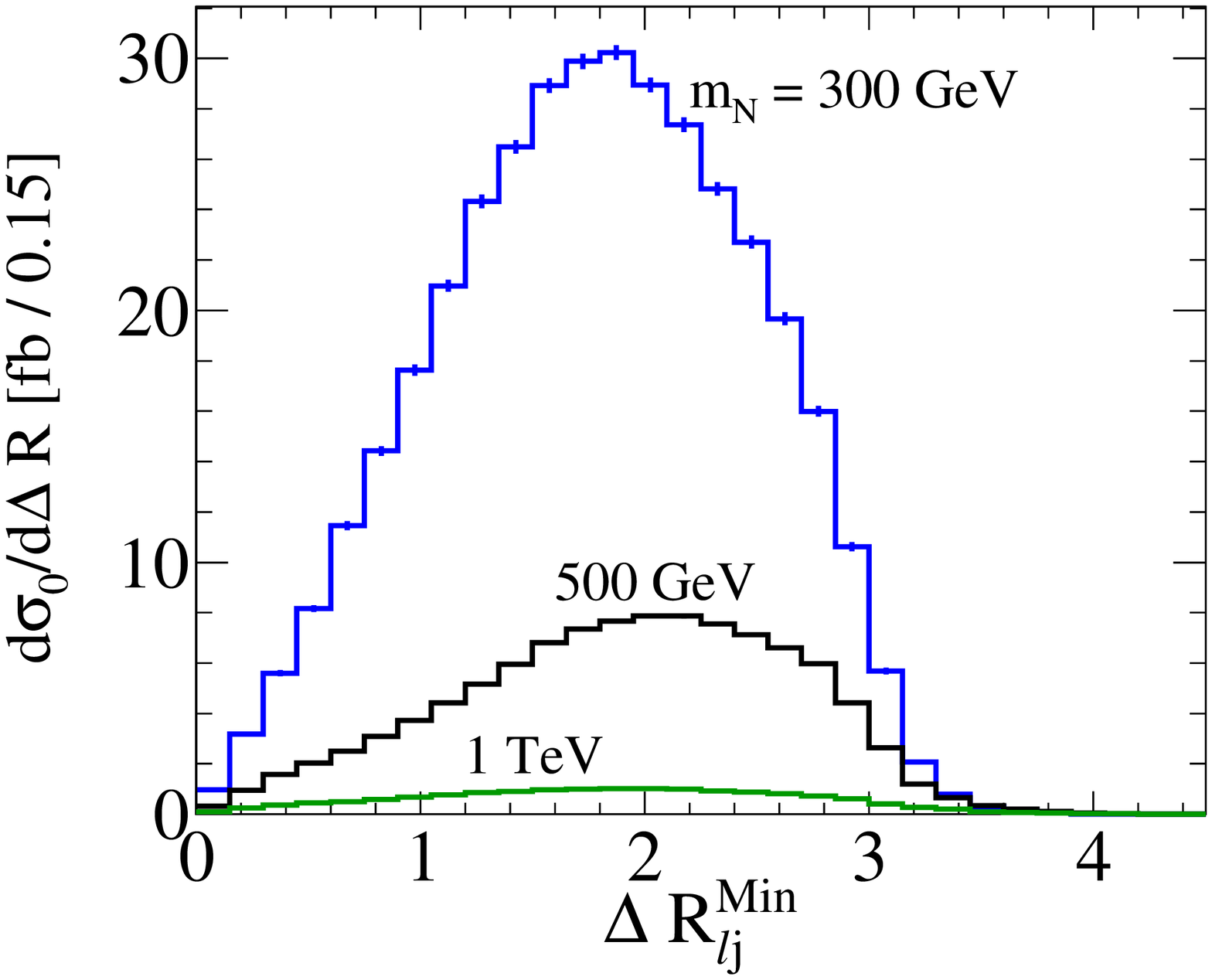}\label{dRljMin_100TeV_MultiGeV.fig}}
\subfigure[]{\includegraphics[scale=1,width=.48\textwidth]{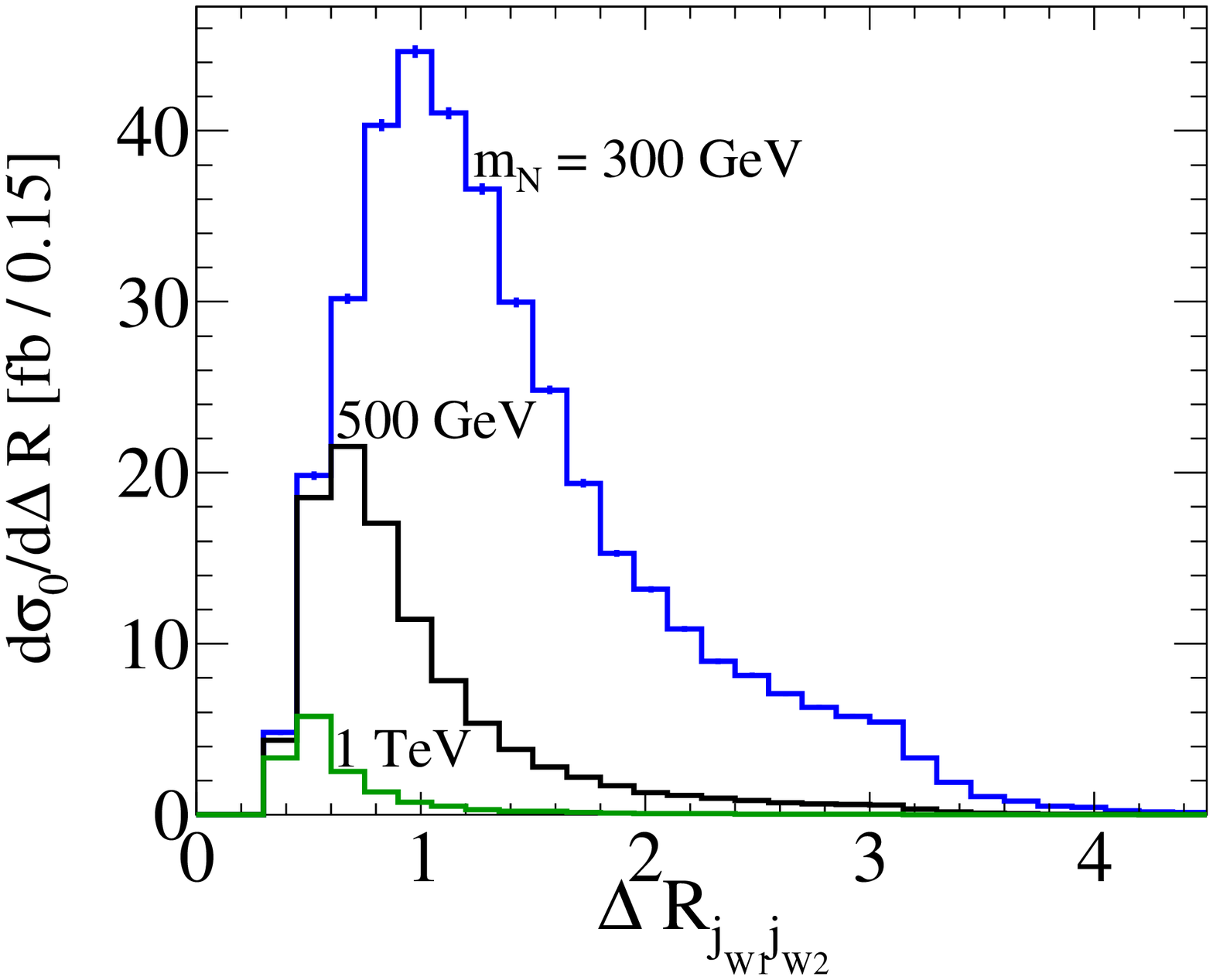}\label{dRjW1jW2_100TeV_MultiGeV.fig}}
\end{center}
\caption[Kinematic distributions of $pp\rightarrow \ell^{\pm}\ell^{'\pm}jjX$ II]{
(a) Maximum jet $p_T$, (b) maximum charged lepton $p_T$, (c) minimum $\Delta R_{\ell j}$, (d) $\Delta R_{j_{W1} j_{W2}}$ distributions for $m_N = 300, \ 500,$ and 1000 GeV.}
\label{maxPt_sep_100TeV.fig}
\end{figure}

We plot the maximum $p_T$ of jets in figure~\ref{ptjMax_100TeV_MultiGeV.fig} and of charged leptons in figure~\ref{ptlMax_100TeV_MultiGeV.fig}, 
for $m_N = 300, ~500,$ and 1000 GeV.
One finds that the $p_T^{j ~\rm Max}$ scale is $m_N/4$ and is set by the $N\rightarrow W \rightarrow jj$ chain.
For the lepton case, $p_T^{\ell ~\rm Max}$ is set by the neutrino decay and scales as $m_N/2$.
In light of these, we apply the following additional selection cuts to reduce background processes:
\begin{equation}
p_{T}^{j~\rm Max} > 40\GeV, \quad p_{T}^{\ell ~\rm Max} > 60\GeV.
\label{leadPTCut.EQ}
\end{equation}
The corresponding rate is given in the second row of Table~\ref{acceptXSec.TB} and we find that virtually all events pass Eq.~(\ref{leadPTCut.EQ}).
As both $p_{T}^{\rm Max}$ are sensitive to $m_N$, searches can be slightly optimized by instead imposing the variable cut 
\begin{equation}
p_{T}^{j~\rm Max} \gtrsim \mathcal{O}\left(\frac{m_N}{4}\right), \quad p_{T}^{\ell ~\rm Max} \gtrsim \mathcal{O}\left(\frac{m_N}{2}\right).
\label{leadPTCutAlt.EQ}
\end{equation}

\begin{figure}[!t]
\begin{center}
\subfigure[]{\includegraphics[scale=1,width=.48\textwidth]{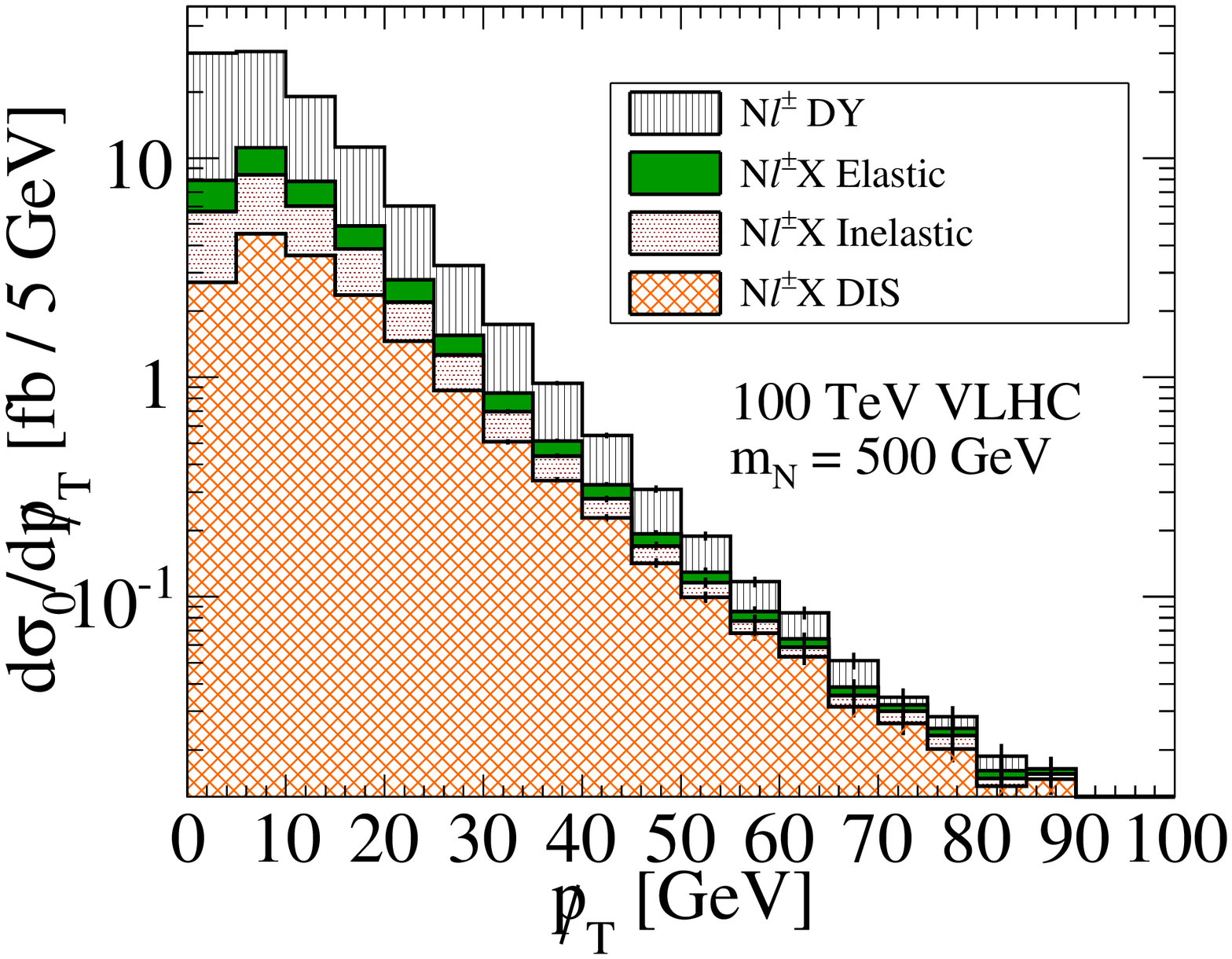}\label{metStacked_100TeV_500GeV.fig}}
\subfigure[]{\includegraphics[scale=1,width=.48\textwidth]{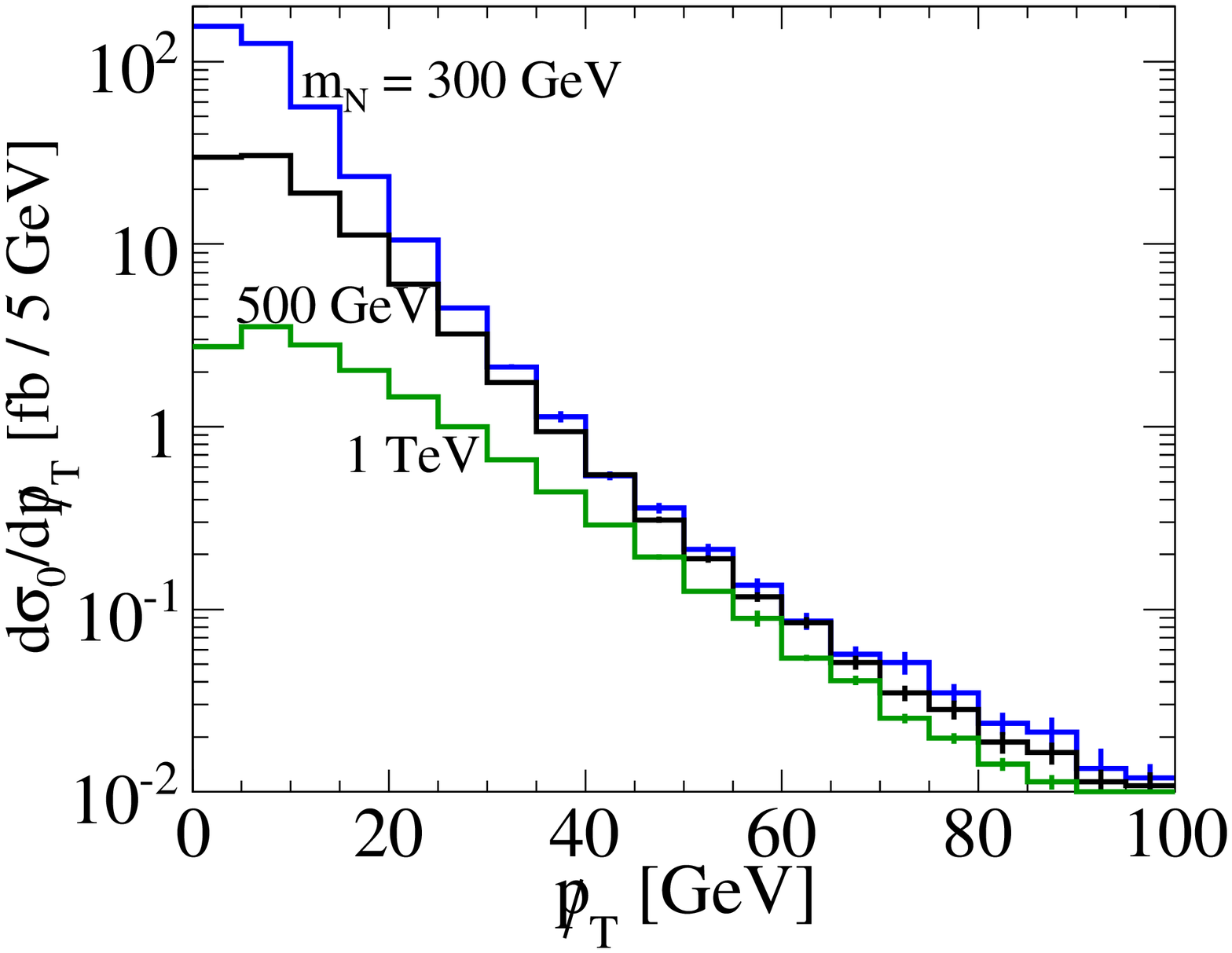}\label{met_100TeV_MultiGeV.fig}}
\end{center}
\caption[Missing transverse momentum in $pp\rightarrow \ell^{\pm}\ell^{'\pm}jjX$]{
(a) $\slashchar{p_T}$  for individual contributions to $pp\rightarrow \ell^{\pm}\ell^{'\pm}jjX$ at $m_N = 500\GeV$.
(b) Total $\slashchar{p_T}$ 
for same $m_N$ as figure~\ref{maxPt_sep_100TeV.fig}.
}
\label{met_100TeV.fig}
\end{figure}

In each of the several production channels, the final-state charged leptons and jets are widely separated in $\Delta R$; 
see figure~\ref{dRljMin_100TeV_MultiGeV.fig}.
With only a marginal effect on the signal rate, we impose the following cut that greatly reduce heavy quarks backgrounds such as $t\overline{t}$ production~\cite{Han:2006ip}:
\begin{equation}
\Delta R_{\ell j}^{\min} > 0.6.
 \label{dRljCut.EQ}
 \end{equation}
The corresponding rate is given in the third row of Table~\ref{acceptXSec.TB}.
If needed, Eq.~(\ref{dRljCut.EQ}) can be set as high as $1.0$ and still maintain a high signal efficiency. 

In figure~\ref{dRjW1jW2_100TeV_MultiGeV.fig}, the separation between the jets in the $N$ decay is shown.
For increasing $m_N$, the separation decreases.
This is the result of the $W$ boson becoming more boosted at larger $m_N$, resulting in more collimated jets.
For TeV-scale $N$, substructure techniques become necessary for optimize event identification and reconstruction.
We reserve studying the inclusive same-sign leptons with at least one (fat) jet for a future analysis.
 
For the signature studied here, no light neutrinos are present in the final state.
For the heavy neutrino widths listed in Eq.~(\ref{widthParam.EQ}), the decay length $\beta c \tau$ is from $10^{-2}$ $-$ 1 fm, indicating that $N$ is very short-lived.
Thus, there is no source of missing transverse momentum (MET) in the same-sign leptons with $(n+2)j$ aside from detector-level mis-measurements, 
which are parameterized by Eqs.~(\ref{jetSmear.EQ})-(\ref{eleSmear.EQ}).
With this smearing parameterization, forward (large $\eta$) jets are observed with less precision than central (small $\eta$) jets.
Due to the naturally larger energies associated with forward jets participating in VBF at 100 TeV, 
the energy-dependent term in Eq.~(\ref{jetSmear.EQ}) provides a potentially large source of momentum mis-measurements in our analysis.
This channel-dependent behavior can be seen in figure~\ref{metStacked_100TeV_500GeV.fig} for $m_N =$500 GeV.
The increase in MET is found to be modest.
In figure~\ref{met_100TeV_MultiGeV.fig}, we plot the combined MET differential distribution for representative $m_N$.
To maximize the contributions to our signal rate, we impose the loose criterion
\begin{equation}
 \slashchar{p_T} < 50\GeV.
 \label{metCut.EQ}
\end{equation}
The corresponding rate is given in the fourth row of Table~\ref{acceptXSec.TB} and show that most events pass.
Though technically difficult, tightening this cut can greatly enhance the signal-to-noise ratio.

\begin{figure}[!t]
\begin{center}
\subfigure[]{\includegraphics[scale=1,width=.48\textwidth]{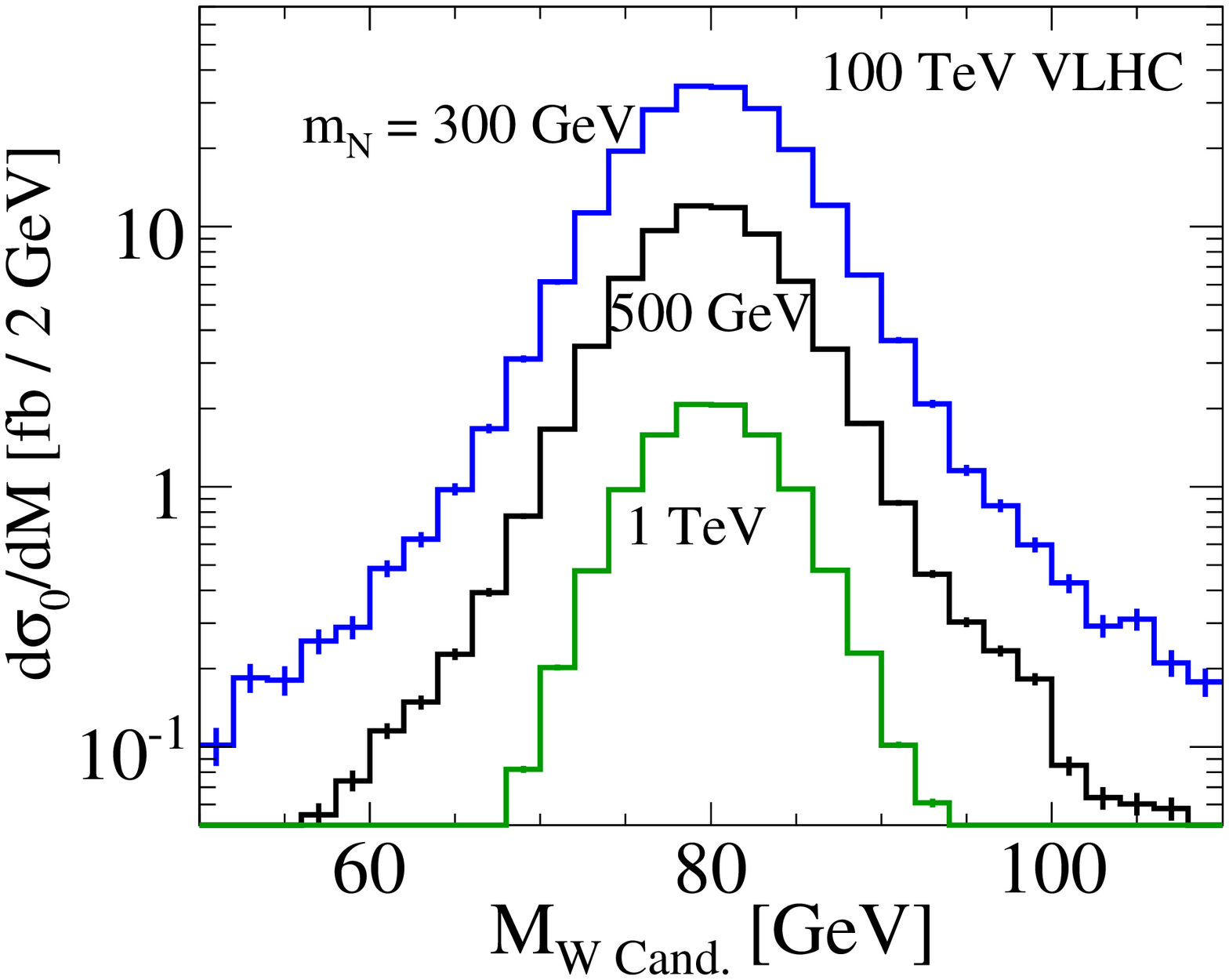}\label{mWReco_100TeV_MultiGeV.fig}}
\subfigure[]{\includegraphics[scale=1,width=.48\textwidth]{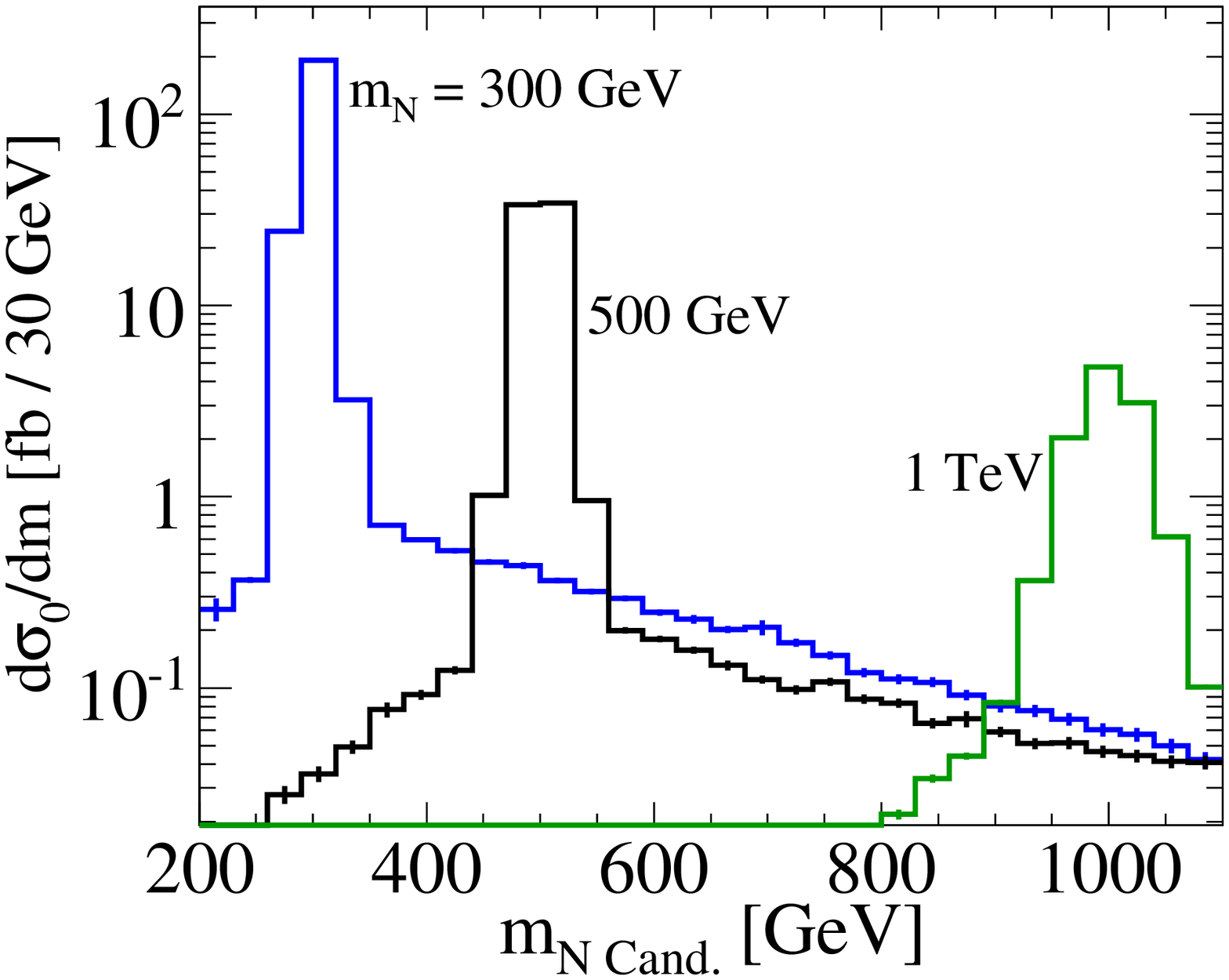}\label{mNReco_100TeV_MultiGeV.fig}}
\end{center}
\caption[Reconstructed invariant mass of $W$ and $N$]{Reconstructed invariant mass of the (a) $W$ boson and (b) heavy $N$ candidates for same $m_N$ as figure~\ref{maxPt_sep_100TeV.fig}.}
\label{mN_100TeV.fig}
\end{figure}	

To identify the heavy neutrino resonance in the complicated $\ell^\pm\ell^\pm+(n+2)j$ topology, 
we exploit that the $N\rightarrow \ell^\pm jj$ decay results in two very energetic jets that remain very central and possess a resonant invariant mass.
In the $4j$ final-state channel, (rare) contributions from $N\ell^\pm W^\mp$ can lead to the existence of a second $W$ boson in our signal.
To avoid identifying a second $W$ (or a continuum distribution) as the $W$ boson from our heavy neutrino decay, we employ the following algorithm: 
(i) First consider all jets satisfying Eq.~(\ref{fidkinsmCut.EQ}) and require that at least one pair possesses an invariant mass close to $M_W$, i.e.,
\begin{equation}
\vert m_{j_m j_n} - M_W \vert < 20\GeV.
 \label{mWCut.EQ}
\end{equation}
(ii) If no such pair has an invariant mass within 20 GeV of $M_W$, then the event is rejected.
(iii) If more than one pair satisfies Eq.~(\ref{mWCut.EQ}), 
including the situation where one jet can satisfy Eq.~(\ref{mWCut.EQ}) with multiple partners, 
we identify the $jj$-system with the highest $p_T$ as the child $W$ boson from the heavy neutrino decay.
This last step is motived by the fact that the $p_{T}$ of neutrino's decay products scale like $p_T \sim m_N/2$, 
and thus at larger values of $m_N$ the $W$ boson will become more boosted.
This is contrary to $N\ell^\pm W^\mp$ and continuum events, 
in which all states are mostly produced close to threshold.
In figure~\ref{mWReco_100TeV_MultiGeV.fig}, we plot the reconstructed invariant mass of the dijet system satisfying this procedure and observe a very clear resonance at $M_W$.
The corresponding rate is given in the fifth row of Table~\ref{acceptXSec.TB} and show most events pass.

To remove background events from $t\overline{t}W$ production, events containing four or more jets with any three jets satisfying 
\begin{equation}
\vert m_{jjj} - m_t \vert < 20\GeV
 \label{mtCut.EQ}
\end{equation}
are rejected. 
As this is a non-resonant distribution in the $N\ell+nj$ channels, its impact on the signal rate is minimal.
The corresponding rate is given in the sixth row of Table~\ref{acceptXSec.TB} and show that nearly all events pass.
A top quark-veto can be further optimized by introducing high-purity anti-$b$-tagging, e.g., Ref.~\cite{Chatrchyan:2012jua}.

We identify $N$ by imposing the $m_N$-dependent requirement on the two $(\ell_i,W_{\rm Cand.})$ pairs 
and choose whichever system possesses an invariant mass closer to $m_N$.
In figure~\ref{mNReco_100TeV_MultiGeV.fig}, we plot the reconstructed invariant mass of this system observing very clear peaks at $m_N$.
It is important to take into account that the width of the heavy neutrino grows like $m_N^3$, and reaches the 10 GeV-level at $m_N = 1\TeV$.
Therefore, we apply the following width-sensitive cut:
\begin{equation}
 \vert m_{N\rm ~Cand.} - m_N \vert < \rm Max(20\GeV,~3\Gamma_N).
 \label{mNCut.EQ}
\end{equation}
The corresponding rate is given in the seventh row of Table~\ref{acceptXSec.TB} and show most events pass.

\begin{table}[!t]
\caption[Expected $\mu^\pm\mu^{\pm}jjX$ (bare) signal and SM background rates at 100 TeV VLHC]{Expected $\mu^\pm\mu^{\pm}jjX$ (bare) signal and SM background rates at 100 TeV VLHC after cuts.
 Number of background events and required signal events for $2\sigma$ sensitivity after 100$\invfb$.}
 \begin{center}
\begin{tabular}{|c|c|c|c|c|c|c|}
\hline\hline  
  $m_N$	[GeV] & $100$	& $200$	& $300$	& $400$	& $500$	&$600$   \tabularnewline\hline\hline
 $\sigma_{0}^{\rm ~All~Cuts}$ [fb]		&205	&588	&244	&118	&64.7	&48.1	\tabularnewline\hline
 $\sigma_{\rm Tot}^{\rm SM}$  [ab]		&16.3	&115	&53.2	&22.2	&11.4	&6.01	\tabularnewline\hline
 $n^{b+\delta_{\rm Sys}}_{2\sigma}(100~\invfb)$	&4	&18	&9	&5	&3	&2	\tabularnewline\hline
 $n^{s}_{2\sigma}(100~\invfb)$			&8	&16	&11	&9	&7	&6	\tabularnewline\hline\hline
  $m_N$	[GeV]	& $700$	& $800$	& $900$	& $1000$	& $1100$	&$1200$   \tabularnewline\hline\hline
 $\sigma_{0}^{\rm ~All~Cuts}$ [fb]		&23.4	&14.4	&10.5	&7.79	&4.61	&4.01	\tabularnewline\hline
 $\sigma_{\rm Tot}^{\rm SM}$  [ab]		&3.47	&1.94	&1.57	&1.25	&0.795	&0.649	\tabularnewline\hline
 $n^{b+\delta_{\rm Sys}}_{2\sigma}(100~\invfb)$	&2	&1	&1	&1	&1	&1	\tabularnewline\hline
 $n^{s}_{2\sigma}(100~\invfb)$			&7	&5	&5	&5	&5	&5	\tabularnewline\hline
 \hline
\end{tabular}
\label{100TeVAnaMuMu.TB}
\end{center}
\end{table}

\begin{table}[!t]
\caption{Same as Table \ref{100TeVAnaMuMu.TB} for $e^\pm\mu^{\pm}jjX$.}
 \begin{center}
\begin{tabular}{|c|c|c|c|c|c|c|}
\hline\hline  
  $m_N$ [GeV]	& $100$	& $200$	& $300$	& $400$	& $500$	&$600$   \tabularnewline\hline\hline
 $\sigma_{0}^{\rm ~All~Cuts}$ [fb]		&408	&1160	&480	&230	&125	&93.2	\tabularnewline\hline
 $\sigma_{\rm Tot}^{\rm SM}$  [ab]		&196	&4000	&578	&82.2	&17.7	&8.20	\tabularnewline\hline
 $n^{b+\delta_{\rm Sys}}_{2\sigma}(100~\invfb)$	&27	&434	&71	&13	&4	&3	\tabularnewline\hline
 $n^{s}_{2\sigma}(100~\invfb)$			&18	&71	&30	&13	&8	&8	\tabularnewline\hline\hline
  $m_N$ [GeV]	& $700$	& $800$	& $900$	& $1000$	& $1100$	&$1200$   \tabularnewline\hline\hline
 $\sigma_{0}^{\rm ~All~Cuts}$ [fb]		&44.9	&27.7	&20.3	&15.1	&8.98	&7.86	\tabularnewline\hline
 $\sigma_{\rm Tot}^{\rm SM}$  [ab]		&4.79	&2.68	&2.07	&1.87	&1.29	&0.932	\tabularnewline\hline
 $n^{b+\delta_{\rm Sys}}_{2\sigma}(100~\invfb)$	&2	&1	&1	&1	&1	&1	\tabularnewline\hline
 $n^{s}_{2\sigma}(100~\invfb)$			&6	&5	&5	&5	&5	&5	\tabularnewline\hline
 \hline
\end{tabular}
\label{100TeVAnaEMu.TB}
\end{center}
\end{table}

The acceptance $\mathcal{A}$ of our signal rate, defined as 
\begin{equation}
 \mathcal{A} = \sigma_{\rm All~Cuts} ~/~ \sigma_{\rm Fidcuial~Cuts + Kinematic~Cuts + Smearing},
\end{equation}
is given in the last row of Table~\ref{acceptXSec.TB}.
The total bare rate for the $\mu\mu$ and $\mu e$ channels at representative values of $m_N$ are given, respectively, 
in the Tables ~\ref{100TeVAnaMuMu.TB} and ~\ref{100TeVAnaEMu.TB}.


\subsection{Background}
\label{sec:background}
Although there are no lepton-number violating processes in the SM, 
there exist rare processes with final-state, same-sign leptons as well as  ``faked'' backgrounds from detector mis-measurement.
Here we describe our estimate of the leading backgrounds to the final-state
\begin{equation}
 p p \rightarrow \ell^\pm \ell^{'\pm} + n\geq2 j + X
 \label{finalState.EQ}
\end{equation}
for the $\mu\mu$ and $e\mu$ channels.
The principle SM processes are $t\overline{t}X$, $W^\pm W^\pm X$, and electron charge misidentification.
We model the parton-level matrix elements of these processes using MG5\_aMC@ NLO~\cite{Alwall:2014hca} and the CTEQ6L PDFs~\cite{Pumplin:2002vw} with
factorization and renormalization scales $Q = \sqrt{\hat{s}}/2.$ We perform the background analysis in the same manner as for the signal-analysis.

\begin{table}[!t]
\caption{Acceptance rates for SM $t\overline{t}$ at 100 TeV $pp$ collider.}
 \begin{center}
\begin{tabular}{|c|c|c|}
\hline\hline
 $\sigma(t\overline{t}W)$ [fb]	& $e\mu $ & $\mu\mu$ \tabularnewline\hline
Fiducial + Kinematics + Smearing [$K = 1.2$]	[Eq.~(\ref{fidkinsmCut.EQ})]	  
									  &20.5 	&10.3~(26\%) \tabularnewline\hline
Leading $p_{T}$ Minimum			[Eq.~(\ref{leadPTCut.EQ})]	  &16.5 	&8.23~(80\%) \tabularnewline\hline
$\Delta R_{\ell j}$ Separation		[Eq.~(\ref{dRljCut.EQ})]    	  &11.8 	&5.91~(72\%) \tabularnewline\hline
$\not\!\! E_{T}$ Maximum		[Eq.~(\ref{metCut.EQ})]   	  &3.58 	&1.78~(30\%) \tabularnewline\hline
$M_{W}$ Reconstruction			[Eq.~(\ref{mWCut.EQ})]	   	  &2.54 	&1.27~(72\%) \tabularnewline\hline
$m_{t}$ Veto				[Eq.~(\ref{mtCut.EQ})] 		  &0.0452 	&0.0213~(2\%) \tabularnewline\hline
\hline
 $\sigma(t\overline{t})$ (Electron Charge Mis-ID) [fb]	&  \multicolumn{2}{|c|}{$e\mu$}  \tabularnewline\hline
Fiducial + Kinematics + Smearing [Eq.~(\ref{fidkinsmCut.EQ})] [$K = 0.96$]
								& \multicolumn{2}{|c|}{94.5 $\times10^{3}$~(21\%)}  	\tabularnewline\hline
Leading $p_{T}$ Minimum		[Eq.~(\ref{leadPTCut.EQ})]	& \multicolumn{2}{|c|}{67.0 $\times10^{3}$~(71\%)} 	\tabularnewline\hline
$\Delta R_{\ell j}$ Separation	[Eq.~(\ref{dRljCut.EQ})]    	& \multicolumn{2}{|c|}{55.2 $\times10^{3}$~(82\%)} 	\tabularnewline\hline
$\not\!\! E_{T}$ Maximum	[Eq.~(\ref{metCut.EQ})]   	& \multicolumn{2}{|c|}{21.4 $\times10^{3}$~(39\%)} 	\tabularnewline\hline
$M_{W}$ Reconstruction		[Eq.~(\ref{mWCut.EQ})]	   	& \multicolumn{2}{|c|}{3.12 $\times10^{3}$~(15\%)} 	\tabularnewline\hline
$m_{t}$ Veto			[Eq.~(\ref{mtCut.EQ})] 		& \multicolumn{2}{|c|}{3.12 $\times10^{3}$~(100\%)}	\tabularnewline\hline
Charge Mis-ID [$\epsilon_{e~ \rm Mis-ID}$]	[Eq.~(\ref{misID.EQ})] 	& \multicolumn{2}{|c|}{10.9 (0.4\%)}     		\tabularnewline\hline
\hline
\end{tabular}
\label{ttBarBkg.TB}
\end{center}
\end{table}

\subsubsection{$t\overline{t}$}
At 100 TeV, radiative EW processes at $\alpha_{\rm s}^{2}\alpha$ such as 
\begin{eqnarray}
 p~p~ \rightarrow ~t ~\overline{t} ~W^\pm 	
    \rightarrow ~b ~\overline{b} ~W^+ ~W^- ~W^\pm \rightarrow ~\ell^\pm ~\ell^{'\pm} ~b ~\overline{b} ~j ~j ~
    \nu_\ell ~\nu_{\ell'},
      \label{ttW.EQ}
\end{eqnarray}
possess non-negligible cross sections. 
At LO, $\sigma(t\overline{t}W \rightarrow \mu^\pm\mu^\pm b\overline{b}jj\nu_\mu\nu_\mu) \approx 40$ fb, and threatens discovery potential.
At 14 TeV, $t\overline{t}W$ possesses a NLO $K$-factor of $K=1.2$~\cite{Campbell:2012dh}. As an estimate, this is applied at 100 TeV.
As shown in Table \ref{ttBarBkg.TB}, the tight acceptance cuts reduce the rate by roughly 75\%.
Unlike the signal process, $t\overline{t}W$ produces two light neutrinos, an inherent source of  MET.
After the MET cut, the background rate is reduced to the 2 fb level. 
Lastly, the decay chain
\begin{equation}
 t ~\rightarrow ~b ~ W~ \rightarrow ~b~j~j
\end{equation}
can be reconstructed into a top quark.
Rejecting any event with a three-jet invariant mass near the top quark mass, i.e., Eq.~(\ref{mtCut.EQ}), 
dramatically reduces this background to the tens of ab level. 
At this point, approximately {0.2\%} of events passing initial selection criteria survive.

At 100 TeV, the NLO $t\overline{t}$ cross section is estimated to be $\sigma(t\overline{t})\approx 1.8\times 10^7$ fb~\cite{Avetisyan:2013onh}.
Hence, rare top quark decays have the potential to spoil our sensitivity, e.g.,
\begin{eqnarray}
 pp ~ \rightarrow ~t ~\overline{t} ~ 
      \rightarrow ~b ~\overline{b} ~ W^+ ~W^- 
      \rightarrow ~b ~\overline{c} ~\ell^{+} ~\ell^{+'} ~\nu_\ell ~\nu_{\ell'} ~W^{-}  + {\rm c.c.},\label{bToc.EQ}
\end{eqnarray}
where a $b$-quark hadronizes into a $B$-meson that then decays semi-leptonically through the $b\rightarrow c\ell \nu_\ell$ subprocess, 
which is proportional to the small mixing $\vert V_{cb}\vert^{2}$.
The MET and $\Delta R_{\ell j}$ cuts render the  rate negligible~\cite{Atre:2009rg}.
Usage of high-purity anti-$b$ tagging techniques~\cite{Chatrchyan:2012jua} can further suppresses this process.
The $b\rightarrow u$ transition offers a similar background but is $|V_{ub}/V_{cb}|^2\sim (0.1)^2$ smaller~\cite{Beringer:1900zz}.


\subsubsection{Electron Charge Misidentification}
An important source of background for the $e^\pm\mu^\pm$ channel is from electron charge misidentification in fully leptonic decays of top quark pairs:
\begin{equation}
p ~p \rightarrow t ~\overline{t} \rightarrow b ~\overline{b} ~W^{+} ~W^{-}\rightarrow b ~\overline{b} ~ e^{\pm} ~\ell^{\mp} ~\nu_e \nu_\ell, \quad \ell = e, ~\mu.
\label{ttBarOS.EQ}
\end{equation}
Such misidentification occurs when an electron undergoes bremsstrahlung in the tracker volume and the associated 
photon converts into an $e^+e^-$ pair. 
If the electron of opposite charge carries a large fraction of the original electron's energy, 
then the oppositely charged electron may be misidentified as the primary electron.
For muons, this effect is negligible due the near absence of photons converting to muons~\cite{Aad:2011vj,Chatrchyan:2011wba}.
At the CMS detector, the electron charge misidentification rate, $\epsilon_{\rm e~ Mis-ID}$, 
has been determined as a function of generator-level $\eta$~\cite{Chatrchyan:2011wba}.
We assume a conservative, uniform rate of 
\begin{equation}
 \epsilon_{e ~\rm Mis-ID} = 3.5\times 10^{-3}.
 \label{misID.EQ}
\end{equation}

To estimate the effect of electron charge mis-ID at 100 TeV, we consider Eq.~(\ref{ttBarOS.EQ}), normalized to NLO.
Other charge mis-ID channels, including $Z+nj$, are coupling/phase space suppressed compared to $t\overline{t}$.
The $t\overline{t}$ rate after selection cuts is recorded  in Table~\ref{ttBarBkg.TB}, and exists at the 100 pb level.
We find that the electron charge mis-ID rate for Eq.~(\ref{ttBarOS.EQ}) can be as large as 11 fb before the $m_{N~\rm Cand}$ cut is applied.
As either electron in the $e^\pm e^\pm$ channel can be tagged, the mis-ID background is the same size as the $e^\pm\mu^\pm$ channel.
Applying the  $m_{N~\rm Cand}$ cut we observe that the background quickly falls off for $m_N\gtrsim 200\GeV$.
As with other backgrounds possessing final-state bottoms, high purity anti-$b$-tagging offers improvements.
We conclude that the effects of charge misidentification are the dominant background in electron-based final states.

\begin{table}[!t]
\caption{Acceptance rates for SM $W^{\pm}W^{\pm}$ at 100 TeV $pp$ collider.}
 \begin{center}
\begin{tabular}{|c|c|c|}
\hline\hline 
 $\sigma(W^{\pm}W^{\pm}+2j)$ [fb]	 & $e\mu $ & $\mu\mu$ \tabularnewline\hline
Fiducial + Kinematics + Smearing 	 [Eq.~(\ref{fidkinsmCut.EQ})]		&11.6	&5.78~(11\%)     \tabularnewline\hline
Leading $p_{T}$ Minimum			[Eq.~(\ref{leadPTCut.EQ})]		&9.45	&4.72~(82\%) \tabularnewline\hline
$\Delta R_{\ell j}$ Separation		[Eq.~(\ref{dRljCut.EQ})]    		&7.46	&3.63~(77\%)  \tabularnewline\hline
$\not\!\! E_{T}$ Maximum		[Eq.~(\ref{metCut.EQ})]   		&2.56	&1.28~(35\%)  \tabularnewline\hline
$M_{W}$ Reconstruction			[Eq.~(\ref{mWCut.EQ})]	  		&0.132	&0.0664~(5\%)	\tabularnewline\hline
$m_{t}$ Veto				[Eq.~(\ref{mtCut.EQ})]			&0.132	&0.0664~(100\%)	\tabularnewline\hline
\hline 
 $\sigma(W^{\pm}W^{\pm}W^{\mp})$ [fb]	&  $e\mu $ & $\mu\mu$ \tabularnewline\hline
Fiducial + Kinematics + Smearing [$K = 1.8$]	[Eq.~(\ref{fidkinsmCut.EQ})]	&3.35 	&1.68~(13\%)   \tabularnewline\hline
Leading $p_{T}$ Minimum			[Eq.~(\ref{leadPTCut.EQ})]			&2.53	&1.26~(75\%) \tabularnewline\hline
$\Delta R_{\ell j}$ Separation		[Eq.~(\ref{dRljCut.EQ})]    			&2.31	&1.11~(87\%)  \tabularnewline\hline
$\not\!\! E_{T}$ Maximum		[Eq.~(\ref{metCut.EQ})]   			&0.754	&0.375~(34\%)  \tabularnewline\hline
$M_{W}$ Reconstruction			[Eq.~(\ref{mWCut.EQ})]	  			&0.735	&0.368~(98\%)	\tabularnewline\hline
$m_{t}$ Veto				[Eq.~(\ref{mtCut.EQ})]				&0.735	&0.368~(100\%)	\tabularnewline\hline
\hline
\end{tabular}
\label{wpwpBkg.TB}
\end{center}
\end{table}


\subsubsection{$W^{\pm}W^{\pm}$}
The QCD and EW processes at orders $\alpha_{\rm s}^2 \alpha^{2}$ and $\alpha^{3}$ , respectively,
\begin{eqnarray}
  p ~p~ &\rightarrow& ~W^\pm   ~W^\pm ~j ~j 	\label{WW2j}\\
  p ~p~ &\rightarrow& ~W^{\pm} ~W^{\pm} ~W^{\mp}	\label{WWW.EQ}
\end{eqnarray}
present a challenging background due to their sizable rates and kinematic similarity to the signal process.
The triboson production rate at NLO in QCD for 14 TeV LHC has been calculated~\cite{Binoth:2008kt}.
As an estimate, we apply the 14 TeV $K$-factor of $K=1.8$ to the 100 TeV LO $W^\pm W^\pm W^\mp$ channel.
After requiring the signal definition criteria, we find the $W^\pm W^\pm$ backgrounds are present at the several fb-level.
Like $t\overline{t}$, the $W^\pm W^\pm X$ final states possess light neutrinos and non-negligible MET. 
Imposing a maximum on the allowed MET further reduces the background by about {35\%}.
As no $W\rightarrow jj$ decay exists in the QCD process, the reconstructed $M_W$ requirement drops the rate considerably.
After the $m_t$ veto, the SM $W^\pm W^\pm X$ rate is {0.4~(0.9) fb} for the $\mu\mu$ ($e\mu$) channel.

\begin{figure}[!t]
\begin{center}
\subfigure[]{\includegraphics[scale=1,width=.45\textwidth]{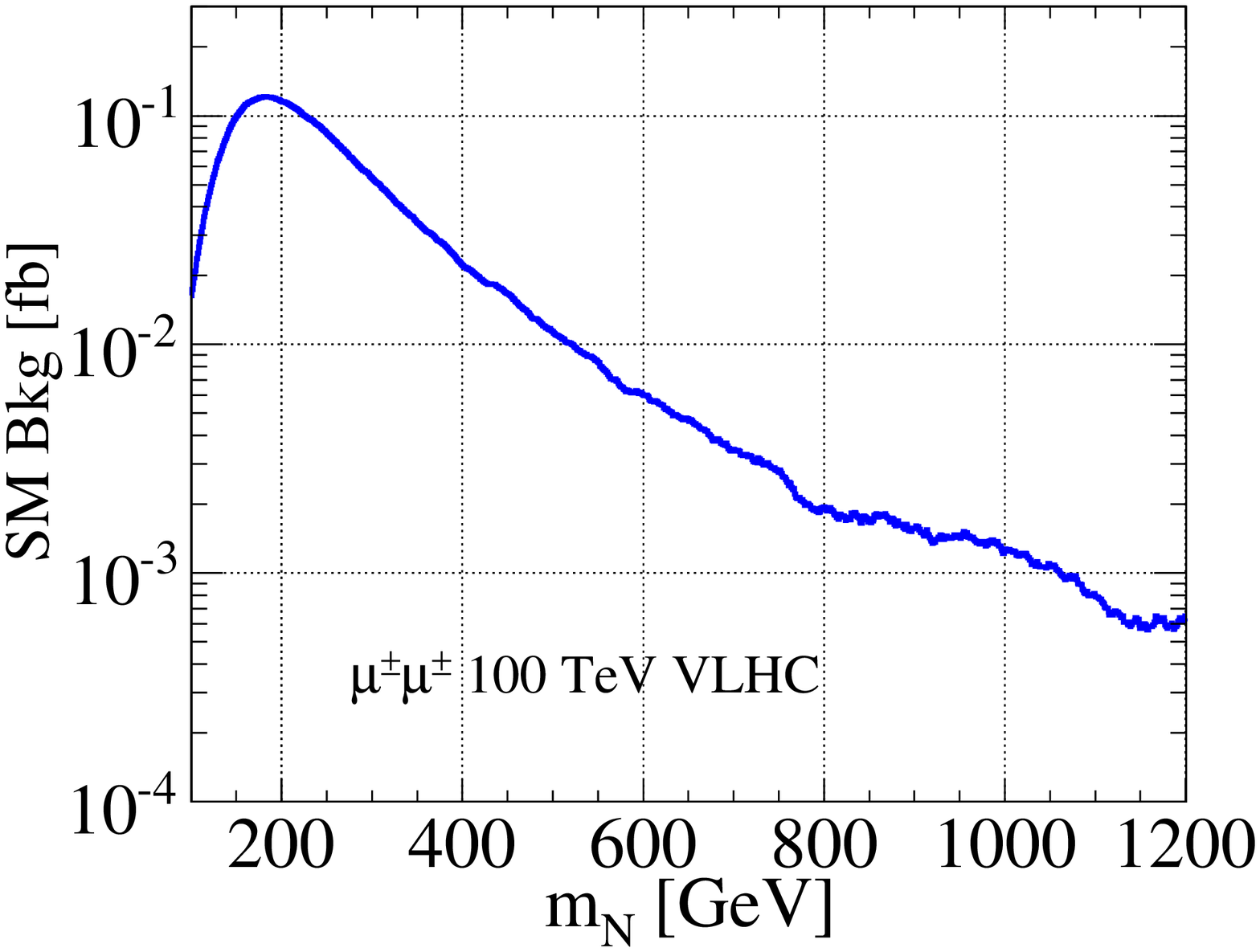}	\label{smBkgMuMu.fig}}
\subfigure[]{\includegraphics[scale=1,width=.45\textwidth]{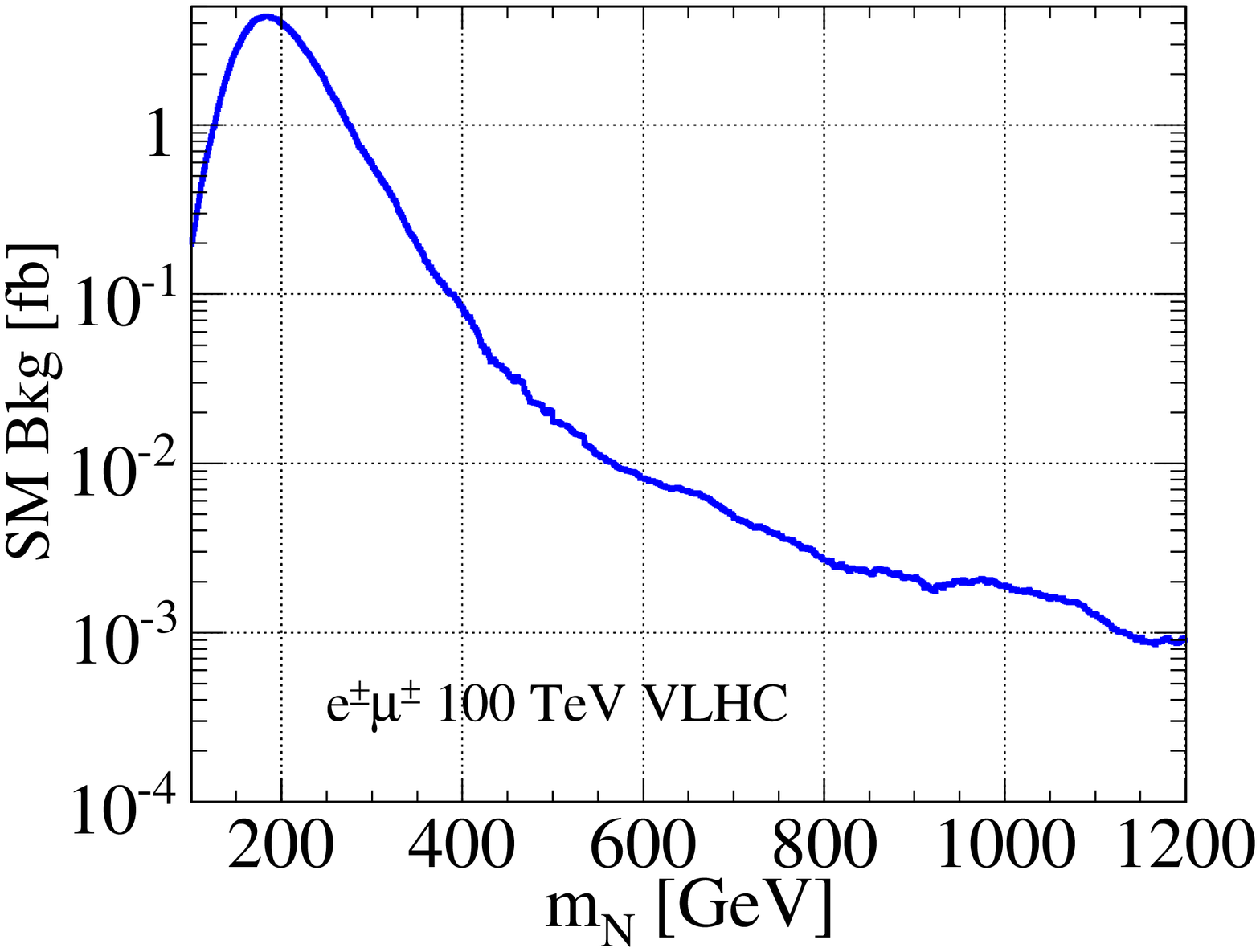}	\label{smBkgEMu.fig}}
\end{center}
\caption{Total SM background versus $m_N$ for (a) $\mu^{\pm}\mu^{\pm}$ and (b) $e^{\pm}\mu^{\pm}$ channels at 100 TeV.}
\label{smBkg.fig}
\end{figure}

For all background channels, 
we apply the $m_N$-dependent cut given in Eq.~(\ref{mNCut.EQ}) on the invariant mass of the reconstructed $W$ candidate with either charged lepton.
The total expected SM background after all selection cuts as a function of $m_N$ are given
for the $\mu\mu$ channel in figure~\ref{smBkgMuMu.fig}, and the $e\mu$ channel in figure~\ref{smBkgEMu.fig}.
The total expected SM background for representative values of $m_N$ are given in Tables~\ref{100TeVAnaMuMu.TB} and ~\ref{100TeVAnaEMu.TB}, respectively.
For these channels, we find a SM background of {$1-115$ ab and $1-4000$ ab} for the neutrino masses considered.
For both channels, the background is greatest for $m_N\lesssim 400\GeV$ and become comparable for $m_N\gtrsim 600\GeV$.

\subsection{Discovery Potential at 100 TeV}
\label{sec:discovery}
\begin{figure}[!t]
\begin{center}
\subfigure[]{\includegraphics[scale=1,width=.48\textwidth]{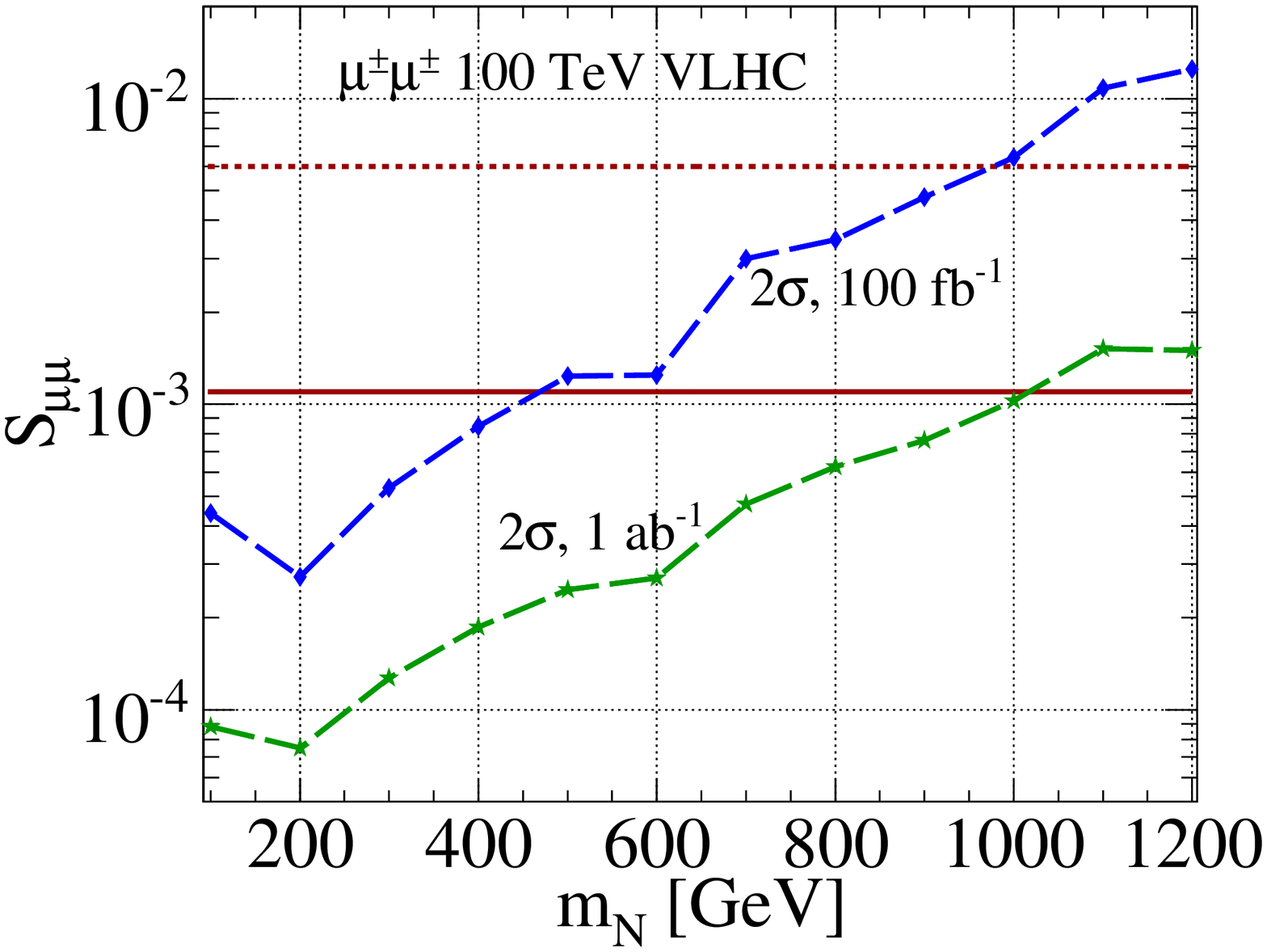}	\label{sMuMuVsMN.fig}}
\subfigure[]{\includegraphics[scale=1,width=.48\textwidth]{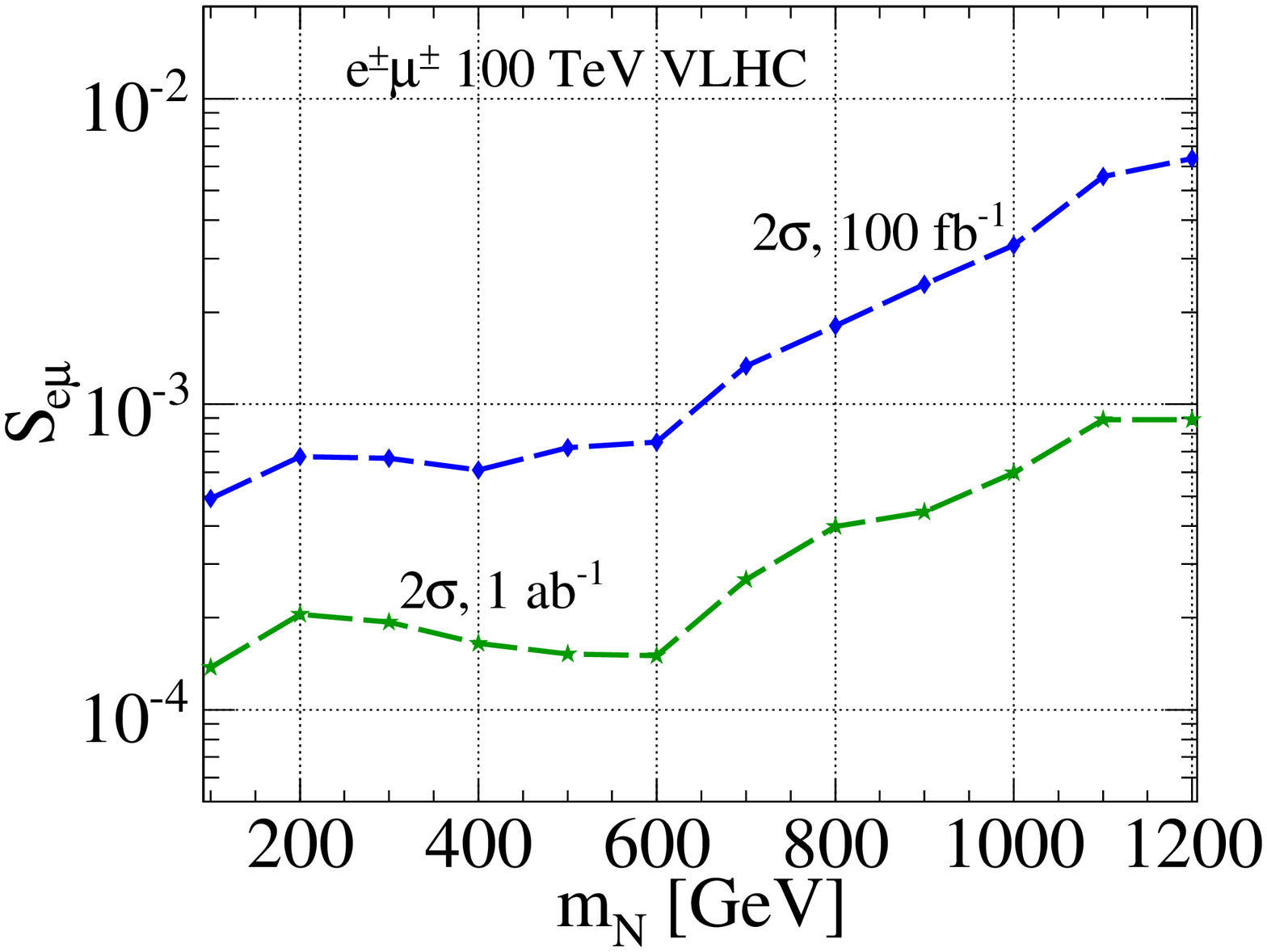}	\label{sEMuVsMN.fig}}
\vspace{.2in}\\
\subfigure[]{\includegraphics[scale=1,width=.48\textwidth]{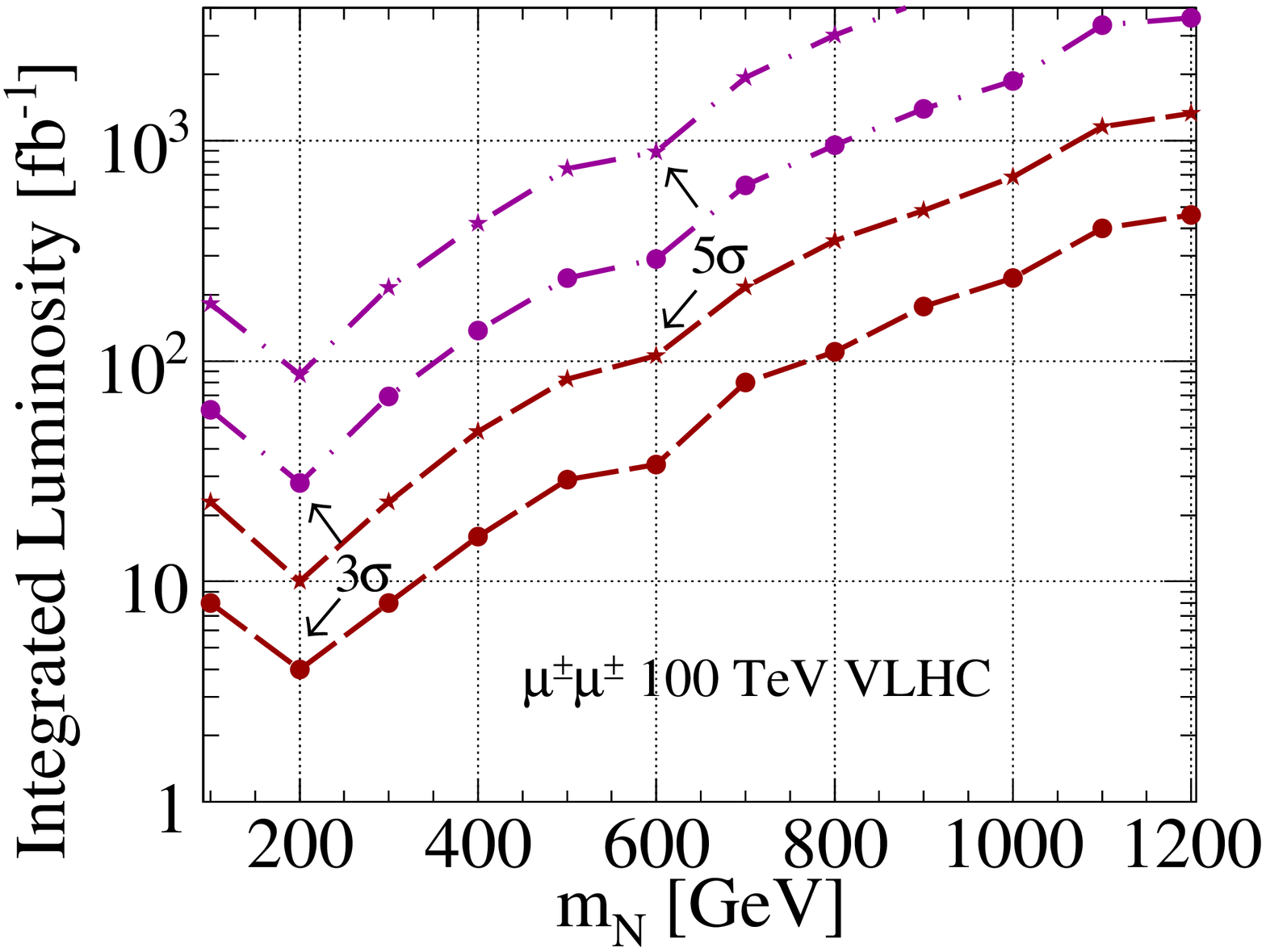}	\label{lumiVsMNMuMu.fig}}
\end{center}
\caption[100 TeV discovery potential of $N$]{
At 100 TeV and as a function of $m_N$, the $2\sigma$ sensitivity to $S_{\ell\ell^{'}}$ after $100\ \invfb$ (dash-diamond) and 1 ab$^{-1}$ (dash-star) 
for the (a) $\mu^{\pm}\mu^{\pm}$ and (b) $e^{\pm}\mu^{\pm}$ channels.
The optimistic (pessimistic) bound is given by the solid (short-dash) horizontal line.
(c) The required luminosity for a 3 (dash-circle) and 5$\sigma$ (dash-star) discovery in the $\mu^{\pm}\mu^{\pm}$ channel}
\label{100TeVdiscovery.fig}
\end{figure}

We now estimate the discovery potential at the 100 TeV VLHC of $L$-violation via same-sign leptons and jets. We quantify this using Poisson statistics. 
Details of our treatment can be found in Section~\ref{sec:stats}.
The total neutrino cross section is related to the total bare cross section by the expression 
\begin{equation}
 \sigma(\ell^\pm\ell^{'\pm}jj+X) = S_{\ell \ell'} ~\times~ \sigma_{0}(\ell^\pm\ell^{'\pm}jj+X).
\end{equation}
We consider two scenarios for $S_{\mu\mu}$, one used by Ref.~\cite{Atre:2009rg}, dubbed the ``optimistic'' scenario,
\begin{equation}
 S_{\mu\mu} = 6\times 10^{-3},
\end{equation}
and the more stringent value obtained in Eq.~(\ref{smumu.EQ}), dubbed the ``pessimistic'' scenario,
\begin{equation}
 S_{\mu\mu} = 1.1\times 10^{-3}.
\end{equation}
For $S_{e\mu}$, we use the $m_N$-dependent quantity obtained in Eq.~(\ref{semu.EQ}), i.e., $10^{-5}-10^{-6}$.
We introduce a $20\%$ systematic uncertainty by making the following scaling to the SM background cross section
\begin{equation}
 \sigma_{\rm SM} \rightarrow \delta_{\rm Sys} \times \sigma_{\rm SM }, \quad  \delta_{\rm Sys}=1.2.
\end{equation}
For the $\mu\mu$ and $e\mu $ channels, respectively, the maximum number of background events
and requisite number of signal events at a 2$\sigma$ significance after 100$\invfb$ are given in Tables ~\ref{100TeVAnaMuMu.TB} and ~\ref{100TeVAnaEMu.TB}.
For the $\mu\mu$ channel, these span $1-18$ background and $5-16$ signal events; for $e\mu$, $1-434$ and $5-71$ events.

We translate this into sensitivity to the mixing parameter $S_{\ell\ell'}$ and
plot the $2\sigma$ contours in $S_{\ell\ell'}-m_N$ space assuming 100 fb$^{-1}$ (dash-diamond) and 1 ab$^{-1}$ (dash-star) for the 
$\mu\mu$ [figure~\ref{sMuMuVsMN.fig}] and $e\mu$ [figure~\ref{sEMuVsMN.fig}] channels.
In the $\mu\mu$ scenario and $m_N = 500\GeV$, a mixing at the level of 
{$S_{\mu\mu}=1.2\times10^{-3}~(2.5\times10^{-4})$} with 100$^{-1}$ (1 ab$^{-1}$) can be probed.
The optimistic (pessimistic) bound is given by the solid (short-dash) horizontal line.
In the $e\mu$ scenario and the same mass, we find sensitivity to {$S_{e\mu}=7.2~(1.5)\times 10^{-4}$}.
For the $e\mu$ channel, the EW+$0\nu\beta\beta$ bound is at the $10^{-6}-10^{-5}$ level.
Sensitivity to $S_{\ell\ell'}$ at 100 TeV is summarized in Table~\ref{mixingReach.TB}.

Comparatively, we observe a slight ``dip'' (broad ``bump'') in the $\mu\mu~(e\mu)$ curve around 200 GeV.
For the $\mu\mu$ channel, this is due to the low signal acceptance rates for Majorana neutrinos very close to the $W$ threshold;
the search methodology for $m_N$ near or below the $M_W$ has been studied elsewhere~\cite{Han:2006ip,Atre:2009rg}.
For $m_N\geq 200\GeV$, the signal acceptance rate grows rapidly, greatly increasing sensitivity.
In the $e\mu$ channel, the electron charge mis-ID background is greatest in the region around 200 GeV and quickly dwindles for larger $m_N$.
In the low-mass regime, we find greater sensitivity in the $\mu\mu$ channel.
However, due to flavor multiplicity and comparable background rates, the $e\mu$ channel has greater sensitivity in the large-$m_N$ regime.

In figure~\ref{lumiVsMNMuMu.fig}, 
we plot as a function of $m_N$ the required luminosity for a $3\sigma$ (circle) and $5\sigma$ (star) discovery in the $\mu\mu$ channel
for the optimistic (red, dash) and pessimistic (purple, dash-dot) mixing scenarios.
With $100\invfb (1\invab)$ and in the optimistic scenario, a Majorana neutrino with {$m_N=580~(1070)\GeV$} can be discovered at $5\sigma$ significance;
with the same integrated luminosity but in the pessimistic scenario, the reach is {$m_N=215~(615)\GeV$}.
In the optimistic (pessimistic) scenario, for a 375 GeV Majorana neutrino, a benchmark used by Ref.~\cite{Atre:2009rg}, 
a $5\sigma$ discovery can be achieved with {$40~(350)\invfb$}; for 500 GeV, this is {$80~(750)\invfb$}.
Sensitivity to $m_N$ at 100 TeV is summarized in Table~\ref{massReach.TB}.

\begin{table}[!t]
\caption{Sensitivity to the mixing parameter $S_{\ell\ell'}$ at the 14 TeV LHC and 100 TeV VLHC}
 \begin{center}
\begin{tabular}{|c|c|c|c|c|}
\hline\hline 
		& $\mathcal{L}$	& $S_{e\mu}(100\TeV) $ & $S_{\mu\mu}(100\TeV) $  	& $S_{\mu\mu}(14\TeV) $ \tabularnewline\hline
\multirow{2}{*}{$2\sigma$} 	& $100\invfb$	& $4.9 \times10^{-4}$	& $2.7 \times10^{-4}$	& $1.4 \times10^{-4}$	\tabularnewline
		& $1\invab$	& $1.4 \times10^{-4}$	& $7.5 \times10^{-5}$	& $3.1 \times10^{-5}$	\tabularnewline\hline
\multirow{2}{*}{$375\GeV$} 	& $100\invfb$	& $6 \times10^{-4}$	& $7.5 \times10^{-4}$	& $3   \times10^{-3}$	\tabularnewline
		& $1\invab$	& $1.7 \times10^{-4}$	& $1.8 \times10^{-4}$	& $5.5 \times10^{-4}$	\tabularnewline\hline
\multirow{2}{*}{$500\GeV$} 	& $100\invfb$	& $7.2 \times10^{-4}$	& $1.2 \times10^{-3}$	& $8   \times10^{-3}$	\tabularnewline
		& $1\invab$	& $1.5 \times10^{-4}$	& $2.5 \times10^{-4}$	& $1.1 \times10^{-3}$	\tabularnewline\hline
\hline
\end{tabular}
\label{mixingReach.TB}
\end{center}
\end{table}

\begin{table}[!t]
\caption{Sensitivity to heavy neutrino production in the $\mu\mu$ channel at 14 and 100 TeV.}
 \begin{center}
\begin{tabular}{|c|c|c|c|c|c|c|}
\hline\hline 
$100\TeV$ & $2\sigma(100\invfb)$  & $5\sigma(100\invfb)$ & $5\sigma(1\invab)$ & $\mathcal{L}_{5\sigma}(375\GeV)$ & $\mathcal{L}_{5\sigma}(500\GeV)$
   \tabularnewline\hline
 Optimistic	& $980 \GeV$		& $580 \GeV$ 		& $1070 \GeV$		& 40$\invfb$	& 80$\invfb$ \tabularnewline\hline
 Pessimistic	& $470 \GeV$		& $215 \GeV$		& $615 \GeV$		& 380$\invfb$	& 750$\invfb$ \tabularnewline\hline
 \hline
$14\TeV$ & $2\sigma(100\invfb)$ & $5\sigma(100\invfb)$ & $5\sigma(1\invab)$ & $\mathcal{L}_{5\sigma}(375\GeV)$ & $\mathcal{L}_{5\sigma}(500\GeV)$
   \tabularnewline\hline
 Optimistic	& $465 \GeV$ 		& $270\GeV$ 		& $530\GeV$		& 300$\invfb$	& 810$\invfb$  \tabularnewline\hline
 Pessimistic	& $255 \GeV$		& $135\GeV$		& $280\GeV$		& 2.6$\invab$	& 6.9$\invab$ \tabularnewline\hline
\hline
\end{tabular}
\label{massReach.TB}
\end{center}
\end{table}

\subsection{Updated Discovery Potential at 14 TeV LHC}
\label{sec:14TeVLHC}
\begin{table}[!t]
\caption{Parton-level cuts on 14 TeV $\mu^\pm\mu^\pm jjX$ Analysis}
 \begin{center}
\begin{tabular}{|c|c|c|}
\hline\hline
Lepton Cuts & Jet Cuts & Other Cuts \tabularnewline\hline\hline
 $\Delta R_{\ell\ell}>0.2$	&$\Delta R_{jj}>0.4$	& $\Delta R_{\ell j}^{\rm Min} > 0.5$	\tabularnewline
 $p_T^\ell ~(p_{T}^{\ell ~\rm Max})> 10~(30)\GeV$ 	& $p_T^j ~(p_{T}^{j~\rm Max})> 15~(40)\GeV$ 	& $\not\!\! E_T < 35\GeV$ \tabularnewline
$\vert \eta^\ell\vert<2.4$ 	& $\vert\eta^j\vert < 2.4$	 	& $\vert m_{N}^{\rm Candidate} - m_N \vert < \rm 20\GeV$ \tabularnewline
				&$\vert M_{W}^{\rm Candidate} - M_W \vert < 20\GeV$ & \tabularnewline	
				&$\vert m_{jjj} - m_t \vert < 20\GeV$ (Veto) &\tabularnewline\hline	
\hline
\end{tabular}
\label{14TeVCuts.TB}
\end{center}
\end{table}

\begin{table}[!t]
\caption{Same as Table \ref{100TeVAnaMuMu.TB} for 14 TeV LHC.}
 \begin{center}
\begin{tabular}{|c|c|c|c|c|c|c|c|}
\hline\hline  
 $\sigma$	$\backslash$ $m_N$ [GeV] & $100$	& $200$	& $300$	& $400$	& $500$	&$600$ &$700$  \tabularnewline\hline\hline
 $\sigma_{0}^{\rm ~All~Cuts}$ [fb]	&576	&132	&36.0	&14.0	&6.28	&3.05  &1.55 \tabularnewline\hline
 $\sigma_{\rm Tot}^{\rm SM}$  [ab]	&14.1	&18.6	&5.62	&2.05	&0.837	&0.393	&0.195 \tabularnewline\hline
 $n^{b+\delta_{\rm Sys}}_{2\sigma}(100~\invfb)$	&4	&4	&2	&1	&1	&0	&0  \tabularnewline\hline
 $n^{s}_{2\sigma}(100~\invfb)$			&8	&8	&6	&5	&5	&4	&4  \tabularnewline\hline
 \hline
\end{tabular}
\label{14TeVAna.TB}
\end{center}
\end{table}

We update the 14 TeV LHC discovery potential to heavy Majorana neutrinos above the $W$ boson threshold decaying to same-sign muons.
Our procedure largely follows the 100 TeV scenario but numerical values are based on Ref.~\cite{Atre:2009rg}.
Signal-wise, we require exactly two same-sign muons (vetoing additional leptons) 
and at least two jets (allowing additional jets) satisfying the cuts listed in Table~\ref{14TeVCuts.TB}.
Differences from the analysis introduced by Ref.~\cite{Atre:2009rg} include:
updated smearing parameterization given in Eqs.~(\ref{jetSmear.EQ}) and (\ref{muSmear.EQ});
an $\not\!\! E_T$ requirement based on the ATLAS detector capabilities given in Ref.~\cite{ATLAS:2012yoa};
cuts on the leading charged lepton and jet; and more stringent requirements on the $W$ and $N$ candidate masses.
These differences sacrifice sensitivity to $m_N\lesssim 100\GeV$ for high-mass reach.
For our NNLO in QCD $K$-factor, we use $K=1.2$, as given in Eq.~(\ref{K.EQ}).
We report the bare heavy neutrino rate after all cuts for representative $m_N$ in the first row of Table~\ref{14TeVAna.TB}.
The total bare rate ranges from {$2-580$} fb for $m_N = 100-700\GeV$.

\begin{figure}[!t]
\begin{center}
\subfigure[]{\includegraphics[scale=1,width=.48\textwidth]{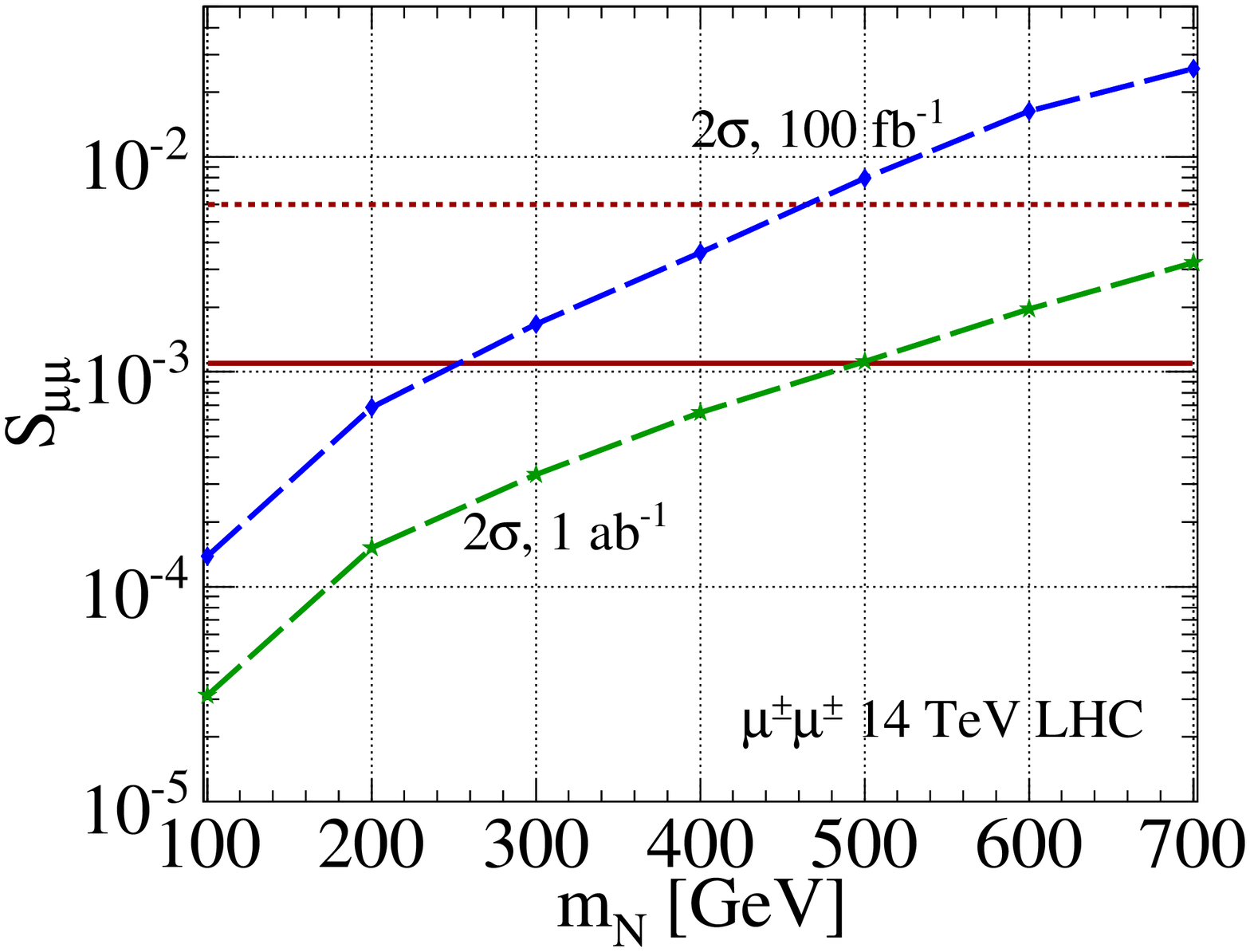}	\label{sMuMuVsMN14TeV.fig}}
\subfigure[]{\includegraphics[scale=1,width=.48\textwidth]{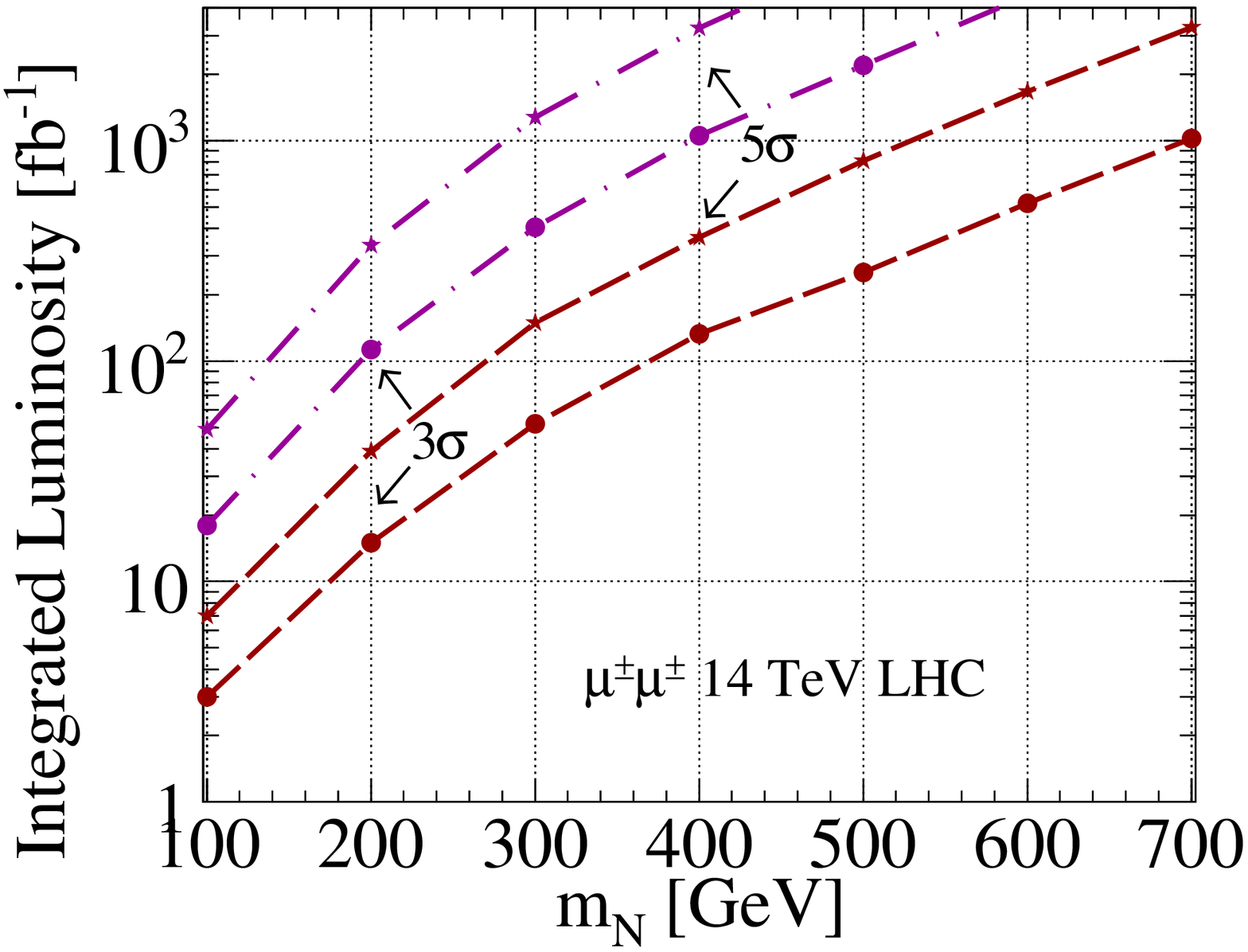}\label{lumiVsMNMuMu14TeV.fig}}
\end{center}
\caption{
At 14 TeV, (a) same as figure~\ref{sMuMuVsMN.fig}; (b) same as figure~\ref{lumiVsMNMuMu.fig}.
}
\label{mumu14TeV.fig}
\end{figure}

As previously discussed or shown, the $t\overline{t}$ background for the dimuon channel is negligible, so we focus on $W^\pm W^\pm$ pairs.
For triboson production, an NLO in QCD $K$ factor of $K=1.8$ is applied~\cite{Binoth:2008kt}.
After all cuts, the expected SM background for representative $m_N$ is given in the second row Table~\ref{14TeVAna.TB}.
After the $m_N$-dependent cut, the expected SM background rate reaches at most {19} ab.
Like the 100 TeV case, a 20\% systematic is introduced into the background.
For the $\mu\mu$ and $e\mu $ channels, respectively, 
The maximum number of background events and requisite number of signal events at a 2$\sigma$ significance after 100$\invfb$ are 
given in the third and fourth rows, respectively, of Table~\ref{14TeVAna.TB}.

%

In figure~\ref{sMuMuVsMN14TeV.fig}, we plot the $2\sigma$ sensitivity to the mixing coefficient $S_{\mu\mu}$ after $100\invfb$ (dash-diamond) 
and 1 ab$^{-1}$ (dash-star).
For the benchmark $m_N = 375\GeV$,
a mixing at the level of {$S_{\mu\mu}=3\times 10^{-3}~(5.5\times10^{-4})$} with 100$^{-1}$ (1 ab$^{-1}$) can be probed;
for $m_N = 500\GeV$, we find sensitivity to be {$S_{\mu\mu}=8\times10^{-3}~(1.1\times10^{-4})$}.
The optimistic (pessimistic) bound is given by the solid (short-dash) horizontal line.
Sensitivity to $S_{\mu\mu}$ at 14 TeV is summarized in Table~\ref{mixingReach.TB}.

In figure~\ref{lumiVsMNMuMu14TeV.fig}, 
we plot as a function of $m_N$ the required luminosity for a $3\sigma$ (circle) and $5\sigma$ (star) discovery in the $\mu\mu$ channel
for the optimistic (red, dash) and pessimistic (purple, dash-dot) mixing scenarios.
With $100\invfb~(1\invab)$ and in the optimistic scenario,
a Majorana neutrino with {$m_N=270~(530)$ GeV} can be discovered at $5\sigma$ significance;
in the pessimistic scenario, the reach is {$m_N=135~(280)$ GeV}. 
In the optimistic (pessimistic) scenario, for the 375 GeV benchmark, a $5\sigma$ discovery can be achieved with {$300~(2600)\invfb$};
for 500 GeV, this is {$810~(6900)\invfb$}.
Sensitivity to $m_N$ at 14 TeV is summarized in Table~\ref{massReach.TB}.


\section{Summary}
\label{sec:summary}
The search for a heavy Majorana neutrino at the LHC is of fundamental importance. 
It is complimentary to the neutrino oscillation programs and, in particular, neutrinoless double-beta decay experiments. 
We have studied the production of a heavy Majorana neutrino at hadron colliders and its lepton-number violating decay as in Eq.~(\ref{ppllnj.EQ}), 
including the NNLO DY contribution, the elastic and inelastic $p\gamma\rightarrow N\ell j$ processes,  and the DIS $pp\rightarrow N\ell jj$ process via $W\gamma^*$ fusion. 
We have determined the discovery potential of the same-sign dilepton signal at a future $100\TeV\,pp$ collider, and updated the results at the 14 TeV LHC.
We summarize our findings as follows:
\begin{itemize}

\item 
Vector boson fusion processes,e.g., $W\gamma \to N \ell$, become increasingly more important at higher collider energies and larger mass scales due to collinear logarithmic enhancements of the cross section. 
At the 14 TeV LHC, the three contributing channels of elastic, inelastic and DIS are comparable in magnitude, 
while at the 100 TeV VLHC, the tendency, in descending importance, is DIS, inelastic, and elastic; see figures~\ref{xsecComb.fig} and ~\ref{xsecComb100TeV.fig}.

 \item We approximately computed the QCD corrections up to NNLO of the DY production of $N\ell$  to obtain the $K$-factor. 
   We found it to span {$1.2-1.5$} for $m_N$ values between  $100\GeV$ and  $1\TeV$ at $14$ and $100~\TeV~pp$ collisions,
   and is summarized in Table~\ref{kFactor.TB}.

   \item The $W\gamma$ fusion processes surpasses the DY mechanism at {$m_{N} \sim 1\TeV \ (770\GeV)$} at the 14 TeV LHC (100 TeV VLHC);
   see figure~\ref{xsecRatio.fig} [\ref{xsecRatio100TeV.fig}].  However, we disagree with the results of Refs.~\cite{Dev:2013wba}, 
 where higher order contributions dominating over the LO DY production at $m_N \geq 200\GeV$ were claimed.
 The discrepancy is attributed to their too low a $p_T^j$ cut that overestimates the contribution of initial-state radiation based on a tree-level calculation.

 \item  We have introduced a systematic treatment for combining initial-state photons from various channels and predict cross sections that are rather stable against the scale choices, typically less than $20\%$.  The exception is the inelastic process, which is rather sensitive to the scale $\lamEl$ where the elastic and inelastic processes are separated. Variation of this scale could lead to about a $30\%$ uncertainty.
  Scale dependence is shown in figure~\ref{scale.fig} and the results summarized in Table~\ref{scale.TB}.
 
\item
We quantified the signal observability by examining the SM backgrounds. 
We conclude that, with the currently allowed mixing {$\vert V_{\mu N}\vert ^2<6\times 10^{-3}$}, 
a $5\sigma$ discovery can be made via the same-sign dimuon channel for {$m_N = 530~(1070)$} GeV at the 14 TeV LHC (100 TeV VLHC) after 1 ab$^{-1}$;
see Table~\ref{massReach.TB}.
Reversely, for $m_N = 500$ GeV and the same integrated luminosity, 
a mixing $\vert V_{\mu N}\vert^2$ of the order {$1.1\times10^{-3}~(2.5\times10^{-4})$} may be probed; see Table~\ref{mixingReach.TB}.
This study represents the first investigation into heavy Majorana neutrino production in 100 TeV $pp$ collisions.
\end{itemize}

%% file: 05_WprimeHeavyN/wprimeHeavyN.tex
\chapter[Lepton Number Violation and $W'$ Chiral Couplings]{Lepton Number Violation and $W'$ Chiral Couplings at the Large Hadron Collider}

\section{Introduction}

Neutrino experiments, over the past decade, have shown undeniably that neutrinos are massive and have large mixing angles
~\cite{Mohapatra:1998rq,Gluza:2002vs,Fukugita:2003en,Barger:2003qi,Eidelman:2004wy,GonzalezGarcia:2007ib,Mohapatra:2006gs,Strumia:2006db}
In the  Standard Model (SM) of particle physics, neutrino masses can be accommodated by a non-renormalizable dimension-5 operator containing left-handed (L.H.) neutrinos, $\nu_{L}$ \cite{dim6}.  Such an operator can be generated at low energy by including heavy right-handed (R.H.) neutrinos, $\nu_{R}$. 
However, the R.H.~neutrinos are gauge singlets and so Majorana mass terms should also be present without violating any gauge symmetry.
The consequences of massive Majorana neutrinos are well-known~\cite{seesaw,TypeIII,Inverse}, 
and have been incorporated into many models, such as left-right symmetric theories~\cite{LRModels}; 
supersymmetric (SUSY) $SO(10)$ grand unified theories (GUTs)~\cite{SO10SUSYGUT} and other GUTs~\cite{MGUT}; 
R-parity violating SUSY~\cite{RparitySUSY}; and extra dimensions~\cite{ExtraDim}.  
A recent review of TeV scale neutrino mass models can be found in Ref.~\cite{Chen:2011de}.

Many of the aforementioned models contain an extended gauge group or Keluza-Klein (KK) excitations of SM gauge bosons.
We refer to additional vector bosons charged under the $U(1)_{EM}$ gauge group collectively as ``$\wpri$''.
If the masses of the $\wpri$ and the lightest heavy neutrino mass eigenstate, $N$, 
are both on the order of a few TeV, then they can be produced in tandem at the Large Hadron Collider (LHC).
As first observed by Ref.~\cite{Keung:1983uu}, 
a $\wpri$ with mass greater than a Majorana neutrino's mass allows the possibility of observing the spectacular lepton number ($L$) violating process 
\begin{equation}
 pp\rightarrow \wpri\rightarrow \ell^\pm N\rightarrow \ell^\pm\ell^\pm jj.
\label{eq:WprN}
\end{equation}

If a $\wpri$ is discovered at the LHC~\cite{schmaltz:2011lrsm}, it is obviously imperative to measure its chiral coupling to fermions.
In a previous work~\cite{Gopalakrishna:2010}, three of the present authors proposed measuring the $W'$ chiral couplings to quarks by studying the process
\begin{equation}
pp\rightarrow\wpri\rightarrow t\bar{b}\rightarrow \ell^+\nu_\ell b\bar{b}.
\end{equation}
It was found that the couplings could be establish as being purely left- or purely right-handed by analyzing the 
polar angle of the charged lepton in the top's rest-frame with respect to the top's direction of motion in the partonic center of momentum (c.m.) frame.

We now extend this prior analysis into the leptonic sector via the $L$-violating cascade decay of Eq.~(\ref{eq:WprN}).
More specifically, by reconstructing the polar angle of the lepton originating from the neutrino decay in the 
neutrino rest-frame and with respect to the direction of motion of the neutrino in the partonic c.m.~frame, 
it can be uniquely determined if the $\wpri$ coupling to leptons is purely left-handed, purely right-handed, 
or a mixture of the two.
We show that the distribution of the angle made between $N$'s production plane and its sequential decay 
plane is sensitive to the $\wpri$ chiral coupling with the initial-state quarks but $independent$ of the $\wpri$ coupling to leptons.  
These results are demonstrated through a combination of analytical calculations and event simulations, assuming nominal LHC parameters.

Majorana neutrinos can decay into either leptons or antileptons, and so $\wpri$ and $N$ may also contribute to the $L$-conserving collider signature
\begin{equation}
 pp\rightarrow \wpri\rightarrow \ell^\pm N\rightarrow \ell^+\ell^- jj.
\label{eq:WprNpm}
\end{equation}
For completeness, we have analyzed the polar angular distributions of the unlike-sign process and comment on the important differences between the $L$-conserving and $L$-violating cases.

This paper is structured as follows:  
First, in section~\ref{ThFrame.SEC}, we present our notation for the $\wpri$ couplings to SM particles and neutrino mass eigenstates, and list current constraints on both $\wpri$'s and $N$'s.
In section~\ref{WpLHC.SEC}, we discuss the production and decay of $\wpri$'s and $N$'s at the LHC.
The like-sign lepton signature, $pp\rightarrow\ell^\pm\ell^\pm jj$, its reconstruction, and suppressed background are fully analyzed in section~\ref{Like.SEC}.
In \ref{WpriAC.SEC}, we propose methods to measure independently the chiral couplings of the $\wpri$ to leptons and to the initial-state quarks.
Finally, in section~\ref{Opp.SEC}, we provide a few comments on the contribution of $\wpri$ and $N$ to the $L$-conserving process $pp\rightarrow \wpri\rightarrow \ell^+\ell^- jj$ regarding the difference between the Majorana and Dirac neutrinos. 
We conclude and summarize our results in section~\ref{Conc.SEC}.
Two appendices are additionally included. 
The first addresses neutrino mass mixing in the context of $W'$ couplings, and the second presents a derivation of the matrix element and angular distributions for our like- and unlike-sign dilepton signals.

\section{Theoretical Framework and Current Constraints}
\label{ThFrame.SEC}
There are many Beyond the Standard Model (BSM) theories containing additional  
vector bosons that couple to SM fermions,
for example: left-right symmetric theories~\cite{Pati:1974yy} with a new $SU(2)_R$ symmetry and an associated $W'_R$; 
Little Higgs models with enlarged gauge symmetries~\cite{ArkaniHamed:2002qy}; 
extra dimensional theories with KK excitations~\cite{Appelquist:2000nn,Csaki:2003zu,Agashe:2003zs}.  
Heavy Majorana neutrinos in BSM theories~\cite{LRModels,SO10SUSYGUT,MGUT,RparitySUSY,ExtraDim}, 
and in particular those with TeV-scale masses~\cite{Dorsner:2006fx,Bajc:2007zf,deGouvea:2006gz,de Gouvea:2007uz}, are just as common.

In this analysis, we assume the existence of a new heavy electrically charged vector boson, $W^{'\pm}$ with mass $M_{W'}$, and a right-handed neutrino, $N_{R}$.
We denote the corresponding heavy neutrino mass eigenstate as $N$ with mass $m_{N}$.
We stipulate that $M_{W'}$ is of the order of a few TeV and  $M_{W'}>m_{N}$ 
 so as the $W'\rightarrow N\ell$ decay is kinematically accessible by the LHC, but do not otherwise tailor to a specific theory.
Regarding the parameterization of mixing between neutrino mass eigenstates with SM flavor eigenstates, we adopt the notation of Ref.~\cite{Atre:2009rg}, 
and extend it to include coupling to a model-independent $W'$ in Section~\ref{appendNeu.APP}.
This parameterization is accomplished with a minimum amount of parameters.

\subsection{Neutrino Mixing with $W'$ Couplings}\label{appendNeu.APP}


\subsubsection{Model-Independent $W'$ Charged Current Couplings}

The goal of this paper is to explore the feasibility of quantifying the properties of a new charged gauge boson, $W'$, at the LHC. 
For this purpose, we relax the $W'$ interactions to include both left-handed and right-handed leptons,
\beq 
L_{aL}
=\left(
\begin{array}{c} \nu_a\\
l_a
\end{array}
\right)_L ,\quad   
R_{b R}
=\left(
\begin{array}{c} N_{b}\\
l_{b}
\end{array}
\right)_R ,
\eeq
with $a,b=1,\:2,\:3$. The L.H.~neutrinos and charged leptons that are members of
SU$(2)_{L}$ doublets in the Standard Model (SM) are denoted by $\nu_{aL}$ and $l_{a}$.
The R.H.~neutrinos, which are SM singlets, and R.H.~charged leptons are denoted by $N_{bR}$ and $l_{b}$.
To grasp the feature of Left-Right symmetric models for a $W'$, we pair $N_{bR}$ and $l_{b}$ into the an SU$(2)_R$ doublet. 
Though there may be more ``sterile'' neutrinos, i.e., $b>3$, we consider only $b=3$ and one new mass eigenstate in our phenomenological presentation.
The mass mixing matrix in Eq.~(\ref{appNuMix.EQ}), in the present case, becomes a $6\times6$ matrix with several repeating entries.

With this assignment, the resulting charged current interactions are 
\begin{eqnarray}
\mathcal{L}=\left( -\frac{g_{L}^{\ell}}{\sqrt{2}}W_{\mu L}^{'+} \
\sum_{a=1}^{3} \overline{\nu_{aL}} \gamma^{\mu}P_{L}l_{a}^{-} -
\frac{g_{R}^{\ell}}{\sqrt{2}}W_{\mu R}^{'+} \ 
\sum_{b=1}^{3} \overline{N_{b R}} \gamma^{\mu}P_{R}l_{b}^{-} \right)
+h.c.
\end{eqnarray}
We have explicitly included the couplings of left- and right-charged currents with new gauge interactions via $W'_{L,R}$. 

The gauge state leptons, $l_{a}$ and $l_{b}$, may be rotated into the mass eigenstates, which are defined to be the flavors eigenstates $\ell=e,\mu,\tau$. 
This amounts to the rotation 
\begin{equation}
 l_{a}^{-}=\sum_{\ell=e}^{\tau}O_{a\ell}\ell^{-}.
\end{equation}
With the SM-like simplest Higgs mechanism, this transformation is trivial and we will make it implicit without loss of generality. 
By simultaneously expanding into the neutrinos' mass basis and into the charged leptons' flavor basis, we obtain 
\begin{eqnarray}
\mathcal{L}=&-&\sum_{\ell=e}^{\tau}\frac{g_{L}^{\ell}}{\sqrt{2}}W_{\mu}^{'+}\left[\sum_{m=1}^{3}\overline{\nu_{m}} U_{m\ell}^{*}+\sum_{m'=4}^{n+3}\overline{N_{m'}^{c}} V_{m'\ell}^{*}\right]\gamma^{\mu}P_{L}\ell^{-} +{\it h.c.}
\nonumber\\
&-&\sum_{\ell=e}^{\tau}\frac{g_{R}^{\ell}}{\sqrt{2}}W_{\mu}^{'+}\left[\sum_{m=1}^{3}\overline{\nu_{m}^{c}} X_{m\ell}+\sum_{m'=4}^{n+3}\overline{N_{m'}} Y_{m'\ell}\right]\gamma^{\mu}P_{R}\ell^{-} +{\it h.c.},
\label{appModIndLag.EQ}
\end{eqnarray}
where
\begin{equation}
U_{m\ell}^{*} \equiv \sum_{a=1}^{3} U_{ma}^{*} O_{a\ell}, \quad V_{m'\ell}^{*} \equiv \sum_{a=1}^{3} V_{m'a}^{*} O_{a\ell}, \quad
X_{m\ell}     \equiv \sum_{b=1}^{3} X_{mb}^{*} O_{b\ell}, \quad Y_{m'\ell}^{*} \equiv \sum_{b=1}^{3} Y_{m'b} O_{b\ell}.
\label{appRotDefs.EQ}
\end{equation}

These are the general couplings for the $W'$ charged currents that we follow in this study.
Leptonic couplings to the SM $W^{\pm}$ boson can be recovered from
Eq.~(\ref{appModIndLag.EQ}) by identifying $W^{'\pm}\to W^{\pm}$ and by setting 
\begin{equation}
g^{\ell}_{L} = g\ {\rm and}\  g^{\ell}_{R} = 0,
\end{equation}
where $g$ is the SU$(2)_{L}$ coupling constant in the SM.
Similarly, we arrive at the SU$(2)_{R}$ charged current coupling by identifying $W' \to W^{\pm}_{R}$ and by setting 
\begin{equation}
g^{\ell}_{L} = 0\ {\rm and}\  g^{\ell}_{R} \neq 0.
\end{equation}

In the quark sector, we do not plan to go through a fully-fledged construction for the charged current couplings.
Instead, we take the simplest approach and just parameterize the model-independent $W'$ Lagrangian by
\begin{equation}
 \mathcal{L}=\frac{-1}{\sqrt{2}}\sum_{i,j=1}^{3}W_{\mu}^{'+}\overline{u_{i}}V_{ij}^{CKM'}\gamma^{\mu}\left[g_{L}^{q}P_{L}+g_{R}^{q}P_{R}\right]d_{j}+{\it h.c.},
\end{equation}
where $V^{CKM'}$ is an unknown flavor mixing matrix.

\subsection{$\wpri$ Chiral Coupling to Fermions}
The model-independent Lagrangian that governs the interaction between SM quarks and a new, massive, electrically charged vector boson, $W'$, is given by
\begin{equation}
 \mathcal{L}=-\frac{1}{\sqrt{2}}\sum_{i,j=1}^{3}W_{\mu}^{'+}\overline{u_{i}}V_{ij}^{CKM'}\gamma^{\mu}\left[g_{R}^{q}P_{R}+g_{L}^{q}P_{L}\right]d_{j}+{\it h.c.},
\end{equation}
where $u_i\ (d_j)$ denotes the  
Dirac spinor of an up-(down-)type quark with flavor $i\ (j)$; 
$V^{CKM'}$ parameterizes the mixing between flavors $i$ and $j$ for the new charged current interactions 
just as the Cabibbo-Kobayashi-Maskawa (CKM) matrix does in the SM;
$g_{R,L}^{q}$ is the $W'$'s universal coupling strength to right-(left-) handed quarks; 
and $P_{R,L}=\frac{1}{2}\left(1\pm\gamma_5\right)$ denotes the $R,L$-handed chiral projection operator.
\label{wpriNcoup.SEC}

We parameterize the new boson's coupling to charged leptons with flavor $\ell$ and neutral leptons with mass $m_{m}$ (for the three light states) or $m_{N}$ (for the heavy state) in the following way:
\begin{eqnarray}
\mathcal{L}=
&-&\sum_{\ell=e}^{\tau}\frac{g_{R}^{\ell}}{\sqrt{2}}W_{\mu}^{'+}\left[\sum_{m=1}^{3}\overline{\nu_{m}^{c}}X_{\ell m} + \overline{N}Y_{\ell N}\right]\gamma^{\mu}P_{R}\ell^{-}\nonumber\\
&-&\sum_{\ell=e}^{\tau}\frac{g_{L}^{\ell}}{\sqrt{2}}W_{\mu}^{'+}\left[\sum_{m=1}^{3}\overline{\nu_{m}}U_{\ell m}^{*} + \overline{N^{c}}V_{\ell N}^{*}\right]\gamma^{\mu}P_{L}\ell^{-}+{\it h.c.}
\label{wpLagrangian.EQ}
\end{eqnarray}
$g_{R}^{\ell}~(g_{L}^{\ell})$ is the $W'$'s coupling strength to R.H. (L.H.) leptons; 
$X_{\ell m}~(U_{\ell m})$ parameterizes the mixing between light neutrino mass eigenstates and R.H. (L.H.) interactions; and
$Y_{\ell N}~(V_{\ell N})$ parameterizes the mixing between the heavy neutrino mass eigenstate and R.H. (L.H.) interactions.
Lastly, $\psi^{c}=\mathcal{C}\overline{\psi}^{T}$ denotes the charge conjugate of the field $\psi$, 
with $\mathcal{C}$ being the charge conjugate operator, and the chiral states satisfy $P_{L}(\psi^{c})=(P_{R}\psi)^{c}.$
In Section~\ref{appendNeu.APP}, our choice of parameterization is discussed in detail. 
From a viewpoint of the model construction as discussed in Refs.~\cite{Mohapatra:1998rq,Gluza:2002vs,Fukugita:2003en,Barger:2003qi,Eidelman:2004wy,
GonzalezGarcia:2007ib,Mohapatra:2006gs,Strumia:2006db,Keung:1983uu,Atre:2009rg}, one may expect that 
$UU^{\dagger},~YY^{\dagger}\sim\mathcal{O}(1)$ and 
$VV^{\dagger},~XX^{\dagger}\sim\mathcal{O}(m_{m}/m_{N}).$ 
Since we prefer a model-independent approach, we will not follow rigorously the above argument and will take the parameters as 
\begin{equation}
UU^{\dagger},~YY^{\dagger}\sim\mathcal{O}(1),\quad\text{and}\quad
VV^{\dagger},~XX^{\dagger}\sim\mathcal{O}(10^{-3}),
\label{mixingFromUnit.EQ}
\end{equation}
which is guided by the current constraints as presented later in this section.

In Eq.~(\ref{wpLagrangian.EQ}), the $\wpri$ is allowed to have both independent right-handed $(g_R^{q,\ell})$ and left-handed $(g_L^{q,\ell})$ couplings.
Subsequently, the pure gauge states $W'_R$ and $W'_L$ are special cases of $W'$ when
\begin{equation}
g_{R}^{q,\ell} \ne 0\quad\text{and}\quad g_{L}^{q,\ell}=0,
\label{pureGuageRH.EQ}
\end{equation}
and
\begin{equation}
g_{R}^{q,\ell}=0\quad\text{and}\quad g_{L}^{q,\ell} \ne 0,
\label{pureGuageLH.EQ}
\end{equation}
respectively. Additionally, the SM $W$ coupling to leptons can be recovered from Eq.~(\ref{wpLagrangian.EQ}) by setting
\begin{equation}
g_{R}^{\ell}=0,\quad\text{and}\quad g_{L}^{\ell}=g.
\end{equation}
Here, $g$ is the usual SM SU$(2)_L$ coupling constant.

\subsection{Current Constraints on $W'$}
\label{Constraints.SEC}
We list only the most stringent, most relevant constraints to our analysis here and refer the reader to Ref.~\cite{Beringer:1900zz,Constraints} for a more complete review.

\begin{itemize}
\item $\bf{Bounds~from~CMS}$: 
The CMS Experiment has searched for $W_{R}$ and heavy $N$, where $M_{W_{R}}>m_{N}$, with the $\ell^{\pm}\ell^{\pm}jj$ collider signature~\cite{Chatrchyan:2011WR}, 
assuming $g_{R}=g$. With 5.0 fb$^{-1}$ of 7 TeV and 3.7 fb$^{-1}$ of 8 TeV $pp$ collisions, the present mass bounds for $W'_{R}$ and $N$ are
\begin{equation}
 M_{W_{R}} > 2.9\text{ TeV }(m_{N}\approx0.8\text{ TeV})\quad\text{and}\quad m_{N}>1.9\text{ TeV }(M_{W_{R}}\approx2.4\text{ TeV}.)
\end{equation}
The search for the sequential SM $W'$, $W'_{SSM}$, decaying into a charged SM lepton plus~$\slash\!\!\!\!E_{T}$, with $g'=g$, has also been performed. 
With 3.7 fb$^{-1}$ of 8 TeV $pp$ collisions \cite{Chatrchyan:2012Wssm}, the present mass bound is
\begin{equation}
 M_{W_{SSM}} > 2.85\text{ TeV.}
\end{equation}

\item $\bf{Bounds~from~ATLAS}$: The ATLAS Experiment has also searched for $W_{R}$ and heavy $N$, under the same stipulations as the CMS Experiment~\cite{Aad:2012WR}. 
With 2.1 fb$^{-1}$ of 7 TeV $pp$ collisions, the present mass bounds for $W'_{R}$ and $N$ are
\begin{equation}
 M_{W_{R}} > 2.5\text{ TeV }(m_{N}\approx0.8\text{ TeV})\quad\text{and}\quad m_{N}>1.6\text{ TeV }(M_{W_{R}}\approx1.8\text{ TeV}.)
\end{equation}
\item $\bf{Global~Fit~Analysis}$: The effects of a generic $Z'$ boson on EW precision observables place bounds~\cite{Cacciapaglia:2006pk} of 
\begin{equation}
 M_{Z'}/g_{Z'}\gtrsim 2.7-6.7~\text{TeV}.
\end{equation}
For $Z'$ and $W'$ bosons originating from the same broken symmetry, we expect similar constraints on $M_{\wpri}/g_{\wpri}$ since
\begin{equation}
 M_{W'}\sim M_{Z'}\times\mathcal{O}(1).
\end{equation}
\item $\bf{Bounds~on~W_{L}-W_{R}~Mixing}$: Non-leptonic Kaon decays~\cite{Donoghue:1982} 
and universality in Weak decays~\cite{Wolf:1984} constrain $W_{L}-W_{R}$ mixing. 
The present bound for the L-R mixing angle $\zeta$~\cite{LRModels} is
\begin{equation}
 \vert \zeta \vert \leq 1\sim4\times 10^{-3}.
\end{equation}
\end{itemize}

\subsection{Current Constraints on $N$}
More complete lists of constraints on low and high mass neutrinos, respectively, are available~\cite{Atre:2009rg,Beringer:1900zz}.
\begin{itemize}
\item $\bf{Bounds~from~0\nu\beta\beta}$: For $m_{N}\gg 1$ GeV, a lack of evidence for neutrinoless double beta decay bounds the mixing between heavy neutrino states and the electron-flavor 
state at~\cite{Olness:1984xb,Langacker:1989xa,Belanger:1995nh,Benes:2005hn}
\bea
\displaystyle\sum_{m'}\frac{|V_{em'}|^2}{m_{m'}}< 5\times10^{-5}\,{\rm TeV}^{-1},
\label{zerovtwobBound.EQ}
\eea
where the sum is over all heavy Majorana neutrinos.
\item $\bf{Bounds~from~EW~Precision~Data}$: A TeV scale singlet neutrino mixing with the SM flavor states is constrained~\cite{delAguila:2008pw} by
\begin{equation}
 |V_{eN}|^2,|V_{\mu N}|^2<0.003\quad\text{and}\quad|V_{\tau N}|^2<0.006
 \label{emutauMixingBounds.EQ}
\end{equation}
\end{itemize}
%


\section{Derivation of Partonic Level Angular Distributions}\label{appendME.APP}
\begin{figure}[!t]
\begin{center}
\includegraphics[clip,width=0.7\textwidth]{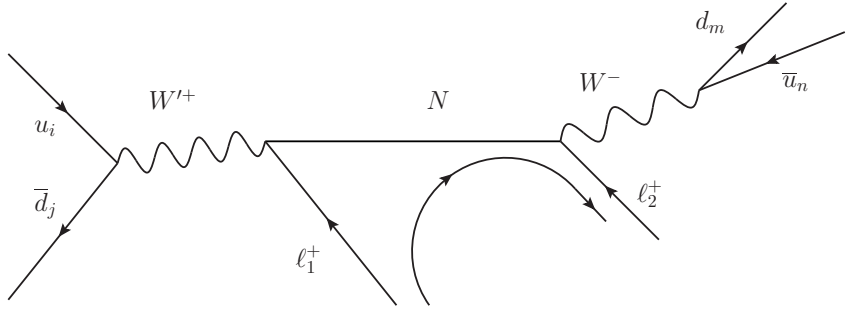}
\caption[Partonic-level process for heavy $W^{'+}$ production and decay into $\ell^\pm\ell^\pm q \overline{q'}$]{The partonic-level process for heavy $W^{'+}$ production and decay into like-sign leptons and quarks in hadronic collisions. 
The longer, black arrow not touching the Feynman diagram denotes fermion flow (FF).}
\label{qqWPllW_Arrow.FIG}
\end{center}
\end{figure}

We strive clarify a few subtleties that arise when calculating observables involving Majorana fermions. 
To do so, we present a detailed derivation of the matrix element for the lepton-number $(L)$ violating process:
\begin{equation}
u_{i}(p_{A})+\overline{d}_{j}(p_{B})\rightarrow W'^{+}\rightarrow\ell_{1}^{+}(p_{1})+\ell_{2}^{+}(p_{2})+q_{m}(p_{3})+\overline{q}_{n}(p_{4}),
\end{equation}
with an intermediate Majorana neutrino of mass $m_{N}$, and 
governed by the Lagrangian given in Section~\ref{ThFrame.SEC}. 
  As discussed in Section \ref{Like.SEC}, and shown in Fig.~\ref{qq2Wp2llqq.FIG}, there are two interfering Feynman diagrams
associated with our $2\ell^{+}2j$ final state.  The interference term may be neglect safely when 
calculating the amplitude squared, $|\mathcal{M}|^2$, since the heavy neutrino's width is very narrow and thus 
the interference is expected to be small. When constructing 
and evaluating $|\mathcal{M}|^2$, we focus on only a single diagram (Fig.~\ref{qqWPllW_Arrow.FIG}) but stress that 
the two diagrams can be treated identically. Additionally, the narrowness of the SM $W$ boson's width allows us to 
further apply the Narrow Width Approximation (NWA). The NWA stipulates that, due to its small width 
compared to its mass, the $W$ boson will dominantly be produced on-shell, and further implies
\begin{equation}
\hat{\sigma}(u_{i}\overline{d}_{j}\rightarrow\ell_{1}^{+}\ell_{2}^{+}q\overline{q}')\approx\hat{\sigma}(u_{i}\overline{d}_{j}\rightarrow\ell_{1}^{+}\ell_{2}^{+}W^{-})\times BR(W\rightarrow q\overline{q}'),
\end{equation}
where $BR(X\rightarrow Y)$ is the branching fraction of $X$ going into $Y$. Since $BR(W\rightarrow q\overline{q}')$ 
is well-known, our work is reduced to determining the analytical expression for 
\begin{equation}
 \hat{\sigma}(u_{i}\overline{d}_{j}\rightarrow\ell_{1}^{+}\ell_{2}^{+}W^{-}).
\end{equation}

\subsection[Summed Squared Matrix Element]{Determination of the Spin-Summed, Polarization-Dependent, Squared Matrix Element}
The usefulness of Feynman rules stems from the ability to assign specific multiplicative 
factors to each component of a Feynman diagram. However, Dirac field Feynman
rules are dependent on Wick's Theorem, which is a statement on field contractions.
For Dirac fields, only combinations of the form $\overline{\psi}\psi$ can contract,
where as for Majorana fields, $\psi\psi$ and $\overline{\psi}\overline{\psi}$ are 
allowed to contract. In short, Feynman rules for Dirac fermions do not account for
all possible Majorana interactions. 

We therefore adopt the Feynman rules developed in Ref.~\cite{Denner:1992} for a two-fold reason. 
The first is that the rules for diagram segments not involving Majorana fermions do not change. 
The second is that for parts that do involve Majorana fermions, 
the new Feynman rules reduce to (a) treating the Majorana fermion like a Dirac fermion 
and modifying the vertex factor for an ordinary Dirac fermion 
with an appropriately placed factor of $-1$, and/or (b) making a single 
$u\leftrightarrow v$ spinor substitution. The placement of the additional 
minus sign and possible spinor substitution is based on the direction of fermion flow (FF)
relative to the traditionally chosen fermion {\em{number}} flow (FNF). 
When the fermion flow and fermion number flow are equal, the newer rules simplify to the usual rules.
Computationally, these rules provide a desirable technique that can be automated in a straight forward manner. 

In the present case, we identify the relevant FF as being identical to the lepton number-changing current. 
The FF current starts at $\ell_{1}$, the charged lepton produced in the $W'$ boson decay, 
and points anti-parallel to $\ell_{1}$'s momentum; the current then continues parallel to the Majorana
neutrino's momentum; and finally terminates at $\ell_{2}$,  the charged lepton produced in the $N$ decay,
and points parallel to $\ell_{2}$'s momentum.
See the curved black arrow in Fig.~\ref{qqWPllW_Arrow.FIG}. With this orientation, the FF is 
parallel to the FNF at the $W'\ell_{1}N$ vertex, and anti-parallel to it at the $N\ell_{2}W$ vertex. 
This change in relative current orientation causes two modifications, the first of which is to the 
spinor of the outgoing lepton originating from the $N\ell_{2}W$ vertex:
\begin{equation}
 \overline{v}_{\ell_2}(p_2)\rightarrow\overline{u}_{\ell_2}(p_2),
\end{equation}
and accounts explicitly for the change in lepton number. The second modification is to the 
$N\ell_{2}W$ vertex itself and occurs in the following way:
\begin{equation}
C^{\rho}_{N\ell_{2}}=\frac{-ig}{\sqrt{2}}V_{\ell_{2}N}\gamma^{\rho}P_{L}\rightarrow C'^{\rho}_{N\ell_{2}}=(-1)^2\frac{ig}{\sqrt{2}}V_{\ell_{2}N}\gamma^{\rho}P_{R},
\label{nChiralFlip.EQ}
\end{equation}
where $g$ is the SM SU$(2)_{L}$ coupling constant, 
$P_{R,L}\equiv\frac{1}{2}(1\pm\gamma_5)$, 
and, as defined in Ref.~\cite{Denner:1992}, the primed-vertex convention indicates
\begin{equation}
\Gamma'\equiv\mathcal{C}\Gamma^{T}\mathcal{C}^{-1}=\eta\Gamma,
\end{equation}
where $\mathcal{C}$ is the charge conjugation operator and for which
\begin{equation}
\eta=\begin{cases}
1, & \Gamma\in\{1,i\gamma^{5},\gamma^{\mu}\gamma^{5}\}\\
-1, & \Gamma\in\{\gamma^{\mu},\sigma^{\mu\nu}\}
\end{cases}.
\label{majEta.EQ}
\end{equation}

As a result, we find that the matrix element describing the $u_{i}\overline{d}_{j}\rightarrow\ell_{1}^{+}\ell_{2}^{+}W_{\lambda}^{-}$
scattering process, for an outgoing SM $W^{-}$ boson with polarization
$\lambda$, and in the Feynman Gauge, is
\begin{eqnarray}
i\mathcal{M}_{\lambda}
&=& \varepsilon_{\lambda\rho}^{*}(p_{W})\cdot\frac{\left[\overline{v}_{Bj}A_{ji}^{\mu}u_{Ai}\right]\cdot\left[\overline{u}_{2}C'^{\rho}_{N\ell_{2}}(\not\! p_{N}+m_{N})B_{\mu\:\ell_{1}N}v_{1}\right]}{\left(\hat{s}-M_{W'}^{2}+i\Gamma_{W'}M_{W'}\right)\left(p_{N}^{2}-m_{N}^{2}+i\Gamma_{N}m_{N}\right)},
\label{udWllME.EQ}
\end{eqnarray}
where the vertex terms are given by
\begin{eqnarray}
A_{ji}^{\mu}&=&\frac{1}{\sqrt{2}}V^{CKM'}_{ji}\gamma^{\mu}~\left[g_{R}^{q}P_{R}+g_{L}^{q}P_{L}\right]\\
B_{\ell_{1} N}^{\nu}&=&\frac{1}{\sqrt{2}}\gamma^{\nu}\left[g_{R}^{\ell}P_{R}Y_{\ell_{1}N}+g_{L}^{\ell}P_{L}V_{\ell_{1}N}^{*}\right]
\end{eqnarray}
To be explicit: $\varepsilon_{\lambda\rho}^{*}(p_W)$ denotes the 
outgoing polarization vector of the on-shell $W$ boson with momentum $p_W$, mass $M_W$, and polarization 
$\lambda$; $\overline{v}_{Bj}$ represents the the spinor $\overline{v}$ of an initial-state 
antiquark of flavor $j$ and momentum $p_B$; similarly, $u_{Ai}$ represents the spinor $u$ of 
an initial-state quark of flavor $i$ and momentum $p_A$; $\overline{u}_{2}$ denotes the spinor 
of our final-state $\it{antilepton}$ with flavor $\ell_{2}$ and momentum $p_2$; and
likewise, $v_{1}$ denotes the spinor of our final-state antilepton with flavor $\ell_{1}$ 
and momentum $p_1$. The $W'$ mass, width, and momentum-squared are respectively given by $M_{W'}$, 
$\Gamma_{W'}$, and the Mandelstam variable 
\begin{equation}
\hat{s}=(p_A+p_B)^2=(p_1+p_2+p_W)^2. 
\end{equation}
The heavy 
neutrino's mass, width, and momentum are similarly given by $m_{N}$, $\Gamma_N$, and 
\begin{equation}
p_N=p_A+p_B-p_1=p_W+p_2.
\end{equation}

After squaring and summing over external spins, diagrams, 
 and colors ($N_{C}$), but not external boson 
polarizations ($\lambda$), the polarization-dependant squared amplitude is
\begin{eqnarray}
\sum\vert\mathcal{M}_{\lambda}\vert^{2} =
\frac{4N^2_{C}~g^{2}~\vert V_{ji}^{CKM'}\vert^{2}~\vert V_{\ell_{2}N}\vert^{2}~\text{Tr}\left[\not\! p_{A}\gamma^{\sigma}\not\!p_{B}\gamma^{\mu}\left(g_{R}^{q~2}P_{R}+g_{L}^{q~2}P_{L}\right)\right]}{2^{3}(1+\delta_{\ell_{1}\ell_{2}})\left[\left(\hat{s}-M_{W'}^{2}\right)^{2}+\left(\Gamma_{W'}M_{W'}\right)^{2}\right]\left[\left(p_{N}^{2}-m_{N}^{2}\right)^{2}+\left(\Gamma_{N}m_{N}\right)^{2}\right]}\nonumber\\
\times \text{Tr}\left[\not\!p_{1}\gamma_{\sigma}\left(\not\! p_{N}+m_{N}\right)\not\!\varepsilon_{\lambda}\not\! p_{2}\not\!\varepsilon_{\lambda}^{*}P_{R}\left(\not\! p_{N}+m_{N}\right)\gamma_{\mu}\left(g_{R}^{\ell~2}P_{R}\vert Y_{\ell_{1}N}\vert^{2}+g_{L}^{\ell~2}P_{L}\vert V_{\ell_{1}N}\vert^{2}\right)\right]\\
=\frac{2^{3}N^2_{C}~g^{2}~\vert V_{ji}^{CKM'}\vert^{2}~\vert V_{\ell_{2}N}\vert^{2}}{(1+\delta_{\ell_{1}\ell_{2}})\left[\left(\hat{s}-M_{W'}^{2}\right)^{2}+\left(\Gamma_{W'}M_{W'}\right)^{2}\right]\left[\left(p_{N}^{2}-m_{N}^{2}\right)^{2}+\left(\Gamma_{N}m_{N}\right)^{2}\right]}\nonumber\\
\times \left[
 \vert Y_{\ell_{1}N}\vert^{2}\left(g_{R}^{q}g_{R}^{\ell}\right)^{2}\mathcal{A}_{\lambda}
+\vert Y_{\ell_{1}N}\vert^{2}\left(g_{L}^{q}g_{R}^{\ell}\right)^{2}\mathcal{B}_{\lambda}
+\vert V_{\ell_{1}N}\vert^{2}\left(g_{L}^{q}g_{L}^{\ell}\right)^{2}\mathcal{C}_{\lambda}
+\vert V_{\ell_{1}N}\vert^{2}\left(g_{R}^{q}g_{L}^{\ell}\right)^{2}\mathcal{D}_{\lambda}\right],
\end{eqnarray}

where
\begin{eqnarray}
\mathcal{A}_{\lambda}&=&2(p_{A}\cdot p_{1})(p_{B}\cdot p_{N})\left[(p_{N}\cdot p_{2})+2(p_{N}\cdot\varepsilon_{\lambda})(\varepsilon_{\lambda}\cdot p_{2})\right]\nonumber\\
&-& m_{N}^{2}(p_{A}\cdot p_{1})\left[(p_{B}\cdot p_{2})+2(p_{B}\cdot\varepsilon_{\lambda})(\varepsilon_{\lambda}\cdot p_{2})\right],\\
\mathcal{B}_{\lambda}&=&2(p_{B}\cdot p_{1})(p_{A}\cdot p_{N})\left[(p_{N}\cdot p_{2})+2(p_{N}\cdot\varepsilon_{\lambda})(\varepsilon_{\lambda}\cdot p_{2})\right]\nonumber\\
&-& m_{N}^{2}(p_{B}\cdot p_{1})\left[(p_{A}\cdot p_{2})+2(p_{A}\cdot\varepsilon_{\lambda})(\varepsilon_{\lambda}\cdot p_{2})\right],\\
\mathcal{C}_{\lambda}&=&m_{N}^{2}(p_{A}\cdot p_{1})\left[(p_{B}\cdot p_{2})+2(p_{B}\cdot\varepsilon_{\lambda})(\varepsilon_{\lambda}\cdot p_{2})\right],\\
\mathcal{D}_{\lambda}&=&m_{N}^{2}(p_{B}\cdot p_{1})\left[(p_{A}\cdot p_{2})+2(p_{A}\cdot\varepsilon_{\lambda})(\varepsilon_{\lambda}\cdot p_{2})\right],
\end{eqnarray}
and $\varepsilon_{\lambda}$ is taken to be real. 

The Majorana neutrino's width, $\Gamma_{N}$, is expected to be very small. 
Therefore, to simplify analytic integration, we again apply the Narrow Width Approximation such that
\begin{equation}
\frac{1}{(p_{N}^{2}-m_{N}^{2})^{2}+(\Gamma_{N}m_{N})^{2}}
\approx \frac{\pi}{\Gamma_{N}m_{N}}~\delta\left(p_{N}^{2}-m_{N}^{2}\right).
\end{equation}
We are motivated to make this additional approximation to highlight and emphasize the 
analyzing power of the angular distributions. Our reported numerical 
results do not reflect this extra stipulation; see Eq.~(\ref{diffXSecNoNWA.EQ}).
Consequentially, the squared and summed amplitude becomes
\begin{equation}
\sum\vert\mathcal{M}_{\lambda}\vert^{2}\approx
\frac{2^{3}\pi N_{C}\: g^{2}~\vert V_{ji}^{CKM'}\vert^{2}~\vert V_{\ell_{2}N}\vert^{2}\:\delta\left(p_{N}^{2}-m_{N}^{2}\right)}{(1+\delta_{\ell_{1}\ell_{2}})\left(\Gamma_{N}m_{N}\right)\left[\left(\hat{s}-M_{W'}^{2}\right)^{2}+\left(\Gamma_{W'}M_{W'}\right)^{2}\right]}
\label{sqME.EQ}
\end{equation}
\begin{equation}
\times \left[
 \vert Y_{\ell_{1}N}\vert^{2}\left(g_{R}^{q}g_{R}^{\ell}\right)^{2}\mathcal{A}_{\lambda}
+\vert Y_{\ell_{1}N}\vert^{2}\left(g_{L}^{q}g_{R}^{\ell}\right)^{2}\mathcal{B}_{\lambda}
+\vert V_{\ell_{1}N}\vert^{2}\left(g_{L}^{q}g_{L}^{\ell}\right)^{2}\mathcal{C}_{\lambda}
+\vert V_{\ell_{1}N}\vert^{2}\left(g_{R}^{q}g_{L}^{\ell}\right)^{2}\mathcal{D}_{\lambda}\right]\nonumber.
\end{equation}

\subsection{Phase Space Volume Element}
\label{PhaseSpace.APP}

We calculate the partonic-level cross section using the usual formula,
\begin{equation}
d\hat{\sigma}=\frac{1}{2\hat{s}}\frac{1}{4N_{C}^{2}}\sum\vert\mathcal{M}\vert^{2}\cdot dPS_{n}.
\label{sigmaHat.EQ}
\end{equation}
Here, the factor of $4N_{C}^{2}$ comes from averaging over initial-state
colors and spins. The factor $dPS_{n}$ represents the
\emph{n}-body phase space volume element, 
\begin{equation}
 dPS_{n}(P;p_{1}\dots p_{n}) = 
\prod_{k=1}^{n}\frac{d^{3}p_k}{(2\pi)^{3}2E_k} ~(2\pi)^{4}\delta^{4}\left(P-p_1-\dots-p_n\right),
\end{equation}
which can be decomposed using the recursion formula
\begin{equation}
dPS_{n}(P;p_{1},\dots,p_{n})=
dPS_{n-1}(P;p_{1},\dots,p_{n-1,n})\times 
dPS_{2}(p_{n-1,n};p_{n-1},p_{n})\times
\frac{d\: p_{n-1,n}^{2}}{2\pi},
\end{equation}
where $P=\sum_{m=1}^{n}p_{m}$ and $p_{i,j}=p_{i}+p_{j}$. In the present case, 
$dPS_{3}$ is expressible as
\begin{equation}
dPS_{3}(p_{A}+p_{B};p_{1},p_{2},p_{W})=
dPS_{2}(p_{A}+p_{B};p_{1},p_{N})\times 
dPS_{2}(p_{N};p_{2},p_{W})\times
\frac{d\: p_{N}^{2}}{2\pi}.
\label{phaseSpace1.EQ}
\end{equation}

Since each $dPS_{k}$ is individually Lorentz invariant,
the two phase space elements in Eq.~(\ref{phaseSpace1.EQ}) can be evaluated in different
reference frames. When $dPS_{2}(p_{1},p_{N})$ is evaluated
in the partonic c.m.~frame and $dPS_{2}(p_{2},p_{W})$
in the neutrino rest-frame, the full volume element is found to be
\begin{equation}
dPS_{3}(p_{A}+p_{B};p_{1},p_{2},p_{W})=d\Omega_{N}
\frac{(1-\tilde{\mu}_{N}^{2})}{2(4\pi)^{2}}\times d\Omega_{\ell_{2}}\frac{(1-\rho_{W}^{2})}{2(4\pi)^{2}}
\times\frac{d\: p_{N}^{2}}{2\pi},
\label{phaseSpace2.EQ}
\end{equation}
with 
\begin{equation}
 \mu_{N}^2=\frac{m_N^2}{\hat{s}},\quad\quad
 \tilde{\mu}_{N}^{2} = \frac{p_N^2}{\hat{s}},\quad\quad
 \rho_{W}^2=\frac{M_{W}}{p_{N}^{2}},
\end{equation}
and, in the on-shell limit,
\begin{equation}
 \mu_{N},\tilde{\mu}_{N}\rightarrow x_N=\frac{m_N}{M_{W'}},\quad\quad
  \rho_{W}\rightarrow y_{W}=\frac{M_{W}}{m_{N}}.
\end{equation}
The solid angle element $d\Omega_{N}$ is defined as the angle made by 
$N$ with respect to the direction of propagation of the initial-state 
quark in the c.m.~frame; $d\Omega_{\ell_{2}}$ is defined as the angle 
made by $\ell_{2}^{+}$ with respect to the heavy neutrino spin 
axis in the neutrino's rest-frame. 

\subsection{Partonic-Level Angular Distributions}
\label{AngDist.APP}

The angular distribution of the charged lepton from the neutrino decay
is most efficiently determined by evaluating $\sum\vert\mathcal{M}\vert^{2}$ 
in the neutrino rest-frame. 
Like individual $dPS_{k}$ volume
elements, $\vert\mathcal{M}\vert^{2}$ is separately Lorentz invariant
and thus can be evaluated in its own reference frame. 

In order to evaluate Eq.~(\ref{sqME.EQ}) in the neutrino rest-frame, we 
must first rotate and boost the four-momenta of the initial-state 
quarks from the c.m.~frame. Without the loss of generality, we assume 
that the initial-state (anti)quark is originally traveling in the 
positive (negative) $\hat{z}-$axis and that the $\ell_{1}^{+}N$ pair propagate in 
$\hat{y}-\hat{z}$ plane. This allows us to rotate the 
entire $2\rightarrow2$ system such that the neutrino's momentum 
is aligned with the $\hat{z}-$axis, and then boost into the neutrino rest-frame. 
Since we are applying the NWA and immediately integrating over $dp_{N}^{2}$, we will take $N$ to be on-shell.
After boosting, our four-momenta are:
\begin{eqnarray}
p_{A}&=&\frac{\hat{s}}{4m_{N}}\left((1-\cos\theta_{N})+\mu_{N}^{2}(1+\cos\theta_{N}),\:0,-2\mu_{N}\sin\theta_{N},\: \mu_{N}^{2}(1+\cos\theta_{N})-(1-\cos\theta_{N})\right),\nonumber\\
p_{B}&=&\frac{\hat{s}}{4m_{N}}\left((1+\cos\theta_{N})+\mu_{N}^{2}(1-\cos\theta_{N}),\:0,~~2\mu_{N}\sin\theta_{N}	,\: \mu_{N}^{2}(1-\cos\theta_{N})-(1+\cos\theta_{N})\right),\nonumber
\end{eqnarray}
\begin{equation}
 p_{N}=(m_{N},0,0,0),\quad\text{and}\quad p_{1}=\frac{\hat{s}}{2m_{N}}(1-\mu_{N}^{2})(1,0,0,-1),
\end{equation}
where $\theta_{N}$ represents the polar angle between $\vec{p}_{N}$
and $\vec{p}_{A}$ in the c.m.~frame. In the neutrino rest-frame, 
the $N\rightarrow\ell_{2}^{+}W^{-}$ decay products' 
momenta are 
\begin{eqnarray}
p_{2}&=&\vert\vec{p}_{2}\vert\left(1,\sin\theta_{\ell_{2}}\cos\phi_{\ell_{2}},\sin\theta_{\ell_{2}}\sin\phi_{\ell_{2}},\cos\theta_{\ell_{2}}\right),\quad\vert\vec{p}_{2}\vert=\vert\vec{p}_{W}\vert=\frac{m_N}{2}(1-y_{W}^{2}),\nonumber\\
p_{W}&=&\vert\vec{p}_{2}\vert\left(\frac{E_{W}}{\vert\vec{p}_{2}\vert},-\sin\theta_{\ell_{2}}\cos\phi_{\ell_{2}},-\sin\theta_{\ell_{2}}\sin\phi_{\ell_{2}},-\cos\theta_{\ell_{2}}\right),\; E_{W}=\frac{m_N}{2}(1+y_{W}^{2}),
\end{eqnarray}
where $\theta_{\ell_{2}}$ and $\phi_{\ell_{2}}$ are defined with respect to the neutrino spin axis in the c.m.~frame.
Explicitly, $\hat{z}=\hat{p}_N$, where $\hat{p}_N =\vec{{p}}_{N} / \vert \vec{p}_{N}\vert$ is measured in the c.m.~frame, and $\phi_{\ell_2}$ w.r.t.~to the $+\hat{y}$ axis.
This is consistent with Eq.~(\ref{phaseSpace2.EQ}). 
The polarization vectors for the SM $W$ boson are subsequently: 
\begin{eqnarray}
\varepsilon_{0}^{\mu}(p_{W})&=&\frac{E_{W}}{m_{W}}\left(\frac{\vert\vec{p}_{2}\vert}{E_{W}},-\sin\theta_{\ell_{2}}\cos\phi_{\ell_{2}},-\sin\theta_{\ell_{2}}\sin\phi_{\ell_{2}},-\cos\theta_{\ell_{2}}\right),\nonumber\\ 
\varepsilon_{T1}^{\mu}(p_{W})&=&\left(0,-\cos\theta_{\ell_{2}}\cos\phi_{\ell_{2}},-\cos\theta_{\ell_{2}}\sin\phi_{\ell_{2}},\sin\theta_{\ell_{2}}\right),\nonumber\\
\varepsilon_{T2}^{\mu}(p_{W})&=&\left(0,\sin\phi_{\ell_{2}},-\cos\phi_{\ell_{2}},0\right).
\end{eqnarray}
Here the labels $0$, $T1,$ and $T2$ denote the longitudinal
and transverse polarizations of the outgoing vector boson. After combining Eqs.~
(\ref{sqME.EQ}), (\ref{sigmaHat.EQ}), (\ref{phaseSpace2.EQ}), and integrating 
over $dp_{N}^{2}$, as well as $d\Omega_{N},$ for the $L$-violating 
process $u_{i}\overline{d}_{j} \rightarrow \ell_{1}^{+}N \rightarrow \ell_{1}^{+}\ell_{2}^{+}W^{-}$ 
with a longitudinally polarized $W^{-}$ boson the angular distribution is
\begin{eqnarray}
\frac{d\hat{\sigma}_{0}}{d\Omega_{\ell_{2}}}&=&\frac{\hat{\sigma}(W_{0})}{2^{4}\pi}\times\{4\left[1+\left(\frac{2-\mu_{N}^{2}}{2+\mu_{N}^{2}}\right)\left(\frac{g_{R}^{\ell\:2}\vert Y_{\ell_{1}N}\vert^{2}-g_{L}^{\ell\:2}\vert V_{\ell_{1}N}\vert^{2}}{g_{R}^{\ell\:2}\vert Y_{\ell_{1}N}\vert^{2}+g_{L}^{\ell\:2}\vert V_{\ell_{1}N}\vert^{2}}\right)\cos\theta_{\ell_{2}}\right]\nonumber\\
&-&\frac{3\pi~\mu_{N}}{\left(2+\mu_{N}^{2}\right)}\left(\frac{g_{R}^{q\:2}-g_{L}^{q\:2}}{g_{R}^{q\:2}+g_{L}^{q\:2}}\right)\sin\theta_{\ell_{2}}\cos\phi_{\ell_{2}}\}.
\end{eqnarray}
Accordingly, for transversely polarized $W$ bosons the angular distributions are
\begin{eqnarray}
\frac{d\hat{\sigma}_{T1}}{d\Omega_{\ell_{2}}}=\frac{d\hat{\sigma}_{T2}}{d\Omega_{\ell_{2}}}&=&\frac{\hat{\sigma}(W_{T})}{2^{5}\pi}\times
\{
4\left[1-\left(\frac{2-\mu_{N}^{2}}{2+\mu_{N}^{2}}\right)\left(\frac
{g_{R}^{\ell\:2}\vert Y_{\ell_{1}N}\vert^{2}-g_{L}^{\ell\:2}\vert V_{\ell_{1}N}\vert^{2}}
{g_{R}^{\ell\:2}\vert Y_{\ell_{1}N}\vert^{2}+g_{L}^{\ell\:2}\vert V_{\ell_{1}N}\vert^{2}}\right)\cos\theta_{\ell_{2}}\right]\nonumber\\
&+&\frac{3\pi~\mu_{N}}{\left(2+\mu_{N}^{2}\right)}
\left(\frac{g_{R}^{q\:2}-g_{L}^{q\:2}}{g_{R}^{q\:2}+g_{L}^{q\:2}}\right)
\sin\theta_{\ell_{2}}\cos\phi_{\ell_{2}}\}.
\end{eqnarray}
In the preceding lines, we have used the following quantities
\begin{eqnarray}
\hat{\sigma}(W_{0})&\equiv&\hat{\sigma}(u\bar{d}\rightarrow \ell^+_1 N\rightarrow \ell^+_1\ell^+_2 W^{-}_{0})\nonumber\\
&=&
\frac{g^{2}~ \vert V^{CKM'}_{ji}\vert^{2}~ \vert V_{\ell_{2}N} \vert^{2}}{3N_{C}~2^{10}~\pi^{2}~(1+\delta_{\ell_{1}\ell_{2}})}
\left(g^{q~2}_{R} + g^{q~2}_{L}\right)
\left(g^{\ell\:2}_{R}\vert Y_{\ell_{1}N}\vert^{2} + g^{\ell\:2}_{L}\vert V_{\ell_{1}N}\vert^{2} \right)\nonumber\\
&\times&
\frac{m_{N}}{\Gamma_{N}}
\frac{\hat{s}}{\left[(\hat{s}-M_{W'}^{2})^{2}+(\Gamma_{W'}M_{W'})^{2}\right]}
(1-\mu_{N}^{2})^{2}(1-y_{W}^{2})^{2}(2+\mu_{N}^{2})
\left(\frac{1}{2y_{W}^{2}}\right)\\
\hat{\sigma}(W_{T})&\equiv&\hat{\sigma}(u\bar{d}\rightarrow \ell^+_1 N\rightarrow \ell^+_1\ell^+_2 W^{-}_{T})\nonumber\\
&=& \hat{\sigma}(W_{0})\times 2y_{W}^{2} 
\label{sigTTil.EQ}
\end{eqnarray}
Integrating over the azimuthal angle, the polar distributions are calculated to be
\begin{equation}
 \frac{d\hat{\sigma}_{0}}{d\cos\theta_{\ell_{2}}}=\frac{\hat{\sigma}(W_{0})}{2}\left[1+\left(\frac{2-\mu_{N}^{2}}{2+\mu_{N}^{2}}\right)\left(\frac
{g_{R}^{\ell\:2}\vert Y_{\ell_{1}N}\vert^{2}-g_{L}^{\ell\:2}\vert V_{\ell_{1}N}\vert^{2}}
{g_{R}^{\ell\:2}\vert Y_{\ell_{1}N}\vert^{2}+g_{L}^{\ell\:2}\vert V_{\ell_{1}N}\vert^{2}}\right)\cos\theta_{\ell_{2}}\right]
\end{equation}
and
\begin{equation}
 \frac{d\hat{\sigma}_{T}}{d\cos\theta_{\ell_{2}}}\equiv\frac{d(\hat{\sigma}_{T1}+\hat{\sigma}_{T1})}{d\cos\theta_{\ell_{2}}}=\frac{\hat{\sigma}(W_{T})}{2}\left[1-\left(\frac{2-\mu_{N}^{2}}{2+\mu_{N}^{2}}\right)\left(\frac
{g_{R}^{\ell\:2}\vert Y_{\ell_{1}N}\vert^{2}-g_{L}^{\ell\:2}\vert V_{\ell_{1}N}\vert^{2}}
{g_{R}^{\ell\:2}\vert Y_{\ell_{1}N}\vert^{2}+g_{L}^{\ell\:2}\vert V_{\ell_{1}N}\vert^{2}}\right)\cos\theta_{\ell_{2}}\right]
\end{equation}
After combining the two, we find that the polarization-summed polar distribution for the full 
$u_{i}\overline{d}_{j}\rightarrow \ell_{1}^{+}\ell_{2}^{+}q\overline{q}'$ 
process is
\begin{eqnarray}
\frac{d\hat{\sigma}_{Tot.}}{d\cos\theta_{\ell_{2}}}
&\equiv&\frac{d(\hat{\sigma}_{0}+\hat{\sigma}_{T})}{d\cos\theta_{\ell_{2}}}\nonumber\\
&=&\frac{\hat{\sigma}_{Tot.}}{2}\left[1+\frac{\hat{\sigma}(W_{0})-\hat{\sigma}(W_{T})}{\hat{\sigma}(W_{0})+\hat{\sigma}(W_{T})}\left(\frac{2-\mu_{N}^{2}}{2+\mu_{N}^{2}}\right)
\left(\frac
{g_{R}^{\ell\:2}\vert Y_{\ell_{1}N}\vert^{2}-g_{L}^{\ell\:2}\vert V_{\ell_{1}N}\vert^{2}}
{g_{R}^{\ell\:2}\vert Y_{\ell_{1}N}\vert^{2}+g_{L}^{\ell\:2}\vert V_{\ell_{1}N}\vert^{2}}\right)\cos\theta_{\ell_{2}}\right],
\label{angDistAppend.EQ}
\end{eqnarray}
where
\begin{equation}
\frac{\hat{\sigma}(W_{0})-\hat{\sigma}(W_{T})}{\hat{\sigma}(W_{0})+\hat{\sigma}(W_{T})}=
\frac{\hat{\sigma}(W_{0})-2y_{W}^{2}\hat{\sigma}(W_{0})}{\hat{\sigma}(W_{0})+2y_{W}^{2}\hat{\sigma}(W_{0})}
=\frac{1-2y_{W}^{2}}{1+2y_{W}^{2}},
\end{equation}
and the total partonic-level cross section is
\begin{eqnarray}
\hat{\sigma}_{Tot.}&\equiv&\hat{\sigma}(u_{i}\overline{d}_{j}\rightarrow \ell_{1}^{+}\ell_{2}^{+}q\overline{q}') \\
&=&(\hat{\sigma}(W_{0})+\hat{\sigma}(W_{T}))\times BR(W\rightarrow q\overline{q}')\\
&=&\hat{\sigma}(W_{0})(1+2y_{W}^{2})\times BR(W\rightarrow q\overline{q}')\\
&=&\frac{g^{2}~ \vert V^{CKM'}_{ji}\vert^{2}~ \vert V_{\ell_{2}N} \vert^{2}}{3N_{C}~2^{10}~\pi^{2}~(1+\delta_{\ell_{1}\ell_{2}})}
\left(g^{q~2}_{R} + g^{q~2}_{L}\right)
\left(g^{\ell\:2}_{R}\vert Y_{\ell_{1}N}\vert^{2} + g^{\ell~2}_{L}\vert V_{\ell_{1}N}\vert^{2}\right)
\left(\frac{m_{N}}{\Gamma_{N}}\right)\nonumber\\
&\times&
\frac{\hat{s}~(1-y_{W}^{2})^{2}(1-y_{W}^{2})^{2}(2+\mu_{N}^{2})}{\left[(\hat{s}-M_{W'}^{2})^{2}+(\Gamma_{W'}M_{W'})^{2}\right]}
\left(\frac{1+2y_{W}^{2}}{2y_{W}^{2}}\right)\times BR(W\rightarrow q\overline{q}').
\label{sig0Til.EQ}
\end{eqnarray}

Having instead chosen to integrate first over the polar angle before the azimuthal angle, 
the polarization-dependent azimuthal distributions for the 
$u_{i}\overline{d}_{j} \rightarrow \ell_{1}^{+}N \rightarrow \ell_{1}^{+}\ell_{2}^{+}W^{-}$ 
process are
\begin{equation}
\frac{d\hat{\sigma}_{0}}{d\phi_{\ell_{2}}}=\frac{\hat{\sigma}(W_{0})}{2\pi}\left[1-\frac{3\pi^{2}}{16}\frac{\mu_{N}}{\left(2+\mu_{N}^{2}\right)}\left(\frac{g_{R}^{q\:2}-g_{L}^{q\:2}}{g_{R}^{q\:2}+g_{L}^{q\:2}}\right)\cos\phi_{\ell_{2}}\right],
\end{equation}
and
\begin{equation}
\frac{d\hat{\sigma}_{T}}{d\phi_{\ell_{2}}} 
\equiv\frac{(d\hat{\sigma}_{T_{1}}+\hat{\sigma}_{T_{2}})}{d\phi_{\ell_{2}}}
=\frac{\hat{\sigma}(W_{T})}{2\pi}\left[1+\frac{3\pi^{2}}{16}\frac{\mu_{N}}{\left(2+\mu_{N}^{2}\right)}\left(\frac{g_{R}^{q\:2}-g_{L}^{q\:2}}{g_{R}^{q\:2}+g_{L}^{q\:2}}\right)\cos\phi_{\ell_{2}}\right].
\end{equation}
Similarly, after combining the azimuthal distributions, the total 
polarization-summed azimuthal distribution for the full 
$u_{i}\overline{d}_{j}\rightarrow \ell_{1}^{+}\ell_{2}^{+}q\overline{q}'$ 
process is
\begin{equation}
\frac{d\hat{\sigma}_{Tot.}}{d\phi_{\ell_{2}}}=\frac{\hat{\sigma}_{Tot.}}{2\pi}\left[1-\frac{3\pi^{2}}{16}\frac{\mu_{N}}{\left(2+\mu_{N}^{2}\right)}\left(\frac{\hat{\sigma}(W_{0})-\hat{\sigma}(W_{T})}{\hat{\sigma}(W_{0})+\hat{\sigma}(W_{T})}\right)\left(\frac{g_{R}^{q\:2}-g_{L}^{q\:2}}{g_{R}^{q\:2}+g_{L}^{q\:2}}\right)\cos\phi_{\ell_{2}}\right].
\end{equation}
Under the definition of the azimuthal angle, $\Phi,$ in Eq.~(\ref{AziDef.EQ}),
we have $\Phi=-\phi_{\ell_{2}},$ and consequentially recover Eq.~(\ref{Azim.EQ}):
\begin{equation}
\frac{d\hat{\sigma}_{Tot.}}{d\Phi}=\frac{\hat{\sigma}_{Tot.}}{2\pi}\left[1-\frac{3\pi^{2}}{16}\frac{\mu_{N}}{\left(2+\mu_{N}^{2}\right)}\left(\frac{\hat{\sigma}(W_{0})-\hat{\sigma}(W_{T})}{\hat{\sigma}(W_{0})+\hat{\sigma}(W_{T})}\right)\left(\frac{g_{R}^{q\:2}-g_{L}^{q\:2}}{g_{R}^{q\:2}+g_{L}^{q\:2}}\right)\cos\Phi\right].
\label{phiDistAppend.EQ}
\end{equation}

Lastly, were the NWA never applied to $N$, the differential cross section 
for the $u_{i}\overline{d}_{j}\rightarrow \ell_{1}^{+}\ell_{2}^{+}W^{-}$
process is
\begin{eqnarray}
\frac{d\hat{\sigma}}{dp_{N}^{2}} &=&
\frac{g^{2}~ \vert V^{CKM'}_{ji}\vert^{2}~ \vert V_{\ell_{2}N} \vert^{2}}{3N_{C}~2^{11}~\pi^{3}~M_{W}^{2}~(1+\delta_{\ell_{1}\ell_{2}})}
\left(g^{q~2}_{R} + g^{q~2}_{L}\right)
\left(p_{N}^{2}g_{R}^{\ell\:2}\vert Y_{\ell_{1}N}\vert^{2} + m_{N}^{2}\vert V_{\ell_{1}N}\vert^{2} g^{\ell~2}_{L}\right)
\nonumber\\
&\times&
\frac{\hat{s}~(1-\tilde{\mu}_{N}^{2})^{2}(2+\tilde{\mu}_{N}^{2})}{\left[(\hat{s}-M_{W'}^{2})^{2}+(\Gamma_{W'}M_{W'})^{2}\right]}
\frac{p_{N}^{2}~(1-\rho_{W}^{2})^{2}(1+2\rho_{W}^{2})}{\left[(p_{N}^{2}-m_{N}^{2})^{2}+(\Gamma_{N}m_{N})^{2}\right]},
\label{diffXSecNoNWA.EQ}
\end{eqnarray}
where $\tilde{\mu}_{N}^{2}\equiv p_{N}^2/\hat{s}$ and $\rho_{W}^{2}\equiv M_{W}^{2}/p_{N}^{2}$.

\subsection{Partonic-Level Angular Distributions: L-Conserving Case}
\label{DiracAngDist.APP}

For comparison, we consider the case where the heavy neutrino decays through the
following $L$-conserving process:
\begin{equation}
 u_{i}\overline{d}_{j}\rightarrow W' \rightarrow \ell_{1}^{+}N\rightarrow\ell_{1}^{+}\ell_{2}^{-}W^{+}.
\end{equation}
Following the identical arguments specified in the preceding section, the subsequent polarization-dependent angular distributions are
\begin{eqnarray}
 \frac{d\hat{\sigma}_{0}}{d\Omega_{\ell_{2}}}&=&\frac{\hat{\sigma}(W_{0})}{2^{4}\pi}\times\{4\left[1-\left(\frac{2-\mu_{N}^{2}}{2+\mu_{N}^{2}}\right)
\left(\frac
{g_{R}^{\ell\:2}\vert Y_{\ell_{1}N}\vert^{2}-g_{L}^{\ell\:2}\vert V_{\ell_{1}N}\vert^{2}}
{g_{R}^{\ell\:2}\vert Y_{\ell_{1}N}\vert^{2}+g_{L}^{\ell\:2}\vert V_{\ell_{1}N}\vert^{2}}\right)\cos\theta_{\ell_{2}}\right]\nonumber\\
&+&\frac{3\pi \mu_{N}}{\left(2+\mu_{N}^{2}\right)}\left(\frac{g_{R}^{q\:2}-g_{L}^{q\:2}}{g_{R}^{q\:2}+g_{L}^{q\:2}}\right)\sin\theta_{\ell_{2}}\cos\phi_{\ell_{2}}\},
\end{eqnarray}
and
\begin{eqnarray}
 \frac{d\hat{\sigma}_{T1}}{d\Omega_{\ell_{2}}}=\frac{d\hat{\sigma}_{T2}}{d\Omega_{\ell_{2}}}&=&\frac{\hat{\sigma}(W_{T})}{2^{4}\pi}\times\{4\left[1+\left(\frac{2-\mu_{N}^{2}}{2+\mu_{N}^{2}}\right)
\left(\frac
{g_{R}^{\ell\:2}\vert Y_{\ell_{1}N}\vert^{2}-g_{L}^{\ell\:2}\vert V_{\ell_{1}N}\vert^{2}}
{g_{R}^{\ell\:2}\vert Y_{\ell_{1}N}\vert^{2}+g_{L}^{\ell\:2}\vert V_{\ell_{1}N}\vert^{2}}\right)\cos\theta_{\ell_{2}}\right]\nonumber\\
&-&\frac{3\pi \mu_{N}}{\left(2+\mu_{N}^{2}\right)}\left(\frac{g_{R}^{q\:2}-g_{L}^{q\:2}}{g_{R}^{q\:2}+g_{L}^{q\:2}}\right)\sin\theta_{\ell_{2}}\cos\phi_{\ell_{2}}\}.
\end{eqnarray}
The polarization-summed distributions for the polar and azimuthal cases are therefore
\begin{equation}
\frac{d\hat{\sigma}_{Tot.}}{d\cos\theta_{\ell_{2}}}=
\frac{\hat{\sigma}_{Tot.}}{2}\left[1-\left(\frac{\hat{\sigma}(W_{0})-\hat{\sigma}(W_{T})}{\hat{\sigma}(W_{0})+\hat{\sigma}(W_{T})}\right)\left(\frac{2-\mu_{N}^{2}}{2+\mu_{N}^{2}}\right)
\left(\frac
{g_{R}^{\ell\:2}\vert Y_{\ell_{1}N}\vert^{2}-g_{L}^{\ell\:2}\vert V_{\ell_{1}N}\vert^{2}}
{g_{R}^{\ell\:2}\vert Y_{\ell_{1}N}\vert^{2}+g_{L}^{\ell\:2}\vert V_{\ell_{1}N}\vert^{2}}\right)\cos\theta_{\ell_{2}}\right],
\end{equation}
and
\begin{equation}
\frac{d\hat{\sigma}_{Tot.}}{d\Phi}
=\frac{\hat{\sigma}_{Tot.}}{2\pi}\left[1+\frac{3\pi^{2}}{16}\frac{\mu_{N}}{\left(2+\mu_{N}^{2}\right)}\left(\frac{\hat{\sigma}(W_{0})-\hat{\sigma}(W_{T})}{\hat{\sigma}(W_{0})+\hat{\sigma}(W_{T})}\right)\left(\frac{g_{R}^{q\:2}-g_{L}^{q\:2}}{g_{R}^{q\:2}+g_{L}^{q\:2}}\right)\cos\Phi\right],
\end{equation}
respectively, where $\sigma_{Tot.}$ is still given by Eq.(\ref{sig0Til.EQ}).
Comparison to Eqs.~(\ref{angDistAppend.EQ}) and~(\ref{phiDistAppend.EQ}) demonstrates that the slopes of the angular distributions differ in sign for the $L$-violating and $L$-conserving cases. 
Consequentially, adding the $L-$conserving and $L-$violating distributions together results in the quantitative feature
\begin{equation}
\hat{\sigma}_{Tot.}
= \frac{d\hat{\sigma}_{Tot.}^{L}}{d\cos\theta_{\ell_{2}}} + \frac{d\hat{\sigma}_{Tot.}^{\not L}}{d\cos\theta_{\ell_{2}}}
= \pi\left[
\frac{d\hat{\sigma}_{Tot.}^{L}}{d\Phi} + \frac{d\hat{\sigma}_{Tot.}^{\not L}}{d\Phi}\right],
\end{equation}
where $L$ ($\not\!\!L$) denotes the lepton number-conserving (violating) angular distributions.

\section{$W'$ and $N$ Production and Decay at the LHC}
\label{WpLHC.SEC}

For the remainder of this analysis, we consider for our various benchmark calculations only the 
pure gauge states $W'_{R}$ and $W'_{L}$, respectively given by Eq.~(\ref{pureGuageRH.EQ}) and Eq.~(\ref{pureGuageLH.EQ}), and with SM coupling strength
\begin{equation}
 g^{q,\ell}_{R,L}=g.
\end{equation}
More general results can be obtained by simple scaling. Unless explicitly stated otherwise, we take
\begin{equation}
\label{benchParam.EQ}
 M_{W'}=3\text{ TeV},\quad 
 m_{N}=500\text{ GeV},\quad
\vert V^{CKM'}_{ud}\vert^{2}=1,
 \end{equation}
 and use the CTEQ6L1 parton distribution functions (pdfs)~\cite{Pumplin:2002vw} for all hadronic-level cross section calculations.
 Explicitly, we consider only the $ud\rightarrow W'$ production mode.

 Regarding our choice of neutrino mixing parameters, for mixing between L.H. gauge states and light mass eigenstates, 
 we use the Pontecorvo-Maki-Nakagawa-Sakata (PMNS) matrix with mixing angles taken from Ref.~\cite{Beringer:1900zz}, which includes recent measurements of 
 $\theta_{13}$, and take $\delta_{CP},\alpha_{1},\alpha_{2}=0$.
 The bounds from $0\nu\beta\beta$ decay are quite severe and discourage collider searches for $L-$violation in the electronic channel.
However, neutrino mixing between the mu- or tau-flavor state and lightest heavy mass eigenstate can still be considerably larger in L.H. interactions.
Therefore, we use 
\begin{equation}
\vert V_{eN}\vert^{2}=2.5\times10^{-5},
\quad
\vert V_{\mu N}\vert^{2}=1\times10^{-3},
\quad\text{and}\quad
\vert V_{\tau N}\vert^{2}=1\times10^{-3}.
\label{mixingHeavyLH.EQ}
\end{equation}
These numerical values are in line with Eqs.~(\ref{zerovtwobBound.EQ}), (\ref{emutauMixingBounds.EQ}), and (\ref{benchParam.EQ}); 
and  furthermore, mimic the observed $\mu-\tau$ symmetry seen in mixing between flavor states and light mass eigenstates. 
Where necessary, for mixing between R.H gauge states and light mass eigenstates, we apply the unitarity condition
\begin{equation}
\sum_{m=1}^{3}\vert X_{\ell m}\vert^{2} = 1 - \sum_{m=1}^{3}\vert U_{\ell m}\vert^{2},
\quad\text{for}\quad
\ell=e,\mu,\tau.
\end{equation}
For mixing between R.H. gauge states and the lightest, heavy mass eigenstate, we apply Eq.~(\ref{mixingFromUnit.EQ}) and take
\begin{equation}
\vert Y_{\ell N}\vert^{2}=1,
\quad\text{for}\quad
\ell=e,\mu,\tau.
\label{mixingHeavyRH.EQ}
\end{equation}
\subsection{$\wpri$ Production and Decay}

\begin{figure}[!t]
\centering
\subfigure[]{	\includegraphics[width=0.45\textwidth]{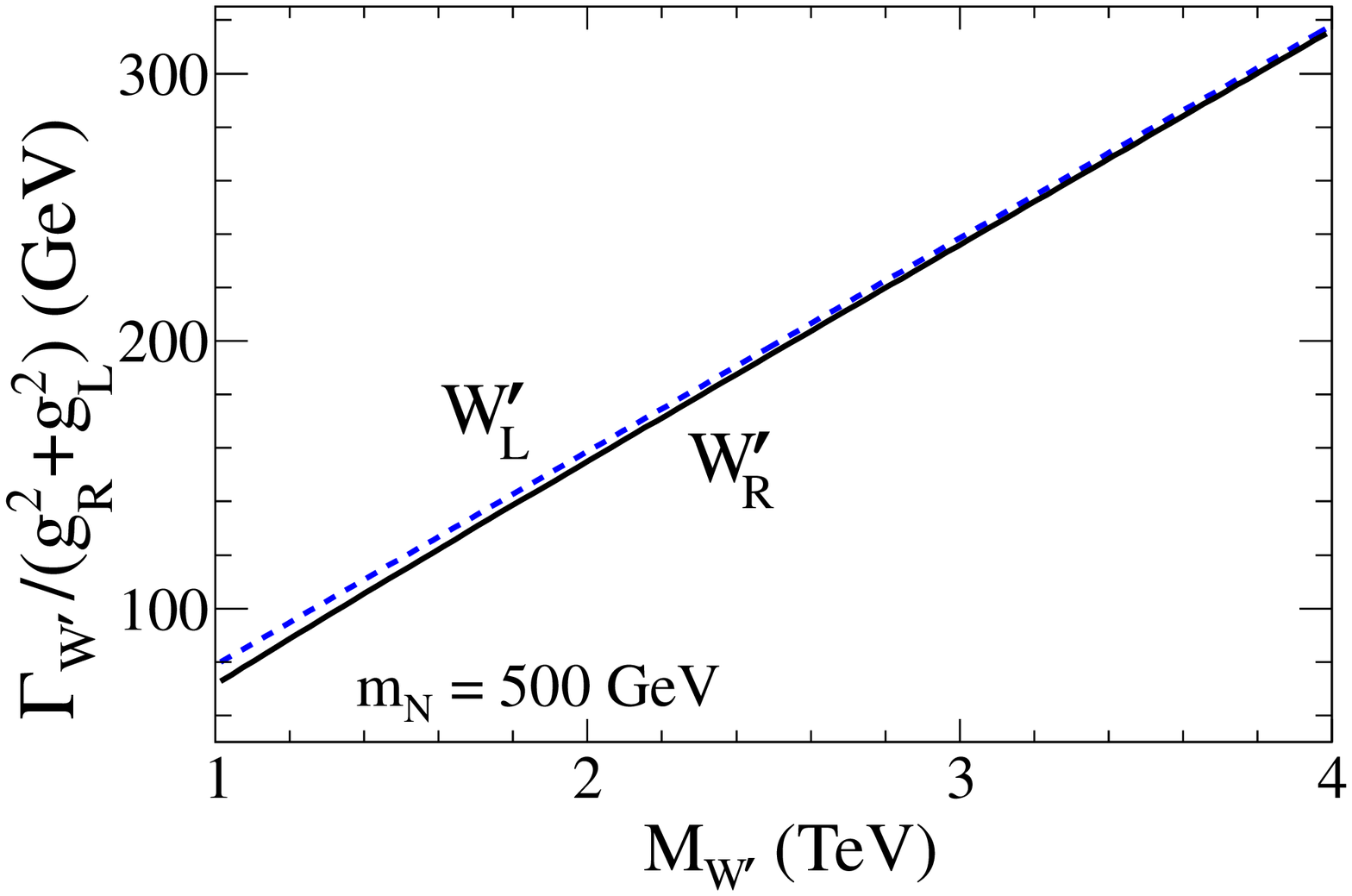}	\label{wpwidth.FIG}
}
\subfigure[]{
	\includegraphics[width=0.45\textwidth]{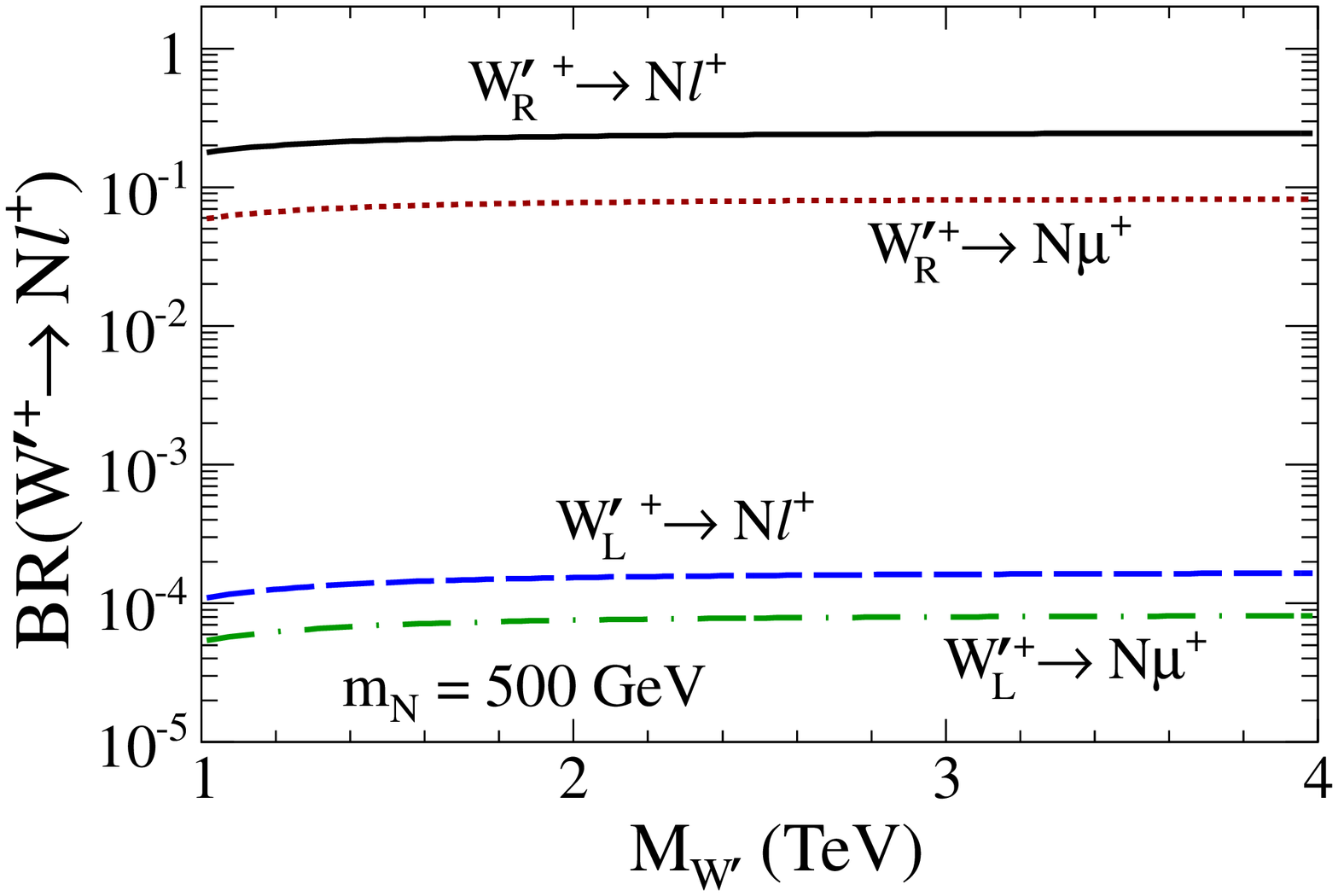}
	\label{BR.FIG}
}\vspace{.15in}\\
\subfigure[]{
	\label{mwprod_8TeV.FIG}
	\includegraphics[width=0.45\textwidth]{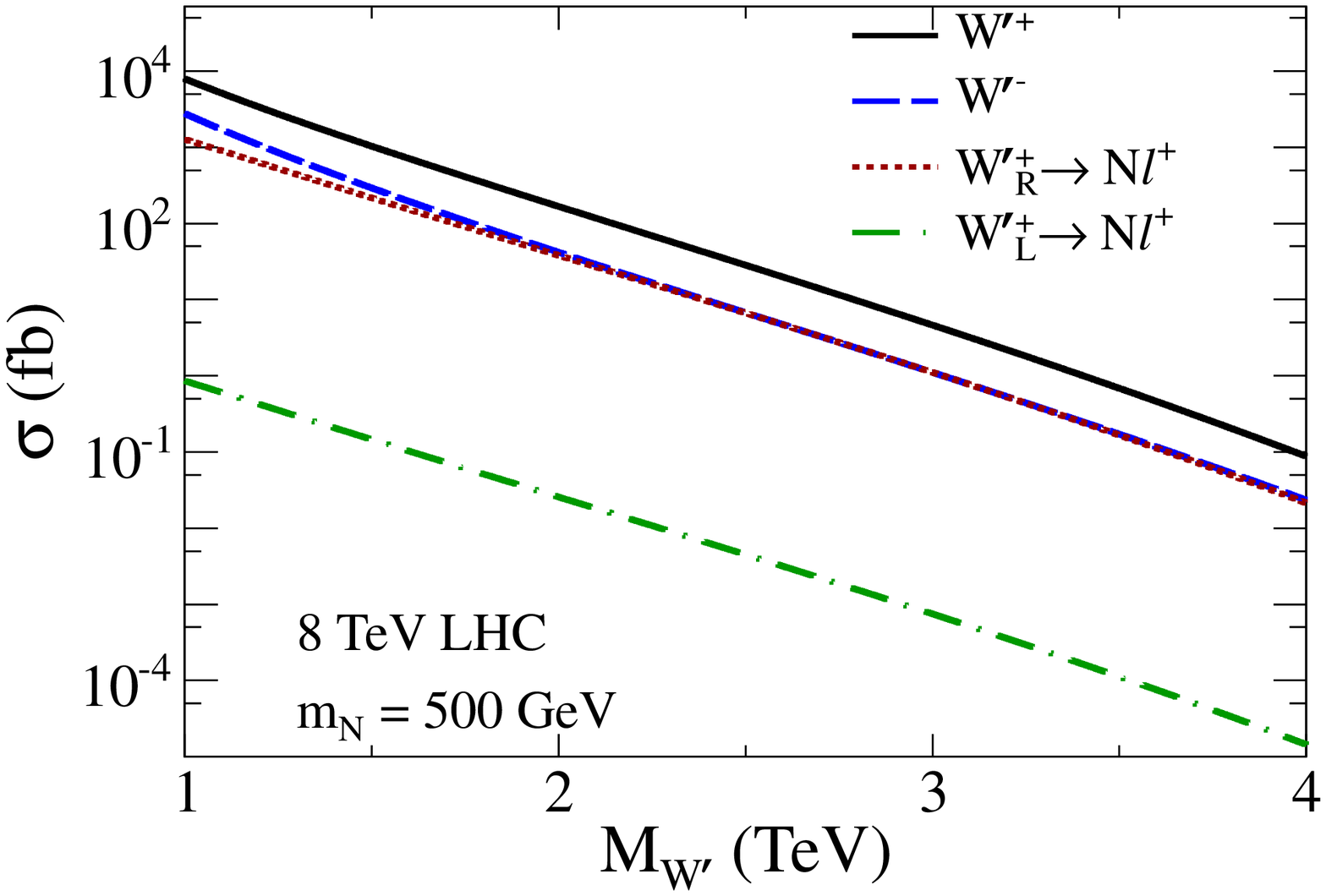}
}
\subfigure[]{
	\label{mwprod_14TeV.FIG}
	\label{mwprod.FIG}
	\includegraphics[width=0.45\textwidth]{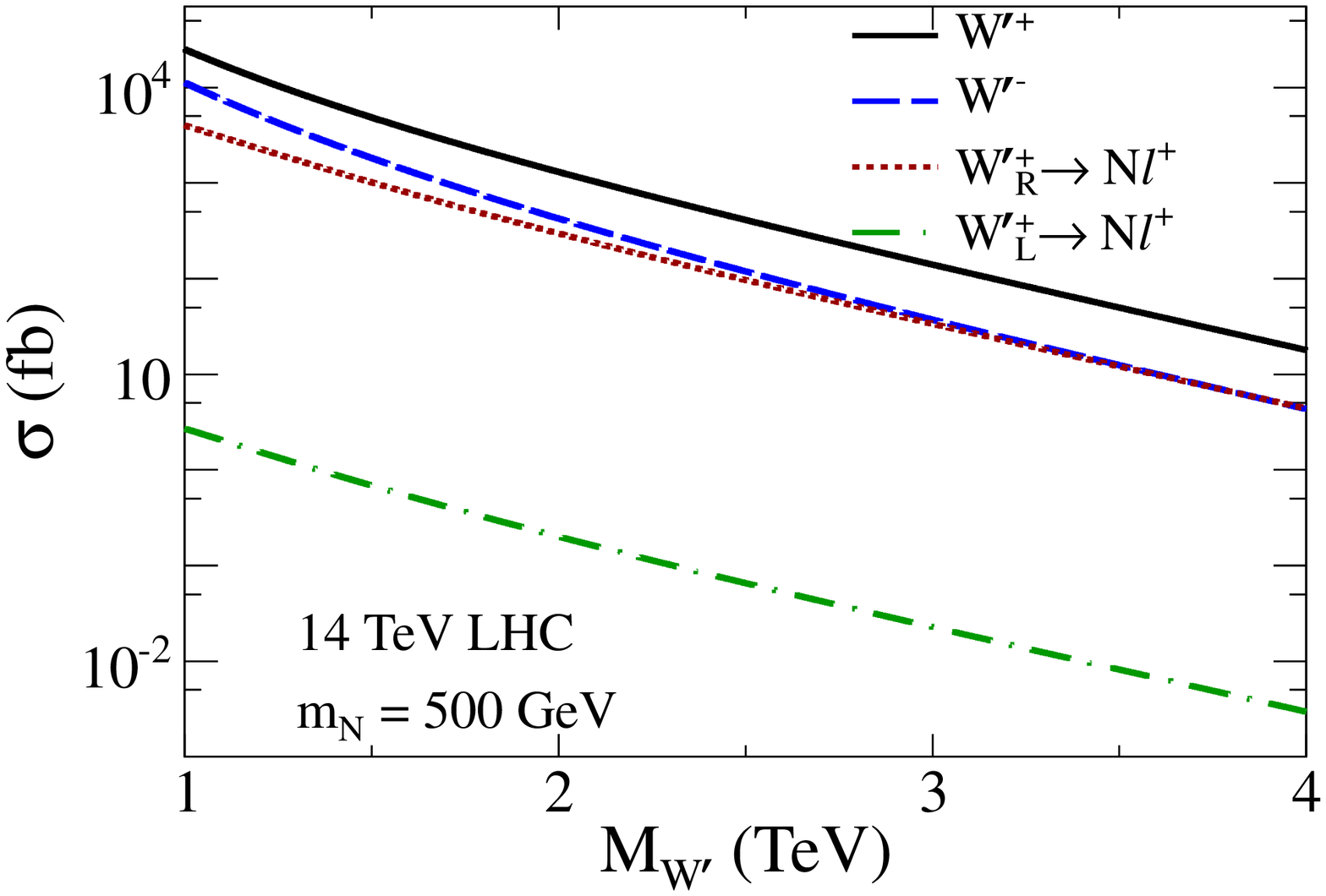}
}	
\caption[The total decay width and branching fractions for $W'_{R}$]{(a) The total decay width for $W'_{R}$ (solid) and $W'_{L}$ (dash); 
(b) the branching ratio of $\wpri_{R,L}\rightarrow N\ell^{+}$,
with subsequent $\wpri_{R}\rightarrow N\mu^{+}$ (dot) and $\wpri_{L}\rightarrow N\mu^{+}$ (dash-dot) ratios;
and the production cross sections at the (c) 8 and (d) 14 TeV LHC of $W'_{R}$ (solid), $W'_{L}$ (dash), 
$W'_{R}\rightarrow N\ell^{+}$ (dot), and $W'_{L}\rightarrow N\ell^{+}$ (dash-dot).}
\label{proddec.FIG}
\end{figure}

Under our parameterization, the partial widths for $\wpri$ decaying into a pairs of quarks are
\bea
\nonumber
\Gamma(W'\rightarrow \bar{q}q') &=& 3|{V^{\rm CKM}_{qq'}}'|^{2}(g_L^{q~2} + g_R^{q~2})\frac{\mwpri}{48\pi},
\qquad \\
\Gamma(W'\rightarrow tb) &=& 3|{V^{\rm CKM}_{tb}}'|^{2}(g_L^{q~2} + g_R^{q~2})\frac{\mwpri}{48\pi}\bigg{(}1-x_{t}^{2}\bigg{)}^2\bigg{(}1+\frac{1}{2}x_{t}^{2}\bigg{)}, \qquad 
\label{parttop.EQ}
\eea
where $x_{i}=m_{i}/M_{W'}$, and the factors of three represent color multiplicity. Likewise, the partial widths of the $W'$ decaying to leptons are
\bea
\Gamma(W'\rightarrow \ell \nu_m)&=&\left(g^{\ell~2}_R |X_{\ell m}|^2 + g^{\ell~2}_L|U_{\ell m}|^{2}\right)\frac{\mwpri}{48\pi},\qquad \\
\Gamma(W'\rightarrow \ell N)        &=&\left(g^{\ell~2}_R |Y_{\ell N}|^2 + g^{\ell~2}_L|V_{\ell N}|^{2}\right)\frac{\mwpri}{48\pi}\bigg{(}1-x_{N}^{2}\bigg{)}^2\bigg{(}1+\frac{1}{2}x_{N}^{2}\bigg{)}. \qquad 
\eea
Summing over the partial widths, the full widths are found to be
\begin{eqnarray}
\Gamma_{W'_{R}} &=& \frac{M_{W'}}{32\pi} \left[ 4 + (1-x_{t}^{2})^2(2+x_{t}^{2}) + (1-x_{N}^{2})^2(2+x_{N}^{2})\frac{1}{3}\sum_{\ell=e}^{\tau}\vert Y_{\ell N}\vert^{2} 
+ \frac{2}{3}\sum_{m=1,\ell=e}^{3,\tau}\vert X_{\ell m}\vert^{2}\right] \\
\Gamma_{W'_{L}} &=& \frac{M_{W'}}{32\pi} \left[ 4 + (1-x_{t}^{2})^2(2+x_{t}^{2}) + (1-x_{N}^{2})^2(2+x_{N}^{2})\frac{1}{3}\sum_{\ell=e}^{\tau}\vert V_{\ell N}\vert^{2}  
+\frac{2}{3}\sum_{m=1,\ell=e}^{3,\tau}\vert U_{\ell m}\vert^{2}\right].
\end{eqnarray}

As a function of $M_{W'}$, Fig.~\ref{proddec.FIG} shows (a) the total $\wpri$ decay width; 
(b) the branding ratio (BR) of $\wpri\rightarrow N\ell$, for $\ell=e,\mu,\tau$, defined as the ratio of the partial width to the total $\wpri$ width, $\Gamma_\wpri$:
\bea
{\rm BR}(\wpri \rightarrow \ell N)=\frac{\Gamma(\wpri \rightarrow \ell N)}{\Gamma_{\wpri}};
\eea
and the production cross sections for the pure gauge eigenstates $W'_{R,L}$, 
along with $pp\rightarrow {\wpri}^+_{R,L}\rightarrow N\ell^+$ in (c) 8 TeV and (d) 14 TeV $pp$ collisions.

The production cross section of the $W'$ and its subsequent decay to $N$ is calculated in the usual fashion~\cite{Denner:1992}.
The treatment of our full $2\rightarrow4$ process, on the other hand, is addressed in Section~\ref{appendME.APP}.
Since the $u$-quark is more prevalent in the proton than the $d$-quark, 
and since the dominate subprocess of $W'^+$ ($W'^-$) production at the LHC is $u\bar{d}\rightarrow {\wpri}^+$ ($d\bar{u} \rightarrow {\wpri}^-$), 
the production cross section of ${\wpri}^+$ is greater than the ${\wpri}^-$ cross section.
In a similar vein, the mixing between L.H. interaction states and heavy neutrino mass eigenstates is suppressed by $\vert V_{\ell N}\vert^{2}\sim\mathcal{O}(10^{-3})$,
whereas the mixing between R.H. interaction states and heavy neutrino mass eigenstates is proportional to $\vert Y_{\ell N}\vert^{2}\sim\mathcal{O}(1)$.
Consequently, the $W'_{L}\rightarrow N\ell$ branching ratio, and hence the $pp\rightarrow W'_{L}\rightarrow N\ell$ cross section, 
is roughly three orders of magnitude smaller than the $W'_{R}$ rates.

\begin{figure}[!t]
\centering
\subfigure[]{
      \includegraphics[clip,width=0.45\textwidth]{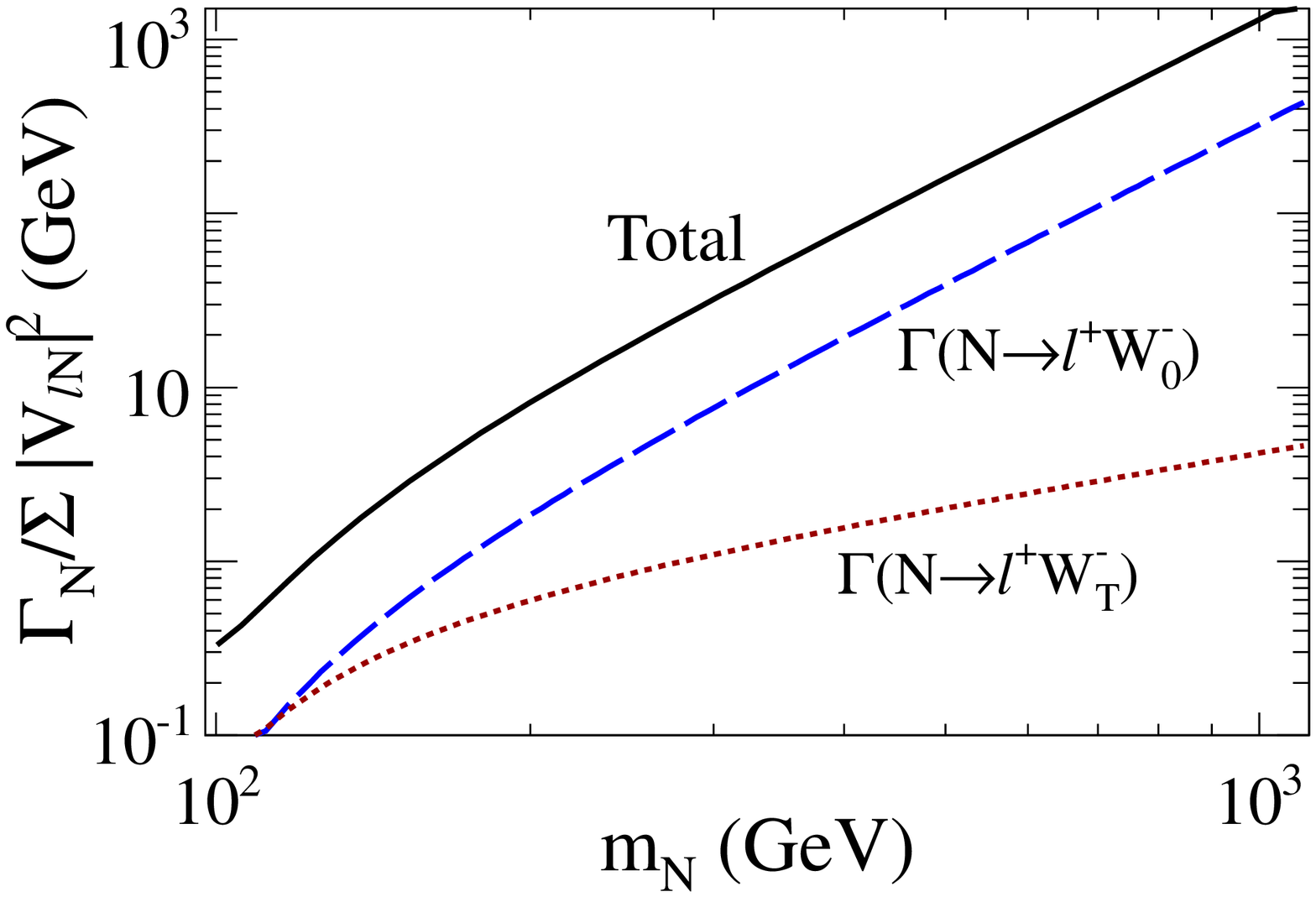}
\label{Nwidth.FIG}
}
\subfigure[]{
      \includegraphics[clip,width=0.45\textwidth]{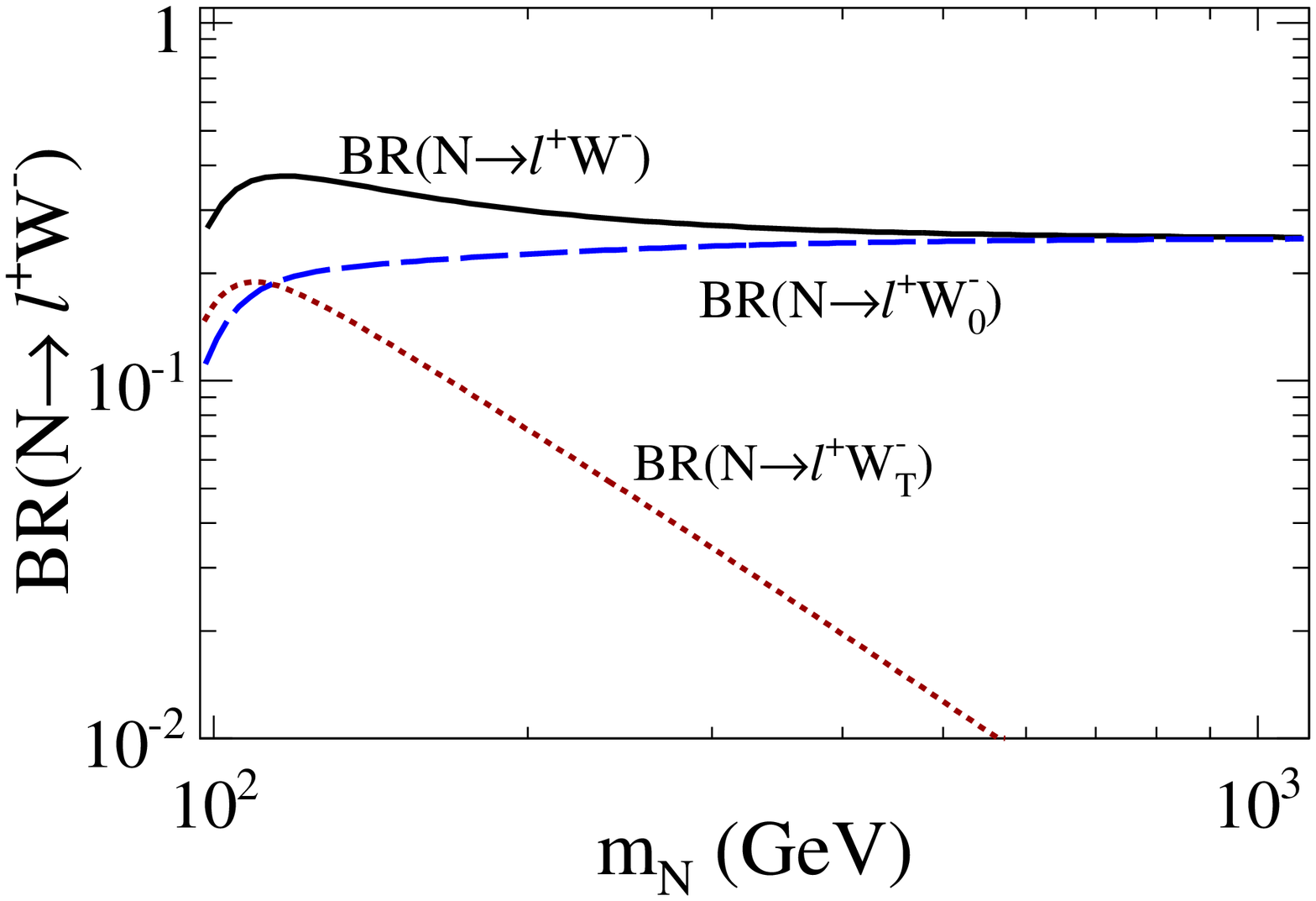}
\label{NBR.FIG}
}
\caption[Heavy Majorana $N$ total width and branching fractions]{As a function of heavy neutrino mass, (a) the total $N$ width and the $N\rightarrow\ell^{+}W^{-}_{\lambda}$ partial widths,
and (b) the combined $N\rightarrow \ell^{+} W^{-}$ and individual $N\rightarrow \ell^{+} W^{-}_{\lambda}$ branching ratios 
for longitudinal $(\lambda=0)$ and transverse $(\lambda=T)$ $W$ polarizations.} 
\label{NW.FIG}
\end{figure}
\subsection{Heavy Neutrino Decay}
\label{NeutDec.SEC}

A heavy neutrino with mass of a few hundred GeV or more can decay through on-shell SM gauge and Higgs bosons.  
The partial widths of the lightest heavy neutrino are
\bea
\Gamma(N\rightarrow \ell^{\pm} W^\mp_0)&\equiv\Gamma_0=&\frac{g^2}{64\pi M^2_W}|V_{\ell N}|^2m^3_N(1-y^2_W)^2\nonumber\\
\Gamma(N\rightarrow \ell^{\pm} W^\mp_T)&\equiv\Gamma_T=&\frac{g^2}{32\pi}|V_{\ell N}|^2m_N\left(1-y^2_W\right)^2\nonumber\\
\Gamma(N\rightarrow \nu_\ell Z)&\equiv\Gamma_Z=&\frac{g^2}{64\pi M_{W}^2}|V_{\ell N}|^2m_N^3(1-y^2_Z)^2\left(1+2y^2_{Z}\right)\nonumber\\
\Gamma(N\rightarrow \nu_\ell H)&\equiv\Gamma_H=& \frac{g^2}{64\pi M^2_W}|V_{\ell N}|^2  m^3_N(1-y^2_H)^2\label{NWidth.EQ}
\eea
where $W_{0,T}$ are longitudinally and transversely polarized $W$'s, respectively, and $y_i=M_i/m_N$.
The decays of the heavy neutrino through a $\wpri$ are not kinematically accessible. 
The total width is
\begin{eqnarray}
\Gamma_{\rm N}=\sum^\tau_{\ell = e}\left(2(\Gamma_0+\Gamma_T)+\Gamma_Z+\Gamma_H\right)
\end{eqnarray}
where the factor of two in front of $\Gamma_{0,T}$ is from the sum over positively and negatively charged leptons.

Figure \ref{Nwidth.FIG} shows the total decay width (solid) and the partial decay widths to positively charged lepton (dashed) normalized to the sum over the mixing matrices.  
For this plot the mass of the SM Higgs boson is set to $125$~GeV.  
The normalized width grows dramatically with mass due to decays into longitudinally polarized $W$'s and $Z$'s and the Higgs boson.  
Although the width appears to be large at high neutrino mass, for mixing angles on the order of a percent or less the width is still narrow.

\begin{figure}[!t]
\centering
\subfigure[]{
      \includegraphics[width=0.45\textwidth,angle=0]{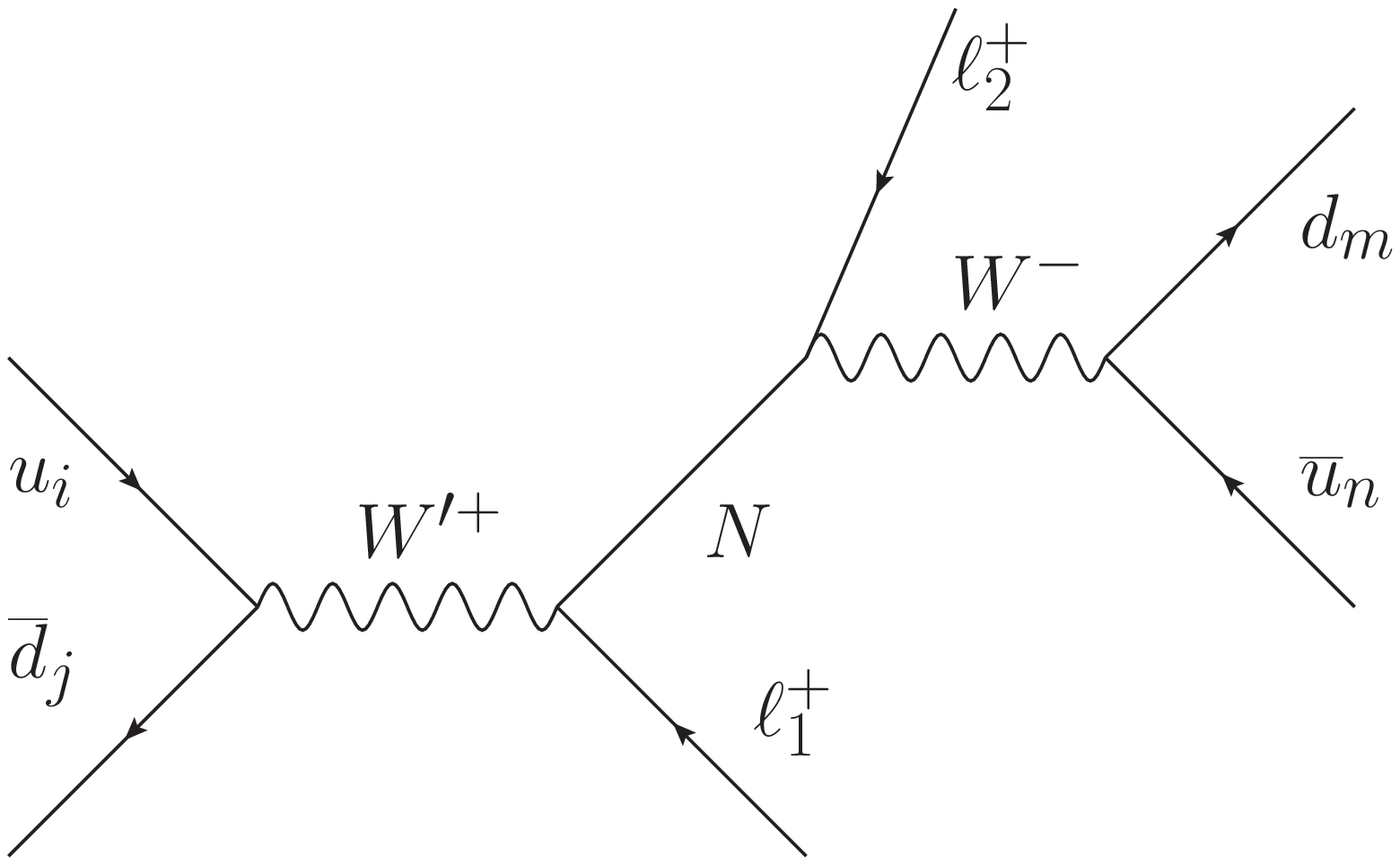}
\label{qq2Wp2llqq1.FIG}
}
\subfigure[]{
    \includegraphics[width=0.45\textwidth,angle=0]{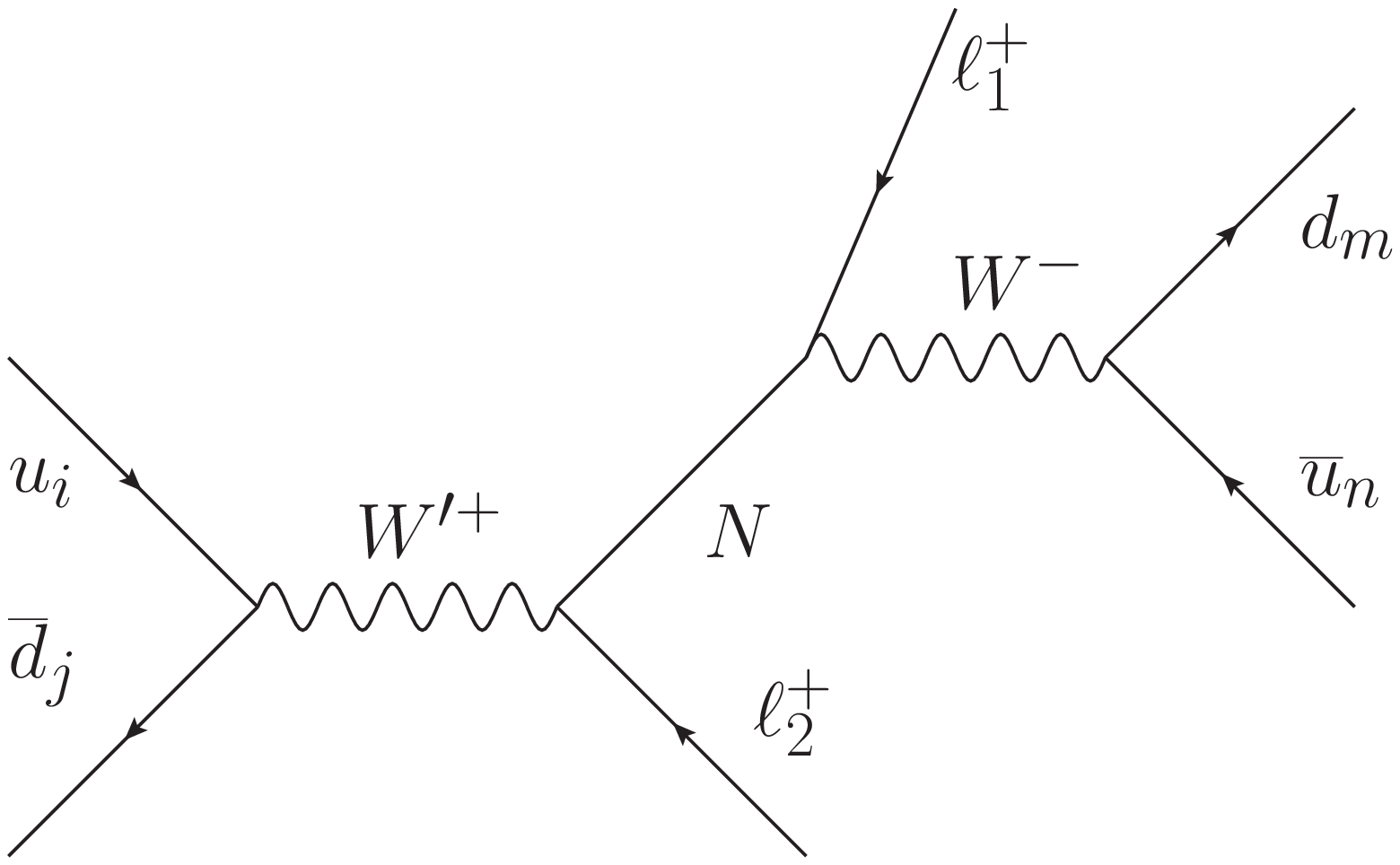}
}
\caption[Partonic level process for a heavy ${\wpri}^+$ production and decay to like sign leptons]{
The partonic level process for a heavy ${\wpri}^+$ production and decay to like sign leptons in hadronic collisions. 
}
\label{qq2Wp2llqq.FIG}
\end{figure}

Also of interest is the branching ratio (BR) of heavy neutrinos into charged leptons:
\bea
\rm{BR}\left(N\rightarrow \ell^\pm W^\mp\right)=\frac{\sum^\tau_{\ell=e} \left(\Gamma_0+\Gamma_T\right)}{\Gamma_{\rm Tot}}
\label{NBR.EQ}
\eea
Figure \ref{NBR.FIG} shows the total BR of the heavy neutrino into positively charged leptons (solid) and individually the BR into longitudinally (dashed) and transversely (dotted) polarize $W$'s as a function of neutrino mass.  The BR's into negatively charged leptons are the same.  As the mass of the neutrino increases the $Z$ and Higgs decay channels open, hence the branching ratio into charged leptons decreases.  Since $\Gamma_0$ grows more quickly with neutrino mass than $\Gamma_T$, for $m_N\gg M_W$ the total BR converges to the BR into longitudinally polarized $W$'s.
  Also, at high neutrino masses
\bea
\Gamma_0\approx\Gamma_H\approx \Gamma_Z
\eea
Hence the total width approaches $4\Gamma_0$ and, from Eq.~(\ref{NBR.EQ}), the branching ratio into a positively charged leptons is approximately $0.25$.
This is a manifestation of the Goldstone Equivalence Theorem when taking $m_N$ and $V_{\ell N}$ as independent parameters.

\section{Like-Sign Dilepton Signature}
\label{Like.SEC}

\begin{figure}[!t]
\centering
\subfigure[]{
      \includegraphics[clip,width=0.45\textwidth]{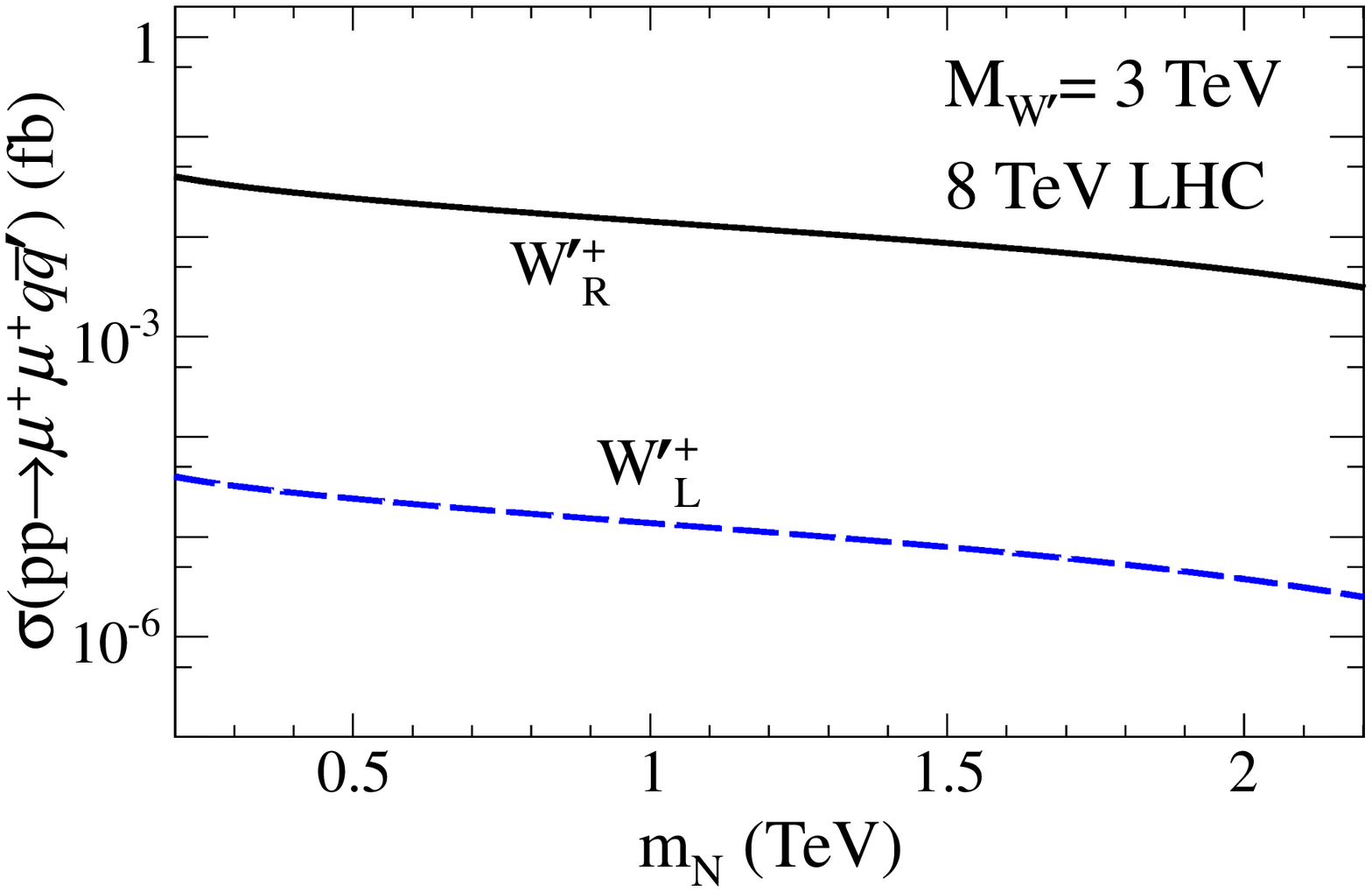}
}
\subfigure[]{
      \includegraphics[clip,width=0.45\textwidth]{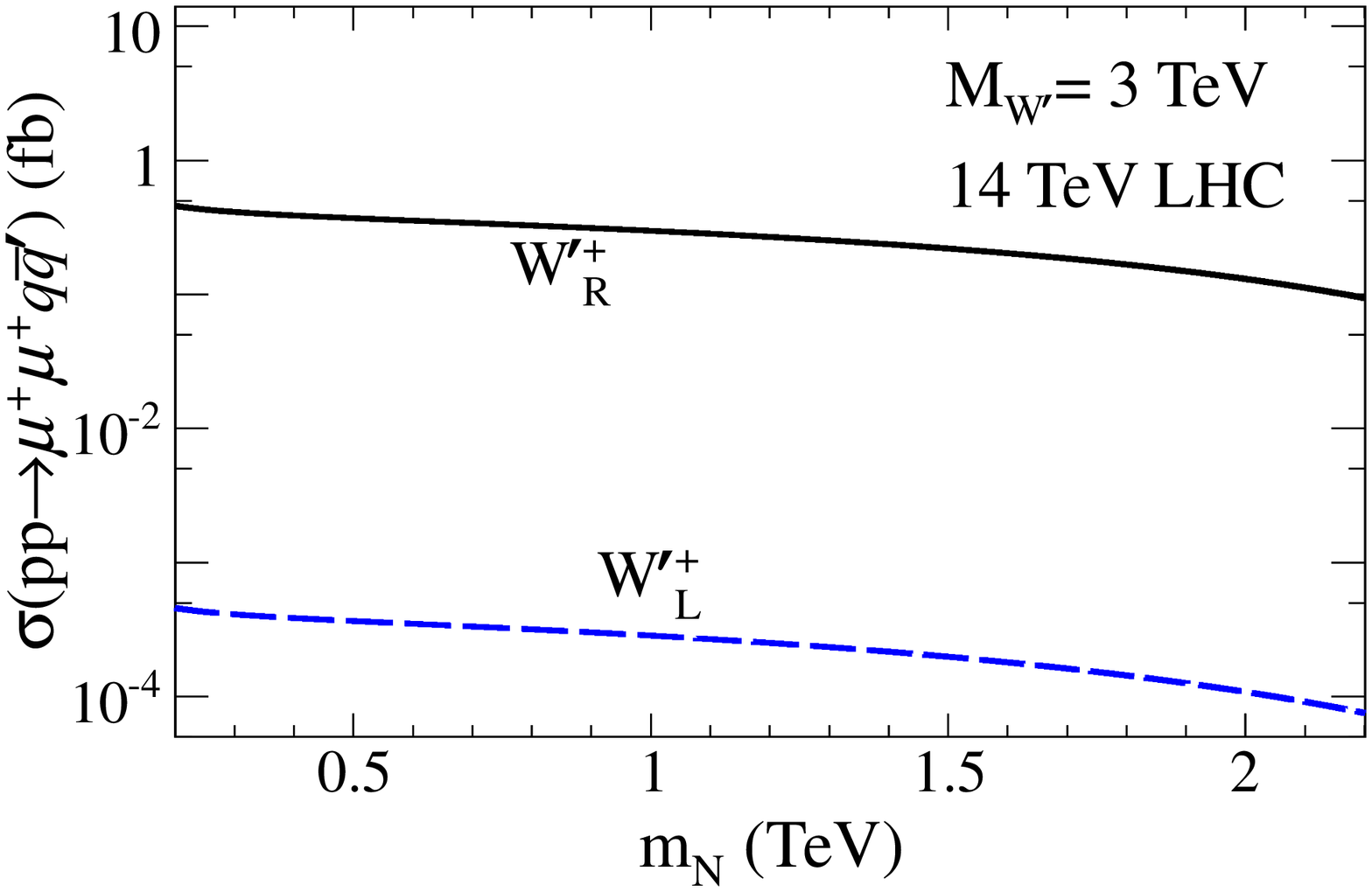}    
}
\caption[Total cross section of $pp\rightarrow {W'}^+\rightarrow \mu^+\mu^+W^{-}$ times $W^{-}\rightarrow q\bar{q}'$ branching ratio]{Total cross section of $pp\rightarrow {W'}^+\rightarrow \mu^+\mu^+W^{-}$ times $W^{-}\rightarrow q\bar{q}'$ branching ratio versus heavy neutrino mass at (a) 8 and  (b) 14 TeV. 
Solid (dashed) line corresponds to $W'_R$ ($W'_L$) gauge state.}
      \label{NeutXsect.fig}
\end{figure}

A distinctive feature of Majorana neutrinos is that they facilitate $L$-violating processes,
and to study this behavior at the LHC we consider the $L$-violating cascade
\begin{eqnarray}
u(p_A)~\bar{d}(p_B)\rightarrow W_{R,L}^{'+} (q)\rightarrow \ell^+_1(p_1)~N(p_{N}) \rightarrow \ell^+_1(p_1)~\ell^+_2(p_2) ~q(p_3)~ \bar{q}'(p_4).
\label{sig.EQ}
\end{eqnarray}
The two diagrams that contribute to this process are shown in Fig.~\ref{qq2Wp2llqq.FIG}. Figure \ref{NeutXsect.fig} 
shows the total production cross section for the like-sign dimuon process as a function of $m_{N}$. 
In it, the solid line denotes the pure $W'_R$ gauge state while the dashed line represents the pure $W'_L$ state.  
Since the $W'_R\rightarrow N\mu$ branching ratio is larger than $W'_L\rightarrow N\mu$ ratio, the cross section for $W'_R$ is systematically larger than for $W'_L$. 
Additionally, as the neutrino mass approaches the $W'$ mass the cross section drops precipitously due to phase space suppression.

In principle, the conjugate process, $\bar{u}d\rightarrow W^{'-}$, should also be possible at the LHC.
However, it will possess a much smaller production rate because the $\bar{u}d$ initial-state has a smaller parton luminosity than $u\bar{d}$.  
Despite this, all reconstruction methods and observables discussed below are applicable to both processes. 

\subsection{Event Selection}
For simplicity, we restrict our study to like-sign muons.
There is no change in the analysis if extended to electrons; however, $\etmiss$ requirements must be reassessed for inclusion of unstable $\tau$'s~\cite{AguilarSaavedra:2012fu,AguilarSaavedra:2012gf}.
Consequently, our signal consists strictly of two positively charged leptons and two jets, a fact that allows for considerable background suppression.
In simulating this like-sign leptons plus dijet signal, to make our analysis more realistic, 
we smear the lepton and jet energies to emulate real detector resolution effects.
These effects are assumed to be Gaussian and parameterized by
\bea
\frac{\sigma(E)}{E}=\frac{a}{\sqrt{E}}\oplus b,
\label{eq:smear}
\eea
where $\sigma(E)/E$ is the energy resolution, $a$ is a sampling term, 
$b$ is a constant term, $\oplus$ represents addition in quadrature, and all energies are measured in GeV.  
For leptons we take $a=5\%$ and $b=0.55\%$, and for jets we take $a=100\%$ and $b=5\%$ \cite{Ball:2007zza}.  

After smearing, we define our candidate event as two positively charged leptons and two jets passing the following 
basic kinematic and fiducial cuts on the transverse momentum, $p_T$, and pseudorapidity, $\eta$:
\begin{eqnarray}
p_T^j\geq 30~{\rm GeV},~ p_T^\ell\geq~{\rm 20~GeV},~ \eta_j\leq 3.0,~  \eta_\ell\leq 2.5.
\label{cuts1.EQ}
\end{eqnarray}
Table~\ref{WpNxsect.TAB} lists the cross sections for Eq.~(\ref{sig.EQ}) assuming the pure $\wpri_{R,L}$ gauge states at the 8 and 14 TeV LHC without smearing or acceptance cuts (row 1), 
and with smearing plus acceptance cuts from Eq.~(\ref{cuts1.EQ}) (row 2).
Here and henceforth, we assume a 100$\%$ efficiency for lepton and jet identification.

 \begin{table}[!t]
 \caption[Cross section for $pp\rightarrow {W'}_{L,R}^+\rightarrow \mu^+\mu^+ q\overline{q}'$]{Cross section for $pp\rightarrow {W'}_{L,R}^+\rightarrow \mu^+\mu^+ q\overline{q}'$ after consecutive cuts for 8 and 14 TeV LHC. 
}
\begin{center}
\begin{tabular}{|c|c|c|c|c|} \hline \hline
                         \multirow{2}{*}{ $\displaystyle\sigma$(fb)}&\multicolumn{2}{c|} {8 TeV}& \multicolumn{2}{c|} {14 TeV} \\\cline{2-5}
                                  &$\wpri_L$             &$\wpri_R$& $\wpri_L$            & $\wpri_R$   \\ \hline \hline
~~Reco.~without Cuts or Smearing  & $4.6\times10^{-5}$   & $0.046$ & $9.3\times10^{-4}$   & $0.95$ \\ \hline
~~~+~Smearing~+~Fiducial~+~Kinematics (Eq.~(\ref{cuts1.EQ})) & $4.0\times10^{-5}$& 0.035 & $8.2\times10^{-4}$ & 0.71 \\ \hline
~+~Isolation~(Eq.~(\ref{cuts2.EQ}))&$2.1\times10^{-5}$   & 0.027   & $3.2\times10^{-4}$   & 0.50 \\ \hline
~~~~~~+$\not\!\!E_{T}$~+~$m_{jj} $~Requirements~(Eq.~(\ref{cuts3.EQ})) & $1.7\times10^{-5}$ & 0.023 & $2.6\times10^{-4}$ & 0.42 \\ \hline 
~~~~+~Mass~Req.~(Eq.~(\ref{cuts4.EQ})) & $7.2\times10^{-6}$ & 0.012 & $2.0\times10^{-4}$ & 0.35  \\ \hline\hline
~$\sigma$(All~Cuts)/$\sigma$(Smearing~+~Fid.~+~Kin.)&  18\% & 35\% & 25\%  & 49\% \\   \hline\hline
\end{tabular} 
\label{WpNxsect.TAB}
\end{center}
\end{table}

The goal of this analysis is to unambiguously determine the properties of $W'$ and $N$.
To do so, our candidate leptons and jets must be well-defined and well-separated, 
that latter of which is measured by
\bea
\Delta R_{ij} = \sqrt{(\Delta \phi_{ij})^2+(\Delta\eta_{ij})^2},
\eea
where $\Delta\phi_{ij}$ and $\Delta\eta_{ij}$ are the difference in the azimuthal angles and rapidities, respectively, of particles $i$ and $j$. 
Subsequently, we apply isolation cuts on our candidate objects: 
\begin{eqnarray}
\Delta R^{\rm min}_{\ell j} \ge 0.4,~~\Delta R_{jj}\ge 0.3
\label{cuts2.EQ}
\end{eqnarray}
for all lepton and jet combinations, where $\Delta R^{\rm min}_{\ell j}$ is defined as
\bea
\Delta R^{\rm min}_{\ell j} = \min_{i=\wpri, N} \Delta R^{\rm min}_{\ell_i j}.
\label{delRMinDef.EQ}
\eea
In Eq.~(\ref{delRMinDef.EQ}), the subscript $i=W',N$ on $\ell_{i}$ denotes the identified parent particle of $\ell_{i}$.
 The effects of the isolation cuts applied at both the 8 and 14 TeV LHC are shown in the third row of Table~\ref{WpNxsect.TAB}.  
To understand the origin of these precise numbers and parent-particle identification, we digress  
to succinctly connect properties of our chiral Lagrangian to the final-state kinematical distributions.

\subsection{Characteristics of Kinematical Distributions}
\label{Recon.SEC}
\begin{figure}[!t]
\centering
\subfigure[]{        \includegraphics[width=0.48\textwidth]{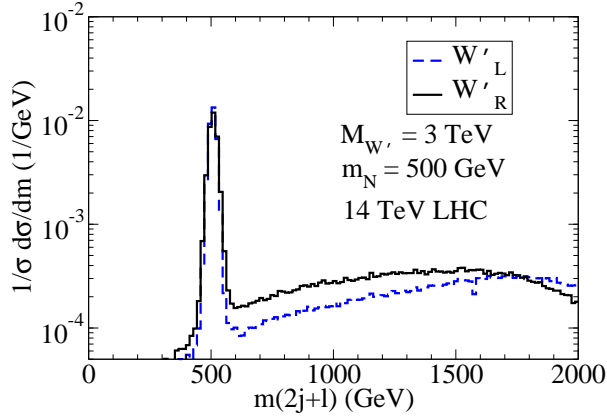}}
\caption[Invariant mass distribution of $m_{\ell_{1}jj}$ and $m_{\ell_{2}jj}$]{Invariant mass distribution of $m_{\ell_{1}jj}$ and $m_{\ell_{2}jj}$, where $\ell_{i}$ for $i=1,2$ originates from either the $W'$ or $N$. 
Cuts from Eqs.~(\ref{cuts1.EQ}) and (\ref{cuts2.EQ}) as well as the energy smearing are applied.}
\label{m2jl.FIG}
\end{figure}

Our signal suffers from a very evident ambiguity: either lepton can originate from the neutrino decay.
The origin of each lepton must thus be determined in order to fully reconstruct an event.
As noted in section~\ref{WpLHC.SEC}, the width of $N$ is narrow.
Consequently, there is a very small probability for the phase space of each diagram in Fig.~\ref{qq2Wp2llqq.FIG} to overlap,
meaning that the interference of the two diagrams is negligible.
In fact, in the $\wpri_R$ case, the interference is exactly zero because the charged lepton from the $N$ decay is left-handed while the charged lepton from the $\wpri_R$ is right-handed.  
Furthermore, since the two diagrams add incoherently, it is reasonable to expect that only one diagram contributes at a time. 
Intuitively, this means that only one of the two following momentum combinations will closely reconstruct the heavy neutrino mass:
\begin{eqnarray}
m^2_{1jj}=(p_1+p_3+p_4)^2~~~{\rm or}~~~m^2_{2jj}=(p_2+p_3+p_4)^2,
\label{mNPermu.EQ}
\end{eqnarray}
where $p_{3}$ and $p_{4}$ are the momenta of our final-state jets.

After calculating both permutations of $m_{N}$ (Fig.~\ref{m2jl.FIG}), the appearance of the $N$ mass peak is stark. 
Using the central value of the mass peak, $m_{N}^{Reco.}$, we identify the charged lepton from the $N$ decay as the 
charged lepton from our candidate event that most closely recovers $m_{N}^{Reco.}$, i.e.,
\begin{equation}
\Delta m_{min} = \min_{i=1,2} \vert m_{ijj} - m_{N}^{Reco.} \vert,
\end{equation}
where $m_{ijj}$ for $i=1,2$ is defined by Eq.~(\ref{mNPermu.EQ}).

\begin{figure}[!t]
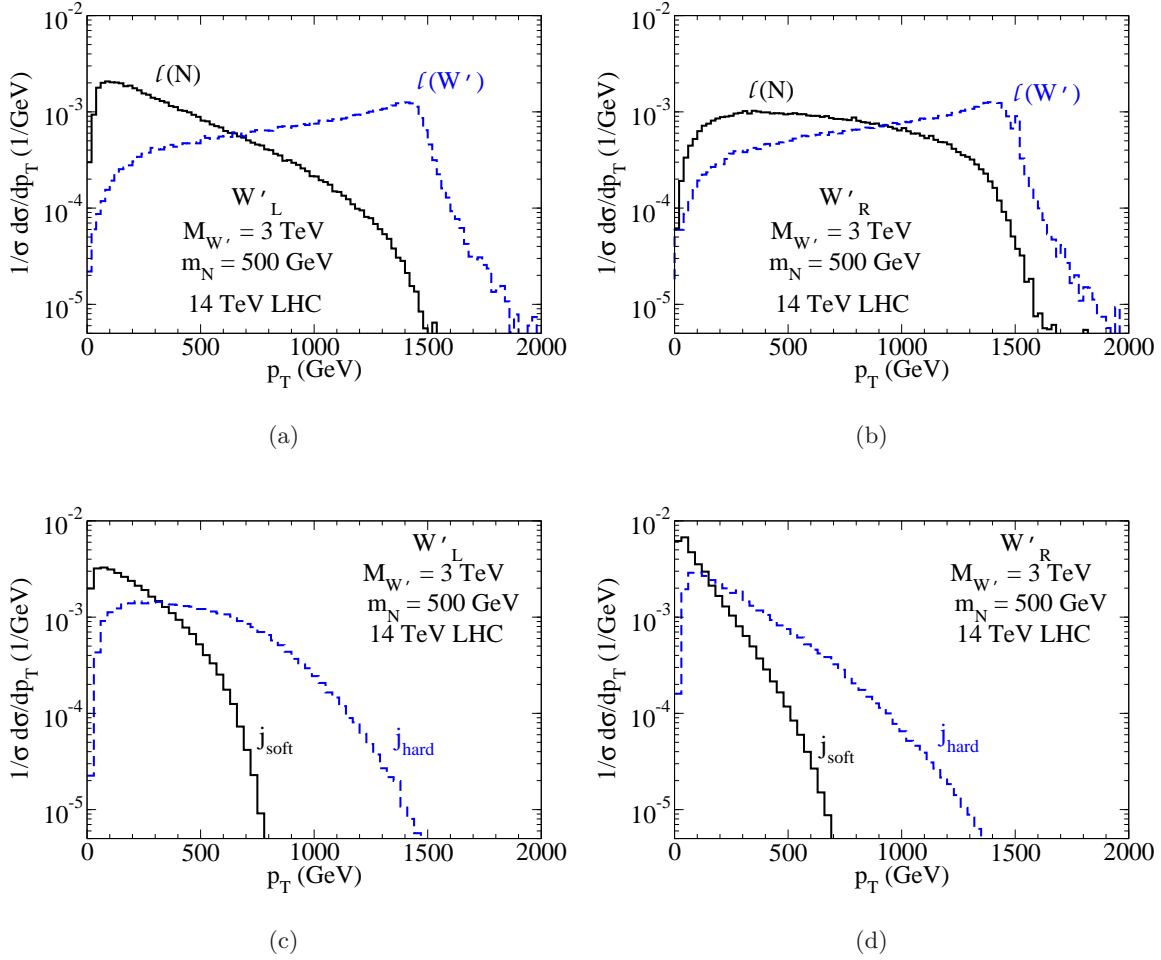

\centering
\subfigure[]{
	\label{ptlLH.FIG}
	\includegraphics[width=0.45\textwidth]{./05_WprimeHeavyN/ptlLH500_3.eps}
	}
\subfigure[]{
	\includegraphics[width=0.45\textwidth]{./05_WprimeHeavyN/ptlRH500_3.eps}
	\label{ptlRH.FIG}
}\vspace{.15in}\\
\subfigure[]{
       \includegraphics[width=0.45\textwidth]{./05_WprimeHeavyN/ptjLH500_3.eps}
       \label{ptjLH.FIG}
}
\subfigure[]{
       \includegraphics[width=0.45\textwidth]{./05_WprimeHeavyN/ptjRH500_3.eps}
}
\caption[Transverse momentum distributions]{ Transverse momentum distributions for (a,b) the lepton identified as originating from the $\wpri$ (dashed) and neutrino (solid), 
and (c,d) the hardest (dashed) and softest (solid) jets in $pp\to W'\to \ell^+ \ell^+ j j$ production.  
The $\wpri_L$ case is represented in (a,c) and the $\wpri_R$ case in (b,d). The energy smearing has been applied. 
}
\label{ptdist.FIG}
\end{figure}

Independent of reconstructing $N$, the charged lepton associated with the $W'$ decay can be identified 
by analyzing the transverse momentum, $p_{T}$, distributions of our final-state objects.
In Fig.~\ref{ptdist.FIG}, the $p_{T}$ distributions of the charged leptons (a,b) and jets (c,d) for the $\wpri_L$ (a,c) and $\wpri_R$ (b,d) gauge states.
As expected, the lepton identified as originating from the $W'$ has a Jacobian peak around $M_{W'}/2$ for both the $\wpri_L$ and $\wpri_R$ cases.   
To understand the other distributions, we consider spin correlations.

\begin{figure}[!t]
\begin{center}
\subfigure[]{
\includegraphics[width=0.35\textwidth,clip=true]{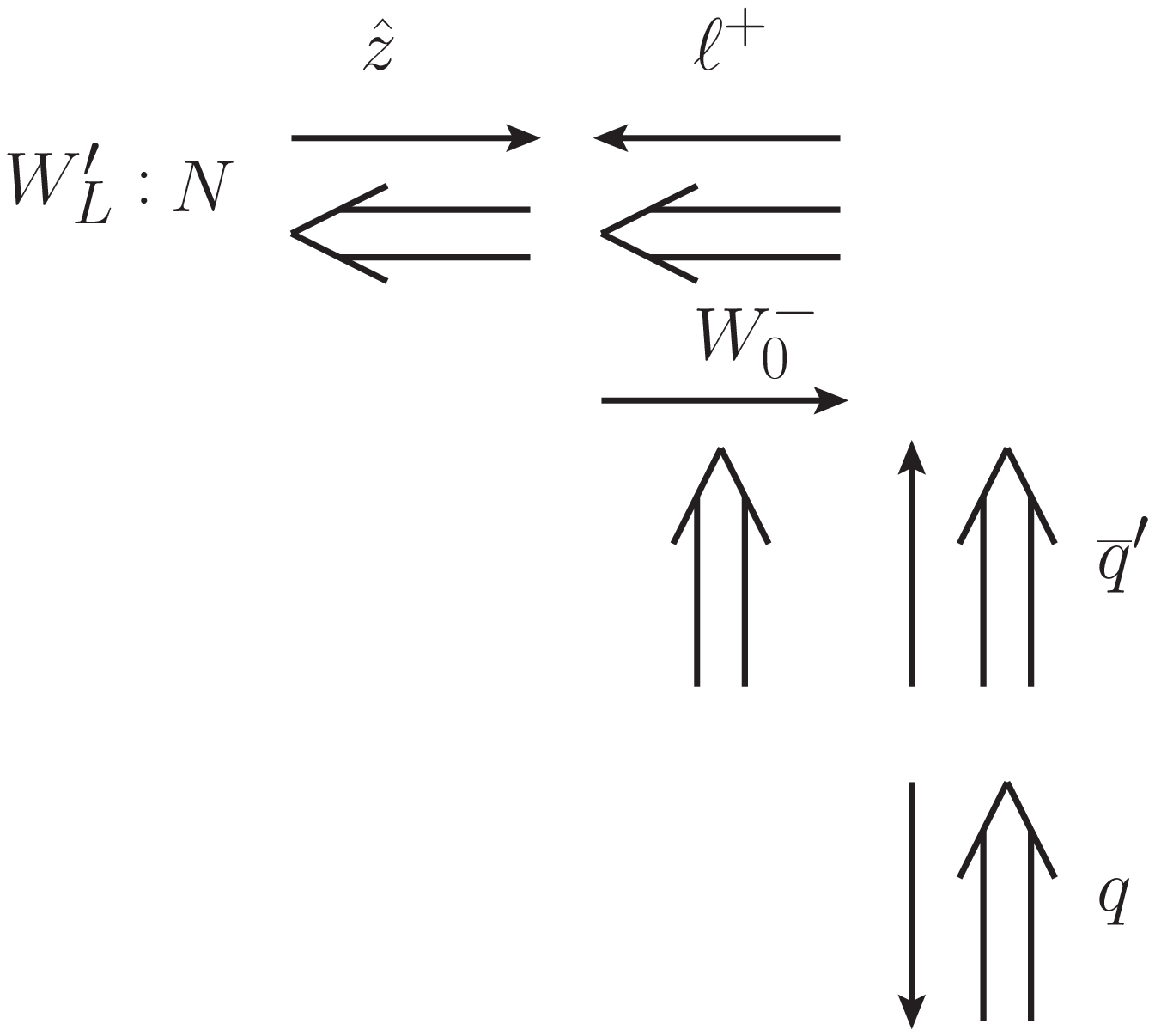}
\label{LspinLH.FIG}
        }
\subfigure[]{
\includegraphics[width=0.35\textwidth,clip=true]{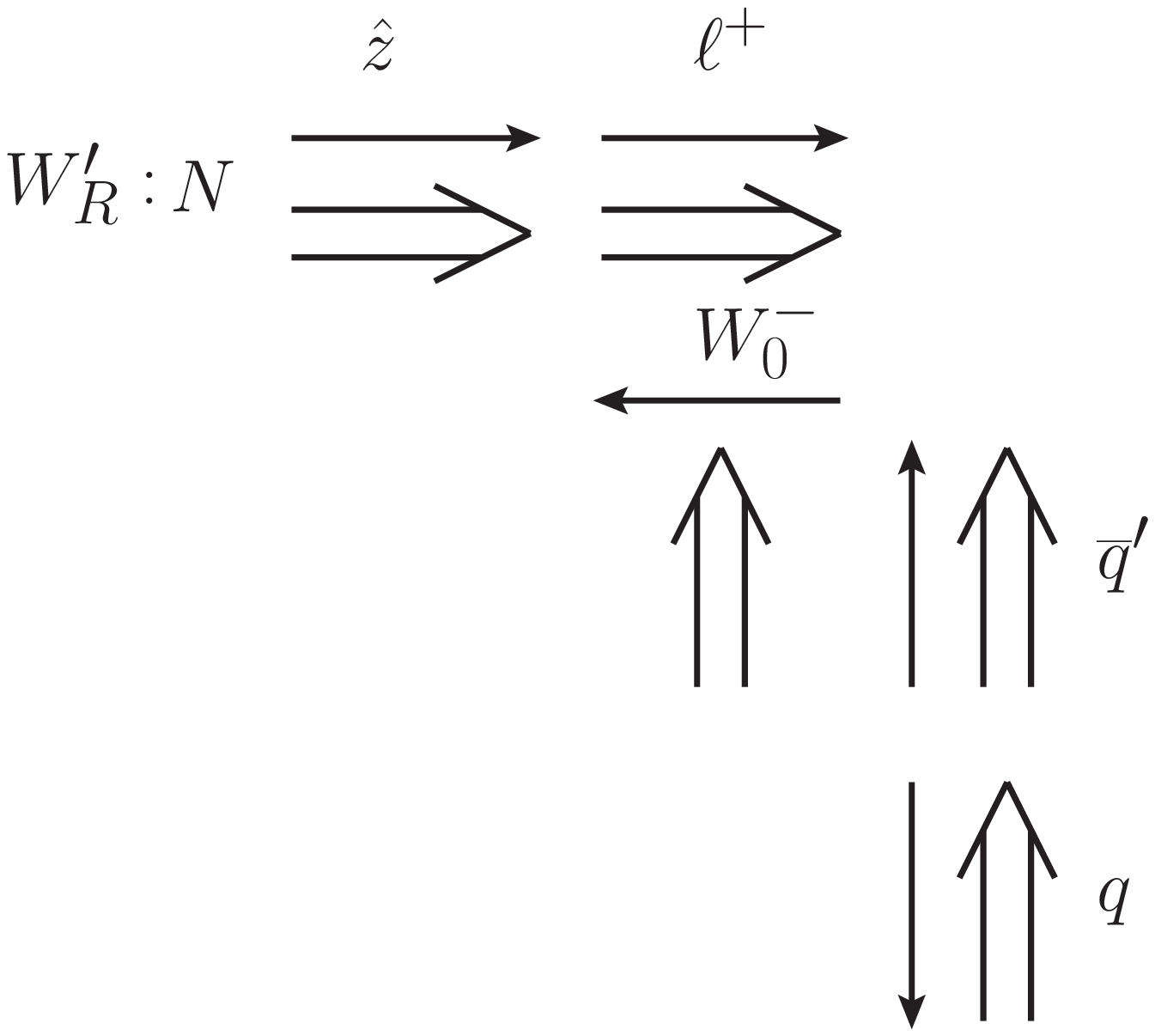}
\label{LspinRH.FIG}
        }\\
\subfigure[]{
\includegraphics[width=0.35\textwidth,clip=true]{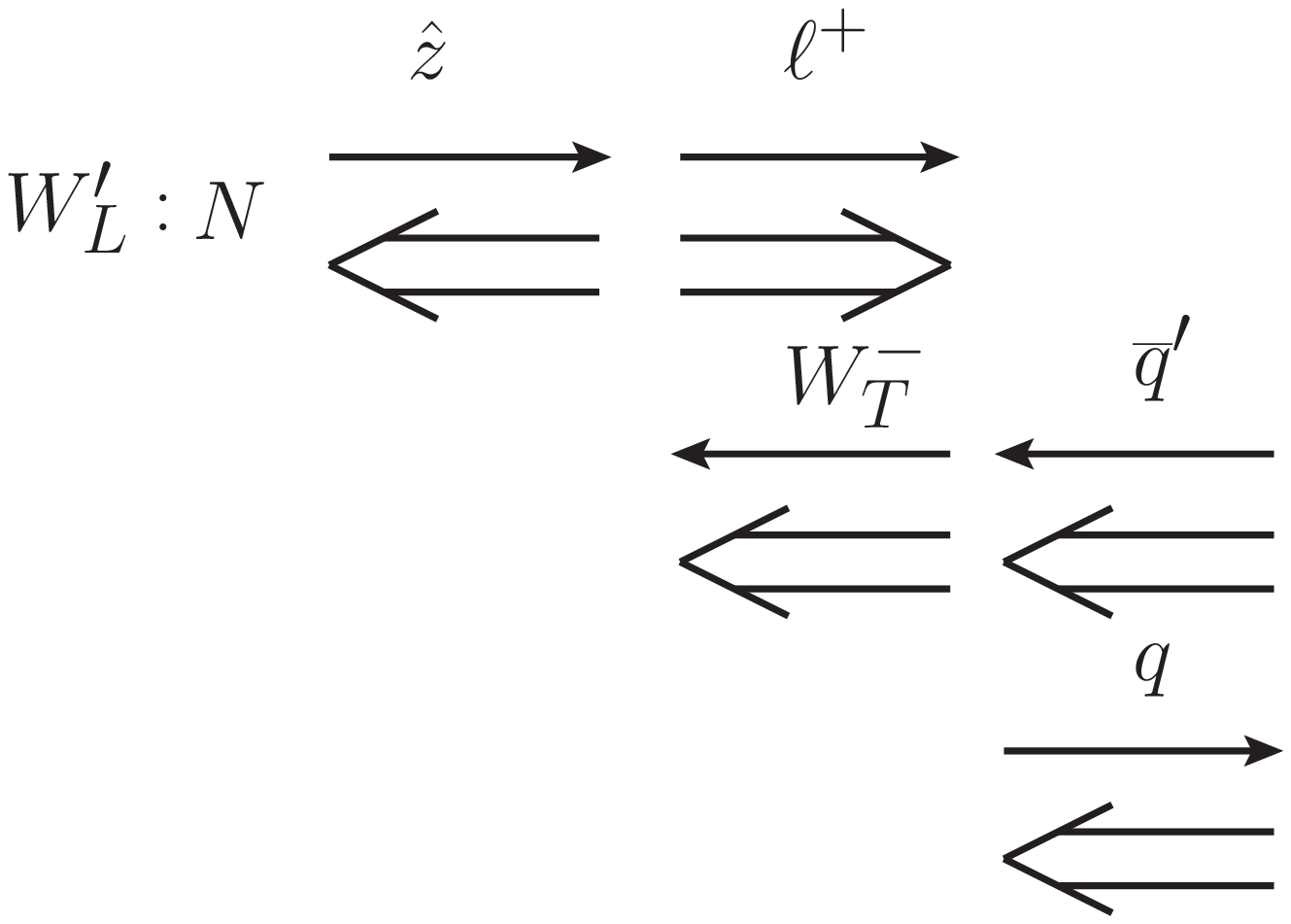}
\label{TspinLH.FIG}
        }
\subfigure[]{
\includegraphics[width=0.35\textwidth,clip=true]{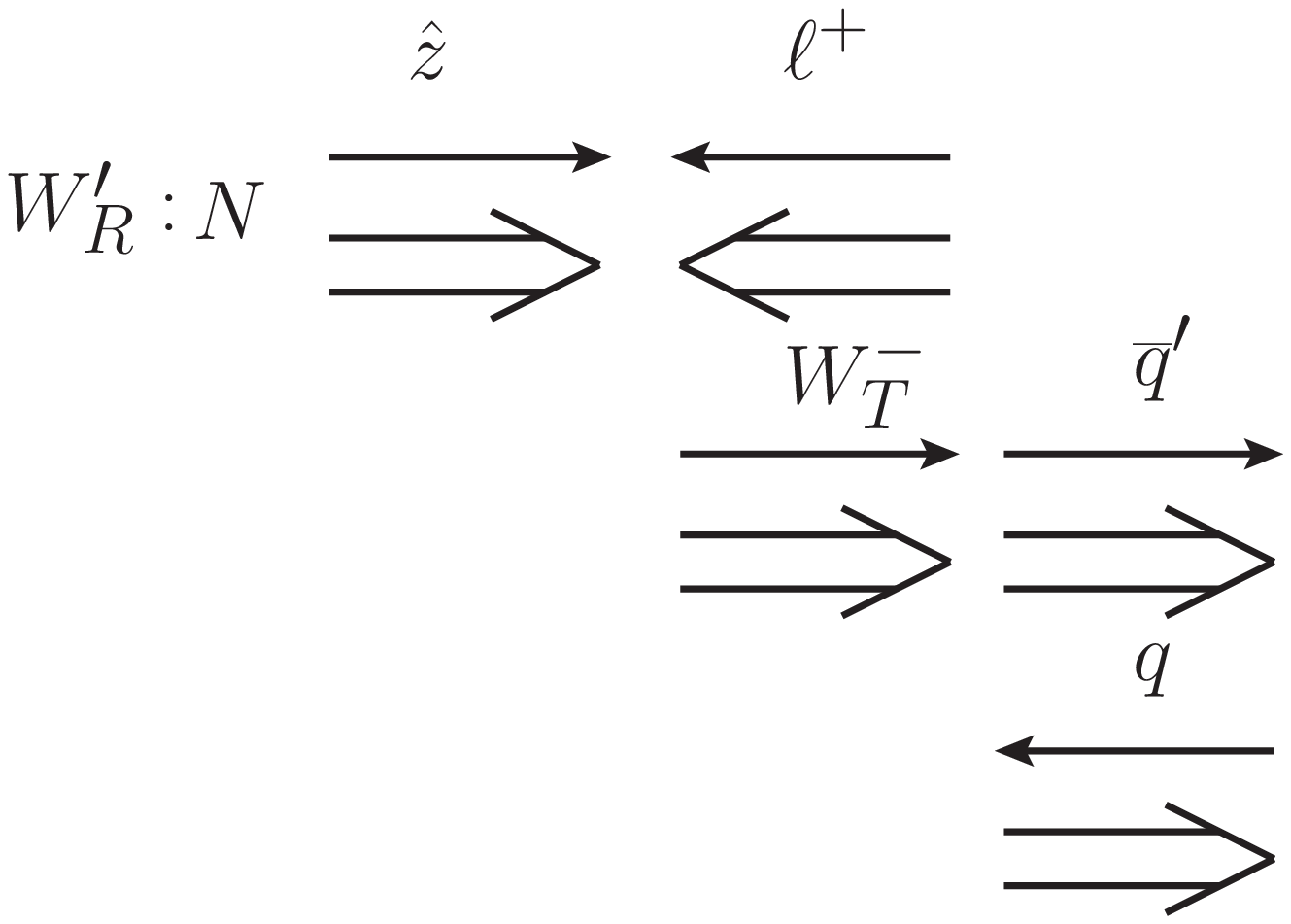}
\label{TspinRH.FIG}
        }
\caption[Helicity and spin correlations in the chains $N_{L,R} \to \ell^+ W^{-} \to \ell^+ q \overline{q}'$ 
from $\wpri_L$ decay]{Helicity and spin correlations in the chains $N_{L,R} \to \ell^+ W^{-} \to \ell^+ q \overline{q}'$ 
from $\wpri_L$ decay in  (a),~(c); and from $\wpri_R$ decay in (b),~(d).
Figures (a) and (b) are for longitudinally polarized SM $W$'s, and Figs.~(c) and (d) are for 
transversely polarized SM $W$'s.  The decay goes from left to right as labeled by the particle names. 
The momenta (single arrow lines) and spins (double arrow lines) are in the parent rest-frame 
in the direction of the heavy neutrino's motion ($\hat{z}$) in the $\wpri$ rest-frame.
}
\label{Lspin.FIG}
\end{center}
\end{figure}

Figure \ref{Lspin.FIG} shows the spin correlations of the process in Eq.~(\ref{sig.EQ}) with the single arrowed lines representing momentum direction and double arrowed lines spin. 
The direction $\hat{z}$ is defined as the direction of motion of the neutrino in the $W'$ rest-frame.  
Each column indicates the spin and momentum of the particles in their parents' rest-frame with the first column in the neutrino rest-frame.  
Note that for the $W'_R$ ($W'_L$) the heavy neutrino is in a mostly right-(left-) handed helicity state.  
Hence, for the $W'_R$ ($W'_L$) the neutrino spin points with (against) the $\hat{z}$ direction.  
The decays of the neutrino through longitudinal $W$ are shown in Fig.~\ref{LspinLH.FIG} and \ref{LspinRH.FIG} for $W'_L$ and $W'_R$, respectively,
 and the decays through a transversely polarized $W$ are shown in Fig.~\ref{TspinLH.FIG} for $\wpri_L$ and Fig.~\ref{TspinRH.FIG} for $\wpri_R$.

  As shown in Fig.~\ref{NW.FIG}, $500$~GeV neutrino preferentially decays into longitudinally polarized $W$'s.  
  We therefore focus on that case.
For the $\wpri_R$, the lepton from the heavy neutrino decay moves preferentially along the $\hat{z}$ direction.  
Hence, the boost into the partonic c.m.~frame will be along the charged lepton's momentum.  
In the $\wpri_L$ case, the charged lepton moves in negative $\hat{z}$ direction and the boost into the partonic 
c.m.~frame is against the lepton's momentum.  
Therefore, the lepton from the heavy neutrino decay is harder in the $\wpri_R$ case than in the $\wpri_L$ case. 
The contribution from decay into transversely polarized $W$'s is in the opposite direction.  
However, as noted previously, this contribution is smaller than the decays into longitudinally polarized $W$'s. 
Similar arguments can be made to explain that the two jets are softer in the $\wpri_R$ case than in the $\wpri_L$ case.

\begin{figure}[!t]
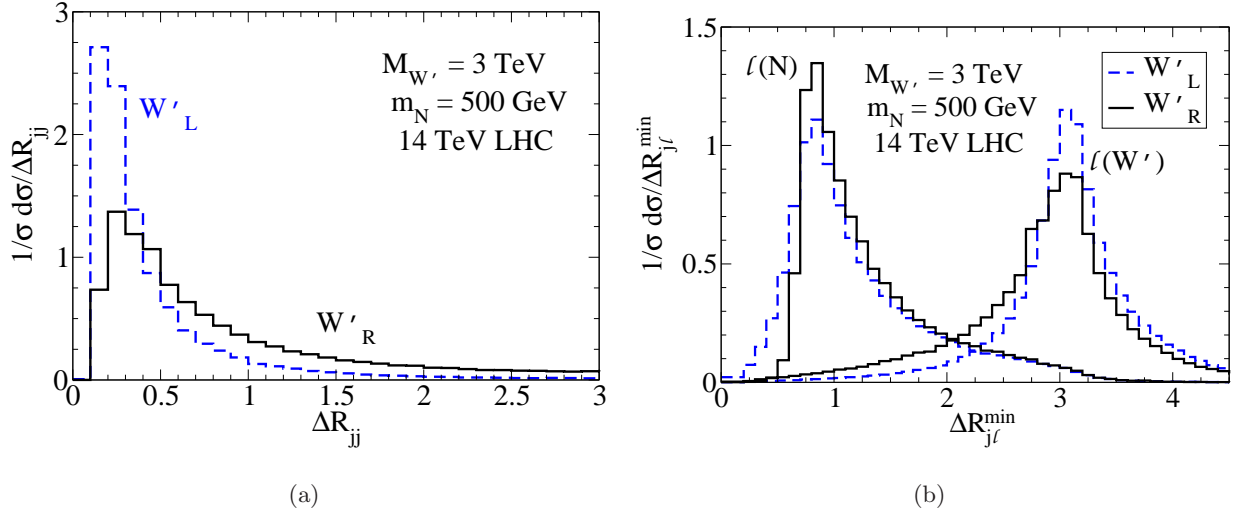

\centering
\subfigure[]{
        \includegraphics[width=0.48\textwidth]{./05_WprimeHeavyN/drjj500_3.eps}
        \label{drjj.FIG}
}
\subfigure[]{
        \includegraphics[width=0.48\textwidth]{./05_WprimeHeavyN/drjlmin500_3.eps}
        \label{drjlmin.FIG}
}
\caption[$\Delta R_{jj}$ distribution and $\Delta R^{\rm min}_{\ell j}$ distributions]{(a) $\Delta R_{jj}$ distribution and 
(b) $\Delta R^{\rm min}_{\ell j}$ distributions for both the lepton identified as originating from $N$ and $\wpri$.  
The solid lines are for the $\wpri_R$ case and dashed lines $\wpri_L$. Energy smearing has been applied.}
\label{deltar.FIG}
\end{figure}

As previously stated, identifying well-separated objects in our event is paramount to measuring our observables.
For 14 TeV LHC collisions, Fig.~\ref{deltar.FIG} shows (a) the separation between the two jets, $\Delta R_{jj}$, 
and (b) the minimum separation between the leptons identifed as originating from the heavy neutrino and $\wpri$ and the two jets defined by 
\begin{eqnarray}
\Delta R^{\rm min}_{\ell_{i} j}=\min_{k=1,2} \Delta R_{\ell_i j_k},
\label{Rmin.EQ}
\end{eqnarray}
where $i=\wpri$ for the lepton coming from the $\wpri$ and $i=N$ for the lepton coming from the neutrino decay.
 The solid lines are for $\wpri_R$ and the dashed lines for $\wpri_L$.  The $\Delta R_{jj}$ distributions peak at low values for both the left- and right-handed cases. 
 This is due to the $W$ from the heavy neutrino decay being highly boosted and its decay products therefore collimated.  
  Also, as can be seen from Fig.~\ref{ptdist.FIG}, in the $\wpri_R$ case the lepton from the neutrino decay is harder and hence the SM $W$ softer than in the $\wpri_L$ case.  
  Since the SM $W$ is less boosted in the right-handed case, the jets are less collimated and the $\Delta R_{jj}$ distribution has a longer tail for $\wpri_R$ than for $\wpri_L$.  
   Also, since the neutrino is highly boosted, its decay products are expected to land opposite in the transverse plane from the lepton from $\wpri$ decay. 
   Hence, $\Delta R^{\rm min}_{\ell_{\wpri} j}$ peaks near $\pi$ for both the the left-handed and right-handed case.  
   Finally, $\Delta R^{\rm min}_{\ell_N j}$ is peaked near $2m_N/E_N\approx 0.7$ for both the $\wpri_L$ and $\wpri_R$ cases.   
 The $\Delta R$ distributions at the 8 TeV LHC are peaked at similar values, but are more narrow than the 14 TeV distributions. 
  Based on these arguments, we define the isolation cuts given by Eq.~(\ref{cuts2.EQ}).

The isolation cuts more severely affect the $\wpri_L$ cross section since the $\Delta R_{jj}$ distribution is strongly peaked at low values for $\wpri_L$.
As the mass of the $\wpri$ increases, the SM $W$ from the heavy neutrino decay will become more boosted.  
Hence, the two jets will become more collimated and the effects of the isolation cuts will be even more significant. 
Since we will only be interested in the angular distributions of the lepton, it is possible to relax the $\Delta R_{jj}$ cut and look for one or two jets with two like sign leptons.  
Also, the separation between the lepton and jets from the heavy neutrino decay depend on the ratio of $m_N/M_{\wpri}$.  
As $m_N/M_{\wpri}$ increases (decreases) the lepton and jets become more (less) well separated.

\subsection{Background Reduction and Statistical Significance}
\label{BkgdRedux.SEC}

\begin{figure}[!t]
\centering
\includegraphics[width=0.55\textwidth]{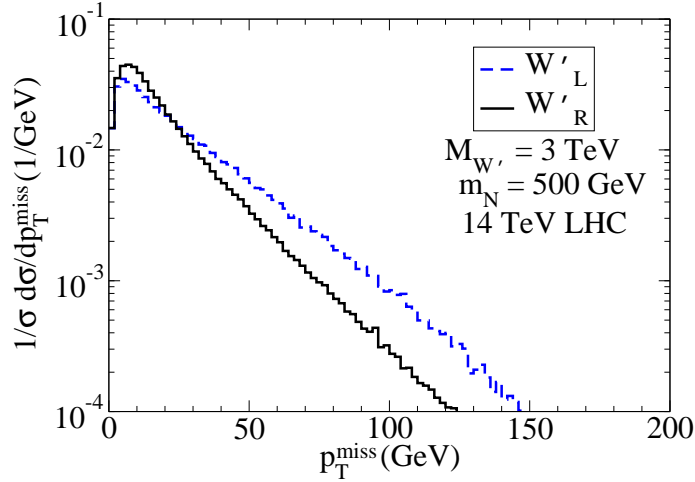}
\caption[Missing energy distribution for $pp\rightarrow {W'}_{L,R}^+\rightarrow \mu^+\mu^+ q\overline{q}'$ at the LHC.]{Missing energy distribution for $pp\rightarrow {W'}_{L,R}^+\rightarrow \mu^+\mu^+ q\overline{q}'$ at the LHC.  
Energy smearing has been applied.}
\label{etmiss.FIG}
\end{figure}

The SM background for our $\ell^{+}\ell^{+}jj$ signature has been thoroughly studied for the 14 TeV LHC by Ref.~\cite{Atre:2009rg}.
The largest background to our process was found to be from $t\bar{t}$ events with the cascade decays,
\begin{equation}
 t\rightarrow W^+b\rightarrow\ell^+\nu_m b,\quad\bar{t}\rightarrow W^-\bar{b}\rightarrow W^-\bar{c}\nu_m \ell^+,
\end{equation}
and was also found to be greatly suppressed by the lepton isolation cuts in Eq.~(\ref{cuts2.EQ}).  
The background can be further suppressed by noting that leptonic $t\bar{t}$ events contain a final state light neutrino and 
therefore a considerable amount of missing transverse energy, $\etmiss$.
This is in direct comparison with our signal where all the $\etmiss$ is due to detector resolution effects.  
The $\etmiss$ for our like-sign leptons + dijet events is shown in Fig.~\ref{etmiss.FIG} for both the right- (solid) and left-handed (dashed) $W'$ cases.  
Furthermore, the two jets in our process originate from a SM $W$ whereas the jets in the top background do not.
Hence $\etmiss$ and dijet invariant mass, $m_{jj}$, cuts are also applied:
\begin{eqnarray}
\etmiss<30~\rm{GeV},~~~60~{\rm GeV}<&m_{jj}&<100~{\rm GeV}.
\label{cuts3.EQ}
\end{eqnarray}
The effect of these cuts on the signal rate are seen in the fourth line of Table~\ref{WpNxsect.TAB}. 

Having obtained a measurement of $m_{N}$ from Eq.~(\ref{mNPermu.EQ}) and $M_{W'}$ from the $W'$'s Jacobian peak, if desired, 
invariant mass cuts on $m_{\ell_{N}jj}$ and $\hat{s}$ can be imposed to further isolate the signal:
\begin{eqnarray}
|m_{\ell_{N}jj}-m_N|\leq 0.1~ m_N~~{\rm and}~~|\hat{s}-M_{W'}|\leq 0.1~M_{\wpri}.
\label{cuts4.EQ}
\end{eqnarray}
The effects of these cuts are shown in the fifth line of Table~\ref{WpNxsect.TAB}.

 As $\sqrt{s}$ increases from 8 TeV to 14 TeV, the percentage of events passing the selection cuts also increases. See the final line of Table~\ref{WpNxsect.TAB}. 
 In particular, we note that relatively fewer events are failing the cuts imposed on the reconstructed masses [Eq.~(\ref{cuts4.EQ})]. 
 To understand this effect, consider that increasing the c.m.~energy also enlarges the phase space. Consequently, our internal propagators are more likely to be on-shell.

 \begin{figure}[!t]
\centering
\subfigure[]{
	\includegraphics[width=0.45\textwidth]{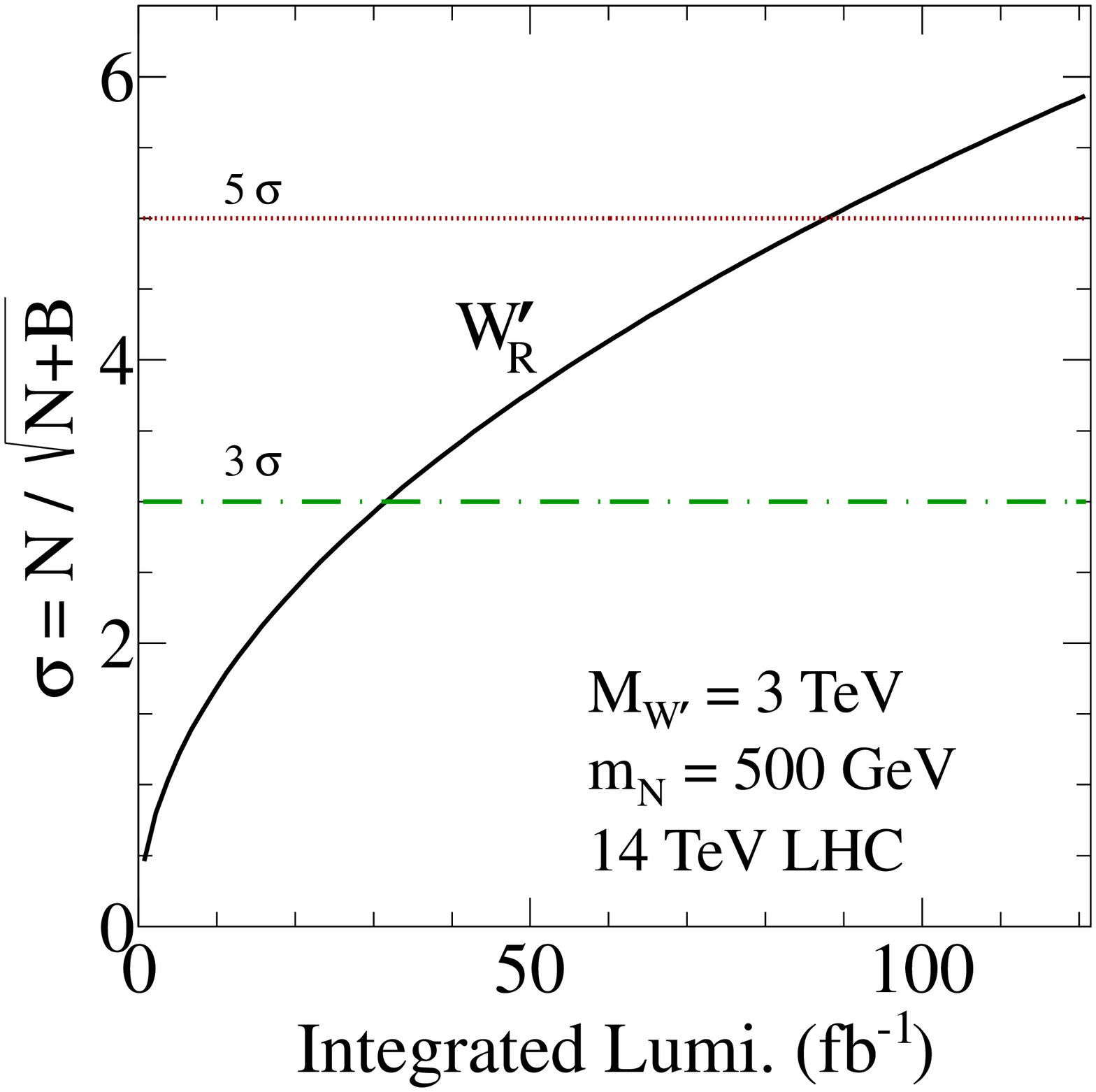}}
\subfigure[]{
		\includegraphics[width=0.45\textwidth]{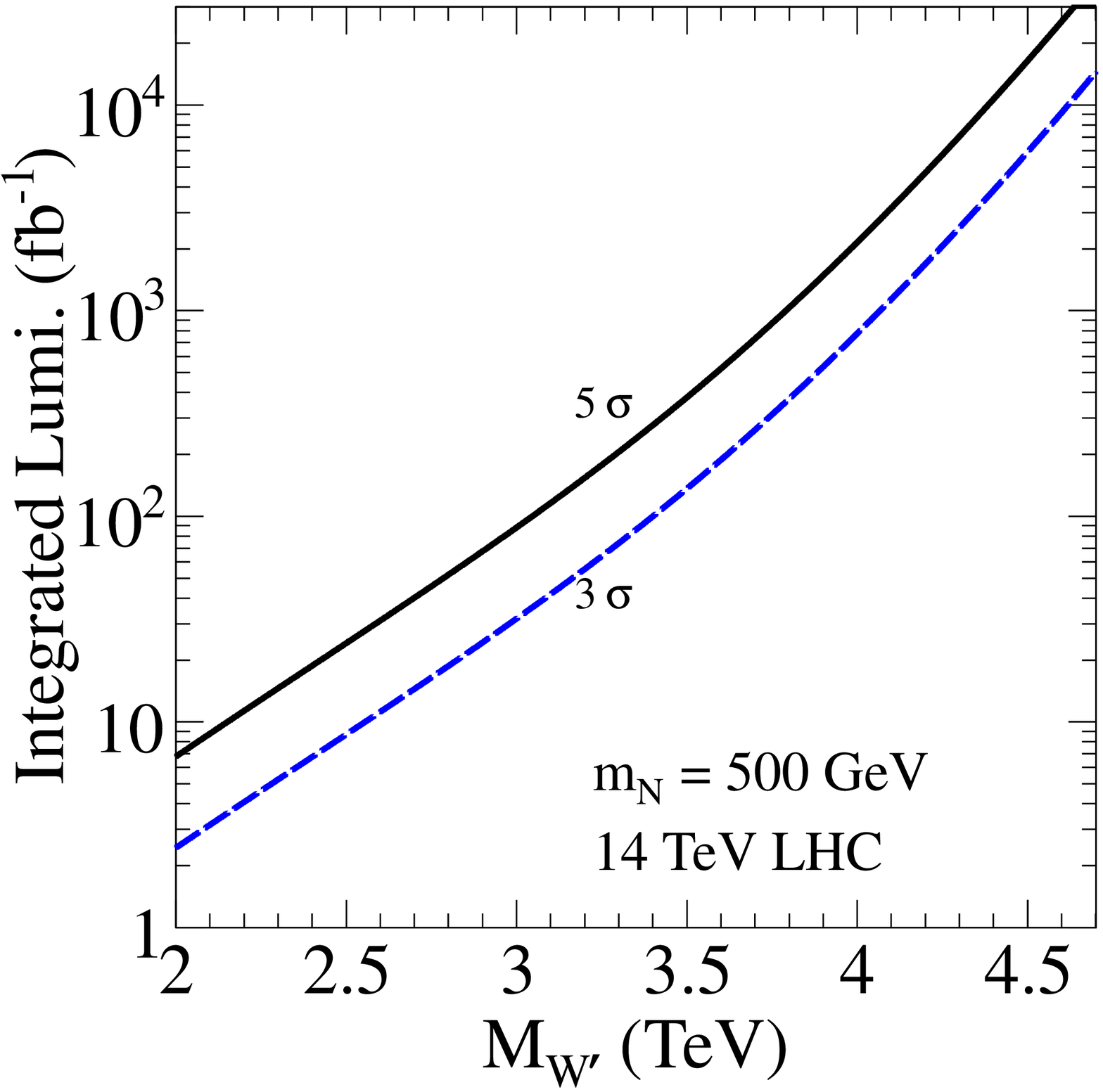}}
\caption[Integrated luminosity needed at 14 TeV LHC for discovery]{Integrated luminosity needed at 14 TeV LHC for (a) achievable statistical significance for $W'_R$ with $M_{W'}=3$ TeV and $m_{N}=500$ GeV, and (b) reachable $W'_R$ mass at $3\sigma$ and $5\sigma$ sensitivity. }
\label{significanceVsLumi.FIG}
\end{figure}

The contribution from the irreducible background for our $\ell^{\pm}\ell^{\pm}jj$ signal,  
\begin{eqnarray}
pp\rightarrow W^{\pm}W^{\pm}W^{\mp},&&pp\rightarrow W^{\pm}W^{\pm}jj,\quad\quad pp\rightarrow t\overline{t}
\end{eqnarray}
events and
\begin{eqnarray}
pp\rightarrow jjZZ,&&pp\rightarrow jjZW,
\end{eqnarray}
wherein leptons from the $Z$ boson escape from a detector, 
are estimated~\cite{Atre:2009rg} to be at most $\sigma=0.08$~fb using a comparable list of selection cuts.
However, this previous analysis does not impose any restriction on the invariant mass of the system as done in Eq.~(\ref{cuts4.EQ}), 
and therefore, realistically, the background will be much less than $0.08$~fb.
In either case, our $W'_{R}$ signal is clearly above background. 
Using $\sigma=0.08$~fb as an estimation for our background, 
we calculate the significance and reachability of our $W'_{R}$ signal at the 14 TeV LHC as shown in 
Fig. ~\ref{significanceVsLumi.FIG}. With 100 fb$^{-1}$ integrated luminosity, 
a $W'_R$ signal via the lepton-number violating process can be observed at a $5\sigma$ level up to a mass of 3 TeV.
As evident, the required integrated luminosity for a discovery at the LHC grows rapidly with increasing $M_{W'_R}$. 
This is expected if we again consider that the $W$ boson becomes increasingly boosted as $M_{W'_R}$ grows.
A more boosted $W$ leads to more collimated jets, which have more difficulty passing the isolation cuts [Eq.~(\ref{cuts2.EQ})]
than their less collimated counterparts.

\section{$W'$ Chiral Couplings From Angular Correlations at the LHC}
\label{WpriAC.SEC}
Once a new gauge boson $W'$ is observed at the LHC, it is of fundamental importance to determined the nature of its coupling to the SM fermions. 
Here, we identify various kinematical quantities that depend on the chiral couplings of the fermions to a $\wpri$.  
Each quantity will have a different dependence on the $\wpri$ chiral couplings and so will provide independent measurements of the chiral couplings.

\begin{figure}[!t]
\begin{center}
\subfigure[]{
      \includegraphics[width=0.29\textwidth,clip=true]{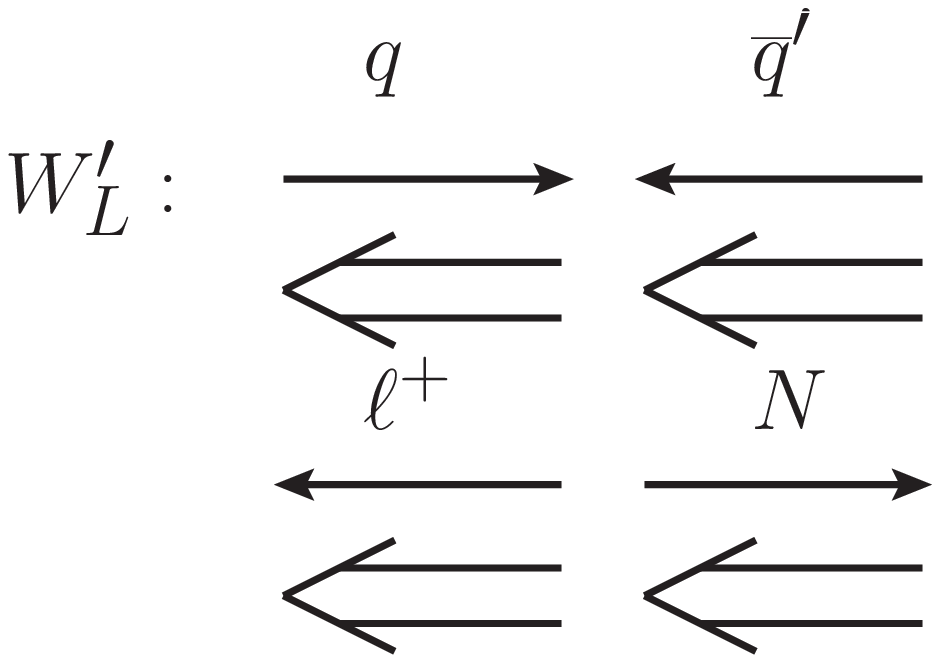}
}
\subfigure[]{
      \includegraphics[width=0.29\textwidth,clip=true]{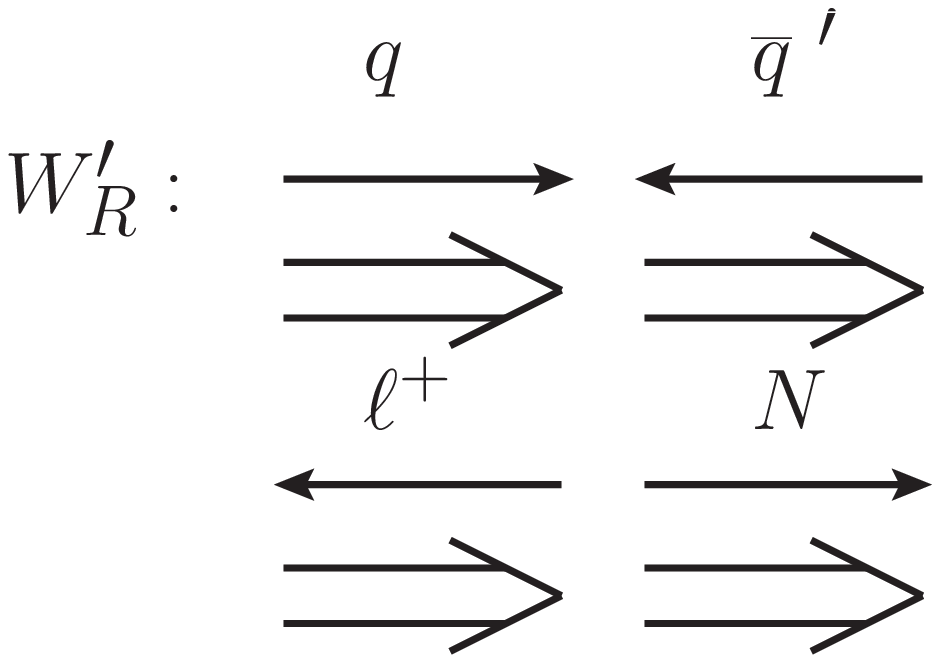}
}
\end{center}
\caption[Spin correlations for $q\bar{q}'\rightarrow W'\rightarrow N\ell^+$]{Spin correlations for $q\bar{q}'\rightarrow W'\rightarrow N\ell^+$ for (a) left-handed and (b) right-handed couplings.  Single arrow lines represent momentum directions and double arrow lines represent spin directions.}
\label{Nspin.FIG}
\end{figure}

\subsection{$W'$ Chiral Couplings To Leptons}
\label{Wptspin.SEC}
Figure \ref{Nspin.FIG} shows the spin correlations for the process $q\bar{q}'\rightarrow W'\rightarrow N\ell^+$ in the partonic c.m.~frame for both 
the (a) left-handed and (b) right-handed cases.  
Double arrowed lines represent spin and single arrowed lines momentum.  As it is well-known, although the 
preferred charged lepton momentum direction leads to a clear distribution of parity violation, it cannot reveal more detailed nature of the chiral coupling. 
On the other hand, the nature of the $W'$ leptonic chiral couplings is encoded in polarization of the heavy neutrino, i.e., 
in the $\wpri_R$ ($\wpri_L$) case the heavy neutrino is preferentially right-handed (left-handed).  
Hence, if the polarization of the neutrino can be determined, the left-handed and right-handed cases can be distinguished. 
Spin observables such as $\left<\hat{s}_N\cdot \hat{a}\right>$, where $s_N$ is the spin of the heavy neutrino and $\hat{a}$ is an arbitrary spin quantization axis, 
are sensitive to the polarization of the heavy neutrino.  
Defining the angle $\theta^*$ between the $\hat a$ and the direction of motion of the charged lepton originating from the heavy neutrino decay, $\hat{p}_{\ell_2},$ 
the angular distribution of the partial width of the neutrino decaying into a charged lepton and two jets is~\cite{Tait:2000sh}
\bea
\displaystyle \frac{1}{\Gamma}\frac{d\Gamma }{d\cos\theta^*}(N\rightarrow \ell^{\pm} jj)=\frac{1}{2}
\left(1+2~A^{\ell^\pm}~\cos\theta^*\right),
\label{Ndec.EQ}
\eea
where $A^{\ell^+}=-A^{\ell^-}\equiv A$ due to the CP invariance.
The coefficient $A$ is related to $\left<\hat{s}_N\cdot \hat{a}\right>$ and is the forward-backward asymmetry of the charged lepton with respect to the direction $\hat a$.  
We will refer to $A$ as the analyzing power.  
The angular distribution of either of the two jets from the neutrino decay will also have a similar linear form and may be used to perform this analysis, 
although uncertainties in jet measurements may cause more complications.  

A highly boosted neutrino from a heavy $W'$ decay will be produced mostly in a helicity state; 
hence, it is natural to choose $\hat a = \hat{p}_N$, 
the direction of motion of the neutrino in the partonic c.m.~frame, and measure $\hat{p}_{\ell_2}$ in the neutrino rest-frame.   
At the partonic level, the angular distribution of the lepton from neutrino decay in the reconstructible neutrino rest-frame is (See App.~\ref{appendME.APP})
\begin{eqnarray}
\label{Ang.EQ}
\frac{d\hat{\sigma}(u\bar{d}\rightarrow\ell^+_1\ell^+_2 W^-)}{d\cos\theta_{\ell_{2}}}
&=&
\\
\frac{\hat{\sigma}_{Tot.}}{2}
\Bigg[
1&+&\left(\frac{\hat{\sigma}(W_{0})-\hat{\sigma}(W_{T})}{\hat{\sigma}(W_{0})+\hat{\sigma}(W_{T})}\right)\left(\frac{2-\mu_{N}^{2}}{2+\mu_{N}^{2}}\right)
\left(\frac
{g_{R}^{\ell\:2}\vert Y_{\ell_1 N}\vert^{2}-g_{L}^{\ell\:2}\vert V_{\ell_1 N}\vert^{2}}
{g_{R}^{\ell\:2}\vert Y_{\ell_1 N}\vert^{2}+g_{L}^{\ell\:2}\vert V_{\ell_1 N}\vert^{2}}\right)\cos\theta_{\ell_{2}}
\Bigg].
\nonumber
\end{eqnarray}
Here $\hat\sigma(W_0)$ and $\hat\sigma(W_T)$ are the partonic level $u\overline{d}\rightarrow W^{'+}\rightarrow \ell^+_1\ell^+_2 W^{-}_{\lambda}$ cross sections 
with $N$ decaying into longitudinally $(\lambda=0)$ and transversely $(\lambda=T)$ polarized $W$'s, respectively. They are given by
\begin{eqnarray}
\hat{\sigma}(W_{0})&\equiv&\hat{\sigma}(u\bar{d}\rightarrow \ell^+_1 N\rightarrow \ell^+_1\ell^+_2 W^-_0)\\
&=&\frac{1}{9}\frac{1}{2^{10}}\frac{g^2}{\pi^2}\frac{ \vert V^{CKM'}_{ud}\vert^{2} \vert V_{\ell_{2}N}\vert^2}{(1+\delta_{\ell_{1}\ell_{2}})}
\left(g^{q~2}_{R} + g^{q~2}_{L}\right)
\left(g^{\ell\:2}_{R}\vert Y_{\ell_{1}N}\vert^{2} + g^{\ell\:2}_{L}\vert V_{\ell_{1}N}\vert^{2} \right)\left(\frac{m_{N}}{\Gamma_{N}}\right)\nonumber\\
&\times&
\frac{\hat{s}}{\left[(\hat{s}-M_{W'}^{2})^{2}+(\Gamma_{W'}M_{W'})^{2}\right]}
(1-y_{W}^{2})^{2}(1-\mu_{N}^{2})^{2}(2+\mu_{N}^{2})
\left(\frac{1}{2y_{W}^{2}}\right)\\
\hat{\sigma}(W_{T})&\equiv&\hat{\sigma}(u\bar{d}\rightarrow \ell^+_1 N\rightarrow \ell^+_1\ell^+_2 W^-_T)\\
&=& \hat{\sigma}(W_{0})\times 2y_{W}^{2}.
\end{eqnarray}
where $\mu_N=m_N/\sqrt{\hat s}$, $y_W=M_W/m_N$, and $\hat{\sigma}_{Tot.}=\left(\hat\sigma(W_0)+\hat\sigma(W_T)\right)\times{\rm BR}(W\rightarrow q\bar{q}')$ 
is the total partonic cross section.
As $W'$ comes on-shell, $\mu_N\rightarrow x_N$.
In this reference frame, $\theta^{*}$ from Eq.~(\ref{Ndec.EQ}) satisfies
\begin{equation}
 \cos\theta^{*} = \cos\theta_{\ell_2} \equiv \hat{p}_{\ell_2} \cdot \hat{p}_{N},
 \label{thetaEll2.Eq}
\end{equation}
where, again, $\hat{p}_{\ell_2}$ is measured in the neutrino rest-frame and $\hat{p}_{N}$ is measured in the partonic c.m. frame.

For an on-shell $\wpri$, the analyzing power at the partonic and hadronic level are the same.  
In such a case, after comparing Eqs.~(\ref{Ndec.EQ}) and (\ref{Ang.EQ}), we find that the analyzing power is
\begin{eqnarray}
A&=&
\frac{1}{2}
\left(\frac{\hat\sigma(W_0)-\hat\sigma(W_T)}{\hat\sigma(W_0)+\hat\sigma(W_T)}\right)
\left(\frac{2-x^2_N}{2+x^2_N}\right)
\left(\frac
{g_{R}^{\ell\:2}\vert Y_{\ell_1 N}\vert^{2}-g_{L}^{\ell\:2}\vert V_{\ell_1 N}\vert^{2}}
{g_{R}^{\ell\:2}\vert Y_{\ell_1 N}\vert^{2}+g_{L}^{\ell\:2}\vert V_{\ell_1 N}\vert^{2}}\right)\nonumber\\
&=&
\frac{1}{2}
\left(\frac{1-2y^2_W}{1+2y^2_W}\right)
\left(\frac{2-x^2_N}{2+x^2_N}\right)
\left(\frac
{g_{R}^{\ell\:2}\vert Y_{\ell_1 N}\vert^{2}-g_{L}^{\ell\:2}\vert V_{\ell_1 N}\vert^{2}}
{g_{R}^{\ell\:2}\vert Y_{\ell_1 N}\vert^{2}+g_{L}^{\ell\:2}\vert V_{\ell_1 N}\vert^{2}}\right).
\label{Apart.EQ}
\end{eqnarray}

The different signs for the analyzing power between the neutrino decays to the two different $W$ polarizations and between the $\wpri_{L,R}$ cases can be understood via the spin correlation in Fig.~\ref{Lspin.FIG}.  
For the $\wpri_R$ case, a heavy neutrino decaying to a longitudinal (transverse) $W$ will have 
the charged lepton preferentially moving with (against) $\hat{p}_{N}$.
For the $\wpri_L$ case the helicity of the neutrino, and therefore the direction of the charged lepton, is reversed. 
Hence the analyzing power is proportional to $(\hat\sigma(W_0)-\hat\sigma(W_T))(g_{R}^{\ell\:2}\vert Y_{\ell_1 N}\vert^{2}-g_{L}^{\ell\:2}\vert V_{\ell_1 N}\vert^{2})$.  

In the analysis of Fig.~\ref{Lspin.FIG}, the left- and right-chiral neutrinos at the $\wpri\rightarrow N\ell^+$ vertex 
are approximated as the left-handed and right-handed helicity states in the partonic c.m.~frame. 
As the neutrino becomes more massive relative to the $\wpri$, the approximation of the chiral basis by the helicity basis begins to break down, 
i.e., the left- (right-) helicity state makes a larger contribution to the right- (left-) chiral state.  
In Eq.~(\ref{Ang.EQ}), this is reflected by the $\cos\theta_{\ell_2}$ ($\cos\theta_\ell$ for simplicity) coefficient  
\begin{equation}
\frac{2-x^2_N}{2+x^2_N} = \frac{2M_{W'}^{2}-m_{N}^{2}}{2M_{W'}^{2}+m_{N}^{2}}.
\end{equation}
As $x_N$ increases, the distribution flattens due to the right-handed (left-handed) neutrino helicity state, 
thereby making a larger contribution to the $\wpri_L$ ($\wpri_R$) distributions.

\begin{figure}[!t]
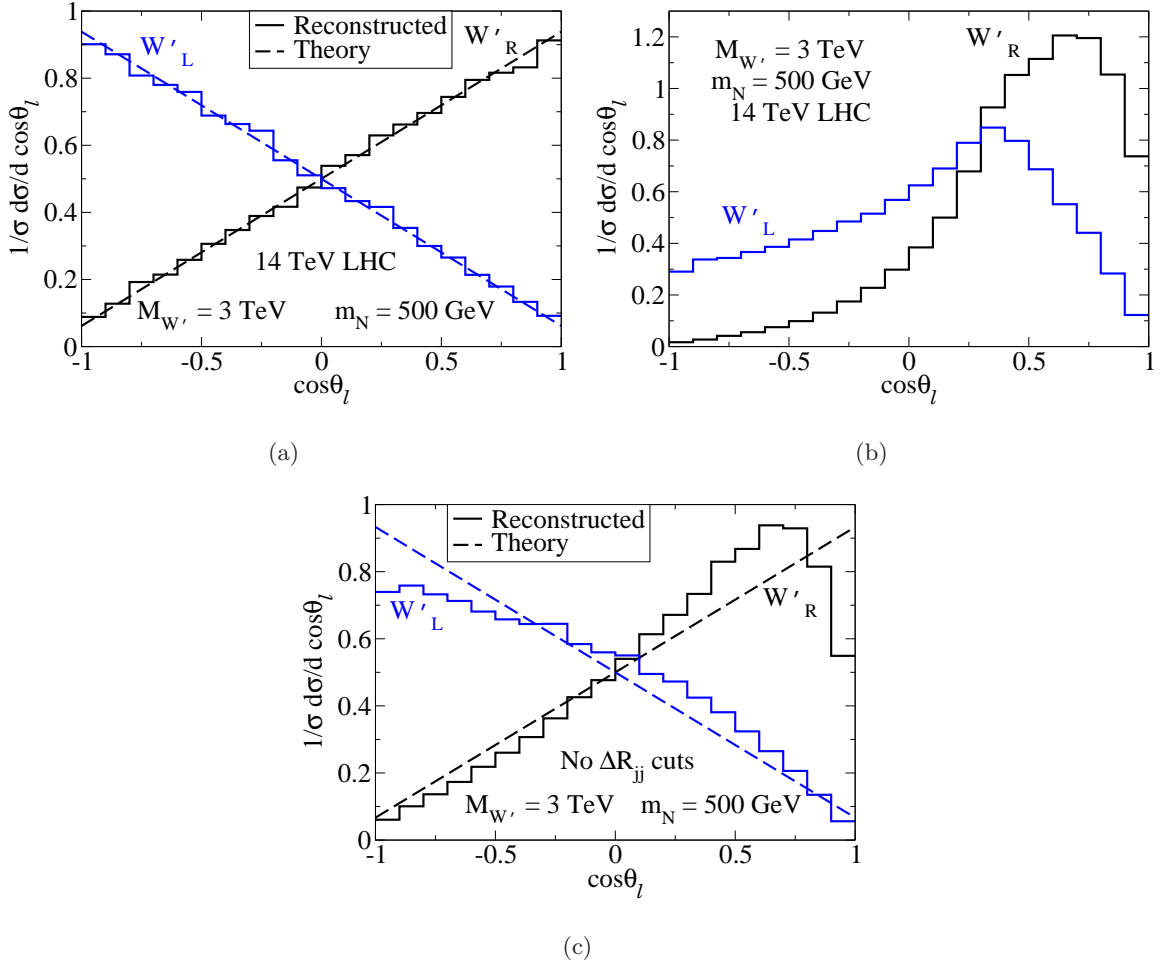

\begin{center}
\subfigure[]{
      \includegraphics[width=0.45\textwidth,clip]{./05_WprimeHeavyN/CthlNCNS500_3.eps}
\label{wpriacNS.FIG}}
\subfigure[]{
      \includegraphics[width=0.45\textwidth,clip]{./05_WprimeHeavyN/CthlAll500_3.eps}
\label{wpriacSM.FIG}}\\
\subfigure[]{
      \includegraphics[width=0.45\textwidth,clip]{./05_WprimeHeavyN/CthlAll500_3NoJIso.eps}
\label{wpriacNoJIso.FIG}}
\caption[Angular distribution of the charged lepton originating from neutrino decay]{
The angular distribution of the charged lepton originating from neutrino decay 
in the heavy neutrino rest-frame with respect to the neutrino moving direction in the partonic c.m.~frame at the LHC with $M_{W'}$, $m_N$ set by Eq.~(\ref{benchParam.EQ}).
Distribution (a) without smearing or cuts, (b) with energy smearing and cuts 
in Eqs.~(\ref{cuts1.EQ}), (\ref{cuts2.EQ}), (\ref{cuts3.EQ}), and (\ref{cuts4.EQ})
, and (c) with all cuts applied to (b) except the $\Delta R_{jj}$ cuts in Eq.~(\ref{cuts2.EQ}). 
The solid lines are for the Monte Carlo simulation results and in (a) and (c) the dashed lines are for the analytical result in Eq.~(\ref{Ang.EQ}).
}
\label{wpriac.FIG}
\end{center}
\end{figure}

Figure \ref{wpriac.FIG} shows the hadronic level angular distribution of the lepton in the neutrino's rest-frame for both $\wpri_L$ and $\wpri_R$ at the LHC.  
The case without smearing or cuts is shown in Fig.~\ref{wpriacNS.FIG}, and contains both the analytical results (dashed line) and Monte Carlo simulation (solid line) histograms.  
As can be clearly seen, the analytical and numerical results are in good agreement.
 Figure \ref{wpriacSM.FIG} shows the leptonic angular distribution after energy smearing and cuts in Eqs.~(\ref{cuts1.EQ}), (\ref{cuts2.EQ}), (\ref{cuts3.EQ}), and (\ref{cuts4.EQ}).  
 Notice that there is a small depletion of events for $\cos\theta_\ell \approx 1$ and a large depletion when $\cos\theta_\ell<0$.  
 First, when $\cos\theta_\ell\approx 1$ the charged lepton is moving with and the jets against the direction of motion of the neutrino in the partonic c.m.~frame. 
 Hence, with boost back to the partonic c.m.~frame, the jets are softest at this point and the jet $p_T$ cuts in Eq.~(\ref{cuts1.EQ}) lead to a depletion of event in this region. 
When $\cos\theta_\ell < 0$, the lepton is moving against and the SM $W$ is moving with the neutrino's direction of motion.  
Hence, with the boost back to the partonic c.m.~frame, the $W$ is boosted and its decay products highly collimated.  
Consequently, the $\Delta R_{jj}$ cuts in Eq.~(\ref{cuts2.EQ}) lead to a large depletion of events.  
Figure \ref{wpriacNoJIso.FIG} shows lepton angular distribution with the same cuts as Fig.~\ref{wpriacSM.FIG} except the $\Delta R_{jj}$ cuts.  
For comparison, both the Monte Carlo simulation with cuts (solid) and analytical results without cuts (dashed) are shown.  
It is clear that the discriminating power of the lepton angular distribution would increases and the Monte Carlo distribution approaches the analytical results if the jet isolation cuts are relaxed.

The analyzing power in Eq.~(\ref{Apart.EQ}) can additionally be related to the forward backward asymmetry
\bea
{\cal A}\, = \, \frac{\sigma(\cos\theta_\ell\geq 0)-\sigma(\cos\theta_\ell < 0)}{\sigma(\cos\theta_\ell\geq 0)+\sigma(\cos\theta_\ell < 0)}.
\eea
Without cuts or smearing, ${\cal A}\,=\,A$; and for the values of $m_N$, $M_{W'}$ stipulated in Eq.~(\ref{benchParam.EQ}), 
\begin{equation}
 A=\left\{
\begin{matrix}
+0.43, & W'=W'_{R}\\ 
-0.43, & W'=W'_{L}
\end{matrix}\right. .
\end{equation}
The simulated values for the forward backward asymmetry with consecutive cuts are shown in Table~\ref{WpNAFB.TAB}. 
Again, simulations are in good agreement with the theoretical prediction for the forward backward asymmetry for no smearing or cuts. 
  As the cuts become more severe, the simulated and theoretical values deviate more, however the $\wpri_L$ and $\wpri_R$ cases can still be distinguished clearly.  
  Furthermore, as shown in the final row, if the $\Delta R_{jj}$ cuts in Eq.~(\ref{cuts2.EQ}) are relaxed, 
  the discriminating power of the asymmetry is greatly increased, and the theory and simulation are in much better agreement.

\begin{table}[!t]
\caption[Forward-backward asymmetry for $pp\rightarrow {W'}_{L,R}^+\rightarrow \mu^+\mu^+ q\overline{q}'$]{Forward-backward asymmetry for $pp\rightarrow {W'}_{L,R}^+\rightarrow \mu^+\mu^+ q\overline{q}'$ with consecutive cuts at 8 and 14 TeV LHC.  The last row has the same cuts applied as the previous row with the removal of the $\Delta R_{jj}$ cuts in Eq.~(\ref{cuts2.EQ}).}
\begin{center}
\begin{tabular}{|l|c|c|c|c|} \hline \hline
                         \multirow{2}{*}{ ~~~~~~~~~~~~~~~~~~~~~~~~~~~~~~~~~~~${\cal A}$ }&\multicolumn{2}{c|} {8 TeV}& \multicolumn{2}{c|} {14 TeV} \\\cline{2-5}
                                   &$\wpri_L$ &$\wpri_R$ & $\wpri_L$ & $\wpri_R$   \\ \hline \hline
~~~~~~~~~~Reco.~without Cuts or Smearing  & $-0.42$  &$0.42$ & $-0.43$ & $0.43$ \\ \hline
+~Smearing~+~Fiducial~+~Kinematics (Eq.~(\ref{cuts1.EQ}))~~& $-0.46$  &$0.33$ & $-0.47$ &$0.34$\\ \hline
~~~~~~~~~~~~~~~~~+~Isolation~(Eq.~(\ref{cuts2.EQ})) & $-0.11$& $0.59$  &$0.083$&$0.72$ \\ \hline
~~~~~~~~~~+$\not\!\!E_{T}$~+~$m_{jj} $~Requirements~(Eq.~(\ref{cuts3.EQ})) & $-0.078$ & $0.62$ & $0.11$ & $0.75$ \\\hline
~~~~~~~~~~~~~~~~~+~Mass~Reco.~(Eq.~(\ref{cuts4.EQ})) & $0.16$ & $0.77$& $0.18$&  $0.77$  \\ \hline
~~~~~~~~~~~~~~~~~$-\Delta R_{jj}$            &$-0.34$  & $0.49$& $-0.34$ & $0.49$ \\\hline \hline 
\end{tabular} 
\label{WpNAFB.TAB}
\end{center}
\end{table}

\subsection{$\wpri$ Chiral Couplings to Initial-State Quarks}

Thus far, we have only presented the results to test the chiral coupling of $W'$ to the final state leptons. It is equally important to examine its couplings to the initial state quarks. 
Define an azimuthal angle 
\begin{equation}
\displaystyle \cos\Phi =
\frac{\hat{p}_N\times{\vec p}_{\ell_{2}}}{|\hat{p}_N\times{\vec p}_{\ell_{2}}|}
\cdot
\frac{\hat{p}_N\times{\vec p}_q}{|\hat{p}_N\times{\vec p}_q|}, 
\label{AziDef.EQ}
\end{equation}
as the angle between the $qq'\rightarrow N\ell^{+}_{1}$ production plane and $N\rightarrow W^-\ell^{+}_{2}$ decay plane in the neutrino rest-frame,  
where ${\vec p}_{\ell_{2}}$ is the three momentum of $\ell_{2}$, the charged lepton identified as originating from the neutrino;
$\hat{p}_N$ is the direction of motion of the neutrino in the partonic c.m.~frame; and ${\vec p}_q$ is the initial-state quark momentum.  
The definition of $\Phi$ is invariant under boosts along $\hat{p}_N$, 
hence the quark and charged lepton momenta can be evaluated either in the partonic c.m.~or the neutrino rest-frame.
The angular distribution between the two planes is thus calculated to be
\bea
\displaystyle\frac{d\hat\sigma}{d\Phi}=\frac{\sigma_{Tot.}}{2\pi}\left[
1+\frac{3\pi^2}{16} \ \frac{\mu_N}{2+\mu^2_N}
\left(\frac{\hat\sigma(W_0)-\hat\sigma(W_T)}{\hat\sigma(W_0)+\hat\sigma(W_T)}\right)
\left(\frac
{g_{R}^{q\:2}-g_{L}^{q\:2}}
{g_{R}^{q\:2}+g_{L}^{q\:2}}\right)
\cos\Phi\right].
\label{Azim.EQ}
\eea
The distribution for $\wpri_L$ is $180^\circ$ out of phase with the $\wpri_R$ distribution and the slope  
only depends on the $W'$ chiral coupling to the initial-state quarks.  
Hence, the phase of this distribution determines the chirality of the initial-state quarks couplings to the $\wpri$ $independently$ of the leptonic chiral couplings to the $\wpri$.

\begin{figure}[!t]
\begin{center}
\subfigure[]{
      \includegraphics[width=0.35\textwidth,clip=true]{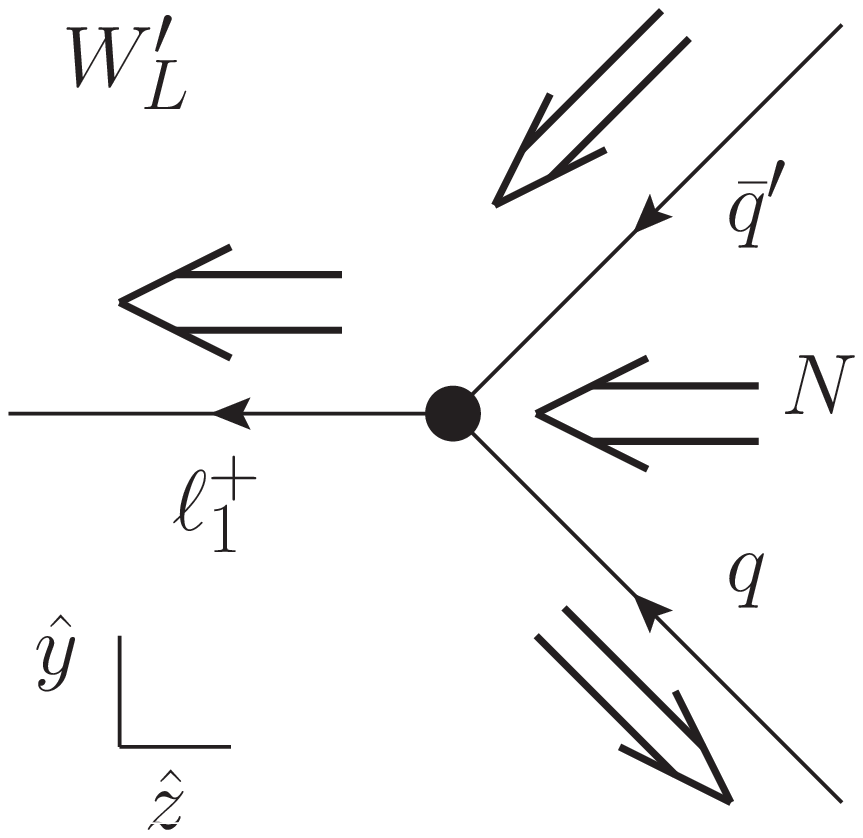}
\label{prodR.FIG}
}
\subfigure[]{
      \includegraphics[width=0.35\textwidth,clip=true]{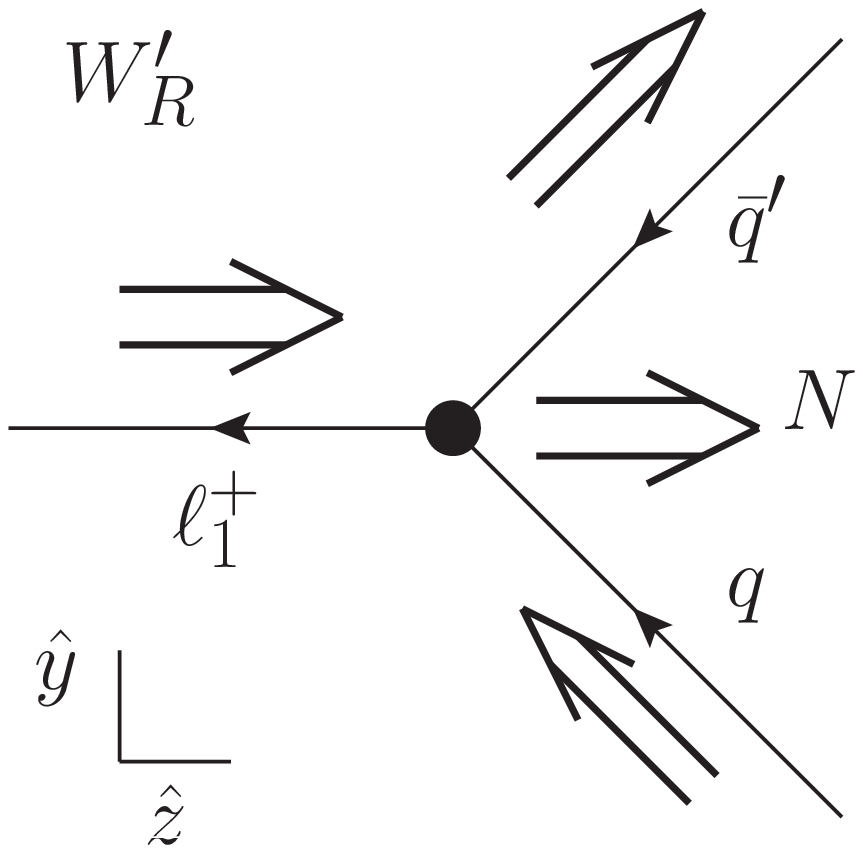}
\label{prodL.FIG}
}
\end{center}
\caption[Spin correlations for neutrino production in the neutrino rest-frame]{Spin correlations for neutrino production in the neutrino rest-frame.  Single arrowed lines represent momentum and double arrowed lines represent spin in the helicity basis.  The $\hat{z}$-axis is defined to be the neutrino's direction of motion in the partonic c.m.~frame and the $\hat{y}$-axis is defined such that $y$-component of the initial-state quark momentum is always positive.}
\label{Kin.FIG}
\end{figure}

To understand the distribution in Eq.~(\ref{Azim.EQ}), we consider the spin correlations between the initial and final states.  
As noted previously, the angle $\Phi$ is invariant under the boosts along $\hat{p}_{N}$.
So for simplicity, we consider the spin correlations in the heavy neutrino rest-frame.  
Figure \ref{Kin.FIG} shows the spin correlations of the neutrino production in the neutrino's rest-frame for both the (a) $\wpri_L$ and (b) $\wpri_R$ cases.  
Like before, single arrowed lines represent momentum directions and double arrowed lines spin in the helicity basis.   
Also, we define the production plane to be oriented in the $\hat{y}-\hat{z}$ plane 
such that the $\hat{y}$-component of the quark momentum always points along the positive $\hat y$-axis and that $\hat{z} = \hat{p}_{N}$. 
With this axis convention, $\Phi=-\phi_{\ell_{2}}$, where $\phi_{\ell_{2}}$ is the azimuthal angle of $\ell_{2}$ as measured from the positive $\hat{y}-$axis.

\begin{figure}[!t]
\centering
\subfigure[]{
      \includegraphics[width=0.48\textwidth,clip=true]{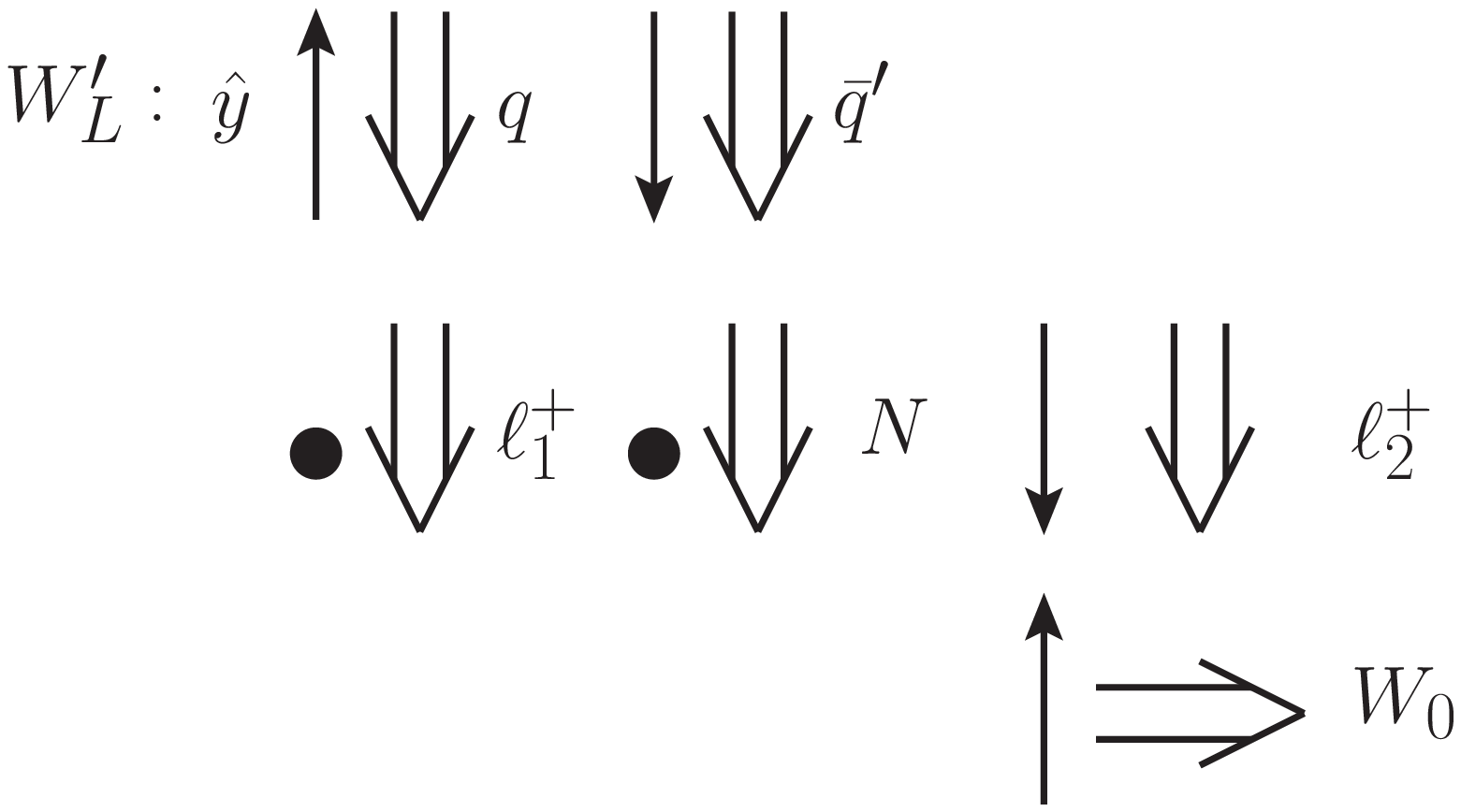}
}
\subfigure[]{
      \includegraphics[width=0.48\textwidth,clip=true]{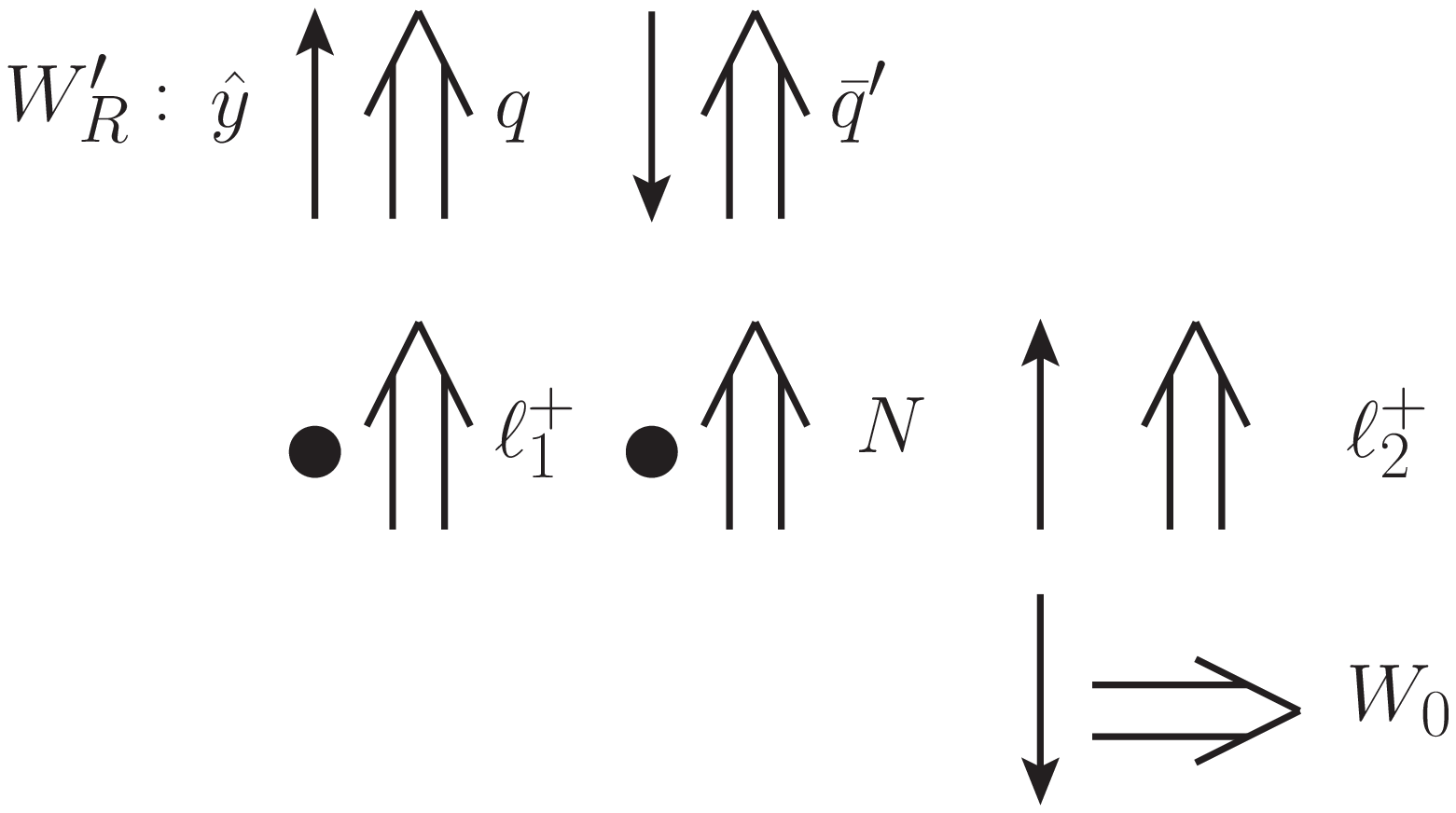}
}\\\vspace{0.3in}
\subfigure[]{
      \includegraphics[width=0.48\textwidth,clip=true]{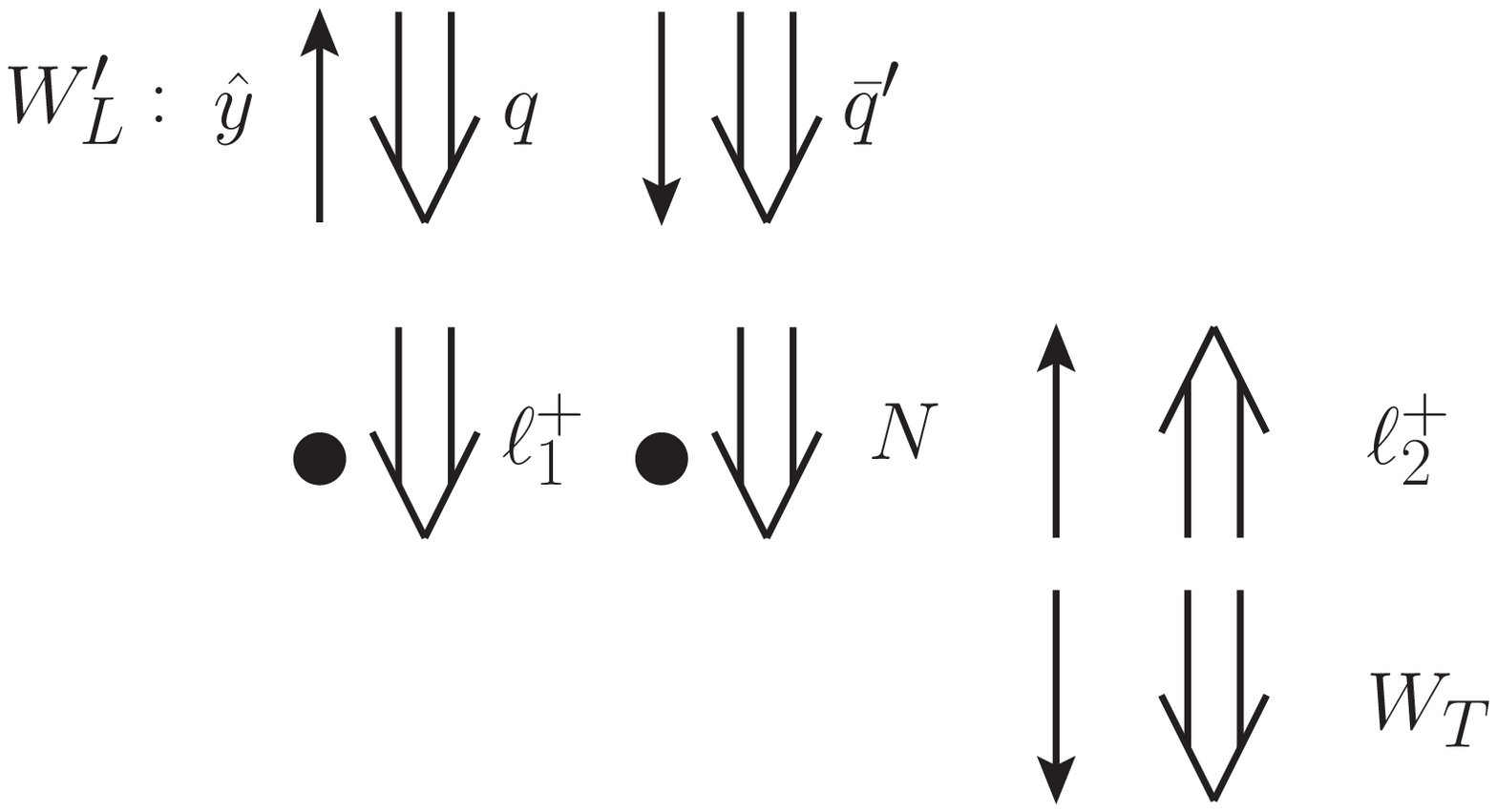}
}
\subfigure[]{
      \includegraphics[width=0.48\textwidth,clip=true]{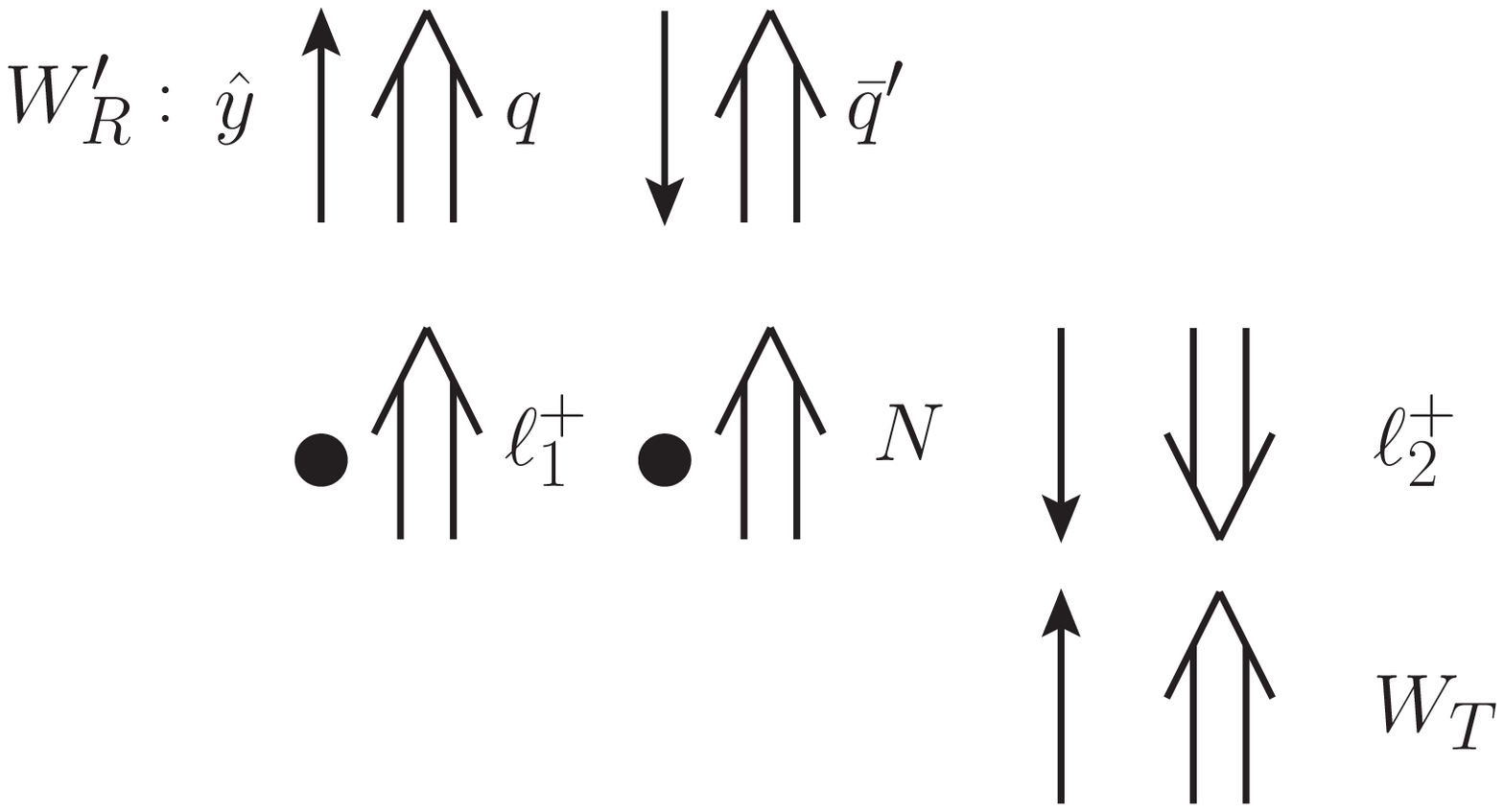}
}
\caption[Spin correlations in the neutrino rest-frame]{Spin correlations in the neutrino rest-frame as described in Fig.~\ref{Kin.FIG}.  Double arrowed lines represent spin with $\hat{y}$ being the quantization axis and single arrowed lines are the $\hat{y}$ component of the particles.}
\label{PhiSpin.FIG}
\end{figure}

Figure \ref{PhiSpin.FIG} shows the spin correlations for the heavy neutrino production and decay with the spin quantization axis chosen to be the $\hat{y}$ direction as defined above.   
The $\wpri_L$ case is shown in Figs.~\ref{PhiSpin.FIG}(a,c) and the $\wpri_R$ case in (b,d). 
The solid dots next to the $N$ and $\ell_1$ indicate that they have no momentum in the $\hat{y}$-direction.  
In the $\wpri_R$ case, the initial-state quark must be right-handed and the initial-state antiquark left-handed.  
Hence, the total spin of the initial-state points in the positive $\hat{y}$-direction, causing the spin of the neutrino to also point in the positive $\hat{y}-$direction.  
When the neutrino decays to a longitudinal or transverse $W$, the lepton from the neutrino decay has spin along or against the $\hat{y}$-axis, respectively.  
For the $\wpri_R$ case, figures \ref{PhiSpin.FIG}(b) and (d) show the decay into longitudinal and transverse $W$'s, respectively.  
Therefore, for the decay into $W_0$ ($W_T$)  case, the lepton prefers to move in the same (opposite) direction as the initial-state quark and $\Phi$ peaks at $0$ ($\pm\pi$).  
In the $\wpri_L$ case, the direction of motion of $\ell_2$ 
relative to the direction of motion of the initial-state quark is reversed and the peaks in the $\Phi$ distribution are shifted by $\pi$.  
This explains the $180^\circ$ phase difference in the angular distribution, Eq.~(\ref{Azim.EQ}), between the $\wpri_L$ and $\wpri_R$ cases, and between the neutrino decay to $W_0$ and $W_T$.  
Also, notice that this argument only relies on the $\wpri-q-q'$ coupling and {\it not} the $\wpri-N-\ell$ chiral couplings.  
Hence, measuring the distribution of the angle between the $qq'\rightarrow N\ell_{1}$ production and the $N\rightarrow {\ell_2}^+ W^-$ decay planes can determine the chiral couplings of a $\wpri$ to light quarks independently from the chiral couplings of the $\wpri$ to leptons.

\begin{figure}[!t]
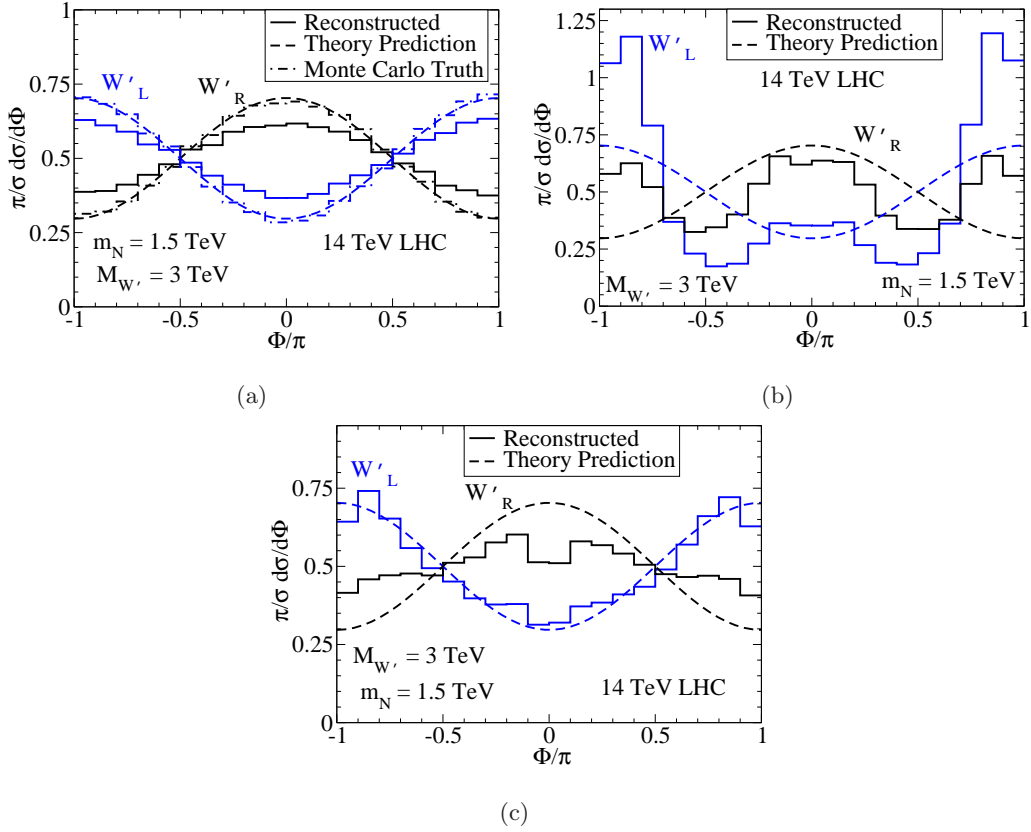

\centering
\subfigure[]{
       \includegraphics[width=0.4\textwidth,clip]{./05_WprimeHeavyN/PhiNCNS1500_3.eps}
\label{PhiNCNS.FIG}}
\subfigure[]{
       \includegraphics[width=0.4\textwidth,clip]{./05_WprimeHeavyN/PhiAll1500_3.eps}
\label{philrecl.FIG}}
\subfigure[]{
       \includegraphics[width=0.4\textwidth,clip]{./05_WprimeHeavyN/PhiAllNoJIso1500_3.eps}
\label{philrecNoJIso.FIG}}
\caption[$\Phi$ distributions at the 14 TeV LHC for fully reconstructed events]{$\Phi$ distributions at the 14 TeV LHC with $M_\wpri=3$~TeV and $m_N=1.5$~TeV for fully reconstructed events (solid), 
the analytical result in Eq.~(\ref{Azim.EQ}) (dashed), and Monte Carlo truth (dash-dot).  
Figure (a) is without energy smearing or cuts, (b) with energy smearing and cuts in Eqs.~(\ref{cuts1.EQ}), (\ref{cuts2.EQ}), (\ref{cuts3.EQ}), and (\ref{cuts4.EQ}),
 and (c) with the same cuts as (b) without the $\Delta R_{jj}$ cut in Eq.~(\ref{cuts2.EQ}).
}
\label{Phi.FIG}
\end{figure}

Most of the angular definition and analysis depend on the initial state quark momentum direction.  
Since the LHC is a symmetric $pp$ machine, this is not known {\it a priori}.  However, at the LHC $u$ and $d$ quarks are valence and antiquarks are sea.  
Hence, the initial-state quark generally has a larger momentum fraction than the initial-state antiquark; 
and the initial-state quark direction can be identified as the direction of motion of the fully reconstructed partonic c.m.~frame.  
Similar techniques have been used for studying forward-backward asymmetries associated with new heavy gauge bosons~\cite{Gopalakrishna:2010,Langacker:1984dc}.

Figure \ref{Phi.FIG} shows the $\Phi$ distributions at the $14$~TeV LHC with $M_\wpri=3$~TeV for both $\wpri_L$ and $\wpri_R$.  
From Eq.~(\ref{Azim.EQ}), the amplitude of the $\Phi$ distribution depends on the ratio $m_N/M_\wpri$, and therefore increase $m_N$ to $1.5$~TeV.  
The solid line is the $\Phi$ distribution with the initial state quark moving direction identified as the partonic c.m.~frame boost direction; 
the dashed lines is the theoretical distribution given in Eq.~(\ref{Azim.EQ}); and in (a) the dash-dot lines are the Monte Carlo truth,~i.e. 
using the known direction of the initial-state quark.  

Figure \ref{PhiNCNS.FIG} does not include cuts or smearing; as can be seen, the Monte Carlo truth and theoretical calculation agree very well.
The reconstructed distribution has a smaller amplitude than the theoretical distribution due to the direction of the initial-state quark being misidentified.  
Figure \ref{philrecl.FIG} shows the theoretical prediction and reconstructed distribution with smearing and the cuts in Eqs.~(\ref{cuts1.EQ},\ref{cuts2.EQ},\ref{cuts3.EQ},\ref{cuts4.EQ}) applied.  
For $\Phi=0$, the SM $W$ is maximally boosted and its decay products are maximally collimated.  
Consequently, the $\Delta R_{jj}$ cut in Eq.~(\ref{cuts2.EQ}) causes a large depletion of events in the central region.  
Figure \ref{philrecNoJIso.FIG} shows the reconstructed distribution with the same cuts as (b) minus the $\Delta R_{jj}$ cut.  
With the relaxation of this cut, the $\wpri_L$ and $\wpri_R$ cases become reasonably discernible with the $\wpri_L$ distribution nearly the same as the theoretical prediction.  
The continued depletion of events at $\Phi=0$ and $\Phi=\pm\pi$ are due to the rapidity cuts on leptons and jets, respectively.

\section{Unlike-Sign Dilepton Angular Distributions}
\label{Opp.SEC}

Intrinsically, Majorana neutrinos can decay to positively or negatively charged leptons, and therefore also contribute to the $L$-conserving process
\bea
pp\rightarrow \wpri \rightarrow \ell_1^+ \ell_2^- jj.
\eea
  These events can be reconstructed similarly to the method described in Section~\ref{Like.SEC}.  
  However, the SM backgrounds for this process, particularly $pp\rightarrow Z jj$, will be larger.  
  Our purpose here is not to do a full signal versus backgrounds study, but to comment on the differences between the like-sign and unlike-sign lepton cases. 
  Again, $u\bar{d}$ has a larger parton luminosity than $d\bar{u}$, so we focus only on $W'^+$ production:
\bea
pp\rightarrow {\wpri}^+\rightarrow N\ell^+_1 \rightarrow \ell_1^+ \ell_2^- jj
\eea

\subsection{$W'$ Chiral Coupling from Angular Distributions}
\label{Same.SEC}

\begin{figure}
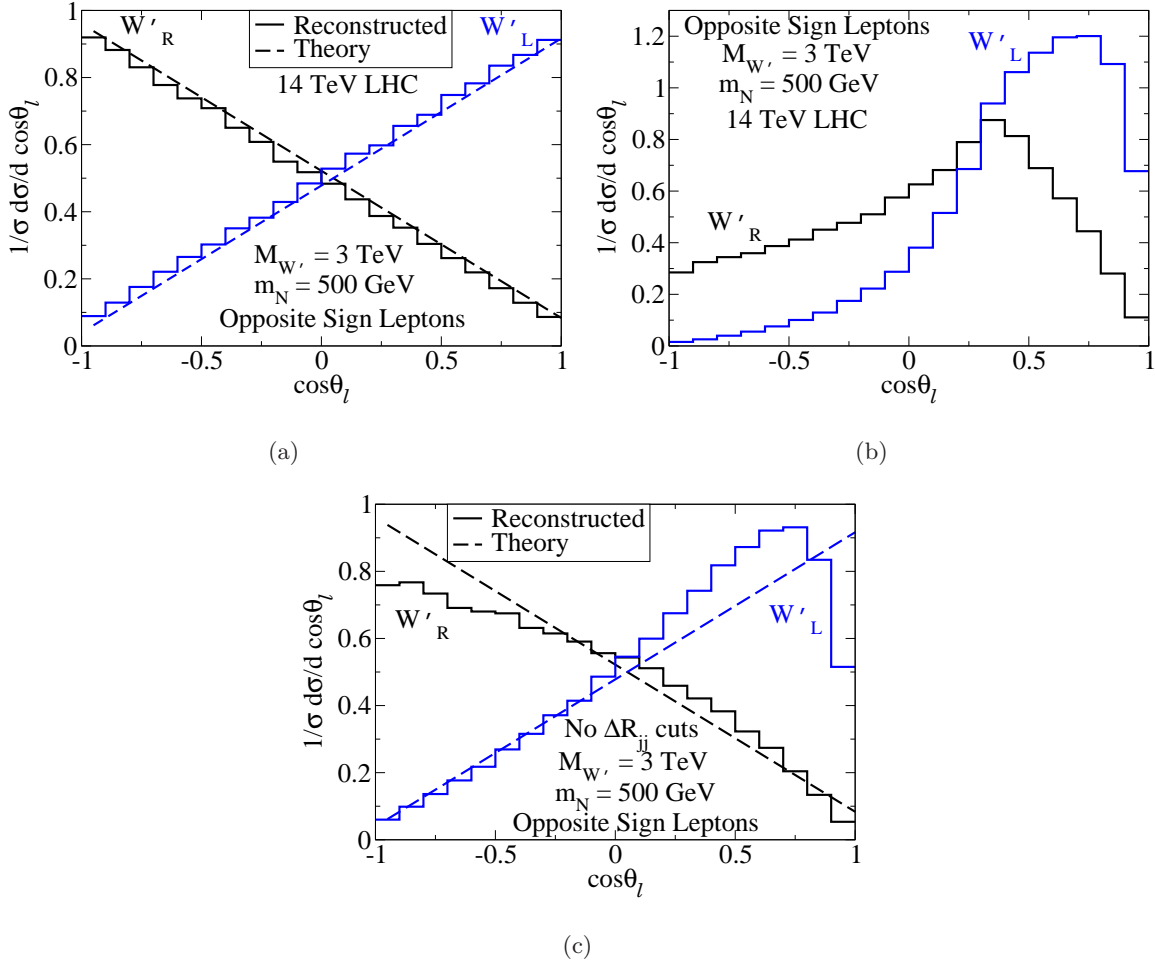

\begin{center}
\subfigure[]{
      \includegraphics[width=0.45\textwidth,clip]{./05_WprimeHeavyN/CthlNCNS500_3_opp.eps}
\label{wpriacNS_opp.FIG}}
\subfigure[]{
      \includegraphics[width=0.45\textwidth,clip]{./05_WprimeHeavyN/CthlAll500_3_opp.eps}
\label{wpriacSM_opp.FIG}}\\
\subfigure[]{
      \includegraphics[width=0.45\textwidth,clip]{./05_WprimeHeavyN/CthlAll500_3NoJIso_opp.eps}
\label{wpriacNoJIso_opp.FIG}}
\caption[Angular distributions of Oppose-Sign $\ell\ell jj$]{
For the opposite sign lepton case, the angular distribution of the charged lepton originating from neutrino decay  
in the heavy neutrino rest-frame with respect to the neutrino moving direction in the partonic c.m.~frame at the LHC with $M_{W'}$, $m_N$ set by Eq.~(\ref{benchParam.EQ}).
Distribution (a) without smearing or cuts, (b) with energy smearing and cuts 
in Eqs.~(\ref{cuts1.EQ}), (\ref{cuts2.EQ}), (\ref{cuts3.EQ}), and (\ref{cuts4.EQ})
, and (c) with all cuts applied to (b) except the $\Delta R_{jj}$ cuts in Eq.~(\ref{cuts2.EQ}). 
The solid lines are for the Monte Carlo simulation results and in (a) and (c) the dashed lines are for the analytical result in Eq.~(\ref{Ang.EQ}).}
\label{wpriac_opp.FIG}
\end{center}
\end{figure}

For the unlike-sign case, we mimic our entire like-sign analysis and 
reconstruct the polar angular distribution of the lepton originating from neutrino decay in the heavy neutrino rest-frame (App.~\ref{DiracAngDist.APP}).
Respectively, the polar and azimuthal distributions are similar to those in 
Eqs.~(\ref{Ang.EQ}) and (\ref{Azim.EQ}) up to a opposite sign in front of the angular dependence.
\begin{equation}
\frac{d\hat{\sigma}}{d\cos\theta_{\ell_{2}}}=
\frac{\hat{\sigma}_{Tot.}}{2}\left[1-\left(\frac{\hat{\sigma}(W_{0})-\hat{\sigma}(W_{T})}{\hat{\sigma}(W_{0})+\hat{\sigma}(W_{T})}\right)\left(\frac{2-\mu_{N}^{2}}{2+\mu_{N}^{2}}\right)
\left(\frac
{g_{R}^{\ell\:2}\vert Y_{\ell_{1}N}\vert^{2}-g_{L}^{\ell\:2}\vert V_{\ell_{1}N}\vert^{2}}
{g_{R}^{\ell\:2}\vert Y_{\ell_{1}N}\vert^{2}+g_{L}^{\ell\:2}\vert V_{\ell_{1}N}\vert^{2}}\right)\cos\theta_{\ell_{2}}\right],
\end{equation}
\begin{equation}
\frac{d\hat{\sigma}}{d\Phi}
=\frac{\hat{\sigma}_{Tot.}}{2\pi}\left[1-\frac{3\pi^{2}}{16}\frac{\mu_{N}}{2+\mu_{N}^{2}}\left(\frac{\hat{\sigma}(W_{0})-\hat{\sigma}(W_{T})}{\hat{\sigma}(W_{0})+\hat{\sigma}(W_{T})}\right)\left(\frac{g_{R}^{q\:2}-g_{L}^{q\:2}}{g_{R}^{q\:2}+g_{L}^{q\:2}}\right)\cos\Phi\right].
\label{AzimUnlikeSig.EQ}
\end{equation}
Figure~\ref{wpriac_opp.FIG} shows the $\Phi$ distributions for the unlike-sign process and follows the identical procedure as for the like-sign case. 
The solid line is the $\Phi$ distribution with the initial-state quark propagation direction identified as the partonic c.m.~frame boost direction; 
the dashed lines are the theoretical distributions given by Eq.~(\ref{AzimUnlikeSig.EQ}); 
and in (a) the dashed-dotted lines are the Monte Carlo truth, i.e., using the known direction of the initial-state quark. 
Figure~\ref{wpriacNS_opp.FIG} does not include cuts or smearing. 
Figure~\ref{wpriacSM_opp.FIG} shows the theoretical prediction and reconstructed distribution with smearing and cuts 
in Eqs.~(\ref{cuts1.EQ}),~(\ref{cuts2.EQ}),~(\ref{cuts3.EQ}), and~(\ref{cuts4.EQ}) applied.
Figure~\ref{wpriacNoJIso_opp.FIG} shows the reconstructed distribution with the same cuts as~\ref{wpriacSM_opp.FIG} minus the $\Delta R_{jj}$ isolation cut.

To understand why the sign of the slope for the $L$-conserving distributions differ from the $L$-violating distributions, we turn to spin correlations.
For $W'^+$, the spin correlations for $u\bar{d}\rightarrow W'^+\rightarrow N\ell^+$ 
are shown in Fig.~\ref{Nspin.FIG} without yet specifying $N$'s decay.
However, we only need to analyze the angular correlation in the neutrino decay.  
 The spin correlations are simply obtained by replacing the right-handed antilepton in Fig.~\ref{Lspin.FIG} with a left-handed lepton.  
 Since the direction of the spin of the lepton is completely determined by the neutrino spin, which is unchanged between the two cases, 
 the effect of the helicity flip is to reverse the direction of the final state lepton momentum relative to the $\hat z$ direction.  
 Therefore, the slopes of the lepton angular distribution are opposite for the like-sign and unlike-sign lepton cases. 
 These same arguments can be made to show that the phases of the 
 $\Phi$ distribution in Eqs.~(\ref{Azim.EQ}) and~(\ref{AzimUnlikeSig.EQ}) differ by $180^\circ$.   
  
The analysis of the two cases also reveals that, unlike the angular distributions, 
the total cross section is independent of having like-sign or unlike-sign leptons in the final state.
This may be understood by recognizing that the difference between the two final states is tantamount to a charge conjugation.
Having integrated out the angular dependence, the total cross section is invariant under parity inversion.
Consequently, by CP-invariance, the total rate is invariant under charge conjugation.
This behavior is evident in Eq.~(\ref{NWidth.EQ}) and Fig.~\ref{NW.FIG}, 
which show that $N$ decays to $\ell^{+}W^{-}$ and $\ell^{-}W^{+}$ equally.

\section{Summary}
\label{Conc.SEC}

The nature of the neutrino mass remains one of most profound puzzles in particle physics. 
The possibility of its being Majorana-like is an extremely interesting aspect since it may have far-reaching consequences in particle physics, nuclear physics and cosmology. 

Given the outstanding performance of the LHC, we are motivated to study the observability for a heavy Majorana neutrino $N$ along with a new charged gauge boson $W'$ at the LHC. 
We first parameterized their couplings in a model-independent approach in Section~\ref{ThFrame.SEC} and presented the current constraints on the mass and coupling parameters. 

We studied the production and decay of $W'$ and $N$ at the LHC, and optimized the observability of the like-sign dilepton signal over the SM backgrounds. 
We emphasized the complementarity of these two particles by exploiting the characteristic kinematical distributions resulting from spin-correlations to unambiguously determine their properties. Our phenomenological results can be summarized as follows.

\begin{itemize}
\item[1.] The heavy neutrino is likely to have a large R.H.~component and thus the $W'_R$ would likely yield a larger signal rate 
than that for $W'_L$, governed by the mixing parameters as discussed in Section~\ref{ThFrame.SEC}.
Under these assumptions, we found that at the 14 TeV LHC a 5$\sigma$ signal,  via the clean channels
$\ell^\pm \ell^\pm  jj$, 
may be reached for $M_{W'_R}=3$ TeV  ($4$ TeV) with 90 fb$^{-1}$ (1 ab$^{-1}$) integrated luminosity, as seen in Fig.~\ref{significanceVsLumi.FIG}.
\item[2.] 
The chiral coupling of $W'$ to the leptons can be inferred by the polar angle distribution of the leptons in the reconstructed neutrino frame, as seen in Fig.~\ref{wpriac.FIG}, 
owing to the spin correlation from the intermediate state $N$. 
\item[3.]
The chiral coupling of $W'$ to the initial state quarks can be inferred by the azimuthal angular distribution of the neutrino production and decay planes, 
as seen in Fig.~\ref{Phi.FIG}.
\item[4.]  
The kinematical distributions for the like-sign and unlike-sign cases have been found to be quite sensitive to spin correlations and are complementary. 
In particular, the angular distributions differ by a minus sign and provide qualitative differences for a Majorana and a Dirac $N$.
Thus in addition to observing final states that violate lepton-number, comparison of the two scenarios provides a means to differentiate the Majorana nature of $N$. 
\end{itemize}

Overall, if the LHC serves as a discovery machine for a new gauge boson $W'$, then  its properties and much rich physics will await to be explored. 
Perhaps a Majorana nature of a heavy neutrino may be first established associated with $W'$ physics.